\documentclass[twoside,12pt]{article}

\usepackage{epsfig}
\usepackage[sort&compress,numbers]{natbib}

\usepackage{graphicx}
\usepackage{latexsym}
\usepackage{amsfonts}
\usepackage{amssymb}
\usepackage{amsmath}
\usepackage{bm}
\usepackage{mathrsfs}
\usepackage{slashed}
\usepackage{ascmac}
\usepackage{comment}
\usepackage{dcolumn}

\usepackage[utf8]{inputenc} 

\allowdisplaybreaks
\graphicspath{{../figs/}}
\usepackage[
bookmarks=true,colorlinks,linkcolor=blue,urlcolor=cyan,citecolor=red]{hyperref}

\usepackage{color} 
\usepackage[normalem]{ulem}  
\newcommand{\com}[1]{{\color[rgb]{0,0,1}{[#1]}}}
\newcommand{\cgd}[1]{{\color[rgb]{1,0,0}{#1}}}
\newcommand{\gray}[1]{{\color[rgb]{0.7,0.7,0.7}{#1}}}

\newcommand{\pp}[1]{{\color[rgb]{0.7,0.3,0.9}{#1}}}

\newcommand{\mass}{m_0}

\renewcommand{\log}{\ln} 

\newcommand{\Ritus}{  {\mathcal R}  }
\newcommand{\qn}{ \chi } 

\newcommand{\pd}{\partial}
\renewcommand{\=}{&=&}
\newcommand{\nnb}{\nonumber \\}
\newcommand{\nn}{\nonumber}
\newcommand{\MSbar}{ \overline{ {\rm MS} }}

\newcommand{\np}{ {n^\prime} }

\newcommand{\cout}[1]{\if 0 {#1} \fi}

\newcommand\eref[1]{Eq.~(\ref{#1})}
\newcommand\fref[1]{Fig.~\ref{#1}}
\newcommand{\sla}{ \slashed }

\newcommand{\LR}[1]{\left\langle {#1} \right\rangle}
\newcommand{\lan}{\langle} 
\newcommand{\ran}{\rangle}

\renewcommand{\a}{\alpha}
\renewcommand{\b}{\beta}

\newcommand{\Lam}{ \Lambda }
\newcommand{\al}{\alpha}
\newcommand{\be}{\beta}
\newcommand{\m}{\mu}

\newcommand{\ep}{ \epsilon }
\newcommand{\vep}{ \varepsilon }
\newcommand{\vphi}{ \varphi }
\newcommand{\Gam}{ \Gamma }
\newcommand{\gam}{ \gamma }
\newcommand{\bgam}{ {\bm \gamma} }

\newcommand{\Lm}{ \Lambda }

\newcommand{\bx}{{\bm{x}}}

\newcommand{\bk}{{\bm{k}}}
\newcommand{\bp}{{\bm{p}}}
\newcommand{\bq}{{\bm{q}}}
\newcommand{\bv}{{\bm{v}}}
\newcommand{\bA}{{\bm{A}}}
\newcommand{\bW}{{\bm{W}}}

\newcommand{\bl}{{\bm{\ell}}}
\newcommand{\bg}{{\bm{\gamma}}}
\newcommand{\bpi}{{\bm{\pi}}}

\newcommand{\bB}{{\bm B}}
\newcommand{\bE}{{\bm E}}
\newcommand{\bj}{{\bm j}}
\renewcommand{\bW}{\bm{W}}
\newcommand{\bP}{\bm{P}}
\newcommand{\bT}{\bm{T}}
\newcommand{\bK}{\bm{K}}

\newcommand{\MeV}{ {\rm MeV} }
\newcommand{\GeV}{ {\rm GeV} }
\newcommand{\fm}{ {\rm fm} }

\newcommand{\order}{ {{\mathcal O} }}
\newcommand{\calA}{ {\mathcal A} }
\newcommand{\calE}{ {\mathcal E } }
\newcommand{\nonA}{ {\mathcal A} }
\newcommand{\nonF}{ {\mathcal{F}} }
\newcommand{\nonB}{ {\mathcal{B}} }
\newcommand{\nonE}{ {\mathcal{E}} }
\newcommand{\F}{ {\mathscr{F}} }
\newcommand{\G}{ {\mathscr{G}} }

\renewcommand{\Finv}{ {\mathscr{F}} }
\newcommand{\Ginv}{ {\mathscr{G}} }
\newcommand{\prj}{ {\mathcal P} }
\newcommand{\Q}{ {\mathcal Q} }

\newcommand{\Ham}{ {\mathcal H} }
\newcommand{\Lag}{ {\cal{L}} }
\newcommand{\M}{ {\mathcal M} }

\newcommand{\T}{ {\mathcal T} }

\newcommand{\tot}{ {\rm tot} }
\newcommand{\ext}{ {\rm ext} }
\newcommand{\vac}{ {\rm vac} }
\newcommand{\EM}{{\rm em}}

\newcommand{\tr}{ {\rm Tr} }
\newcommand{\diag}{ {\rm diag} }
\newcommand{\sgn}{ {\rm sgn} }
\newcommand{\Hall}{ {\rm Hall} }
\newcommand{\eff}{ {\rm eff} }

\newcommand{\quark}{ {\rm quark} }
\newcommand{\gluon}{ {\rm gluon} }
\newcommand{\ghost}{ {\rm ghost} }

\newcommand{\UV}{ {\rm UV} }
\newcommand{\IR}{ {\rm IR} }

\newcommand{\q}{ {\rm q} }
\newcommand{\YM}{ {\rm YM} }
\newcommand{\dyn}{ {\rm dyn} }
\newcommand{\LLL}{ {\rm LLL} }
\newcommand{\hLL}{ {\rm hLL} }
\newcommand{\QCD}{ {\rm QCD} }
\newcommand{\QED}{ {\rm QED} }

\newcommand{\CME}{{\rm CME}}
\newcommand{\Ohm}{{\rm Ohm}}

\newcommand{\integ}{\int\!\!}

\newcommand{\id}{\mbox{1}\hspace{-0.25em}\mbox{l}}
\newcommand{\1}{\mbox{1}\hspace{-0.25em}\mbox{l}}

\newcommand{\rp}{{r_\parallel^2}}
\newcommand{\rt}{{r_\perp^2}}

\newcommand{\para}{ \parallel}

\newcommand{\zero}{ {(0)} }
\newcommand{\one}{ {(1)} }

\newcommand{\Br}{B_{\rm r}}

\newcommand{\mq}{m}

\newcommand{\xc}{ { x_c}  } 
\newcommand{\yc}{ {y_c}  }

\newcommand{\FS}{{\rm FS}}

\newcommand{\J}{ {\mathcal J} }
\newcommand{\D}{ {\mathcal D} }

\newcommand{\ab}{ {(a)} }

\newcommand{\ar}{{ar}}
\newcommand{\ra}{{ra}}
\newcommand{\rr}{{rr}}

\newcommand{\temp}{ {\rm med} }

\newcommand{\idx}{ { {}_{\ell n} } }

\newcommand{\eq}{ {\rm eq} }

\newcommand{\del}{\partial}

\newcommand{\e}{{\rm e}}
\newcommand{\aaa}{\mathfrak{a}}
\newcommand{\bbb}{\mathfrak{b}}
\newcommand{\eee}{\mathfrak{e}}
\newcommand{\A}{\mathcal{A}}

\newcommand{\beq}{\begin{eqnarray}}
\newcommand{\eeq}{\end{eqnarray}}
\newcommand{\bseq}{\begin{subequations}}
\newcommand{\eseq}{\end{subequations}}

\def\simge{\mathrel{%
   \rlap{\raise 0.511ex \hbox{$>$}}{\lower 0.511ex \hbox{$\sim$}}}}
\def\simle{\mathrel{
   \rlap{\raise 0.511ex \hbox{$<$}}{\lower 0.511ex \hbox{$\sim$}}}}
\def\bigs{\mathrel{
   \rlap{\raise 0.531ex \hbox{$>$}}{\lower 0.531ex \hbox{$<$}}}}

\begin{document}

\title{Strong-Field Physics in QED and QCD: 
\\
From Fundamentals to Applications}


\author{
Koichi Hattori,$^{1,2 }$\footnote{koichi.hattori@outlook.com} \ 
Kazunori Itakura,$^{3, 4}$\footnote{kazunori.itakura@kek.jp} \ 
and 
Sho Ozaki$^{5}$\footnote{sho.ozaki@hirosaki-u.ac.jp}
\\
\\
$^1$Zhejiang Institute of Modern Physics, Department of Physics, 
\\
Zhejiang University, Hangzhou, Zhejiang 310027, China
\\
$^2$Research Center for Nuclear Physics, Osaka University, 
\\
10-1 Mihogaoka, Ibaraki, Osaka 567-0047, Japan
\\
$ ^3 $Nagasaki Institute of Applied Science, Nagasaki 851-0193, Japan
\\
$^4$KEK Theory Center, Institute of Particle and Nuclear Studies,\\
High Energy Accelerator Research Organization, \\1-1, Oho, Ibaraki, 305-0801, Japan
\\
$ ^5$ Graduate School of Science and Technology, Hirosaki University, \\3 Bunkyo, Hirosaki, Aomori, 036-8561, Japan
}

\maketitle 

\begin{abstract} 
We provide a pedagogical review article on fundamentals and applications 
of the quantum dynamics in strong electromagnetic fields in QED and QCD. 
The fundamentals include the basic picture of the Landau quantization and the resummation techniques applied to the class of higher-order diagrams 
that are enhanced by large magnitudes of the external fields. 
We then discuss observable effects of the vacuum fluctuations 
in the presence of the strong fields, which consist of 
the interdisciplinary research field of nonlinear QED. 
We also discuss extensions of the Heisenberg-Euler 
effective theory to finite temperature/density 
and to non-Abelian theories with some applications. 
Next, we proceed to the paradigm of 
the dimensional reduction emerging 
in the low-energy dynamics in the strong magnetic fields. 
The mechanisms of superconductivity, the magnetic catalysis of the chiral symmetry breaking, 
and the Kondo effect are addressed from a unified point of view in terms of the renormalization-group method. 
We provide an up-to-date summary of the lattice QCD simulations in magnetic fields for the chiral symmetry breaking 
and the related topics as of the end of 2022. 
Finally, we discuss novel transport phenomena 
induced by chiral anomaly and the axial-charge dynamics. 
Those discussions are supported by a number of appendices. 
\end{abstract}

\clearpage
\tableofcontents

\section{Introduction}

\label{sec:intro} Since the seminal paper by Heisenberg and Euler \cite{Heisenberg:1935qt}, 
the strong-field physics, quantum dynamics in external strong electromagnetic fields, 
has been studied in various research fields which include high-energy physics, condensed matter physics, high-intensity laser physics, and astrophysics. 
It is still growing as one of the most exciting interdisciplinary research fields in modern physics. 


Recent progress in laboratory experiments and astronomical observations has been driving intensive and extensive theoretical studies on the strong-field physics. 
To name a few, 
as proposed in classic papers \cite{Fermi, Weizsacker, Williams, Breit:1934zz}, highly accelerated nuclei can be used as a source of strong electromagnetic fields. 
Ultrarelativistic heavy-ion collisions with 
Relativistic Heavy Ion Collider (RHIC) 
and the Large Hadron Collider (LHC) create 
the ever strongest electromagnetic fields \cite{Skokov:2009qp, Voronyuk:2011jd, Bzdak:2011yy, Deng:2012pc, Deng:2014uja} 
(see also Refs.~\cite{Huang:2015oca, Hattori:2016emy} for 
reviews on the estimates of the strengths). 
We have basically two types of opportunities. 
When the accelerated nuclei smash each other, 
the quark-gluon plasma is created under the strong 
electromagnetic fields, enabling us to study 
the interplay between QCD and QED \cite{Kharzeev:2013jha, Kharzeev:2013ffa, Liao:2014ava, Kharzeev:2015znc, 
Huang:2015oca, Hattori:2016emy}. 
In ultraperipheral collision events where 
the nuclei pass by each other, 
electromagnetic processes in vacuum play dominant roles 
without the matter effects 
(see recent experimental progress 
at the LHC \cite{ATLAS:2017fur, CMS:2018erd, ATLAS:2019azn, ATLAS:2020hii} and RHIC \cite{STAR:2019wlg}). 
Also, observations of compact stars, especially magnetars, 
have been providing us with an opportunity to investigate 
effects of strong magnetic fields \cite{Harding:2006qn, Enoto:2019vcg} that are stable in time 
and have an extension over a macroscopic spatial scale. 
The existence of a strong magnetic field 
in the early universe has been also suggested by observations and there are hot discussions about 
its generation mechanism \cite{Grasso:2000wj, Giovannini:2003yn, Kandus:2010nw, Durrer:2013pga, Subramanian:2015lua, Kamada:2022nyt}.  
The world record of the laser intensity 
has been increased, aiming at probing the nonperturbative  modifications of the vacuum properties such as the Schwinger pair creation and the vacuum birefringence~\cite{Mourou:2006zz, DiPiazza:2011tq, Ejlli:2020yhk, Zhang:2020lxl, Fedotov:2022ely}.

%
%
%
%
%
%
%
%



This review paper is composed of two parts to provide 
a consistent picture of the strong-field physics in fundamentals and applications. 
There are great review papers and textbooks that are useful 
even beyond the scope of each research field 
such as a basic textbook \cite{sokolov1986radiation}, 
those in astrophysics \cite{1977FCPh, meszaros1992high}, 
and in condensed matter physics \cite{yoshioka2013quantum, Tong:2016kpv}. 
We will try to provide a complementary review 
that starts with fundamentals but includes 
more recent developments. 
Below, we give some more detailed survey of this paper 
and then remarks on general properties of constant electromagnetic fields and conventions.


\subsubsection*{Fundamentals}

In the first part from Sec.~\ref{sec:Q} to Sec.~\ref{sec:resum}, 
we will provide fundamental materials that we think important to understand the basic picture of the dynamics in external fields and to be prepared for applications discussed in the later sections. 

In Sec.~\ref{sec:Q}, we first address the most fundamental fact that the spectrum of a charged particle is  
quantized in magnetic fields as known as the Landau quantization. 
To provide the simplest demonstration, 
we start with nonrelativistic quantum mechanics. 
We put an emphasis on the gauge-invariant formulation 
and the physical meaning of the quantum numbers 
specifying the energy eigenstates which include not 
only the principal quantum numbers 
but also the other auxiliary quantum numbers. 
We then discuss a relativistic extension for a Dirac fermion. 


Bearing this in mind, we proceed to the resummation technique 
by means of the proper-time method in Sec.~\ref{sec:resum}. 
After suggestive papers by Nambu~\cite{Nambu:1950rs},  Feynman~\cite{Feynman:1950ir}, 
and Schwinger's seminal paper \cite{Schwinger:1951nm}, 
this method is widely used in the strong-field physics and 
allows us to work on the resummation of the higher-order diagrams 
that are enhanced by large magnitudes of external fields. 
Namely, 
they are equally important as the leading-order diagram in the naive perturbation theory.

We first explain the resummation in QED where the charged fermions 
are interacting with external Abelian electromagnetic fields. 
Treatment of photons does not require the same resummation, since 
they do not have self-interactions with the external fields. 
We apply the same technique to scalar QED. 
Then, we explain an extension of the resummation technique to QCD. In contrast to QED, the external chromo-electromagnetic fields interact with all the constituents, i.e., quarks, gluons, and ghosts. 
In general, this resummation is a quite tough task because of the non-Abelian nature of QCD. 
We therefore confine ourselves to the simplest extension, that is, 
the so-called covariantly constant fields. 
In this specific configuration of an external chromo-electromagnetic field, we only need to investigate the diagonal sector of the  SU($N_c$) color symmetry.


\subsubsection*{Applications to classic but unsolved problems}

We review classic problems which have been addressed since the very early times 
of Heisenberg and Euler~\cite{Heisenberg:1935qt}. 
We will, however, encounter unestablished aspects of the nonlinear QED. 

As mentioned above, photons do not directly interact with the external fields. 
Nevertheless, we first observe that photons can interact with the external fields through 
the virtual fermion pairs in vacuum, 
and, importantly, those quantum processes are not suppressed 
in sufficiently strong external fields as compared to the process in the absence of the external fields. 
This fact poses a fundamental question about the vacuum properties in the strong fields. 
This is the subject discussed in Sections~\ref{sec:HE} and \ref{sec:photons}. 

We are able to construct the effective theory 
by integrating out the fermion loop. 
This is called the Heisenberg-Euler effective theory 
and is understood as the low-energy effective theory 
of photons in the zero frequency and momentum limits. 
We derive the effective Lagrangian as an application 
of the resummation technique established in the preceding sections, 
and discuss the consequences of the nonlinear interactions 
with the external fields, such as the Schwinger mechanism.

We next discuss one of the drastic changes of the photon properties in external fields. 
We will see that photons acquire nontrivial refractive indices in external fields, 
so that the photon propagation in external fields is far nontrivial 
and reflects the modifications of the vacuum as if propagating in dielectric substances. 
When the photon energy is small, one may use the low-energy effective theory in computing refractive indices, 
which offers another application of the Heisenberg-Euler Lagrangian. 
However, when the photon carries a hard momentum, we need to explicitly include the photon momentum 
and encounter computation of the two-point function, 
that is, the vacuum polarization tensor. 
We review a recent progress on the general analysis of 
the photon's refractive index. 
Also, we briefly discuss the photon splitting in magnetic fields 
arising from the three-point functions. 

Those phenomena are expected to be observed 
with the developments of laser fields, 
astrophysical observations, 
and relativistic heavy-ion collisions.


\subsubsection*{Developments of further interdisciplinary fields}

From Sec.~\ref{sec:HE_QCD}, we move on to the topics that are newly proposed and have been intensively studied 
in the last two decades.

In Sec.~\ref{sec:HE_QCD}, we begin with discussing 
extensions of the Heisenberg-Euler effective theory 
to non-Abelian theories as well as to finite temperature/density. 
By using the results from the covariantly constant fields, 
we are able to obtain the low-energy effective theory of QCD 
in the coexisting Abelian and non-Abelian fields. 
In the last decade, there appeared a number of suggestive 
results from lattice QCD simulations 
and ultrarelativistic heavy-ion collisions 
that have been providing driving force for the study of QCD 
in the presence of the external fields.

In Sec.~\ref{sec:dmr}, we focus on the low-energy dynamics 
in strong magnetic fields. 
The central topics of this section is the quantum many-body effect induced by the dimensional reduction in the LLL 
and its analogy to the low-energy dynamics 
in the vicinity of the large Fermi surface 
where a similar dimensional reduction occurs. 
We discuss the chiral symmetry breaking in strong magnetic fields 
induced by the pairing between fermions and antifermions 
in parallel to the Cooper pairing in superconductor. 
Also, we discuss another pairing phenomenon 
occurring between light fermions 
and a heavy and dilute impurity embedded in 
a bulk of conducting light fermions. 
Such a situation is known occurs in alloys 
and was recently suggested to occur in quark matter 
with heavy-quark impurities, 
serving as platforms of the Kondo effect.

We provide a unified point of view of all 
these systems in terms of the renormalizaton-group (RG) method. 
It is noteworthy that the Kondo effect was one of main  motivations for Wilson to develop his concept of the renormalization group and effective field theories \cite{Wilson:1974mb}. 
The first basic observation in this section is the fact that the four-Fermi interaction becomes a marginal operator irrespective of the magnitude of the coupling constant thanks to the dimensional reduction. 
With this suggestive result, we construct the RG equations for weak-coupling theories at zero temperature. 
This allows us to understand the very basic concept of the dimensional reduction and how it manifests itself in all of the aforementioned phenomena. 
We then proceed to the low-energy QCD in strong magnetic fields 
with a body of summary for recent lattice QCD simulations 
and discuss interpretations of the interesting results 
that are induced by the interplay between 
nonperturbative QCD and strong magnetic fields.

In Sec.~\ref{sec:transport}, 
we discus the novel transport phenomena induced 
by quantum anomaly that opened a new interdisciplinary avenue 
toward applications to ultrarelativistic heavy-ion collisions, 
the Weyl/Dirac semimetals in condensed matter physics, 
and astrophysics/cosmology. 
We discuss a diagrammatic derivation of 
the anomalous currents and an exponential amplification 
of a helical magnetic field as a consequence of 
chiral anomaly.

\subsubsection*{Concluding remarks and appendices}

Finally, we devote Sec.~\ref{sec:summary} to our summary 
and brief remarks on recent developments 
which are not covered in earlier sections. 
In appendices, we provide details of the basic formulation 
which we think are useful as the machinery 
for the future studies.

\subsubsection*{Preparations: Generalities of constant electromagnetic fields and conventions} 

In this review paper, we mostly focus on the dynamics in external constant electromagnetic fields. 
The field strength tensor and its dual are, respectively, given by 
\begin{subequations}
\begin{eqnarray}
F^{\mu\nu} &=& \partial^\mu A^\nu - \partial^\nu A^\mu
\, ,
\\
\tilde F^{\mu\nu} &=& \frac{1}{2} \epsilon^{\mu\nu\alpha\beta} F_{\alpha\beta}
\, .
\end{eqnarray}
\end{subequations}
We use the mostly minus convention 
for the Minkowski metric $g^{\mu\nu} = {\rm diag}(1,-1,-1,-1)$ 
and the completely antisymmetric tensor with $ \ep^{0123} =+ 1 $. 
An electric and magnetic field can be expressed as 
$ E^i = - \nabla^i A^0 - \dot A^i = F^{i0} $ and 
$ B^i = \sum_{j,k}\epsilon^{ijk}  \nabla^j A^k =  -  \sum_{j,k} \epsilon^{ijk }F^{jk} /2 $ 
with $ \nabla^i = - \partial^i $ and the antisymmetric tensor $ \epsilon^{123} = +1 $, respectively. 
%
%
Their components are explicitly given as 
\begin{subequations}
\label{eq:field-strength}
\begin{eqnarray}
&&
F ^{\mu\nu} = 
\begin{pmatrix}
0& - E_x  & - E_y  & - E_z
\\
E_x& 0& - B_z &  B_y
\\
E_y&   B_z &0  & - B_x
\\
E_z& - B_y & B_x  & 0
\end{pmatrix}
\, , \quad
F _{\mu\nu} = 
\begin{pmatrix}
0&  E_x  &  E_y  &  E_z
\\
-E_x& 0& - B_z &  B_y
\\
-E_y&   B_z &0  & - B_x
\\
-E_z& - B_y & B_x  & 0
\end{pmatrix}
\, ,
\\
&&
\tilde F ^{\mu\nu} =
\begin{pmatrix}
0& - B_x  & - B_y  & -B_z
\\
B_x&0 &  E_z &  - E_y
\\
B_y&   - E_z & 0 &  E_x
\\
B_z&  E_y & - E_x  & 0
\end{pmatrix}
\, , \quad
\tilde F _{\mu\nu} = 
\begin{pmatrix}
0&  B_x  &  B_y  & B_z
\\
-B_x&0 &  E_z &  - E_y
\\
-B_y&   - E_z &0  &  E_x
\\
-B_z&  E_y & - E_x  & 0
\end{pmatrix}
\, ,
\end{eqnarray}
\end{subequations}
%
%
Since the field strength tensors have antisymmetric Lorentz indices, 
one can only construct two Lorentz invariants of mass-dimension four: 
\begin{subequations}
\label{eq:fg}
\begin{eqnarray}
\Finv &\equiv& \frac{1}{4} F_{\mu\nu} F^{\mu\nu} = \frac{1}{2} ( \bB^2 - \bE^2 ) \label{eq:f} 
\, ,\\
\Ginv &\equiv& \frac{1}{4} F_{\mu\nu} \tilde F^{\mu\nu} = - \bB \cdot \bE \label{eq:g}
\, ,
\end{eqnarray}
\end{subequations}
where 
the three-dimensional vectors $\bE$ and $\bB$ are 
electric and magnetic fields in an arbitrary Lorentz frame. 
Regarding $F_\mu^{\ \, \nu}$ as a matrix, 
one finds that the equation $F^{\mu}_{\ \nu}\phi^\nu=\lambda \phi^\mu$ has four independent eigenvalues,\footnote{
This is because the matrix $F_\mu^{\ \, \nu}$ is not antisymmetric 
in the sense that $F_\mu^{\ \, \nu} \not= - F_\nu^{\ \, \mu}$.
}
of which the pairwise forms $\pm a $ and $ \pm i b$ are written in terms of the invariants (\ref{eq:f}) and (\ref{eq:g}) as 
\begin{subequations}
\label{eq:ab}
\begin{eqnarray}
a &=& \sqrt{ \sqrt{ \F^2 + \G^2 } - \F } \label{eq:a} \, , \\
b &=& \sqrt{ \sqrt{ \F^2 + \G^2 } + \F } \label{eq:b} \, .
\end{eqnarray}
\end{subequations}
The field configurations are characterized by these Lorentz invariants.  

As long as the electromagnetic field has a finite invariant $ \Ginv $, 
there exists a Lorentz frame such that the invariant is given by $ \Ginv = \mp |\bB_0|  |\bE_0| $. 
This means that the electric and magnetic fields are parallel or antiparallel to each other 
depending on the sign of $  \Ginv$ which is, of course, invariant under the proper Lorentz transform. 
Therefore, the appropriate choice of the Lorentz frame, in general, 
simplifies analyses of the strong-field physics. 
In such a frame, one finds $ a =  |\bE_0|$ and $b = |\bB_0|  $.\footnote{ 
Occasionally, the definitions of $a$ and $b$ are interchanged in the literature. 
However, the definitions in Eqs.~(\ref{eq:a}) and (\ref{eq:b}) may be more convenient 
in the sense that the $ b $ reduces to the $ B $ field by the Lorentz transform.
}

If the invariant is vanishing $ \Ginv =0 $, i.e., the electric and magnetic fields are 
orthogonal to each other in a Lorentz frame, they are also orthogonal in any other frame. 
In this case, we have $ a =0 $ and $  b = \sqrt{|\bB|^2-|\bE|^2}  $ 
when the other invariant is positive $ \Finv >0 $, 
or $ a = \sqrt{|\bE|^2-|\bB|^2}   $ and $ b =  0  $ when negative $ \Finv<0 $. 
Therefore, there exists a Lorentz frame in which either an electric or magnetic field vanishes. 
We shall denote the electric and magnetic fields in such a frame as $ |\bE_0| $ and $|\bB_0|  $, respectively. 
In the former case ($ |\bB| > |\bE| $), 
one can take a vanishing electric field ($  |\bE_0|=0 $) in the above expression, 
leaving only a magnetic field with a magnitude $   |\bB_0| = b  $. 
Similarly, in the latter case ($ |\bE| > |\bB| $), 
one finds that $ |\bB_0| = 0   $ and  $  |\bE_0| = a  $. 

A pair of vanishing invariants $ \Finv=\Ginv=0 $ corresponds to not only the absence of the field 
but also the so-called ``crossed field'' such that $   |\bE | = | \bB| $ and $ \bE \cdot\bB=0 $. 
This feature is invariant under the Lorentz transform.

In the second rank of mass-dimension four, one can again construct two tensors by the use of the field strength tensors. 
In the parallel configuration with $ E $ and $ B $ being the magnitudes of the fields, their explicit forms read 
\begin{subequations}
\begin{eqnarray}
F^{\mu\lambda} F^\nu_{\ \, \lambda} &=& - E^2 g_\para^{\mu\nu} + B^2 g_\perp^{\mu\nu}
\label{eq:FF_mn}
\, ,
\\
F^{\mu\lambda} \tilde F^\nu_{\ \, \lambda} &=& - E B g^{\mu\nu}
\label{eq:FFtilde_mn}
\, .
\end{eqnarray}
\end{subequations}
Therefore, we are naturally led to introduce two additional metrics 
for the longitudinal and transverse subspaces with respect to the fields. 
They are, respectively, defined as 
\begin{eqnarray}
\label{eq:metrics}
g_\para^{\mu\nu}  = {\rm diag} (1,0,0,-1) 
\quad
{\rm and}
\quad 
g_\perp^{\mu\nu} = {\rm diag} (0,-1,-1,0)
\, .
\end{eqnarray}
Without loss of generality, we have taken the directions of the fields 
in the positive $ z $ direction when $E,B > 0  $.\footnote{
It is also useful to note frame-independent identities 
$ F^{\mu \lambda} F^\nu_{\ \, \lambda} - \tilde F^{\mu \lambda} \tilde F^\nu_{\ \, \lambda} 
= 2 \F g^{\mu\nu}$ and $  F^{\mu \lambda} \tilde F^\nu_{\ \, \lambda} = \G g^{\mu\nu}  $. 
}
Accordingly, the momenta in these subspaces are introduced 
as $ p_\parallel^\mu =g_\para^{\mu\nu} p_\nu  $ 
and $ p_\perp ^\mu =g_\perp^{\mu\nu} p_\nu  $. 
Also, for the gamma matrices obeying the Clifford algebra $\{\gam^\mu , \gam^\nu \} = 2 g^{\mu\nu}  $, 
we introduce $ \gam_\parallel^\mu =g_\para^{\mu\nu} \gam_\nu  $ 
and $ \gam_\perp ^\mu =g_\perp^{\mu\nu} \gam_\nu  $. 
We will often use these notations throughout this review paper 
together with $ \gam^5 = i \gam^0 \gam^1 \gam^2 \gam^3 $.


\section{Landau quantization in a constant magnetic field}

\label{sec:Q} In this section, we provide a pedagogical discussion about 
the energy spectrum of charged particles in magnetic fields. 
We first analyze a nonrelativistic Hamiltonian 
to demonstrate the essence of the Landau quantization  \cite{Landau1930, Landau:1991wop} (see also Sec.~2.5 
in Ref.~\cite{hoddeson1992out} for an anecdotal review) 
and then proceed to QED in Sec.~\ref{sec:Ritus-Feynman}.

\begin{figure}[t]
     \begin{center}
           \includegraphics[width=0.5\hsize]{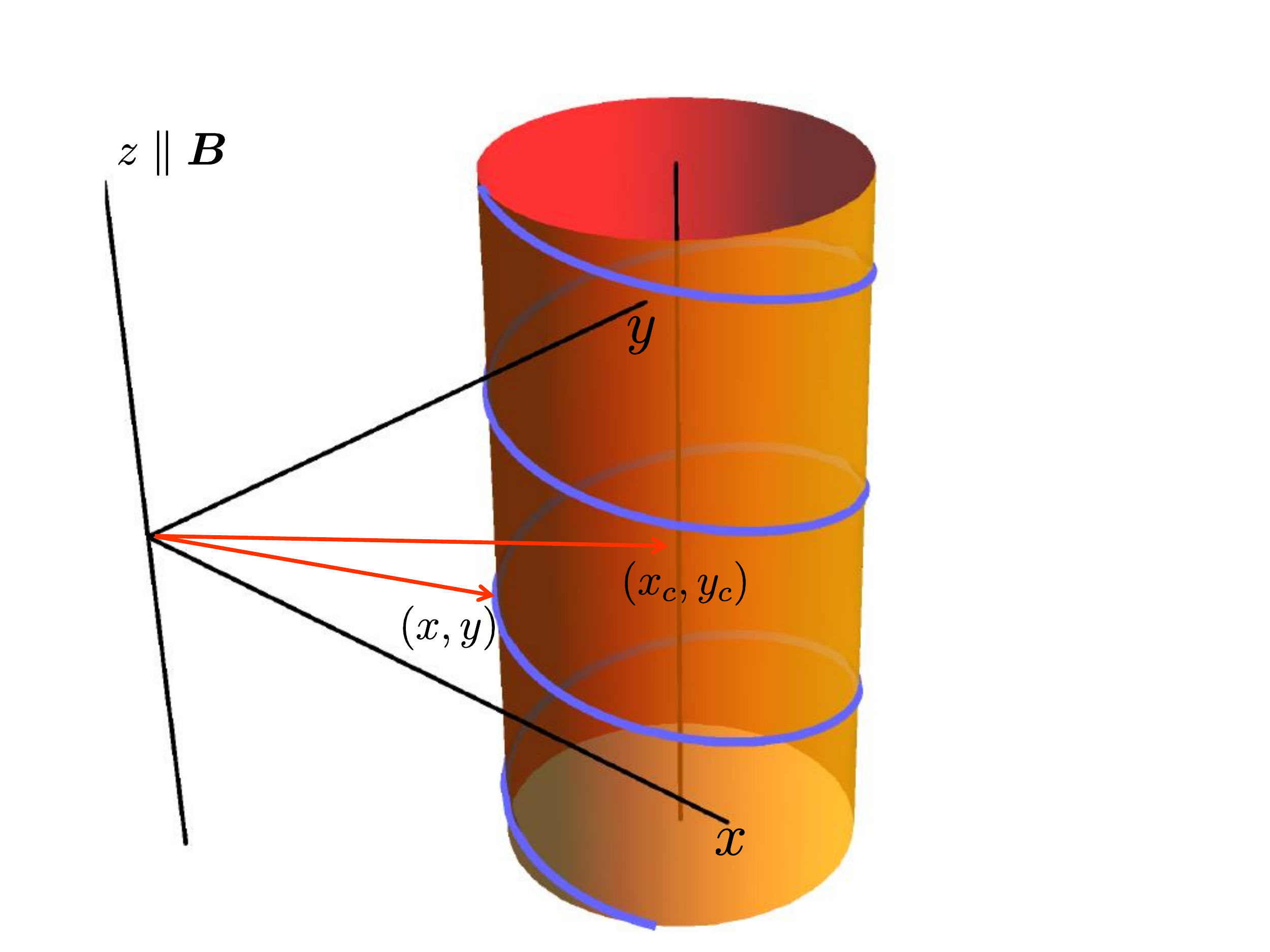}
     \end{center}
\caption{Cyclotron motion in a magnetic field $( \bB \parallel \hat z )$ with a finite longitudinal momentum $  p_z$. 
The transverse coordinates of the classical trajectory (blue curve) and 
of the center of the cyclotron orbit is given by $ (x,y) $ and $  (x_c, y_c)$, respectively.}
\label{fig:center}
\end{figure}

The quantum mechanical Hamiltonian is given as 
\begin{eqnarray}
\hat \Ham = \hat \Ham_\perp + \frac{\hat p_z^2}{2m} - \hat  \mu_B B
\label{eq:Ham}
\, .
\end{eqnarray}
Without loss of generality, we apply a constant magnetic field 
$ \bB =(0,0,B) $ in the $z$ direction, 
so that the longitudinal momentum $  \hat p_z$ is a constant of motion and can be replaced by a c-number. 
The last term is responsible for the Zeeman effect for 
the nonrelativistic magnetic moment $ \hat \mu_B \equiv q_f g \frac{\hbar}{2m} \hat s_z $, 
which is given by the $  g$-factor, an electrical charge $q_f$, 
and the spin operator $ \hat s_z$ along the magnetic field. 
The transverse part of the Hamiltonian is given as 
\begin{eqnarray}
\hat \Ham_\perp = \frac{1}{2 m} \{ \hat \bp_\perp - q_f   \bA_\perp (\hat \bx_\perp) \}^2 
= \frac{1}{2m} (\hat \pi_x^2 + \hat \pi_y^2)
\label{eq:H_perp}
\, ,
\end{eqnarray}
where the gauge field $\bA_\perp(\hat \bx_\perp)  $ 
generates the constant external magnetic field. 
The subscript $  \perp$ denotes the components transverse 
to the external magnetic field, that is, 
the $x $ and $ y$ components.

The commutation relation is imposed on 
the {\it canonical momentum} $ \hat \bp$ as
\begin{eqnarray}
[\hat x^i, \hat p^j] = i \hbar \delta^{ij}
\label{eq:can}
\, .
\end{eqnarray}
It is very important to note that, 
in the presence of the gauge field, 
the canonical momentum $ \hat \bp $ is different from 
the {\it kinetic momentum} defined by 
\begin{eqnarray}
\hat \bpi \equiv \hat \bp - q_f  \bA (\hat \bx)
\, .
\end{eqnarray}
This momentum corresponds to the covariant derivative. 
Its kinetic nature, relation to the mass and velocity, 
is suggested by the Heisenberg equation 
for the coordinate, i.e., 
$ \dot \bx = (i \hbar )^{-1} [ \hat \bx, \hat \Ham ]
= \hat \bpi / m $. 
Also, the Heisenberg equation for the kinetic momentum reads 
$ \dot \bpi = (i \hbar)^{-1} [ \hat \bpi, \hat \Ham ]
=   q_f \hat \bpi \times \bB /m $, indicating that 
the kinetic momentum evolves in time 
independently of the gauge choice for $\bA_\perp $ 
and that the transverse components $\hat \bpi_\perp $ 
are not conserved quantities in a magnetic field. 
The transverse components of 
the canonical momentum $\hat \bp_\perp $ are, in general, 
not conserved quantities either, 
i.e., $ [ \hat \bp_\perp , \hat \Ham] \not = 0$, 
because of the coordinate dependence of 
the gauge field $ \bA_\perp(\bx) $ under 
the canonical commutation relation (\ref{eq:can}): 
Whether or not the canonical momentum is conserved 
depends on the gauge choice.

\begin{figure}[t]
     \begin{center}
              \includegraphics[width=0.85\hsize]{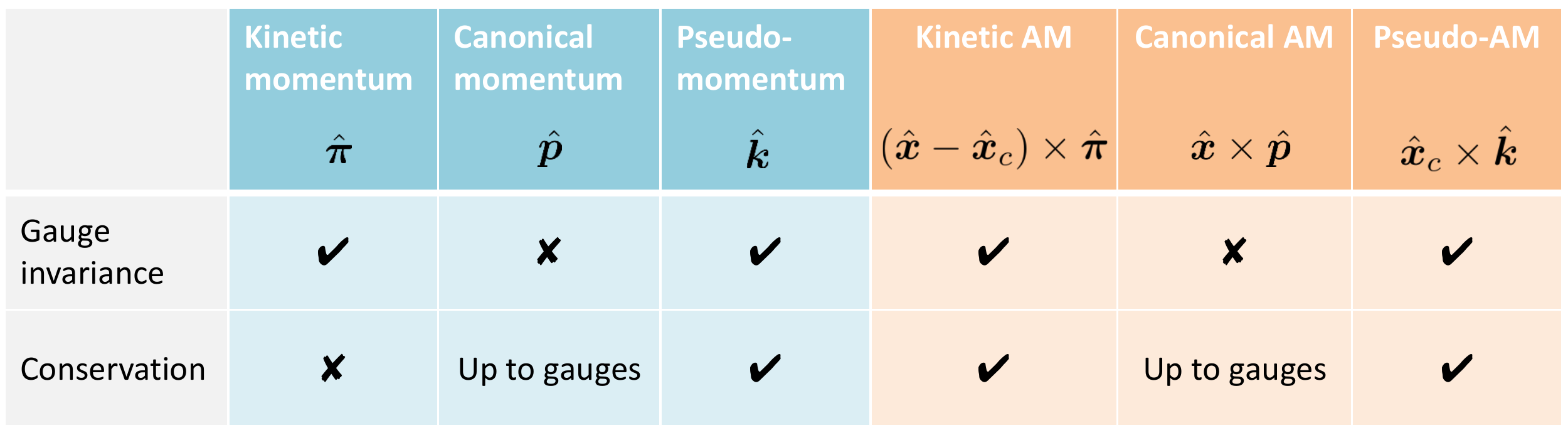}
     \end{center}
\vspace{-0.5cm}
\caption{Properties of the three linear momenta and the associated angular momenta (AM). }
\label{fig:conservation}
\end{figure}

The presence of a conserved and gauge-invariant momentum 
is suggested by the above Heisenberg equations of motion 
that describe the effects of the Lorentz force. 
Integrating the both sides of the equation for $ \dot \bpi$, 
we define the {\it pseudomomentum} as 
\begin{eqnarray}
\hat \bk \equiv  \hat \bpi - q_f \, \hat \bx \times {\bm B} 
\label{eq:pseudo-momentum}
\, .
\end{eqnarray}
The pseudomomentum commutes with the Hamiltonian, 
i.e., $ [ \hat \bk , \hat \Ham]  = 0$, 
independently of the gauge choice. 
Figure~\ref{fig:center} shows a classical trajectory 
of the cyclotron motion in a constant magnetic field. 
The classical solution for the center coordinate of 
a cyclotron orbit is found to be  
\begin{eqnarray}
\label{eq:center-coordinate}
( \hat x_c , \hat y_c) = ( \hat x + \frac{ \hat \pi_y}{q_f B}  ,  \hat y - \frac{\hat \pi_x}{q_f B})
\, .
\end{eqnarray}
We lifted $\hat \bx_\perp $ and  $ \hat \bpi_\perp$ 
to quantum operators on the right-hand side, 
which defines the center coordinate $(\hat x_c, \hat y_c)$ 
as a quantum operator. 
This coordinate, which is sometimes called 
the {\it guiding center}, is also a constant of motion 
both in classical and quantum theories, 
i.e., $ [ \hat x_c , \hat \Ham]  = 0 
=  [ \hat y_c , \hat \Ham]$. 
In fact, those two conserved quantities 
are not independent of one another and are related as 
\begin{eqnarray} 
\label{eq:ps-mom-center}
( \hat k_x , \hat k_y) 
= q_fB ( - \hat y_c ,  \hat x_c)
\, .
\end{eqnarray}
Figure~\ref{fig:conservation} summarizes 
the properties of the three distinct momenta 
as well as the associated angular momenta 
that we will discuss in detail. 
Based on the above observations, we will carefully examine 
quantum picture of the cyclotron motion.

\cout{
NOTE: Hamilton's equations read 
\begin{subequations}
\label{eq:}
\begin{eqnarray} 
\dot p_\perp^i \= - \frac{ \pd \Ham_\perp }{\pd x^i_\perp} 
= -\frac{1}{m} \sum_j \pi^j
 (- q_f \frac{ \pd A^j_\perp }{\pd x^i_\perp}) 
\\
\dot \bx \=  \frac{ \pd \Ham }{\pd \bp} 
=  \frac{1}{m} \{ \hat \bp_\perp - q_f   \bA_\perp (\hat \bx_\perp) \} = \frac{\bpi}{m}
\end{eqnarray}
\end{subequations}
On the other hand, 
\begin{eqnarray}
\dot \do \pi^i \= m \frac{d \ }{d t} \frac{\pd \Ham}{\pd p}
= \dot p^i - q_f \sum_j \frac{ \pd A^i}{ \pd x^j } \dot x^j
\nnb
\= q_f \sum_j \dot x^j \left( 
 \frac{ \pd A^j_\perp}{\pd x^i_\perp} 
 -   \frac{ \pd A^i}{ \pd x^j } \right)
\nnb
\= q_f \sum_j \dot x^j ( - \sum_k \epsilon^{k j i } B^k )
\nnb
\= q_f ( \dot \bx \times \bB)^i
\end{eqnarray}

\begin{eqnarray}
i \hbar \dot \pi^i_\perp \= [ \pi^i_\perp, \Ham_\perp ]
\nnb
\= \frac{1}{2m} [ \pi^i_\perp,( \pi_\perp^j)^2 ] 
\nnb
\= \frac{2 i \hbar q_f }{2m} \sum_{j,k} \epsilon^{ijk} \pi^j B^k 
\nnb
\= \frac{ i \hbar q_f }{m} (\bpi \times \bB)^i 
\end{eqnarray}

\begin{eqnarray}
i \hbar \dot \bx \= [ \hat \bx, \hat \Ham_\perp ]
\nnb
\= \frac{ 1}{m} ( \hat \bp - q_f \bA) [\hat \bx, \hat \bp]
\nnb
\=  \frac{ i \hbar }{m} ( \hat \bp - q_f \bA)  
\nnb
\= i \hbar \frac{ \hat \bpi}{m}
\end{eqnarray}
 
}

\subsection{Gauge-invariant consequences}

\label{sec:LL_1}

We first discuss the gauge-invariant consequences 
that can be extracted from the Hamiltonian (\ref{eq:H_perp}) 
and gauge-invariant algebra. 
We will not choose a particular gauge 
for $\bA_\perp(\hat \bx_\perp) $
until  Sec.~\ref{sec:wavefunctions} 
where we look for explicit forms of the wave functions. 
As mentioned above, neither kinetic nor canonical momentum 
is a conserved quantity in general. 
The canonical momentum, even if it is conserved, does not provide 
a gauge-invariant formulation of the Landau quantization. 
We formulate the Landau quantization in a gauge-invariant way 
by the use of eigenstates 
of the center coordinates $(\hat x_c, \hat y_c) $, 
and thus of the pseudomomentum $(\hat k_x, \hat k_y)$, 
that are the simultaneous eigenstates of 
the Hamiltonian $\hat \Ham_\perp $. 
We will, however, find that $ \hat x_c$ and $\hat y_c $ 
do not commute with each other in quantum theory, 
i.e., $[\hat x_c, \hat y_c] \not = 0 $. 
Until Sec.~\ref{sec:zeeman-effect}, 
we focus on the transverse part of the Hamiltonian $\Ham_\perp $.

\subsubsection{Landau quantization}

According to the canonical commutation relation (\ref{eq:can}), 
one can immediately show that the kinetic momentum satisfies a commutation relation
\begin{eqnarray}
[\hat \pi_x, \hat \pi_y] = i \hbar q_f ( \frac{\partial  A_y}{\partial  x} 
- \frac{\partial  A_x}{\partial y} ) = i \hbar q_f  B
\, .
\label{eq:canonical_B}
\end{eqnarray}
This is a gauge-invariant relation and can be regarded 
as a canonical commutation relation 
after an appropriate normalization is applied. 
Motivated by this observation, one can define 
a set of ``creation and annihilation operators'' 
\begin{eqnarray}
\hat a = \frac{\ell_f }{ \hbar \sqrt{2} } 
(\hat \pi_x + i s_f \hat \pi_y) 
\, \quad
\hat a^\dagger = \frac{\ell_f }{ \hbar \sqrt{2}} 
(\hat \pi_x - i s_f \hat \pi_y)
\label{eq:CA}
\, ,
\end{eqnarray}
which satisfies the commutation relation $ [\hat a, \hat a^\dagger] = 1$. 
Here, we defined a sign function $s_f = \sgn(q_f B) $ 
and $ \ell_f = \sqrt{\hbar/\vert  q_f B \vert}$ called the magnetic length. 
One can diagonalize the Hamiltonian (\ref{eq:H_perp}) 
in the same manner as that for the harmonic oscillator. 
That is, we get 
\begin{eqnarray}
\hat \Ham_\perp = \hbar \omega_f \left( \hat a^\dagger  \hat a + \frac{1}{2} \right)
\, .
\end{eqnarray}
The cyclotron frequency $ \omega_f = \vert q_f B \vert/m$ naturally appears in the spectrum 
since a cyclotron motion is a sort of harmonic motion. 
We have found that the energy spectrum is quantized 
in constant magnetic fields, 
and is specified by the eigenvalue $ n$ of the ``number operator $ \hat a^\dagger \hat a $'' as 
\begin{eqnarray}
\ep_n = \hbar \omega_f \left( n+ \frac{1}{2} \right)
\label{eq:energy}
\, .
\end{eqnarray}
This is called the Landau levels. 
As usual, the corresponding eigenvectors are constructed as 
\begin{eqnarray}
\hat a \vert 0 \rangle = 0 
\, , \quad
\vert n \rangle =  \frac{ ( \hat a^\dagger)^n}{\sqrt{n!}} \vert 0 \rangle
\label{eq:states}
\, ,
\end{eqnarray}
which are orthonormal among themselves, i.e., 
\begin{eqnarray}
\langle n \vert n^\prime \rangle = \delta_{n n^\prime}
\, .
\end{eqnarray}
Below, we investigate the properties and the expectation values 
of the linear and angular momentum operators 
with the eigenvectors (\ref{eq:states}).

\subsubsection{Kinetic momentum}

The kinetic momentum does not commute with the Hamiltonian, 
i.e., $[\hat \pi_{x,y} , \hat \Ham_\perp] \not = 0 $, 
and does not serve as a good quantum number. 
This is expected from classical intuition for a cyclotron motion  
where the direction of the kinetic momentum is changing in time.

The expectation value of the kinetic momentum reads 
\begin{eqnarray}
\label{eq:pi-average}
\langle n \vert \hat \pi_{x, y} \vert n \rangle \propto \langle n \vert  ( \hat a \pm \hat a^\dagger) \vert n \rangle = 0
\, ,
\end{eqnarray}
where the upper and lower signs are for the $ x$ and $y $ components, respectively. 
The expectation value of the velocity 
$\hat \bv =  \hat \bpi /m $ is also vanishing. 
This is again expected from the classical motion 
on a cyclotron orbit due to 
the alternating direction of motion.\footnote{
The velocity will have a finite expectation value 
when there is an electric field applied perpendicularly to the magnetic field because of the drift motion. 
This leads to the (classical) Hall effect, but not quantum Hall effect without effects of disorder (see, e.g., Ref.~\cite{girvin2002quantum, yoshioka2013quantum}). 
} 
The mean square is also easily obtained as 
\begin{eqnarray}
\label{eq:pi-square}
\langle n \vert \hat \pi_{x, y} ^2\vert n \rangle =
\frac{\hbar^2}{\ell_f^2} \left( n+ \frac{1}{2} \right)
=  \hbar |q_f B| \left( n+ \frac{1}{2} \right)
\, ,
\end{eqnarray}
where both the components take the same value. 
Then, we find an uncertainty relation 
$ \langle (\Delta \hat \pi_x)^2 \rangle\langle (\Delta \hat \pi_y)^2 \rangle 
\geq | \langle [ \hat \pi_x, \hat \pi_y] \rangle |^2/4 =\hbar^2 |q_fB|^2$ 
with $ \Delta \hat \order := \hat \order - \langle  \hat \order \rangle $ 
(cf., e.g., Ref.~\cite{sakurai_napolitano_2017}). 
Therefore, the ground state, called the lowest Landau level (LLL), satisfies the minimal uncertainty 
as anticipated from the analogy with the harmonic oscillator.

\subsubsection{Pseudomomentum, guiding center, and the magnetic translation}

\label{sec:pseudomomentum}

We have obtained the energy spectrum (\ref{eq:energy}) 
from the manifestly gauge-invariant algebra. 
One would, however, wonder if there is a mismatch in 
the number of degrees of freedom. 
Whereas we had two components of the transverse momenta in the transverse Hamiltonian (\ref{eq:H_perp}), 
the energy spectrum is only specified by one quantum number $ n$ in \eref{eq:energy}. 
To resolve this mismatch, we examine the eigenstates 
of the pseudomomentum \cite{PhysRev.76.828, 1977FCPh, 
meszaros1992high, AlHashimi:2008hr}.

As mentioned in Eq.~(\ref{eq:ps-mom-center}), 
the pseudomomentum is related to the center coordinate 
of the cyclotron orbit called the guiding center. 
It should be noticed that there is no preferred position for 
a center of cyclotron orbit in a constant magnetic field, 
so that the energy spectrum is expected to be 
independent of the center coordinate. 
Namely, there must be energy degeneracy.

One can easily show commutation relations for 
the center coordinate $[\hat x_c^i, \hat \pi^j]=0 $ for any combination of the components
and other two sets of important commutation relations 
\begin{eqnarray}
\label{eq:H-center}
[\hat x_c, \hat \Ham ]=0 = [\hat y_c,\hat \Ham] 
\, ,
\end{eqnarray}
and 
\begin{eqnarray}
[\hat x_c , \hat y_c] = - i s_f \ell_f^2
\label{eq:xcyc}
\, .
\end{eqnarray}
According to Eq.~(\ref{eq:ps-mom-center}), 
these relations also imply the corresponding relations 
for the pseudomomentum, $[\hat k^i, \hat \pi^j]=0 $ and so on. 
The first relation (\ref{eq:H-center}) indicates that the center coordinate is a constant of motion in quantum theory. 
However, this quantum number does not appear 
in the energy spectrum (\ref{eq:energy}), explicitly 
indicating that the Landau levels are degenerated with respect to 
the center coordinate (or the pseudomomentum). 
One can therefore use the center coordinate 
to label the degenerate energy eigenstates. 
However, the second relation (\ref{eq:xcyc}) indicates 
quantum nature that one cannot simultaneously determine 
both components of the center coordinate, meaning that 
only one of the components, or any one of functions of 
$\hat x_c, \hat y_c $, serves as a simultaneous eigenstate 
of the Hamiltonian. 
One can choose, for example, one of the followings 
\begin{eqnarray}
\label{eq:label}
\hat x_c \, , \quad 
\hat y_c \, , \quad 
\hat r_c^2 \equiv \hat x_c^2 + \hat y_c^2 \, , 
\end{eqnarray}
and so on. 
The existence of this one simultaneous quantum number 
resolves the aforementioned mismatch 
in the number of degrees of freedom. 
In Sec.~\ref{sec:wavefunctions}, we discuss 
two of an infinite number of possible choices.

\begin{figure}[t]
\begin{center}
   \includegraphics[width=0.6\hsize]{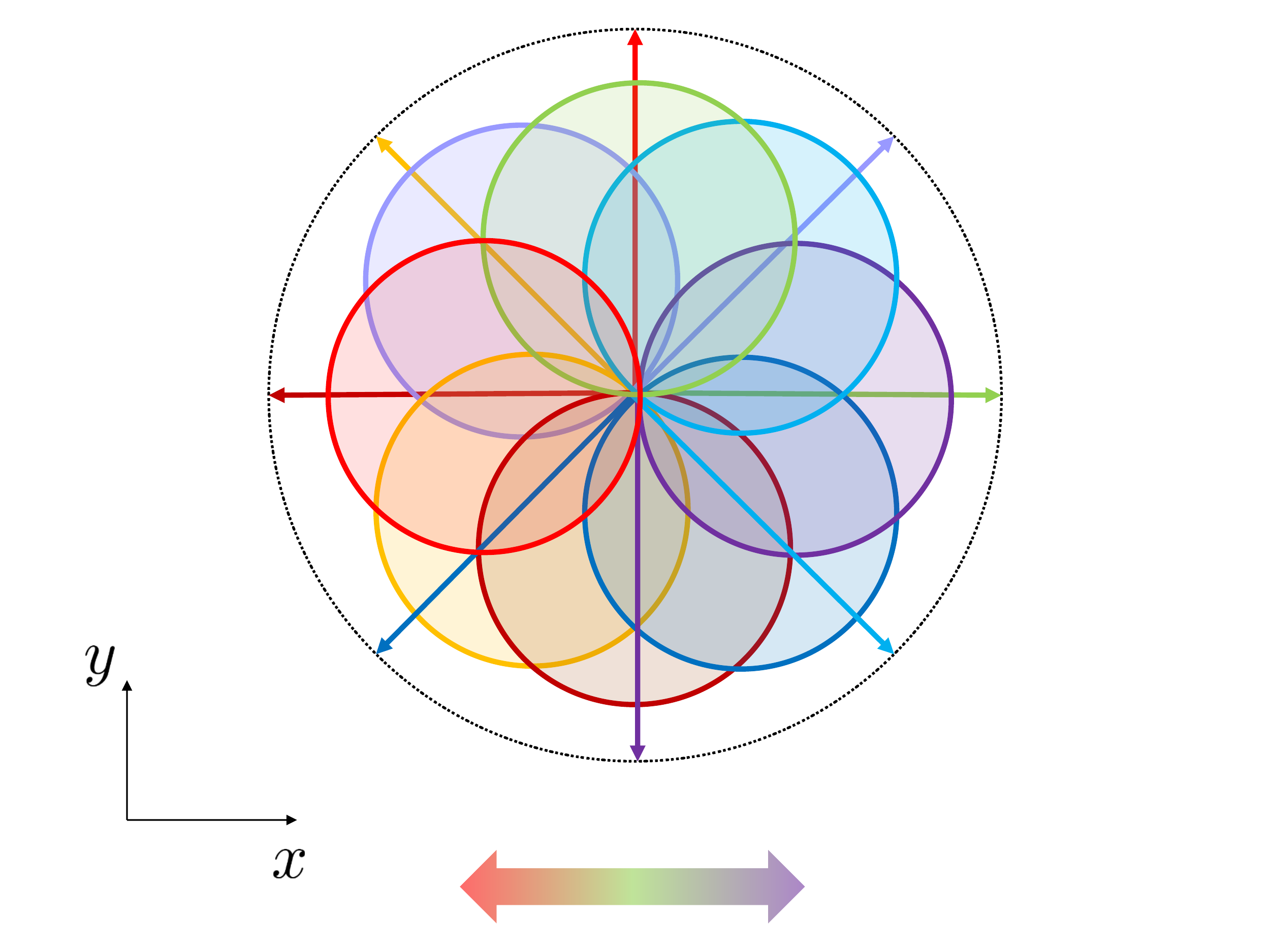}
\end{center}
\caption{
Superposition of cyclotron orbits that all pass through 
a given coordinate position. 
The arrows show the uncertainty in the direction of 
the kinetic momentum at a given energy eigenvalue (shown by dotted circle) and the circles in the same colors 
show the corresponding cyclotron orbits tangential 
to the arrows. 
The two-headed arrow in the bottom shows the magnitude of the standard deviation of the order of the cyclotron radius 
in the $ x$ direction; The same holds in the $ y$ direction. 
}
  \label{fig:superposition}
\end{figure}

The commutation relation (\ref{eq:xcyc}) leads to 
an uncertainty relation $ \langle (\Delta \hat x_c)^2 \rangle\langle (\Delta \hat y_c)^2 \rangle \geq
| \langle [ \hat x_c, \hat y_c] \rangle |^2/4 = \ell_f^4/4 $.  
The existence of the uncertainty in the center coordinate 
can be understood by convincing oneself 
that an energy eigenstate is not 
corresponding to one (classical) cyclotron orbit but 
to superposition of cyclotron orbits of the same radius. 
To see this, recall the existence of the uncertainty 
in the kinetic momentum (\ref{eq:canonical_B}). 
In the energy eigenstate $|n \rangle $, 
the magnitude of the kinetic momentum is  
determined via the Hamiltonian (\ref{eq:H_perp}). 
However, the direction of the kinetic momentum 
is uncertain in quantum theory. 
As a consequence, there are infinitely many cyclotron orbits 
that pass through a given coordinate position 
with the same magnitude of the kinetic momentum. 
See Fig.~\ref{fig:superposition} drawn in the coordinate space 
where the arrows and circles in the same colors show 
the uncertainty in the directions of the kinetic momentum 
and the corresponding cyclotron orbits 
tangential to the kinetic momentum, respectively. 
The uncertainty in the center coordinate 
originates from the superposition of the cyclotron orbits, 
and thus the magnitude of the uncertainty is 
of the order of the cyclotron radius. 
We find that the LLL satisfies the lower bound 
in the uncertainty relation. 
The lower bound decreases as we increase the magnetic field strength since the cyclotron radius 
itself shrinks with an increasing magnetic field strength.



\cout{
Knowing the $y $ component of the kinetic momentum at $\bar x_c $ 
does not uniquely specifies a cyclotron orbit that 
the particle is belonging to since the $x $ component of 
the kinetic momentum is uncertain at 
a given position $\bar x_c $. 
Consequently, the energy eigenstate is inherently given by 
the superposition of possible cyclotron orbits, leading to 
a nonzero standard deviation of $\hat x_c $ 
of the order of the cyclotron radius in quantum theory (see Fig.~\ref{fig:superposition}). 
Computing the same quantities for $\hat y_c $, 
we find that the LLL satisfies the lower bound 
in the uncertainty relation. 
}



As we increase the magnetic-field strength, 
more and more orbits are packed in the transverse plane. 
The density of degenerate states for a given $ n $ can be obtained from 
the phase space volume of the canonical pair in Eq.~(\ref{eq:xcyc}). 
When the system size is given by 
$ 0 \leq \xc, \yc \leq L_{x,y} $, 
one finds the density of states 
often called the Landau degeneracy factor 
\begin{eqnarray}
 \frac{1}{L_x L_y} \left(\frac{ L_x L_y |q_fB|} { 2\pi \hbar}\right)
 = \frac{ |q_f B|}{ 2\pi \hbar }
 \label{eq:DoS-0}
 \, ,
\end{eqnarray}
where the factor of $  |q_fB|$ comes form the normalization of the canonical pair in Eq.~(\ref{eq:xcyc}). 
The density of states is common to all the Landau levels.\footnote{
This may not be a rigorous statement for finite-volume systems. 
Charged particles feel boundary effects when their orbits touch the boundary, 
and, actually, the cyclotron radius depends on the Landau levels [cf. Eq.~(\ref{eq:R})]. 
Nevertheless, one may neglect the contribution of such states near the boundary 
as compared to a large number of states in the bulk as long as $L_x L_y/ \ell_f^2 \gg1$ and we focus on bulk properties. 
One should note that boundary contributions can be crucial 
when bulk contributions are suppressed for some reason 
like in quantum Hall effect. 
} 
Note also that there could be further spin degeneracies depending on particle species 
as discussed in Sec.~\ref{sec:zeeman-effect}.

We remark on an interesting property of $\hat r_c^2 $ 
defined in Eq.~(\ref{eq:label}) that is 
the distance between a cyclotron orbit 
and the coordinate origin. 
The eigenvalue of $\hat r_c^2 $ is quantized 
independently of the gauge choice, 
and the degenerate states are labeled by integers 
(when we decide to choose this label among other choices). 
Led by the canonical commutation relation (\ref{eq:xcyc}) 
between $ \hat \xc$ and $\hat \yc $, we can construct 
a set of ``creation and annihilation operators'' 
\begin{eqnarray}
\label{eq:b-operators}
\hat b = \frac{1}{\sqrt{2}\ell_f} ( \hat x_c - i s_f \hat y_c) 
\, , \quad 
\hat b^\dagger = \frac{1}{\sqrt{2}\ell_f} ( \hat x_c + i s_f \hat y_c) 
\, ,
\end{eqnarray}
which satisfy the commutation relation $ [ \hat b , \hat b^\dagger] = 1$. 
They commute with $\hat a$ and $\hat a^\dagger $. 
The radial coordinate is expressed by the ``number operator'' as 
\begin{eqnarray}
\hat r_c^2 = 2 \ell_f^2 \left( \,\hat  b^\dagger \hat b + \frac{1}{2} \, \right)
\label{eq:bbdagger}
\, .
\end{eqnarray}
Therefore, the degenerate states are labeled by 
an integer $ m $ that is the eigenvalue of $\hat  b^\dagger \hat b $.\footnote{
We follow this frequently used notation for $ m $, 
which should not be confused with a mass parameter. 
} 
The above algebraic properties are gauge-invariant ones. 
We will use those properties in Sec.~\ref{sec:S-gauge}.

In classical mechanics, 
the system in a constant magnetic field has 
a translational invariance in the transverse plane 
that manifests itself in a shift of cyclotron orbit. 
In quantum theory, neither the kinetic nor canonical momentum 
is a gauge-invariant generator of such a translation operator. 
Nevertheless, we expect the existence of 
a translational invariance since it is the very origin 
of the Landau degeneracy discussed above. 
The reason for the apparent breaking is that 
the usual translation operator, 
generating a linear coordinate shift 
$\hat \bx_\perp \to \hat \bx_\perp'= \hat \bx_\perp 
+ \Delta \bx_\perp $, transforms the vector potential as 
$\bA_\perp(\hat \bx_\perp) \to  \bA_\perp(\hat \bx_\perp') $, 
which breaks the translational invariance of the Hamiltonian. 
This translation, however, does not contradict with 
the translation invariance of the system 
since the magnitude of the magnetic field is invariant, 
i.e, $\nabla\times \bA_\perp(\hat \bx_\perp) = \bB = \nabla \times  \bA_\perp(\hat \bx_\perp') $. 
That is, the translational invariance is hidden 
behind the gauge dependence induced by 
the translation of the vector potential. 
Below, we find that the pseudomomentum serves as the generator 
of the translation operator connecting two degenerate states 
at different center coordinates 
with an appropriate phase rotation. 
Likewise, the rotational symmetry with respect to 
the direction of the magnetic field should imply 
the existence of an associated conserved angular momentum 
as we will indeed find below.


We introduce the ``magnetic translation operator'' generated 
by the pseudomomentum as \cite{Zak:1964zz, brown1964bloch}\footnote{See also Ref.~\cite{kohmoto1985topological} 
for an application to quantum Hall effect.}
\begin{eqnarray}
T(\bx) = 
\exp\left(-\frac{i}{\hbar}\bx \cdot  \hat \bk\right)
\label{eq:magneticT}
\, .
\end{eqnarray}
One can show that the pseudomomentum is canonical conjugate 
to the cyclotron center, i.e., 
\begin{eqnarray}
\label{eq:xc-k}
[\hat x_c^i , \hat k ^j] = i \hbar \delta^{ij} 
\, ,
\end{eqnarray}
as well as to $ \hat \bx $. 
Therefore, this translation connects the degenerate energy eigenstates labeled by 
different eigenvalues of $ \hat \bx_c $.\footnote{
For notational brevity, we use the vector form of $ \hat \bx_c $. 
However, remember that the two transverse components 
do not commute with each other 
and cannot be simultaneous eigenstates of the Hamiltonian 
at the same time as shown in Eq.~(\ref{eq:xcyc}). 
Also, we do not include the coordinate $z $ 
into the label as it is not a good quantum number. 
} 
Namely, when a state $ | n, \bx_c \rangle   $ is an energy eigenstate of the Landau level $ n $, 
so is a new state $| n, \bx_c + \bx \rangle  \equiv T(\bx) | n, \bx_c \rangle   $. 
Note, however, that the two components of $ \hat \bk $ do not commute with each other\footnote{
The pseudomomentum can be generalized to 
a center-of-mass momentum of multi-body systems 
in constant magnetic fields, 
and is conserved if the system is charge neutral in total \cite{avron1978, herold1981}. 
This generalized pseudomomentum commutes with the Hamiltonian 
as long as the interaction Hamiltonian has a translational invariance, 
so that it 
serve as a good quantum number of bound states 
in a magnetic field. 
This conserved momentum was, for example, used in the study of quarkonium spectrum in a magnetic field \cite{Alford:2013jva, Bonati:2015dka, Suzuki:2016kcs, Yoshida:2016xgm} (see Ref.~\cite{Hattori:2016emy, Iwasaki:2021nrz} for reviews). 
However, if a bound state carries a total nonzero charge, 
the transverse components do not commute with each other. 
Therefore, one of the components cannot persist 
as a good quantum number, 
and, instead, an integer specifies a bound-state spectrum  \cite{Hattori:2015aki}. 

}
\begin{eqnarray}
\label{eq:k-k}
[\hat k_x, \hat k_y] = - i \hbar q_f B
\, .
\end{eqnarray}
Accordingly, the translation operators do not commute either, 
and interchanging the order of the magnetic translation operators 
gives rise to a phase factor as 
\begin{eqnarray}
\label{eq:AB-magnetic_translataion}
T(\bx_2) T(\bx_1) = T(\bx_1) T(\bx_2) \, 
e^{\frac{1}{\hbar^2} [ \bx_1 \cdot \hat \bk, \, \bx_2 \cdot \hat  \bk]}
= T(\bx_1) T(\bx_2) \, e^{ - i \frac{ q_f}{\hbar} \Phi_B}
\, .
\end{eqnarray}
The phase is proportional to $ \Phi _B = (\bx_1 \times \bx_2)\cdot \bB$, which is the magnetic flux penetrating the area spanned by $\bx_1$ and $\bx_2$, 
and vanishes only when the magnetic flux 
is an integer multiple of the magnetic flux quantum, 
i.e., $\Phi_B = (h /q_f )n_B $ with $ n_B $ being an integer. 
This phase is essentially the Aharonov-Bohm phase 
(though the magnetic field is directly interacting 
with a charged particle in the present case).

\cout{

\begin{eqnarray}
T(\bx) \= \exp\left(-\frac{i}{\hbar}\bx \cdot  \hat \bk\right)
\nnb
\= \exp\left(-\frac{i}{\hbar}\bx \cdot  \hat \bp \right)
\exp\left( + q_f \frac{i}{\hbar}\bx \cdot 
( \hat \bA(\hat \bx) + \hat \bx \times \bB )\right)
\exp\left(  - \frac12 [-\frac{i}{\hbar}\bx \cdot  \hat \bp,
q_f \frac{i}{\hbar}\bx \cdot 
( \hat \bA(\hat \bx) + \hat \bx \times \bB )] \right)
\nnb
\= \exp\left(-\frac{i}{\hbar}\bx \cdot  \hat \bp \right)
\exp\left( + q_f \frac{i}{\hbar}\bx \cdot 
( \hat \bA(\hat \bx) + \hat \bx \times \bB )\right) 
\, .
\end{eqnarray}
where $ [\nabla \times (\bx \times \bB)]_z
= \pd_x (\bx \times \bB)_y - \pd_y (\bx \times \bB)_x
= \pd_x (-xB) - \pd_y (yB) = -2 B$. 

\begin{eqnarray}
[-\frac{i}{\hbar}\bx \cdot  \hat \bp,
q_f \frac{i}{\hbar}\bx \cdot 
( \hat \bA(\hat \bx) + \hat \bx \times \bB )] 
&\propto& x^i x^j [\hat p^i,  \hat A^j(\hat \bx) 
+ \epsilon^{jlm} \hat x^l B^m ] 
\nnb
\= x^i x^j [\hat p^i,  \hat A^j(\hat \bx) 
+ \epsilon^{jlm} \hat x^l B^m] 
\nnb
\= -i x^i x^j \nabla^i \hat A^j(\hat \bx) 
\end{eqnarray}

\begin{eqnarray}
\hat \bA(\hat \bx + \Delta \bx_c) - \hat \bA(\hat \bx)
=  \Delta \bx_c \cdot \nabla  \hat \bA(\hat \bx + \Delta \bx_c)
+ ( \Delta \bx_c^2) \, .
\end{eqnarray}

\begin{eqnarray}
\nabla \left( \hat \bA(\hat \bx) + \hat \bx \times \bB \right)
= \nabla   \hat \bA(\hat \bx) + 0
\end{eqnarray}

}

\subsubsection{Angular momenta}

We focus on the rotational motion of a cyclotron orbit 
and associated angular momenta. 
At abstract level, an angular momentum is introduced 
as the generator of the spatial rotations. 
We examine whether or not there is any representation of 
such generators in the Hilbert space of energy eigenstates 
in the current situation. 
For this purpose, it is useful to examine concrete forms 
of angular momenta 
that can be introduced as the exterior product of 
a coordinate and momentum in analogy with classical mechanics. 
In general, properties of as such introduced angular momenta 
depend on the reference point in the coordinate system. 
Besides, one can here define distinct angular momenta 
by the use of the three linear momenta introduced above 
which are the kinetic momentum, canonical momentum, 
and pseudomomentum. 
Among those various options, there are, however, few 
conserved angular momenta since a magnetic field 
breaks the rotational symmetries except for the one 
with respect to the direction of the magnetic field 
and the vector potential can completely break 
the rotational symmetries depending on a gauge choice. 
Not all the components of an angular momentum 
are conserved in any definition introduced below. 
Recall also that the classical picture of cyclotron motion 
is a rotational motion around the center 
of a cyclotron orbit $ (\hat x_c , \hat y_c) $ 
distributed in the transverse plane 
that is different from the coordinate origin. 
Such a classical intuition is useful below.

Motivated by the classical picture, 
we first define the {\it kinetic angular momentum} around 
the center coordinate of a cyclotron orbit as 
\begin{eqnarray}
\label{eq:kinetic-AM}
\hat {\bm  M} \equiv  
\hat {\bm \xi} \times \hat {\bm \pi}
\, ,
\end{eqnarray} 
with the kinetic momentum $  \hat {\bm \pi}$. 
We take 
$\hat {\bm \xi} = (\hat x - \hat x_c, \hat y - \hat y_c, \hat z  ) $ 
that is canonical conjugate to the kinetic momentum, i.e., 
$[ \hat  \xi ^i  ,\hat \pi^j]  = i\hbar \delta^{ij}$. 
By the use of Eq.~(\ref{eq:center-coordinate}), 
one can immediately rewrite the Hamiltonian as 
\begin{eqnarray}
\label{eq:Ham-rotation}
\hat \Ham_\perp 
= - \frac12 \omega_f s_f \hat M_z
\, .
\end{eqnarray} 
Clearly, the $z $ component along the magnetic field 
is a conserved quantity, i.e., 
$ [\hat M_z, \hat \Ham] = 0 $, 
which is anticipated from the rotational symmetry 
around the center of cyclotron orbit. 
Kinetic nature is made clear if one writes it 
by the use of Eq.~(\ref{eq:center-coordinate}) as 
\begin{eqnarray}
\hat M_z   = - s_f I_z \omega_f 
\, , \quad 
\hat \Ham_\perp = \frac12 I_z \omega_f^2
\, ,
\end{eqnarray}
where one can define the ``moment of inertia'' of 
a point-like particle as 
$  I _z \equiv m  \hat R^2
$ 
with the distance from the center coordinate $ \hat R^2 
\equiv (\hat  x- \hat  x_c)^2 + (\hat  y- \hat  y_c)^2 $. 
The average radius of 
the cyclotron orbit depends on the Landau level as 
\begin{eqnarray}
\langle n \vert \hat R^2 \vert n \rangle 
=\ell_f^2 (2n+1)
\, .
\label{eq:R}
\end{eqnarray}
The average radius is proportional to 
the magnetic length $\ell_f$ and shrinks as we increase 
the magnetic-field strength. 
The average radius increases with 
an increasing $n$ for the higher Landau levels (hLLs). 
The cyclotron motion gives rise to a nonzero expectation value 
of the kinetic angular momentum 
\begin{eqnarray}
\label{eq:Mz}
\langle n \vert \hat M_z \vert n \rangle 
= - s_f \hbar (2n+1)
\, .
\end{eqnarray}
The magnitude depends on the Landau level, 
but is independent of the magnetic-field strength. 
The sign of the expectation value depends on the direction of 
the cyclotron motion via the sign function $  s_f$. 
Moreover, due to the overall minus sign in Eq.~(\ref{eq:Mz}), 
the magnetic moment of the orbital current is 
opposite to the applied magnetic field, 
implying occurrence of the {\it Landau diamagnetism}~\cite{Landau1930}.  
None of the other components of $  \hat {\bm M} $ nor  
the kinetic angular momenta around the coordinate origin, 
$ \hat \bx \times \hat {\bm \pi} $ 
and $ \hat \bx_c \times \hat {\bm \pi} $, 
are conserved in a magnetic field 
due to the absence of associated rotational symmetries.

Recalling that $\hat {\bm \xi}$ and $\hat \bpi $ are 
canonical conjugate to each other, one can show 
the following commutation relations 
\begin{eqnarray}
\label{eq:MM}
[\hat M_y, \hat M_z] = 2 i\hbar \hat M_x
\, , \quad 
[\hat M_z, \hat M_x] = 2 i\hbar \hat M_y
\, .
\end{eqnarray}
These relations indicate that the kinetic angular momentum 
satisfies a part of the algebra for the rotation generators 
when it is normalized as $\hat {\bm M}/2 $. 
The factor of $ 2$ stems from the fact that 
$\hat \pi_x $ and $\hat \pi_y $ do not commute with each other. 
For the same reason, we find that 
the algebra is not closed, i.e.,  
$[\hat M_x, \hat M_y] \not = 2 i\hbar \hat M_z $. 
One of the components $\hat M_z/2 $ generates 
a rotation of the coordinate $ \hat {\bm \xi} $ 
with respect to the direction of the magnetic field. 
That is, we can define a rotation operator 
\begin{eqnarray}
\hat D_z(\theta) = 
e^{ - \frac{i}{\hbar} \theta \frac{ \hat M_z}{2} }
\, ,
\end{eqnarray}
by an angle $ \theta$ around the center coordinate 
of a cyclotron orbit. 
Then, the coordinate operator $\hat {\bm \xi}$
is rotated in the transverse plane as 
\begin{eqnarray}
\hat D_z^{-1}(\theta) \hat \xi_x \hat D_z(\theta) 
= \hat \xi_x \cos\theta - \hat \xi_y \sin \theta
\, .
\end{eqnarray}
Applying a rotation by an angle $\theta = 2\pi $ 
to the energy eigenstate, we find a phase factor 
\begin{eqnarray}
\label{eq:kinetic-rotation}
\hat D_z(2\pi) | n \rangle 
= e^{ i\frac{q_f}{\hbar}  \hat \Phi_B  } | n \rangle
= e^{ i s_f  (2n+1) \pi } | n \rangle
= - | n \rangle
\, ,
\end{eqnarray}
where $\hat \Phi_B = B \pi \hat R^2 $ is the magnetic flux 
penetrating through a cyclotron orbit 
$ \hat \Phi_B |n\rangle 
= s_f \frac{\hbar}{q_f} (2n+1) \pi|n\rangle$. 
We obtained a negative sign generated by a $2\pi $ rotation 
similar to that appearing in the spin precession (see, e.g., 
Sec.~3.2 in a textbook~\cite{sakurai_napolitano_2017}).

One can also define the {\it canonical angular momentum} 
\begin{eqnarray}
\label{eq:canonical-AM}
\hat {\bm L} \equiv\hat \bx \times \hat \bp 
\, .
\end{eqnarray}
This angular momentum satisfies the algebra 
$[ \hat L_i, \hat L_j] =i\hbar \sum_k \epsilon_{ijk} \hat L_k$ 
required for the rotation generators 
and serves as the representation in 
the three-dimensional coordinate space. 
This property is solely guaranteed by 
the canonical commutation relation (\ref{eq:can}) 
and is intact in the presence a magnetic field. 
However, the canonical angular momentum 
does not serve as a representation in the Hilbert space 
of the energy eigenstates 
since it is, in general, not a conserved quantity 
unless the gauge configuration $ \bA_\perp(\hat \bx)$ 
holds a rotational symmetry. 
Whether or not it is conserved depends on the gauge choice 
(see Eq.~(\ref{eq:Lz1}) and below for the ``symmetric gauge'' 
where it becomes a conserved quantity).

Finally, we define another angular momentum 
by the pseudomomentum as 
\begin{eqnarray}
\hat \bK \equiv   \hat \bx_c \times \hat \bk 
\, ,
\end{eqnarray}
where 
we introduced a three dimensional vector 
$ \hat \bx_c \equiv (\hat x_c, \hat y_c, \hat z )$. 
This is just for the notational convenience\footnote{
While the transverse components $(\hat x_c , \hat y_c) $ 
are originally introduced as conserved quantities, 
$\hat z$ and thus $\hat \bx_c $ as a three-vector are 
of course not conserved quantities. 
} 
and the z components of $\hat \bx_c $ and $\hat {\bm \xi} $ 
are the same. 
Then, $  \hat \bx_c $ and $ \hat \bk$ are canonical conjugate 
to each other as shown in Eq.~(\ref{eq:xc-k}) 
and imposed in Eq.~(\ref{eq:can}). 
Here, we call $ \hat \bK$ the {\it pseudo angular momentum}. 
This is a gauge-invariant quantity. 
Notice that its component along the magnetic field 
is proportional to the radius defined in Eq.~(\ref{eq:label}) as 
\begin{eqnarray}
\label{eq:Kz-rc}
\hat K_z =   q_f B \hat r_c^2
\, ,
\end{eqnarray}
and is a conserved quantity, i.e., 
$[\hat K_z, \hat \Ham ] \propto 
[\hat r_c^2, \hat \Ham ] =0 $ as discussed there. 
As clear in Eq.~(\ref{eq:Kz-rc}), 
$ \hat K_z $ does not have kinetic nature 
since it is independent of particle mass and velocity 
and only depends on the magnetic-field strength and the center coordinate. 
Neither the other possible quantities, 
$ \hat \bx \times \hat \bk $ nor $ (\hat \bx - \hat \bx_c) \times \hat \bk $, 
is a conserved angular momentum because we have 
$\hat \bx \times \hat \bk = \hat \bx \times \hat \bpi   $
which is not conserved as mentioned above.

One can immediately show that $\hat K_z $ satisfies 
the following commutation relations 
\begin{eqnarray}
\label{eq:commutators-KK}
[ \hat K_y, \hat K_z]  = 2i \hbar  \hat K_x
\, , \quad 
[ \hat K_z, \hat K_x] = 2 i \hbar \hat K_y
\, .
\end{eqnarray}
Similar to Eq.~(\ref{eq:MM}) for the kinetic angular momentum, 
these commutation relations stem from 
the canonical commutation relation between $ \hat \bx_c $ and $ \hat \bk $. 
Due to the non-commutative property (\ref{eq:k-k}) 
between $\hat k_x $ and $ \hat k_y$, 
there is an additional factor of $ 2$ on the right-hand side 
as compared to the algebra for the rotation generators 
and the other combination does not satisfy 
the algebra required for the rotation generators, i.e., 
$[ \hat K_x, \hat K_y] \not = i \hat K_z $. 
The normalized operator $\hat K_z/2 $ serves as the generator of 
spatial rotation with respect to 
the direction of the magnetic field 
\begin{eqnarray}
\hat R_z(\theta) 
= e^{ - \frac{i}{\hbar} \theta \frac{\hat K_z}{2} }
\, ,
\end{eqnarray}
by an angle $ \theta$ around the coordinate origin.  
According to the canonical commutation between 
$ \hat \bx_c$ and $ \hat \bk$, 
this operator rotates $ \hat \bx_c$ in the transverse plane as 
\begin{eqnarray}
\hat R_z^{-1}(\theta) \hat x_c \hat R_z (\theta)
= \hat x_c \cos\theta - \hat y_c \sin \theta
\, .
\end{eqnarray}
Since the center of rotation is the coordinate origin, 
this operator also rotates $\hat \bx$ as expected from 
the canonical commutation relation 
$[\hat x^i,\hat k^j ] = i\hbar \delta^{ij} $. 
When we label the degenerate states 
with $ \hat r_c^2$ in Eq.~(\ref{eq:label}) 
and apply a rotation by an angle $\theta = 2\pi $, 
we again find a phase factor (see Eq.~(\ref{eq:AB-magnetic_translataion}) for the phase factor from the magnetic translation along a closed path)
\begin{eqnarray}
\label{eq:AB-rotation}
\hat R_z(2\pi) | n, r_c^2 \rangle
= e^{ - i\frac{q_f}{\hbar} \Phi_B } | n, r_c^2 \rangle
=e^{ - i s_f ( 2m + 1)\pi  } | n, r_c^2 \rangle
= - | n, r_c^2 \rangle
\, ,
\end{eqnarray}
where the magnetic flux penetrating through the area $ \pi  r_c^2$ is quantized as $\Phi_B = (\pi r_c^2) B 
= \hbar s_f/q_f  ( 2m + 1)\pi $ according to 
Eq.~(\ref{eq:bbdagger}). 
Also, in analogy with the familiar theory of angular momentum, 
we define the ladder operators 
\begin{eqnarray}
\label{eq:K-ladder}
\hat K_\pm = \frac12 ( \hat K_x \pm i s_f \hat K_y )
\, ,
\end{eqnarray}
which satisfy the commutation relations 
$ [ \hat K_z/2 , \hat K_\pm] = \pm s_f \hbar \hat K_\pm$. 
According to the relation (\ref{eq:Kz-rc}), 
these are equivalent to 
$ [ \hat r_c^2 , \hat K_\pm] = \pm 2 \ell_f^2  \hat K_\pm $. 
Therefore, these ladder operators shift 
the eigenvalue of $\hat r^2_c $ as 
\begin{eqnarray}
\label{eq:rc-ladder}
\hat r_c^2 \hat K_\pm |n, r_c^2 \rangle 
= ( r_c^2 \pm 2 \ell_f^2  ) \hat K_\pm |n, r_c^2 \rangle 
\, .
\end{eqnarray} 
This relation indicates that the ladder operators $\hat K_\pm $ 
connect the degenerate states in a given Landau level $n $ 
and that the eigenvalue of $\hat r_c^2 $ is quantized 
as we have already seen in Eq.~(\ref{eq:bbdagger}). 
Accordingly, the $2\pi $ rotation gives rise to 
a non-vanishing phase in Eq.~(\ref{eq:AB-rotation}) 
that flips the sign of the state vector. 
Here, we stress that the algebraic structures for $\hat \bK $ 
have been constructed in a gauge-invariant way.
In Sec.~\ref{sec:S-gauge}, 
we will use the quantized eigenvalue of $\hat r_c^2  $ 
to label the degenerate states.

\cout{

In the beginning of Sec.~\ref{sec:S-gauge}, 
we will come back to the quantization 
and find that the phase factor does not vanish 
in Eq.~(\ref{eq:AB-rotation}). 
Here, we stress that the above algebraic structures 
have been constructed in a gauge-invariant way. 


}

\cout{
NOTE: 

\begin{eqnarray}
[\hat K_\pm, \hat K_\mp] \= 
[ \hat K_x \pm i s_f \hat K_y , \hat K_x \mp i s_f \hat K_y ]
\nnb
\=  \mp i s_f  [ \hat K_x  ,  \hat K_y ]
 \pm i s_f  [ \hat K_y , \hat K_x   ]
\nnb
\=  \mp 2 i s_f  [ \hat K_x  ,  \hat K_y ]
\end{eqnarray}

}

\cout{
Rotation by $ 2\pi$ should be identical 
\begin{eqnarray}
2\pi m_K = - \frac{q_f}{\hbar} (2\pi) K_z
\end{eqnarray}
Therefore, the eigenvalue of $\hat K_z $ is quantized as 
\begin{eqnarray}
K_z = - \frac{\hbar}{q_f} m_K
\end{eqnarray}
This means that the eigenvalue of $\hat r_c^2 $ is quantized as 
\begin{eqnarray}
r_c^2 = \frac{2}{q_f B} K_z = - 2 s_f \ell_f^2 m_K
\end{eqnarray}

\begin{eqnarray}
2\pi m = - \frac{q_f}{\hbar} \Phi_B
= - \frac{ s_f}{ \ell_f^2} \pi r_c^2
\end{eqnarray}
Thus,
\begin{eqnarray}
r_c^2 = 2  \ell_f^2  s_f {\mathfrak m}
\end{eqnarray}
with an integer $s_f {\mathfrak m}$. 
For a sequence of the eigenvalues $r_c^2 $ 
terminate at a non-negative eigenvalues 
in Eq.~(\ref{eq:rc-ladder}), 
$r_c^2 $ should not take a vanishing eigenvalue. 
Therefore, we should have $ s_f {\mathfrak m} = 2m +1 $.


$\hat K_z $ is the simultaneous eigenstate of $\hat K^2 $ 
whereas the other components are not, i.e., 
$ [ \hat K^2 , \hat K_z] =0$ and 
$ [ \hat K^2 , \hat K_{x,y}] \not =0 $. 

}

\subsubsection{Zeeman effect and resultant energy spectrum} 

\label{sec:zeeman-effect}

\begin{figure}[t]
     \begin{center}
           \includegraphics[width=0.7\hsize]{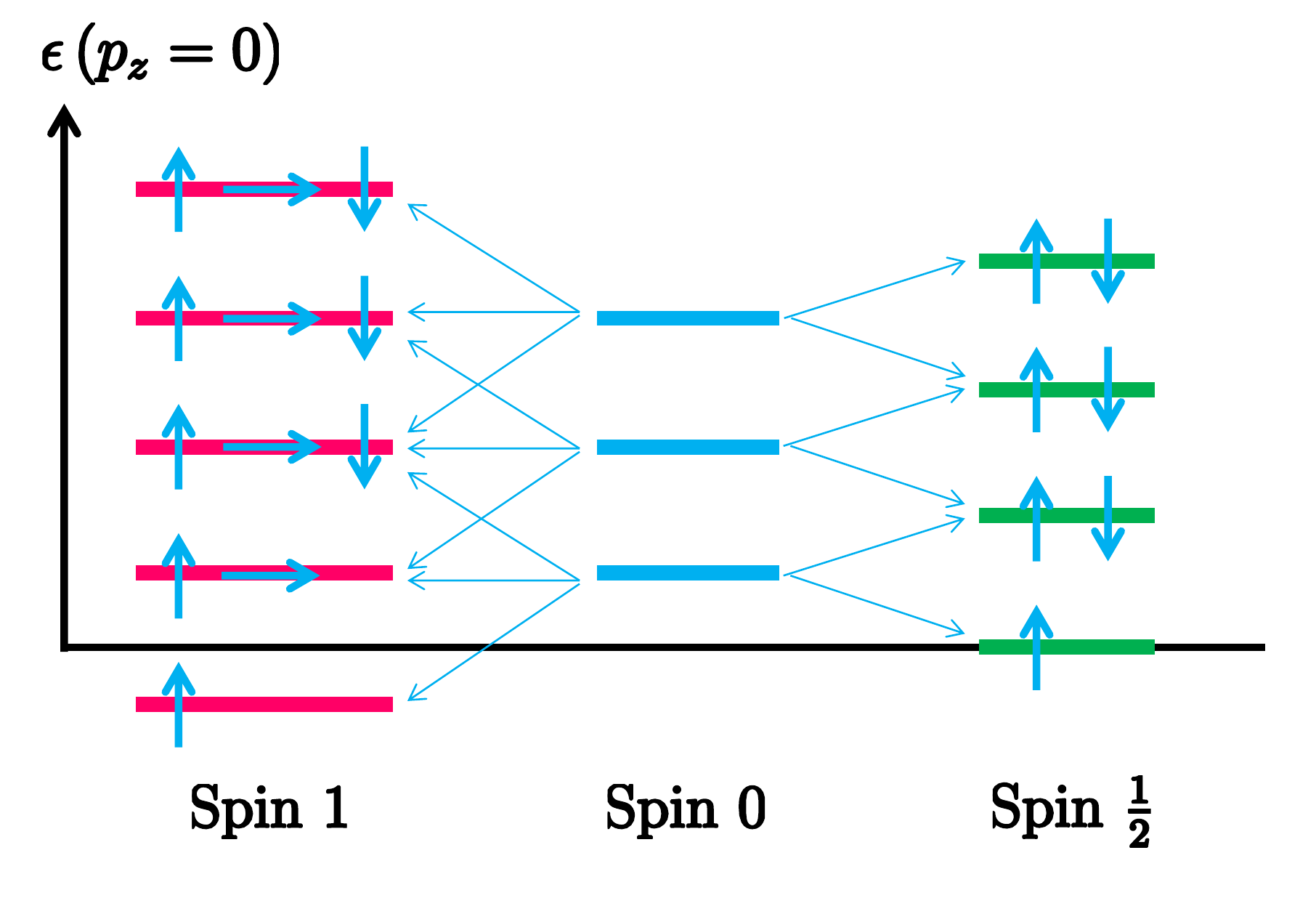}
     \end{center}
\vspace{-1cm}
\caption{ 
Landau-level discretization and Zeeman splitting with 
the vanishing longitudinal momentum $ p_z = 0 $. 
The ground states of spin-$  \frac{1}{2}$ and spin-$  1$ particles have unique spin directions, 
of which the spectra could become zero modes and tachyonic, respectively. 
}
\label{fig:Zeeman}
\end{figure}

We now include the Zeeman effect given by the last term in \eref{eq:Ham}, 
which resolves the spin degeneracy in each Landau level. 
Including the Zeeman splitting and longitudinal momentum in \eref{eq:Ham}, 
we obtain the energy spectrum 
for a spin-$  \frac{1}{2}$ particle 
\begin{eqnarray}
\epsilon_{\rm{spin}-\frac{1}{2}} =  \frac{p_z^2}{2m_f} + \hbar \omega_f n
\label{eq:energy-levels}
\, ,
\end{eqnarray}
with an integer $ n \geq0 $. 
Figure~\ref{fig:Zeeman} shows the resultant energy levels in Eq.~(\ref{eq:energy-levels}). 
The LLL has a unique spin direction, 
while there is still a spin degeneracy in each hLL due to the coincidence between the Landau-level spacing 
and the magnitude of the Zeeman effect for $  g =2$. 
The Zeeman energy cancels the ``zero-point energy'' 
in the Landau quantization when $g=2 $.

As for a spin-$  1$ particle, we have 
\begin{eqnarray}
\label{eq:spin1}
\epsilon_{\rm{spin}-1} =  \frac{p_z^2}{2m} + \hbar \omega_f \left( n - \frac{1}{2} \right)
\, ,
\end{eqnarray}
with an integer $ n \geq0 $. 
Similar to the spin-$ 1/2 $ particles, 
higher states are degenerated 
with respect to the spin directions, 
while the ground state is a unique spin eigenstate. 
The lowest energy takes a negative value, $ \epsilon_{\rm{spin}-1} = - \hbar \omega_f /2$.\footnote{ 
As we will see in Sec.~\ref{sec:gluon-prop} with the field-theoretical framework, 
the relativistic form of the energy spectrum is given by 
$  \epsilon_{\rm{spin}-1} =  \sqrt{ p_z^2  + m^2 + \hbar |q_fB| ( 2n -1 ) }$. 
When $ m^2 \gg p_z^2, |q_fB| $, the mass expansion leads to the real-valued spectrum (\ref{eq:spin1}). 
On the other hand, when $  \hbar |q_fB| > (p_z^2  + m^2) $, the low-energy spectrum becomes an imaginary number, 
$  \epsilon_{\rm{spin}-1} = i  \sqrt{  \hbar |q_fB| - (p_z^2  + m^2)  }$. 
A well-known example of such unstable modes 
is the Nielsen-Olesen instability \cite{Nielsen:1978rm} 
which occurs due to the self-interactions between 
the (massless) gauge bosons and external magnetic fields 
in non-Abelian theories, 
e.g., electroweak theory and QCD (cf. Sec.~\ref{sec:HE_QCD}). 
}

Summarizing this subsection, we have introduced 
the quantum-mechanical concept of the cyclotron center 
and the magnetic translation in the gauge-invariant way. 
One can draw two gauge-invariant conclusions. (i) The energy level is quantized into the Landau levels. 
Each Landau level is infinite-fold degenerated 
with respect to the center coordinate of the cyclotron motion 
with the density (\ref{eq:DoS-0}). 
(ii) The degenerate states can be labeled by any function of 
the conserved $\hat \bx_c$ according Eq.~(\ref{eq:H-center}). 
Those degenerate states are connected by the magnetic translation introduced in Eq.~(\ref{eq:magneticT}). 
However, the commutation relation (\ref{eq:xcyc}) indicates 
that only one of such functions can be a simultaneous energy eigenstate. 
The simplest choice would be the eigenstate of 
either $ \hat x_c$ or $\hat y_c$. 
Alternatively, one can take the eigenstate of 
the radius $ \hat r_c^2 $ defined in Eq.~(\ref{eq:label}). 
We will examine those two cases separately 
in the next subsection.

\subsection{Wave functions for the Landau levels} 

\label{sec:wavefunctions}

Here, we look for explicit forms of the wave functions, 
that is, the coordinate representation of 
the eigenvectors (\ref{eq:states}). 
As discussed in Sec.~\ref{sec:pseudomomentum}, 
the degenerate states are labeled 
by any function of $\hat \xc$ and $\hat \yc$. 
However, $\hat \xc$ and $\hat \yc$ do not commute 
with each other [cf. Eq.~(\ref{eq:xcyc})]. 
Specifying a function of the center coordinates is intimately, 
though {\it not} necessarily, related to the gauge choice. 
We will examine two choices of such functions, 
and choose a convenient gauge in each case 
to get an explicit form of the wave function.

\subsubsection{Landau gauge}

\label{sec:Landau-g}

We first choose $ \hat \xc$ as the label of 
the degenerate energy eigenstates that satisfy 
\begin{eqnarray}
\label{eq:center-xc}
\hat x_c \vert n,  \xc \rangle = \xc \vert n, \xc \rangle
\, ,
\end{eqnarray}
where $\xc $ is an eigenvalue. 
Note that making this choice itself is independent of the gauge choice 
and that the $y_c$ cannot be determined due to the quantum uncertainty (\ref{eq:xcyc}). 
One can write $\hat \xc $ as 
$ \hat x_c = \hat x + \frac{ 1}{q_f B} 
\{ \hat p_y - q_f   A_y (\hat \bx) \}$. 
This form suggests that a suitable gauge choice 
will be the ``Landau gauge'': 
\begin{eqnarray}
A_y = B \hat x 
\, , \quad A_t = A_x = A_z = 0
\label{eq:Landau-g}
\, ,
\end{eqnarray}
where the above relation reduces to 
\begin{eqnarray}
\hat x_c =   \frac{\hat p_y}{q_f B}
\, .
\end{eqnarray}
The canonical momentum $ \hat p_y$ is a constant of motion, 
while $\hat p_x$ is not conserved. 
Therefore, the operator $\hat p_y$ can be replaced by its eigenvalue as 
\begin{eqnarray}
\hat p_y \vert n, \xc \rangle = q_f B \xc \vert n, \xc \rangle
\label{eq:py}
\, .
\end{eqnarray}
This equality only holds in the Landau gauge. 
In this gauge, the energy eigenstate is labeled by the principle quantum number $  n$ and the conserved canonical momentum $ p_y$ 
that is equivalent to $ x_c$ chosen in Eq.~(\ref{eq:center-xc}). 
On the other hand, the gauge configuration apparently breaks the rotational symmetry 
around the direction of the magnetic field. 
The rotational symmetry should be restored 
in gauge-invariant quantities 
in the end of the day after correct computations.

\begin{figure}[t]
     \begin{center}
           \includegraphics[width=0.6\hsize]{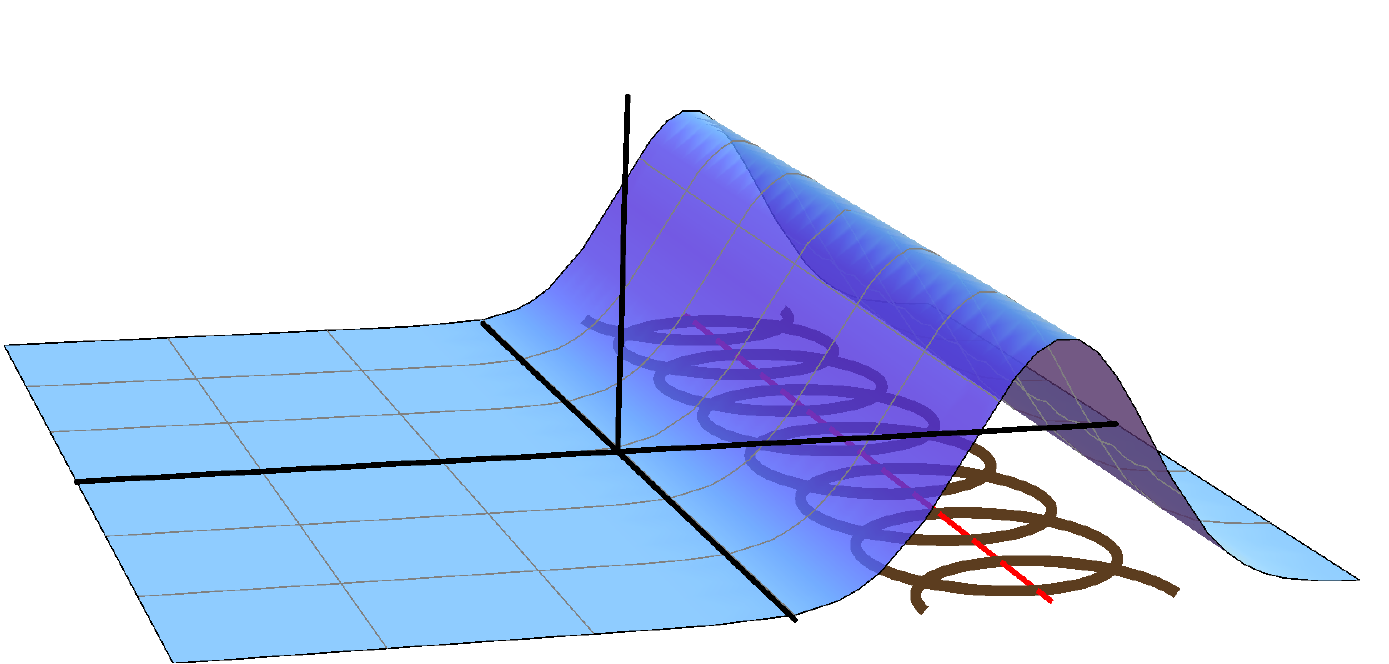}
     \end{center}
\caption{The LLL wave function in the Landau gauge composed of 
the superposition of cyclotron motions (circles) 
of which the center coordinates are aligned on a red line at $ x_c = p_y/( q_f B ) $. 
A similar picture holds in the higher Landau levels ($ n \geq1 $). 
}
\label{fig:L-gauge}
\end{figure}

Now that we have completely specified 
the labels of the energy eigenstates, 
we can obtain the wave function $ \phi_n (x, y) \equiv \langle x \vert n, x_c \rangle$. 
First, according to \eref{eq:py}, 
we can factorize the plane-wave component as 
\begin{eqnarray}
\phi_{n , p_y} (x, y) = e^{i \frac{ p_y y}{\hbar} } \tilde \phi_n(x)
\, ,
\end{eqnarray}
where $ p_y =  q_f B \xc$.  
In the previous subsection, we have already constructed the complete set of the eigenstates in \eref{eq:states},  
so that the remaining part $  \tilde \phi(x)$ can be obtained 
by using the explicit representations of the creation and annihilation operators in the Landau gauge. 
Following the detailed account in Appendix~\ref{sec:wf_Landau}, 
we obtain the wave function for the general Landau level $  n$: 
\begin{eqnarray}
\phi_{n , p_y} (\xi,y) = e^{i\frac{p_y y}{\hbar}} i^n \Ham_n (\xi) 
\label{eq:WF_Landau}
\, ,
\end{eqnarray}
where we have defined $ \xi = (x - x_c)/\ell_f$ and the normalized Hermite function 
\begin{eqnarray}
\label{eq:Hermite-f-p}
\Ham_n(\xi) &=& \sqrt{ \frac{1}{ 2^n n! \pi^{\frac{1}{2}} \ell_f} } e^{- \frac{\xi^2}{2} } H_n (\xi)
\, ,
\end{eqnarray}
with the Hermite polynomial $ H_n (x)  $ is defined as 
\begin{eqnarray}
\label{eq:H_poly}
 H_n(\xi) = ( -1)^n e^{ \xi^2 }  \frac{\partial^n \ }{\partial \xi^n } e^{ - \xi^2 }
 \, .
\end{eqnarray} 
This means that the wave function in the Landau gauge 
is given by that of the displaced harmonic oscillator with the center located at $ x_c$ 
(see Fig.~\ref{fig:L-gauge}). 
Some useful properties of the Hermite function are summarized as  
\begin{subequations}
\label{eq:Hermite-orthogonal}
\begin{eqnarray}
&&
\int \!\!  dx \, \Ham_m (\xi) \Ham_n (\xi) = \delta_{m,n}
\, ,
\\
&&
\sum_n  \Ham_n (\xi) \Ham_n (\xi^\prime) = \ell_f^{-1} \delta( \xi - \xi^\prime)
\, ,
\\
&&
\int \!\!  dx \, x\, \Ham_m (\xi) \Ham_n (\xi) 
= \ell_f \Big[ \sqrt{\frac{n}{2}} \delta_{m,n-1} + \sqrt{\frac{n+1}{2}} \delta_{m,n+1}  \Big]
\, ,
\\
&&
\partial_x \Ham_n(\xi) = \ell_f^{-1} \Big[  \sqrt{\frac{n}{2}} \Ham_{n-1} (\xi) 
- \sqrt{\frac{n+1}{2}} \Ham_{n+1} (\xi)   \Big]
\, .
\end{eqnarray}
\end{subequations}
Note a difference between $  x$ and $\xi  $ which have different mass dimensions.

As already discussed in the gauge-invariant manner, 
the density of degenerate states can be counted in a finite box with the side $ L_x $ in the $  x$ direction. 
Since $0 \leq  \xc \leq L_x $, we have $ 0 \leq p_y \leq q_f B L_x$ when $s_f >0  $, 
and $ - | q_f B| L_x \leq p_y \leq 0$ when $ s_f <0 $. 
Therefore, we get the density of states 
\begin{eqnarray}
\frac{1}{ L_x} \int_0^{|q_f B| L_x} \!\! \frac{dp_y}{2\pi \hbar }  = \frac{|q_f B|}{2\pi\hbar}
\, .
\label{eq:DoS}
\end{eqnarray}
This expression agrees with that in Eq.~(\ref{eq:DoS-0}). 
Here, there is no $ p_x $ integral since $ p_x $ is not a quantum number specifying our phase space.

\subsubsection{Symmetric gauge}

\label{sec:S-gauge}

\cout{ 

We shall examine another way of labeling 
the degenerate states with the radius of 
the center coordinate $\hat r_c^2 $ 
defined in Eq.~(\ref{eq:label}). 
This is a distance between the origin of the coordinate system 
and the center of the cyclotron orbit. 
As in Eq.~(\ref{eq:center-xc}), making this choice itself is independent of the gauge choice 
since the commutation relations (\ref{eq:H-center}) with the Hamiltonian are gauge-invariant properties. 
Led by the canonical commutation relation (\ref{eq:xcyc}) 
between $ \hat \xc$ and $\hat \yc $, 
we construct another set of the creation and annihilation operators 
\begin{eqnarray}
\label{eq:b-operators}
\hat b = \frac{1}{\sqrt{2}\ell_f} ( \hat x_c - i s_f \hat y_c) 
\, , \quad 
\hat b^\dagger = \frac{1}{\sqrt{2}\ell_f} ( \hat x_c + i s_f \hat y_c) 
\, ,
\end{eqnarray}
which satisfy the commutation relation $ [ \hat b , \hat b^\dagger] = 1$. 
They commute with $\hat a$ and $\hat a^\dagger $. 
The radial coordinate is expressed by the ``number operator'' as 
\begin{eqnarray}
\hat r_c^2 = 2 \ell_f^2 \left( \,\hat  b^\dagger \hat b + \frac{1}{2} \, \right)
\label{eq:bbdagger}
\, .
\end{eqnarray}
Therefore, the degenerate states are labeled by 
an integer $ m $ that is the eigenvalue of $\hat  b^\dagger \hat b $.\footnote{
We follow this frequently used notation for $ m $, 
which should not be confused with a mass parameter. 
} 
One can construct the complete set of the energy eigenstates by starting out 
with the ground state annihilated by both of the annihilation operators 
\begin{eqnarray}
 \hat a \vert 0 , 0 \rangle = 0 
 =  \hat b \vert 0 , 0 \rangle 
 \label{eq:S-00}
 \, .
\end{eqnarray}
The other states can be obtained by operating the creation operators as 
\begin{eqnarray}
\phi_{nm} \equiv \langle \bx \vert n , m \rangle = 
\langle \bx \vert \frac{\big(\hat a^\dagger\big)^n }{\sqrt{n!}}
\frac{\big(\hat b^\dagger\big)^m}{\sqrt{m!}} \vert 0 , 0 \rangle
\label{eq:S-nm}
\, .
\end{eqnarray}
The above algebraic properties are gauge-invariant ones. 

} 

We shall examine another way of labeling 
the degenerate states by the eigenvalue of $\hat r_c^2 $ 
which is the distance between the cyclotron center 
and the coordinate origin as defined in Eq.~(\ref{eq:label}). 
In Eq.~(\ref{eq:bbdagger}), we have found that 
the eigenvalue of $\hat r_c^2 $ is quantized with 
the creation and annihilation operators, 
$\hat b $ and $\hat b^\dagger $, in Eq.~(\ref{eq:b-operators}). 
Therefore, one can construct the complete set of the energy eigenstates by starting out with the ``ground state'' 
annihilated by both of the annihilation operators 
\begin{eqnarray}
 \hat a \vert 0 , 0 \rangle = 0 
 =  \hat b \vert 0 , 0 \rangle 
 \label{eq:S-00}
 \, .
\end{eqnarray}
The other states can be obtained by operating the creation operators as 
\begin{eqnarray}
\phi_{nm} \equiv \langle \bx \vert n , m \rangle = 
\langle \bx \vert \frac{\big(\hat a^\dagger\big)^n }{\sqrt{n!}}
\frac{\big(\hat b^\dagger\big)^m}{\sqrt{m!}} \vert 0 , 0 \rangle
\label{eq:S-nm}
\, .
\end{eqnarray}
Note that operating $\hat b, \hat b^\dagger $ does not change 
the Landau level $n $ as $\hat a, \hat a^\dagger $ 
and $\hat b, \hat b^\dagger $ commute with each other. 
Also, operating $\hat a, \hat a^\dagger $ does not change 
the degeneracy label $m $. 
As in Eq.~(\ref{eq:center-xc}), making this choice itself is independent of the gauge choice 
since the commutation relations (\ref{eq:H-center}) with the Hamiltonian are gauge-invariant properties. 
The algebraic properties for $\hat b , \hat b^\dagger $ 
as well as for $\hat a , \hat a^\dagger $ are 
gauge-invariant as well.  
Remember also that, in the gauge-invariant language of 
the pseudo angular momentum $\hat \bK $, 
the quantization of $\hat r_c^2 $, which is 
found to be proportional to $ \hat K_z$, 
is led by the ladder operators (\ref{eq:K-ladder}) 
that connects the degenerate states. 
One needs to specify the gauge only when getting 
an explicit form of the wave function $ \phi_{nm}  $ 
from the coordinate representations of 
the creation and annihilation operators.  
The coordinate representations of the creation and annihilation 
operators depend on the gauge choice.

\begin{figure}[t]
     \begin{center}
           \includegraphics[width=0.5\hsize]{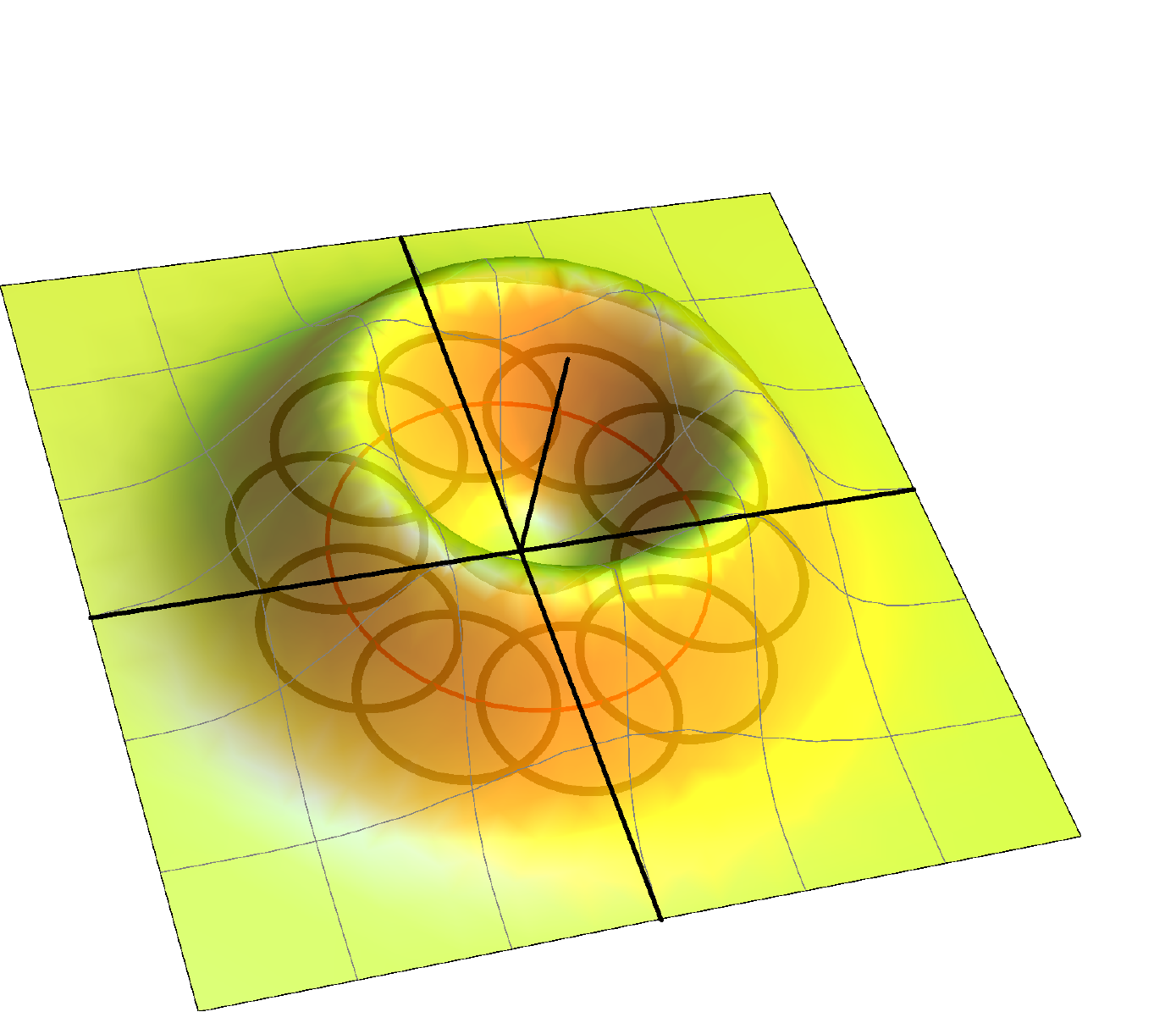}
     \end{center}
\caption{The LLL wave function in the symmetric gauge composed of 
superposition of cyclotron motions (circles), of which the center coordinates are aligned on a red circle. 
The radius of each cyclotron motion and the location of the center coordinate 
are given by $R^2 = (2n+1) \ell_f^2 $ and $r_c^2 = (2m+1) \ell_f^2  $, respectively.}
\label{fig:S-gauge}
\end{figure}

It is also straightforward to show that the canonical angular momentum operator (\ref{eq:canonical-AM}), 
$\hat L_z = \hat x \hat  p_y - \hat y \hat p_x$, 
can be written by $\hat  \bx_c$ and $ \hat \bpi$ 
in a general gauge as 
\begin{eqnarray}
\hat L_z = \frac{ s_f}{ \omega_f} \left(  \frac{1}{2} \mass  (\hat r_c \omega_f)^2 -  \hat \Ham_\perp \right) 
+q_f  \left[ \,  \{ \, \hat x A_y (\hat \bx) - \hat y A_x(\hat \bx) \, \}  - \frac{1}{2} B \hat r^2 \, \right]
\label{eq:Lz1}
\, ,
\end{eqnarray}
where $\hat r^2 \equiv \hat x^2 + \hat y^2$ 
and a mass parameter is denoted as $ \mass $ in this subsection 
(to avoid a possible confusion with the quantum number $  m$). 
The first two terms between the parentheses are diagonalized 
by the creation and annihilation operators as discussed above. 
However, those terms between the square brackets do not 
commute with the Hamiltonian in a general gauge.
Nevertheless, it is clear from the symmetry argument that, if one takes a rotationally symmetric gauge, 
the canonical angular momentum is also a simultaneous eigenstate of the Hamiltonian. 
Therefore, one can make a suitable choice with the so-called ``symmetric gauge'' 
\begin{eqnarray}
A_x = - \frac{1}{2} B \hat y \, , \quad A_y =  \frac{1}{2} B \hat x 
\, , \quad A_t = A_z = 0
\label{eq:symmetric-g}
\, .
\end{eqnarray}
In this gauge, the last term in Eq.~(\ref{eq:Lz1}) cancels away,
and the canonical angular momentum is diagonalized as 
\begin{eqnarray}
\hat L_z = s_f \hbar \left( \hat b^\dagger \hat b - \hat a^\dagger \hat a \right) 
\, .
\end{eqnarray}
The canonical angular momentum is a gauge-dependent quantity, and is not an observable quantity 
as clear from its definition by the canonical momentum. 
We find an eigenvalue of the angular momentum as 
\begin{eqnarray}
\hat L_z \vert n, m \rangle = s_f  \hbar   ( m - n) \vert n, m \rangle 
\label{eq:Lz2}
\, .
\end{eqnarray}
For each Landau level, 
the angular momentum $\ell \equiv  m - n $ takes 
an integer from $- n$ to the positive infinity $( \ell \geq - n) $, 
according to the lower bounds, $m, n \geq 0 $. 
Since the integer $ m$ specifies the center of the cyclotron motion, 
a larger angular momentum at a fixed $ n$ corresponds to an outer orbit 
that is more distant from the origin of the coordinate system (cf. \fref{fig:S-gauge}).


Following detailed descriptions in Appendix~\ref{sec:wf_symm}, 
the coordinate representation of the wave function is obtained as 
\begin{eqnarray}
\label{eq:wf-sym}
\phi_{nm} (r, \theta) 
&=& C_S   (- i)^{ n  }  e^{i s_f \ell \theta}  
\sqrt{\frac{n!}{m!}} e^{- \frac{r^2}{4\ell_f^2} } \left(\frac{r^2}{2\ell_f^2}\right)^{\frac{ \ell }{2}}  
L_{ n } ^{ \ell } \Big(\frac{r^2}{2\ell_f^2} \Big)
\nn
\\
&=&C_S \,   i^n   (-1)^{ \min(n,m) } e^{i s_f \ell \theta} 
\sqrt{\frac{n!}{m!}}^{ \, \sgn(\ell)}  
e^{- \frac{r^2}{4\ell_f^2} } \left(\frac{r^2}{2\ell_f^2}\right)^{\frac{\vert \ell \vert}{2}} 
L_{ \min(n,m ) } ^{\vert \ell \vert} \Big(\frac{r^2}{2\ell_f^2} \Big)
\label{eq:wavefunction-symmetric}
\, ,
\end{eqnarray}
with the associated Laguerre polynomial $ L_k^\alpha(z) $ 
defined as 
\begin{eqnarray}
\label{eq:Laguerre}
L_k^\alpha(\rho) = \frac{\rho^{-\alpha} e^\rho }{k!} \frac{ \partial^k \ }{\partial  \rho^k } 
\left( \, \rho^{k+\alpha} e^{-\rho} \, \right)
\, .
\end{eqnarray}
The radial coordinate and the normalization constant are denoted as $ r =\sqrt{x^2+y^2} $ 
and $  C_S= (2\pi\ell_f^2)^{-\frac{1}{2}} $, respectively. 
In the second line of Eq.~(\ref{eq:wavefunction-symmetric}), 
we used the formula (\ref{eq:Laguerre-negative}) 
to have a positive index in the upstairs of the Laguerre polynomial 
and promise to take the smaller integer in $ \min(n,m ) $. 
Especially, the LLL wave function is found to be 
\begin{eqnarray}
\phi_{0m}(r, \theta) =  C_S \, e^{i s_f m \theta}  \sqrt{ \frac{1}{ m!} } \, 
e^{ - \frac{r^2}{4\ell_f^2}}  \left( \frac{r^2 }{ 2\ell_f^2 } \right)^{\frac{m}{2}} 
\, .
\end{eqnarray}
Notice that the last factor can be any positive power of $ r $. 
Therefore, any linear combination of those wave functions, 
that is a product of any polynomial of $ r$ and the Gaussian, 
serves as the wave function of the LLL.

When the transverse area of the system is given by $  S_\perp$, 
the quantum number $  m$ is bounded according to an inequality $ S_\perp \geq \pi r_c^2  $. 
From this inequality, we immediately find the density of degenerate states 
\begin{eqnarray}
\frac{m_{\rm max}}{S_\perp} \sim \frac{|q_f B|}{2\pi \hbar}
\, .
\end{eqnarray}
Here, we assumed $ m_{\rm max} \gg 1 $ or, in other words, $ S_\perp \gg \pi(2 \ell_f^2) $ 
where the latter is the area enclosed by each cyclotron orbit. 
Although the spatial profiles of the wave functions and the labels of 
the degenerate states strongly depend on the gauge choice, 
the density of states is a gauge-invariant quantity [cf. Eqs.~(\ref{eq:DoS-0}) and (\ref{eq:DoS})].

\cout{

\subsection{Drift motion in crossed magnetic and electric fields}

\label{sec:E-effect}

\begin{figure}[bt]
     \begin{center}
           \includegraphics[width=0.5\hsize]{drift}
     \end{center}
\caption{
Classical trajectories of positive and negative charges that are drifting in the orthogonal electric and magnetic fields 
applied in the positive $ x $- and $  z$-directions, respectively. 
The initial velocity is taken to be $ v_x(0)=0 $ and $ v_y(0) >0 $. 
The direction of the drift motion is independent of the signs of electric charges, 
and only depends on the direction of the cross product $ \bE \times \bB $. 
}
\label{fig:drift}
\end{figure}

Including a potential term for an electric field, the Hamiltonian is given by 
\begin{eqnarray}
\hat \Ham_\perp = \frac{1}{2m} ( \hat \pi_x^2 + \hat \pi_y^{ 2} ) - q_f E \hat x
\label{eq:Ham_E}
\, ,
\end{eqnarray}
where the momenta $ \hat \pi_{x,y} $ are the same as in the previous subsections. 
When the electric field is applied in perpendicular to the magnetic field, 
a charged particle is subject to the drift motion, 
of which the velocity is given by $ \bv_{\rm drift} = \bE \times \bB/ |\bB|^2  $ 
independently of the electric charge (cf. Fig.~\ref{fig:drift}).

Since the Hamiltonian (\ref{eq:Ham_E}) explicitly depends on 
one of the transverse coordinate $ \hat x $, 
the number operator (\ref{eq:CA}) and $ \hat y_c $ no longer 
commute with the Hamiltonian, $ [\hat a^\dagger \hat a, \hat \Ham_\perp] \not = 0$ 
and $ [\hat y_c, \hat \Ham_\perp] \not = 0$. 
The latter implies that the degenerate states having been specified by 
the eigenvalues of $  \hat y_c$ are resolved by effects of the electric field. 
Nevertheless, one still observes the cyclotron motion in the comoving frame of the drift motion. 
Therefore, let us subtract the momentum of the drift motion as ($ v_{\rm drift} = - E/B $) 
\begin{eqnarray}
\hat \pi_y ^\prime = \hat \pi_y - m   v_{\rm drift}
\label{eq:drift-shift}
\, .
\end{eqnarray}
Then, the Hamiltonian reads 
\begin{eqnarray}
\hat \Ham_\perp = \frac{1}{2m} ( \hat \pi_x^2 + \hat \pi_y^{\prime\, 2} ) + \frac{1}{2}m v_{\rm drift}^2 
+ q_f E (  \hat x_c - \frac{ m  v_{\rm drift}}{q_f B}  )
\, .
\end{eqnarray}
The cyclotron center $ \hat x_c $ still commutes with both of 
$ \hat \pi_x $ and $\hat \pi_y^{\prime}  $, and thus with the Hamiltonian.  
Furthermore, the transverse momenta $ \hat \pi_x $ and $\hat \pi_y^{\prime}  $ serve as a canonical pair. 
Therefore, the Fock space can be spanned by 
the eigenstates of the $ \hat x_c $ and of the number operator $\hat a^{\prime \dagger} \hat a^\prime  $ 
constructed with $ \hat \pi_x $ and $\hat \pi_y^{\prime}  $, 
so that the gauge-invariant discussion in Sec.~\ref{sec:LL_1} still holds 
with the replacement of $ \hat \pi_y $ by $\hat \pi_y^\prime  $. 
By the use of those new creation and annihilation operators, 
the energy eigenvalues are obtained as 
\begin{eqnarray}
\langle  n^\prime , x_c   | \hat \Ham_\perp | n^\prime , x_c \rangle 
=  \omega_f ( n^\prime + \frac{1}{2} ) + \frac{1}{2}m v_{\rm drift}^2  + q_f E x_c^\prime  
\, ,
\end{eqnarray}
where a non-negative integer $ n^\prime $ is an eigenvalue of 
the number operator $\hat a^{\prime \dagger} \hat a^\prime  $ 
and the cyclotron center is shifted as $  x_c^\prime = x_c -m v_{\rm drift}/(q_fB) $. 
The first term is for the discrete spectrum of the Landau levels, 
and the second and third terms are for the kinetic and potential energies of the drift motion, respectively. 
As already clear in the above discussion, 
the momentum perpendicular to the electric field acquires a finite expectation value 
\begin{eqnarray}
\langle n^\prime | \hat \pi_y | n ^\prime\rangle
= \langle n^\prime | (\hat \pi_y^\prime  + m v_{\rm drift}) | n^\prime \rangle
=  m v_{\rm drift} 
\, ,
\end{eqnarray}
while it was vanishing in Sec.~\ref{sec:Landau-g}. 
The expectation value of the other component, which is perpendicular to the drift motion, 
is still vanishing, i.e., $ \langle n^\prime | \hat \pi_x | n ^\prime\rangle = 0$. 
Those expectation values are independent of the Landau level $ n' $.

\cgd{
We have not chosen any gauge choice in the above discussion. 
A specific gauge needs to be chosen only for an explicit form of the wave function. 
}
Clearly, the electric field specified above breaks 
the rotational symmetry with respect to the direction of the magnetic field 
and the translational symmetry along the electric field. 
On the other hand, the system still has a translational invariance in 
the direction perpendicular to both of the fields. 
We thus choose the Landau gauge (\ref{eq:Landau-g}) that preserves the remaining translational invariance. 
In this gauge, the shift of the momentum (\ref{eq:drift-shift}) 
can be absorbed into a shift of the cyclotron center $ x_c \to x_c^\prime$ 
in the creation and annihilation operators. 
Therefore, the wave function is obtained in the same form as in Eq.~(\ref{eq:WF_Landau}) 
with the replacement of $ x_c $ by $ x_c^\prime = (p_y -m v_{\rm drift})/(q_fB) $.


Those observations are important for understanding the (integer) quantum Hall effect~\cite{PhysRevB.23.5632, PhysRevB.25.2185}. 
The interested reader is referred to the literature, e.g., Refs.~\cite{yoshioka2013quantum, Tong:2016kpv}. 
\com{[Expand this paragraph a little bit.]}

}

\subsection{Relativistic fermions in a magnetic field}

\label{sec:Ritus-Feynman}

In the previous subsection, 
we have examined the Landau quantization 
in terms of nonrelativistic quantum mechanics. 
In this section, we proceed to a field theoretical framework 
for relativistic fermions in magnetic fields. 
Specifically, we will introduce 
the Ritus-basis method \cite{Ritus:1972ky, Ritus:1978cj} 
to find the eigenspinor of the Dirac operator in magnetic fields. 
Hereafter, we work in natural units.

\subsubsection{Dirac equation in magnetic fields}

\label{sec:relativistic-fermion}

Interactions between relativistic fermion and an external electromagnetic field are described by the spinor QED Lagrangian: 
\begin{eqnarray}
\Lag = \bar \psi \left( i \slashed D - m \right) \psi
\label{eq:Lqed}
\, .
\end{eqnarray}
Here, we use the following convention of the covariant derivative 
\begin{eqnarray}
D^\mu = \partial ^\mu + i q_f A^\mu(x)
\label{eq:covariantD-QED}
\, .
\end{eqnarray}
The Abelian gauge field $  A^\mu(x)$ is for an external field, and we do not include a dynamical photon (see, e.g., Ref.~\cite{Hattori:2020htm}). 
The Euler-Lagrange equation results in the Dirac equation in the external field: 
\begin{eqnarray}
\left( i \slashed D - m \right) \psi = 0
\label{eq:Dirac}
\, .
\end{eqnarray}
By using an identity, 
$\gamma^\mu \gamma^\nu = \frac{1}{2} [\gamma^\mu, \gamma^\nu] 
+ \frac{1}{2} \{ \gamma^\mu, \gamma^\nu \}$, 
we get 
\begin{eqnarray}
\left( D^2 + m^2 + \frac{q_f}{2} F^{\mu\nu} \sigma_{\mu\nu}\right) \psi = 0
\label{eq:K-GandZeeman}
\, ,
\end{eqnarray}
with $\sigma_{\mu\nu}=\frac{i}{2}[\gamma_\mu, \gamma_\nu]$.

In the presence of a magnetic field in the $  z$-direction, 
we have a non-vanishing commutation relation for the transverse components of the covariant derivative 
\begin{eqnarray}
[ i D^1 , iD^2] = - i q_f F^{12} = iq_fB
\, .
\end{eqnarray}
One can use the covariant derivative 
as a pair of the ``canonical variables'' 
as in Eq.~(\ref{eq:canonical_B}) for nonrelativistic quantum mechanics. 
Therefore, one can apply the techniques discussed 
in the previous subsection. 
Identifying the variables in the present and previous discussions as $ i D^i \leftrightarrow \hat \pi^i $, 
one can define the ``creation and annihilation operators'' for the Landau levels [cf. Eq.~(\ref{eq:CA})]: 
\begin{eqnarray}
\label{eq:CA-rela}
\hat a = \frac{1}{ \sqrt{2 |q_f B|}} ( i D^1 - s_f  D^2)
\, , \quad
\hat a^\dagger = \frac{1}{\sqrt{2 |q_f B|}} ( i D^1 + s_f  D^2)
\, .
\end{eqnarray}
Then, we find a simple relation 
\begin{eqnarray}
D^2 + m^2  =  \partial_t^2 - \partial_z^2 + \left(2 \hat a^\dagger \hat a +1\right)  \vert q_f B \vert + m^2 
\, .
\end{eqnarray}
Here and below, the creation and annihilation operators, $ \hat a^\dagger $ and $ \hat a $, should be understood 
as their coordinate representations and act on the wave function $\phi_n(x)  $ in the coordinate representation. 
The above Klein-Gordon operator represents 
the relativistic Landau quantization 
for charged scalar particles.

One may easily guess that the remaining term $  \frac{q_f}{2} F^{\mu\nu} \sigma_{\mu\nu}$ 
in Eq.~(\ref{eq:K-GandZeeman}) is responsible for the Zeeman effect. 
To explicitly see this, we shall introduce 
spin projection operators\footnote{
These operators have useful properties: 
$  \prj_\pm^\dagger = \prj_\pm$, $ \prj_+ + \prj_- = 1 $, 
$ \prj_\pm \prj_\pm = \prj_\pm $, and $\prj_\pm \prj_\mp =0  $. 
Therefore, one also finds that $\prj_\pm \gam^\mu \prj_\pm = \gam_\para^\mu \prj_\pm  $ 
and $\prj_\pm \gam^\mu \prj_\mp = \gam_\perp^\mu \prj_\mp  $, 
which will be useful for diagrammatic calculations. 
} 
\begin{eqnarray}
\label{eq:spin-projection}
 \prj _\pm = \frac{1}{2} \left(1 \pm  i s_f \gam^1 \gam^2 \right)
 \, .
\end{eqnarray} 
In the standard (Dirac or Weyl) representation, they are expressed as 
\begin{eqnarray}
 \prj _\pm 
    = \frac{1}{2}
 \begin{pmatrix}
 {1 \pm s_f \sigma_z} & {} 
 \\
 {} & {1 \pm s_f \sigma_z} 
 \end{pmatrix}  
 \, ,
\end{eqnarray} 
with the Pauli matrix $  \sigma_z = {\rm diag}(1,-1) $. 
Then, we indeed find a spin decomposition 
\begin{eqnarray}
\frac{q_f}{2} F^{\mu\nu} \sigma_{\mu\nu}  =  \vert q_f B \vert ( - \prj_+ + \prj_-)
\, .
\end{eqnarray}
Therefore, the (squared) energy of the spin state $\psi_+ \equiv \prj_+ \psi$ 
decreases by $ \vert q_f B \vert $, 
while that of the opposite spin state $ \psi_- \equiv \prj_- \psi $ increases by the same amount. 
This is nothing but the Zeeman effect for a spin-1/2 particle. 
The energetically favored spin direction depends on 
the direction of the magnetic field 
and the electric charge, 
which is thus encoded in the sign function $ s_f =\sgn(q_f B)$ in $ \prj_\pm $. 
Including the Zeeman splitting term, we have 
\begin{eqnarray}
 \left[\, \partial_t^2 - \partial_z^2 + \left(2 \hat a^\dagger \hat a  + 1 \mp 1\right)  \vert q_f B \vert  + m^2 \, \right] \psi_\pm = 0
 \, .\label{eq:rel_EoM}
\end{eqnarray}
Note that the $ g$-factor in the Dirac equation is $ g=2$, 
up to higher-order corrections by interactions. 
The zero-point energy in the LLL is canceled by the Zeeman shift as in Fig.~\ref{fig:Zeeman} for nonrelativistic fermions.

In the previous subsections, we have already obtained the transverse wave functions in magnetic fields such that 
$\phi_{n  ,\qn} (x_\perp)=\langle x_\perp | n , \qn \rangle$ with $\hat a^\dagger \hat a |n , \qn \rangle =n |n, \qn\rangle$.\footnote{ 
Recall that the wave functions were denoted as $ \phi_{n,p_y} $ and $\phi_{nm}  $ 
in the Landau and symmetric gauges, respectively. 
Here, $ \phi_{n  ,\qn}  $ is a generalized notation for an arbitrary gauge. 
The quantum numbers specifying a state $  |n , \qn \rangle $ are denoted as subscripts of the wave function $ \phi_{n  ,\qn}  $ 
regardless of whether they are discrete or continuous ones, while the coordinate basis is denoted as the argument. 
}
Recall that this ``number operator'' has been defined in a gauge-invariant manner 
and its eigenvalue $ n $ corresponds to the Landau levels. 
The longitudinal part is the same as the (1+1)-dimensional free theory.  
Therefore, one can write the eigenfunction in a factorized form 
$ \psi \propto e^{-ip_\para  x} \phi_{n, \qn}(x_\perp) $ 
to find the relativistic dispersion relation 
\begin{eqnarray}
\ep_n = \pm \sqrt{ p_z^2 + 2  n \vert q_f B \vert + m^2  }
\label{eq:fermion-rela}
\, .
\end{eqnarray}
The upper and lower signs are for the positive and negative energy solutions, respectively, 
and the non-negative integer $ n $ is the resultant quantum number 
after the sum of the Landau level and the Zeeman shift. 
Similar to the nonrelativistic fermions discussed in Sec.~\ref{sec:zeeman-effect}, 
the LLL has a unique spin direction, 
while the higher levels ($n\geq1  $) are two-fold degenerated 
with respect to the spin directions (cf. Fig.~\ref{fig:Zeeman}). 
One can reproduce the nonrelativistic expression 
as an expansion in the limit, $m^2 \gg p_z^2, \ |q_fB|  $. 
While the Landau-level spacing is given by 
the cyclotron frequency $ \omega_c $ in the nonrelativistic case, 
the relativistic energy levels are not equally spaced. 
The level spacing is now of the order of $ \sqrt{q_fB} $.

\subsubsection{Mode expansion with the Ritus basis}

Next, we examine the Dirac spinor structure of $ \psi $. 
As clear from the above discussion, the spin projection operators 
provides an appropriate basis for the solution of the Dirac equation as 
\begin{eqnarray}
\Ritus_{n, \qn} (x_\perp) =  \phi_{n, \qn} (x_\perp) \prj_+ +  \phi_{n-1 ,\qn} (x_\perp) \prj_- 
\label{eq:Ritus}
\, ,
\end{eqnarray}
where $  \phi_{-1} \equiv 0 $ is understood. 
This expression reflects the two-fold spin degeneracy in the hLLs, 
while the uniqueness in the LLL is guaranteed by the prescription $\phi_{-1} = 0$. 
This is called the {\it Ritus basis}, which was proposed for computation of 
the fermion self-energy in external fields \cite{Ritus:1972ky, Ritus:1978cj}. 
By the use of the Ritus basis, one may put an Ansatz 
\begin{eqnarray}
\psi (x) =   e^{ - i p_\para x} \Ritus_{n,\qn} (x_\perp)  \, u
\label{eq:Dirac-Ritus}
\, ,
\end{eqnarray}
where $ u $ is a four-component spinor to be determined by the Dirac equation. 
In both the Landau and symmetric gauges, 
we have $A_0=A_3=0$, so that $i \sla D =i \sla \partial_\para -iD^1 \gamma^1 -iD^2 \gamma^2$. 
Then, by the use of the $\hat a, \ \hat a^\dagger  $ and $ \prj_\pm $, the Dirac operator is cast into a more convenient form 
\begin{eqnarray}
\label{Dirac_Ritus}
i \sla D  =  i \sla \partial _\para- \sqrt{2|q_fB|} \  \gam^1 \left(\hat a \prj_+ + \hat a^\dagger \prj_- \right)
\, . 
\end{eqnarray}
Acting the Dirac operator on the above Ansatz, we find that 
\begin{eqnarray}
i \sla D   \psi(x) = e^{ - i p_\para x}   \Ritus_{n,\qn} (x_\perp)  
\Big( \sla p _\para - \sqrt{2n |q_fB| }\ \gam^1  \Big) \, u
\, .\ \label{Ritus_u}
\end{eqnarray}
The ansatz (\ref{eq:Dirac-Ritus}) provides us  
with a solution for the Dirac equation (\ref{eq:Dirac}) 
if the spinor $ u $ satisfies the ``free'' Dirac equation 
\begin{eqnarray}
 ( {\sla p}_n -m ) u ( p_n) = 0
 \label{eq:free}
 \, ,
\end{eqnarray}
with the four momentum $ p^\mu_n \equiv ( \ep_n , \sqrt{2n |q_fB| },0,p^3) $. 
This is one of the most fundamental equations in quantum field theory, 
and the solutions are available in numerous textbooks, e.g., Ref.~\cite{Peskin:1995ev}. 
This equation has two solutions corresponding to 
the two ``spin states'' labeled as $ u^\kappa ( p_n)  $.\footnote{ 
The general solution for the Dirac equation (\ref{eq:free}) can be written as 
$ u^\kappa(p_n) = ( \sqrt{p_n \cdot \sigma} \, \xi^\kappa,  \sqrt{p_n \cdot \bar \sigma}  \, \xi^\kappa) $ 
with $  \sigma^\mu = (1, {\bm \sigma})$, $\bar  \sigma^\mu = (1, - {\bm \sigma})  $, 
and a still arbitrary spinor $ \xi^\kappa $, according to $ (p_n \cdot \sigma)(p_n \cdot \bar \sigma) = p_n^2 = m^2 $ \cite{Peskin:1995ev}. 
Precisely speaking, the ``spin label $\kappa$'' in the notations of Ref.~\cite{Peskin:1995ev} refers to an arbitrary spinor basis $ \xi^\kappa$. 
While $  \xi^\kappa$ can be chosen as an eigenstate of 
arbitrary operator, e.g., a spin operator $ \sigma_z $, 
the solutions $  \sqrt{p_n \cdot \sigma} \, \xi^\kappa$ and $  \sqrt{p_n \cdot \bar \sigma}  \, \xi^\kappa $ 
do not remain eigenstates of the same operator if it does not commute with $ {\bm \sigma} $. 
In the present case, the spinors are eventually projected with $ \prj_\pm $ in the Ritus basis $ \Ritus_{n,\qn}(x;p) $ 
whatever the basis is chosen. 
}
Similarly, we can put an ansatz for the negative-energy solution: 
\begin{eqnarray}
\psi (x) =    e^{ i p_\para x}  \Ritus_{n,\qn} (x_\perp)    \, v
\label{eq:Dirac-Ritus_v}
\, ,
\end{eqnarray}
where the sign of the longitudinal momentum $p_\para ^\mu $ is flipped. 
Inserting it into the Dirac equation (\ref{eq:Dirac}), we have 
\begin{eqnarray}
i \sla D \psi(x) =  
-  e^{ i p_\para x}  \Ritus_{n,\qn} (x_\perp)   \Big(  \sla p _\para +  \sqrt{2n |q_fB| }\ \gam^1 \Big) \, v
\, .\ \label{Ritus_v}
\end{eqnarray} 
Therefore, the ansatz (\ref{eq:Dirac-Ritus_v}) provides us  
with the solution for the Dirac equation (\ref{eq:Dirac}) 
if the spinor $ v $ satisfies the ``free'' Dirac equation 
\begin{eqnarray}
 ( \bar {\sla p}_n + m ) v (\bar p_n) = 0
 \label{eq:free_v}
 \, .
\end{eqnarray}
We defined $ \bar p^\mu_{ n} \equiv ( \ep_n , - \sqrt{2n |q_fB| },0,p^3) $ that needs to be distinguished from $ p_n^\mu $ in Eq.~(\ref{eq:free}): 
Since the sign of $ p_\para^\mu $ has been flipped in the above, 
$ \bar p^1_{ n}  $ has a relative minus sign against those longitudinal components. 
Those spinors satisfy the useful relations \cite{Peskin:1995ev}
\begin{eqnarray}
\label{eq:spin-sum}
\sum_{\kappa=\pm}  u^\kappa (p_n)  \bar u^\kappa (p_n) = ( \sla p_n + m)
\, , \quad
\sum_{\kappa=\pm}    v^\kappa(\bar p_n)  \bar v^\kappa (\bar p_n)  = ( \bar {\sla p}_n - m)
\, .
\end{eqnarray}

Using the obtained eigenspinors, 
we can organize a mode expansion. 
For the notational brevity, we hereafter take a specific gauge, i.e., the Landau gauge (\ref{eq:Landau-g}), 
where $ p_y $ plays the role of $ \qn $. 
In the Landau gauge, the mode expansion with the Ritus basis reads\footnote{
The creation and annihilation operators for particle states 
should not be confused with $ \hat a $ and $ \hat a^\dagger $ defined in Eq.~(\ref{eq:CA-rela}). 
} 
\begin{subequations}
\label{eq:Ritus-mode}
\begin{eqnarray}
\psi(x) &=& \sum_{\kappa=\pm} \sum_{n=0}^{\infty} \int \frac{d p_z}{2\pi} \int  \frac{dp_y}{2\pi}  \frac{1}{\sqrt{2 \epsilon_n}}
\Ritus_{n,p_y} (x_\perp)   \left[ \, 
 a_{p_n, p_y}^\kappa   e^{ - i p_\para x}   u ^\kappa (p_n)+  b^{\kappa\dagger}_{\bar p_n, p_y}  e^{ i p_\para x}  v^\kappa (\bar p_n) \,\right]
 ,
 \\
\bar  \psi(x) &=&\sum_{\kappa=\pm}  \sum_{n=0}^{\infty} \int \frac{d p_z }{2\pi} \int  \frac{dp_y}{2\pi} \frac{1}{\sqrt{2 \epsilon_n}}
\left[ \, 
  b_{\bar p_n, p_y}^\kappa  e^{-i p_\para x}  \bar v ^\kappa (\bar p_n)  + a_{p_n, p_y}^{\kappa\dagger}  e^{ i p_\para x}   \bar u ^\kappa (p_n) \,\right] 
\Ritus_{n,p_y}^\dagger(x_\perp)  
 ,
\end{eqnarray}
\end{subequations}
where $ \kappa $ is for the ``spin sum'' 
and $ p^0 = \epsilon_n $. Note also that $ \prj_\pm^\dagger = \prj_\pm $. 
Each mode is specified by $ n $ and $p_z  $ in $  p_n$ or $\bar p_n  $ as well as the gauge-dependent quantum number $ p_y $, 
which are indicated explicitly on the creation and annihilation operators. 
The operators $ a_{p_n, p_y}^\kappa$ and $ b_{p_n, p_y}^\kappa$ annihilate the vacuum, 
i.e., $ a_{p_n, p_y}^\kappa | 0\rangle =0= b_{\bar p_n, p_y}^\kappa | 0\rangle $ 
for any $ p_n $, $\bar p_n  $, and $  p_y $. 
The creation and annihilation operators satisfy the anticommutation relations  
$ \{ a_{p_n, p_y}^\kappa,  a_{p'_\np, p_y}^{\kappa' \dagger} \} = \{ b^\kappa_{p_n, p_y},  b_{p'_\np, p_y}^{\kappa' \dagger} \} 
= (2\pi)^2 \delta(p_y-p'_y) \delta( p_z - p'_z)  \delta_{nn'} \delta_{\kappa\kappa'}$, 
while those for all the other combinations vanish. 
The anticommutation relations 
imposed on the creation and annihilation operators 
are consistent with the equal-time anticommutation relation 
for the fermion field, 
$ \left.\{ \psi (x), \psi^\dagger (x') \}\right|_{x^0 = x^{\prime 0} } =   \delta^{(3)} ( \bx - \bx') $.

One can show orthogonal relations 
\begin{subequations}
\label{eq:}
\begin{eqnarray} 
\label{eq:orthogonal-x}
&&
\sum_{n=0}^{\infty} \int \frac{d p_y}{2\pi}  
   \Ritus _{n,p_y}  (x_\perp)   \Ritus_{n,p_y}^\dagger (x'_\perp) 
=
\delta^{(2)}(x_\perp - x'_\perp) 
\, ,
\\
&&
\int d^2 x_\perp    \Ritus _{n,p_y}  (x_\perp)   \Ritus_{\np,p_y^\prime}^\dagger (x_\perp) 
= 2\pi \delta( p_y - p'_y) \delta_{nn'} I_n
\label{eq:orthogonal-p}
\, ,
\end{eqnarray}
\end{subequations}
where a ``unit matrix'' is introduced as 
\begin{eqnarray}
\label{eq:id-magnetic}
I_n = \left\{
\begin{array}{ll}
 \prj_+ & (n=0)
\\
\id_{\rm spinor} & (n \geq1)
\end{array}
\right.
\, .
\end{eqnarray} 
\cout{
The first relation (\ref{eq:orthogonal-x}) is a consequence 
of the completeness of the spin projection operators $ \prj_+ + \prj_- = 1 $ 
and of the Landau-level wave function: 
\begin{eqnarray}
\sum_{n=0}^\infty\int \frac{d p_y}{2\pi}  \phi_{n,p_y} (x_\perp)  \phi_{n,p_y}^\ast (x'_\perp) 
= \sum_{n=0}^\infty\int \frac{d p_y}{2\pi}  \langle x_\perp | n , p_y \rangle  \langle n, p_y | x'_\perp \rangle 
=  \delta^{(2)}(x_\perp - x'_\perp) 
\, .
\end{eqnarray}
The second relation (\ref{eq:orthogonal-p}) is a consequence 
of the orthogonality among the Landau levels: 
\begin{eqnarray}
 \int d^2 x_\perp \phi_{n,p_y} (x_\perp)  \phi_{\np,p_y^\prime}^\ast (x_\perp) 
= \int d^2 x_\perp  \langle x_\perp | n , p_y \rangle  \langle n', p'_y | x_\perp \rangle 
= 2\pi\delta(p_y-p'_y) \delta_{nn'} 
\, .
\end{eqnarray}  
}
Note that the explicit forms of those orthogonal relations depend on 
the choice of the label for the Landau degeneracy as discussed in Sec.~\ref{sec:LL_1}. 
However, as long as the Landau degeneracy is specified by a Hermite operator 
composed of $ \hat x_c $ and $\hat y_c  $, one can organize an orthonormal set of eigenstates 
in a corresponding choice of gauge, which serves as good a choice as the Landau gauge with Eq.~(\ref{eq:py}). 
Mode expansions in other choices of the label and gauge can be organized in a similar way.

In Sec.~\ref{sec:Ritus-CME}, we will apply the Ritus-basis method 
to computation of currents (one-point functions). 
The reader is also referred to the literature, e.g., Ref.~\cite{Lee:1997zj, Elizalde:2000vz, 
Ayala:2006sv, Ferrer:2006vw, Noronha:2007wg, Warringa:2012bq, Kojo:2012js, 
Watson:2013ghq, Mueller:2014tea, Hattori:2015aki, Hattori:2016lqx,
 Hattori:2016cnt, Hattori:2017qih, Hattori:2020htm}, for more examples. 
In Appendix~\ref{sec:relation}, 
we discuss a fermion propagator in the Ritus-basis method.

\cout{

This can be straightforwardly verified as follows. 
Plugging the Ritus-basis mode expansion (\ref{eq:Ritus-mode}) into the equal-time anticommutator, we have 
\begin{eqnarray}
\{ \psi (x), \psi^\dagger (x') \} 
&=& 
\sum_{\kappa=\pm}  \sum_{n=0}^{\infty} \int \frac{d p_z}{2\pi} \int  \frac{dp_y}{2\pi}   
 \frac{1}{2\epsilon_n}
 \\
 && \times 
\Ritus_{n,p_y} (x_\perp)   \big[ \,  e^{ - i p_\para (x-x') }     u^\kappa (p_n)  \bar u^\kappa (p_n) 
+  e^{ i p_\para (x-x')} v^\kappa (\bar p_n) \bar v^\kappa (\bar p_n)  
\, \big ] \Ritus_{n,p_y}^\dagger  (x'_\perp)  \gam^0
\nn
\, .
\end{eqnarray}
Then, according to Eq.~(\ref{eq:spin-sum}) and the completeness of the wave function, 
the Dirac spinors can be arranged as 
\begin{eqnarray}
\label{eq:canonical-psi}
\{ \psi (x), \psi^\dagger (x') \} 
  &=& 
 \sum_{n=0}^{\infty} \int \frac{d p_z}{2\pi} \int  \frac{dp_y}{2\pi}    \frac{1}{2\epsilon_n}
  \nn
 \\
 && \times 
  \Ritus_{n,p_y}  (x_\perp)   
 \big[ \,  e^{- i p_\para (x-x')}      ( \sla p_n + m)   +  e^{  i p_\para (x-x') }      ( \bar {\sla p}_n - m )  
\,  \big]   \Ritus_{n,p_y}^\dagger  (x'_\perp)  \gam^0
  \nn
  \\
  &=& 
   \sum_{n=0}^{\infty} \int \frac{d p_y}{2\pi}  
   \Ritus _{n,p_y}  (x_\perp)   \Ritus_{n,p_y}^\dagger (x'_\perp) 
    \int  \frac{dp_z}{2\pi}  e^{ - i p_\para (x-x') }   
     \nn
  \\
  &=& 
  \delta^{(3)} ( \bx - \bx') 
  \, .
\end{eqnarray} 
To reach the last line, we used the orthogonal relation 
\begin{eqnarray}
\label{eq:orthogonal-x}
\sum_{n=0}^{\infty} \int \frac{d p_y}{2\pi}  
   \Ritus _{n,p_y}  (x_\perp)   \Ritus_{n,p_y}^\dagger (x'_\perp) 
 &=&
\sum_{n=0}^{\infty} \int \frac{d p_y}{2\pi}  
  \big[ \, \phi_{n,p_y} (x_\perp)  \phi_{n,p_y}^\ast (x'_\perp) \prj_+ +  \phi_{n-1,p_y} (x_\perp)  \phi_{n-1,p_y}^\ast (x'_\perp) \prj_-  \, \big]
 \nn
 \\
  &=&
\delta^{(2)}(x_\perp - x'_\perp) 
\, .
\end{eqnarray} 
This is a manifestation of the completeness in the Landau quantization: 
\begin{eqnarray}
\sum_{n=0}^\infty\int \frac{d p_y}{2\pi}  \phi_{n,p_y} (x_\perp)  \phi_{n,p_y}^\ast (x'_\perp) 
= \sum_{n=0}^\infty\int \frac{d p_y}{2\pi}  \langle x_\perp | n , p_y \rangle  \langle n, p_y | x'_\perp \rangle 
=  \delta^{(2)}(x_\perp - x'_\perp) 
\, ,
\end{eqnarray}
and of the spin projection operators $ \prj_+ + \prj_- = 1 $. 
Incidentally, one can show another orthogonal relation  
\begin{eqnarray}
\label{eq:orthogonal-p}
\int d^2 x_\perp    \Ritus _{n,p_y}  (x_\perp)   \Ritus_{\np,p_y^\prime}^\dagger (x_\perp) 
 &=&
\int d^2 x_\perp 
  \big[ \,  \phi_{n,p_y} (x_\perp)  \phi_{\np,p_y^\prime}^\ast (x_\perp) \prj_+ 
  +  \phi_{n-1,p_y} (x_\perp)  \phi_{\np-1,p_y^\prime}^\ast (x_\perp) \prj_- \, \big]
 \nn
 \\
  &=& 2\pi \delta( p_y - p'_y) \delta_{nn'} I_n
\, ,
\end{eqnarray}
according to the orthogonality among the Landau levels: 
\begin{eqnarray}
 \int d^2 x_\perp \phi_{n,p_y} (x_\perp)  \phi_{\np,p_y^\prime}^\ast (x_\perp) 
= \int d^2 x_\perp  \langle x_\perp | n , p_y \rangle  \langle n', p'_y | x_\perp \rangle 
= 2\pi\delta(p_y-p'_y) \delta_{nn'} 
\, .
\end{eqnarray}  
We introduced a ``unit matrix'' 
\begin{eqnarray}
\label{eq:id-magnetic}
I_n = \left\{
\begin{array}{ll}
 \prj_+ & (n=0)
\\
\id_{\rm spinor} & (n \geq1)
\end{array}
\right.
\, .
\end{eqnarray} 
Note that the explicit forms of those orthogonal relations depend on 
the choice of the label for the Landau degeneracy as discussed in Sec.~\ref{sec:LL_1}. 
However, as long as the Landau degeneracy is specified by a Hermite operator 
composed of $ \hat x_c $ and $\hat y_c  $, one can organize an orthonormal set of eigenstates 
in a corresponding choice of gauge, serving as good a choice as the Landau gauge with Eq.~(\ref{eq:py}). 
Therefore, mode expansions in other choices of the label and gauge may be organized in a similar way.

In Sec.~\ref{sec:Ritus-CME}, we will apply the Ritus-basis Feynman rules 
to computation of currents (one-point functions), 
and find that an intriguing dynamics in the LLL induces the so-called chiral magnetic/separation effect. 
The reader is also referred to the literature, e.g., Ref.~\cite{Lee:1997zj, Elizalde:2000vz, 
Ayala:2006sv, Ferrer:2006vw, Noronha:2007wg, Warringa:2012bq, Kojo:2012js, 
Watson:2013ghq, Mueller:2014tea, Hattori:2015aki, Hattori:2016lqx,
 Hattori:2016cnt, Hattori:2017qih, Hattori:2020htm}, for more examples.

 }

\subsubsection{Massless fermions, spectral flow, and chiral anomaly}

\label{sec:massless}


\begin{figure}
\begin{minipage}{0.45\hsize} 
	\begin{center} 
		\includegraphics[width=0.75\hsize]{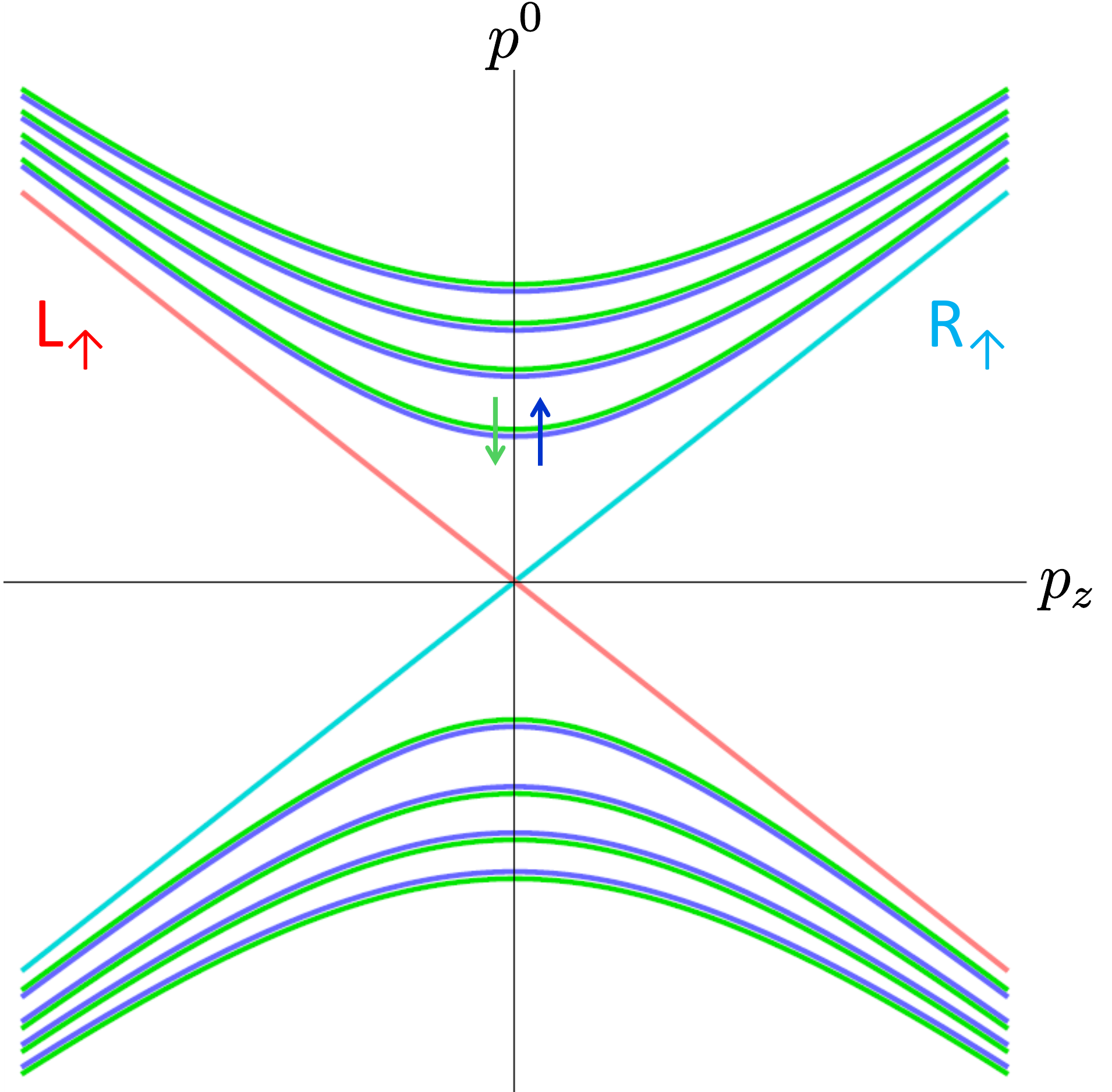}
	\end{center}
\end{minipage}
\begin{minipage}{0.55\hsize}
	\begin{center}
\includegraphics[width=0.99\hsize]{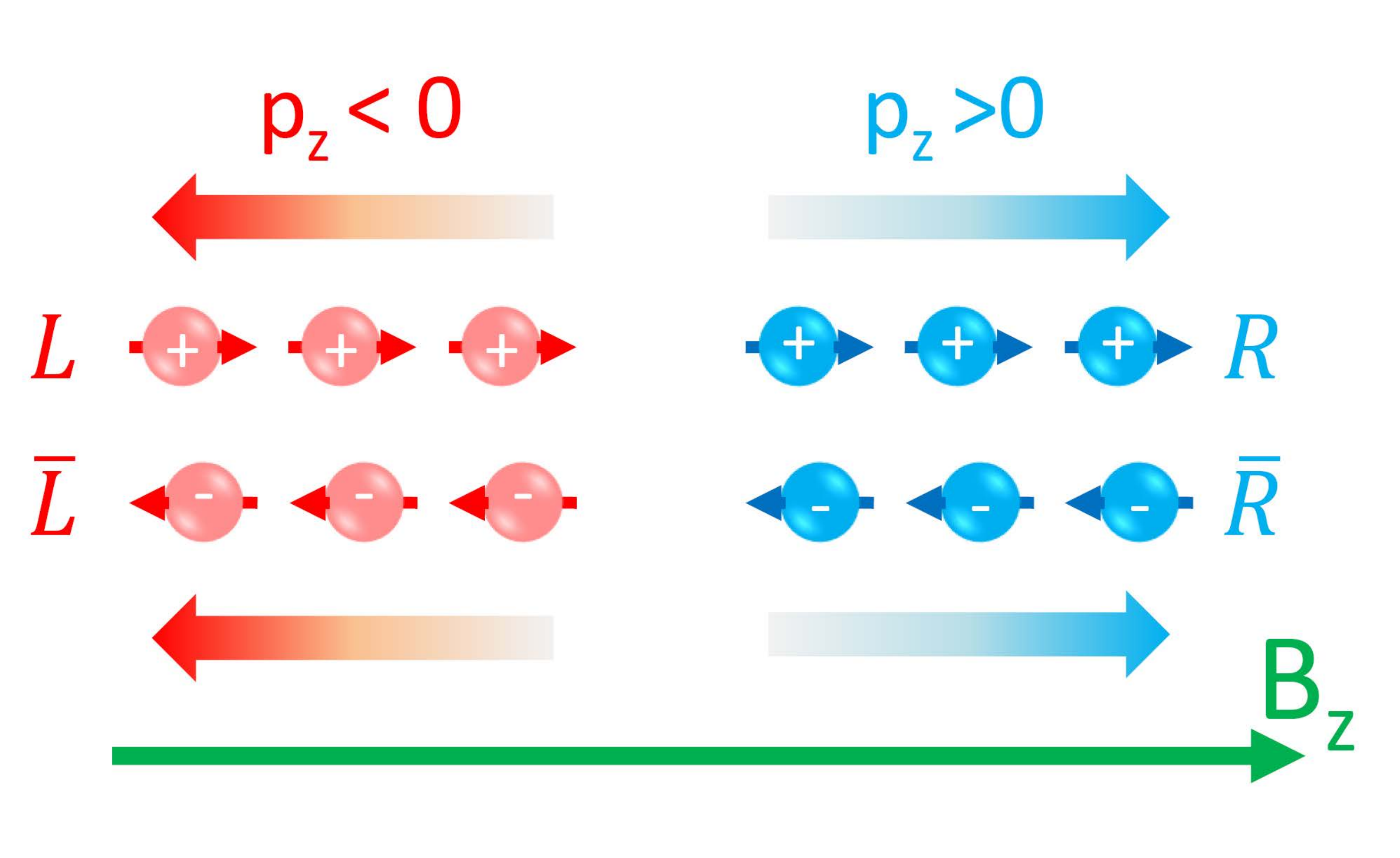}
	\end{center}
\end{minipage}
\caption{
Dispersion relations of the massless fermions in a constant magnetic field when $ s_f >0 $ (left). 
Arrows denote the spin directions. 
The right figure shows the momentum directions (long arrows), the spin directions (short arrows) 
and the electric charges $  (\pm)$. 
Bars on $ \bar R/\bar L $ denote the antiparticle states contained in the $ R/ L $ spinors. 
The particles and antiparticles contained in the spinor of a given chirality, say the blues, 
obey the one-way motion in the {\it same} direction 
(cf. the color correspondence between the left and right figures). 
}
\label{fig:LL}
\end{figure}

In this transient subsection, we discuss 
some special properties of massless fermions in the LLL. 
In the massless limit $ (m=0) $, the solutions for 
the free Dirac equations (\ref{eq:free}) and (\ref{eq:free_v}) 
are given by the chirality eigenstates, i.e., $ \gam^5 u = \pm u $, 
and so is the solution for $ \psi $ 
because the $ \prj_\pm $ commutes with 
$ \gam^5 = i \gam^0 \gam^1 \gam^2 \gam^3 = \diag( -\id, \id) $. 
As shown in Fig.~\ref{fig:LL}, the hLLs still have the quadratic dispersion relations 
and are degenerated with respect to the spin directions or now the chirality.

On the other hand, the massless LLL exhibits 
some special properties originating from the fact that 
the positive- and negative-energy states are
directly connected with each other on each linear dispersion relations (see blue and red diagonal lines in Fig.~\ref{fig:LL}). 
Such a diagonal dispersion relation often induces 
interesting phenomena as well as confusions, 
so that we will carefully identify the quantum numbers of 
the physical excitations and discuss the spectral flow, 
an adiabatic acceleration along the dispersion lines 
in response to an external electric field, 
which serves as an intuitive interpretation of 
the chiral anomaly \cite{Nielsen:1983rb, Ambjorn:1983hp} 
(see also Ref.~\cite{Creutz:2000bs} for a review article).


To get the correct dispersion relation, 
it is important to note that Eq.~(\ref{eq:rel_EoM}) 
from the squared Dirac operator does not determine the relative sign between $p^0  $ and $ p_z  $ in the dispersion relation, 
which could be $ p^0 = \pm |p_z| $, $ p^0 = \pm p_z $, or anything else. 
Thus, one should refer the original Dirac equation that can be arranged as 
\begin{eqnarray}
\label{eq:Dirac-LLL}
 (p^0 - s_f  p_z \gam^5  ) \psi _\LLL = 0
 \, ,
\end{eqnarray}
where we used the facts that the LLL spinor satisfies $ \hat a \psi_\LLL = 0 $ 
and $ (i \gam^1 \gam^2 ) \psi _\LLL=  s_f \psi _\LLL $. 
We denote the right/left chirality eigenstate as 
$ \gam^5 \psi_\LLL^{R/L} = \pm \psi_\LLL^{R/L} $. 
The dispersion relation for each chirality is found to be 
\begin{eqnarray}
\label{eq:gappless}
 p^0 = \pm s_f p_z 
  \, ,
\end{eqnarray} 
{\it without} the symbol of the absolute value. 
Here, the upper and lower signs are for the right- and left-handed chirality eigenstates, respectively, 
and should not be mixed up with the signs for positive- and negative-energy solutions. 
The sign of $ p^0 $ depends on that of $ p_z $, 
so that each chirality eigenstate contains 
both positive- and negative-energy states connected 
by the diagonal dispersion relation in Fig.~\ref{fig:LL}. 
The correct dispersion relation (\ref{eq:gappless}) 
is different from the naive massless limit 
of Eq.~(\ref{eq:fermion-rela}) that erroneously results in 
an absolute value of $ p_z $ and thus separate Dirac cones 
in the positive- and negative-energy regions.

Below, by antiparticles, 
we mean the physical excitations carrying positive energies. 
In each of $\psi_\LLL^{R/L} $,  
the momentum direction of {\it antiparticles}, as holes of 
the negative-energy particle states, is the same as 
that of particles (cf. the right panel of Fig.~\ref{fig:LL}). 
This is because, 
in the linear dispersion relation (\ref{eq:gappless}), 
the positive- and negative-energy states 
have the opposite momenta 
(cf. each diagonal line in the left panel of Fig.~\ref{fig:LL}). 
On the other hand, in each of $\psi_\LLL^{R/L} $, 
the particles and {\it antiparticles} carry 
the opposite helicity $h $ and 
the opposite axial charge $n_A $,\footnote{
The axial charge density, or the chiral charge density, is defined as the temporal component of 
the axial current $n_A =  j_A^{\mu=0} = \bar \psi \gam^0 \gam^5 \psi 
=  \psi_R ^\dagger \psi_R -    \psi_L ^\dagger \psi_L$. 
The particles in $\psi_\LLL^R$ ($\psi_\LLL^L$) carry the axial charge $n_A = + 1$ ($n_A = - 1$ ), 
while the antiparticles in $\psi_\LLL^R$ ($\psi_\LLL^L$) 
carry the opposite axial charge $n_A = - 1$ ($n_A = + 1$ )
} 
meaning that those quantities for the antiparticles 
have the opposite signs to the chirality, 
i.e., the eigenvalue of $ \gam^5 $. 
To explicitly check the correspondence between 
the chirality and the helicity, 
one can arrange the Dirac equation (\ref{eq:Dirac-LLL}) as 
\begin{eqnarray}
\label{eq:h-5}
 \hat h \, \psi_\LLL = \frac{1}{2} \sgn(p^0)  \gam^5 \psi_\LLL 
\, ,
\end{eqnarray}
where we used $ (\gam^5)^2 =1 $ and defined the helicity operator $\hat h = (i \gam^1 \gam^2/2)  (p_z/|p^0|)  $ 
with the spin operator $ (i \gam^1 \gam^2/2) $ 
and momentum direction along the magnetic field. 
According to Eq.~(\ref{eq:h-5}), 
the relative sign between the chirality and the helicity 
depends on the sign of energy $p^0 $. 
Therefore, $\psi_\LLL^R$ ($\psi_\LLL^L$) contains the right-handed helicity particles and left-handed helicity antiparticles 
(the left-handed helicity particles and right-handed helicity antiparticles). 

%
%



\begin{figure}[t]
     \begin{center}
              \includegraphics[width=0.8\hsize]{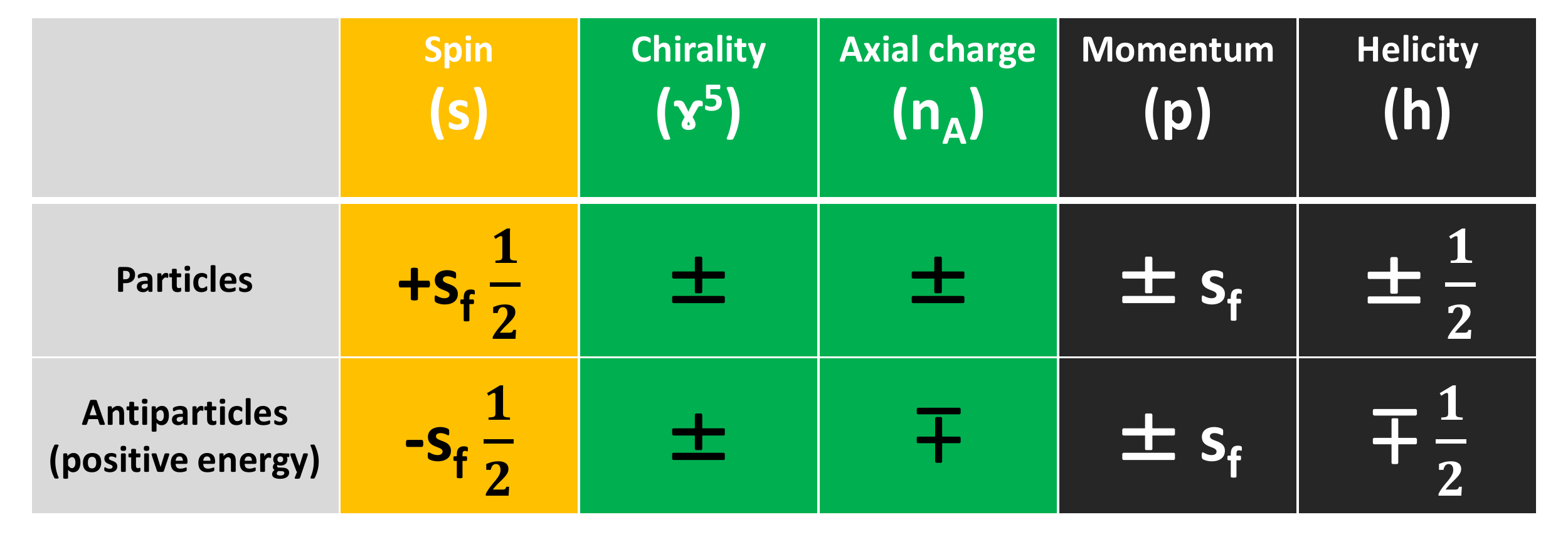}
     \end{center}
\vspace{-0.7cm}
\caption{
Quantum numbers characterizing the massless fermions in the LLL. 
All upper (lower) signs are for the right (left) chirality. 
The momentum and helicity are automatically assigned 
once the spin direction and the chirality are specified. 
}
\label{fig:chirality}
\end{figure}


In Fig.~\ref{fig:chirality}, we summarize 
the correspondences among the quantum numbers. 
All upper and lower signs are for $\psi_\LLL^R $ 
and $\psi_\LLL^L $, respectively. 
The momentum direction and the chirality (or the helicity)  
are locked with each other in the LLL 
(cf. the right panel of Fig.~\ref{fig:LL}): 
One needs two quantum numbers 
out of chirality, spin, and, momentum 
to specify the massless excitations 
and the spin direction is frozen in 
the unique direction along a magnetic field, 
leaving only one independent quantum number. 
The spin direction depends on the sign function $ s_f $, 
and so does the momentum direction according to the locking. 
The massless LLL fermions are equivalent to 
the chiral fermions in purely (1+1) dimensions 
that do not have spin degrees of freedom.\footnote{
We should keep it in mind that the wave function of the LLL fermions have 
a transverse part and that charge-neutral particles such as photons and gluons 
are not confined in the (1+1) dimensions, that makes big differences in, e.g., the spontaneous symmetry breaking 
and bound-state problems (or the confinement). 
}

\begin{figure}[t]
     \begin{center}
              \includegraphics[width=0.8\hsize]{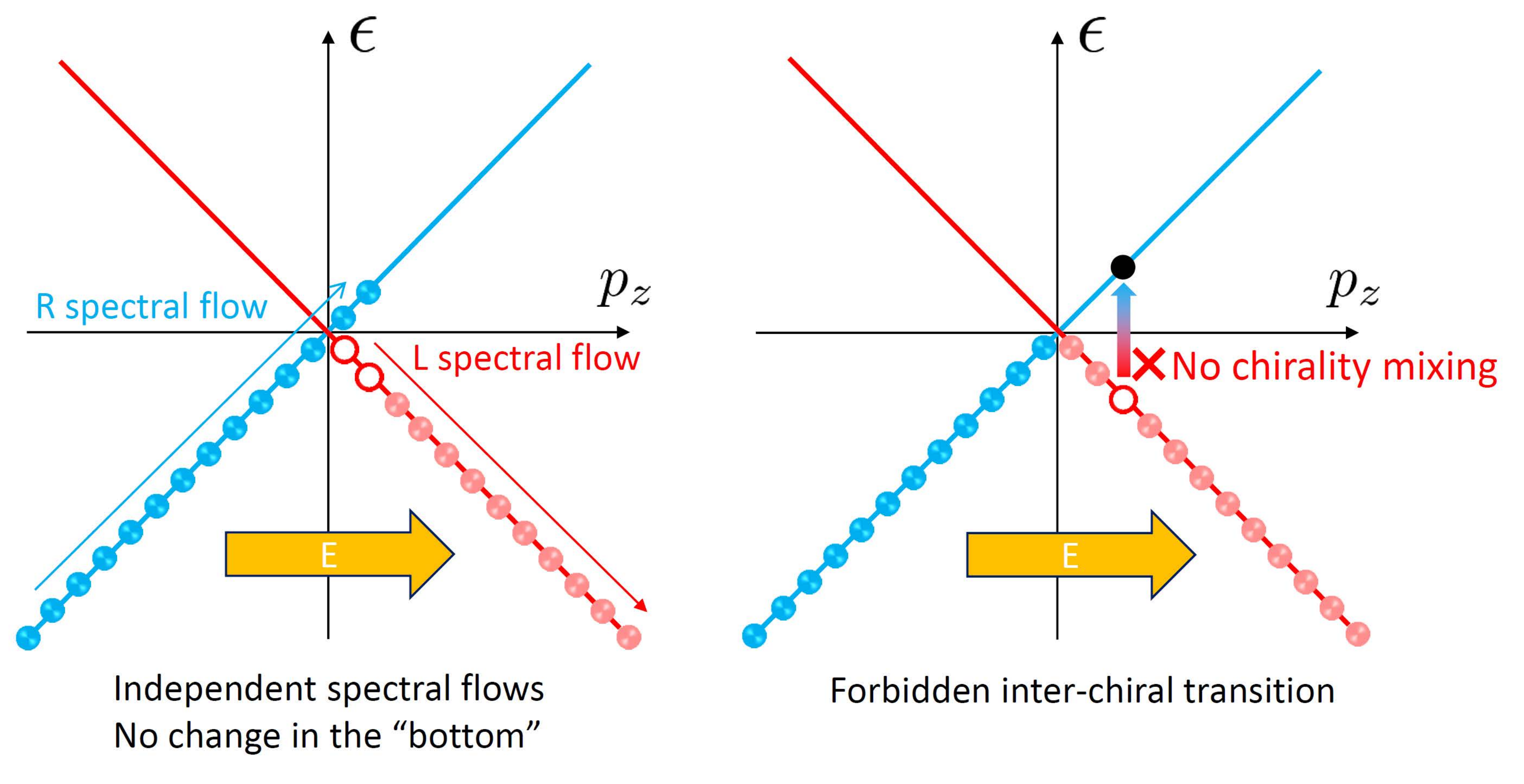}
     \end{center}
\vspace{-0.8cm}
\caption{Spectral flow (left) and a forbidden inter-chiral-state transition (right) 
in an electric and magnetic field applied in a parallel direction. 
We take a case $  q_f >0$, $ B>0 $, and $q_f E >0  $ to draw a figure. 
}
\label{fig:spectral_flow}
\end{figure}

\cout{

We focus on the following two important properties 
of the diagonal dispersion relations (\ref{eq:gappless}). 
(i) The two branches are not directly coupled to each other 
at perturbative interaction vertices since they are belonging to 
the different chirality eigenstates. 
This means that the right-moving and left-moving particles 
do not interact with each other, 
and the right-moving (left-moving) particles are 
only coupled to the right-moving (left-moving) antiparticles. 
(ii) In each chirality 
or in a single (1+1)-dimensional Weyl fermion, 
the positive- and negative-energy states are 
smoothly connected with each other 
across the surface of the Dirac sea 
along the gapless dispersion relations. 
This implies that an infinitesimal electric field 
in parallel to the magnetic field 
can create particle or antiparticle 
out of the Dirac sea, i.e., vacuum. 
Such an adiabatic shift along the dispersion curves 
is often referred to as the {\it spectral flow}. 
}

Now, we consider applying an external electric field 
along the magnetic field adiabatically, 
which will in general accelerate particles 
along dispersion curves. 
Such an adiabatic shift along dispersion curves 
is often referred to as the spectral flow. 
Here, an infinitesimal electric field can create 
particles and antiparticles out of the Dirac sea, i.e., vacuum, 
since the positive- and negative-energy states are 
smoothly connected with each other 
on the diagonal dispersion lines. 
However, the creation of particles or antiparticles would possibly violate the charge conservation. 
This anomalous charge violation is offset only if particles and antiparticles are created with the same amount. 
Nevertheless, the total energy of the system seems to increase even if the total charge is somehow conserved. 
To understand how this issue is cured and if there is any consequence of the deposited energy, 
we should recall the quantum numbers of the massless LLL fermions discussed above 
and identify the quantum numbers created by an electric field.

Since particles and antiparticles are accelerated in opposite directions in an electric field, 
we immediately notice that they are belonging to the {\it different} chirality eigenstates (cf. Fig.~\ref{fig:LL}). 
Therefore, each chirality eigenstate gains the same amount of charge with a relative minus sign, i.e., $ n_R = - n_L $ 
(cf. Fig.~\ref{fig:chirality}). 
This means that the net charge remains zero $ n_R +  n_L = 0 $ thanks to the cancellation. 
However, the particles and antiparticles in the final state carry 
the same amount of chiral charges created out of vacuum, 
providing a nonzero axial charge $n_A =  n_R -  n_L  \not = 0 $. 
This clearly indicates violation of the chirality conservation, and of the chiral symmetry. 
Namely, the spectrum flow across the surface of the Dirac sea, 
shown in the left panel of Fig.~\ref{fig:spectral_flow}, 
serves as an intuitive interpretation of the chiral anomaly 
\cite{Adler:1969gk, Bell:1969ts} in massless fermion systems
as discussed in Refs.~\cite{Nielsen:1983rb, Ambjorn:1983hp} 
(see also Ref.~\cite{Creutz:2000bs} for a review article).


Notice that the creation of particles and antiparticles 
occurs independently of each other along each diagonal line 
(cf. left panel in Fig.~\ref{fig:spectral_flow}). 
During the spectral flow, the occupied states shift altogether 
since the electric field accelerates all of them. 
In this way, the occupied states in the infrared (IR) region 
near the surface is entangled with those 
in the ultraviolet (UV) region in the bottom of the Dirac sea. 
The UV behavior of a theory is still crucial 
though the axial charge creation reveals itself 
as a consequence in the IR region \cite{Ambjorn:1983hp}. 
Counting of the created quantum numbers depends 
on whether the Dirac sea is bottomless or not 
and whether there is, for example, a periodicity 
in the finite cases like lattice systems. 
That is, regularization of the UV region matters.

Chiral anomaly is not explained just by processes 
in the infrared region such as an inter-chirality transition 
from an occupied negative-energy state to 
a positive-energy state in the other chirality state 
(cf. right panel in Fig.~\ref{fig:spectral_flow}). 
This transition is prohibited due to 
the absence of chirality mixing in massless QED Lagrangian. 
In fact, adiabatic creation of the axial charge 
is not induced by a vertical transition 
from the lower to upper half of 
the $ p_z - \epsilon $ plane 
but by a horizontal transition between 
the left- and right-half planes 
due to acceleration of particles across 
a vanishing momentum $ p_z = 0 $. 
At this instant, the helicity is flipped. 
In the massless LLL systems, those transitions occur simultaneously in the diagonal way. 
For massive cases, the spectral flow goes along 
the parabolic dispersion curves. 
In vacuum, the spectral flow under the Dirac sea does not 
create an axial charge \cite{Ambjorn:1983hp}. 
However, an axial charge can be created 
when there are pre-occupied positive-energy states 
in medium at finite temperature and/or density 
where the helicity of massive particles can be flipped due to 
acceleration by an electric field \cite{Hattori:2022wao}. 
Things are more intricate when one considers 
diabatic processes such as the Schwinger pair production 
induced by nonperturbatively strong electric fields 
(see, e.g., Refs.~\cite{Ambjorn:1983hp, Copinger:2018ftr} 
and Sec.~\ref{sec:pair-production}) 
and/or transitions induced by 
finite-frequency electric fields 
(see Ref.~\cite{Hattori:2022wao} and Appendix~\ref{sec:screening}). 
We will discuss how the created axial charge manifests itself 
in physical quantities in Sec.~\ref{sec:transport}. 


\cout{

Notice that the net creation of chirality can be put forward 
only when there are no changes in the deep interior of the Dirac sea. 
For instance, one cannot put forward the net creation 
if a finite number of negative-energy states flows 
in a shallow valence sea along an electric field. 
In contrast, continuous quantum field theories have a {\it bottomless} Dirac sea, 
and an {\it infinite} number of negative-energy states flow collectively without leaving a difference 
in the ultraviolet regime between the initial and final states. 
In this sense, a particle is pulled out from the ``bottom'' of the bottomless Dirac sea by an electric field, 
and its consequence manifests itself in a difference at the surface of the Dirac sea. 
Therefore, the boundary condition at the ultraviolet regime plays a crucial role 
and is connected to the consequence in the infrared regime in such a curious way. 
Therefore, the chiral anomaly is not a consequence of inter-chiral-state transitions just in the vicinity of the infrared regime 
from an occupied negative-energy state to a positive-energy state of the opposite chirality 
(cf. right panel in Fig.~\ref{fig:spectral_flow})~\cite{Ambjorn:1983hp}. 
Indeed, as clear from the first point (i), the absence of the chirality mixing forbids 
such a transition in the vertical direction in the $ p_z - \epsilon $ plane, 
and the creation of particle or antiparticle should occur independently of each other along each diagonal line. 
Notice also that the forbidden vertical transition, which leaves a hole state behind, 
does not create any net quantum number of the system except for an increase of the total energy, even if it occurred. 
Namely, the crucial process for the axial charge creation 
is not a transition between the upper and lower half of the $ p_z - \epsilon $ plane but by a transition between the left- and right-half planes due to acceleration of particles across a vanishing momentum $ p_z = 0 $. 
At this instant, the helicity is flipped. 
In the massless fermion systems, 
the vertical and horizontal transitions occur 
simultaneously in the diagonal way. 

}

\section{Resummation by the proper-time method} 

\label{sec:resum} In this section, we review the proper-time method 
in external electromagnetic fields. 
After describing generalities of the method, 
we will apply it to the construction of the propagators of 
spin-$\frac{1}{2}$ and spinless particles in external Abelian fields. 
In particular, we will elaborate the Schwinger phase which is responsible for the gauge dependence of the propagators 
and the Landau levels encoded in the propagator. 
We will, then, discuss an extension to the case of external 
chromo-electromagnetic fields that is straightforwardly performed 
when we focus on the so-called covariantly constant 
field configurations. 
We discuss the gluon propagator in the external chromo-electromagnetic field in detail. 
In Appendix~\ref{sec:relation}, 
we discuss an equivalence between the proper-time method 
and the Ritus-basis method introduced in the previous section.

\subsection{Generalities of the proper-time method}

\label{sec:PP_general}

Nambu \cite{Nambu:1950rs} and Feynman \cite{Feynman:1950ir} discussed quantum field theories 
on the basis of the idea of Fock \cite{Fock:1937dy} where 
a proper-time variable is introduced to parametrize the worldline in the four dimensional spacetime. 
Schwinger applied the proper-time method to reformulation of the Heisenberg-Euler effective action 
and further investigated fundamental properties of vacuum in strong electromagnetic fields~\cite{Schwinger:1951nm}. 
He established a systematic way of studying the strong-field physics 
with the proper-time method which makes it simpler to perform the resummation of propagators 
or effective actions in terms of external fields. 
Here, we first explain the essence of the proper-time method 
by taking scalar field theory as a simple example \cite{Schwartz, Schubert:2001he}.

Consider the Feynman propagator in the coordinate space:
\beq
G(x,y)=\int \frac{d^4p}{(2\pi)^4}\ {\rm e}^{ip(x-y)} \frac{i}{p^2-m^2+i\epsilon}
\, .
\label{Feynman_prop}
\eeq
We rewrite this by using a mathematical identity
\beq
\frac{i}{X+i\epsilon}=\int_0^\infty ds \ {\rm e}^{is(X+i\epsilon)}\, ,
\label{propertime1}
\eeq
where $X$ is a real quantity. 
An infinitesimal small number $\epsilon >0$, 
which serves as a damping factor ${\rm e}^{-\epsilon s}$, 
ensures the convergence of integral at infinity. 
Integrating both sides of Eq.~(\ref{propertime1}) with respect to $  X$ from $ B $ to $  A$, 
we find another formula 
\beq
\ln\frac{A+i\epsilon}{B+i\epsilon} = 
- \int_0^\infty \frac{ds}{s} \left({\rm e}^{is(A+i\epsilon)}-{\rm e}^{is(B+i\epsilon)}\right)
\, .
\label{propertime2}
\eeq
We will later find that this identity is useful for evaluating the effective action. 
The integration variable $s$ introduced in these formulas is called the proper time for the reason explained later. 
By using the first formula (\ref{propertime1}), the propagator (\ref{Feynman_prop}) is rewritten as 
\beq
G(x,y)=\int \frac{d^4p}{(2\pi)^4}\, {\rm e}^{ip(x-y)} \int_0^\infty ds\ {\rm e}^{is(p^2-m^2+i\epsilon)}\, .
\label{propagator_propertime1}
\eeq
Introduction of the proper time makes the inverse operator $i(p^2-m^2+i\epsilon)^{-1}$ (where the external fields will appear) more tractable with an exponential form ${\rm e}^{is(p^2-m^2+i\epsilon)}$. 

Instead of performing the Gaussian integral over the momentum $p$ in Eq.~(\ref{propagator_propertime1}), we examine another way.  
Let us introduce the Hilbert spaces spanned by $|x\rangle$ and $|p\rangle $ where $x$ and $p$ are {\it four-dimensional} vectors 
$x^\mu$ and $p^\mu$, respectively. 
Namely, we define $|x\rangle$ and $|p\rangle $ as eigenstates of the ``operators'' $\hat x$ and $\hat p$:
\beq
\hat x^\mu |x\rangle = x^\mu |x\rangle\, ,\qquad 
\hat p^\mu |p\rangle = p^\mu |p\rangle\, .
\eeq
These states are normalized as 
\beq
 \langle p'|p\rangle = \delta^4(p'-p)\, ,\qquad 
\int \frac{d^4p}{(2\pi)^4}\, |p\rangle \langle p|=1\, ,
\eeq
and so on, and the inner product between $|x\rangle$ and $|p\rangle$ is defined as 
\beq
\langle p | x\rangle = {\rm e}^{ip^\mu x_\mu}.
\eeq
By using these properties, one can rewrite Eq.~(\ref{propagator_propertime1}) 
in a compact and suggestive form:
\beq
G(x,y)=\int_0^\infty ds \, {\rm e}^{-s\epsilon} {\rm e}^{-ism^2} 
\langle y | {\rm e}^{-i\hat H_0 s} |x\rangle  \, ,
\label{propagator_propertime2}
\eeq
where we have defined an operator $\hat H_0=-\hat p^2$ and used ${\rm e}^{isp^2}\langle p|x\rangle=\langle p|{\rm e}^{-is\hat H_0} |x\rangle $. As is evident from this expression, we notice that the matrix element $\langle y | {\rm e}^{-i\hat H_0 s} |x\rangle $ can be regarded as a kind of transition amplitude with $\hat H_0$ being the ``Hamiltonian" and $s$ being the ``time". In this picture, the four-vectors $x^\mu$ and $y^\mu$ are parametrized by $s$. 
Recall that, in the special relativity, the proper time $s$ of a particle is defined by the worldline element $ds^2=g_{\mu\nu}dx^\mu dx^\nu$ along the trajectory of a particle, and is conjugate to an operator $\hat H=g_{\mu\nu}\del^\mu \del^\nu $. Therefore, we can indeed regard $s$, introduced above, as the proper time and $\hat H_0=-\hat p^2$ as the Hamiltonian generating the ``proper-time" evolution of a state $|x(s)\rangle = {\rm e}^{-i\hat H_0 s}|x(0)\rangle$ (it is customary to define the Hamiltonian without the mass contribution).

Then we are left with the problem to compute the transition matrix element $\langle y | {\rm e}^{-i\hat H_0 s} |x\rangle $ \cite{Schwinger:1951nm}. Similar to the ordinary quantum mechanics, this transition amplitude satisfies 
a counterpart of the Schr\"odinger equation for the proper-time evolution. 
Alternatively, we can work in the Heisenberg picture with the Heisenberg equations of motion 
for the operators $\hat x^\mu$ and $\hat p^\mu$.\footnote{
With the proper-time evolution operator, the Heisenberg equation of motion is given by $d \hat x^\mu/ds= i[\hat H, \hat x^\mu]$, etc. 
We impose the canonical commutator at the equal ``proper time" $[\hat x^\mu(s), \hat p^\nu(s)]= - ig^{\mu\nu}$. 
} 
In any case, the result turns out to be $ \langle y | {\rm e}^{-i\hat H_0 s} |x\rangle=- i/(16\pi^2 s^{2} )\exp\{-i (x-y)^2/(4s) \}$ and reproduces the result directly obtained from the Gaussian integral in Eq.~(\ref{propagator_propertime1}) (See, e.g., Ref.~\cite{Schwinger:1951nm, Hattori:2020guh} for more detailed discussions).

In interacting field theories, we are able to apply the proper-time method for the computation of propagators. Then, in place of the transition amplitude defined by $\hat H_0$ in Eq.~(\ref{propagator_propertime2}), we will encounter a transition amplitude $\langle y | {\rm e}^{-i\hat H s}|x\rangle$ with a Hamiltonian $\hat H$ which includes the interaction effects. 
Nevertheless, we can compute the transition amplitude in a similar way as discussed above by solving the Schr\"odinger equation or the Heisenberg equation of motion. 
An application to the spinor QED was suggested by Nambu \cite{Nambu:1950rs}. 
Soon later, Feynman also discussed a similar idea for Klein-Gordon particles \cite{Feynman:1950ir}. 
An advantage of the proper-time method is that 
the amplitude can be canst into a path-integral form 
in analogy with quantum mechanics \cite{Feynman:1950ir, Affleck:1981bma, Bern:1990cu, Fradkin:1991ci, Strassler:1992zr, Schmidt:1993rk, Reuter:1996zm}. 
Then, one can apply conventional techniques 
that have been accumulated for evaluation of the path integral 
such as the Monte Carlo sampling and the saddle-point method with the so-called ``worldline instanton'' solutions (see Ref.~\cite{Schubert:2001he, Schubert:2007xm} for reviews and more references therein for applications). 


\cout{
Once we obtain the transition amplitude $\langle y | {\rm e}^{-i\hat H s} |x\rangle$, we are able to calculate the effective action.
In the present simple example of a scalar field theory with a potential $V(\phi)$, 
the one-loop effective potential $\Gamma[\phi]$ is given by \cite{ItzyksonZuber}
\beq
\Gamma[\phi]=-\frac{i}{2} {\rm Tr}\, \ln \left[ \frac{\del^2 + m^2 + V''(\phi)}{\del^2 + m^2} \right]\, , \label{eff_action_scalar}
\eeq
where $V''(\phi)=\frac{\delta^2 V(\phi)}{\delta\phi \delta\phi}$. 
We already subtracted the contribution of a free scalar field. 
By using the formula (\ref{propertime2}), one finds the proper-time expression:
\beq
\Gamma[\phi]=\frac{i}{2}  \int d^4x 
\int_0^\infty \frac{ds}{s}\ {\rm e}^{-s\epsilon}{\rm e}^{-ism^2} \ \left[ 
\langle x | {\rm e}^{-i\hat H s}|x \rangle
-
\langle x | {\rm e}^{-i\hat H_0 s}|x \rangle
\right] \, ,
\eeq
where $\hat H=\del^2 +V''(\phi)$ and $\hat H_0=\del^2$ 
and the trace was taken over the coordinate basis. 
Because of the trace, the world-line trajectory forms a loop in the spacetime 
after the proper-time evolution, i.e., $ x^\mu(s) = x^\mu(0) $. 
The looped transition amplitude corresponds to the loop diagram as expected from the above construction. 
Notice that the proper-time integral diverges as $s \to 0  $, 
which corresponds to a large momentum regime 
according to the dimensionless combination $ \bar s = p^2 s $ appearing in Eq.~(\ref{propagator_propertime1}). 
Therefore, this divergence can be identified with the ultraviolet divergence in quantum field theory, 
and can be regularized by simply introducing a cutoff. 
This procedure violates neither Lorentz nor gauge invariance (when we consider gauge theories), 
and provides a useful regularization scheme called the proper-time regularization. 
The original work by Schwinger was indeed motivated by the gauge invariant formulation 
of the quantum field theory~\cite{Schwinger:1951nm} (see also Ref.~\cite{ZinnJustin:2002ru}). 
We will use the proper-time regularization in Sec.~\ref{sec:HE_QCD} to compute the effective actions.


Interpretation of the matrix element $\langle y | {\rm e}^{-i\hat H s}|x \rangle=\langle y (0)|x(s)\rangle$ as a transition amplitude leads to an intriguing representation of the propagator called the world-line representation. Similar to the ordinary quantum mechanics, we can express the transition matrix element $\langle y (0)|x(s)\rangle$ in a path-integral form \cite{Schubert:2001he}. In the simplest case with a (relativistic) scalar field theory, the Hamiltonian $\hat H=\del^2+V''(\phi)$ can be re-interpreted as a non-relativistic Hamiltonian\footnote{The kinetic energy becomes positive definite if one  goes into the Euclidean representation. $-\hat p^2/2\tilde m=\hat p_{\rm E}^2/2\tilde m$.} $\hat H=-\hat p^2/2\tilde m + U(\phi)$ with $\tilde m=1/2$ and $U(\phi)=V''(\phi)$, and we rewrite the transition amplitude as follows: 
\beq
\langle y | {\rm e}^{-i\hat H s} |x\rangle = \int_{x^\mu(0)=x}^{x^\mu(s)=y} {\cal D}x^\mu(\tau)\, {\rm e}^{-i\int_0^s d\tau \left[ \frac14 \dot{x}(\tau)^2 + V''(\phi(x(\tau)))\right]} \, ,
\eeq
where the dot over $x$ is a derivative with respect to the proper time, $\dot{x}=dx/d\tau$ and we sum over the paths connecting the initial and final spacetime points $x^\mu(0)=x$ and $x^\mu(s)=y$. Even though we are not able to exactly solve the Schr\"odinger equation for the amplitude, we can apply various approximation schemes in the path-integral expression, such as the semi-classical (WBK) approximation. This approach is in particular useful when the external fields are inhomogeneous in spacetime. 

}

\subsection{Spin-$\frac{1}{2}$ particles}

\label{sec:prop_half}


In this subsection, we elaborate the proper-time method for 
spin-$\frac{1}{2}$ particles in constant external electromagnetic fields. 
We derive the proper-time representation of the fermion propagator $ S(x,y|A)  $, 
where the argument $A$ indicates the presence of an external electromagnetic field. 
In the presence of the external field $A^\mu (x) $, 
we encounter an issue when transforming the propagator into the Fourier space: 
Namely, the gauge field 
and, thus the propagator $ S(x,y|A)$, do not have a manifest translational invariance, 
which spoils the Fourier transform by a single momentum. 
Nevertheless, we have already learned in Sec.~\ref{sec:LL_1} that 
the translational invariance is merely hidden behind the gauge dependence since constant electric and magnetic fields 
do not break the translational invariance. 
Putting it the other way round, one can take an advantage of 
the gauge choice to bypass the above issue as follows.

Notice that, as long as the gauge field is a linear function of the coordinate 
[cf. Eqs.~(\ref{eq:Landau-g}) and (\ref{eq:symmetric-g})], 
the choice of the origin of the coordinate is arbitrary; 
A gauge field $A^\mu(x- x_0) $ shifted by $ x_0$ produces the same external field as $A^\mu(x) $ does. 
Setting the origin to be $ x_0 = y$, 
the Green's function will satisfy the equation of motion in a translation-invariant form 
\begin{eqnarray}
\left( i \slashed D_x - m \right) S(x-y|A) = i \, \delta^4(x-y)
\, ,
\label{eq:Gx}
\end{eqnarray}
where the covariant derivative (\ref{eq:covariantD-QED}) 
includes the gauge field $A^\mu (x- y) $. 
This gauge choice enables us to perform the Fourier transform 
of the Green's function by a single momentum. 
Then, we need to restore a general gauge 
by examining the gauge-transformation property of 
the obtained propagator 
to guarantee that there is no conflict of gauge choices 
among all the fermion lines in Feynman diagrams.


Below, we will introduce a useful gauge 
called the Fock-Schwinger gauge~\cite{Fock:1937dy, Schwinger:1951nm} 
and will carefully examine the gauge covariance of the propagator 
that is expressed in the form of the so-called Schwinger phase.
There are several ways to solve Eq.~(\ref{eq:Gx}). 
Here, we employ the strategy given in Refs.~\cite{Brown:1975bc, Dittrich:1975au, Dittrich:1985yb}
and extend the discussions to more general cases. 
As an alternative way, 
one may evaluate the inverse Dirac operator 
by using the complete set of the wave functions discussed in the previous section. 
In Appendix~\ref{sec:relation}, we explicitly show the equivalence between the proper-time method and the Ritus basis method. 
The reader is also referred to, e.g., appendices in Refs.~\cite{Hayata:2013sea,Miransky:2015ava} for 
this alternative derivation of the Schwinger phase and the resummed propagator. 
It is also interesting to look up the original derivation by Schwinger on the basis of 
the ``equations of motion'' for the proper-time evolution~\cite{Schwinger:1951nm} (see also Refs.~\cite{Dittrich:2000zu, Hattori:2020guh} for reviews). 



\subsubsection{The Fock-Schwinger gauge and the Schwinger phase}

\label{sec:FS}

We introduce the Fock-Schwinger (FS) gauge for a constant field strength tensor $ F^{\mu\nu} $~\cite{Fock:1937dy, Schwinger:1951nm}: 
\begin{eqnarray}
A^\mu_{\FS} (x) 
= - \frac{1}{2} F^{\mu\nu}  (x_\nu - x_{0\nu})
\, ,
\label{eq:FS}
\end{eqnarray}
where $ x_0^\mu $ is a free parameter. 
By choosing $ x_0^\mu = y^\mu $, we can get 
the translation-invariant form (\ref{eq:Gx}). 
This gauge is sometimes called the fixed-point gauge 
because $ x_0^\mu $ specifies a particular point in spacetime \cite{Novikov:1983gd, Reinders:1984sr}. 
This is an extension of the symmetric gauge (\ref{eq:symmetric-g}) 
that we have used for a constant magnetic field in Sec.~\ref{sec:S-gauge}. 
The FS gauge allows us to include an electric field as well as a magnetic field 
in terms of the gauge-invariant field strength tensor $ F^{\mu\nu} $. 
While we here focus on constant fields, 
this gauge can be extended to inhomogeneous cases \cite{Novikov:1983gd, Reinders:1984sr}. 
The general gauge condition is given by 
\begin{eqnarray}
\label{eq:FS-condition}
 (x^\mu - x_{0}^{\mu}) A^{\FS}_\mu (x) = 0
 \,  .
\end{eqnarray}
If the field strength tensor has a weak spacetime dependence, one can solve Eq.~(\ref{eq:FS-condition}) on an order-by-order basis in a derivative expansion. 
Such a gauge configuration has been used 
for calculations of the Wilson coefficients in 
the operator product expansion \cite{Novikov:1983gd, Reinders:1984sr} 
and for constructions of the Heisenberg-Euler Lagrangian. 
We will discuss the latter in Sec.~\ref{sec:HE_inhom}.

To get the fermion propagator in a general gauge, 
we carefully examine how the fermion propagator is 
transformed in a gauge transformation. 
Under the $  U(1)$ gauge transformation by a local angle $\alpha(x)$, we have 
\begin{eqnarray}
\psi(x) &\rightarrow& \psi^{\prime} = {\rm e}^{i\alpha(x)} \psi(x)
\label{eq:gauge-transf}
\, ,
\\
A^\mu(x) &\rightarrow & A^{\prime\mu} (x) = A^\mu(x) - \frac{1}{q_f} \partial^\mu \alpha(x)
\label{eq:gauge-transf-A}
\, .
\end{eqnarray}
According to the definition~(\ref{eq:Gx}), the Green's function $S(x,y|A')$ with the transformed field $A'_\mu(x)$ 
should satisfy $( i \slashed D'_x - m ) S(x,y|A') = i\,\delta^4(x-y)$ with $D'_\mu=\del_\mu +iq_f A_\mu'$. By using the covariance of the Dirac operator, this equation  can be rewritten as 
\begin{eqnarray}
i\, \delta^4(x-y) &=&   {\rm e}^{i\alpha(x) } \left( i \slashed D_x  - m \right)  {\rm e}^{-i\alpha(x) } \ S(x,y|A^\prime) 
\nonumber \\
&=&\left( i \slashed D_x  - m \right) {\rm e}^{-i\left\{\alpha(x) - \alpha(y) \right\} } \ S(x,y|A^\prime) 
\label{eq:Gx3}
\, \ ,
\end{eqnarray}
where we have replaced  $ \alpha(x) $ by $ \alpha(y) $ 
because the overall exponential factor is relevant only when $x=y$ as implied by the delta function. 
Comparing Eq.~(\ref{eq:Gx3}) with Eq.~(\ref{eq:Gx}),  
the propagator transforms under the gauge transformation from the FS gauge (\ref{eq:FS}) to an arbitrary gauge $ A(x)$ as 
\begin{eqnarray}
S(x,y|A) &= & {\rm e}^{ i \Phi_A(x,y) } \ S(x-y|A_\FS) 
\label{eq:transf}
\,  .
\end{eqnarray}
The transformation property is solely encoded 
in the Schwinger phase $ \Phi_A(x,y) $ given by 
\begin{eqnarray}
\Phi_A(x,y) \equiv - \{ \alpha(x) - \alpha(y)  \}
=  - q_f  \int_y^x \left( A_\mu(\xi) 
+\frac{1}{2}F_{\mu \nu} (\xi^\nu - y^\nu ) \right) d\xi^ \mu
\, .
\label{eq:Phi_S}
\end{eqnarray}  
It is important to remember that the origin of the coordinate was taken at $ x_0 = y$ in the FS gauge (\ref{eq:FS}) for {\it each} propagator. 
This choice is reflected in the integrand of the Schwinger phase 
and apparently breaks the translational invariance. 
The rest of the propagator $ S(x-y|A_\FS)  $ is expressed with the field strength tensor $ F^{\mu\nu} $, and 
has a manifest gauge invariance. 
Thus, we conclude that the gauge dependence and the translation-breaking part are both factorized as the Schwinger phase. 
In short, insertion of a gauge link between the fermion fields at $  x$ and $ y $ 
is necessary for the gauge covariance of the two-point function. 
The Schwinger phase guarantees that 
there is no conflict of gauge choices among fermion lines. 
The total Schwinger phase in Feynman diagrams 
should be automatically organized 
into a gauge- and translation-invariant form 
when one computes a gauge-invariant quantity.

\begin{figure}[t]
     \begin{center}
           \includegraphics[width=0.9\hsize]{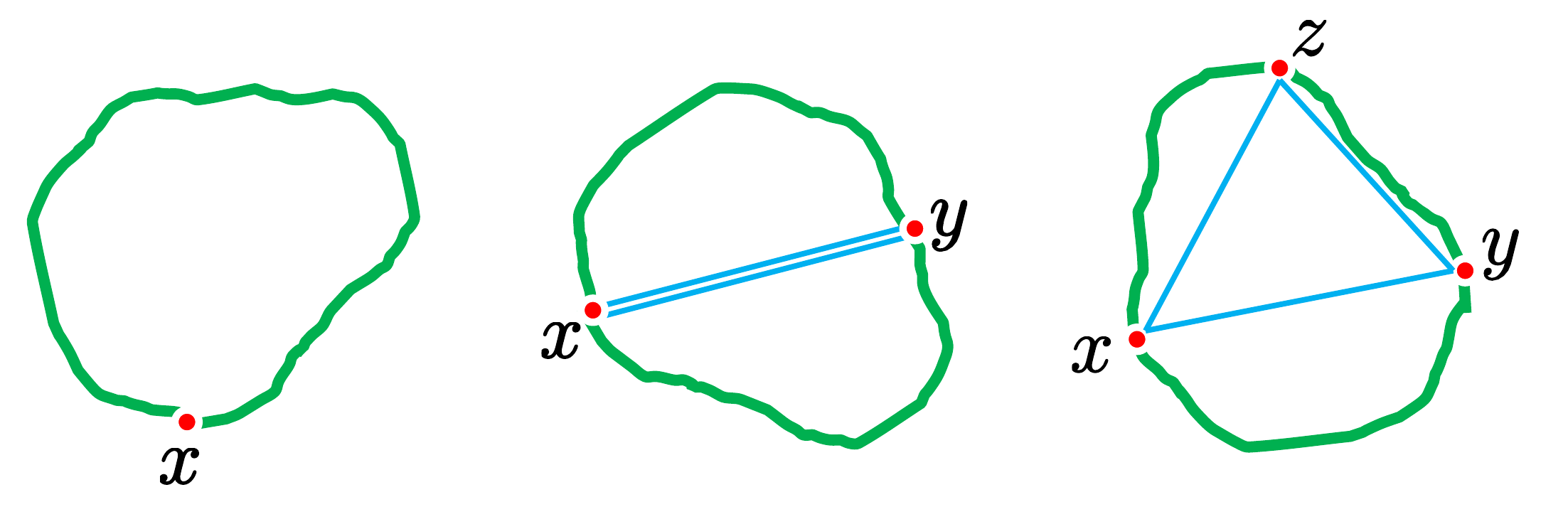}
     \end{center}
\caption{Schwinger phases on closed paths. 
}
\label{fig:Sphase}
\end{figure}

In prior to getting the explicit form of $ S(x-y|A_\FS)  $ in the next subsection, 
we summarize some basic properties of the Schwinger phase. 
We find in Eq.~(\ref{eq:Phi_S}) that the integrand of the Schwinger phase $[\, \equiv f_{\mu}(\xi; y) \,]$ 
is curl-free, i.e., $ \partial^\mu_\xi f^\nu (\xi; y) - \partial^\nu_\xi f^\mu (\xi; y)  = 0$. 
Therefore, first of all, the result of the integral is independent of the path, 
and the integral along a closed path vanishes\footnote{
This statement is true only when the closed path can be continuously squeezed to a point. 
For example, in the imaginary-time formalism of thermal field theory, 
the integral around the compact temporal direction results in a finite winding number (see Sec.~\ref{sec:HE_QCD}).
}
\begin{eqnarray}
\Phi_A(x,x) =   - q_f  \oint 
 f_{\mu}(\xi; x)  d\xi^ \mu= 0
 \, .
 \label{eq:vanish}
\end{eqnarray}
It is important to note that this means neither $\Phi_A(x,y)  = \Phi_A(x^\prime,y^\prime)  $ 
for $ x \not = x^\prime$ and  $ y \not = y^\prime$ nor 
the absence of the Schwinger phase on an arbitrary closed loop when there are more than two vertices. 
This is because the integrand $ f_{\mu}(\xi; y)$ does not have a translational invariance since we have chosen the different values of the parameter $ x_0^\mu $ for different fermion lines in a Feynman diagram.

In Fig.~\ref{fig:Sphase}, we show closed loops with different numbers of vertices, $ n_{\rm v} = 1$, $2 $, and $ 3$. 
When $ n_{\rm v} = 1$ which includes the case of the Heisenberg-Euler effective action discussed in the subsequent section, 
one finds the vanishing phase as in Eq.~({\ref{eq:vanish}). 
However, when $ n_{\rm v} \geq 2$, 
the Schwinger phase on a closed loop is nonvanishing 
due to the breaking of translational invariance.\footnote{As shown in Appendix~\ref{sec:polygon}, we find that $ n_{\rm v}=2 $ is an exceptional case where the Schwinger phase identically vanishes.} 
Nevertheless, 
one can show that the Schwinger phase for a closed fermion loop is a gauge-invariant quantity and can be cast into a translation-invariant form (\ref{eq:Schwinger_n}) as shown in Appendix~\ref{sec:polygon}, 
though the Schwinger phase on each side of polygons 
is not a gauge- or translation-invariant quantity.

When considering derivative couplings, 
one needs to take the derivative of the Schwinger phase at the interaction vertices. 
This is required not only in a theory/model with derivative couplings 
but also in the correlators of the energy-momentum tensor which may contain derivatives. 
One can show that (see Appendix~\ref{sec:polygon})
\begin{eqnarray}
\label{eq:D-Phi}
D_x^\mu {\rm e}^{i\Phi_A(x,y)} = {\rm e}^{i\Phi_A(x,y)} \{\, \partial_x^\mu -  \frac{i}{2} q_f F^{\mu\nu} (x_\nu - y_\nu) \, \}
=  {\rm e}^{i\Phi_A(x,y)} 
\{\, \partial_x^\mu + i q_f A^\mu _\FS (x_0 \to y) \,\}
\, .
\end{eqnarray}
After the derivative went through the Schwinger phase, 
the terms between the braces are given in a manifestly gauge- and translation-invariant form 
and are the covariant derivative expressed in the Fock-Schwinger gauge (\ref{eq:FS}) with $ x_0^\mu = y^\mu $. 
This result is expected from the covariance of the derivative operator. 
When the Dirac operator $ (i \sla D_x -m) $ is operated on the both sides of Eq.~(\ref{eq:transf}), 
the derivative relation (\ref{eq:D-Phi}) also implies that 
\begin{eqnarray}
 (i \sla D_x -m) S(x,y|A) 
 =  e^{ i\Phi_A(x,y) } i \delta^{(4)} ( x-y)
 \, ,
\end{eqnarray}
where we used Eq.~(\ref{eq:Gx}). Recalling that $ \Phi_A(x,x)=0 $, we notice that 
the right-hand side reads $ i \delta^{(4)} ( x-y) $ when $ x=y $ and, otherwise, vanishes. 
Therefore, we confirm that the resummed propagator $ S(x,y|A) $ in a general gauge 
is the Green's function of the Dirac operator, 
although it does not have a translation invariance like in familiar cases.

\subsubsection{Proper-time representation}

\label{sec:prop-time}

By using the trick discussed below Eq.~(\ref{eq:Gx}), 
one can Fourier transform the both sides of \eref{eq:Gx} as 
\begin{eqnarray}
\left( \,  \slashed p - q_f \slashed A - m \, \right) S(p|A) = i
\, ,
\label{eq:Gp}
\end{eqnarray}
with the single momentum $p $. 
The coordinate dependence in the gauge field is understood 
to be replaced by the derivative operator with respect to the momentum. 
The above equation in the momentum space can be solved in a formal way 
\begin{eqnarray}
S(p|A) &=& \frac{i}{\slashed p - q_f \slashed A - m} 
= i ( \slashed p - q_f \slashed A  + m ) \ \Delta(p|A)
\, ,
\label{eq:GDp}
\end{eqnarray}
where 
\begin{eqnarray}
\Delta(p|A) &=& \frac{1}{i}
\int_0^\infty \!\! ds \ {\rm e}^{ is \{ (\slashed p - q_f \slashed A )^2 - m^2 + i\epsilon  \} }
\label{eq:Dp} 
\, .
\end{eqnarray}
As mentioned in Sec.~\ref{sec:PP_general}, the integral variable $s$ is called the proper time, 
and this integral is convergent at the infinity owing to an infinitesimal imaginary part, $i \epsilon$. 
As clear from the appearance of the Dirac operator in Eq.~(\ref{eq:GDp}), 
the $ \Delta(p|A) $ transforms in precisely the same way as 
the $ S(p|A) $ does in Eq.~(\ref{eq:transf}) by the Schwinger phase. 
When the external field is weak, one may perform a perturbative expansion of the full propagator (\ref{eq:GDp}) 
with respect to the coupling constant $ e $ (in $ q_f $). 
However, when $e A \gtrsim m  $, the higher-order terms, 
which is of the order of $ (eA)^n $ ($ n \geq 1 $), 
are all as important as the leading term without the external field ($ n=0 $), 
and a naive perturbative expansion breaks down (cf. Fig.~\ref{fig:resummed_prop}). 
In such a case, the proper-time method is useful for organizing 
the full propagator into the compact form (\ref{eq:Dp}).

\begin{figure}
     \begin{center}
              \includegraphics[width=0.9\hsize]{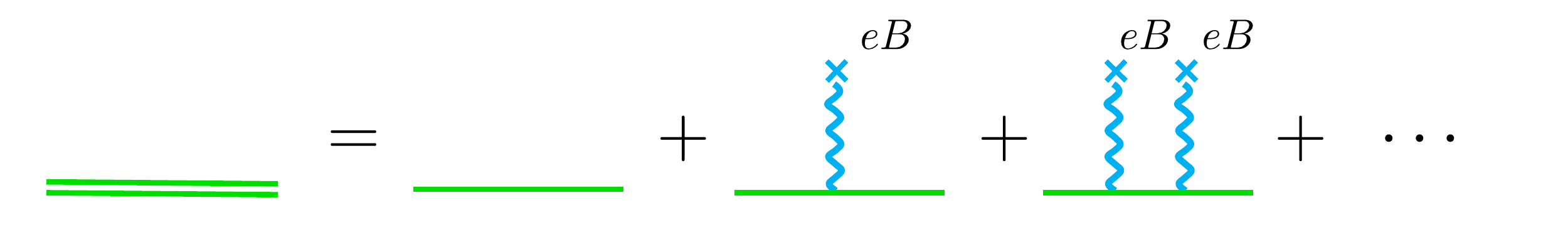}
     \end{center}
\caption{Resummed propagator in a strong external field (double line). 
The higher-order terms with respect to the coupling constant are not suppressed 
when the smallness of the coupling constant is compensated 
by the large field strength of the external field appearing in the combination $ q_f F^{\mu\nu} $.
}
\label{fig:resummed_prop}
\end{figure}

Following Eq.~(\ref{eq:Gp}), we obtain the equation for the $\Delta(p|A)$: 
\begin{eqnarray}
&&\left[ \, \left( \, \slashed p - q_f \slashed A \, \right)^2 - m ^2 \, \right] \Delta(p|A) = 1 
\label{eq:eqDp} 
\, .
\end{eqnarray}
By using the FS gauge and an identity, 
$\gamma^\mu \gamma^\nu = \frac{1}{2} [\gamma^\mu, \gamma^\nu] + \frac{1}{2} \{ \gamma^\mu, \gamma^\nu \}$, 
we have 
\begin{eqnarray}
 \left[ \, p^2  + \frac{q_f^2}{4} 
F^{\mu\alpha}F_\alpha^{\ \nu} \frac{\partial}{\partial p^\mu} \frac{\partial}{\partial p^\nu} 
- \kappa^2 
\, \right ] \Delta(p|A_\FS) = 1
\label{eq:DELp}
\, ,
\end{eqnarray}
where $\kappa^2 :=(m^2- i\epsilon) + \frac{q_f}{2} F^{\mu\nu} \sigma_{\mu\nu}$. 
We have dropped a term proportional to $F_{\mu\nu}  p^\mu \partial_p^\nu $, 
which is the generator of the Lorentz transform by an ``angle matrix $F^{\mu\nu}$". 
Since the operator in Eq.~(\ref{eq:DELp}) has a manifest Lorentz invariance, 
the solution for $ \Delta(p|A_\FS) $ is expected to be a Lorentz-scalar function, 
and thus the increment in an infinitesimal transform $\sim   p^\mu \partial_p^\nu \Delta(p|A_\FS)$ should vanish.

To obtain the resummed propagator, one needs to solve Eq.~(\ref{eq:DELp}). 
This can be carried out by using an ansatz, which is explained in detail in Appendix~\ref{sec:resum_prop}. 
After some computations, one finds the following solution:
\begin{subequations}
\begin{eqnarray}
\label{eq:Dscalar}
\Delta (p|A_\FS) &=&- i \int _0^\infty \!ds \exp \left[ \,
- i \kappa^2 s + i p_\mu X^{\mu\nu}(s)  p_\nu + Y(s) \, \right]
\, ,
\\
\label{eq:DscalarX}
X^{\mu\nu}(s) &=& \left[(q_f F)^{-1} \tanh(q_f Fs) \right]^{\mu\nu}
\, ,
\\
\label{eq:DscalarY}
Y(s) &=& - \frac{1}{2} \tr \big[ \ln\{ \cosh( q_f F s ) \} \big]
\, ,
\end{eqnarray}
\end{subequations}
where $ F^{-1}$ is the inverse matrix of the field strength tensor, 
and the Lorentz indices are suppressed for 
notational simplicity. 
Inserting $\Delta(p|A_\FS)$ back to Eq.~(\ref{eq:GDp}) with the gauge field in the FS gauge (\ref{eq:FS}), 
we find the proper-time representation of the resummed propagator 
\begin{eqnarray}
S(p|A_\FS) 
=  \int _0^\infty \!\! ds 
\left[  \slashed p +  q_f  \gamma^\mu F_{\mu \nu} X^{\nu \lambda}(s) p_\lambda + m \right] 
{\rm e}^{-i\kappa^2s + i p^\mu X_{\mu \nu}(s) p^\nu + Y(s) }
\, .
\label{eq:electron} 
\end{eqnarray}
It is straightforward to confirm that the free propagator is reproduced in the vanishing field limit 
and to obtain the order-by-order expansion with respect to the external field in the weak field limit (see Appendix~\ref{sec:prop-weak}). 
As discussed in Sec.~\ref{sec:FS}, 
one should remember that the above result is obtained in the FS gauge, 
and transforms with the Schwinger phase under a gauge transformation as in Eq.~(\ref{eq:transf}). 





\subsubsection{Spectrum and spin polarizations}

\label{sec:parallel}

In this subsection, 
we shall more closely look into the proper-time representation (\ref{eq:electron}) 
and clarify the physical meaning of the spinor structures and the energy spectrum encoded therein. 
To this end, we first specify an explicit form of the field strength tensor. 
As discussed below \eref{eq:b}, one can choose the Lorentz frame 
in which the external electric and magnetic fields are parallel/antiparallel to each other. 
Therefore, without loss of generality, we work with this field configuration 
where the external electric and magnetic fields are both oriented in the $  z$ direction. 
This configuration also covers the cases 
where either of the electric or magnetic field vanishes 
as one can easily take such a limit.

In this Lorentz frame, the square of the field strength tensor has the diagonal form (\ref{eq:FF_mn}), 
so that the following calculations are greatly simplified. 
Especially, since the $  X^{\mu\nu}$ and $  Y$ in Eqs.~(\ref{eq:DscalarX}) and (\ref{eq:DscalarY}) 
are even functions of $ F^{\mu\nu} $, they also have diagonal forms: 
\begin{eqnarray}
X^{\mu \nu}(s)  &=& 
\frac{1}{q_f E} \tanh(q_f E s )\, g^{\mu\nu}_\parallel + \frac{1}{q_f B} \tan(q_f Bs)\, g^{\mu\nu}_\perp
\label{eq:Xpara} 
\, ,
\\ 
Y(s) &=& - \ln \big\{ \cosh(q_fEs) \cos(q_fBs) \big\}
\label{eq:Ypara}
\, .
\end{eqnarray}
The remaining ingredient is the spin coupling 
$ F^{\mu\nu} \sigma_{\mu\nu} = 2 i ( E \gamma^0 \gamma^3 - B \gamma^1 \gamma^2)$ in \eref{eq:Dscalar}. 
One can decompose the exponential factor there as 
\begin{eqnarray}
{\rm e}^{-i \kappa^2 s} &=& {\rm e}^{-i m^2 s}
\Big(  \cosh(q_f Es ) + \gamma^0 \gamma^3 \sinh(q_f Es)  \Big)
\Big( \cos(q_f Bs) - \gamma^1 \gamma^2 \sin(q_f Bs)  \Big)
\label{eq:exp_decomp}
\, ,
\end{eqnarray}
where we have utilized the fact that $\gamma^0 \gamma^3$ and $\gamma^1 \gamma^2$ commute with each other. 
The signs of $ E$ and $B $ only matter in this exponential factor related to spin configurations: 
The other parts of the propagator are even functions of 
$ E$ and $B $ as seen above. 
The decomposition in Eqs.~(\ref{eq:Xpara})--(\ref{eq:exp_decomp}) casts the resummed propagator (\ref{eq:electron}) into 
the following form 
\begin{eqnarray}
S(p|B, E) &=&
\int_0^\infty \!\!\!\! ds \Bigl[ \, 
 \frac{ 1- \gamma^1 \gamma^2 \tan(q_f B s) }{\cosh^{2}(q_f E s) } \slashed p _\parallel 
+ 
\frac{1+ \gamma^0 \gamma^3 \tanh(q_fEs) }{\cos^{2}(q_f Bs) } \slashed p _\perp  
\nonumber
\\
&& \hspace{1.5cm} 
+
m 
\big( 1- \gamma^1 \gamma^2 \tan(q_f Bs) \big)
\big( 1+ \gamma^0 \gamma^3 \tanh(q_f Es) \big)
\, \Bigr]
\nonumber
\\
&& \hspace{1.cm} \times
\exp \Big( \,
- i m^2 s 
+ i \frac{ p_\perp^2}{q_f B } \tan( q_f B s)
+ i \frac{ p_\parallel^2}{q_f E} \tanh( q_f E s)
\, \Big)
\label{eq:prop-q}
\, .
\end{eqnarray}
It should be noticed that there is a duality under 
the following simultaneous interchanges: 
\begin{eqnarray}
(B  , \, p_\perp^\mu , \,  \gamma^1 \gamma^2 ) &\longleftrightarrow  &
( iE , \, p_\parallel^\mu  , \, i \gamma^0 \gamma^3)
\label{eq:duality_q}
\, ,
\end{eqnarray} 
whereas the ordinary electromagnetic duality under the interchange $ B \longleftrightarrow E$ 
is broken due to the coupling to the matter fields. 


To investigate the basic properties of the resummed propagator (\ref{eq:prop-q}), 
we focus on a purely magnetic-field case 
without an electric field $(B \not = 0 ,\  E=0) $. 
The expression for a purely electric-field case $(E \not = 0 ,\  B=0) $
can be obtained by using the extended duality (\ref{eq:duality_q}). 
Taking the limit of the vanishing electric field in \eref{eq:prop-q}, one finds that 
\begin{eqnarray}
\label{eq:S_B}
S(p|B, E=0) &=& 
\int_0^\infty \!\!\!\! ds
\left[ \,
( \,  \slashed p_\parallel + m \, )
\big( 1- s_f \gamma^1 \gamma^2 \tan(\vert q_f B\vert s) \, \big) 
+ \slashed p _\perp   \cos^{-2}( \vert q_f B \vert s) 
\, \right]
\nonumber
\\
&& \hspace{1cm} \times
\exp \Big( \, 
- i m^2 s + i p_\parallel^2 s
+ i \frac{ p_\perp^2}{\vert q_f B\vert} \tan(\vert q_f B \vert s)
\, \Big)
\, .
\end{eqnarray}
As discussed in Sec.~\ref{sec:S-gauge}, the wave functions of charged particles in magnetic fields 
are given by the associated Laguerre polynomial $ L_k^\alpha(x)$ in the symmetric gauge 
that is nothing but the FS gauge when $ E=0 $. 
Therefore, it is natural to expect that the resummed propagator can be 
decomposed in terms of $ L_k^\alpha(x)$. 
To use the generating function of the associated Laguerre polynomial 
\begin{eqnarray}
(1-z)^{-(1+\alpha)} \exp \left( \frac{xz}{z-1} \right) 
= \sum_{n=0}^{\infty} L_n^\alpha (x) z^n
\label{eq:g_Laguerre}
\, ,
\end{eqnarray}
we put 
\begin{eqnarray}
z = - {\rm e}^{-2i \vert q_f B \vert s}
\label{eq:zzz}
\,  .
\end{eqnarray}
Then, the tangent in the exponential is rewritten in a desired form 
\begin{eqnarray}
\exp\Big( \, i \frac{ p_\perp^2 }{ \vert q_f B \vert } \tan( \vert q_f B \vert s) \, \Big)
= \exp\Big(\, - \frac{ u_\perp }{ 2 } \, \Big) \exp\Big(\, \frac{ u_\perp z}{z-1} \, \Big)
\, ,
\end{eqnarray}
where $ u_{\perp}= -2 p_\perp^2 /\vert q_f B \vert$. 
The second factor is identified with the exponential factor in the generating function (\ref{eq:g_Laguerre}), 
and thus can be decomposed into the polynomial of $  z$ by the use of \eref{eq:g_Laguerre}.
The other trigonometric functions in \eref{eq:S_B} are also rewritten in terms of $ z$.

After all, the fermion propagator is written by the polynomial of $  z$, 
leading to a simple expression in terms of the exponential factor of $  s$: 
\begin{eqnarray}
S(p|B) &=& 
2 \, e^{ -\frac{u_\perp}{2} }  
\sum_{n=0}^\infty (-1)^n \hat B_n (u_\perp) 
\, \int_0^\infty\!\!\! ds \, {\rm e}^{i ( p_\parallel^2 - m^2+ i \epsilon - 2n \vert q_f B \vert   ) s}
\, ,
\end{eqnarray}
where
\begin{eqnarray}
\hat B_n (x) &=& 
( \slashed p _\parallel + m ) \big\{ 
\prj_+ L_n(x) - \prj_-  L_{n-1} (x) \big\}
- 2 \slashed p _\perp L_{n-1}^1 (x)
\, .
\end{eqnarray}
Here, $ L_n(x) = L_n^0(x) $ and $ L_{-1} (x)=0 = L^1_{-1} (x)$ are understood. 
Now, the proper-time integral can be easily performed, 
and the resummed propagator is represented 
in the form of summation 
over the Landau levels \cite{Chodos:1990vv, Gusynin:1995nb}
\begin{eqnarray}
S(p|B) &=&
2 i 
\, e^{ \frac{ p_\perp^ 2}{|q_fB|} } 
 \sum_{n=0}^\infty 
\frac{ (-1)^n \hat B_n (u_\perp)  }{ p_\parallel^2 - m^2   - 2n \vert q_f B \vert   }
\label{eq:fermion_prop-LL}
\, ,
\end{eqnarray}
where $ p_\perp^2 = - \vert \bp_\perp\vert^2$. 
The origin of the spin projection operators $ \prj_\pm = (1\pm i s_f \gam^1 \gam^2)/2$ is traced back to the spin coupling $ F^{\mu\nu}\sigma_{\mu\nu}$ in Eq.~(\ref{eq:exp_decomp}) 
as encountered earlier in Eq.~(\ref{eq:spin-projection}). 
Note that the sign of $ q_f B$ only appears in the spin projection operator. 
As discussed below Eq.~(\ref{eq:prop-n}), 
the first two terms in $ \hat B_n (u_\perp) $ correspond to the degenerate spin eigenstates, 
while the last term originates from the overlap between the wave functions of those states. 
The last point is more clearly seen in 
the Ritus basis method discussed in Appendix~\ref{sec:relation}. 


There are an infinite number of poles located at $p_\parallel^2 - m^2 - 2n \vert q_f B  \vert =0$. 
The dispersion relation $p^0 = \pm \sqrt{ (p^3)^2 + m^2 + 2 n \vert q_f B \vert }$ represents nothing but 
the resultant energy levels from 
the Landau quantization and the Zeeman effect (cf. \fref{fig:Zeeman}). 
The propagator in the LLL is given by a particularly simple form 
\begin{eqnarray}
S_{\rm LLL} (p|B) 
=  2i\, {\rm e}^{ \frac{p_\perp^2 }{\vert q_f B \vert} } \frac{ \sla p_\para + m }{ p_\parallel^2 - m^2  }   \prj_+
\label{eq:prop-LLL}
\, ,
\end{eqnarray}
with $L_0(x)=1$. 
The LLL propagator only has one term because of 
the absence of the spin degeneracy. 
Notice that this expression is the same as 
the free fermion propagator in (1+1) dimensions 
up to two residual factors due to the presence of 
(frozen) spin $  \prj_+$ and the wave function 
in the transverse plane given by the Gaussian factor. 
This propagator is useful to compute physical quantities in the strong field limit.

\cout{
Applying the extended duality (\ref{eq:duality_q}) 
to Eq.~(\ref{eq:fermion_prop-LL}), 
we get the propagator in a constant electric field: 
\begin{eqnarray}
\label{eq:propagator-E}
S(p|E) &=&
2 i  \, e^{ -i \frac{ p_\para^ 2}{|q_fE|} } 
 \sum_{n=0}^\infty 
\frac{ (-1)^n \hat E_n ( i \frac{ 2 p_\para^ 2}{|q_fE|} )  }
{ p_\perp^2 - m^2   - 2 i n \vert q_f E \vert   }
\, ,
\end{eqnarray}
where
\begin{eqnarray}
\hat E_n (ix) &=& 
( \slashed p _\perp + m ) \big\{ 
\prj_+^E  L_n(ix) - \prj_-^E  L_{n-1} (ix) \big\}
- 2 \slashed p _\para L_{n-1}^1 (ix)
\, .
\end{eqnarray}
The spin projection operators $\prj_\pm $ are transformed 
to $\prj_\pm^E = (1 \mp s_f^E \gam^0 \gam^3)/2 $ with $ s_f^E = \sgn(q_f E)$. 
These operators can be rewritten as 
$\prj_\pm^E = (1 \mp s_f^E (i \gam^1 \gam^2) \gam^5 )/2 $
with the chirality and 
the spin component along the electric field. 
Polynomials $L_n(ix) $ do not have poles in the $p^0 $ plane, 
and neither does the propagator (\ref{eq:propagator-E}). 
This may be because charged particles in an electric field 
are not in definite energy eigenstates. 

}


\subsection{Charged scalar particles}

\label{sec:charged}

The proper-time representation of the propagator for charged scalar particles can be obtained in a similar manner 
as we have done for the fermion propagator in the preceding subsections. 
The Lagrangian of scalar QED is given by 
\begin{eqnarray}
\Lag = (D^\mu \phi)^\ast ( D_\mu \phi ) - m^2 \phi^\ast \phi
\label{eq:Lqed_s}
\, ,
\end{eqnarray}
where the covariant derivative (\ref{eq:covariantD-QED}) 
contains an external field. 
The propagator of the scalar particles 
obeys the equation of motion 
\begin{eqnarray}
( D^2 + m^2 )  G(x,y|A) = i \delta^4(x-y)
\, .
\label{eq:Gx_s}
\end{eqnarray}
Similar to the procedure in the preceding sections, 
this equation can be solved formally as 
\begin{eqnarray}
G(p|A) 
&=& i \times \frac{1}{i}
\int_0^\infty \!\! ds \ {\rm e}^{ is \left\{ (p - e A )^2 - m^2 + i \epsilon \right\} } 
\equiv \ \Delta_{\rm scalar} (p|A)
\label{eq:GDp_s}
\, .
\end{eqnarray}
Here, we notice a correspondence between $ \Delta (p|A) $ in Eq.~(\ref{eq:Dp}) for spin-$  \frac{1}{2}$ particles and 
$ \ \Delta_{\rm scalar} (p|A)$ in Eq.~(\ref{eq:GDp_s}) for scalar particles. 
They only differ by the spinor structure $ F^{\mu\nu} \sigma_{\mu\nu}$ 
which arises from the square of the Dirac operator $\slashed D \slashed D$ 
and thus is absent for spinless particles. 
We find a useful relation in a formal limit 
at $\sigma_{\mu\nu} \to 0 $ as 
\begin{eqnarray}
\Delta_{\rm scalar} (p|A) = \Delta (p|A) \, \vert_{\sigma^{\mu\nu}=0}
\, ,
\end{eqnarray} 
where one should get rid of an implicit unit matrix 
in the spinor space on the right-hand side. 
The absence of the spinor structure 
should result in the absence of the Zeeman effect 
in the energy spectrum of spinless particles 
as explicitly seen below. 
Also, we can apply the discussion around Eq.~(\ref{eq:transf}) to the Klein-Gordon equation (\ref{eq:Gx_s}) 
using the covariance of the Klein-Gordon operator 
to find the gauge-transformation property 
\begin{eqnarray}
G(x,y|A) =  {\rm e}^{ i \Phi_A(x,y) } \, G(x-y|A_\FS) 
\label{eq:transf-scalar}
\,  ,
\end{eqnarray}
where $ \Phi_A(x,y)$ is the Schwinger phase (\ref{eq:Phi_S}).

Borrowing the result for spin-1/2 fermions shown in Eqs.~(\ref{eq:Dscalar})--(\ref{eq:DscalarY}), 
one can immediately obtain the propagator of charged scalar particles 
\begin{eqnarray}
G(p|A) &=& \int _0^\infty \!\! ds \, 
{\rm e}^{-i (m^2-i\vep) s + i p_\mu X^{\mu \nu}(s) p_\nu + Y(s) } 
\,  .
\label{eq:scalar}
\end{eqnarray}
In the vanishing and weak field limits, 
the free propagator $ G(p|B = E=0) = i/(p^2 - m^2 ) $ 
and the perturbative expansion are reproduced from \eref{eq:scalar} (see Appendix~\ref{sec:prop-weak}). 
When $ E=0$, the resummed propagator is simplified and decomposed into the Landau levels as  
\begin{subequations}
\begin{eqnarray}
G(p|B, E=0) &=& \int_0^\infty \!\!\!\! ds
\frac{1}{\cos(q_f Bs)}
\exp \Big\{
- i (m^2-i\epsilon) s + i p_\parallel^2 s
+ i \frac{ p_\perp^2}{q_f B} \tan(q_f Bs)
\Big\}
\label{eq:G_B_scalar}
\\
&=&
2i\, {\rm e}^{- \frac{\vert \bp_\perp \vert^2 }{\vert q_f B \vert}}  
\sum_{n=0}^\infty (-1)^n L_n \left( \frac{\vert \bp_\perp \vert^2 }{2\vert q_f B \vert} \right) 
\frac{ 1 }{  p_\parallel^2 - m^2 - (2n+1)  \vert q_f B \vert   }
\label{eq:G_B_scalar-LL}
\, .
\end{eqnarray}
\end{subequations}
As mentioned in the previous section, 
the proper-time integral is convergent thanks to the infinitesimal imaginary part $i\epsilon$. 
In the second line,  the decomposed propagator 
has an infinite number of poles precisely at the Landau levels 
$\omega^2 = m^2 + p_z^2 + (2n+1)  \vert q_f B \vert  $ without the Zeeman effect. 
Again, the expression in a constant electric field $ E$ is obtained by using the extended duality (\ref{eq:duality_q}). 

\subsection{Simplest extension to QCD: Covariantly constant fields}

\label{sec:QCD_prop}

In this subsection, we further extend the resummation techniques discussed in the previous subsections for QED. 
Here, we will obtain the resummed propagators for quarks, gluons, and ghosts in the presence of an external chromo-electromagnetic field in QCD. 
In general, QCD contains much richer and more complex dynamics than QED, 
so that we confine ourselves to the so-called {\it covariantly constant fields}. 
As shown below, the covariantly constant chromo-electromagnetic field is decomposed into three Abelian fields, 
and thus one can apply the resummation techniques to QCD without a technical leap. 
We will use those propagators to derive 
the non-Abelian Heisenberg-Euler 
effective action in Sec.~\ref{sec:L_YM}.

We first briefly recapitulate the QCD Lagrangian in an external chromo-electromagnetic field, following the ``background field method.'' 
We shall start with the full QCD action of the SU$(N_{c})$ gauge group: 
\beq
S_{\rm QCD}
&=& \int d^{4}x \, \Big[
\bar{\psi} \left( i \sla D_{\mathcal{A}} - m \right) \psi 
-\frac{1}{4} \mathcal{F}_{\nonA \mu \nu}^{a} \mathcal{F}^{a \mu \nu}_\nonA  
\Big] \, ,
\label{QCDaction}
\eeq
where we use the following conventions of the covariant derivative 
\beq
D^{\mu}_{\mathcal{A}} = \partial^{\mu} - ig\mathcal{A}^{a \mu} t^{a} 
\, .
\label{QCDcovderivative}
\eeq
The associated field strength tensor is given by
$
\mathcal{F}^{a \mu \nu}_\nonA 
= \partial^{\mu} \mathcal{A}^{a \nu} - \partial^{\nu} \mathcal{A}^{a \mu} + gf^{abc} \mathcal{A}^{b \mu} \mathcal{A}^{c \nu} $. 
The generator of the non-Abelian gauge symmetry obeys the algebra 
$[ t^a, t^b] = i f^{abc} t^c$ and the normalization $\tr [ t^a t^b ] = C \delta^{ab}$ 
with $C =1/2  $ and $C= N_c=3  $ for the fundamental and adjoint representations, respectively. 
While we consider one-flavor case for notational simplicity, 
extension to multi-flavor cases is straightforward.

We divide the non-Abelian gauge field into a dynamical and external fields as 
\begin{eqnarray}
\nonA^{a\mu} = a^{a\mu} + \nonA_\ext^{a\mu}
\,  .
\label{eq:aA}
\end{eqnarray} 
Accordingly, the coupling between the quark and gluon fields in the covariant derivative 
is also divided as 
\begin{eqnarray}
\label{eq:pdp}
\bar \psi (i \slashed D_{\mathcal{A}}) \psi 
=
\bar \psi (i \slashed D) \psi - ig \bar \psi \slashed a^a (t^a) \psi
\,  .
\end{eqnarray}
Hereafter in this section, we frequently use the covariant derivative defined 
with the external chromo-field (without any suffix)
\begin{eqnarray}
D^{ \mu} \equiv  \partial^\mu - ig \nonA_\ext^{a\mu} t^a
\,  .
\label{eq:Dext}
\end{eqnarray}
Based on the decomposition (\ref{eq:aA}), one can also decompose the Yang-Mills part. 
As summarized in Appendix~\ref{sec:bgd}, this procedure provides 
the kinetic and interaction terms of the fluctuation field $a^a  $ in the presence of the external field. 
Then, the kinetic terms, from which we get the propagators in the external field, are obtained as 
\begin{eqnarray}
\Lag_{\rm kin}  &=& 
\bar \psi (i \slashed D-m) \psi  - \bar c^a (D^2)^{ac} c^c 
\nn
\\
&& 
- \frac{1}{2} a_\mu^a \left( 
- (D^2)^{ac} g^{\mu\nu}  + ( 1-  \frac{1}{\xi_{g} } )  D^{ ab \mu} D^{ bc\nu} 
+ i g ( \nonF_{\alpha\beta}^b \J^{\alpha\beta})^{\mu\nu} f^{abc} 
\right) a_\nu^c
 \, ,
 \label{eq:Lkg_d}
\end{eqnarray}
where the ghost field and the gauge parameter (for the dynamical gauge field) are denoted as $ c^a $ and $ \xi_g $, respectively. 
We also introduced the field strength tensor of the external field 
\begin{eqnarray}
\label{eq:Fext}
\nonF^{a \mu\nu}\equiv \partial^\mu \nonA^{a \nu }_\ext - \partial^\nu \nonA^{a\mu}_\ext
 - i g (t^b)^{ac} \nonA^{b\mu}_\ext \nonA^{c\nu}_\ext
 \, ,
\end{eqnarray}
and the generator of the Lorentz transformation $\J_{\alpha\beta}^{\mu\nu} 
 = i ( \delta_\alpha ^\mu \delta_\beta^\nu - \delta_\beta^\mu \delta_\alpha^\nu )$. 
Using those quantities, we can write the spin interaction term as 
$( \nonF_{\alpha\beta}^b \J^{\alpha\beta})^{\mu\nu}= \nonF^{b\alpha\beta} \J_{\alpha\beta}^{\mu\nu}=2i\nonF^{b\mu\nu}$. 
The interaction vertices in the external field are summarized in Appendix~\ref{sec:bgd}.

While we have not assumed any specific configuration of the external field in the above arrangement, 
we now focus on the covariantly constant external field 
which is introduced as an extension of constant Abelian fields that satisfy $\partial_\lambda F_{\mu\nu} = 0 $. 
For this extension, we require that a covariant derivative of the field strength tensor vanishes as follows 
\cite{Batalin:1976uv, Yildiz:1979vv, Ambjorn:1982bp, 
Gyulassy:1986jq, Suganuma:1991ha, Nayak:2005yv, Nayak:2005pf, Tanji:2010eu, Ozaki:2013sfa} 
\begin{eqnarray}
D^{ab}_\lambda \nonF^b_{\mu\nu} = 0
\label{eq:CCF}
\,  .
\end{eqnarray}
To find a solution for this condition, 
we evaluate a quantity $[ D_\lambda, D_\sigma] ^{ab} \nonF_{\mu\nu}^b$ in two ways. 
First, the above condition immediately leads to $  [ D_\lambda, D_\sigma] ^{ab} \nonF_{\mu\nu}^b = 0$. 
On the other hand, the commutator can be written by the field strength tensor 
and the structure constant. 
Therefore, the covariantly constant field satisfies a condition 
\begin{eqnarray}
f^{abc} \nonF_{\mu\nu}^b \nonF_{\lambda \sigma}^c = 0
\label{eq:fGG}
\,  .
\end{eqnarray}
Since the four Lorentz indices are arbitrary, 
this condition is satisfied only when 
the contractions of the color indices vanish. 
Therefore, we find the solution in a factorized form 
\begin{eqnarray}
\nonF^a_{\mu\nu} = \nonF_{\mu\nu} n^a
\label{eq:fact}
\, ,
\end{eqnarray}
where $n^a$ is a vector in the color space 
and is normalized as $n^a n^a = 1$. 
On the right-hand side of \eref{eq:fact}, the vector $ n^a $ represents the color direction, 
while an Abelian-like field $\nonF^{\mu\nu}$ does not carry the color index 
and only quantifies the magnitude of the external field.

Accordingly, the external gauge field 
in the covariant derivative (\ref{eq:Dext}) is also 
factorized into the color direction and the magnitude. 
As explained in Appendix~\ref{sec:cc}, the color structures in 
the covariant derivatives are diagonalized as 
\begin{subequations}
\begin{eqnarray}
D^{ij \mu} &=&  \delta^{ij} \left(  \partial^\mu - i w_i  { \nonA}_\ext ^\mu \right)
\label{eq:fund_ccf}
\, ,
\\
D^{ab \mu} &=&  \delta^{ab}  \left(  \partial^\mu - i v^{a} \nonA^\mu_\ext  \right) 
\label{eq:adj_ccf}
\, ,
\end{eqnarray}
\end{subequations}
where the first and second lines are for the fundamental and adjoint representations, respectively. 
Note that the Einstein summation convention is not applied to the color indices represented with the Roman alphabets. 
For $N_c=3$, the effective color charges $ w_k $ have the three components 
\begin{eqnarray}
w^k
= \frac{ g }{ \sqrt{3} }  \sin \left (  \frac{2}{3}k \pi - \theta\right)
\, , \ \ \ k=1,2,3
\label{eq:angle-fund-rep}
\, .
\end{eqnarray}
As shown in \eref{eq:angle-fnd-rep}, the color direction $ \theta $ is specified by the second Casimir invariant.  
The effective color charges $  v^{a}  $ for the adjoint representation are given by 
\begin{eqnarray}
\begin{array}{ll}
\displaystyle
v^a = \frac{g}{2  } \sin \left ( \frac{2}{3} a \pi -   \theta_{\rm ad}  \right ) \, , & a=1,2,3 \, , \\
\displaystyle
v^a = - \frac{g}{2  } \sin \left ( \frac{2}{3} a \pi -   \theta_{\rm ad} \right ) \, , & a=5,6,7 \, , \\
\displaystyle
v^a = 0 \, , & a=4,8 \, .
\end{array}
\label{eq:angle-adj-rep}
\end{eqnarray}
Again, the color direction $ \theta_{\rm ad} $ is given by a gauge-invariant form (\ref{eq:angle_adj}). 
Similarly, we also get the diagonal form of the spin-interaction term 
\begin{eqnarray}
ig (\nonF_{\alpha\beta}^b \J^{\alpha\beta})^{\mu\nu} f^{abc}  
= 
v^a \delta^{ac}(\nonF_{\alpha\beta} \J^{\alpha\beta})^{\mu\nu} 
\label{eq:spin-gluon}
\, .
\end{eqnarray}
With the help of these simplifications, 
we will obtain the resummed propagators in the covariantly constant fields below, 
which goes along the procedure in the previous sections for the Abelian field.

\subsubsection{Resummed quark propagator}

\label{sec:quark-cc}

In the covariantly constant field, the kinetic term of the quark field is given as 
\begin{eqnarray}
\bar \psi^i  (i \slashed D^{ij} -m ) \psi^j = 
\bar \psi^i i \delta^{ij} \! \left(  \slashed \partial - i w^i\, {/\!\!\!\! {\nonA}}_{\ext} -m \right) \psi^j
\, .
\end{eqnarray}
This means that the interaction with the external field does not 
cause a rotation of the color carried by the fermion, 
and thus that the color index only provides additive degrees just like the flavor index (within QCD). 
We can still use the Fock-Schwinger gauge for 
the Abelian-like field $\nonF^{\mu\nu}$ in Eq.~(\ref{eq:fact}). 
Therefore, the resummed propagator is obtained simply 
by using Eq.~(\ref{eq:electron}) with the replacement of the coupling constant (\ref{eq:angle-fund-rep}) as 
\begin{eqnarray}
q_f \rightarrow -w^i
\, .
\end{eqnarray} 
Explicitly, the Fourier transformation of the quark propagator ${\mathcal S}^{ij} (x,y|\nonA_\FS) =\langle 0 | T \psi^i (x) \bar\psi^j(y)|0\rangle$ reads  
\begin{eqnarray}
{\mathcal S}^{ij} (p|\nonA_\FS) = \delta^{ij} S^i (p|\nonA_\FS)
\, ,
\end{eqnarray}
where $  S^i (p|\nonA_\FS)$ denotes the fermion propagator (\ref{eq:electron}) with the effective charge, $  -w^i$. 
The Einstein notation is not applied to the index $  i$. 
We should introduce the Schwinger phase as in the Abelian case [cf., Eq.~(\ref{eq:transf})], 
which now has the color index in the diagonal form. 
Therefore, we have 
\begin{eqnarray}
{\mathcal S}^{ij} (x,y|\nonA_\ext) = \delta^{ij}\, {\rm e}^{i \Phi_A^i} \, {\mathcal S}^i (x-y|\nonA_\FS)
\, ,
\end{eqnarray}
where $ {\mathcal S}^i (x-y|\nonA_\FS)$ is the coordinate representation of $ {\mathcal S}^i (p|\nonA_\FS)$ 
and $  \Phi_A^i$ is the Schwinger phase (\ref{eq:Phi_S}). 

\subsubsection{Resummed ghost propagator}

\label{sec:ghost-prop}

The covariant derivative in the adjoint representation, 
is diagonal in the color structure as well. 
Therefore, each diagonal component of the ghost propagator 
is also obtained from the propagator (\ref{eq:scalar}) 
for charged scalar particles in the Abelian field. 
This amounts to the replacement of 
the coupling constant (\ref{eq:angle-adj-rep}) as 
\begin{eqnarray}
  q_f \to - v^a
  \, .
\end{eqnarray} 
Explicitly, we have 
\begin{eqnarray}
{\mathcal G}_{\rm ghost}^{ab} (p|\nonA_\FS) = \delta^{ab} G^a (p|\nonA_\FS)
\, ,
\end{eqnarray}
where $ G^a (p|\nonA_\FS)$ denotes the scalar propagator (\ref{eq:scalar}) 
with the effective charge $  -v^a$. 
According to the discussion around Eq.~(\ref{eq:transf-scalar}), the Schwinger phase should be introduced. 
This is simply done with the replacement of the charge as in the fermion propagator mentioned just above (but for the adjoint representation).

One should remember that the ghost and the Klein-Gordon fields have different statistics: 
The ghost fields anticommute with each other. 
Therefore, one should put a minus sign to each ghost loop 
as in the familiar Feynman rules without external fields.

\subsubsection{Resummed gluon propagator}

\label{sec:gluon-prop}


Finally, we examine the gluon propagator in the Feynman gauge (for the dynamical fields). 
Setting $\xi_g = 1$ in the Lagrangian (\ref{eq:Lkg_d}), the kinetic term reads 
\begin{eqnarray}
\Lag_{\rm Feynman}
=
- \frac{1}{2} a_\mu^a \delta^{ac} \left\{
- (D^2) g^{\mu\nu} +  v^{a} \left( \nonF_{\alpha\beta} \J^{\alpha\beta} \right)^{\mu\nu}
\right\} a_\nu^c
\,  .
\end{eqnarray}
Interactions with the covariantly constant external field do not induce color rotations. 
Therefore, the color structure of the resummed gluon propagator is factorized as 
\begin{eqnarray}
 \D_{\mu\nu}^{ac} (p|\nonA_\ext) = \delta^{ac} \D_{\mu\nu}^{(a)} (p|\nonA_\ext)
 \, .
\end{eqnarray}  
By using the trick discussed below \eref{eq:Gx}, 
the equation of motion for the gluon propagator in the momentum space is obtained as 
\begin{eqnarray}
\left[\ 
- ( p - v^a \nonA_\ext (p) )^2 g^{\mu\nu} 
-  v^{a} \left( \nonF_{\alpha\beta} \J^{\alpha\beta} \right)^{\mu \nu}
\ \right] \D^{(a)}_ {\nu \sigma} (p|\nonA_\ext) = i \delta^\mu_\sigma 
\label{eq:green}
\, ,
\end{eqnarray}
where summation of the color index is not assumed. 
Because of the covariance in the this expression, 
the gluon propagator again transforms with the same Schwinger phase as the quark and ghost propagators do. 

The propagator is formally given by the inverse of the operator in Eq.~(\ref{eq:green}), i.e., 
\begin{eqnarray}
\D_{ \mu \nu}^{(a)} (p|\nonA_\ext) = - i
\left[ \ 
 \left( p - v^a \nonA_\ext (p) \right)^2 \1 -v^a \left( \nonF_{\alpha\beta} \J^{\alpha\beta} \right)
\ \right] ^{-1}  _{\mu\nu}
= i \Delta_{\mu\nu}^{(a)} (p|\nonA_\ext)
\,  .
\label{eq:D0}
\end{eqnarray}
Similar to the case of the fermion propagator, 
we have exponentiated the inverse operator 
by the use of the proper-time integral 
\begin{eqnarray}
\Delta_{\mu \nu}^{(a)} (p|\nonA_\ext) = \frac{1}{i} \int_0^\infty \!\! ds \, {\rm e}^{ -is ( p - v^a \nonA_\ext(p) )^2} 
\left[ e^{- is v^a  \nonF_{\alpha\beta} \J^{\alpha\beta}} \right]_{\mu\nu}
\label{eq:D-gluon}
\, .
\end{eqnarray}
At this point, one can easily confirm that the above propagator reduces to the free propagator 
$\D^{(a) \mu \nu}_ {\rm free} (p) =  - \frac{i}{p^2}  g^{\mu\nu}$ in the vanishing field limit. 
Now, we adopt the FS gauge for the Abelian part of the covariantly constant field $  \nonA_\ext^\mu$. 
Then, the resummed gluon propagator obeys an equation 
\begin{eqnarray}
- \left[ \, \left( p^2  
+ \frac{(v^a)^2}{4}  \nonF^{\alpha \beta} \nonF_\beta^{\ \gam} \frac{\partial}{\partial p^\alpha} \frac{\partial}{\partial p^\gam} 
\right) g^{\mu\nu}
+ v^a ( \nonF_{\alpha\beta} \J^{\alpha\beta} )^{\mu\nu}
\, \right ] \Delta_{\nu\sigma}^{(a)}(p| \nonA_\FS) = \delta^\mu_\sigma
\label{eq:D-g}
\, .
\end{eqnarray}
As explained in Appendix~\ref{sec:resum_prop-gluon}, this equation can be solved by using an ansatz 
in the same manner as that we have done for the fermion propagator in Sec.~\ref{sec:prop-time}. 
In fact, after the replacement of the effective color charge $ q_f \to - v^a $, 
the gluon propagator can be written in terms of $X_{\mu\nu} (s)$ and $Y(s)$ 
already given in Eqs.~(\ref{eq:DscalarX}) and (\ref{eq:DscalarY}), respectively. 
Namely, the resummed gluon propagator in the Feynman gauge is obtained as 
\begin{eqnarray}
\D_{\mu\nu}^{(a)}(p|\nonA_{\rm FS}) = i \Delta_{\mu\nu}^{(a)}(p|\nonA_{\rm FS}) =
\int_0^\infty \!\!\! ds \ 
{\rm e}^{ - i p_\alpha X^{\alpha \beta} (s) p_\beta + Y(s) }
\exp \!\! \left[ - i s v^a \left( \nonF_{\alpha\beta} \J^{\alpha\beta} \right)
\right]_ {\mu\nu}
\label{eq:prop-gluon}
\,  .
\end{eqnarray}
Compared with the fermion propagator in \eref{eq:Dscalar}, 
the exponential factor, $\exp ( - i s q_f F_{\alpha\beta} \sigma^{\alpha\beta}/2 )$, 
is replaced by $\exp ( - i s v^a (\nonF_{\alpha\beta} \J^{\alpha\beta} ) )_{\mu \nu}$. 
This replacement originates from the difference between 
the spin degrees of freedom of fermions and the gluons, 
which should manifest themselves in the Zeeman effects. 
We will explicitly see this point below 
after the decomposition of the gluon propagator 
into the Landau levels. 



We shall closely look into the gluon spectrum 
and the spin structure encoded in the propagator (\ref{eq:prop-gluon}) as we did in Sec.~\ref{sec:parallel} for fermions. 
We again work in the Lorentz frame where the electric and magnetic fields 
are parallel/antiparallel to each other. 
Then, the expressions of $ X^{\mu\nu}(s) $ and $ Y(s) $ are simplified to 
those in Eqs.~(\ref{eq:Xpara}) and (\ref{eq:Ypara}) with the replacement of the charge $q_f \to - v^a  $. 
To decompose the exponential factor in \eref{eq:prop-gluon}, 
we introduce two antisymmetric tensors 
\begin{eqnarray}
&&
\epsilon_{\parallel }^{\mu \nu} 
=
\left(
\begin{array}{llll}
 & & & -1
\\
 & &0 &
\\
 & 0 & & 
 \\
1  & & &
\end{array}
\right) \, ,\ \ \ \  
\epsilon_{\perp }^{\mu \nu} 
=
\left(
\begin{array}{llll}
 & & & 0
\\
 & & 1 &
\\
 &- 1 & & 
 \\
0  & & &
\end{array}
\right)
\, ,
\end{eqnarray}
and separate the electric and magnetic components of the field strength tensor as $ \nonF^{\mu \nu} 
= \nonE \epsilon_{\parallel }^{\mu \nu} - \nonB \epsilon_{\perp}^{\mu \nu}$. 
Since those antisymmetric tensors are orthogonal to each other and have properties 
$\epsilon_{\parallel\sigma}^\mu \epsilon_{\parallel}^{ \sigma \nu} = g_{\parallel}^{ \mu \nu}$ 
and $\epsilon_{\perp \sigma}^\mu \epsilon_{\perp}^{ \sigma \nu} = - g_{\perp}^{ \mu \nu}$, 
one can arrange the even- and odd-power terms in the form 
\begin{eqnarray}
\exp \!\! \left[ - i s v^a \left( \nonF_{\alpha\beta} \J^{\alpha\beta} \right) \right]^{\mu\nu}
=
\cosh ( 2 v^a \nonE s ) g_{\parallel}^{ \mu \nu} + \sinh ( 2 v^a \nonE s ) \epsilon_{\parallel}^{ \mu\nu} 
+ \cos ( 2 v^a \nonB s ) g_{\perp}^{ \mu \nu} - \sin ( 2 v^a \nonB s ) \epsilon_{\perp}^{ \mu \nu} 
\label{eq:moment}
\, .
\end{eqnarray}
The gluon propagator is then given by 
\begin{eqnarray}
\D_{ \mu\nu}^{(a)} (p) = 
\int_0^\infty \!\!\! ds
\frac{ \exp \!\! \left[ - i s v^a \left( \nonF_{\alpha\beta} \J^{\alpha\beta} \right) \right]_{\mu\nu}}
 {\cos( v^a \nonB s)\cosh( v^a \nonE s)}
\exp \Big( \, 
- i \frac{ p_\perp^2}{v^a \nonB} \tan( v^a \nonB  s) - i \frac{ p_\parallel^2}{v^a \nonE} \tanh( v^a \nonE s) \, \Big)
\label{eq:reprop-glu}
\, ,
\end{eqnarray}
where the exponential factor is understood to be decomposed as 
already shown in \eref{eq:moment}. 
Similar to the case of the fermion propagator, 
there is an extended duality between the electric and magnetic fields [cf. Eq.~(\ref{eq:duality_q})]. 
Namely, the gluon propagator (\ref{eq:reprop-glu}) is invariant under the simultaneous interchanges 
\begin{eqnarray}
(\nonB, \,  \, p_\perp^\mu , \, g_{\perp}^{ \mu\nu} , \, \epsilon_{\perp}^{ \mu \nu}  ) 
&\longleftrightarrow  &
( i \nonE ,  \, p_\parallel^\mu, \, g_{\parallel}^{ \mu \nu} ,  \, i \epsilon_{\parallel}^{ \mu \nu})
\label{eq:duality_g}
\, .
\end{eqnarray}

Below, we shall elaborate the gluon propagator with a magnetic component without an electric component. 
The propagator with an electric component can be deduced from the expression with a magnetic component 
owing to the extended duality (\ref{eq:duality_g}). 
To use the generating function of the Laguerre polynomial (\ref{eq:g_Laguerre}), 
we introduce a variable\footnote{In \eref{eq:zzz} for the fermion propagator, 
we chose $ z=- {\rm e}^{-2i \vert q_f B \vert s }$ with an opposite sign in the exponential. 
This is because the bilinear $ p _\alpha X^{\alpha \beta} p_\beta$ had an opposite sign 
stemming from the overall sign conventions 
in the free propagators. 
} 
\begin{eqnarray}
z = - {\rm e}^{2i \vert v^a \nonB \vert s }
\, .
\end{eqnarray}
Then, the tangent appearing in \eref{eq:reprop-glu} is rewritten as 
\begin{eqnarray}
\exp\Big( - i \frac{ p_\perp^2 }{ v^a \nonB }  \tan( v^a \nonB s ) \, \Big)
=
\exp\Big( - \frac{u_\perp}{2} \Big) \exp \Big(  \frac{ u_\perp z}{z-1} \Big)
\label{eq:eqeq1}
\, ,
\end{eqnarray}
where $u_\perp = - 2 p_\perp^2 / \vert v^a \nonB \vert$ as in the case of the fermion propagator. 
Then, the last exponential factor in \eref{eq:eqeq1} can be decomposed into polynomials of $  z$ 
by the use of the generating function (\ref{eq:g_Laguerre}). 
In addition, all of the trigonometric functions appearing 
in the gluon propagator (\ref{eq:reprop-glu}) can be rewritten by $  z$. 
Therefore, after some arrangements, 
the gluon propagator in the covariantly constant magnetic field is found to be 
\begin{eqnarray}
\D^{\ab  \mu \nu} (p) 
&=& 
 2 \, 
{\rm e}^{-\frac{u_\perp}{2}} \sum_{n=0}^\infty
 (-1)^n \, \hat \nonB_{n}^{ \mu \nu}  
 \int_0^\infty ds  \, \e^{ - is (p_\parallel^2  - (2n-1) \vert v^a \nonB\vert) - \epsilon s}
\nn
\\
&=& 
 2 i \, 
{\rm e}^{-\frac{u_\perp}{2}} \sum_{n=0}^\infty
\frac{ (-1)^n \, \hat \nonB_{n}^{ \mu \nu} }{ p_\parallel^2 - (2n-1) \vert v^a \nonB\vert } \, ,
\label{eq:g-prop-B}
\\
\hat \nonB_{n}^{\mu  \nu} &=& L_{n}  (u_\perp ) \Q_+^{\mu\nu}
- L_{n-1} (u_\perp ) g_{\parallel}^{\mu \nu} 
+ L_{n-2}  (u_\perp ) \Q_-^{\mu\nu} 
\label{eq:B-g}
\, .
\end{eqnarray}
We promise again that the Laguerre polynomials with negative indices vanish, i.e., $L_{-1} = 0= L_{-2} $, 
and introduce new operators $  \Q_\pm^{\mu\nu} =  ( g_\perp^{\mu\nu} \pm i s^a \ep_\perp^{\mu\nu} )/2$ 
with $ s^a = \sgn(- v^a \nonB) $.\footnote{
The sign of the spin interaction depends on that of the (color) charge 
specified by the convention of the covariant derivative (\ref{QCDcovderivative}). 
In our conventions following those in a standard textbook \cite{Peskin:1995ev}, a particle carries a color charge, $ - g$, 
while a fermion carries an electric charge, $ +q_f $, 
as specified in the covariant derivative (\ref{eq:covariantD-QED}). 
} 

The pole positions in \eref{eq:g-prop-B} 
specify the gluon dispersion relations as 
$ (p^0)^2 = (p^3)^2 + (2n-1) \vert v^a \nonB \vert $. 
This is nothing but the spectrum of spin-1 particles 
in a magnetic field that results from 
the Landau quantization and the Zeeman effect (cf. \fref{fig:Zeeman}). 
Especially, the ground state $(n=0)$ takes a negative value, 
$ (p^0)^2 = p_z^2 -\vert v^a \nonB \vert $, when $ p_z^2 < \vert v^a \nonB \vert $. 
In this case, the energy $p^0 = \sqrt{p_z^2 - \vert v^a \nonB \vert }$ is a pure imaginary number, 
indicating the presence of a tachyonic mode called the Nielsen-Olesen instability \cite{Nielsen:1978rm, Nielsen:1978nk, Ambjorn:1980ms}.

The operators $  \Q_\pm^{\mu\nu} $ work as 
the helicity-projection operators for 
the transverse polarization modes. 
Those operators have 
the eigenvectors $ \vep^\mu_\pm = (0,1, \pm is^a ,0) /\sqrt{2}$ 
that have unit eigenvalues $\Q_\pm^{\mu\nu} \vep_{\pm \nu} = \vep_\pm^\mu $ and are orthogonal to the other eigenvectors $ \Q_\pm^{\mu\nu} \vep_{\mp \nu}  =0$. 
The eigenvectors are nothing but the circular polarization vectors 
with the spin directions oriented along the magnetic field. 
This helicity projection is also evident in the energy levels: 
The terms proportional to $\Q_\pm^{\mu\nu}  $ correspond to 
the lower and higher energy levels in the Zeeman splitting 
for a given $ n $ (as schematically shown in Fig.~\ref{fig:Zeeman} {\it albeit} for the nonrelativistic case). 
The second term in \eref{eq:B-g} corresponds to the longitudinal mode. 
The spin direction is perpendicular to the magnetic field, so that 
the energy level is not shifted by the Zeeman effect. 
When computing gauge-invariant quantities, 
this term is expected to be canceled out 
with the ghost contribution given in Sec.~\ref{sec:ghost-prop}.
We will explicitly confirm this cancellation in the derivation of the non-Abelian Heisenberg-Euler 
effective action in Sec.~\ref{sec:L_YM}.

\section{Heisenberg-Euler effective action}

\label{sec:HE} 

The previous sections were devoted to a review of the fundamental techniques of quantum field theories under strong electromagnetic fields. From this section, we are going to discuss applications of the machinery to various physical phenomena. As the first and important application, we discuss the Heisenberg-Euler effective action for low-energy electromagnetic fields by including quantum corrections from the one-loop vacuum fluctuations. 
We then discuss the Schwinger mechanism signaled 
by an imaginary part of the effective action, 
which indicates the electron-positron pair production from vacuum driven by electric fields.  
We have already seen that the effect of external fields on propagation of charged particles 
is in general nonlinear in nature and becomes non-negligible even if the coupling to the external field is small 
when the magnitude of the external field is strong.


It was Heisenberg and Euler who for the first time constructed the effective action for the electromagnetic fields \cite{Heisenberg:1935qt}. They obtained the effective action through the evaluation of the vacuum energy shift induced by electromagnetic fields before the systematic machinery of diagrammatic calculations or renormalization is established. 
Later, Schwinger applied the proper-time formalism, discussed in Sec.~\ref{sec:resum}, 
to the field-theoretical derivation of the Heisenberg-Euler effective action \cite{Schwinger:1951nm}. 
Since then, quantum field theories in external fields 
have been developed in various directions.

This includes, for example, analyses for the fields with spacetime inhomogeneity, computation of the effective action beyond the one-loop level, and application to the other gauge theories such as QCD. 
Note also that the physical processes induced by an electric field are not static phenomena but are inevitably dynamical ones 
because of the energy injection to the system by the electric field. 
There have been a number of studies addressing the real-time dynamics of the pair production 
including the back-reactions \cite{Cooper:1989kf, Kluger:1991ib, 
BialynickiBirula:1991tx, Kluger:1992gb, Tanji:2008ku, Gelis:2013oca, Taya:2016ovo, Copinger:2018ftr} 
(see also Refs.~\cite{Dunne:2004nc,Dunne:2012vv,Gelis:2015kya,Fedotov:2022ely} for reviews 
on the Heisenberg-Euler effective action, the Schwinger mechanism, and beyond).

Below, we first present a derivation of the Heisenberg-Euler effective action by utilizing the resummed propagators already obtained in the previous section (see also Ref.~\cite{Hattori:2020guh} for a review of the original derivation by Schwinger \cite{Schwinger:1951nm}). 
An analog of the Heisenberg-Euler effective action in QCD (and in the presence of both QCD and QED fields) will be discussed in Sec.~\ref{sec:HE_QCD}.  
For now, we only briefly mention that the quark-loop contribution in the chromo-electric field was applied to 
the description of the quark and antiquark pair production in the color flux tubes \cite{Casher:1978wy, Casher:1979gw} 
and the particle production mechanism in the relativistic heavy-ion collisions \cite{Biro:1984cf, Kajantie:1985jh, Gyulassy:1986jq}. 
See a review article \cite{Gelis:2015kya} for more developments in the particle production 
in the early-time dynamics of the relativistic heavy-ion collisions that leads to 
the creation of the quark-gluon plasma 
and a recent review article \cite{Fedotov:2022ely} 
for particle production in various systems.

\subsection{Resummed effective action}

\label{sec:resummed-action}

\begin{figure}[t]
\begin{center}
   \includegraphics[width=0.8\hsize]{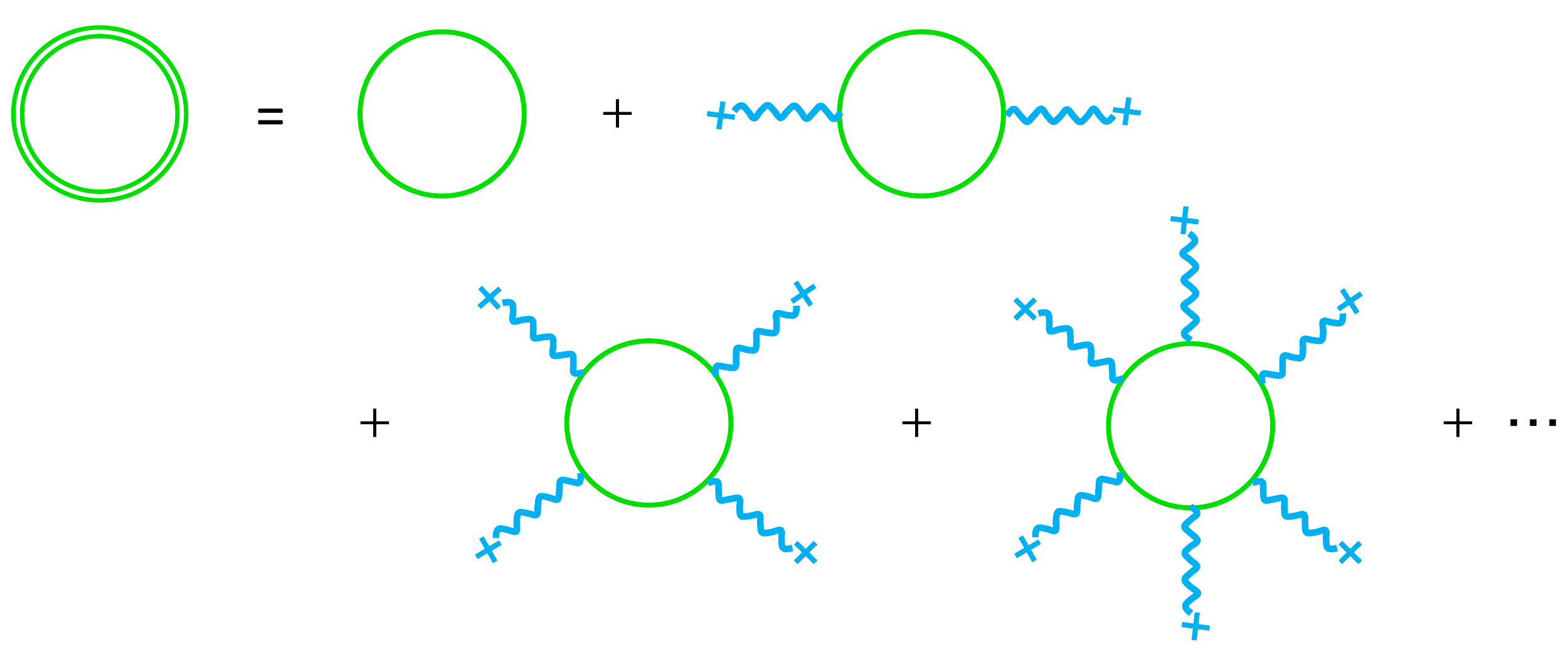}
\end{center}
\vspace{-0.5cm}
\caption{One-loop diagrams contributing to the Heisenberg-Euler effective action. 
The first diagram without external legs is subtracted in Eq.~(\ref{effectiveaction}). 
}
  \label{fig:OneLoopHE}
\end{figure}

The effective action for the gauge fields $S_{\rm eff}=\int d^4x\, {\cal L}_{\rm eff}[A_\mu]$ is formally obtained 
through the path integration of the QED action with respect to the fermion field: 
\beq
S_{\rm eff} [A_\mu] = -i \ln \det (i\sla D-m)=-\frac{i}{2}\ln \det (\sla D^2 + m^2)
\, ,
\label{Lag_original}
\eeq 
with $D_\mu=\del_\mu+iq_f A_\mu $ for the external field $ A^\mu $. 
The electrical charge is negative for electrons ($q_f=-|e|$). 
Without the dynamical photon field, this corresponds to a series of 
the one-loop contributions (see Fig.~\ref{fig:OneLoopHE}).\footnote{
As mentioned in the beginning of Sec.~\ref{sec:photon-splitting}, 
Furry's theorem states that the odd-order diagrams vanish identically 
in charge-conjugation even environments 
and that only even-order diagrams contribute to the series \cite{Furry:1937zz}. 
} 
Thus, together with the original Maxwell part ${\cal L}^{(0)}=-\frac14 F_{\mu\nu}F^{\mu\nu}$, 
we write the total effective Lagrangian as ${\cal L}_{\rm eff}={\cal L}^{(0)}+{\cal L}^{(1)}$. 
In the last equality of Eq.~(\ref{Lag_original}), we have used a relation $\det (i \sla D-m)= (-1)^d \det (i \sla D+m)$ 
obtained from the operation of the charge conjugation $C \sla D  C^{-1}=- \sla D ^T$ 
and $ \det(-1) = (-1)^4 =1 $ in the four dimensions.

As discussed in Sec.~\ref{sec:PP_general}, we can rewrite Eq.~(\ref{Lag_original}) by using the formula (\ref{propertime2}) with the proper time $s$: 
\beq
{\cal L}^{(1')}=\frac{i}{2}\int_0^\infty \frac{ds}{s}\, {\rm e}^{-\epsilon s} {\rm e}^{-ism^2} 
{\rm tr}\, \left[ 
\langle x | {\rm e}^{-i \hat Hs}|x\rangle - \langle x|{\rm e}^{-i \hat H_0 s}|x\rangle 
\right] \, ,
\label{effectiveaction}
\eeq
where we have recovered an infinitesimal parameter $\epsilon>0$. 
We factorized the trace over the coordinate space in the form of the corresponding integral measure 
and will take the remaining trace over the Dirac spinor space denoted as ``tr". 
The ``Hamiltonian" is defined as (see also Sec.~\ref{sec:PP_general}) 
\beq
\hat H= D^2 + \frac{q_f}{2}F^{\mu\nu}\sigma_{\mu \nu}
\label{Ham_fermion}
\, ,
\eeq
where we defined the Hamiltonian without including the mass term. 
The mass dependence only appears in the overall exponential factor. 
We subtracted the contribution in the absence of external fields, 
which corresponds to the second term in Eq.~(\ref{effectiveaction}) with the free Hamiltonian $\hat H_0=\del^2$. 

So far, there is no restriction to the configuration of the external electromagnetic fields, 
and thus the action (\ref{effectiveaction}) is valid for electromagnetic fields with arbitrary spacetime dependence. 
Here, we first focus on a constant electromagnetic fields without a spacetime dependence, 
which yields the original Heisenberg-Euler effective action. 
The non-constant cases will be discussed in Sec.~\ref{sec:HE_inhom}.

\subsubsection{Constant electromagnetic fields}

\label{sec:resummed-action-constant}

Consider the transition matrix element in the proper-time formalism:
\beq
K(x,x';s|A_{\rm FS})\equiv \langle x | {\rm e}^{-i \hat H s}|x'\rangle
\label{eq000}
\, ,
\eeq 
whose special case with $x=x'$ appears in Eq.~(\ref{effectiveaction}). As we discussed in Sec.~\ref{sec:PP_general}, we can compute this quantity by solving the Schr\"odinger-like equation with the proper time $s$ being the evolution parameter, or in the path-integral formalism. 
This is the strategy taken by Schwinger (see Sec.~III of Ref.~\cite{Schwinger:1951nm} and a review part in Ref.~\cite{Hattori:2020guh}). 
However, we can in fact utilize the fermion propagator under external fields 
which was already computed in Sections \ref{sec:prop-time} and \ref{sec:parallel}. 
While this derivation of the Heisenberg-Euler effective action is somewhat different from those known in the literature, 
we can take the shortest path by the efficient use of our previous results.

One finds that a proper-time integral of $K(x,x';s|A_{\rm FS}) $  corresponds to the coordinate representation of $\Delta(p|A)$ defined in Sec.~\ref{sec:prop-time} as 
\beq
\Delta (x,x'|A_{\rm FS})=\frac{1}{i}\int_0^\infty ds\, {\rm e}^{-im^2s} K(x,x';s|A_{\rm FS})
\label{eq111}
\, .
\eeq
For constant electromagnetic fields in the Fock-Schwinger gauge, $\Delta(x,x'|A_{\rm FS})$ becomes a function of the coordinate difference, and can be Fourier-transformed as 
\beq
\Delta(x-x'|A_{\rm FS})=\int\frac{d^4p}{(2\pi)^4}\, {\rm e}^{ip(x-x')}\Delta(p|A_{\rm FS})
\, . 
\label{eq222}
\eeq
Consider the parallel field configuration as defined in Sec.~\ref{sec:parallel}. Since we already know the explicit form of the propagator for this configuration, let us utilize it for computation of the effective action. Recall that $\Delta(p|A_{\rm FS})$ is defined by Eqs.~(\ref{eq:DscalarX})--(\ref{eq:DscalarY}). 
Using the explicit forms for the parallel field configurations 
in Eqs.~(\ref{eq:Xpara}) and (\ref{eq:Ypara}), we find that 
\beq
\Delta(p|A_{\rm FS})&=&\frac{1}{i}\int_0^\infty ds\, {\rm e}^{-im^2 s}
\big(1-\gamma^1\gamma^2 \tan (q_fBs)\big)\left(1+\gamma^0\gamma^3 \tanh (q_fEs)\right)\nonumber\\
&&\qquad \times \exp\Big(
i\frac{p_\perp^2}{q_fB}\tan(q_fBs) + i\frac{p_\parallel^2}{q_fE}\tanh(q_fEs)
\Big)\, .
\eeq
Now, comparing Eqs. (\ref{eq111}) and (\ref{eq222}), 
it is straightforward to read off
the expression for tr $\langle x | {\rm e}^{-i\hat Hs}|x\rangle $:
\beq
{\rm tr}\, \langle x | {\rm e}^{-i\hat Hs}|x\rangle &=& {\rm tr}\, K(x,x;s|A_{\rm FS})
\nonumber \\
&=& -\frac{i}{4\pi^2}\frac{q_fB}{\tan (q_fBs)}\cdot \frac{q_fE}{\tanh (q_fEs)}
\label{eq:K-fermion}
\, .
\eeq
The imaginary unit $i$ in the last line comes from the Gaussian integration over 
$p_\parallel^\mu=(p^0,0,0,p^3)$. 
Remember that the gauge dependence of $\Delta(p|A)$ 
has been factorized in the form of the Schwinger phase 
as mentioned below Eq.~(\ref{eq:Dp}). 
Therefore, the transition amplitude $ \langle x^\prime | {\rm e}^{-i\hat Hs}|x\rangle $ 
has the manifest gauge invariance in the coincidence limit $ x^\prime \to x $, 
and so does the Heisenberg-Euler effective Lagrangian 
represented by the closed loop diagrams in Fig.~\ref{fig:OneLoopHE} (see also discussions in Sec.~\ref{sec:FS}).

Plugging the matrix element (\ref{eq:K-fermion}) into Eq.~(\ref{effectiveaction}), we arrive at the Heisenberg-Euler effective action ${\cal L}_{\rm HE}={\cal L}^{(0)}+{\cal L}^{(1')}$ for the parallel field configurations: 
\beq
{\cal L}_{\rm HE}=\frac{1}{2}(E^2-B^2)+
\frac{1}{8\pi^2}\int_0^\infty \frac{ds}{s}\, {\rm e}^{-i(m^2-i\epsilon)s} 
\Big[\,
 \frac{ (q_f E  )(q_f B )}{   \tanh(q_fEs) \, \tan (q_fBs) }
-\frac{1}{s^2}
\, \Big]
\, , \label{HE_parallel}
\eeq 
where the second term in the bracket corresponds to the term in Eq.~(\ref{effectiveaction}), which was added to subtract the one-loop contribution without the external fields. The first term in the bracket is nonlinear with respect to the electromagnetic fields $E$ and $B$, which are multiplied by the electric charge $ q_f $ 
as they originate from the interactions with the fermion loop in Fig.~\ref{fig:OneLoopHE}. 
The effective action is invariant under the interchange $B \leftrightarrow iE$.


Recall that the quantities $a$ and $b$, which are the eigenvalues of a field strength tensor $F_\mu^{\ \, \nu}$, correspond to $E$ and $B$ in the parallel configurations. 
Therefore, by replacing $E$ and $B$ in the previous results by $a$ and $b$, we can easily obtain the Heisenberg-Euler effective action 
for arbitrary constant configurations as a function of 
those two Lorentz invariants [see Eqs.~(\ref{eq:a}) and (\ref{eq:b})]:
\beq
\hspace{-5mm}
{\cal L}_{\rm HE}=-{\mathscr F}&+&
\frac{1}{8\pi^2}\int_0^\infty \frac{ds}{s}\, {\rm e}^{-i(m^2-i\epsilon)s} 
\Big[\, 
 \frac{ (q_f a  )(q_f b )}{   \tanh(q_f a s) \, \tan (q_f b s) }
-\frac{1}{s^2}
\, \Big]
\, .\label{HE_general1}
\eeq
We can find several different representations of the Heisenberg-Euler effective action. Among them, the following expression is useful in the sense that the action looks obviously real (we will discuss the appearance of an imaginary part for $E\neq 0$ in the next subsection): 
\beq
{\cal L}_{\rm HE}=-{\mathscr F}-\frac{1}{8\pi^2}\int_0^\infty \frac{ds}{s}\, {\rm e}^{-m^2s} \left[q_f^2{\mathscr G}\, \frac{\Re e \left[\cosh \left\{q_f s \sqrt{2({\mathscr F}+i{\mathscr G})}\right\}\right]}{\Im m\left[\cosh \left\{q_f s \sqrt{2({\mathscr F}+i{\mathscr G})}\right\}\right]}-\frac{1}{s^2}-\frac23 q_f^2 {\mathscr F}\right]\, .\label{HE_general2}
\eeq 
Here we have replaced the proper time $s$ by the ``imaginary" proper time $s\to -is$ on the basis of the Wick rotation.

The last term in Eq.~(\ref{HE_general2}) has been added in order to renormalize the effective action: 
The last two terms cancel the divergent terms $1/s^2+(2/3)q_f^2{\mathscr F}$ that appear 
in the expansion of the first term with respect to $q_f$ (or weak fields) [cf. Fig.~\ref{fig:OneLoopHE}]. 
The term $1/s^2$ is the divergent vacuum energy density in the absence of external fields, 
which comes from a fermion loop without external fields attached. 
The term $(2/3)q_f^2{\mathscr F}$ has a quadratic dependence on the external fields 
that comes from the diagram with two insertions of the external field in Fig.~\ref{fig:OneLoopHE}. 
This diagram contains a logarithmic divergence and requires the charge renormalization. 
The higher-order diagrams do not yield UV divergences.\footnote{
The actual degrees of divergences are lower than 
the superficial ones due to the gauge symmetry~(see e.g., Ref.~\cite{Peskin:1995ev}). 
} 
To regularize the ultraviolet divergences, one can simply insert 
a cutoff as the lower boundary of the integral region.\footnote{Note that the proper time $s $ has a mass dimension ``$ -2$'', and the small $s $ regime corresponds to the ultraviolet regime.} 
This simple regularization, called the proper-time regularization, is a gauge-invariant one, 
which serves as one of the merits of the proper-time method. 
We will discuss renormalization in more detail in Sec.~\ref{sec:vacuum_magnetism}.

All the calculations discussed so far can be equally applied to scalar QED defined by Eq.~(\ref{eq:Lqed_s}). 
By using the resummed propagator (\ref{eq:scalar}) for scalar fields, one can easily obtain the effective action for scalar QED \cite{Schwinger:1951nm}.  
The result is obtained as 
(see also Ref.~\cite{Hattori:2020guh} for a comparison between scalar and spinor QED) 
\begin{eqnarray}
\label{eq:HE-scalar}
{\cal L}_{\rm scalar \ QED} &=& -
\frac{1}{16\pi^2}\int_0^\infty \frac{ds}{s}\,  {\rm e}^{-is(m^2-i\epsilon)} 
\left[\frac{ (q_f a )(q_f b ) }{\sinh (q_f a s) \sin (q_f b s)} - \frac{1}{s^2} \right] 
\, .
\end{eqnarray}


\subsubsection{Beyond the Heisenberg-Euler effective action}

\label{sec:HE_inhom}

The Heisenberg-Euler effective action is obtained for constant electromagnetic fields. However, as mentioned before, the very first expression (\ref{effectiveaction}) in the proper-time formalism is valid for any non-constant field. In many actual situations where strong fields appear, we encounter time-dependent and/or spatially inhomogeneous electromagnetic fields. Except for a few special cases some of which we will discuss later, 
it is difficult to exactly compute the effective actions 
in non-constant electromagnetic fields. 
We thus discuss some approximations.

A useful  method is the derivative expansion with respect to the electromagnetic fields. Since we treat the spacetime variation of the electromagnetic fields (instead of the gauge potentials), it is again convenient to work in the Fock-Schwinger gauge introduced in Sec. \ref{sec:FS}. 
According to the gauge condition (\ref{eq:FS-condition}), one finds an identity  
\begin{eqnarray}
x_\alpha^\prime  F^{\alpha \mu} (x)
= (1+x_\alpha^\prime \partial^\alpha ) A^\mu_\FS (x)
\,  , 
\end{eqnarray}
where $ F^{\alpha \mu} (x) = \partial^\alpha A^\mu_\FS (x) - \partial^\mu A^\alpha_\FS (x) $ 
and $ x^{\prime\mu} := x^\mu - x_0^\mu $. 
Performing a scale transformation $ x^{\prime\mu} \to \sigma x^{\prime\mu }$, 
we have  
\begin{eqnarray}
\sigma x_\alpha^\prime F^{\alpha \mu} (\sigma x^\prime+x_0)
= \frac{d}{d\sigma} [ \, \sigma A_\FS^\mu(\sigma x^\prime + x_0) \, ]
\,  .
\label{eq:FSdiffeq}
\end{eqnarray}
If the field strength tensor takes a constant value $  F^{\alpha \mu} (\sigma x^\prime+x_0) = F^{\alpha \mu}  $, 
one can easily integrate the both sides as 
\begin{eqnarray}
A_\FS^\mu( x) = \int_0^1 d\sigma \sigma (x_\alpha- x_{0\alpha}) F^{\alpha \mu} 
= - \frac{1}{2} F^{ \mu \alpha} (x_\alpha- x_{0\alpha}) 
\,  .
\label{eq:FSdiffeq-const}
\end{eqnarray}
This reproduces the constant case (\ref{eq:FS}). 
When the spacetime dependence is weak, one can organize a derivative expansion for the field strength tensor.  
Then, we are able to solve Eq.~(\ref{eq:FSdiffeq}) on an order-by-order basis of the derivative expansion 
to express the gauge potential in terms of the electromagnetic fields as 
\begin{eqnarray} 
A^{\rm FS}_\nu(x)&=&\frac{1}{2}(x^\mu-x_0^\mu) F_{\mu\nu}(x_0)+\frac13 (x^\mu-x_0^\mu)(x^\sigma - x_0^\sigma) \del_\sigma F_{\mu\nu} (x_0)+\cdots\nonumber\\
&=&\sum_{n=0}^\infty \frac{(x^\mu-x_0^\mu)(x^{\sigma_1}-x_0^{\sigma_1})\cdots 
(x^{\sigma_n}-x_0^{\sigma_n})}{n!\, (n+2)}\del_{\sigma_1}\del_{\sigma_2}\cdots \del_{\sigma_n}F_{\mu\nu}(x_0)\, .
\end{eqnarray}  
The leading term corresponds to the constant case (\ref{eq:FSdiffeq-const}). 
Since the spacetime derivative $\del_\mu$ has a mass dimension, it must be accompanied by a characteristic scale in the system, the electron mass $m_e$. The derivative expansion is a good approximation when the ratio $\del_\mu/m_e$ is small enough. In other words, the Heisenberg-Euler effective action is valid when the electromagnetic fields are almost constant over the Compton wavelength $\lambda=1/m_e$, which means $\del_\mu/m_e \ll 1$. 
The effective action in the derivative expansion will read 
\begin{eqnarray}
{\cal L}_{\rm eff}={\cal L}_{\rm HE}(F_{\mu\nu}) + \del_\sigma F_{\alpha\beta}\del_\rho F_{\gamma\delta}\, {\cal L}_1^{\sigma\alpha\beta\gamma\delta\rho}(F_{\mu\nu})+\cdots 
\end{eqnarray}
Calculation of the first correction to the Heisenberg-Euler action was explicitly done in Refs. \cite{Lee:1989vh,Gusynin:1995bc,Gusynin:1998bt}. 
In Ref.~\cite{Dunne:1998ni}, the result of the derivative expansion is compared with that from a solvable model.


\begin{figure}[t]
\begin{center}
   \includegraphics[width=0.4\hsize]{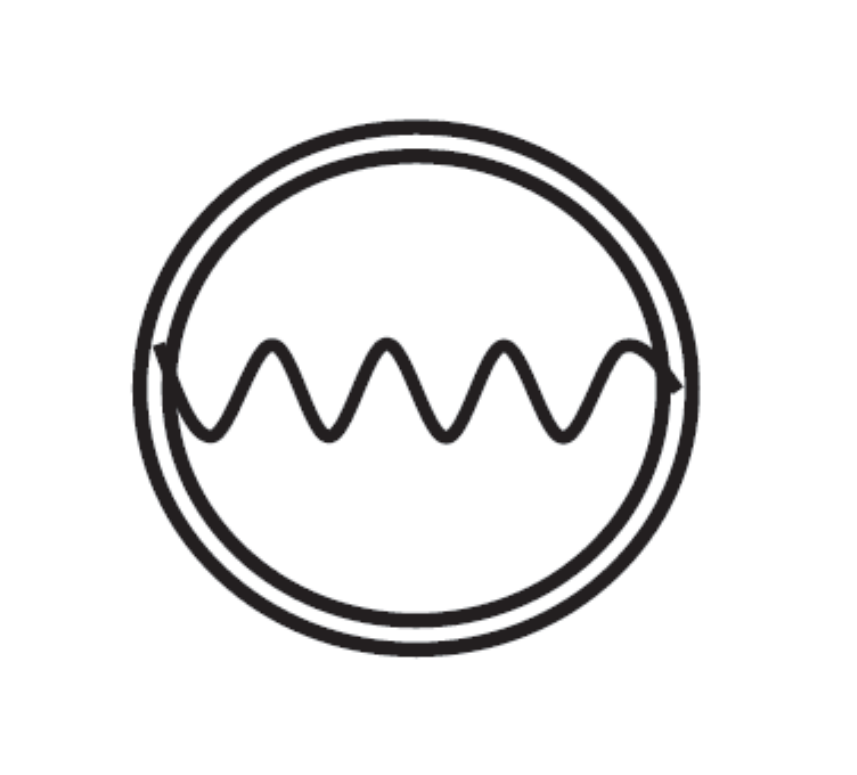}
\end{center}
\vspace{-10mm}
\caption{Two-loop diagram contributing to the Heisenberg-Euler effective action.
}
  \label{fig:TwoLoopHE}
\end{figure}

Another direction to go beyond the Heisenberg-Euler effective action is evaluation of the higher-loop contributions. In deriving the Heisenberg-Euler effective action, we treated the electromagnetic field as an external background field and did not treat it as a dynamical field. Let us allow a photon field to fluctuate around the external classical field: $A^\mu=A^\mu_{\rm ext}+a^\mu$. Then the path-integration over a photon fluctuation field $a_\mu$ in addition to the electron field $\psi ,\ \bar \psi$ formally defines the effective action including higher-loop effects. The two-loop diagram is just an electron's one-loop diagram with a virtual photon propagator connecting two points on the loop (see Fig.~\ref{fig:TwoLoopHE}). Compared to the one-loop diagram, this contribution is down by a factor of $e^2$ due to the coupling between a photon and an electron loop (at two vertices).\footnote{In QED with a small coupling constant, the higher-loop contributions are of course small compared to the one-loop contribution. However, when the coupling is relatively large as in QCD near the QCD scale, the higher-loop effects will become important.} It is in general difficult to exactly evaluate the higher-loop contributions, but a conjecture about the imaginary part of the effective action (corresponding to the pair production rate as discussed in the next subsection) has been suggested based on the results of strong-coupling and weak-field case \cite{Affleck:1981bma}. A concise review about higher-loop effects including relevant references is available in Ref. \cite{Huet:2017ydx} (see also Refs.~\cite{Gies:2016yaa, Karbstein:2019wmj}).

Since the physical degrees of freedom of the effective action are the low-energy electromagnetic fields, the effective action cannot be directly applied to the situations involving energetic particles. For example, if we wish to understand how energetic electrons or photons propagate in strong electromagnetic fields, we need to evaluate the self-energy diagrams 
that have two external lines with a large energy/momentum flow 
together with infinitely many insertions of external fields. 
Such an investigation also goes beyond the Heisenberg-Euler effective action. We will discuss this direction in Sec.~\ref{sec:photons}.

\subsection{Schwinger mechanism in constant fields}

\label{sec:pair-production}

\subsubsection{Pair production from purely electric fields}

As we will see shortly, the Heisenberg-Euler effective action acquires 
an imaginary part in a non-vanishing electric field $(E\neq 0)$. 
According to the Cutkosky's cutting rule, the presence of an imaginary part implies 
creation of on-shell particles out of, otherwise, virtual states forming bubble diagrams in vacuum. 
While the real part of the Heisenberg-Euler effective action describes electromagnetism, 
the production of electron-positron pairs in the electric fields is 
signaled by the emergence of an imaginary part \cite{Sauter:1931zz, Heisenberg:1935qt}. 
This is often called the Schwinger mechanism \cite{Schwinger:1951nm}.

In the presence of an external background field $A_\ext$, 
the vacuum state in the infinite past $\vert \Omega_{\rm in}\rangle $ is 
in general not equal to the one in the infinite future  $\vert \Omega_{\rm out}\rangle$. 
In the present case, an electric field will inject energy to the system in the form of the acceleration of the particles 
and antiparticles, which causes the mixing between the positive- and negative-energy states. 
Therefore, the definition of vacuum does not uniquely persist throughout the time evolution, 
and the vacuum persistence probability (VPP) $ \left| \langle \Omega _{\rm out} | \Omega _{\rm in} \rangle_{A_\ext } \right| ^2$ is not equal to unity. In fact, the transition amplitude defines the effective action 
\begin{eqnarray}
\label{eq:VPA}
\langle \Omega _{\rm out} | \Omega _{\rm in} \rangle_{A_\ext }
={\rm e}^{iW[A_\ext]}=\exp\Big\{ i\int d^4x {\cal L}_{\rm eff}[A_\ext]\Big\}
\, .
\end{eqnarray}
Accordingly, the origin of the transition from the initial to final vacua is identified 
with the particle-antiparticle pair production process from the vacuum, 
and is signaled by the emergence of the imaginary part in the effective action. 
Then, we define the complement of the VPP as  
\begin{eqnarray}
\Gamma \equiv 1 - \left| \langle \Omega _{\rm out} | \Omega _{\rm in} \rangle_{A_\ext } \right| ^2 
= 1 - {\rm e}^{ -2 \, \Im  m  \, W [A_\ext] }
\, .  \nonumber
\end{eqnarray}
When the imaginary part of the effective action is small, we have 
\begin{eqnarray}
\Gamma \simeq  2 \ \Im {m}\,  W [A_\ext]    
= 2 \integ d^4x \ \Im {m}\,  {\cal L}_\eff [A_\ext] 
\label{eq:Gam-Im}
\, .
\end{eqnarray}
Note that the VPP is by definition different from the pair production rate (PPR) 
that should be defined with the expectation value of the number operator for the produced pairs. 
Indeed, the direct computation of the PPR has shown that those quantities have 
similar but different forms \cite{Nikishov:1969tt, Nikishov:1970br}. 
We will come back to this issue in Sec.~\ref{sec:tunneling}.


Here, we compute the imaginary part of the Heisenberg-Euler effective action (\ref{HE_parallel}) 
in a constant electromagnetic field at the one-loop level. 
In Eq.~(\ref{HE_parallel}), the integrand for the proper-time integral 
has the form $ ds s^{-1} f(s) $ with $ f(s) $ being a real-valued and even function of $  s$. 
Thus, its imaginary part may be written as 
\begin{eqnarray}
\label{eq:L-imag}
\Im m {\cal L}_{\rm HE} &=&
\frac{1}{2i}\cdot \frac{1}{8\pi^2}\left[
\int_0^\infty \frac{ds}{s}\, {\rm e}^{-i (m^2-i\epsilon) s}f(s) 
- \int_0^\infty \frac{ds}{s}\, {\rm e}^{i (m^2+i\epsilon) s}f(s)\right]\nonumber\\
&=&\frac{1}{2i}\cdot \frac{1}{8\pi^2}\int_{-\infty}^\infty \frac{ds}{s}\, {\rm e}^{-i (m^2 - i \sgn(s)\epsilon ) s}f(s)
\, .
\end{eqnarray}
We can easily perform this integral with the help of the residue theorem. 
A positive-definite mass term $( m^2>0) $ suggests closing the contour in the lower-half plane (see Fig.~\ref{fig:contour_electric}). 
When analytically continued to the complex $s$-plane, 
the function $ f(s) $ has poles on the real and imaginary axes. 
The pole on the origin is common to the free-theory contribution, 
and is nothing to do with the Schwinger mechanism. 
Since the poles on the real axis are avoided due to 
the inclined contour specified by the $  \epsilon$ parameter, 
we should only pick up the poles on the imaginary axis at $s=-i  n\pi/(q_fE)$ $(n=1,2,\cdots)$.\footnote{ 
The poles in the lower half plane should be denoted by $s= - i  n\pi/|q_fE| \, (n=1,2,\cdots)$ when $ q_f E<0 $.
We will, however, suppress the symbol of absolute value for notational simplicity. 
}
Note that the poles on the real and imaginary axes are induced by a magnetic and electric field, respectively. 
This means that only an electric field is responsible for the emergence of the imaginary part. 
Therefore, we first consider the case only with an electric field. 
Effects of a magnetic field on the pair production will be examined in Sec.~\ref{sec:Effects_of_magnetic_field} 
for QED and in Sec.~\ref{sec:HE_QCD} for QCD.

\begin{figure}[t]
\begin{center}
   \includegraphics[width=0.6\hsize]{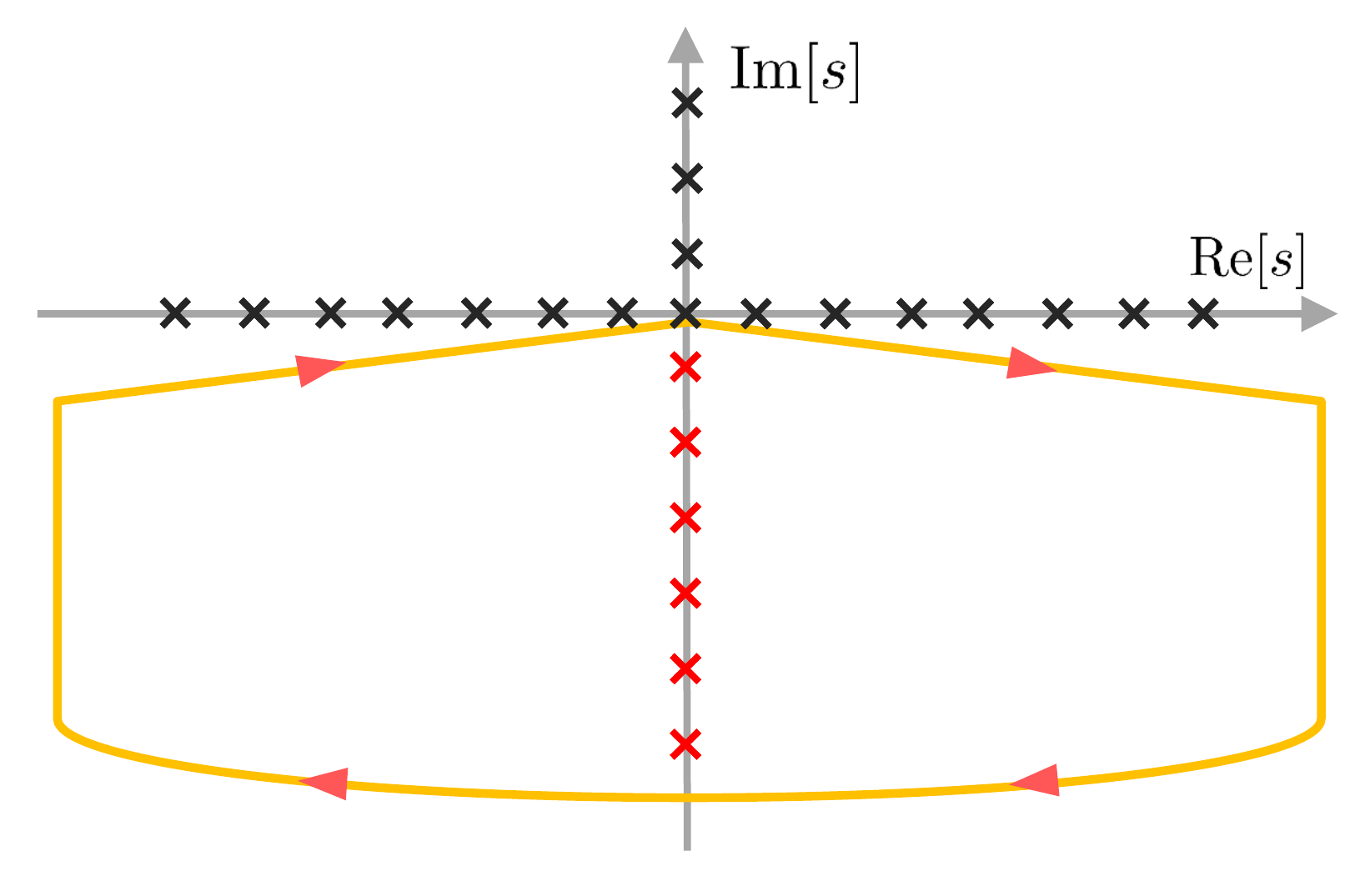}
\end{center}\vspace{-5mm}
\caption{
Contour on the complex $s$-plane with the poles on the real and imaginary axes 
induced by a magnetic and electric field, respectively. 
The path along the real axis is inclined with the $\epsilon  $ parameter. 
}
  \label{fig:contour_electric}
\end{figure}

The effective action in the presence of an electric field is 
easily obtained from the previous result (\ref{HE_parallel}) by taking the limit $B\to 0$: 
\beq
&&
{\cal L}_{\rm HE}(E\neq 0,B=0)
=\frac{1}{2}E^2+\frac{1}{8\pi^2}\int_0^\infty \frac{ds}{s^3}\, {\rm e}^{-i m^2 s}f_e(s)
\, ,
\label{HE_pureE}
\\
&&
f_e(s) \equiv s^2 \lim_{B\to0} f(s) = q_fEs \coth (q_f Es)-1  
\nn
\, .
\eeq
At this point, we took the limit $\epsilon\to 0$ (though this is not mandatory), 
since the integrand damps fast enough $f_e(s)/s^3\to q_fE/s^2$ as $s\to \infty$. 
Then, we obtain the integral value as the summation of the residues 
\begin{eqnarray} 
\Im m {\cal L}_{\rm HE}(E\neq 0,B=0)&=&
\frac{1}{2i}\cdot \frac{1}{8\pi^2}\cdot (-2\pi i) \sum_{n=1}^\infty {\rm Res} 
\left[\frac{{\rm e}^{-im^2 s}}{s^3}\, f_e(s)\right]_{s=-i\frac{n\pi}{q_fE}}
\nonumber
\\
& =& \frac{|q_fE|^2}{8 \pi^3} \sum_{n=1}^\infty \frac{1}{n^2 }\, {\rm e}^{-\frac{E_c}{E}n\pi}
\label{eq:SCH} 
\, ,
\end{eqnarray}
where the summation over $n$ can be represented by the dilogarithm function ${\rm Li}_2(z)=\sum_{n=1}^\infty {z^n}/{n^2}$ as $\Im m {\cal L}_{\rm HE}(E) = q_f^2E^2/(8\pi^3){\rm Li}_2({\rm e}^{-\frac{E_c}{E}\pi}) $. 
Here, we defined the critical electric field \cite{Heisenberg:1935qt} 
\beq
E_c \equiv \frac{m^2}{|q_f|}
\label{eq:E_cr}
\, .
\eeq
The first term of this series $ (n=1) $ in Eq.~(\ref{eq:SCH}) was obtained by Sauter in the form of the transmission rate 
between the negative- and positive-energy states \cite{Sauter:1931zz}. 
Then, Heisenberg and Euler recognized the correspondence between the Sauter's result 
and the imaginary part of their effective Lagrangian at $ n=1 $ 
as well as the presence of the other sequential poles (though an explicit result was not shown) \cite{Heisenberg:1935qt}. 
Sometime later, the complete form of the series was systematically derived by Schwinger \cite{Schwinger:1951nm}.

The imaginary part of the effective Lagrangian is a monotonically increasing function of $E$ 
and finally increases quadratically without the exponential suppression when the electric field exceeds the critical value $E_c$. 
Then, the VPP decreases significantly. 
The exponential factor crucially controls the visibility of the Schwinger mechanism in real systems. 
Nevertheless, even an infinitesimally weak electric field 
induces a finite imaginary part, though it is exponentially suppressed, which implies that the Schwinger mechanism is a quantum phenomenon. 
The Schwinger mechanism is also an inherently non-perturbative phenomenon because the exponential factor 
$\exp(-\frac{E_c}{E}n\pi)=\exp(-n\pi \frac{m^2}{q_fE})$ behaves non-analytically 
when the coupling constant $ q_f $ is formally sent to zero, 
which cannot be obtained from an finite order of the perturbative theory with respect to $q_f$. 
In other words, the pair production can be captured only when the infinitely many insertions of external fields are summed.

%
%

One may wonder whether the imaginary part of the effective Lagrangian 
provides the pair production rate (PPR) per unit spacetime volume. 
Since the PPR is essentially an expectation value of the number operator, 
there is no {\it a priori} reason that it coincides with the imaginary part representing the overlap of the vacua. 
It should be emphasized that an explicit comparison between those quantities has 
shown that they are indeed given by different expressions~\cite{Nikishov:1969tt, Nikishov:1970br}. 
The PPR is actually given by the first term ($ n=1 $) of the Schwinger's formula (\ref{eq:SCH}). 
Therefore, the two quantities approximately agree with each other 
only when the exponential factors are small. 
Those facts also suggest that one cannot interpret the $n $-th term in the summation as a probability for production of $n $ pairs: 
If it were so, the expectation value computed with 
this would-be probability 
should agree with the PPR. 
The difference between the VPP and PPR has been discussed repeatedly in pedagogical ways 
\cite{Ambjorn:1982nd, Holstein:1999ta, haro2003schwinger, Cohen:2008wz, Tanji:2008ku}, 
although there are yet confusions in some literature. 
We will come back to this issue in terms of the quantum-tunneling picture in the next subsection.

While the pair production has been discussed in the seminal papers \cite{Heisenberg:1935qt, Schwinger:1951nm} 
on the basis of the effective-action formalism as above, 
it actually may not be a suitable framework for explicitly describing the real-time dynamics of the pair production process 
since the fermionic degrees of freedom are already integrated out. 
To define the particle number and discuss the dynamical issues 
including the (kinetic) momentum distribution of the produced particles, 
one needs to begin with defining the particle and antiparticle states 
by the Bogoliubov transformation at each step of the time evolution \cite{Nikishov:1969tt, Nikishov:1970br, Ambjorn:1982bp, Ambjorn:1982nd, Ambjorn:1983ne}, 
since an amount of energy injected by the electric field inevitably causes 
the mixing between the positive- and negative-energy states. 
In other words, one needs to define the vacuum that varies dynamically. 
For those dynamical issues, the reader is referred 
to, e.g., Refs.~\cite{Tanji:2008ku, Gelis:2015kya, Hidetoshi:2017uok, Fedotov:2022ely} and references therein.

One can similarly compute the imaginary part for scalar QED \cite{Schwinger:1951nm}. 
The imaginary part of the effective action (\ref{eq:HE-scalar}) reads 
\begin{eqnarray}
\Im m \Lag_{\rm scalar\, QED}=\frac{|q_fE|^2}{16 \pi^3} \sum_{n=1}^\infty \frac{(-1)^{n-1}}{n^2 }\, {\rm e}^{-\frac{E_c}{E}n\pi}
\, .
\label{eq:SM-scalar}
\end{eqnarray}
Here, $q_f$ and $m$ in the critical field $E_c=m^2/|q_f|$ are the charge and mass of the complex scalar field. 
The result looks similar to Eq.~(\ref{eq:SCH}) for spinor QED except for 
an overall numerical factor and the alternating sign of each term in the summation over $n$. 
The statistics in the Gaussian path integral, which is now for the two degrees of freedom 
in a complex scalar field without the Grassmann nature, is reflected in the numerical factor, 
while the absence of the spin-interaction term in the Hamiltonian (\ref{Ham_fermion}) is reflected in 
both the numerical factor and the alternating sign (see Ref.~\cite{Hattori:2020guh} for the comparisons). 
As we mention in Sec.~\ref{sec:Effects_of_magnetic_field}, a difference between scalar and spinor QED is more prominent 
when there is a coexisting magnetic field \cite{Hattori:2020guh}. 
This is because the ground-state energy is lowered in spinor QED owing to the Zeeman shift 
that reduces the exponential suppression.

\subsubsection{Schwinger mechanism as quantum tunneling}

\label{sec:tunneling}

\begin{figure}[t]
\begin{center}
   \includegraphics[width=0.8\hsize]{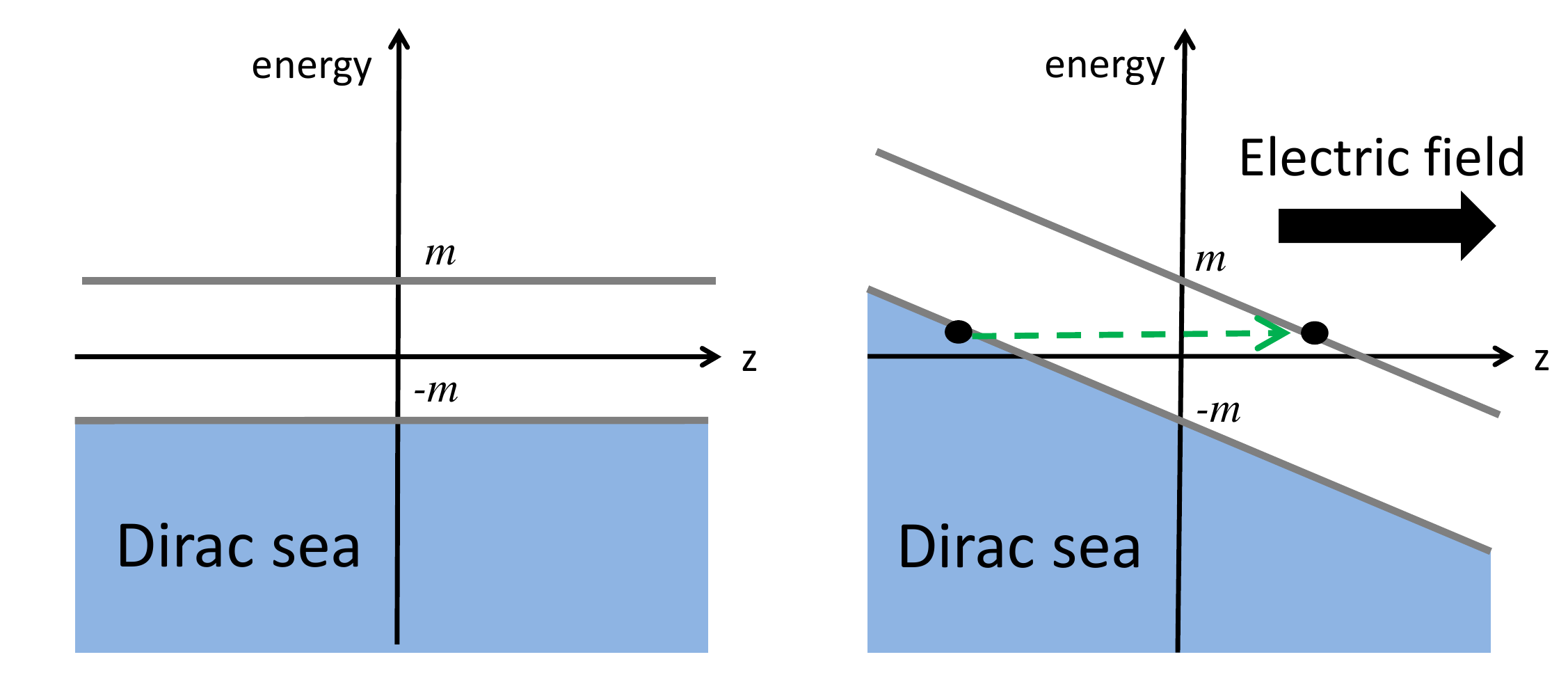}
\end{center}
\caption{
The static Dirac sea at $E=0$ (left) and the Schwinger mechanism as tunneling 
through the mass gap at $E\neq 0$ (right). 
}
  \label{fig:Schwinger}
\end{figure} 

The result in Eq.~(\ref{eq:SCH}) can be interpreted in an intuitive way as quantum tunneling. First of all, the critical electric field $E_c$ in Eq.~(\ref{eq:E_cr}) is consistent with the minimum work necessary for exciting a virtual pair of a 
fermion and an antifermion to become a real excitation: 
Namely, the $ E_c $ satisfies a relation $2|q_f|E_cd \sim 2m$ where $d\sim 1/m$ is the Compton length of a fermion and $2m$ on the right-hand-side is the threshold energy.\footnote{This is consistent with the validity condition for the Heisenberg-Euler effective action that the electromagnetic field should be almost constant over the scale of the Compton wavelength.} Next, the result is understood as the tunneling probability for a fermion in the Dirac sea going out across the mass gap~\cite{Casher:1978wy}. 
When an electric field is generated by a potential $A^0=-Ez$, 
the dispersion relation of a charged fermion is tilted in the coordinate space 
as $(\epsilon + q_fEz)^2-|{\bp}|^2=m^2$. 
For the notational simplicity, we take $ q_f E > 0 $. 
For fixed values of $\epsilon , |\bp_\perp| \not = 0$ and $p_z = 0$, 
one can define the tilted band gap $z_- < z < z_+ $ with $z_\pm= ( - \epsilon  \pm m_T)/q_fE$ and 
the transverse mass $m_T=\sqrt{m^2+|{\bp}_\perp|^2}$ (see Fig.~\ref{fig:Schwinger}). 
This band is the mas gap region and plays a role of a potential barrier for the pair production.
The Dirac sea is occupied in the vacuum and the quantum tunneling will take place 
from the top of the occupied state at $z=z_-$ to the vacant state at $z=z_+$. 
Therefore, the tunneling probability in the WKB approximation, for a fermion 
carrying a transverse momentum $\bp_\perp$, is given by 
\beq
P(\bp_\perp)=\exp\left( - 2\int_{z_-}^{z_+} dz \sqrt{m_T^2 - ( \epsilon+q_fEz)^2} \right) 
= \exp\left(-\frac{\pi m_T^2}{q_fE} \right)
\, .
\eeq

The tunneling probability defines the pair production rate (PPR) $ \kappa $ and 
vacuum persistence probability (VPP) in the summation and product forms, respectively, as 
\begin{subequations}
\begin{eqnarray}
&&
\kappa = \frac{1}{TL^3} \sum_{\rm spin}\sum_{t,z}\sum_{{\bp}_\perp } P(\bp_\perp)
\, ,
\\
&&
\label{eq:VPP-sum}
|\langle \Omega_{\rm out}| \Omega_{\rm in}\rangle_{A_{\rm ext}} |^2
=\prod_{\rm spin}\prod_{t,z}\prod_{{\bp}_\perp } \Big[1-P(\bp_\perp)\Big]
=\exp\Big[ \, \sum_{\rm spin}\sum_{t,z}\sum_{{\bp}_\perp } \ln \left[1-P(\bp_\perp)\right]  \, \Big]
\, ,
\end{eqnarray}
\end{subequations}
where $  T L^3  $ is the spacetime volume. 
The summation over transverse coordinates $x,y$ is replaced by the summation over $\bp_\perp=(p_x, p_y)$. 
Then, we take summation over small cells $(\Delta p_x, \Delta p_y, \Delta z, \Delta t)$ and replace each sum as\footnote{
One can evaluate the summation as follows \cite{Casher:1978wy, Glendenning:1983qq}: 
A pair creation occurs in a volume $ \Delta t \Delta z $. 
According to the uncertainty principle, we have $\Delta t = 2\pi/ \omega =\pi / m_T $, where 
the frequency $\omega $ counts how often the tunneling occurs 
and the threshold energy is given as $2m_T$. 
The longitudinal extension $\Delta z$ is given by 
the penetration length through the barrier $\Delta z =z_+ - z_-=2m_T/q_fE$.
}
\begin{eqnarray}
\sum_{\bp_\perp}=L^2\int\frac{d^2\bp_\perp}{(2\pi)^2}
\, , \qquad 
\sum_{t}=\frac{m_T}{\pi}\int {dt}
\, , \qquad 
\sum_{z}=\frac{q_fE}{2m_T}\int dz
\, .
\end{eqnarray}
On the basis of these replacements, we immediately get the PPR 
\begin{eqnarray}
\kappa 
=  \frac{|q_f E|^2}{4\pi^3} \exp\left(-\frac{\pi m^2}{q_fE} \right)
\, .
\end{eqnarray}
As for the VPP, we define $ \gamma $ 
such that $|\langle \Omega_{\rm out}| \Omega_{\rm in}\rangle_{A_{\rm ext}} |^2=\exp( - T L^3 \gamma )$. 
Then, we have 
\begin{eqnarray}
\gamma
= \frac{|q_fE|^2}{4\pi^3}\sum_{n=1}^\infty \frac{1}{n^2}\exp\left(-\frac{\pi m^2}{q_fE} n \right)\, ,
\label{eq:gamma-4}
\end{eqnarray}
where the summation appears from an expansion $\ln (1-x)=-\sum_{n=1}^\infty \frac{1}{n}x^n$. 
The tunneling picture reproduces the VPP obtained from the Heisenberg-Euler effective action, i.e., 
$\gamma = 2 \Im m {\cal L}_{\rm HE}(E\neq 0,B=0)$ [cf. Eqs.~(\ref{eq:VPA}) and (\ref{eq:SCH})] 
\cite{Casher:1978wy, Glendenning:1983qq, Holstein:1999ta, Kim:2000un, Kim:2003qp}. 
As mentioned earlier, the PPR $ \kappa $ coincides with the first term 
of the series ($  n=1$) in the VPP $ \gamma $, 
which confirms the previous observations \cite{Nikishov:1969tt, Nikishov:1970br, Holstein:1999ta, 
haro2003schwinger, Cohen:2008wz, Tanji:2008ku}. 
This point is already clear in Eq.~(\ref{eq:VPP-sum}) 
where the expansion of the logarithm results in the summation over $ n $: 
\begin{eqnarray} 
\gam = \sum_{n=1}^\infty \frac1n 
\Big[ \, \frac{ 1} {T L^{d}} \sum_{\rm spin}\sum_{t,z}\sum_{{\bp}_\perp }   P(\bp_\perp)^n \, \Big]
\, .
\end{eqnarray}
This expression is valid for a general (d+1)-dimensional case 
with the (d-1)-dimensional transverse-momentum integral $ \sum_{{\bp}_\perp }   $. 
Thus, $\kappa $ is given by the first term in $ \gamma $ in any dimensions. 
For example, the corresponding expressions in the (1+1)-dimension are obtained as 
\begin{subequations}
\begin{eqnarray}
&&\kappa 
 = \frac{q_f E}{2\pi} e^{ - \frac{\pi m^2}{q_f E} }
 \, ,
\\
&& \gam 
 = \frac{q_f E}{2\pi}  \sum _{n=1}^\infty \frac1n e^{ -  \frac{\pi m^2}{q_f E} n } 
 \, ,
\end{eqnarray}
\end{subequations}
where there is no spin degrees of freedom or transverse-momentum integral 
in the pure (1+1) dimensions. 
Without the transverse-momentum integral, the order of the inverse $  n$ factor 
is lower than that in the four dimensional case (\ref{eq:gamma-4}). 
This makes a prominent difference in the strong-field limit or the small-mass limit 
where the exponential factor reduces to unity as $m^2/(q_fE) \to 0  $. 
In such a limit, the summation over $  n$ diverges in the (1+1) dimensions \cite{Cohen:2008wz, Hidaka:2011dp, Hidaka:2011fa}, while it remains finite in the (3+1) dimensions 
where $ \sum_{n=1}^\infty 1/n^2 = \zeta(2) = \pi^2/6 $. 
Note that $ P(\bp_\perp) $ itself depends on the spatial dimensions via 
the transverse momentum in $ m_T $. 
Creating pairs with nonzero transverse momentum causes 
an additional energy cost for filling up the transverse phase space. 
Accordingly, a suppression of $ P(\bp_\perp)^n $ by a Gaussian factor $ \exp[ - n \pi |\bp_\perp|^2/(q_fE)] $ 
results in a faster convergence of the series in a higher-dimensional system. 
Once again, one should note that the $  n$-th term cannot be interpreted as 
the probability of $ n $-pair production \cite{Brezin:1970xf}. 
In the context of the worldline instanton, the summation over $ n $ is reproduced 
from a periodic motion along a classical path parametrized by the proper time \cite{Dunne:2005sx, Dunne:2006st} (see also Ref.~\cite{Affleck:1981bma}).

\begin{figure}[t]
\begin{center}
   \includegraphics[width=0.8\hsize]{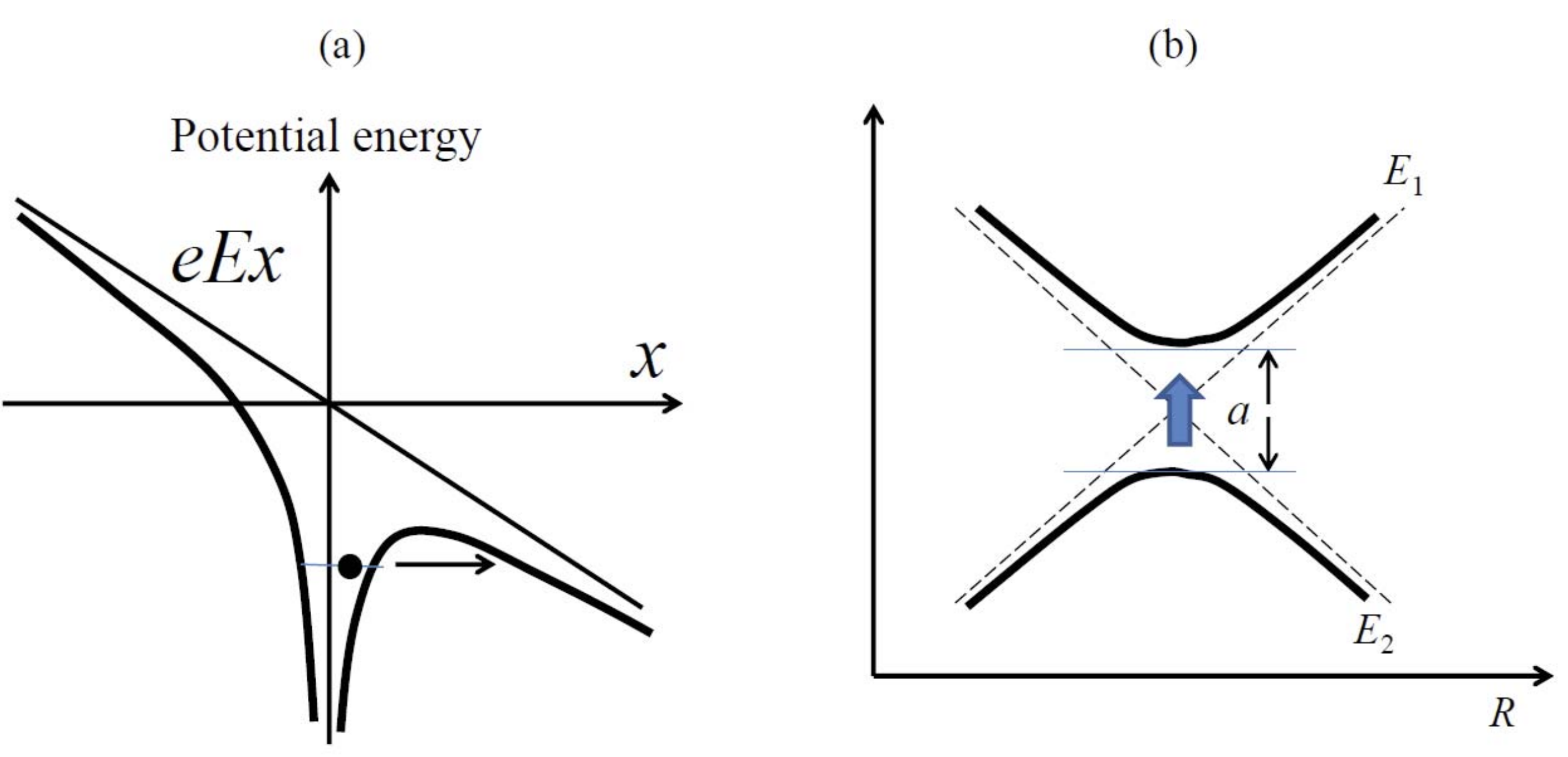}
\end{center}
\caption{(a) Ionization of a hydrogen atom by a static electric field. 
A bound electron can escape from the atom due to quantum tunneling. 
(b) 
Diabatic transition between two states at the avoided crossing point with an energy gap $a$. 
Figure shows two energy levels that typically appear in a system of two molecules at a distance $R$. 
} 
\label{fig:Ionization-LZ}
\end{figure}

The tunneling interpretation of the Schwinger mechanism reminds us of 
the essential similarity with 
the other related phenomena. 
For example, as shown in Fig.~\ref{fig:Ionization-LZ}, 
the PPR is very similar to the ionization rate of a hydrogen atom by a static electric field $E$ 
\cite{Oppenheimer:1928zz, Keldysh:1965ojf}
(see also Refs.~\cite{popov2004tunnel, Dunne:2004nc, keldysh2016dynamic} for more details) 
\begin{eqnarray}
P_{\rm ionization}=\exp\left(-\frac23 \frac{m^2e^5}{E}\right)
\, ,
\end{eqnarray}
and the Landau-Zener formula for non-adiabatic transition \cite{Zener:1932ws, zener1934theory} 
(see also section 90 in Ref.~\cite{Landau:1991wop})
\begin{eqnarray}
P_{\rm LZ}=\exp \left(-\frac{2\pi a^2}{\left|\frac{d}{dt}(E_1-E_2)\right|} \right)
\, ,
\end{eqnarray}
where $a$ is the closest energy gap between two levels (corresponding to the mass gap in the Schwinger mechanism), and $\frac{d}{dt}(E_1-E_2)$ is a velocity of the energy-level difference under the change of time dependent parameters (corresponding to the applied electric field). 
The interested reader is referred to Ref.~\cite{Oka:2011kf} 
for further discussions about similarities to condensed matter systems.

\subsubsection{Effects of magnetic fields on the pair production}
\label{sec:Effects_of_magnetic_field}

When both electric and magnetic fields are applied in vacuum, 
we need to generalize the formula (\ref{eq:SCH}) to include effects of a magnetic field. This extension can be simply carried out by the use of the Lorentz invariants $ a $ and $b  $. Therefore, we start with the expression in Eq.~(\ref{HE_general1}).

The integrand has poles both on the real and imaginary axes in the $s$ plane. 
Nevertheless, the actual integral contour is displaced downward from the real 
axis because of the infinitesimal imaginary part $ i \epsilon $, as discussed 
below Eq.~(\ref{eq:L-imag}). Therefore, we can again close the integral 
contour in the lower half plane as shown in Fig.~\ref{fig:contour_electric}, 
and only need to pick up the residues on the imaginary axis. 
Then, we obtain the the imaginary part of the effective Lagrangian and the PPR $  \kappa$: 
\begin{subequations}
\label{eq:SCHab}
\begin{eqnarray}
&&
 2{\Im}m\, {\cal L}_{\rm HE}(a,b)
= \frac{q_f^2}{4 \pi^2} \sum_{n=1}^\infty \left[ \, \frac{ab}{n} \coth \left(\frac{b}{a}n\pi\right) \, \right]
{\rm e}^{-\frac{E_c}{a}n\pi}
\, ,
\\
&&
\kappa (a,b) = \frac{q_f^2}{4 \pi^2} \left[ \, ab\coth \left(\frac{b}{a}\pi\right) \, \right]
{\rm e}^{-\frac{E_c}{a}\pi}
\, .
\end{eqnarray}
\end{subequations}
As discussed above, the PPR $  \kappa$ coincides with the first term of the series ($ n=1 $) in $  2{\Im}m\, {\cal L}_{\rm HE}(a,b) $. 
Of course, this result reproduces the previous result (\ref{eq:SCH}) for purely electric fields 
in the limit $ a \to E,\, b \to 0 $, 
and the imaginary part vanishes when $ a\to0 $. 

%
%

Now, we fix the magnitude of the electric field $ |\bE| $, and investigate 
how a magnetic field modifies the PPR as compared to the one 
in a purely electric field. To see a dependence on the relative direction 
between the electric and magnetic fields, we first consider two particular 
configurations in which those fields are applied in parallel/antiparallel 
and orthogonal to each other.

When a magnetic field is applied in {\it parallel/antiparallel} to 
the electric field, we have $a=|\bE|$ and $b=|\bB|$. Compared with 
the purely electric field configuration, we get a finite $b$ without 
changing the value of $ a $. Thus, there is no modification in the exponent. 
The prefactor, shown between the square brackets, should be compared with 
$E^2/\pi n^2$ in the purely electric case. It has a positive first derivative 
with respect to $ b $ for any positive value $ a,b > 0 $. 
One can also show that the prefactor is always larger than $E^2/\pi n^2$. 
This means that the PPR (\ref{eq:SCHab}) is enhanced 
as we increase the magnetic-field strength (with a fixed electric field). 
Therefore, the {\it parallel/antiparallel} magnetic field ``catalyzes'' the pair production. 
We will discuss a more intuitive picture below on the basis of the Landau-level decomposition.

When a magnetic field is applied in {\it orthogonal} to the electric field, i.e., when $\G=0$ (with $ \F \not = 0 $), 
any field configuration reduces to either a purely electric or magnetic field by a Lorentz transform. 
Since the production rate is a Lorentz scalar, 
we may evaluate the PPR in such a particular frame 
and then go back to the original frame in which we have the orthogonal field configuration. 
When $\F<0$, i.e., $|\bB| < |\bE|$, we have $a = \sqrt{|\bE^2 - \bB^2|} $ and $ b=0 $. 
Therefore, we observe a pair production induced by a purely electric field with a strength $  a$. 
Because of a smaller value  $a < |\bE|$, the PPR is suppressed by the magnetic field. 
As we increase the magnetic field (with the electric field fixed), 
we have a smaller $ a $ and observe a stronger suppression. 
As we further increase the magnetic field, we reach the ``crossed-field'' configuration specified by $ \F = \G =0 $, 
where the electric and magnetic fields have the same strengths. 
In this case, we have $a=b=0$, and thus observe no pair production eventually. 
With a stronger magnetic field, we have $\F>0$, i.e., $a=0$, 
so that we again find a vanishing PPR, $\kappa = 0  $. 
Therefore, the {\it orthogonal} magnetic field suppresses the PPR 
and eventually prohibits the pair production when $|\bE| \leq |\bB|$.

In the presence of an orthogonal magnetic field, a fermion and antifermion drift in the same direction 
perpendicular to both the electric and magnetic fields. 
This cyclic motion prevents the pair from receding from each other along the electric field, 
which may cause a suppression of the PPR. 
In the beginning of Sec.~\ref{sec:tunneling}, we discussed a semi-classical picture 
in which the critical electric field strength is interpreted as a minimum strength 
that can separate a pair over the Compton length. 
Therefore, it is reasonable to compare the Compton length ($ = 1/\sqrt{|q_f E_c|} $) 
and the cyclotron radius ($ = 1/\sqrt{|q_fB|} $). 
In the presence of a magnetic field stronger than the electric field, i.e., $ |B| > |E_c| $, 
the cyclotron radius is smaller than the Compton length. 
Therefore, there is little chance for a pair to separate over the Compton length. 
However, in the opposite case, i.e., $ |B| < |E_c| $, 
the cyclotron radius is larger than the Compton length. 
Therefore, a pair can be separated from each other along the electric field 
in a cycle of the cyclotron motion, which may lead to a finite, {\it albeit} suppressed, production rate.

\begin{figure}
     \begin{center}
              \includegraphics[width=0.6\hsize]{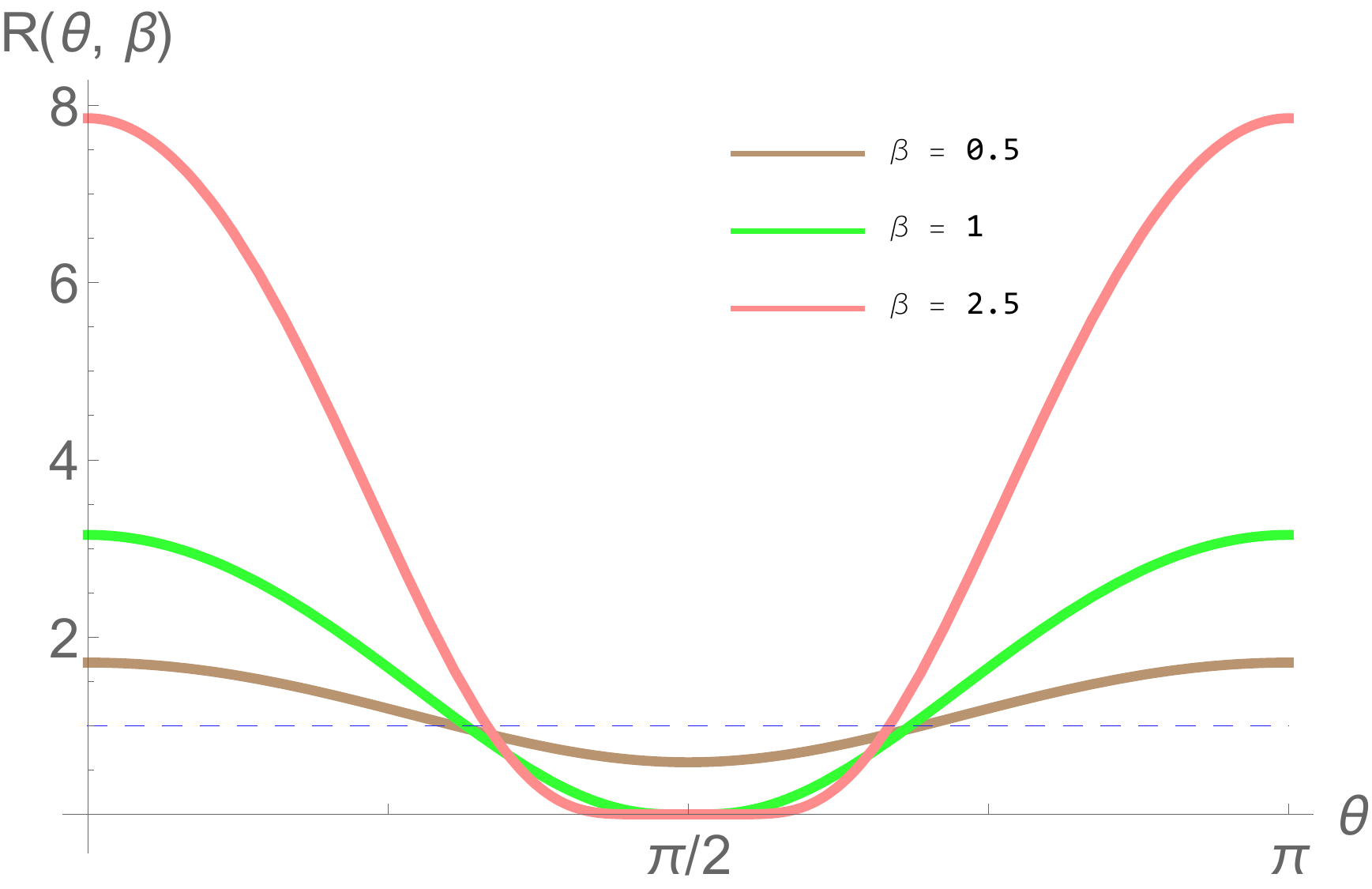}
     \end{center}
\vspace{-0.5cm}
\caption{Effects of a magnetic field on the pair production rate as a function 
of the relative strength $ \beta \equiv |B/E| $ and the relative angle $ \theta $.}
\label{fig:EandB}
\end{figure}

For a general configuration with an arbitrary angle, we shall see a numerical plot. 
We focus on the PPR at the critical field strength $ |E| = E_c $ which may be of our interest. 
Nevertheless, qualitative features shown below do not depend on the magnitude of the electric field. 
We investigate the ratio 
\begin{eqnarray}
R(\theta, \beta) \equiv 
\frac{\kappa(a,b) }{ \kappa(E,0) } 
=
\frac{ a  b\coth \left(\frac{ b}{a} \pi\right) e^{-\frac{|E|}{ a} \pi}}
{ (2\pi)^{-1} E^2  e^{- \pi}}
\, ,
\end{eqnarray}
where we set $ |E|=E_c $. 
This ratio can be regarded as a function of a normalized magnetic-field strength $ \beta \equiv |B/E|$ and of the relative angle $  \theta$ between the directions of the electric and magnetic fields. 
In Fig.~\ref{fig:EandB}, all the aforementioned behaviors at $ \theta = 0, \pi$ and $ \theta=\pi/2$ are confirmed, and are interpolated for a general value of $ \theta$.

\subsubsection{Landau-level decomposition}

\label{sec:Sch-LL}

In the presence of a parallel/antiparallel magnetic field, we could pursue the origin of the enhanced PPR 
by performing the Landau-level decomposition as we have done in Sec.~\ref{sec:parallel}. 
To see this, we apply the formula (\ref{eq:g_Laguerre}) to the transition amplitude (\ref{eq:K-fermion}) 
before performing the transverse-momentum integral~\cite{Hattori:2020guh}. 
Then, we get 
\beq
{\rm tr}\, \langle x | {\rm e}^{-i\hat Hs}|x\rangle 
&=& 2 \sum_{n=0}^{\infty} g_n (q_f B)\, (-  z)^n
\left[  \frac{ i }{(2\pi)^2} \frac{\pi q_f E }{ i \tanh(q_f Es)} \right]
\label{eq:K-fermion-Landau}
\, ,
\eeq
where we defined $ z = - \exp \left( -2i  q_f B  s \right) $. 
The longitudinal-momentum integral has been performed, and the transverse-momentum integral is left in the form 
\begin{eqnarray}
g_n(q_f B) = 2
\int \frac{d^2p_\perp}{(2\pi)^2} \, {\rm e}^{ \frac{p_\perp^2}{q_f B} } \, (-1)^n L_n^{-1}\left( - \frac{2p_\perp^2}{|q_f B|}  \right)
\, .
\end{eqnarray}
Note that the transition amplitude (\ref{eq:K-fermion-Landau}) has been completely 
factorized into the longitudinal (1+1)-dimensional part and the residual transverse part. 
The integral in the transverse part can be performed as\footnote{
Putting $I_n := \int_0^\infty d\zeta e^{ -  \zeta/2 }  L_{n}^{-1}\left(  \zeta  \right) $, 
one can find a simple recursive relation $ I_{n+1} = - I_{n} $ for $n \geq1$. 
Use the derivative formula $ d L_n^\alpha(\zeta)/d\zeta = - L_{n-1}^{\alpha+1}(\zeta) $ 
and the recursive relation $  L_{n+1}^{\alpha-1} (\zeta)=  L_{n+1}^{\alpha}  (\zeta) -  L_{n}^{\alpha}  (\zeta) $ 
to show this relation, and then $ L_0^{-1}(\zeta) =1  $ and $ L_1^{-1}(\zeta) = - \zeta $ to get $ I_0 = 2 $ and $I_1 = -4  $, respectively \cite{AssociatedLaguerrePolynomial}. 
} 
\begin{eqnarray}
g_n(q_f B) 
=  \frac{|q_f B|}{ 4 \pi}  (-1)^n \int_0^\infty d\zeta \, {\rm e}^{ -  \zeta/2 }  L_{n}^{-1}\left(  \zeta  \right)
= \kappa^{}_n\frac{|q_f B|}{2\pi} 
\label{eq:g-transverse}
\, .
\end{eqnarray}
Interestingly, the dependence on the integer index appears only in the prefactor $  \kappa^{}_n =2 - \delta_{n0}  $. 
This result for the integral is actually anticipated, since this is nothing but the Landau degeneracy factor for each Landau level 
with $  \kappa^{}_n  $ being the spin degeneracy factor for the two-fold degenerated hLLs.

Inserting those expressions into Eq.~(\ref{effectiveaction}), we obtain the Heisenberg-Euler Lagrangian 
in the Landau-level representation: 
\beq
{\cal L}_{\rm HE}=\frac{1}{2}(E^2-B^2) 
+ \sum_{n=0}^\infty \left[ \kappa^{}_n \frac{|q_f B|}{2\pi} \right]
\frac{i}{4\pi}\int_0^\infty \frac{ds}{s} {\rm e}^{-i(m_n^2-i\epsilon)s}   \frac{ q_f E }{ \tanh(q_f E s) }
\, .
\label{HE_parallel-Landau}
\eeq
We did not include the counter term since there is no divergence in the imaginary part. 
We have defined the effective mass $ m_n^2 = m^2 + 2 n |q_f B|  $ 
where the integer $ n $ specifies the Landau level as foreseen from our prior experience 
that the Landau levels are eigenstates of the Dirac operator. 
The covariant form is obtained by simple replacements, $ |\bE| , \, |\bB| \to a , \, b  $ (see Ref.~\cite{Hattori:2020guh}). 
In fact, one can directly obtain the same result from the Heisenberg-Euler action 
for the parallel field configuration (\ref{HE_parallel}) by the use of an identity 
\cite{Karbstein:2019wmj, Hattori:2020guh}: 
\begin{eqnarray}
\label{eq:cot-exp}
\cot x 
= i\left[1+ \frac{ 2 \e^{-2i  x} }{1-\e^{-2i  x} }  \right]
= i \sum_{n=0}^\infty \kappa_n \, \e^{-2in x}  
\, .
\end{eqnarray}
We can also apply this formula to $\cot (q_f b s)$ in the HE effective action (\ref{HE_general1}) 
for arbitrary configuration to get an infinite-series expansion. 
As long as $  b \not = 0$, there exists such a Lorentz frame where 
it reduces to the magnetic-field strength $ b = |\bB| $. 
Accordingly, we can identify the Landau levels in that frame. 
One may not consider the vanishing limit $ b \to 0$ naively, 
since this limit and the summation over $ n$ do not commute with each other. 
The physical origin of this non-commutativity comes from the fact that 
the infinite tower of the discrete Landau levels collapses into the ground state 
if a finite value of $  b $ is not maintained in the summation form. 
Therefore, the summation must be performed before the limit $ b \to 0$.

As before, one can obtain the imaginary part of the effective Lagrangian 
by picking up the residues of the poles on the imaginary axis: 
\begin{eqnarray}
2 \Im m {\cal L}_{\rm HE} 
&=& \frac{2}{2i}\cdot  \frac{i}{4\pi} 
\sum_{n=0}^\infty  \left[ \kappa^{}_n \frac{|q_f B|}{2\pi} \right]
\Im m \int_0^\infty \frac{ds}{s} 
{\rm e}^{-i(m_n^2-i\epsilon)s} \frac{ q_f E }{ \tanh(q_f E s) }
\nn
\\
&=&   \sum_{n=0}^\infty 
\left[ \kappa^{}_n \frac{|q_f B|}{2\pi} \right] 
\left[  \frac{|q_f E|}{2\pi}  \sum_{\sigma=1}^\infty  \frac{1}{\sigma}\, {\rm e}^{-  \frac{E_c^n }{| E|} \sigma \pi }  \right]
 \label{imaginary_E-Landau}
\, .
\end{eqnarray}
Alternatively, one can again apply the expansion (\ref{eq:cot-exp}) to the previous results (\ref{eq:SCHab}). 
Here, we have defined $ E_c^n = m_n^2/|q_f| $ for each Landau level. 
Therefore, there is an infinite number of the ``Landau-Schwinger limits'' specifying 
the critical field strengths for the pair production. 
It is quite natural that the exponential suppression is stronger for 
the higher Landau level that has a larger energy gap measured from the Dirac sea. 
Once we overcome the exponential suppression with a sufficiently strong electric field, 
the PPR is enhanced by the Landau degeneracy factor \cite{Hidaka:2011dp, Hidaka:2011fa, Hattori:2020guh}.  
This is because an energy provided by the external electric field can be consumed only to fill up the 
one-dimensional phase space along the magnetic field, and the degenerated transverse phase space can be filled without an additional energy cost. 
Although we have an infinite tower of the Landau levels, the hLL contributions are exponentially suppressed in the strong magnetic field limit. 
In contrast, the LLL contribution is still enhanced by the Landau degeneracy factor as long as $ E \gtrsim E_c^{n=0} $. 
The simple point is that the lowest critical field strength, $ E_c^{n=0} = m^2/|q_f| $, is independent of the magnetic field, 
while the higher levels are gapped out in a strong magnetic field. 
This one-dimensional structure of the phase space is the origin of the catalysis phenomenon, 
leading to the enhancement of the PPR found in Fig.~\ref{fig:EandB}.\footnote{In Sec.~\ref{sec:dmr}, 
we will discuss consequences of the effective dimensional reduction in more detail. 
One can get a covariant form by replacing $ E,B $ by $ a,b $ in Eq.~(\ref{imaginary_E-Landau}), respectively. 
The quickest way to get the LLL contribution is 
to simply take the strong magnetic field limit $  |B/E| \to \infty $ in Eq.~(\ref{eq:SCHab}). 
} 
Consistent with those observations, the expression between the second brackets 
in the Landau-Schwinger formula (\ref{imaginary_E-Landau}) is nothing but the (1+1)-dimensional Schwinger formula. 
The pair production in the parallel electric and magnetic fields was recently studied 
with the kinetic equation from the Wigner-function formalism as well \cite{Sheng:2018jwf}.

Now, it is a straightforward exercise to get the imaginary part for scalar QED \cite{Hattori:2020guh}: 
\begin{eqnarray}
\label{eq:scalarQED-Landau}
2 \Im m {\cal L}_{\rm scalar \, QED} 
&=&  \left[ \frac{|q_f B|}{2\pi} \right]  \sum_{n=0}^\infty \left[  \frac{|q_f E|}{2\pi} 
 \sum_{\sigma=1}^\infty  \frac{(-1)^{\sigma-1}}{\sigma}\, {\rm e}^{-  \frac{E_c^n }{| E|} \sigma \pi }  \right]
\, .
\end{eqnarray}
In addition to the alternating sign mentioned previously, 
there appear two-fold differences between spinor and scalar QED 
because of the absence of spin degrees of freedom. 
First, there is not a factor of $ \kappa_n $ without a spin degeneracy in each Landau level. 
Second, the critical field strength is now given by 
$ E_c^n = \{ m^2 + (2 n +1 ) |q_fB| \}/|q_f| $ 
due to the absence of the Zeeman shift. 
The ``zero-point energy'', that increases with $ |q_fB| $, yields a huge difference 
in the exponential suppression factor. 
Thus, the Schwinger mechanism is suppressed in scalar QED 
as we increase the strength of the parallel magnetic field.

\subsection{Schwinger mechanism beyond constant fields}

\subsubsection{Pair production from time-dependent electric fields}

We have discussed the Schwinger mechanism in a constant electric field as a nonperturbative phenomenon. 
Recall that the pair production is described by the imaginary part of the Heisenberg-Euler effective action, and thus is diagrammatically represented as a one-loop diagram with a cut. 
Since there is no energy-momentum flow at each interaction vertex from a constant electric field, 
massive particles have a chance of becoming on-shell excitations 
only when their dispersion relations are modified by the infinitely many insertions. 
This is the rough understanding for the reason why the Schwinger mechanism is a nonperturbative process. When the electric field is time-dependent, however, the situation is qualitatively different. A generic time-dependent electric field $E(t) )$ can be Fourier-transformed into a power spectrum $\tilde E(\omega) $. Thus, if the power spectrum has a large support above the threshold energy $\omega \simge 2m$, particle production becomes possible with only a few insertions of electric fields. In other words, the pair production in a generic time-dependent electric field contains both nonperturbative and perturbative contributions.

If the characteristic energy $\omega_0$ in the spectrum $\tilde E(\omega)$ is much {\it smaller} than $m$ (meaning that we can ignore the time-dependence) and the strength of electric field $E_0$ at that energy is {\it larger} than the critical electric field $E_c$, 
the pair production predominantly occurs with the nonperturbative mechanism. 
On the other hand, if the characteristic energy $\omega_0$ is much {\it larger} than $m$ and the electric field $E_0$ at that energy is {\it smaller} than $E_c$, the pair production is mainly realized with the perturbative mechanism, i.e., a few insertions of electric fields. 
These two limiting cases are distinguished by the ``Keldysh parameter" $\gamma^{\rm (time)}_{\rm K}$ defined by
\begin{equation}
\gamma_{\rm K}^{\rm (time)}\equiv \frac{\omega_0}{m}\cdot \frac{E_c}{E_0}=\frac{m\omega_0}{eE_0}\, , \label{eq:Keldysh_time}
\end{equation}
where $\gamma_{\rm K}^{\rm (time)}\ll 1$ and $\gamma_{\rm K}^{\rm (time)}\gg 1$ correspond to nonperturbative and perturbative regimes, respectively. This dimensionless parameter, sometimes also called the adiabaticity parameter, was first introduced by Keldysh in the analysis of ionization of an atom by laser \cite{keldysh1965ionization}. As mentioned before, ionization can take place in a static electric field due to quantum tunneling. Ionization is also possible by absorption of multiple photons which assists a bound electron to be liberated. These two cases correspond to nonperturbative and perturbative mechanisms, respectively. Such an interplay between the perturbative and nonperturbative mechanisms in the scalar and spinor QED was demonstrated by Brezin and Itzykson \cite{Brezin:1970xf}. They computed particle creations from an oscillating electric field $E_z(t)=E_0\, {\rm e}^{i\omega_0 t}$ and obtained an approximate formula connecting the perturbative and nonperturbative regimes. Interestingly, they found that the exponential suppression in the production ratio (the leading $n=1$ contribution in Eq.~(\ref{eq:SCH})) is now modified as 
\begin{eqnarray}
\exp \left[-\frac{\pi m^2}{eE_0}g(\gamma_{\rm K}^{\rm (time)})\right]
\, ,
\end{eqnarray}
where $g(\gamma_{\rm K}^{\rm (time)})$ is a smooth decreasing function with a boundary condition $g(0)=1$. This reproduces the Schwinger's formula and the perturbative result for $\gamma_{\rm K}^{\rm (time)}\ll 1$ and $\gamma_{\rm K}^{\rm (time)}\gg 1$, respectively. In fact, at large $\gamma_{\rm K}^{\rm (time)}$, the $g(\gamma_{\rm K}^{\rm (time)})$ behaves as $g(\gamma_{\rm K}^{\rm (time)})\sim (4\pi/\gamma_{\rm K}^{\rm (time)})\ln (2\gamma_{\rm K}^{\rm (time)})$, so that the nonperturbative dependence $\propto 1/eE_0$ in the exponent is canceled and a perturbative power dependence is reproduced. 

More recently, the authors of Ref.~\cite{Taya:2014taa} have carefully investigated this interplay by using a pulsed electric field 
called the Sauter-type field. The electric field is oriented to the $z$-direction $[{\bm E}=(0,0,E(t))]$ and its time-dependence has a pulsed shape: 
\begin{equation}
E(t)=\frac{E_0}{{\rm cosh}^2 (t/\tau )}\, ,
\end{equation}
where $E_0$ is the peak value of the field and $\tau$ is the duration of the pulse. The Sauter-type field is one of the few examples with non-constant electromagnetic fields which allow us to solve the equation of motion to exactly compute the pair production rate.\footnote{In fact, this is an alternative (and more direct) method for the calculation of the pair production rate. It is not easy to compute the effective action itself for electric fields with generic time dependences.} The Fourier transformation of the electric field (the power spectrum) is given by
\begin{equation}
\tilde E(\omega)
=\frac{i\pi E_0 \tau^2 \omega}{\sinh \frac{\pi \tau \omega}{2}}\, ,
\end{equation}
which is plotted in Fig.~\ref{fig:Sauter} for a normalized frequency (putting $\pi \tau/2 =1$). 
The characteristic spectrum $|\omega| \simle \frac{2}{\pi \tau}$ is determined by the time duration of the pulse $\tau$. 
Therefore, a shorter pulse involves a wider spectrum that could contain the frequency above the threshold value $2m$ 
and could induce the pair production perturbatively. 
On the other hand, an electric field with a large $\tau$ has a sharp spectrum near $\omega \sim 0$ 
and will induce the nonperturbative pair production (the Schwinger mechanism).  
This expectation is explicitly verified by analytic results for the number density of produced pairs:
\begin{eqnarray}
\frac{(2\pi)^3}{V}\, \frac{d^3N}{dp^3}\to \left\{
\begin{array}{ll}
\left(1-\frac{p_z^2}{p_0^2}\right) \left|\frac{eE_0}{p_0^2}\right|^2 \frac{(\pi p_0 \tau)^4}{\pi^2 |{\rm sinh} (\pi p_0 \tau)|}  &\quad (\gamma_{\rm K}^{\rm (time)}\gg 1)\\
\exp\left( -\frac{\pi m_T^2}{|eE_0|}  \right) &\quad (\gamma_{\rm K}^{\rm (time)}\ll 1)\, ,
\end{array}
\right.
\end{eqnarray}
where the Keldysh parameter is now given by $\gamma_{\rm K}^{\rm (time)}=\frac{m}{|eE_0|\tau}$. The result with $\gamma_{\rm K}^{\rm (time)}\gg 1$ corresponds to the perturbative pair production from one photon, and is a good approximation when the pulse duration is small enough $\tau \ll 1/m$.  

\begin{figure}[t]
\begin{center}
   \includegraphics[width=0.5\hsize]{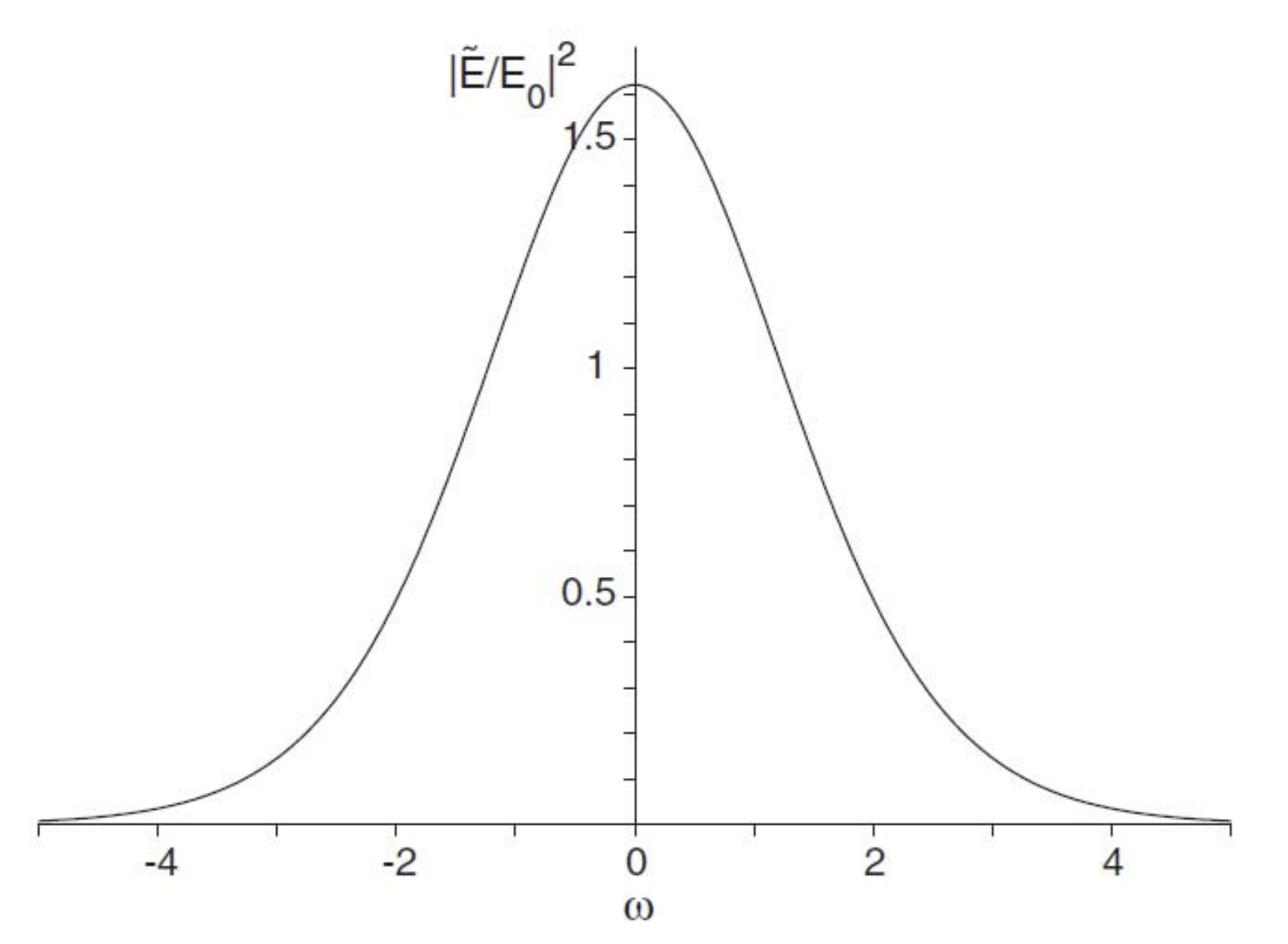}
\end{center}
\vspace{-5mm}
\caption{Spectrum $|\tilde E(\omega)/E_0|^2$ of the Sauter-type electric field for $\tau=2/\pi$. 
} \label{fig:Sauter}
\end{figure}

In recent years, another interesting development has been seen. Based on the observation discussed above, 
we can also think of superposition of two pulsed electric fields. In particular, it is interesting to consider the mixed situation where one pulsed field induces nonperturbative pair production while the other perturbative one:
\begin{eqnarray}
E(t)=\frac{E_1}{{\rm cosh}^2(t/\tau_1)}+\frac{E_2}{{\rm cosh}^2(t/\tau_2)}\, .
\end{eqnarray}
Here, $E_1\gg E_2$ and $\tau_1\gg \tau_2$ and thus the first (second) term is for the strong and long (weak and short) pulse. In this case, even if $E_1$ has a subcritical strength $E_1\ll E_c$, we can gain a large number of particle production due to the assist of the weaker short pulse. 
This kind of pair production is called the  ``dynamically assisted Schwinger mechanism" and was first discussed in Ref.~\cite{Schutzhold:2008pz}. Physically, the tunneling barrier is reduced thanks to the absorption of perturbative photons, and the pair production rate is significantly enhanced \cite{DiPiazza:2009py}. It was explicitly demonstrated that in the perturbative treatment of the short pulse around the long pulse, the pair production rate is indeed enhanced \cite{Torgrimsson:2017pzs}. This opened up an interesting question what is the optimized pulse shape that maximizes the pair-creation rate. For example, it was recently found that the pair-creation rate strongly depends on the shape of the short (perturbative) pulse \cite{Linder:2015vta}, effects of envelope profiles of laser fields were investigated in Ref.~\cite{Aleksandrov:2017owa}, and lastly, optimization of the pulse shape was discussed for rotating fields in Ref.~\cite{Fillion-Gourdeau:2017uss}. 

In fact, there is a condensed matter analog of the dynamically assisted Schwinger mechanism. In semiconductor physics, photo absorption plays a fundamental role since it directly reflects the band structure of a bulk material. It has been known both theoretically and experimentally that the photo absorption rate is significantly modified if a strong slow electric field is applied to the semiconductor in such a way that the rate becomes finite even below the band gap energy and oscillates above the gap energy. 
This phenomenon, called the Franz-Keldysh effect \cite{franz1958einfluss, keldysh1958effect} (see also a review article \cite{RevModPhys.90.021002}), 
has a similar situation as the dynamically assisted Schwinger mechanism since the absorptive photons correspond to perturbative electric fields which coexist with a slow strong electric field. 
Very recently, the Franz-Keldysh effect was formulated in QED and it was found that the similar interplay between the perturbative and nonperturbative pair production indeed occurs~\cite{Taya:2018eng}.

\subsubsection{Pair production from spatially inhomogeneous electric fields}

As we discussed above, the crucial point of time-dependent fields is that those fields are able to directly inject 
a significant amount of energy to the pairs of virtual excitations 
so that they can become real excitations over a smaller energy barrier. 
Those ``hard'' processes can occur within a short spacetime scale. 
Contrary, time-independent electric fields provide energies to the pairs only 
by consuming the electric energies for the acceleration of the pairs. 
This process, as a sequence of the ``soft'' interactions, requires a quantum coherence 
over a certain spatial length as well as the time duration 
before the virtual pairs acquire energy enough to become real particles. 
Therefore, 
the spatial profile of the electric fields is crucial for the realization of the pair production. 
This observation is consistent with the demand for the resummation of external-field insertions, 
which leads us to the Heisenberg-Euler effective action. 


To understand the effects of spatial inhomogeneity, let us reexamine the condition for pair production to occur. For simplicity, we assume that the electric field is oriented to the $z$ direction and has only $z$-dependence, $\bE =(0,0,E(z))$. The condition for the pair production to occur is that the work done by the electric field for the virtual pair is larger than the mass gap: 
\begin{eqnarray}
2\int_0^\ell |q_f|E(z)dz \ge 2 m\, .
\label{eq:condition_paircreation}
\end{eqnarray}
For a given $E(z)$, we can determine the minimum length $\ell_{\rm min}$ by the length $\ell$ that satisfies the equality. The $\ell_{\rm min}$ will be short when $E(z)$ is large over a wide range of space, while it may be long for a weak electric field. However, the length $\ell$ cannot be too long because the pair is a (virtual) quantum fluctuation before the condition is satisfied. Typically, the formation length $\ell$ will be shorter than or of the order of the Compton wavelength $\lambda =1/m$. In other words,  in the tunneling picture, the Compton wavelength roughly corresponds to a typical penetration length in a mass gap region.  Therefore, the pair production will be realized when the condition (\ref{eq:condition_paircreation}) is satisfied for the length scale $\ell \simle \lambda=1/m$. 
When the electric field is spatially constant and present in the whole region of $z$, $E(z)=E_0\, (>0)$, the threshold condition reads $E_0 \ell = m/|q_f|$. For the length scale $\ell \simle \lambda=1/m$, one finds that there is a minimum strength for the electric field: $E_0 \simge E_c=(m/|q_f|)(1/\lambda)=m^2/|q_f|$. This is the critical field strength which we previously obtained. 
Notice that the minimum length $\ell_{\rm min}$ necessary for the pair production is given by $\ell_{\rm min}=(m/|q_f|)(1/E_0)$ 
and thus becomes shorter than the Compton wavelength for a super-critical field strength $E_0\simge E_c$.

Now let us turn to inhomogeneous electric fields. The simplest example is the case where the electric field is applied only in a finite extent as in a capacitor: $E(z)=E_0>0$ only for $-L<z<L$. Consider a strong (super-critical) electric field $E_0\simge E_c$. If the electric field is applied in a whole space, then the length $\ell$ for the pair production to occur is $\ell_{\rm min}=m/(|q_f|E_0)\simle \lambda=1/m$. Therefore, if the extent $L$ is larger than $\ell_{\rm min}$, there will be a spatial region where the condition (\ref{eq:condition_paircreation}) will be satisfied and thus the pair production occurs. On the other hand, when $L$ is shorter than $\ell_{\rm min}$, it is not possible to satisfy the condition (\ref{eq:condition_paircreation}) and thus the pair production does not occur. This means that there is a critical extent $L_{\rm crit}$ for the onset of pair production and it is simply given by 
\begin{eqnarray}
L_{\rm crit}=\frac{m}{|q_f|E_0}\, .
\label{eq:crit_length}
\end{eqnarray} 
In fact, by exactly solving the Dirac equation with a finite-extent electric field, one can confirm the presence of the critical extent as explicitly demonstrated in Ref.~\cite{Wang:1988ct} (see also Ref.~\cite{Martin:1988gr} for the calculation in the worldline formalism and Ref.~\cite{Gavrilov:2015zem} for a recent analysis). When $E_0=E_c$, the critical extent coincides with the Compton wavelength $L_{\rm crit}=1/m=\lambda$.

Next, consider the electric field with a generic but mild $z$-dependence \cite{Tuchin:2014hza}. Assuming that the electric field takes its maximum value at $z=0$ and slowly decays to zero at large $|z|$, we may use the following expansion of the electric field (for $z>0$) 
\begin{eqnarray}
E(z)=E_0 + z \left.\frac{d E(z)}{d z}\right|_{z\sim 0} + \cdots 
\, .
\end{eqnarray}
The coefficient of the first derivative term is roughly evaluated as 
$$
\left.\frac{d E(z)}{d z}\right|_{z\sim 0}\sim - \frac{E_0}{\sigma}
\, ,
$$
with $\sigma$ being a typical width of the profile. By using this expression, the work done by the electric field ${\mathscr{W}}(\ell)\equiv 2 |q_f|\int_0^\ell E(z)dz $ is also expanded as 
${\mathscr{W}}(\ell)={\mathscr{W}}_0(\ell)+{\mathscr{W}}_1(\ell)+\cdots $. 
Now if we take a ratio between the zeroth term ${\mathscr{W}}_0(\ell)$ and the first-derivative term ${\mathscr{W}}_1(\ell)$, we find that the result is expressed by the ratio between two characteristic length scales:
\begin{eqnarray}
\left|\frac{{\mathscr{W}}_1(\ell)}{{\mathscr{W}}_0(\ell)}\right|= \frac{\frac{E_0}{\sigma}\frac{\ell^2}{2}}{E_0\ell}=\frac{\ell}{2\sigma}\, .
\end{eqnarray}
When the ratio is small enough, the pair production occurs predominantly by the zeroth term, and we are able to ignore the effects of inhomogeneity. In this case, we can approximate $\ell$ as $\ell\sim \ell_{\rm min}$ which is determined for the electric field with an infinite extent. On the other hand, if the ratio is large enough, the derivative expansion breaks down and the effects of inhomogeneity is significant. Therefore, it is natural to define an inhomogeneity parameter in analogy with the Keldysh parameter for adiabaticity \cite{Dunne:2005sx}: 
\begin{eqnarray}
\gamma_{\rm K}^{\rm (space)}\equiv \frac{\ell_{\rm min}}{\sigma}\, .
\end{eqnarray}
Plugging the explicit form $\ell_{\rm min}=m/(|q_f|E_0)$ and introducing an inverse width $k=1/\sigma$, one finds 
\begin{eqnarray}
\gamma_{\rm K}^{\rm (space)}= \frac{mk}{|q_f|E_0}\, ,
\label{eq:Keldysh_space}
\end{eqnarray}
which has the same representation as the Keldysh adiabaticity parameter (\ref{eq:Keldysh_time}) under the replacement of the typical frequency $\omega_0$ by the typical momentum scale $k$. 
If we take $E_0$ strong enough, the inhomogeneity parameter can become small even though the electric field has a narrow width. In such a case, we can forget about the effects of inhomogeneity and can use the ordinary formula of the Schwinger mechanism. 

In the above case with  mild $z$-dependences, the inhomogeneity parameter $\gamma_{\rm K}^{\rm (space)}$ in Eq.~(\ref{eq:Keldysh_space}) makes sense only for $\gamma_{\rm K}^{\rm (space)}\ll 1$. However, it can be applied to more generic $z$-dependent electric fields. Similar to the case with time-dependent electric fields, the Sauter profile 
$$
E(z)=\frac{E_0}{ {\rm cosh}^2 kz}
$$ 
can be exactly treated \cite{Nikishov:1970br}, and the inhomogeneity parameter indeed plays the role of distinguishing the homogeneous case from the strongly inhomogeneous case \cite{Dunne:2005sx, Kim:2007pm}. Since the Sauter profile has a width $1/k$, 
we define the condition for the pair production to occur as 
\begin{eqnarray}
|q_f|\int_{-\infty}^\infty E(z)dz \ge 2m\, . 
\label{eq:condition_infinity}
\end{eqnarray}
The left-hand side yields $2|q_f|E_0/k$, and we find that the condition is rewritten as $\gamma_{\rm K}^{\rm (space)}\le 1$ with $\gamma_{\rm K}^{\rm (space)}$ defined by Eq.~(\ref{eq:Keldysh_space}).  Therefore, the Sauter profile with $\gamma_{\rm K}^{\rm (space)}>1$ does not allow pair production, and  $\gamma_{\rm K}^{\rm (space)}=1$ corresponds to the critical value. Incidentally, the critical extent (\ref{eq:crit_length}) for the capacitor also gives $\gamma_{\rm K}^{\rm (space)}=1$ under the identification $k=1/L_{\rm crit}$. 

So far we discussed three examples of inhomogeneous electric fields: the capacitor-like profile, the mild decaying profile, and the Sauter profile. All these cases allow for the pair production depending on parameters characterizing the profiles (the magnitude at the peak $E_0$ and the inverse width $k$). In particular, the onset of pair production is equally specified by the inhomogeneity parameter $\gamma_{\rm K}^{\rm (space)}=1$. This brings us into awareness that there are in fact infinitely many profiles that can satisfy the condition (\ref{eq:condition_infinity}). 
This is, of course, simply because the condition is written in terms of the integration over the space, 
and only the integral value matters. 
Then, one may ask if the spatial profiles obey any classification. 
Recently, the authors of Refs.~\cite{Gies:2015hia,Gies:2016coz} found a similarity between the pair production and phase transitions in critical phenomena. Recall that the onset of pair production is expressed as a nonzero value of the imaginary part of the effective action $\Im m {\mathcal L}_{\rm eff}$. The authors of Refs.~\cite{Gies:2015hia,Gies:2016coz} regarded the imaginary part of the effective action as an order parameter for the pair production, and claimed that the notion of universality appears at the ``critical point" $\gamma_{\rm K}^{\rm (space)}=1$. 
We shall write the electric fields as $E(z)=E_0 f'(u)$, where the normalized potential function 
$f$ is an antisymmetric and monotonic function of $u=kz$. 
Then, this field indeed yields the imaginary part of the effective action around the critical point 
\begin{eqnarray}
\Im m {\mathcal L}_{\rm eff}\sim \left[1-\big(\gamma_{\rm K}^{\rm (space)}\big)^2\right]^\beta
\, ,
\end{eqnarray}
with $\beta$ being a critical exponent. 
The generic profiles $E(z)$ are subdivided into several ``universality classes" depending on numerical values of $\beta$. 
Interestingly, the exponent $\beta$ only depends on the large-scale behavior of the profile, and is insensitive to microscopic details.  
We also point out that the Landau-Zener transition, 
mentioned above as an analog of the Schwinger mechanism, 
was analyzed in terms of the Kibble-Zurek mechanism \cite{Damski:2004bz, Damski:2005by} 
that has been originally developed to connect 
defect production during the phase transition 
to the universal critical exponents  \cite{Kibble:1976sj, Kibble:1980mv, Zurek:1985qw} (see Ref.~\cite{delCampo:2013nla} for a review). 
It is quite interesting to pursue the connection 
between the production rate and the ``universality class'' 
for the inhomogeneous electric field by the Kibble-Zurek mechanism.

\section{Vacuum fluctuations and nonlinear effects}

\label{sec:photons} 
Many studies in physics have addressed how the systems respond to externally applied fields in various ways. 
When an external field is weak, one may focus on the linear responses to the external field. 
In the strong-field physics, we however explore the nonlinear regime beyond the linear order. 
For example, ``nonlinear optics" is the research field where one studies nonlinear responses of media \cite{Y.R.Shen,LandauLifshitz_contmedia}.
An electric polarization $\bm{P}$ of a medium in the {\it linear} optics is assumed to be proportional 
to the applied electric field $\bm{P}=\epsilon_0 \chi \bm{E}$ where $\epsilon_0$ is the vacuum permittivity and $\chi$ is the electric susceptibility. 
However, it is called the {\it nonlinear} optics when the same quantity is subject to 
significant nonlinear corrections like $\bm{P}=\epsilon_0 [\chi^{(1)} \bm{E}+ \chi^{(2)} \bm{E}^2 + \chi^{(3)} \bm{E}^3 + \cdots ]$. 
Such nonlinear dependences can be assembled in the form of the nonlinear response function, e.g., 
the nonlinear electric susceptibility $\chi=\chi(\bm{E})$ in this example.

In quantum field theories, vacuum accommodates perpetual fluctuations of the matter fields, 
and is mixed with the states having a nonzero number of particle-antiparticle pairs. 
Moreover, due to the presence of interactions between (charged) particles and gauge fields, 
the external fields are coupled with such fluctuations and induce nontrivial responses of the {\it vacuum} 
as if the vacuum plays a role of a medium. 
This observation was one of the motivations that drove Heisenberg and Euler 
to compute the effective action of the electromagnetic fields \cite{Heisenberg:1935qt}, 
and was clearly recognized by Weisskopf \cite{Weisskopf:1996bu, Weisskopf}. 
With the vacuum behaving like a medium, we can expect various phenomena to occur in parallel to the nonlinear optics. 
Such examples include ``birefringences''  (cf. a photo in Fig.~\ref{fig:biref-pic}),\footnote{Birefringence of a medium induced by an electric (magnetic) field is called the optical Kerr effect (the Cotton-Mouton effect). When both $\bm{E}$ and $\bm{B}$ are present, birefringences induced by the parallel configuration $\bm{E}\parallel \bm{B}$ and perpendicular configuration $\bm{E}\perp \bm{B}$ are, respectively, called Jones and magneto-electric birefringences. All of these are already observed in experiments \cite{Battesti:2012hf}.} 
``wave mixing'' and ``high harmonic generation.''

In this section, we study how the properties of photons are modified by strong electromagnetic fields. 
As we discussed in Sec.~\ref{sec:HE}, electromagnetism of slowly varying fields 
is described by the Heisenberg-Euler (HE) effective action (\ref{HE_general2}) \cite{Heisenberg:1935qt}, 
which contains nonlinear terms in addition to the Maxwell term. 
In fact, similar to the nonlinear optics, one can compute an electric polarization $\bm{P}$ and a magnetization $\bm{M}$ of the vacuum 
from the HE action ${\cal L}_{\rm HE}$ as $\bm{P}=\bm{D} - \bm{E}$ and $\bm{M}=\bm{H}-\bm{B}$ with $\bm{D}\equiv \del {\cal L}_{\rm HE}/\del \bm{E}$ and $\bm{H}\equiv -\del {\cal L}_{\rm HE}/\del\bm{B}$ (see, for example, \S 129 in Ref.~\cite{Berestetsky:1982aq}). 
In other words, one can define the electric and magnetic permeability tensors $(\bar\epsilon^{ij}, \bar\mu^{ij})$ of the vacuum as $ {D^i}=\bar\epsilon^{ij} E^j$ and $B^i = \bar\mu^{ij} H^j$. Since they are just unity $(\bar\epsilon^{ij}=\epsilon_0\delta^{ij}=\delta^{ij},\ \bar\mu^{ij}=\mu_0\delta^{ij}=\delta^{ij})$ in the ordinary vacuum without a strong electromagnetic field, 
deviations from unity signals the polarization and magnetization of the vacuum.

Recall that the combination of the electric and magnetic permeability 
provides 
the speed of light $c_0$ 
which is, of course, unity $c_0^{-1}=\sqrt{\epsilon_0\mu_0}=1$ 
in the natural unit according to the Lorentz and gauge symmetries in the ordinary vacuum.\footnote{
As is well-known, photons become massive in a gauge invariant way 
in the (1+1)-dimensional QED called the Schwinger model \cite{Schwinger:1962tn, Schwinger:1962tp}. 
A similar situation occurs in a strong magnetic field (see also Appendix~\ref{sec:VP_vac}). 
} 
However, when an external field is applied, the symmetries of the ``vacuum'' will be changed. 
The external field breaks the Lorentz and full spatial rotational symmetries, 
so that the speed of light, as consequences of the response from the vacuum, 
will acquire a nontrivial modification and moreover becomes not only anisotropic but also polarization-mode dependent. 
This is the {\it vacuum birefringence}. 
In the soft-photon regime, we can compute its velocity from the HE action through $\bm{P}$ and $\bm{M}$, 
which will be obtained in the tensor forms consistent with the residual spatial symmetries. 
In this way, properties of propagating photons are intimately related to the properties of the vacuum under strong fields. 
It is also important to notice that a photon having no electric charge gets modified by the electromagnetic fields 
only through quantum fluctuations creating a virtual electron-positron pair. 
Without a strong field, such effects are higher order in the perturbation theory. 
However, as we have already seen, the coupling between the fluctuating matter field and the strong external field 
can be enhanced by a large field strength.

Nonlinear physics in vacuum under strong electromagnetic fields has a long history. After the pioneering work by Heisenberg and Euler \cite{Heisenberg:1935qt}, Schwinger re-derived the effective action from a modern view of quantum field theory 
by using the proper-time method \cite{Schwinger:1951nm} as we have seen in the previous section. 
Then people started discussing the vacuum birefringence and the photon splitting for low-energy photons as a result of the effective theory including nonlinear corrections 
to the Maxwell part \cite{Toll:1952rq, Klein:1964zza, Baier:1967zza, Baier:1967zzc,BialynickaBirula:1970vy,Brezin:1971nd,Adler:1971wn}. 
Analysis beyond the first nonlinear correction to the Maxwell theory was also performed in order to treat higher energy photons 
and/or stronger electromagnetic fields \cite{Toll:1952rq,Adler:1971wn,batalin1971green,Tsai:1974fa, Tsai:1975iz,Tsai:1974ap,Melrose:1977ftf,Urrutia:1977xb,Heyl:1996dt, Heyl:1997hr}. 
There are many studies which discussed the photon polarization tensor in limited kinematical regimes 
with appropriate approximations 
(see \fref{fig:scales-polarization} for summary of the representative regimes in external magnetic fields). 
The reader is also referred to Ref.~\cite{Dittrich:2000zu} for more references until 2000. 
Most recently, an analytic evaluation of the double parameter integral for the polarization tensor 
was performed in Refs.~\cite{Hattori:2012je, Hattori:2012ny}. 
Below, we discuss how to describe the vacuum birefringence and the photon splitting with relevant references.

\begin{figure}[t]
\begin{minipage}{0.5\hsize} 
	\begin{center}\hspace*{-5mm}
		\includegraphics[width=0.9\hsize]{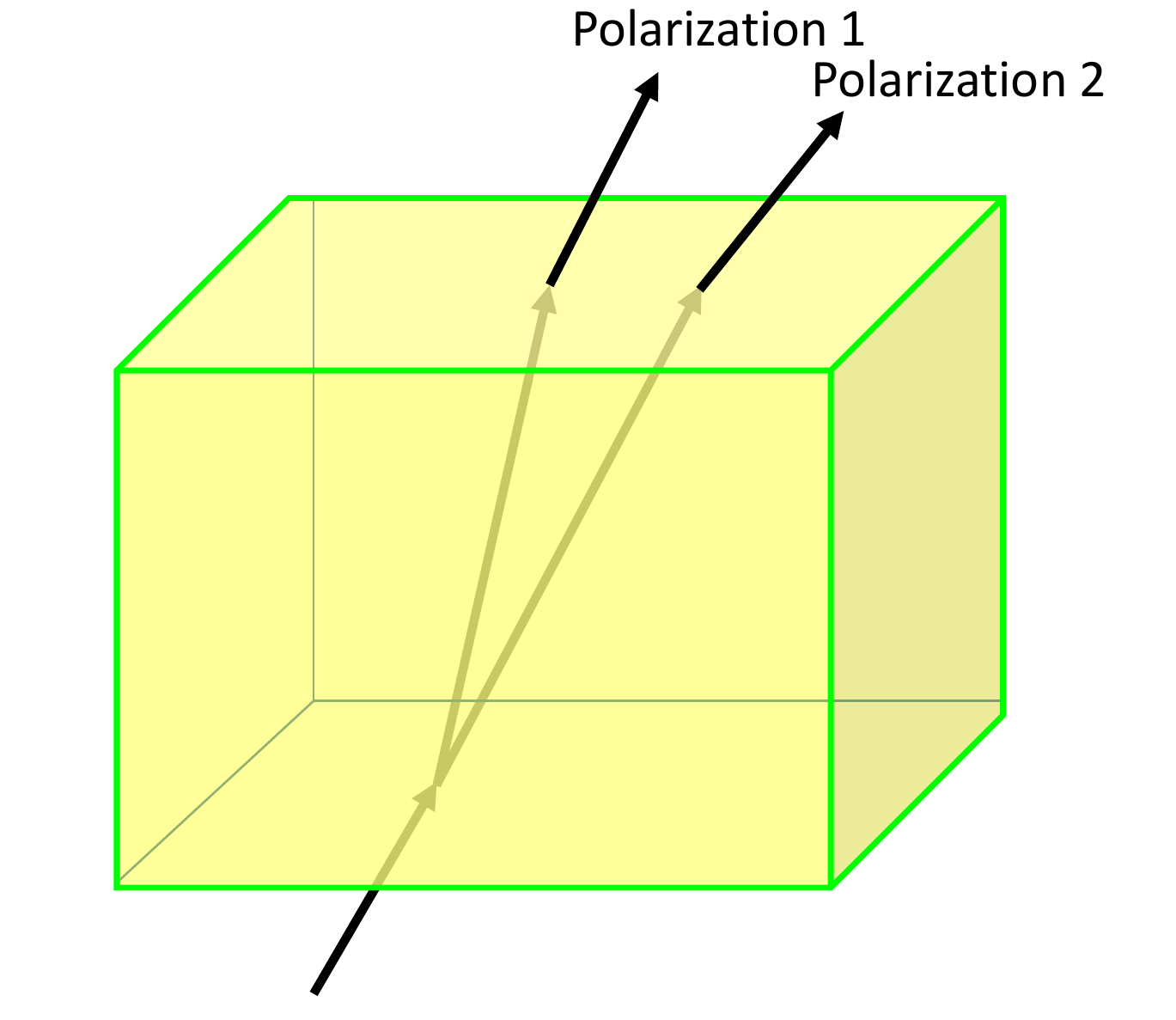}
	\end{center}
\vspace{-0.8cm}
\end{minipage}
\begin{minipage}{0.5\hsize}
	\begin{center}
\includegraphics[width=0.9\hsize]{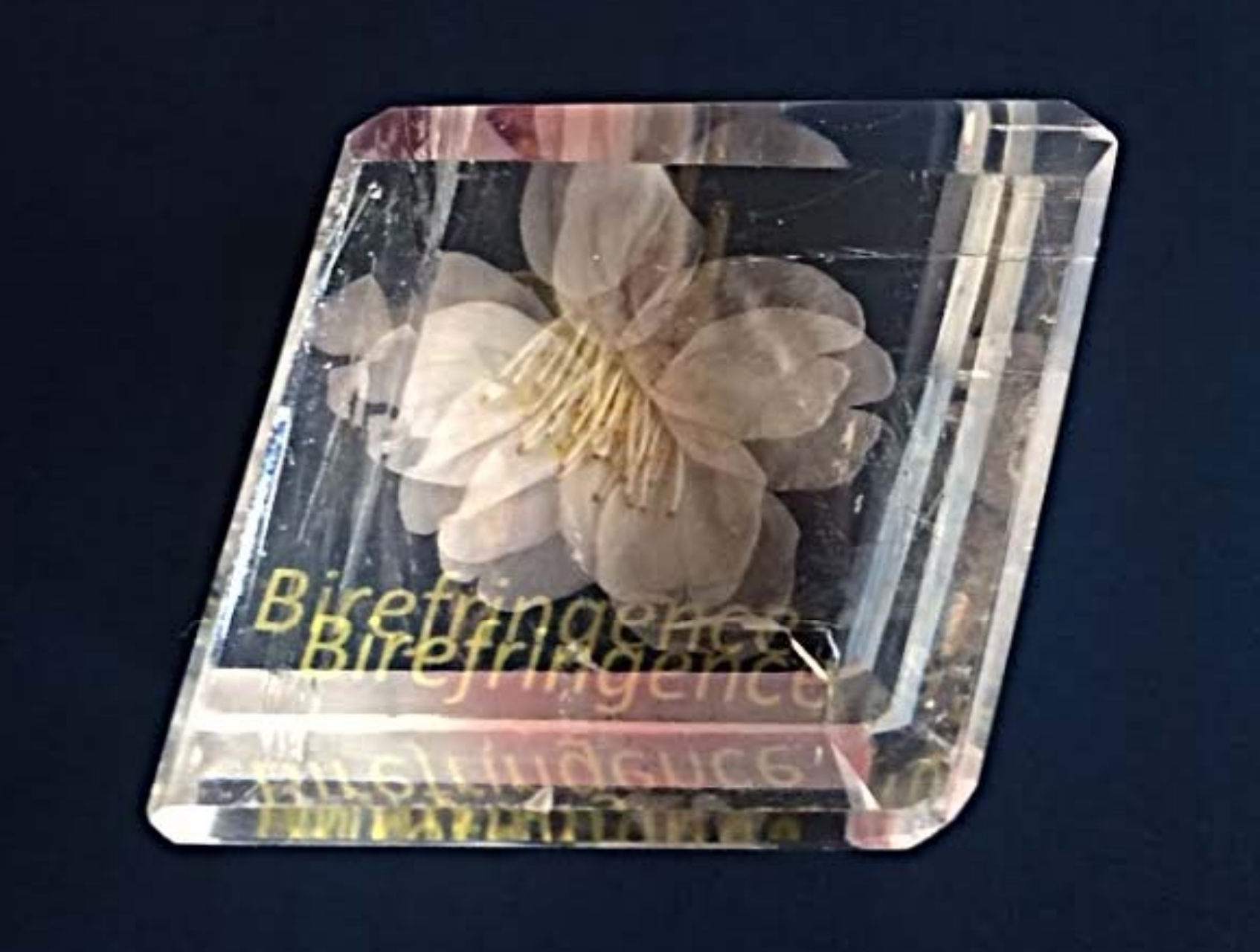}
	\end{center}
\end{minipage}
\caption{Ray splitting on the faces of a calcite caused by the birefringence.}
\label{fig:biref-pic}
\end{figure}

\begin{figure}
     \begin{center}
              \includegraphics[width=0.7\hsize]{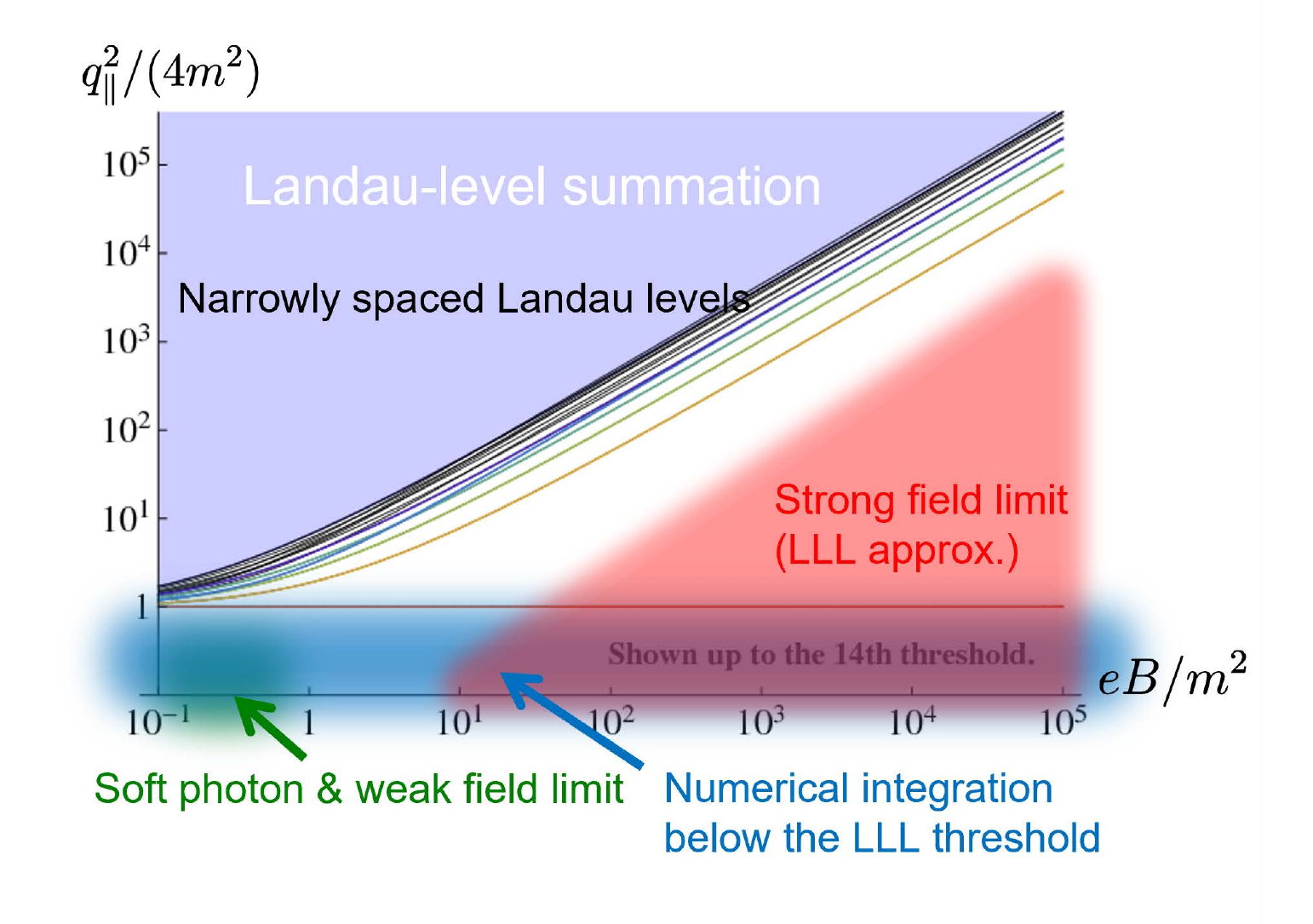}
     \end{center}
\vspace{-0.7cm}
\caption{Summary of scales involved in the analyzes of the vacuum fluctuations in magnetic fields. 
The photon momentum $  q_\parallel^2$ and the magnetic field strength $ eB $ 
are normalized by the mass scale of the fluctuation.}
\label{fig:scales-polarization}
\end{figure}

\subsection{Vacuum polarization tensor in strong fields}

There are mainly two different ways to compute the vacuum polarization tensor of a photon in external fields. 
The first one is to extract the vacuum polarization tensor from the effective action for the electromagnetic fields, and the second one is to directly compute it from the polarization diagram. In both cases, the proper-time method turns out to be very useful. In the present review, we explain the latter approach in detail, while the former is only briefly outlined below.

One can use the effective Lagrangian (\ref{effectiveaction}) to derive the polarization tensor for a dynamical photon field. 
Since we have not specified the gauge field $A_\mu(x)$ at the level of Eq.~(\ref{effectiveaction}), we can regard it as consisting of an external field and a dynamical field: $A^\mu = A^\mu_{\rm ext}+ a^\mu$. Using this decomposition and expanding Eq.~(\ref{effectiveaction}) with respect to the dynamical gauge field, we obtain ${\cal L}^{(1')}[A^\mu]={\cal L}^{(1')}[A^{\mu}_{\rm ext}]+\delta{\cal L}^{(1')}[A^{\mu}_{\rm ext},a^\mu]$ where the second term is quadratic in $a_\mu$: 
\beq
\delta {\cal L}^{(1')}[A^\mu_{\rm ext}, a^\mu] = - \frac{1}{2}\int d^4x' a_\mu(x)\, \Pi^{\mu\nu}( x,x'|F_{\rm ext})\, a_\nu(x')\, . \label{eq:fluctuation_pol}
\eeq
Here, the polarization tensor of a propagating photon is obtained from the second-order variation of the action 
\begin{eqnarray}
%
\Pi^{\mu\nu}(x,x'|F_{\rm ext}) = - \frac{\delta^2  }{\delta a_\mu(x)\delta a_\nu(x')}
\int d^4 x \, \delta {\cal L}^{(1')}[A^\mu_{\rm ext}, a^\mu]
\, .
\end{eqnarray}
It has a nonlinear dependence on the field strength of 
the external field $F_{\rm ext}^{\mu\nu}=\del^\mu A^{\nu}_{\rm ext}-\del^\nu A_{\rm ext}^{\mu}$. 
This approach was adopted, for example, by Tsai and Erber \cite{Tsai:1974fa, Tsai:1975iz}. 
When the spacetime dependence of an electromagnetic field is smaller 
than the electron's Compton wavelength $|\nabla F|/F \ll m$, 
the effective action can be explicitly evaluated and yields the HE effective action. 
In this ``soft photon limit,'' one may use the previous result on the HE effective action.

One can also explicitly compute the same quantity from the polarization diagrams. 
At the one-loop level, the diagram is drawn in Fig.~\ref{fig:photon_vp}. 
Accordingly to the standard Feynman rules, its momentum representation reads 
\begin{eqnarray}
i\, \Pi^{\mu \nu}(q|F_{\rm ext}) = 
 (-1) (-ie)^2 \integ \frac{d^4p}{(2\pi)^4} 
{\rm tr} \Big[ \ \gamma^\mu S(p|F_{\rm ext}) \gamma^\nu 
S(p+q|F_{\rm ext}) \ \Big] 
\label{eq:prop}
\, ,
\end{eqnarray}
where ``tr" is taken for the spinor indices and $S(p|F_{\rm ext})$ 
is (the gauge-invariant part of) the electron propagator in a strong field as defined in Eq.~(\ref{eq:eqDp}). 
As we discussed in Sec.~\ref{sec:FS}, the Schwinger phase identically vanishes 
for the two-point diagram in Fig.~\ref{fig:photon_vp} 
and we are left with the gauge-invariant part, implying the gauge invariance of the polarization tensor.\footnote{Here, we mean the gauge invariance of $\Pi^{\mu\nu}$ with respect to the external gauge field $A^{\mu}_{\rm ext}$. 
We discuss the gauge for the dynamical gauge field $a^\mu$ just below.} 
In contrast to the effective-action approach, there is no {\it a priori} restriction on 
the momentum scale $ q $ of the incoming photon. 
Since the resummed propagator is expressed by the use of the proper-time method, 
the explicit form (\ref{eq:prop}) contains two proper times corresponding to the two electron propagators. This approach was taken, for example, 
in Refs.~\cite{Hattori:2012je, Hattori:2012ny} (see also the literature therein). 
In the next subsection, we will give its explicit form and discuss how to analytically perform the proper-time integrals.

\begin{figure}[t]
	\begin{center}
		\includegraphics[scale=0.3]{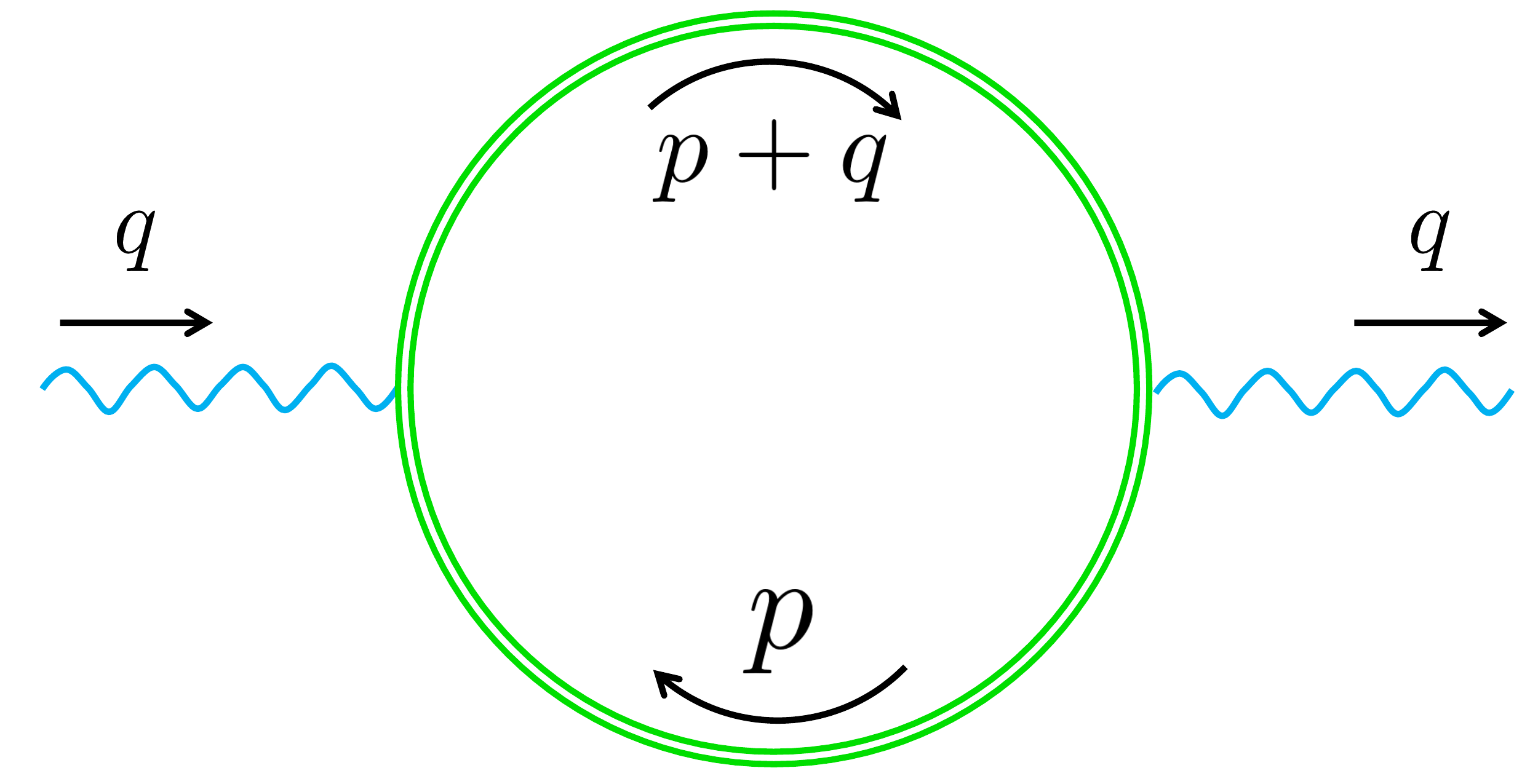}
	\end{center}
	\vspace{-0.5cm}
\caption{One-loop diagram of the vacuum polarization tensor in a strong magnetic field. 
Double lines denote the dressed electrons interacting with the external field 
treated by the proper-time method.}
\label{fig:photon_vp}
\end{figure}

\subsection{Photon's vacuum birefringence and decay into an $e^+e^-$ pair}

\label{sec:photon-vp}

Consider the case where the external field only consists of a magnetic field. As we discussed in previous sections,  the direction of the magnetic field can be taken along the $z$ axis without loss of generality. 
In the ordinary vacuum, the Lorentz structure of the polarization tensor is 
proportional to the transverse projection $P^{\mu\nu}_0(q)= q^2 g^{\mu\nu}-q^\mu q^\nu$ 
so that the Ward identity is satisfied, $q_\mu  \Pi_0^{\mu\nu}(q)=0$. 
On the other hand, in the presence of a magnetic field, 
we need to distinguish the longitudinal and transverse directions with respect to the magnetic field, which can be performed  with the help of 
the additional metrics $ g_\para^{\mu\nu} $ and $g_\perp^{\mu\nu}  $ defined in Eq.~(\ref{eq:metrics}). 
Thus, in addition to $P_0^{\mu\nu}$, we can construct two additional tensor structures $P_1^{\mu\nu}= q^2_\parallel  g^{\mu\nu}_\parallel  - q^\mu_\parallel  q^\nu_ \parallel$ and $P_2^{\mu\nu}= q^2_\perp g^{\mu\nu}_\perp - q^\mu_\perp q^\nu_\perp$. 
By using them, 
the general form of the vacuum polarization tensor is given as\footnote{
When there are both electric and magnetic fields, there is one additional tensor structure (see, e.g., Ref.~\cite{Dittrich:2000zu}).}
\begin{eqnarray}
&&
 \Pi^{\mu\nu} (q|B) = - \big( \,   \chi_0  P_0^{\mu\nu} + \chi_1 P_1^{\mu\nu} + \chi_2  P_2^{\mu\nu} \, \big)
\, .
\label{eq:Pex}
\end{eqnarray}
While we have suppressed arguments of the Lorentz-scalar functions $\chi_i \ (i=0,1,2)$, 
they are in general functions of $ q_\para^2 $ and $ q_\perp^2 $ 
according to the boost invariance and rotational symmetry with respect to the direction of the magnetic field. 
By construction, the Ward identity is satisfied $q_\mu  \Pi^{\mu\nu}=0$ 
thanks to the transversality $q_\mu P_i^{\mu\nu}=0$ for $i=0,1,2$.

In the presence of an external magnetic field, we need to analyze the equation of motion for the dynamical photon field including the effect of the vacuum polarization. Namely, with the quantum correction to the Lagrangian (\ref{eq:fluctuation_pol}), 
the Maxwell equation 
is modified as 
\begin{equation}
\left[ \ q^2 g^{\mu\nu} - \left( 1-\frac{1}{\xi_g} \right) q^\mu q^\nu 
- \Pi^{\mu\nu}(q_\parallel, q_\perp | B) \ \right] a_\nu(q)  =  0\, , 
\label{modifiedMaxwell}
\end{equation}
where $\xi_g$ is a parameter in the gauge-fixing term, 
$\Lag_{\rm GF} = - \frac{1}{2 \xi_g} (\partial^\mu a_\mu)^2$. 
Substituting the general form of the vacuum polarization tensor (\ref{eq:Pex}), one finds 
\begin{eqnarray}
\left[ \  (1+\chi_0) P_0^{\ \mu\nu} + \chi_1 P_1^{\ \mu\nu} + \chi_2 P_2^{\ \mu\nu} + \frac{1}{\xi_g}  q^\mu q^\nu  \ \right] a_\nu(q) = 0
\, .
\label{eq:Maxwell_2}
\end{eqnarray}
We shall examine modifications in a refractive index of photon 
\beq
n^2=\frac{|\bm{q}|^2}{\omega^2}
\, ,
\label{def:epsilon}
\eeq
where $ \omega $ and $ \bq $ are the frequency and spatial momentum of the propagating photon, respectively. 
We relate the dielectric constant $ \epsilon $ to the refractive index as 
$  \epsilon = n^2$ assuming that the modification of the magnetic permeability is small. 
By selecting an appropriate basis for the photon polarization modes, we find two independent dispersion relations\footnote{The dynamical photon field $a^\mu$ can be decomposed into two modes having different polarizations. They are specified by vectors $\varepsilon^\mu_{(1)}\propto (0,q^2,-q^1,0)$ and $\varepsilon^\mu_{(2)}\propto (q^3,0,0,q^0)$ so that $a^\mu = N(\varepsilon^\mu_{(1)}{\rm e}^{-i(\omega_{(1)}t-\bm{q\cdot x})}+\varepsilon^\mu_{(2)}{\rm e}^{-i(\omega_{(2)}t-\bm{q\cdot x})})$ with $N$ being a normalization constant. The polarization modes $\varepsilon^\mu_{(1)}$ and $\varepsilon^\mu_{(2)}$ correspond to $\epsilon_\perp$ and $\epsilon_\parallel$, respectively. One can also read off the dispersion relation and corresponding polarization from the resummed photon propagator (see Appendix~C and D of Ref.~\cite{Hattori:2012je} and Ref.~\cite{Hattori:2017xoo}).} 
in terms of the distinct dielectric constants: 
\begin{subequations}
\begin{eqnarray}
&&\epsilon_\perp = \frac{1+\chi_0}{1+\chi_0+\chi_2 \sin^2 \theta }  \, , 
\label{eq:eps1bperp}
\\
&&\epsilon_\parallel = \frac{1+\chi_0+\chi_1}{1+\chi_0+\chi_1 \cos^2 \theta } 
\, ,
\label{eq:eps1bparallel}
\end{eqnarray}
\end{subequations}
where $\theta$ is the angle between the magnetic field ${\bm B}$ and the photon momentum $\bm{q}$. 
With a nonzero $\chi_i$, the $\epsilon_\perp$ and $\epsilon_\parallel$ are in general not equal to unity 
and are also different from each other. 
Thus, there are two propagating modes whose velocities are not equal to the speed of light in the ordinary vacuum. 
This phenomenon is called the {\it vacuum birefringence} after a similar phenomenon 
in the birefringent substances such as calcite crystal. 
The birefringence in a material causes, for example, splitting of 
the polarization modes of the transmitting ray and then induces a double image as shown in Fig.~\ref{fig:biref-pic}. 
Intuitively, the vacuum birefringence originates from the anisotropic response of charged particles 
to the photon field in the presence of an external magnetic field. 
As we already saw in Sec.~\ref{sec:Q}, motions of charged particles 
in the directions parallel and transverse to the magnetic field are completely different. 
Especially, the (1+1)-dimensional motion of the LLL fermions behaves like a polarizer 
which modifies the propagation of the parallel mode ($ \epsilon_\para \not =1 $) 
but transmits the perpendicular mode ($ \epsilon_\perp=1 $) [see Eq.~(\ref{eq:chi02_LLL}) and below 
for the analysis in the LLL approximation]. 
It should be emphasized that, while the modified Maxwell equation (\ref{modifiedMaxwell}) is linear in the dynamical field $a_\mu$, the vacuum birefringence is a result of the nonlinear effects with respect to the external fields. The polarization tensor carries the information about the interaction between the virtual fluctuations and the external fields. 
Here are a few comments in order: 

\begin{itemize}
\item Both of the two dielectric constants $\epsilon_\perp$ and $\epsilon_\parallel$ are independent of the gauge fixing parameter $\xi_g$, meaning that they are gauge invariant quantities (the same result is indeed available in the radiation gauge). 
The dependence on $\xi_g$ only appears in the unphysical mode. 

\item In the ordinary vacuum, both $\epsilon_\perp$ and $\epsilon_\parallel$ are unity in accordance with the  Lorentz symmetry. As mentioned before, quantum fluctuations in the ordinary vacuum without an external field induces nonzero $\chi_0$ while the other two are vanishing ($\chi_0\neq 0, \chi_1=\chi_2=0$). 
In this ordinary case, we confirm that $\epsilon_\perp = \epsilon_\parallel = 1$ as it should be. 


\item Due to the breaking of the Lorentz and rotational symmetries 
by an external magnetic field, the dielectric constants explicitly depend on the zenith angle with respect to the magnetic field direction, while they do not depend on the azimuthal angle according to the residual rotational symmetry. On the other hand, the system maintains a boost invariance in the direction of the constant external magnetic field, and thus photon propagations in the directions at $\theta =0, \pi$ are special. Substituting these angles into Eqs.~(\ref{eq:eps1bperp}) and (\ref{eq:eps1bparallel}), we find that both of the dielectric constants reduce to unity $\epsilon_\perp(\theta=0,\pi)=\epsilon_\parallel(\theta=0,\pi) =1$, as a consequence of the boost invariance. 


\end{itemize}

Let us assume that the scalar functions $\chi_i$ have imaginary parts in general. 
Then, the dielectric constants also have imaginary parts: $\epsilon = \epsilon_{\rm real} + i \, \epsilon_{\rm imag}$. Since the dielectric constants and refractive indices are related with each other via  $n^2 = \epsilon$, we can similarly define real and imaginary parts of the refractive indices: $ n = n_{\rm real} + i \, n_{\rm imag}$. 
Then, one can easily find relations 
\begin{eqnarray}
 n_{\rm real} =  \frac{1}{\sqrt{2}} \sqrt{  | \epsilon | + \epsilon_{\rm real}  }
 \, ,  \quad 
  n_{\rm imag} =   \frac{1}{\sqrt{2}} \sqrt{  | \epsilon | - \epsilon_{\rm real}  } \, , 
\end{eqnarray}
where $   | \epsilon |=( \epsilon_{\rm real} ^2 + \epsilon_{\rm imag} ^2 )^{1/2}$. 
As is well explained in standard textbooks of optics \cite{Hecht}, 
refractive indices have clear physical meaning: 
The real and imaginary parts of the refraction index provide 
a {\it phase velocity} $v_{\rm phase}=1/n_{\rm real}$ and an {\it extinction coefficient} $\kappa = n_{\rm imag}$ of a propagating photon, respectively. 
In the photon field $\Psi(t,\bx)$, they appear in a phase factor and damping factor, respectively, 
\begin{eqnarray}
\Psi (t,\bx) \propto 
\exp \{  - i \omega ( t - v_{\rm phase}^{-1}\, \hat \bq \cdot \bx ) \} \ 
\exp\{ -  \omega \, \kappa \, \hat \bq \cdot \bx \}
\, ,
\label{eq:photon_field}
\end{eqnarray}
where $\hat \bq$ is a unit vector along the photon momentum. 
Thus, it is natural to define a {\it decay length} 
at which the intensity of the photon field falls off 
\begin{eqnarray}
\lambda \equiv \frac{1}{2  \omega \kappa}= \frac{1}{ 2  \omega  n_{\rm imag}}
\, .
\label{eq:decay_length}
\end{eqnarray}
As we will explain later, the coefficient functions $\chi_i$ indeed acquire imaginary parts, 
and there emerges complex refractive indices, which originates from the fact 
that even a real photon can decay into an electron-positron pair in a strong magnetic field.\footnote{
We know that the cyclotron radiation occurs in a magnetic field and is 
a representative example of the 1-to-2 processes. 
Those 1-to-2 (2-to-1) processes are, however, prohibited in the absence of a magnetic field for the kinematical reason. 
}


\subsubsection{Low-energy photons (I): Weak fields}

\label{sec:photons_HE-1}

Let us first consider the case where the energy of propagating photon is small enough. More precisely, the photon energy is assumed to be much smaller than the electron mass $\omega \ll m\simeq 0.5$~MeV so that the HE effective Lagrangian is valid. This condition is satisfied by visible lights (with a few eV) and X-rays (with 1-100 keV), and thus typically matters when we consider photon propagation in a high-intensity laser or in the magnetospheres of neutron stars or magnetars. The refractive indices are always real quantities for such a low-energy photon (well below the threshold of the pair production). 
Here we compute the refractive indices from the HE effective Lagrangian.


When the external fields are not stronger than the critical value $|F^{\mu\nu}|\ll m^2/|e|$, 
we are able to expand the HE effective Lagrangian with respect to the external fields. 
The expansion with respect to the gauge-invariant quantities $\F$ and $\G$, 
defined in Eqs.~(\ref{eq:fg}), reads 
\beq
{\cal L}_{\rm HE}={\cal L}^{(0)}_{\rm HE}+{\cal L}^{(1)}_{\rm HE}+\cdots
\, .
\eeq
with the Maxwell term ${\cal L}^{(0)}_{\rm HE}=-\F$. 
The nonlinear part in the HE effective Lagrangian (\ref{HE_general2}) is evaluated as 
(we take $q_f=e <0 $ for electrons) 
\beq
(es)^2 \G \frac{{\Re e} \left[\cosh \big\{es \sqrt{2(\F+i\G)}\big\}\right]}{{\Im m} \left[\cosh \big\{es\sqrt{2(\F+i\G)}\big\}\right]}=\frac{1+(es)^2 \F + \frac{1}{6}(es)^4(\F^2-\G^2)+\cdots}{1+\frac13 (es)^2 \F +\frac{1}{90}(es)^4 (3\F^2-\G^2)+\cdots}
\, .
\eeq
Note that the discrete symmetries of electromagnetic fields are the following: 
$(C,P,T) = (-,-,+)$  for the electric field $\bm{E}$ and $(C,P,T) = (-,+,-)$ for the magnetic field $\bm{B}$. 
Consequently, $\F$ is $(C,P,T) = (+,+,+)$ and $\G$ is $(C,P,T) = (+,-,-)$. 
Therefore, the effective Lagrangian 
can have $\F$ with odd powers, but $\G$ must always appear with even powers 
unless the theory has any parameter that breaks those symmetries. 
Therefore, we find the leading nonlinear corrections~\cite{Schwinger:1951nm} 
\beq
{\cal L}^{(1)}_{\rm HE}&=&-\frac{1}{8\pi^2}\int_0^\infty \frac{ ds }{ s^{3} } {\rm e}^{-m^2s}\left[
-\frac{(es)^4}{45} (4\F^2+7\G^2)
\right]\nonumber\\
&=&\frac{2\alpha^2}{45m^4} (4\F^2+7\G^2)\, .\label{EulerKockel}
\eeq
This is called the Euler-Kockel Lagrangian \cite{Euler:1935zz} 
as they first derived it before the paper by Heisenberg and Euler \cite{Heisenberg:1935qt}.

As we mentioned in the begining of this section, we are able to compute the electric and magnetic permeability tensor $\bar\epsilon^{ij}$ and $\bar\mu^{ij}$ through $D^i=\del {\cal L}_{\rm HE}/\del E^i = \bar \epsilon^{ij}E^j$ and $H^i=-\del {\cal L}_{\rm HE}/\del B^i=(\bar\mu^{-1})^{ij}B^j$.  
By using the Euler-Kockel Lagrangian (\ref{EulerKockel}), 
the leading nonlinear corrections to those quantities are obtained as 
\begin{subequations}
\label{eq:leading-nonlinear}
\beq
\bar \epsilon^{ij}&=&\delta^{ij}+\frac{4\alpha^2}{45m^4}\left[2(\bm{E}^2-\bm{B}^2)\delta^{ij}+7B^i B^j\right]\, ,\\
\bar \mu^{ij}&=&\delta^{ij}+\frac{4\alpha^2}{45m^4}\left[-2(\bm{E}^2-\bm{B}^2)\delta^{ij}+7E^i E^j\right]\, .
\eeq
\end{subequations}
This result implies that the polarization $\bm{P}=\bm{D}-\bm{E}$ and the magnetization $\bm{M}=\bm{H}-\bm{B}$ 
of the vacuum acquire the nonlinear corrections 
\begin{subequations}
\beq
P^{i}&=&\frac{4\alpha^2}{45m^4}\left[2(\bm{E}^2-\bm{B}^2)\delta^{ij}+7B^i B^j\right]E^j\, ,\\
M^{i}&=&\frac{4\alpha^2}{45m^4}\left[2(\bm{E}^2-\bm{B}^2)\delta^{ij}-7E^i E^j\right]B^j\, .
\eeq
\end{subequations}
This result is similar to that in (nonlinear) optics. 
However, unlike optics, the above vacuum result does not 
have a linear term in the electromagnetic fields and starts from the third order. 
Note also that there is mixing between $\bm{E}$ and $\bm{B}$.

Consider the case where a low-energy photon propagates in vacuum in the presence of only a magnetic field $\bm{B}=\bm{B}_{\rm ext}=(0,0,B)$. To compute the refractive indices, we can again use the Euler-Kockel Lagrangian in which we substitute $\bm{B}=\bm{B}_{\rm ext}+ \bm{b}$ and $\bm{E}=\bm{e}$ with $\bm{b}$ and $\bm{e}$ being the propagating modes. From the modified Maxwell equations, we can find two independent polarization modes: the ``$\perp$-mode" is that the propagating electric field ${\bm e}$ is perpendicular to the plane spanned by the photon momentum ${\bm k}$ and the magnetic field ${\bm B}_{\rm ext}$ (${\bm b}$ is in the plane), and the ``$\parallel$-mode" is that the propagating magnetic field ${\bm b}$ is perpendicular  to the same plane (${\bm e}$ is in the plane).\footnote{
Note that these definitions are interchanged in some literature, e.g., Ref.~\cite{Adler:1970gg}.} 
The refractive indices of these two modes are found to be 
\begin{subequations}
\begin{eqnarray}
n_\perp&=&1+\frac{8\alpha^2}{45m^4}B^2 \sin^2 \theta\, , \label{nperp-weak}\\
n_\parallel&=&1+\frac{14\alpha^2}{45m^4}B^2 \sin^2 \theta \, ,\label{nparallel-weak}
\end{eqnarray}
\end{subequations}
where $\theta$ is the angle between the direction of propagation and the magnetic field. 
Those results are valid only when $B\ll B_c=m^2/e$. 
Therefore, it is useful to rewrite them with respect to the ratio $B/B_c$ 
in the form $n_{\perp,\para} = 1+\Delta n_{\perp,\para}$. 
The corrections are given by  $\Delta n_{\perp}= (7/4) \Delta n_{\para}= (2\alpha /45\pi)(B/B_c)^2\sin^2\theta$, 
so that the deviations of the refractive indices from unity are quite small $\Delta n_{\perp,\para}\ll 1$. 
This regime corresponds to the ``soft-photon and weak-field limit'' in Fig.~\ref{fig:scales-polarization}.

One of the possible experimental setup to measure the vacuum birefringence is to use the ultra-high-intensity laser\footnote{Another setup is to measure the rotation angle of the polarization direction of laser photons that travel in a static homogeneous magnetic field. The merit of this setup is that the deflecting angle should be proportional to the propagation length so that a large effect is expected if one could prepare a homogeneous magnetic field of a large size. Typical experiment is the PVLAS collaboration, which also searches for axions in the same facility. Although it was once reported that they observed a small rotation of the photon polarization, it turned out later that it was due to an instrumental artifact and was not of physical origin \cite{Zavattini:2005tm, Zavattini:2007ee}. See also Ref.~\cite{DellaValle:2015xxa} for current status of the experiment.} such as to be built in Extreme-Light Infrastructure (ELI) in Europe \cite{ELI,TopicalReviewLASER,Heinzl:2006xc}. A probe laser will get affected during its propagation in a high-intensity laser, yielding a small amount of change in the polarization direction. This change could be detected as an interference pattern between the original and modified probe laser fields. Some of the recent attempts are discussed in Ref.~\cite{Battesti:2012hf}. Intensity of the laser field in ELI will reach the current highest value $I=E^2=B^2\sim 10^{25}\, $W/cm$^2$, which is however far below the intensity of the critical field $I_c=E_c^2=B_c^2=4.4\times 10^{29}\, $W/cm$^2$. Thus, even with such a high-intensity laser, the ratio $B/B_c$ is small $\sim 10^{-2}$, and we should treat it as a weak field. If one uses Eqs.~(\ref{nperp-weak}) and (\ref{nparallel-weak}),  one finds that deviation of the refractive indices from unity will be quite tiny $\Delta n\sim 10^{-9}$. Extremely high precision measurement will be necessary to detect this small deviation. Nevertheless, considering that the vacuum birefringence as a dispersive effect occurs in external fields whose strengths are well below the critical value, we expect that, in laboratory experiments, it is much easier to detect the vacuum birefringence than the Schwinger mechanism which at least requires strong electric fields of the order of the critical field $E_c=m^2/e$.

\subsubsection{Low-energy photons (II): Wrenchless fields}

\label{sec:photons_HE-2}

Low-energy photon propagation in a strong field should be investigated 
with the full nonlinear expression of the HE effective Lagrangian. 
While it is still tough to analytically treat the HE Lagrangian for arbitrary values of $\F$ and $\G$, 
one could find other expansion parameters instead. 
One such example is the ``wrenchless" fields specified as $\F\neq 0,\ \G=0$ 
and its first correction with respect to a nonzero $\G$ 
as was shown by Heyl and Hernquist \cite{Heyl:1996dt, Heyl:1997hr}. 
The wrenchless fields cover the weak- to strong-field regimes within the constraint that $ \G $ is a small quantity.

The HE Lagrangian for the wrenchless fields and its corrections are simply given 
in the series expansion with respect to $\G^2$:
\beq
{\cal L}_{\rm HE}={\cal L}_{\rm HE}(\F\neq 0,\G=0)\, +\, \left. \frac{\del{\cal L}_{\rm HE}}{\del \G^2}\right|_{\G^2=0}\G^2 +\cdots
\, ,
\eeq 
where each term is given as a function of a dimensionless variable $\rho \equiv \sqrt{B_c^2/(2\F)}$ with $B_c=m^2/|e|$. 
It is quite straightforward to perform the expansion and obtain explicit forms of the first two terms\footnote{
Note correspondences between the notations here and in Refs.~\cite{Heyl:1996dt, Heyl:1997hr}: 
$ \F = I/4$, $ \G = K/4$, and $\rho  =  \xi^{-1} $. 
Also, the effective Lagrangian therein is smaller by an overall factor of $ 1/(8\pi^2) $ 
according to the difference in electromagnetic units. 
} 
\begin{subequations}
\begin{eqnarray}
{\cal L}_{\rm HE}(\F,\G=0)&=&-\F \left[1-2 \frac{ e^2}{ 8\pi^2} X_0\left( \rho  \right)\right]
\, ,
\\
\left.\frac{\del{\cal L}_{\rm HE}}{\del \G^2}\right|_{\G^2=0} 
&=&
\frac{e^2}{ 8\pi^2}  \frac{1}{4 \F} X_1\left( \rho  \right)
\, ,
\end{eqnarray}
\end{subequations}
with 
\begin{subequations}
\begin{eqnarray}
\label{eq:X0-reg}
X_0 (\rho) &=& 
\int_0^\infty \frac{ ds}{ s } e^{ - i (1-i\epsilon) \rho s} 
\Bigg[ \ \frac{1}{s} \cot s - \frac{1}{s^2} + \frac{1}{3}  \Bigg]
\, ,
\\ 
X_1(\rho)  &=& 
\int_0^\infty \frac{ ds}{ s^2 } e^{ - i (1-i\epsilon) \rho s} 
\frac{1}{3}  [ \,  (  3  +  2s^2  )  \cot s -3 s \csc^2 s \, ] 
 \, .
\end{eqnarray}
\end{subequations}
Both of those integrals are convergent. 
Notice that $ X_1(\rho) $ can be expressed with $ X_0(\rho) $ by integrating by parts as 
\begin{eqnarray}
X_1(\rho)  = -2 X_0(\rho)+\rho X_0^{(1)}(\rho)+\frac23 X_0^{(2)}(\rho)-\frac{2}{9 \rho^2}
\, ,
\end{eqnarray}
where $X_0^{(n)}(\rho) \equiv d^n X_0(\rho) /d\rho^n$.

Now we evaluate the coefficient function $ X_0(x) $ that appears in the purely electric or magnetic case ($ \G=0 $), 
and has been well investigated in the classic literature after the seminal papers  \cite{Euler:1935zz, Heisenberg:1935qt}. 
One can also find the literature in the modern languages \cite{Salam:1974xe, Dittrich:1975au, Dittrich:1978fc} 
as well as Refs.~\cite{Heyl:1996dt, Heyl:1997hr} (see also 
review articles~\cite{Dittrich:1985yb, Dunne:2004nc, Andersen:2014xxa}). 
Assuming a real-valued $ \rho $ for a positive value of $ \F>0 $, i.e., the purely magnetic-field case, 
one can rotate the integral contour as  
\begin{eqnarray}
X_0 (\rho)  &=&   \lim_{\delta \to 0}  \mu^{2\delta} \int_0^\infty \frac{ ds }{ s^{2-\delta } } e^{ -  \rho s } 
 \Bigg[ \ -  \coth s + \frac{1}{s }  + \frac{s}{3}  \Bigg]
 \, .
\end{eqnarray}
While this integral is finite in total (as the counter terms have been already included), 
each term in the integrand diverges at the lower boundary $ s = 0 $. 
It is, thus, helpful to regularize each term in the intermediate steps of the computation. 
One such way is to introduce a cut-off to the proper-time integral as mentioned earlier. 
In the above expression, we employed an alternative way via 
the dimensional regularization~\cite{Salam:1974xe, Dittrich:1975au, Dittrich:1978fc} 
and introduced a displacement $ \delta $ that is assumed to be $ \delta > 2 $ and is sent to zero in the end of computation. 
Accordingly, we have introduced an overall factor composed of a dimension-one parameter $ \mu $ 
since the proper-time $ s $ is a dimensionful variable.\footnote{
Nevertheless, this (renormalization) scale does not appear in the final result of $ X_0(\rho) $ 
which has been already made a finite quantity by the subtraction of the second and third terms in its integrand.  
For the same reason, we did not explicitly include the factor of $ (-1)^d $ mentioned below Eq.~\eqref{Lag_original} 
which depends on the dimension $  d$. We will restore these factors in Sec.~\ref{sec:gluon-condensate} 
to discuss renormalization in QCD. 
} 
These schemes provide gauge-invariant regularizations and finally a finite result. 
One can immediately identify the second and third terms with the integral representation of the gamma function $ \Gam(z) $ 
and the the first term with the Hurwitz zeta function $ \zeta(z,a) = \sum_{n=0}^\infty (n+a)^{-z}$ after some arrangements. 
The integral representation of $ \zeta(z,a) $ is given as \cite{HurwitzZetaFunction} 
\begin{eqnarray}
\label{eq:Hurwitz-zeta}
\zeta (z,a ) =  \frac{1}{\Gamma(z)} \int _0^\infty\frac{ ds}{ s^{1-z}}  \frac{ e^{- a s}  }{ 1 - e^{-s}}
\, , \quad \Re e \, [z] >1 \,, \ \Re e \, [a] >0
\, .
\end{eqnarray}
Therefore, one can find a useful formula 
\begin{eqnarray}
\label{eq:zeta-fermion}
\int _0^\infty\frac{ ds}{ s^{1-z}} e^{- a s} \coth s
=  \frac{\Gamma(z)} {2^{z-1}} \left[ \, \zeta (z,  \frac{a}{2} ) - 2^{z-1} a^{-z} \, \right]
 \, ,
\end{eqnarray}
and arrive at a well-known result (see Ref.~\cite{Dittrich:1978fc} for alternative expressions)
\begin{eqnarray}
\label{eq:X0-result}
X_0(\rho) 
 &=&  \lim_{\delta\to 0} \mu^{2\delta}
 \left[ \, -  \Gam(-1+\delta) \{ \, 2^{2-\delta} \zeta(-1+\delta,\frac{\rho}{2}) -  \rho^{1-\delta}  \, \}
 + \rho^{2-\delta} \Gam(-2+\delta) + \frac{1}{3} \rho^{-\delta} \Gam(\delta)
 \, \right]
  \nn
 \\
 &=&  \left(  \frac{\rho}{2} \right)^2 - \frac13
+ 4 \zeta(-1, \frac{\rho}{2}) \ln \frac{\rho}{2} +  4 \zeta'( - 1, \frac{\rho}{2} ) 
\, ,
\end{eqnarray}
where we have the first derivative of the Hurwitz zeta function $ \zeta'(z,a) \equiv \partial  \zeta(z,a)/\partial z  $ 
and a polynomial form $ \zeta(-1,a) = - ( 2a^2 - 2a + 1/3) /4 $. 
Since we do not have constraints on the magnitude of $ \F $, 
we may consider both the weak- and strong-field limits such that $ \rho\gg1 $ and $ \rho\ll1 $, respectively. 
In the weak-field limit ($ \rho\gg1 $), 
the above results correctly reproduce the Euler-Kockel Lagrangian (\ref{EulerKockel}) 
and systematically yields higher-order corrections in $ 1/\rho $.  
In the strong-field limit ($ \rho\ll1 $), the leading behavior of $ X_0(\rho) $ reads 
$ X_0(\rho) \sim \frac13  \ln (2/\rho) $ with constants under the logarithm. 
On the other hand, we get $ X_1(\rho) \sim 2 /( 3 \rho) $, 
and the coefficient of the $ \G^2 $ term reads $ \left. \del{\cal L}_{\rm HE}/\del \G^2\right|_{\G^2=0} 
\propto (e^4/m^4) \rho^2 X_1(\rho) \sim (e^4/m^4) \rho $ which becomes less and less important 
as we increase $ \F $ with fixed values of $ \G $, $ e $, and $m  $. 
Therefore, one finds the leading behavior of the HE effective Lagrangian  
$  {\cal L}_{\rm HE} \sim -  \F ( 1-  \frac{  e^2  }{12\pi^2 }  \ln \frac{2\sqrt{2 e^2 \F} }{ m^2 }  ) $.  
While we assumed a positive value of $ \F>0 $ in the above, 
we can obtain expressions for the other case with a negative value of $ \F<0 $, i.e., the purely electric case, 
with the help of analytic continuation applied to the above results. 
In this case, an imaginary part of the effective Lagrangian should appear from analytic continuation 
in accordance with the Schwinger mechanism discussed in the previous section~\cite{Dittrich:1975au, Dittrich:1978fc, 
Heyl:1996dt, Heyl:1997hr, Dunne:2004nc}.

By using the above result up to the first order with respect to $\G^2$, 
one obtains the electric and magnetic permeability tensors as ($B_c=E_c=m^2/|e|$)
\begin{subequations}
\begin{eqnarray}
\bar\epsilon^{ij}&=&\delta^{ij} 
    +\frac{\alpha}{2\pi}
     \left[
     \left\{-2X_0\left(\rho \right)+ \rho X_0^{(1)}\left(\rho \right)\right\}\delta^{ij} 
     - \rho^2 X_1\left(\rho \right)\frac{B^iB^j}{B_c^2}
     \right]  \, ,\\
(\bar \mu^{-1})^{ij}&=& \delta^{ij} 
    +\frac{\alpha}{2\pi}
     \left[
     \left\{-2X_0\left(\rho \right)+ \rho  X_0^{(1)}\left(\rho \right)\right\}\delta^{ij} 
     + \rho^2  X_1\left(\rho \right)\frac{E^iE^j}{E_c^2}
     \right] \, .
\end{eqnarray}
\end{subequations}
To extract the photon polarization tensor, 
we need the second derivative of the HE Lagrangian with respect to $E^i$ or $B^i$. 
We can obtain a modified Maxwell equation for the propagating fields, and again find two independent modes. When there is only a magnetic field, we find the following refractive indices ($\F=\bm{B}^2/2$ and thus $\rho=B_c/|\bm{B}|$): 
\begin{subequations}
\begin{eqnarray}
n_\perp&=& 1 -\frac{\alpha}{4\pi}X_1\left( \rho \right)\sin^2 \theta\, , \\
n_\parallel&=& 1+\frac{\alpha}{4\pi}\left\{ \rho^2  X_0^{(2)}\left(\rho \right) 
- \rho X_0^{(1)}\left( \rho \right)\right\}\sin^2 \theta\, .
\end{eqnarray}
\end{subequations} 
The deviations from unity could be much larger than those in Eqs.~(\ref{nperp-weak}) and (\ref{nparallel-weak}) for the weak magnetic fields.

It has been discussed in the literature \cite{Heyl:1999dw, Heyl:2003kt, vanAdelsberg:2006uu} that the vacuum birefringence in strong magnetic fields could give rise to a visible effect on the low-energy photon radiation emitted from neutron stars or magnetars, where the strength of magnetic fields would be close to or beyond the critical value $B_c=m^2/e$ (see a review \cite{Harding:2006qn} for more references). In particular, it is expected that the polarization properties of radiation emitted by isolated neutron stars will be strongly influenced by strong magnetic fields of stars. To minimize possible background effects originating from the QED plasma, ``isolated" neutron stars have been thought to be the best candidates for the observation of the vacuum birefringence.  Recently, it was reported that radiation from the famous isolated neutron star, RX J1856.5-3754, having a dipolar surface magnetic field $B \sim 10^{13}$-$10^{14}$~Gauss is substantially polarized, which cannot be explained by known effects and thus would suggest other new effects such as the vacuum birefringence. Indeed, the measured values of polarization are consistent with the picture that they are from the vacuum birefringence\cite{Mignani:2016fwz}. The polarization measurement of neutron stars and magnetars is very important and one of the main targets in future experiments such as PRAXyS (the Polarimeter for Relativistic Astrophysical X-ray Sources) \cite{PRAXyS}. More precise measurement in a wide range of radiation energies will be necessary for definitive identification of the effects with that of the vacuum birefringence, which will also improve our understanding of the magnetic structure of the neutron stars or magnetars. 
There is also an attempt to detect the vacuum birefringence 
by relativistic heavy-ion collisions in recent years \cite{STAR:2019wlg, Brandenburg:2022tna}. 
The maximum magnitudes of the electromagnetic fields can be 
as large as the QCD scale $\sim 200 \ \MeV $ that is 
much larger than the critical field strength defined by electron mass. 
The above expansion may be relevant for 
the field configurations created there.

\subsubsection{High-energy photons: Beyond Heisenberg-Euler effective theory}

\label{sec:vp}

The HE Lagrangian is no longer a useful effective theory 
when the photon field varies in spacetime at scales shorter than the electron's Compton wavelength. 
Propagation of such a high-energy photon in a slowly varying magnetic field is a typical situation which requires to go beyond the HE Lagrangian. Such situation will be relevant when we consider high-energy photons emitted in heavy-ion collisions or $\gamma$ ray propagation in astrophysical jets, both of which are accompanied by strong magnetic fields. 
In order to study such problems, we need to go back to the original representation (\ref{effectiveaction}) of the effective action for arbitrary gauge fields or directly evaluate the polarization tensor (\ref{eq:prop}) as shown in the diagram, Fig.~\ref{fig:photon_vp}. 
While the formal expression of the polarization tensor (at the one-loop level) had been known for a long time \cite{batalin1971green,Tsai:1974ap,Urrutia:1977xb, Baier:2007dw}, analytic insights in a whole kinematical region have not been available until recently due to technical difficulties in the expression in terms of the double-parameter integral \cite{Karbstein:2011ja}. Here, based on Refs.~\cite{Hattori:2012je, Hattori:2012ny}, we outline the way to reach an analytic representation of the polarization tensor under magnetic fields with arbitrary strength.

\subsubsection*{Double-integral representation of the polarization tensor}

Again, let us focus on the case with only a homogeneous magnetic field $\bm{B}=(0,0,B)$ directed to the $z$ direction. The polarization tensor (\ref{eq:prop}) has integrals with respect to the internal momentum $p$, and two proper times 
which we denote as $\tau_1$ and $\tau_2$ originating from the two resummed propagators. 
The integration over $p$ is just a Gaussian integral and is straightforwardly carried out. 
We are left with the remaining proper-time integrals which can be rewritten by using two dimensionless variables 
$\tau = eB ( \tau_1 + \tau_2 )/2 $ and $\beta = eB ( \tau_1 - \tau_2 )/\tau$. 
Notice that the scalar functions $\chi_i$ $(i=0,1,2)$ defined in Eq.~(\ref{eq:Pex}) are dimensionless, and thus we further introduce three dimensionless variables, 
$\Br = {B}/{B_c}=eB/m^2$, 
$r_\parallel^2 =  { q_\parallel^2 }/{ (4m^2) }$ 
and 
$r_\perp^2 = { q_\perp^2 }/{ (4m^2) } = - { | \bq_\perp | ^2 }/{ (4m^2) }$, 
which are taken as the axes of Fig.~\ref{fig:scales-polarization}. 
Then, the scalar functions $\chi_i$ are expressed as 
\begin{eqnarray}
\chi_i (r_\parallel^2, r_\perp^2; \Br) 
&=& 
\frac{\alpha}{4\pi} \int_{-1}^1 \!\!\! d\beta  \int_0^\infty \!\!\!\! d\tau \ 
\frac{ \Gamma_i (\tau, \beta) }{ \sin \tau } \ 
{\rm e}^{-i u \cos (\beta \tau) } \ 
{\rm e}^{ i \eta \cot \tau } \ 
{\rm e}^{-i\phi_\parallel \tau} \, .
\label{eq:vp0} 
\end{eqnarray}
Here, we have introduced two shorthand notations, $\eta \equiv - 2 { r_\perp^2 }/{ \Br } $ and $u \equiv { \eta }/{\sin\tau} $, 
which are useful to proceed to the analytic calculation below. 
Lastly, $\phi_\parallel$ and $\Gamma_i$ are known functions given by the following forms 
\cite{Adler:1971wn, batalin1971green, Tsai:1974fa, Tsai:1975iz, Tsai:1974ap, Melrose:1977ftf, 
Urrutia:1977xb, Schubert:2000yt, Dittrich:2000wz} (see also 
Refs.~\cite{Dittrich:1985yb,Dittrich:2000zu} for details):
\begin{eqnarray}
&&\phi_\parallel (r_\parallel^2, \Br) 
=  \frac{1}{\Br}\left\{ 1 - (1-\beta^2) \ r_\parallel^2  \right\}\, , 
\label{eq:phi0} 
\end{eqnarray}
and
\begin{eqnarray}
&&\Gamma_0  (\tau, \beta)
= \cos( \beta \tau) - \beta \sin( \beta \tau) \cot \tau \, ,
\nonumber
\\
&&\Gamma_1  (\tau, \beta) 
= (1-\beta^2) \cos \tau   - \Gamma_0 (\tau, \beta)
\label{eq:Gam0}
\, ,\\
&&\Gamma_2  (\tau, \beta)
= 2 \frac{\cos( \beta \tau ) - \cos \tau}{\sin^2 \tau} - \Gamma_0 (\tau, \beta)
\nonumber
\, .
\end{eqnarray}
In Eq.~(\ref{eq:vp0}), the coupling constants in the overall factor $\alpha=e^2/4\pi$ 
come from the two vertices at which the propagating photon is coupled to the electron one-loop. 
The others are from the interactions between the electron-positron fluctuations and the external magnetic field,  
as is evident from the fact that they always appear with the magnetic field in the form of $eB$. 

\begin{figure}
     \begin{center}
              \includegraphics[width=0.6\hsize]{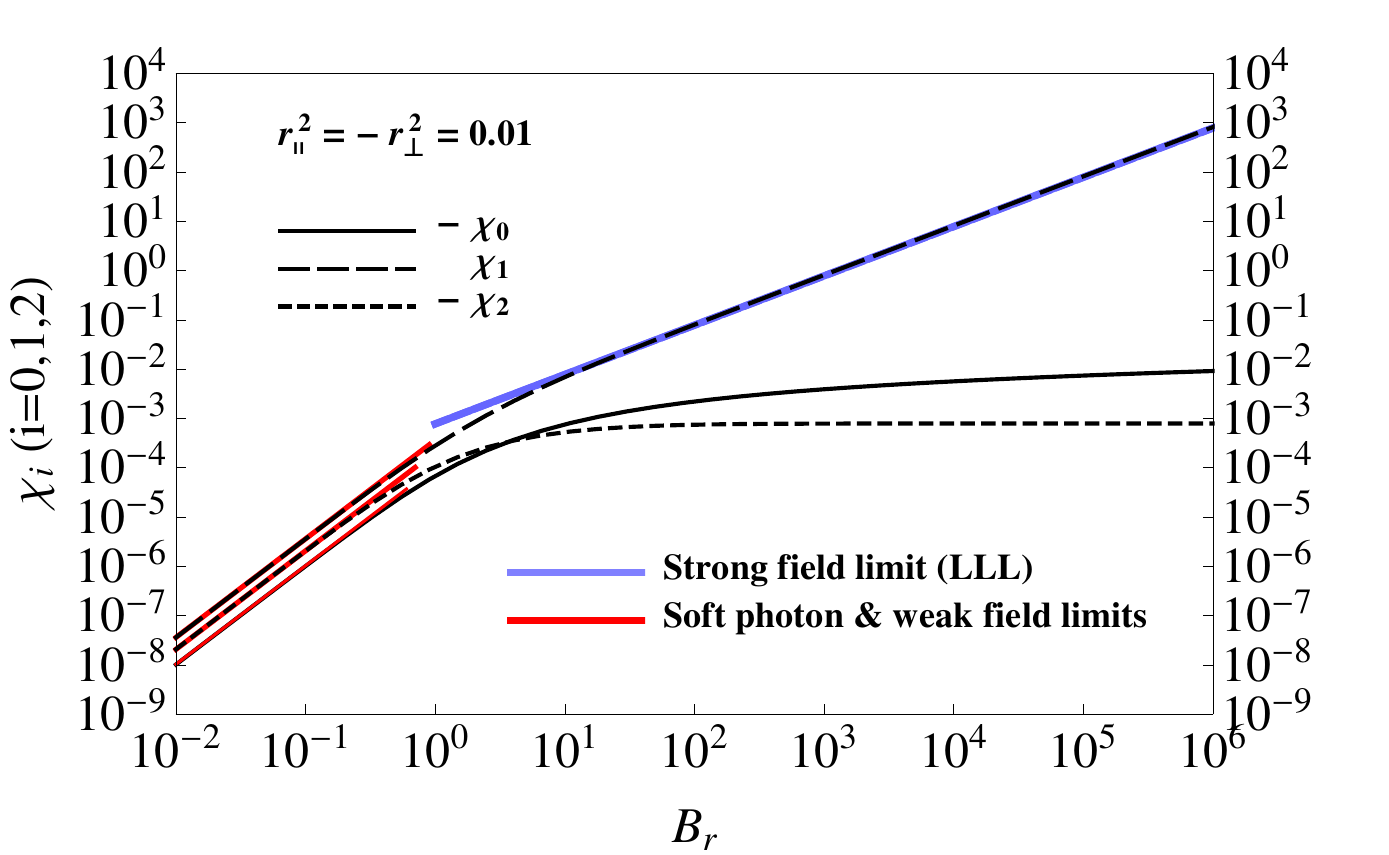}
     \end{center}
\vspace{-0.5cm}
\caption{Numerical results of $ \chi_i $ in the soft-photon regime, compared with 
the analytic expressions both in the weak and strong field limits, taken from \cite{Hattori:2012ny}.}
\label{fig:chi_B-dep}
\end{figure}

This representation contains involved integration with respect to $\beta$ and $\tau$. 
In fact, this complexity has prevented previous studies from complete analytical understanding of the vacuum birefringence \cite{Adler:1971wn,Tsai:1974fa, Tsai:1975iz,Melrose:1977ftf}, 
and even from performing numerical computation except in a limited kinematical region 
where there is no imaginary part (cf. Fig.~\ref{fig:scales-polarization}) \cite{Kohri:2001wx}. 
All the studies started from this representation and had to resort to some kind of approximations 
depending on interested kinematics and conditions.

\subsubsection*{Soft-photon and weak-field limit}

It would be instructive to first look at one of these approximations in the soft-photon and weak-field limit, 
which turns out to reproduce the results in Eqs.~(\ref{nperp-weak}) and (\ref{nparallel-weak}) 
obtained earlier from the HE Lagrangian. 
Consider the simultaneous limits for the soft photon $\omega^2 \ll 4m^2$ and weak field $\Br \ll 1$. 
In Eq.~(\ref{eq:phi0}), the integrands of the $\tau$-integral is damped 
within a scale $ \sim \Br $ when $  r_\parallel^2 \ll 1$. 
Therefore, when $\Br \ll 1$, we can expand the integrands with respect to the small value of $\tau$. 
In this limit, the exponential factor is approximated as 
\begin{eqnarray}
\exp(-i\phi_\para \tau) 
&\sim& 
\left[ 1 + (1-\beta^2) r^2 \frac{\tau}{\Br} \right]  e^{-\frac{\tau}{\Br}}
\, ,
\end{eqnarray}
where $ r^2 = q^2/(4m^2) $. The $\Gamma_i$ are also expanded as 
\begin{subequations}
\begin{eqnarray}
\frac{1}{\sinh\tau} \left( \Gamma_0 - \Gamma_0^{\rm free} \right) & = & - \frac{ 1 }{ 6 } (1-\beta^2)^2 \  \tau 
\, ,
\\
\frac{1}{\sinh\tau} \Gamma_1  &=& \frac{1}{6} (1-\beta^2)(3-\beta^2) \ \tau
\, ,
\\
\frac{1}{\sinh\tau}  \Gamma_2   &=& - \frac{1}{12} (1-\beta^2)(3+\beta^2) \ \tau
\, .
\end{eqnarray}
\end{subequations}
A function $ \Gamma_0^{\rm free}  $ is the contribution in the absence of the magnetic field 
that is equal to the vanishing $ B $ limit of $ \Gamma_0 $, 
and removes the logarithmic singularity $ \sim1/s $ coming from the ultraviolet divergence.\footnote{
In case of the two-point function, the vacuum contribution, 
without any insertion of the external field, has the logarithmic divergence, 
while all the other diagrams with insertions of the external magnetic field are convergent.}
After performing the elementary integrals, we obtain 
\begin{subequations}
\begin{eqnarray}
\chi_0 &=& - \frac{2\alpha}{315 \pi} \Br^2 ( 7 + 12 r^2 )
\, ,
\\
\chi_1 &=& \frac{\alpha}{315 \pi} \Br^2 ( 49 + 80 r^2 )
\, ,
\\
\chi_2 &=& - \frac{4\alpha}{315 \pi} \Br^2 ( 7 + 11 r^2 )
\, .
\end{eqnarray}
\end{subequations}
In \fref{fig:chi_B-dep}, the above analytic results are shown with the numerical results in the soft-photon limit. 
They agree with each other in the soft-photon and weak-field limit, 
which is indicated as an overlap region in Fig.~\ref{fig:scales-polarization}. 
The analytic expression for the strong field limit, the LLL approximation, will be discussed below. 
Plugging the weak-field results into the refractive indices given in Eq.~(\ref{def:epsilon})--(\ref{eq:eps1bparallel}), 
we find that the results from the HE Lagrangian (\ref{nperp-weak}) and (\ref{nparallel-weak}) are reproduced 
in the vanishing momentum limit $ r^2 \to 0 $. 
The above results provide the leading corrections with respect to the photon momentum $ r^2 $.

These approximates are, however, valid only in the limited kinematical regions 
and the other regions were not fully investigated. 
After almost 40 years since the expression (\ref{eq:vp0}) was obtained, 
an analytic result for the double integral 
was obtained in Ref.~\cite{Hattori:2012je}, 
and the result was soon confirmed numerically\footnote{
The authors of Ref.~\cite{Ishikawa:2013fxa} also gave a renormalization prescription 
when $ \Gamma_0 $ is given by the discrete sum of contributions from the Landau levels, 
while the subtraction term $ \Gamma_0^{\rm free} $, which is not affected by the magnetic field, 
is a continuous function.} \cite{Ishikawa:2013fxa}. 
The final analytic representation for the scalar functions $\chi_i$ obtained in Ref.~\cite{Hattori:2012je} 
is given by an infinite series of functions composed of the associated Laguerre polynomials $L_\ell^n(x)$. 
The appearance of the associated Laguerre polynomials is quite natural 
because they appear in the wave functions of charged fermions in a magnetic field 
as we saw in Sec.~\ref{sec:S-gauge}.\footnote{
Nevertheless, the vacuum polarization diagram is invariant with respect to the gauge of the external magnetic field 
as mentioned earlier.
} 

\subsubsection*{Full analytic result and the Landau levels}

Below, we give the outline to reach the analytic expression 
and explain the physical meaning of the result. 
The main difficulties in performing the double integral in the scalar functions $\chi_i$ (\ref{eq:vp0}) come from the fact 
that trigonometric functions appear in the exponentials in the integrands. 
Actually, those double-exponential forms generate arbitrarily higher harmonic modes due to nonlinear nature. 
Those difficulties are overcome as follows: First of all, the factor ${\rm e}^{-iu\cos(\beta \tau)}$ can be ``expanded" by using the partial wave decomposition which is an established technique in quantum mechanics. 
This exponential factor is expressed as an infinite series including the modified Bessel functions $I_n(-iu)$. 
Next, we can further rewrite the product of the modified Bessel function $I_n(-iu)$ 
with the other exponential factor ${\rm e}^{i\eta \cot \tau}$ 
by using another infinite series including the associated Laguerre polynomials $L_\ell^n(x)$. 
After those decompositions, all the exponential factors have simple shoulders linear in $ \tau $, 
and we are able to exactly perform the double integrals with respect to $\tau$ and then $\beta$ to obtain the final result. 
The details are written in Ref.~\cite{Hattori:2012je}.

After all, we have two integer indices $n$ and $\ell$ to be summed from zero to infinity. 
Although both of them were originally introduced in the above decompositions for the technical purposes, 
they actually correspond to the Landau-level indices of the fermion-antifermion pair. 
This interpretation is justified by the following observation. Each term in the infinite summation contains the function:
\begin{eqnarray}
I_{\ell \Delta}^n (r_\parallel^2)
&\equiv&
\int_{-1}^1  
\frac{ d\beta }{ \ r_\parallel^2 \beta^2 - n \Br \beta + (1-r_\parallel^2) + (2\ell+ n) \Br \ }\, .
\end{eqnarray}
Its kinematical property is specified by a discriminant for the second-order function of $\beta$ in the denominator: 
$
{\mathcal D}\equiv  ( - n \Br )^2 - 4\rp\, \{(1-\rp)+(2\ell +n)\Br \}\, .
$
Depending on the sign of $\mathcal D$, the integral $I^n_{\ell \Delta}(\rp)$ 
takes either real or complex value. Therefore, the solutions to an equation, 
$\mathcal D(r_\parallel ) = 0$, specify the threshold for the emergence of imaginary part:
\begin{eqnarray}
\rp =  \frac{1}{4} \left[
\sqrt{ 1 + 2 \ell \Br } \pm \sqrt{ 1 + 2 ( \ell + n ) \Br }  \right] ^2 
\equiv s_\pm^\idx
\, .
\label{eq:s_pm}
\end{eqnarray} 
One can prove that an imaginary part appears only when $ s_+^\idx < r_\parallel^2$. 
In terms of dimensionful quantities, the threshold condition $\rp=s_+^\idx$ is rewritten as 
\begin{eqnarray}
q_\parallel^2 = \left[ \sqrt{  m^2 + 2 \ell eB \ } + \sqrt{  m^2 + 2 ( \ell + n ) eB \ }  \right]^2
\, .
\label{eq:thrs}
\end{eqnarray}
Recall that the fermion dispersion relation in a magnetic field is given as $\varepsilon_n(p_z)=\sqrt{m^2+p_z^2+2neB}$ with $n\ge 0$. 
The right-hand side of Eq.~(\ref{eq:thrs}) exactly agrees with 
the invariant mass of a fermion-antifermion pair with a vanishing longitudinal momentum, 
and the integers $\ell$ and $\ell+n$ specify the Landau levels. 
Therefore, the infinite summations over $n$ and $\ell$ correspond to 
summing all the Landau levels associated with the fermion-antifermion pair constituting the one-loop. 
The lowest threshold is given by the lowest Landau levels $n=\ell=0$, 
which agrees with the strong-field limit obtained in other methods \cite{Melrose:1977ftf, Fukushima:2011nu,Hattori:2012je, Hattori:2012ny}.

The appearance of an imaginary part indicates that a {\it real photon} decays 
into an electron-positron pair in the magnetic field above the threshold energy. 
Clearly, this contrasts to the case without a magnetic field 
where the on-shell conditions for the fermion-antifermion pair require a nonzero invariant mass of photon, $ q^2 \geq (2m)^2 $. 
The kinematics in a magnetic field can be understood as follows. 
If we regard the Landau level and the transverse photon momentum as 
an ``effective mass'' of the fermion and the photon, respectively, 
their dispersion relations may be interpreted as (1+1)-dimensional ones. 
Then, the kinematics in a magnetic field can be identified with 
that for the decay of a ``massive'' gauge boson to a fermion-antifermion pair in the (1+1) dimensions: 
There is no kinematical prohibition of decay of a massive boson. 
Consistent to this observation, the threshold condition (\ref{eq:thrs}) constrains 
only the longitudinal part of the photon momentum $ q_\para^2 $.\footnote{
Remember also that the transverse components of the kinetic momentum are not good quantum numbers 
and that those of the canonical momentum are not conserved either 
in a general gauge (cf. Sec.~\ref{sec:Ritus-Feynman}). 
}  
Setting $q_z=0$ in Eq.~(\ref{eq:thrs}),\footnote{ 
One can always move to such a Lorentz frame where the longitudinal photon momentum vanishes $q_z=0$, 
thanks to the boost invariance along the magnetic field. 
} 
we find the threshold photon energy $\omega_{\rm th}(n,\ell)=\sqrt{  m^2 + 2 \ell eB \ } + \sqrt{  m^2 + 2 ( \ell + n ) eB \ }$. 
In the LLL, the threshold does not depend on $B$, and is simply given by $\omega_{\rm th}(0,0)=2m$. 
Therefore, even in a magnetic field, a real photon does not decay if its energy is less than $2m$.


\begin{figure}
\begin{minipage}{0.5\hsize} 
	\begin{center}\hspace*{-5mm}
		\includegraphics[width=0.99\hsize]{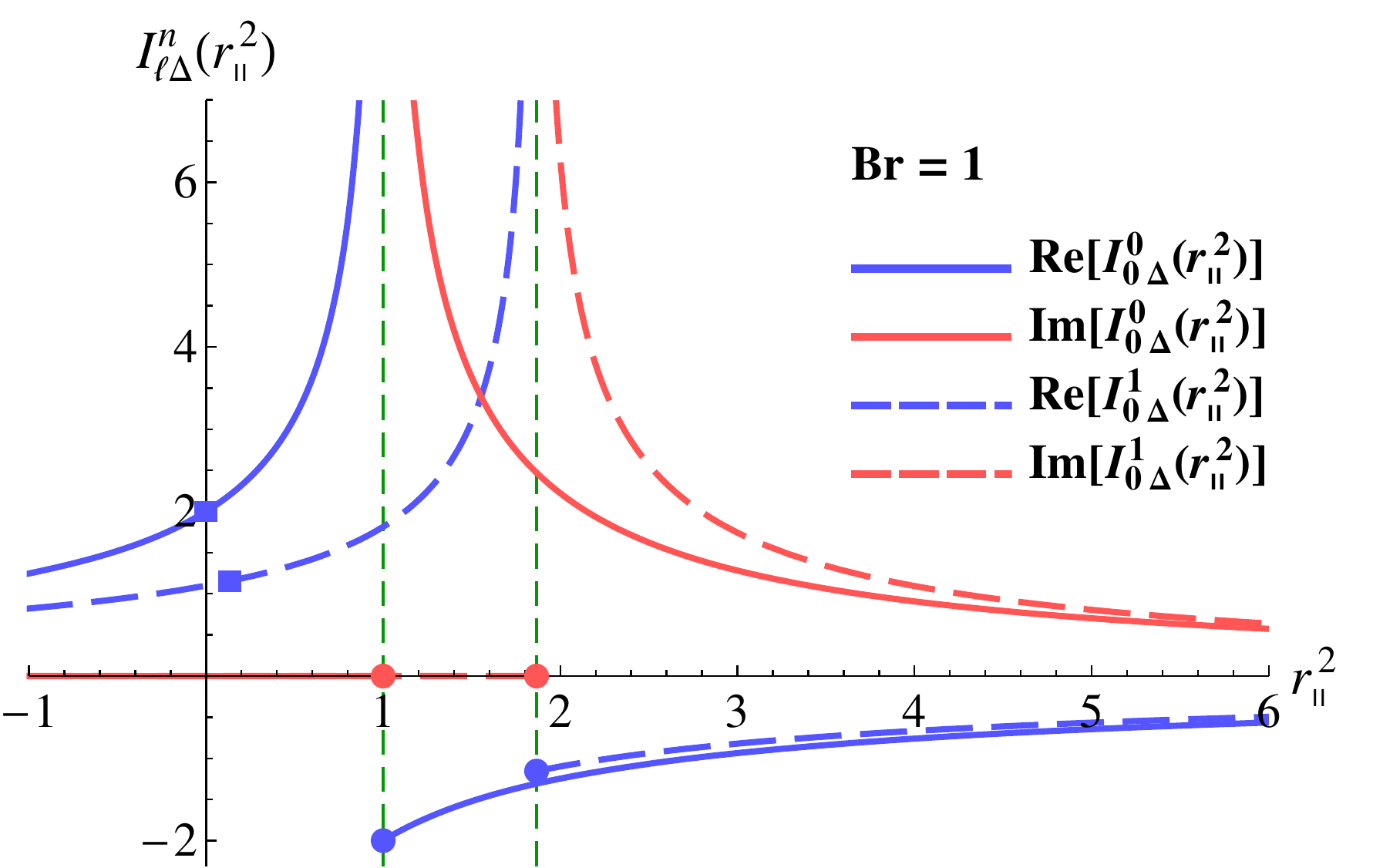}
	\end{center}
\end{minipage}
\begin{minipage}{0.5\hsize}
	\begin{center}
\includegraphics[width=0.99\hsize]{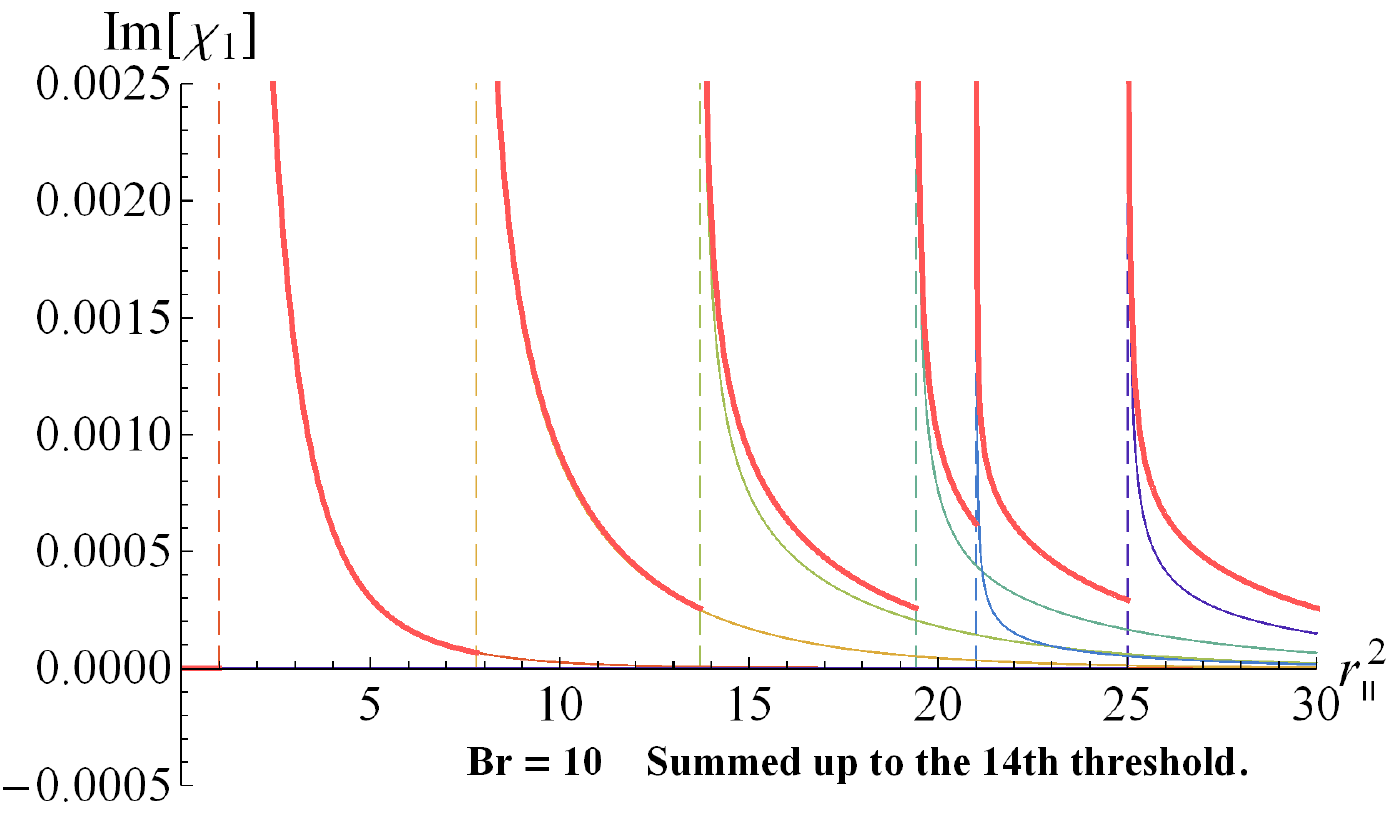}
	\end{center}
\end{minipage}
\caption{Real and imaginary parts of the function $I_{\ell \Delta}^n(\rp)$ 
for the lowest and first Landau levels ($n=\ell=0$ and $n=1,\ell=0$) [left]. 
The imaginary part of $ \chi_1 $ with many thresholds 
which show contributions of each Landau level (thin lines) and the sum (solid line) [right]. 
Both figures are taken from Ref.~\cite{Hattori:2012je}. 
}
  \label{fig:I}
\end{figure}

The function $I_{\ell \Delta}^n(r_\parallel^2)$ has a characteristic behavior around the threshold $r_\parallel^2=s_+^{\ell n}$, which is common to any pair of the Landau levels $(\ell,\ell+n)$. In Fig.~\ref{fig:I}, we show threshold structures of the real and imaginary parts of $I_{\ell \Delta}^n(\rp)$ for the lowest and first Landau levels ($n=\ell=0$ and $n=1,\ell=0$). 
The real part of $I_{\ell \Delta}^n(r_\parallel^2)$ increases as $r_\parallel^2$ approaches 
the threshold from below, and diverges at the threshold. 
On the other hand, the imaginary part is zero below the threshold, 
but has an inverse-square-root dependence $ \sim 1/(r_\para^2-s_+^{\ell n})^{1/2} $ above the threshold. 
This is the typical threshold behavior in the (1+1) dimensions, determine by the phase-space volume. 
As a result of such a divergent behavior of $I_{\ell\Delta}^n(r_\parallel^2)$ at each threshold, the scalar coefficients $\chi_i(r_\parallel^2)$ 
have divergences at infinitely many points, 
exhibiting a comb-shaped dependence on $r_\para^2 $ (see \fref{fig:I}). 
It is important to notice that these divergences are harmless for the physical quantities, i.e., 
the dielectric constants $\epsilon_{\perp,\parallel}$ and the refractive indices $n_{\perp, \parallel}$. 
This is easily understood from the explicit representation of the dielectric
constants,  Eqs.~(\ref{eq:eps1bperp}) and (\ref{eq:eps1bparallel}). 
First, all the coefficients $\chi_i\, (i=0,1,2)$ contain the function  $I_{\ell \Delta}^n (r_\parallel^2)$, and thus they have divergences of the same order at the same $ r_\para^2$. Second, 
these coefficients appear both in the denominators and numerators of Eqs.~(\ref{eq:eps1bperp}) and (\ref{eq:eps1bparallel}). 
Therefore, the singularities are canceled to give finite values of the dielectric constants and the refractive indices.

\subsubsection*{Refractive indices in a strong magnetic field}

Having obtained the explicit forms of the scalar functions $\chi_i$, 
we are able to compute the dielectric constants $\epsilon_{\perp, \parallel}$ or equivalently the refractive indices $n_{\perp,\parallel}$. The formulas are already given in Eqs.~(\ref{eq:eps1bperp}) and (\ref{eq:eps1bparallel}). However, it should be noticed that these equations must be solved self-consistently with respect to $\epsilon_\perp$ and $\epsilon_\parallel$ since $\chi_i$ are functions of $\rt$ and $\rp$ \cite{Hattori:2012ny}. In fact, we have not yet specified any dispersion relation for the external photon momentum when we computed the scalar coefficient functions $\chi_i$. For an on-shell photon, the dispersion relations should be determined by Eqs.~(\ref{eq:eps1bperp}) and (\ref{eq:eps1bparallel}), which do contain the dielectric constants on the right-hand sides through the photon momenta, $\rp$ and $\rt$. According to the definition of the dielectric constant (\ref{def:epsilon}), those photon momenta are rewritten in terms of $\epsilon, \omega$ and $\theta$ as (by using $q_z^2=|\bm{q}|^2\cos^2 \theta=\epsilon\, \omega^2 \cos^2\theta,\, \bm{q}_\perp^2=|\bm{q}|^2-q_z^2=|\bm{q}|^2(1-\cos^2\theta)=\epsilon\, \omega^2 \sin^2 \theta$), 
\begin{subequations}
\begin{eqnarray}
\rp &=& \tilde\omega^2 (1- \epsilon \, \cos^2 \theta)\, ,\label{rp}\\
\rt &=& - \epsilon \, \tilde\omega^2 \sin^2 \theta\, ,\label{rt}
\end{eqnarray}
\end{subequations}
where we introduced a scaled photon energy, $\tilde \omega^2 = \omega^2/(4m^2)$. This fact indicates that we have to solve these relations in a self-consistent way with respect to the dielectric constant appearing on the both sides. Physically, self-consistent treatment corresponds to including the backreaction of the vacuum against the incident photon field.

Such procedure will demand careful treatments of $\chi_i$ with the infinitely many divergences. 
However, analysis is greatly simplified if we take the limit of a strong magnetic field and focus on the region of photon energies around the first threshold. 
In this limit, we are allowed to approximate $\chi_i$ by the lowest Landau levels (LLL) $n=\ell=0$. 
This is because the higher Landau levels are all gapped by the energy scale $ \sim \sqrt{eB} $ and go far away from the LLL 
as the magnetic field increases. 
With such a simplification, we are able to explicitly study how the self-consistent treatment affects the solution \cite{Hattori:2012ny}. 
Since $\chi_i$ behaves in the same manner in the vicinity of all the thresholds, 
which goes like the inverse square root as mentioned above, 
the behaviors of the refractive indices near the threshold regions 
should share qualitatively the same character as that in the LLL.

\begin{figure}[t]
\begin{center}
 \begin{minipage}[t]{0.45\hsize}
   \includegraphics[width=\hsize]{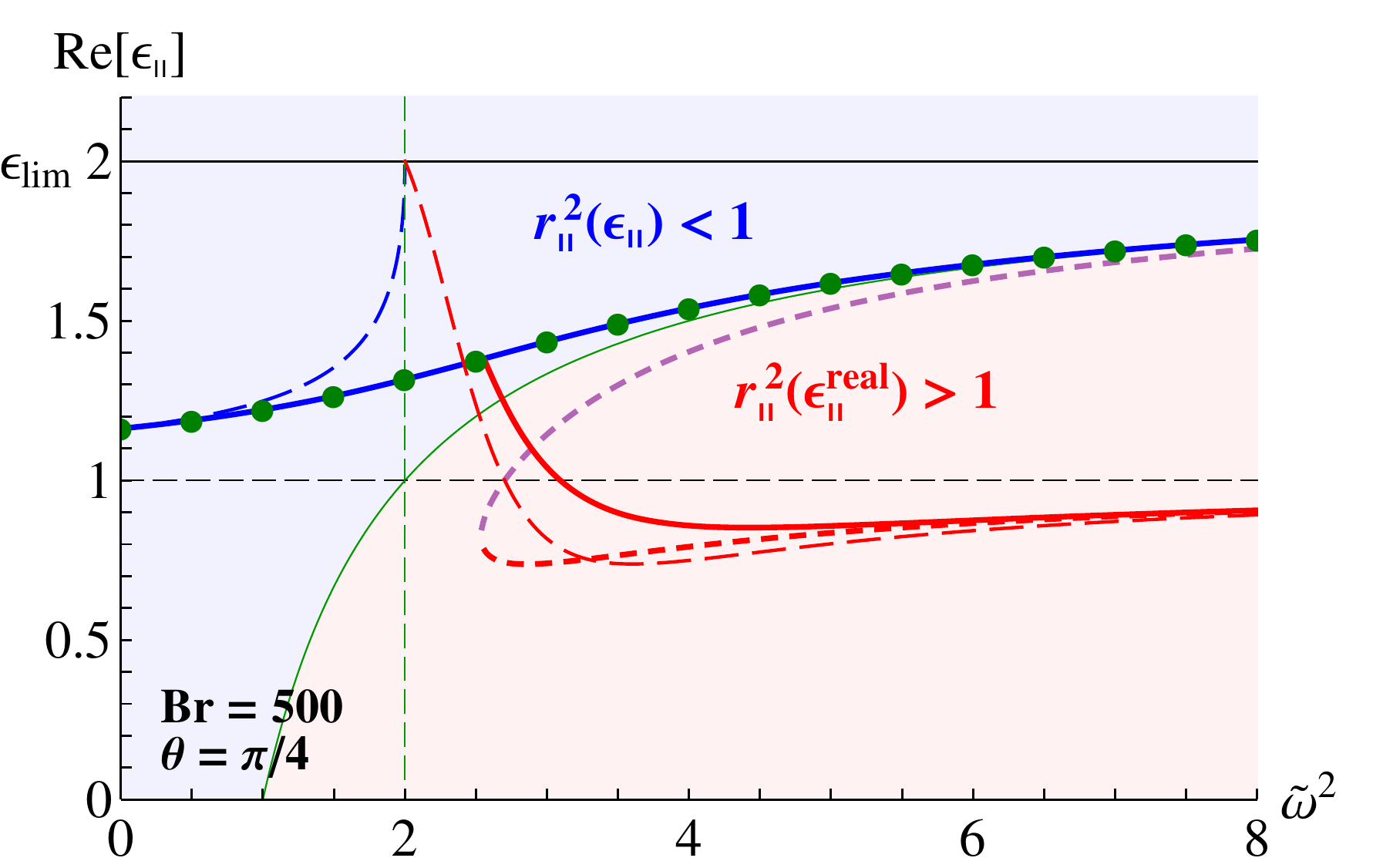}
 \end{minipage}
  \hspace{0.01\hsize}
 \begin{minipage}[t]{0.45\hsize}
   \includegraphics[width=\hsize]{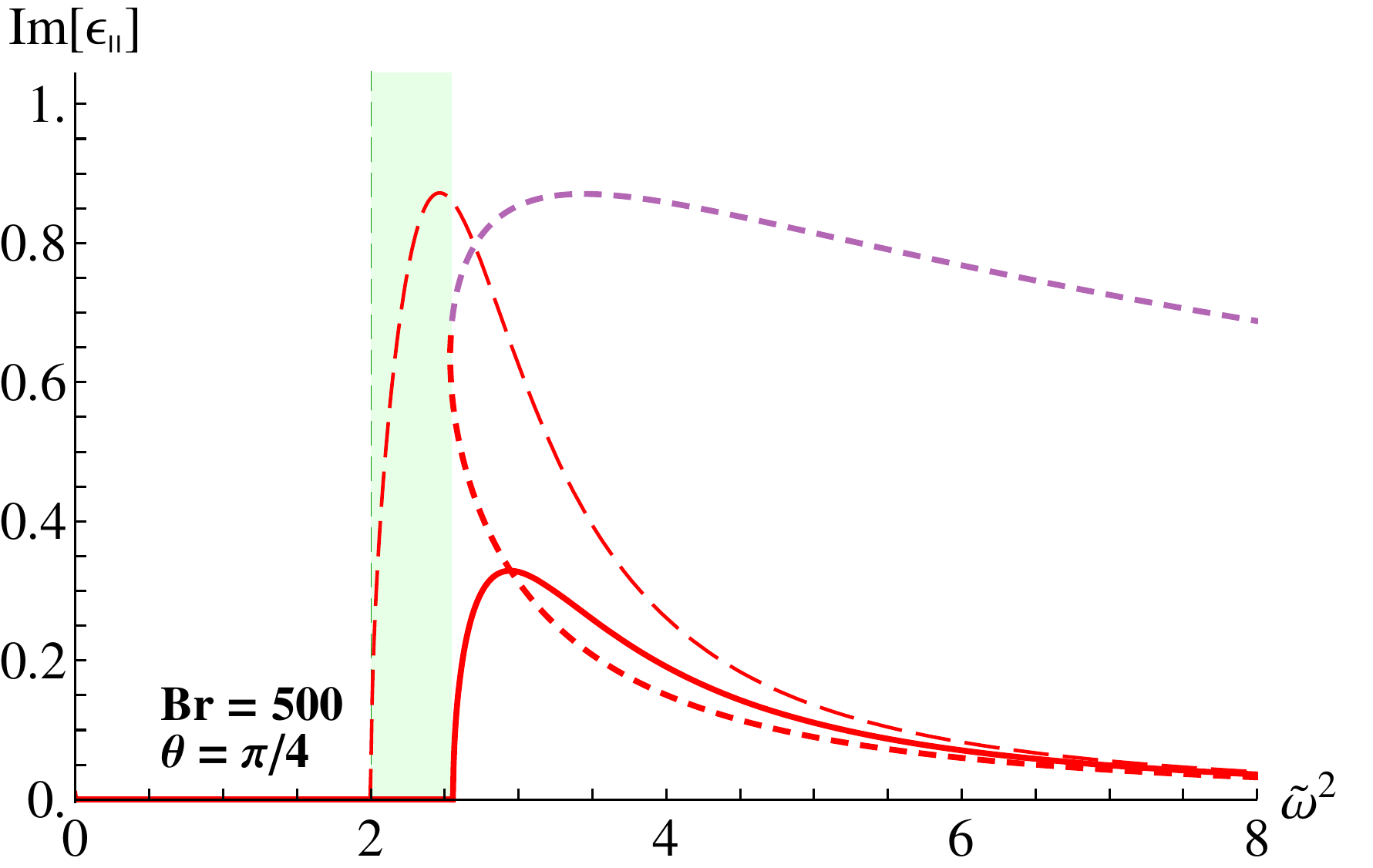}
 \end{minipage}
\end{center}
\vspace{-0.5cm}
\caption{
Energy dependence of the dielectric constant $\epsilon_\parallel$ in the LLL approximation. The angle $\theta$ between the photon momentum and the magnetic field is taken as $\theta=\pi/4$ and $\Br = B/B_c=500$. Taken from Ref.~\cite{Hattori:2012ny}. 
}
  \label{fig:full500}
\end{figure}

The scalar coefficient functions in the LLL approximation are given as 
(see Appendix~\ref{sec:screening} for more details) 
\begin{subequations}
\begin{eqnarray}
&&\chi_0^{\rm LLL} = \chi_2^{\rm LLL} = 0
\label{eq:chi02_LLL}
\, ,\\
&&\chi_1^{\rm LLL}(\rp, \rt ; \Br) 
= \frac{\alpha \Br }{4\pi } \ {\rm e}^{-\eta}  
\times \frac{1}{r_\parallel^2} \left[ I_{0 \Delta}^0 (r_\parallel^2) - 2  \right]
\label{eq:chi1_LLL}
\, .
\end{eqnarray}
\end{subequations}
In Fig.~\ref{fig:chi_B-dep}, we confirm the agreement between this analytic expression and the numerical result, 
and clearly see the dominance of $ \chi_1 $. 
The linear growth of $ \chi_1 $ with respect to $eB $ 
is attributed to the Landau degeneracy factor. 
Then, the dielectric constants are obtained as 
\begin{subequations}
\begin{eqnarray}
&&
\epsilon_\perp^{\rm LLL} = 1\, ,
\label{eq:eps_LLL_t}
\\
&&
\epsilon_\parallel^{\rm LLL} (\tilde \omega, \theta; \Br) 
= \frac{ 1 + \chi_1^{\rm LLL} (\rp,\rt; \Br) }{ 1 +  \chi_1^{\rm LLL}
(\rp,\rt; \Br) \times \cos^2 \theta}
\, .
\label{eq:eps_LLL}
\end{eqnarray}
\end{subequations}
Here, we explicitly write the arguments of $\epsilon_\parallel^{\rm LLL}$ and $\chi_1^{\rm LLL}$. The photon momenta $\rp$ and $\rt$ on the right-hand side should be expressed as Eqs.~(\ref{rp}) and (\ref{rt}), respectively, with $\epsilon=\epsilon_\parallel$.

Even with the above simplified case, we need to solve Eq.~(\ref{eq:eps_LLL}) numerically. The result for $\Br=500$ and $\theta=\pi/4$ is shown in Fig.~\ref{fig:full500}. Solid lines correspond to the self-consistent solutions, and the other lines are all incomplete solutions where some kind of approximations are adopted. First of all, look at the green solid line on the left panel. This corresponds to the threshold line $\rp=1$ which is rewritten as $\epsilon=(1-1/\tilde\omega^2)/\cos^2\theta$. In the high energy limit $\tilde \omega \to \infty$, we find $\epsilon \to 1/\cos^2\theta\equiv \epsilon_{\rm lim}$. This is the largest value for $\epsilon$, and can be also obtained from Eq.~(\ref{eq:eps_LLL}) in the limit $\chi_1^{\rm LLL}\to \infty$. The blue shaded region on the left-hand side of the green solid line corresponds to the region below the threshold $\rp<1$. Second, notice that the blue thick solid line (the real part of $\epsilon_\parallel$) largely deviates from unity and increases with increasing photon energies. At some energy, another red solid line appears, which is accompanied by a nonzero imaginary part (the right panel). Therefore, these red lines correspond to unstable decaying states. Lastly, comparing with the other lines, one finds that the effect of the self-consistent treatment is rather large, in particular around the threshold. We can also study the magnetic field dependence and 
angle dependence which are all available in Ref.~\cite{Hattori:2012ny}.

We can perform the same calculation for the refractive index $n_\parallel^{\rm LLL}$. The result is qualitatively the same as for $\epsilon_\parallel^{\rm LLL}$. In particular, $n_\parallel^{\rm LLL}$ is clearly larger than unity (for the same $\Br$ and $\theta$), and its largest value is given by $n_\parallel^{\rm LLL}=\sqrt{\epsilon_{\rm lim}}\equiv n_{\rm lim}$. For example, at the angle $\theta=\pi/4$, the refractive index $n_\parallel$ keeps increasing with increasing magnetic fields, and 
approaches the limiting value 
$n_{\rm lim}=\sqrt2$. This is comparable to the values which we encounter 
in ordinary life. To name a few, atmosphere of the earth 
($1$ atm, $0\ {}^\circ\mathrm{C}$) and water ($20\ {}^\circ\mathrm{C}$) 
have refractive indices, 
$n_{\rm air} = 1.000293$ and $n_{\rm water} = 1.333$, 
respectively, and ``calcite" known as a representative birefringent 
material shown in Fig.~\ref{fig:biref-pic} has refractive indices 
$n_{\rm o}  = 1.6584$ and $n_{\rm e}  = 1.4864$ 
for ordinary and extraordinary modes, respectively. 
At a magnetic field strength $\Br \sim (m_\pi/m_e)^2 \sim 10^{4-5} $ 
which could be realized in the ultrarelativistic heavy-ion collisions\footnote{
$m_\pi$ and $m_e$ are masses of a pion $m_\pi \sim 140$ MeV and 
an electron $m_e \sim 0.5$ MeV, respectively.} 
(see Ref.~\cite{Hattori:2016emy} and references therein), 
the refractive index is close to the limiting value $n_{\rm lim}=\sqrt2$, 
and thus can be larger than that of gas, and even comparable to those of liquid and solid.

\begin{figure}
     \begin{center}
              \includegraphics[width=0.6\hsize]{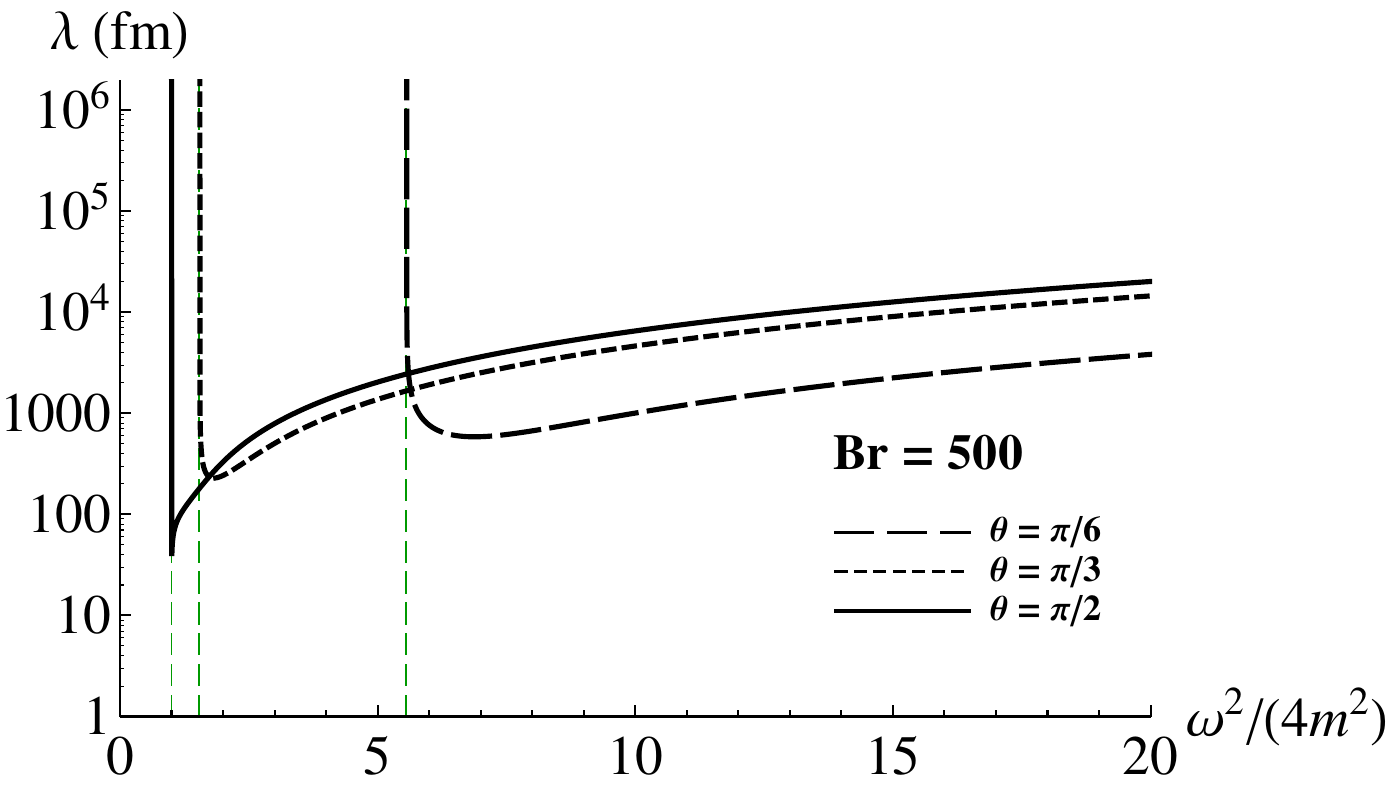}
     \end{center}
\vspace{-0.5cm}
\caption{The decay length of real photons against the photon energy, taken from Ref.~\cite{Hattori:2012ny}. 
$ \theta $ is an angle between the photon momentum and the magnetic field.}
\label{fig:optical}
\end{figure}

In Fig.~\ref{fig:optical}, we show the decay length (\ref{eq:decay_length}) 
obtained from the imaginary part of the polarization tensor in the LLL approximation. 
The magnitude of the magnetic field is taken to be supercritical $ \Br = 500 $, 
which may be relevant for physics in the magnetosphere near magnetars (see, e.g., Refs.~\cite{Harding:2006qn, Enoto:2019vcg} for reviews). 
The locations of the thresholds depend on the angle $ \theta $ 
between the photon momentum and the magnetic field. 
We find that the photons having energy larger than the threshold value decay into $ e^+ e^- $ within a microscopic scale $ \sim $ 0.1-10 pm near the thresholds. 
This implies an importance of the real-photon decay in the magnetic fields of magnetars 
which extend over the macroscopic scales. 
Corresponding to the large value of the imaginary part near the threshold (see Fig.~\ref{fig:I}), 
the decay length is small near the threshold and increases with an increasing photon energy. 
Then, approaching the threshold for the next Landau level, we will again find a minimum of the decay length. 
As mentioned before, the behaviors of the imaginary parts near the thresholds 
are essentially common to all Landau levels, 
so that the decay length scales as $ \sim 1/\omega $ in Eq.~(\ref{eq:decay_length}). 
When the photon energy is of the order of GeV, the decay length falls in 0.1-10 fm. 
Therefore, the real-photon decay could be an interesting effect 
in the relativistic heavy-ion collisions \cite{Tuchin:2010gx, Tuchin:2013ie, Hattori:2012ny, Hattori:2020htm}. On the other hand, photons below the threshold energy can survive without decaying into $e^+e^-$ pairs. Photon spectrum after traversing the magnetic regions will be modified so that higher energy modes are cut off. 
 
A couple of the next directions will be the inclusion of electric fields and/or inhomogeneities (depending on the systems). 
Recently, the refractive indices were investigated in a monochromatic plane-wave configuration of 
external electromagnetic field \cite{Yatabe:2018hhs}. 
The inhomogeneity was included on the basis of the derivative expansion.  

\subsection{Photon splitting}

\label{sec:photon-splitting}

Another interesting phenomenon induced by strong fields is the {\it photon splitting} $\gamma \to \gamma+\gamma$ for on-shell photons. 
As well-known as Furry's theorem \cite{Furry:1937zz}, 
diagrams having an odd number of external photon lines do not contribute in charge-conjugation even systems, 
because contracting the fermion fields in all possible ways 
results in the pairwise diagrams with the opposite charge flows 
that turn out to exactly cancel each other in the total amplitude.\footnote{
This can be easily verified by using the charge-conjugation property $ C \gam^\mu  C^{-1} =- ( \gam^\mu)^T$ with $C $ and $ T $ being the charge conjugation operator and the transpose of the spinor indices, respectively. 
Since the trace (for the spinor index) is invariant under the transpose, one can invert the ordering of the vertices by the use of the above property, and finds that the pairwise diagrams have the same amplitudes but with a relative sign $ (-1)^{n_v} $ for the number of vertices $  n_v$ (see also Ref.~\cite{Greiner_QED} for a pedagogical explanation). 
When the charge-conjugation symmetry is broken by, e.g., a finite chemical potential, 
the exact cancellation may not occur. 
} 
Therefore, in vacuum without external fields, it does not make any sense to discuss a diagram having three external photon legs. However, in the presence of an external field, it becomes possible to have a nonzero amplitude 
which persists with the help of an additional odd number of legs for external fields. 
This means that a real photon can split into two photons in the presence of an external field.

Let us see this in a higher order QED diagram. The lowest order diagram for the effective interactions among photons is given by the box diagram shown in Fig.~\ref{diagram} (a). 
Its cross section is proportional to $\alpha^4$, 
and experimental detection of such processes is challenging.\footnote{  
When all the photon legs are on-shell, 
this process is called the light-by-light scattering.
There are intensive experimental studies of this process 
at the Large Hadron Collider (LHC) in recent years 
\cite{ATLAS:2017fur, CMS:2018erd, ATLAS:2019azn, ATLAS:2020hii} 
(see Ref.~\cite{dEnterria:2013zqi} for 
a pre-experiment estimate of signal efficiency). 
This measurement is made possible by an enhancement of 
the cross section by a large photon flux induced 
by highly accelerated nuclei and a suppression of medium-induced 
backgrounds in the ultraperipheral collision events 
where the nuclei pass by each other 
and the medium, or the quark-gluon plasma, is not created. 
}
But if we replace one photon leg by a strong external field such as a Coulomb field of an atomic nucleus having a charge number $Z$ [Fig.~\ref{diagram} (b)], the cross section is enhanced by a factor of $Z^2$. This corresponds to the lowest order photon splitting diagram in the perturbation theory.\footnote{Similarly, if we replace two photon legs by the atomic Coulomb fields, the cross section gets a huge enhancement factor $Z^4$ for a large $Z$. This process is called the Delbr\"uck scattering, and is already observed in experiments. See Ref.~\cite{Milstein:1994zz} for the current status of experiments and theoretical description including higher order corrections.} In fact, such an enhancement and $Z^2$-scaling of the cross section were already observed long time ago in the scattering of synchrotron lights off atomic targets with various charge numbers \cite{Jarlskog:1974tx}. More recently, direct observation of the same process by back-scattered laser lights was also reported \cite{Akhmadaliev:2001ik}. When the external field is much stronger than the atomic Coulomb field and approaches the critical field, we need to sum up all the diagrams having multiple interactions with the external field, and the photon splitting shown in Fig.~\ref{diagram} (c) is expected to occur. There, the double line corresponds to a dressed electron in strong external field that was introduced in previous sections. 

\begin{figure}[t]
\begin{center}
\includegraphics[width=0.8\hsize]{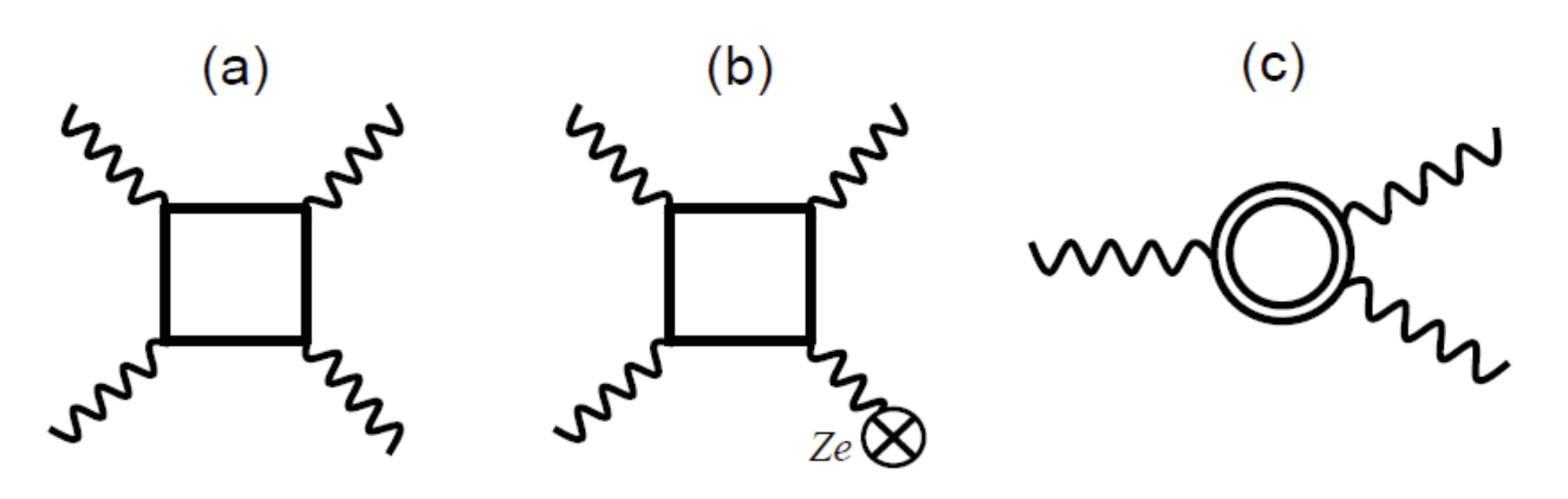}
\end{center}
\caption{(a) Photon-photon scattering at lowest order in QED, (b) photon splitting in atomic Coulomb field, (c) photon splitting in strong external fields.}
\label{diagram}
\end{figure}

Similar to the vacuum birefringence in strong fields, if photon's energy is small enough $\omega \ll m$, we can study the photon splitting by using the HE effective theory. This was done by Adler, et al.  \cite{Adler:1970gg, Adler:1971wn}, who discussed a weak-field approximation. 
They found that the lowest contribution comes from the hexagonal diagram with six photon fields attached,\footnote{
To draw this conclusion, one assumes the light-like dispersion relation of photons: 
Otherwise, the square diagrams have finite contributions. 
This implies an interplay between the vacuum birefringence and the photon splitting changes the splitting rate. 
In nonlinear optics, it is well-known 
that the ``phase matching'' is important to observe the nonlinear processes such as high-harmonic generation 
and can be achieved by use of birefringent materials. 
} 
among which three are propagating photons and the rest are external fields. 
The absorption coefficient $\kappa$ for the photon splitting in the magnetic field is computed as $\kappa\sim \int {dk_1 dk_2} |{\cal M}(\gamma(k) \to \gamma(k_1)+ \gamma(k_2))|^2 \delta^4(k-k_1-k_2)$ with ${\cal M}$ being the photon splitting amplitude, and is estimated as 
\beq
\kappa = c\frac{\alpha^3}{60\pi^2} \left(\frac{\omega}{m}\right)^5 \left(\frac{B\sin \theta}{B_c}\right)^6 m\, , 
\eeq
where $c$ is a numerical constant which depends on the polarization, $\omega$ is the energy of an incoming photon $k^\mu=(\omega,{\bm k})$,  and $\theta$ is the angle between the incoming photon momentum ${\bm k}$ and the external magnetic field. The hexagonal diagram having six vertices should amount to $(e^6)^2=e^{12}$ in total. In the above result, a half of them appears in $\alpha^3=(e^2/4\pi)^3$ and the other half in $(B/B_c)^6=(eB/m^2)^6$. 
Being suppressed by small numbers $\alpha\ll 1$, $\omega/m \ll 1$, and $B/B_c\ll 1$,  this cross section is quite small.

However, for the photon energy below the threshold of the decay into an $e^+ e^-$ pair, 
the photon splitting is the only process at work. 
Besides, when the magnetic field is strong enough, 
we can again use the analytic representation of the HE effective theory for the wrenchless fields \cite{Heyl:1996dt, Heyl:1997hr}. The result shows that the coefficient $\kappa$ increases with an increasing $B$, and saturates at a high value for extremely strong magnetic fields beyond the critical field. Thus, we may expect to have observable effects, for example, around neutron stars or magnetars: There, a lot of photons will traverse a strong magnetic fields over a large spatial scale, which will compensate the smallness of the absorption coefficient to make the effects sizable. Summary of applications to neutron stars and magnetars is available in Refs.~\cite{Baring:2008aw, Enoto:2019vcg}.

When the photon energy is close to or larger than the electron mass, one needs to compute 
the triangle diagrams with the resummed fermion propagators (see, e.g., 
Refs.~\cite{Baring:1997gs, Chistyakov:1998wf, baring1998radio, Weise:1998qu, Baier:1986cv, Baier:1996bq, Adler:1996cja, Baring:2000cr, Baring:2000fc, Chistyakov:2012ms, Hattori:2015jia}). 
In this case, one would need to examine the competition between the photon splitting and the decay into an $e^+e^-$ pair 
possibly in the magnetosphere of magnetars~\cite{Baring:1997gs, baring1998radio, Baring:2000cr}.

%



\section{Heisenberg-Euler effective action revisited: Extensions to finite temperature/density and QCD}

\label{sec:HE_QCD}

There is a long history in the investigations of low-energy effective actions in pure Yang-Mills theories and QCD. One of the motivations for studying this problem is to determine the ground state of QCD. While this may require nonperturbative analyses of strongly coupled quantum dynamics, perturbative analyses 
have also been playing some roles in understanding and modeling the QCD ground state. 
Of course, a simple fixed-order perturbative calculation may not suffice because 
a class of diagrams would be enhanced by nonperturbative physics underlying in the QCD vacuum. 
Assuming that the physical degrees of freedom relevant for the low-energy dynamics 
are the non-Abelian electromagnetic fields which we call the {\it chromo}-electromagnetic fields, 
one may perform resummed perturbative calculations of the effective actions 
in a similar way that we analysed the HE effective action in Sec.~\ref{sec:HE}. 
In this sense, the early investigations of 
the low-energy effective actions can be understood 
as the {\it non-Abelian extensions of the HE effective action}.

The first attempt in this direction was performed for the pure Yang-Mills theory with SU($N$) symmetry \cite{Duff:1975ue}, 
and the result was expressed with the proper-time integral. Explicit evaluation of the proper-time integration was later done in the pure SU(2) Yang-Mills theory \cite{Batalin:1976uv, Savvidy:1977as, Matinyan:1976mp}. 
In particular, the authors computed the effective potential in homogeneous chromo-magnetic fields 
and found that the effective potential has a local minimum at a nonzero chromo-magnetic field, 
suggesting the realization of paramagnetism in the ground state of 
the Yang-Mills theories known as the Savvidy vacuum~\cite{Batalin:1976uv, Savvidy:1977as} 
(see Ref.~\cite{Savvidy:2019grj} for a recent retrospective review).

It was important to notice that this local minimum 
is induced by the logarithmic term arising from the gluon-loop contribution to the effective action 
(see Ref.~\cite{Vanyashin:1965ple} for an early observation), 
and that the prefactor of the logarithm has the opposite sign as compared to those both in spinor and scalar QED 
(remember the logarithmic behavior mentioned below Eq.~(\ref{eq:X0-result}) for spinor QED). 
The authors explicitly showed that this logarithm leads to the renormalization-group equation 
with the negative beta function~\cite{Batalin:1976uv, Savvidy:1977as, Matinyan:1976mp}, 
which had been known to give rise to the asymptotic freedom~\cite{Gross:1973id, Politzer:1973fx}. 
Under the constraint of the Lorentz symmetry $\epsilon  \mu  = 1 $, 
the charge antiscreening effect $(\epsilon < 1  )$ and the paramagnetism $ (\mu>1) $ imply each other, 
and one may attribute the asymptotic freedom to a consequence of 
the vacuum paramagnetism~\cite{
Hughes:1980ms, Nielsen:1980sx, Hughes:1981nw, RevModPhys.77.837, Grozin:2008yd}.\footnote{
We discuss recent lattice QCD studies on magnetization 
in Sec.~\ref{sec:MC-lattice}.  
}

After those effective-potential studies, Nielsen and Olesen found the presence of a tachyonic unstable mode 
in fluctuations around the non-Abelian magnetic fields~\cite{Nielsen:1978rm} (see also Refs.~\cite{
Yildiz:1979vv, Dittrich:1983ej, Elizalde:1984zv}). 
Namely, the spin-1 Zeeman effect in a chromo-magnetic field causes 
a negative value of the energy square in the lowest Landau levels. 
We have already seen a similar behavior in Fig.~\ref{fig:Zeeman} with a (hypothetical) electrically charged spin-1 boson 
in an Abelian magnetic field (see Ref.~\cite{Tsai:1972iq} for an early observation). 
As we will explicitly see in this section, the unstable mode provides about a half of the total logarithmic contributions 
to the effective action among the infinite tower of the Landau levels, 
and also gives rise to an imaginary part of the effective action (even in the absence of an electric field).  
It is worth mentioning that similar implications were obtained in an earlier study~\cite{Vanyashin:1965ple} 
where the authors considered the vacuum fluctuation of a spin-1 vector boson in an Abelian electromagnetic field. 
The authors observed a charge antiscreening effect in their concluding remarks, 
which implies the essential role of the spin interaction, 
although the presence of such a vector boson was not 
located in the Abelian theory.

Since a nonzero chromo-magnetic field having specific color and spatial directions 
breaks the gauge and Lorentz symmetries, it cannot be regarded as the true vacuum by itself. 
However, Ambjorn and Olesen later proposed that the ground state of pure Yang-Mills theories would be realized by the so-called ``Spaghetti vacuum" which consists of nonzero non-Abelian magnetic fields having a nontrivial spatial structure (Kagome-like configuration in the transverse plane) so that the gauge and Lorentz symmetries are recovered on average at large scales \cite{Ambjorn:1979xi, Ambjorn:1980ms}. On the other hand, to determine the ground state in QCD, we need to include the contribution of quarks. Actually, as was done in Ref.~\cite{Claudson:1980yz}, this procedure is just a straightforward extension of the HE effective action. 
Calculation of the effective action of the Yang-Mills theory or QCD at finite temperature also followed in Refs.~\cite{Dittrich:1980nh, Muller:1980kf, Kapusta:1981nf, Gies:2000dw} (see also Refs.~\cite{Dittrich:1979ux, Loewe:1991mn, Elmfors:1993wj, Elmfors:1993bm, Elmfors:1994fw, Gies:1998vt} for the QED effective action at finite temperature). These studies based on the perturbative effective actions provided significant insights into the understanding of the QCD ground state, which then followed by various nonperturbative analyses~\cite{Cea:1987ku, Pagels:1978dd, Adler:1981as, Adler:1982rk, Cho:2002iv, Kondo:2004dg, Kondo:2006ih, Kondo:2013cka}.



Recently, we have seen a revival of interest in strong non-Abelian electromagnetic fields motivated by a modern picture of relativistic heavy-ion collisions. 
When the collision energy is high enough, colliding nuclei can be described as ``color glass condensates" (weakly interacting high-density gluonic states) \cite{Iancu:2003xm, Gelis:2010nm} which, after the collision, produce color electromagnetic fields whose strength is very large compared with the quark mass: $g{\cal F}\gg m_q^2$. This strong color electromagnetic field is a transitional state in between the color glass condensate and a quark-gluon plasma, and is called the ``glasma" \cite{Kovner:1995ja,Kovner:1995ts,Lappi:2006fp}.  While the glasma is generated by color sources made of valence-like partons of each colliding nucleus (partons carrying larger fractions of momentum), we are able to treat it as a source-free field in the forward light cones (i.e., in the region between two receding nuclei). Therefore, to describe the particle production from the glasma we are able to use the non-Abelian analog of the HE effective action. Moreover, it has been also recognized that the heavy-ion collisions of two electrically charged nuclei produce extremely strong electromagnetic fields \cite{Kharzeev:2007jp, Skokov:2009qp, Voronyuk:2011jd, Bzdak:2011yy, Deng:2012pc}. Thus, there comes a theoretical interest in the interplay between the electromagnetic fields and the color electromagnetic fields. For example, analytic representation of the HE effective action in the presence of both fields was recently obtained at zero temperature~\cite{Galilo:2011nh, Ozaki:2013sfa} and at finite temperature~\cite{Ozaki:2015yja}. At finite temperature, the effective action also contains the Polyakov loop as well as the color electromagnetic fields. Since the Polyakov loop is an order parameter for the deconfinement phase transition, the generalized HE effective action allows us to study the effects of electromagnetic fields on the deconfinement phase transition~\cite{Ozaki:2015yja}.

This section is devoted to the outline of such recent extensions of the HE effective action. 
After we explain the derivation of general form of the effective action at zero and finite temperatures, we discuss physical applications of the results to the Schwinger mechanism and the analyses of the QCD effective potential for color electromagnetic fields at zero temperature and the Polyakov-loop effective action at finite temperature. 

\subsection{Extension to finite temperature/density}

\label{sec:finiteT}

We extend the HE effective action discussed in Sec.~\ref{sec:resummed-action} to finite temperature/density 
by the use of the imaginary-time formalism. 
This extension was carried out for the effective action in a magnetic field in an early work~\cite{Dittrich:1979ux} 
and that for electric and magnetic fields in Ref.~\cite{Elmfors:1994fw}. 
It was further elaborated by Gies in Ref.~\cite{Gies:1998vt}, 
and the technical details were reviewed in Ref.~\cite{Dittrich:2000zu}. 


First of all, the Lorentz symmetry is broken in the presence of medium, 
and our reference frame may be specified by the medium flow vector $ u^\mu $. 
This means that configurations of constant electromagnetic fields are not completely classified 
by the Lorentz invariants $a$ and $b$ like in vacuum. 
Therefore, one cannot simply deduce the general form of the effective action by starting out from the parallel/antiparallel configuration, 
which is now not uniquely connected to the class of field configurations specified by $ a, \, b $ by a Lorentz transformation. 
We will need to introduce the covariant form of the electric field $E^\mu \equiv F^{\mu \nu} u_{\nu}   $ 
to express the effective action in a general Lorentz frame. 
Moreover, since the temporal component of the gauge field is distinguished from the spatial components, 
the extended duality (\ref{eq:duality_q}) 
under the interchange between the electric and magnetic components, $ i a$ and $b $ 
does not hold in the presence of medium.

Next, one should be careful of the gauge transformation  
in a compact spacetime. 
In the imaginary-time formalism for thermal field theory,
the temporal coordinate is compactified with the periodic and antiperiodic boundary conditions $ \psi(\tau=0,\bx) = \pm \psi(\tau=\beta,\bx) $ for bosons in scalar QED 
and fermions in spinor QED; 
We introduce the imaginary time $t \to \tau = - i t $, 
and the period is set by an inverse temperature, $ \beta = 1/T $. 
Those boundary conditions are, however, not invariant 
under general gauge transformations. 
Thus, one should keep track of the pair of 
the covariant derivative and the boundary conditions 
when a gauge transformation is performed.

We recapitulate a general discussion with 
a covariant derivative under an external field $ A_\ext^\mu$: 
\begin{eqnarray}
i D_\tau =  \pd_\tau - q_f A^0_\ext \  \ {\rm with} \ \
\psi(\tau=0,\bx) = \pm \psi(\tau=\beta,\bx)
\, .
\end{eqnarray}
Here, we focus on the temporal component 
since the spatial directions are the same as in vacuum. 
We examine whether $A^0_\ext $ can be gauged away as 
\begin{eqnarray}
\label{eq:gauge-transf-T}
\psi (\tau,\bx) \to \psi' (\tau,\bx) 
= e^{ q_f \int_0^\tau d\tau ' A^0_\ext(\tau')   } \psi(\tau, \bx)
\, .
\end{eqnarray}
As we should expect, we cannot simply remove 
$A^0_\ext $ from the theory 
and find a modification of the boundary condition as 
\begin{eqnarray}
\label{eq:gauged-away}
i D_\tau \to  \pd_\tau   \  \ {\rm with} \ \ 
\psi(\tau =0 , \bx)= \pm e^{ q_f \int_0^\beta d\tau ' 
 A^0_\ext(\tau')  } \psi(\beta, \bx)
\, .
\end{eqnarray}
The modification of the boundary condition gives rise to 
a shift of the Matsubara frequency 
\begin{eqnarray}
\label{eq:Matsubara-replacement-0}
p^0  = i \omega_k^{B/F} +  q_f T 
\int_0^\beta d\tau '  A^0_\ext(\tau')  
\, ,
\end{eqnarray}
where $\omega_k^{B/F} = 2k \pi T, \ (2k+1)\pi T $ 
are the Matsubara frequency 
for a boson and fermion field, respectively. 
This is just similar to the shift by 
a chemical potential: $ p^0 \to i \omega_k + \mu $ 
(see, e.g., Ref.~\cite{Kapusta:2006pm, Bellac:2011kqa}).\footnote{
The shift in Eq.~(\ref{eq:Matsubara-replacement-0}) 
is also analogous to the periodic energy shift 
due to the Aharonov-Bohm phase on a circle; 
The energy degeneracy between the clockwise and counterclockwise 
propagation is lifted due to the AB phase (see, e.g., Ref.~\cite{griffiths2003consistent}). 
One could interpret the above gauge-invariant integral 
(or a chemical potential) in a similar way 
(though the Matsubara frequency is 
not an energy spectrum in the imaginary world). 
The shift in Eq.~(\ref{eq:Matsubara-replacement-0}) is 
periodic for an Euclidean gauge field 
when the integral value coincides with 
an integer multiple of $2\pi/q_f $, i.e, 
$\int_0^\beta d\tau ' A^0_\ext(\tau',\bx)  
= i (2 \pi /q_f) \lambda $ with an integer $\lambda $; 
Such a shift can be absorbed into the Matsubara frequency. 
}
The integral over the temporal period is invariant 
under a periodic gauge transformation where 
the gauge parameter is taken to be the same at 
$\tau =0 $ and $ \tau=\beta$. 
Namely, if we perform another gauge transformation, 
the integral in Eq.~(\ref{eq:Matsubara-replacement-0}) 
gives a vanishing surface term that is a vanishing difference between the gauge parameter at $\tau = 0$ and $ \beta $.

Below, we discuss the case where 
$A_\ext^0 $ can contain not only a constant external field 
but also other external fields. 
To include the constant fields with the FS gauge as in vacuum, 
we perform a gauge transformation 
\begin{eqnarray}
\label{eq:gauge-transf-T}
\psi (\tau,\bx) \to \psi' (\tau,\bx) 
= e^{ q_f \int_0^\tau d\tau ' 
\{ A^0_\ext(\tau') - A^0_\FS(\tau')\} } \psi(\tau, \bx)
\, .
\end{eqnarray}
Accordingly, the pair of the covariant derivative and 
the boundary condition is transformed to 
\begin{eqnarray}
\label{eq:BC-modified}
i D_\tau =  \pd_\tau - q_f A^0_{FS} \  \ {\rm with} \ \ 
\psi(\tau =0 , \bx)= \pm e^{ q_f \int_0^\beta d\tau ' 
\{ A^0_\ext(\tau') - A^0_\FS(\tau')\} } \psi(\beta, \bx)
\, .
\end{eqnarray}
The modification of the boundary condition (\ref{eq:BC-modified}) 
gives rise to the shift of the Matsubara frequency 
\begin{eqnarray}
\label{eq:Matsubara-replacement}
p^0  = i \omega_k^{B/F} +  q_f T \int_0^\beta d\tau ' 
\{ A^0_\ext(\tau') - A^0_\FS(\tau')\}
\, .
\end{eqnarray}
Since only the FS gauge component is left in the covariant derivative (\ref{eq:BC-modified}) 
after the gauge transformation, 
the temporal component does not contribute to 
the Schwinger phase (\ref{eq:Phi_S}). 
Instead, one should include the shift 
in Eq.~(\ref{eq:Matsubara-replacement}). 
The constant-field component in $A^0_\ext$ 
is subtracted by $A^0_\FS $; Namely, if $A^0_\ext$ 
is just gauge-equivalent to $A^0_\FS $, 
the integral gives a vanishing surface term 
under a periodic gauge transformation. 
For notational simplicity, we denote the shift as 
$ p^0  = i \omega_k^{B/F} + q_f \varphi$ with 
\begin{eqnarray}
\label{eq:vphi}
 \varphi := T \int_0^\beta d\tau ' 
\{ A^0_\ext(\tau') - A^0_\FS(\tau')\}
\, .
 \end{eqnarray} 
A non-Abelian analog of the shift is 
the Polyakov loop discussed in Sec.~\ref{sec:Polyakov}.

\cout{
Next, one should be careful of the gauge transformation  
in a compact spacetime. 
In the imaginary-time formalism for thermal field theory,
the temporal coordinate is compactified with the periodic and antiperiodic boundary conditions $ \psi(\tau=0,\bx) = \pm \psi(\tau=\beta,\bx) $ for bosons and fermions, respectively; 
We introduce the imaginary time $t \to \tau = - i t $, 
and the period is set by an inverse temperature, $ \beta = 1/T $. 
Those boundary conditions are, however, not invariant 
under general gauge transformations. 
Thus, one should keep track of the pair of 
the covariant derivative and the boundary conditions 
when a gauge transformation is performed. 
We start with a covariant derivative with 
an external field $ A_\ext^\mu$ in an arbitrary gauge: 
\begin{eqnarray}
i D_\tau =  \pd_\tau - q_f A^0_\ext \  \ {\rm with} \ \
\psi(\tau=0,\bx) = \pm \psi(\tau=\beta,\bx)
\, .
\end{eqnarray}
Here, we focus on the temporal component 
since the spatial directions is the same as in vacuum. 
To use the FS gauge as in vacuum, 
we perform a gauge transformation 
\begin{eqnarray}
\label{eq:gauge-transf-T}
\psi (\tau,\bx) \to \psi' (\tau,\bx) 
= e^{ q_f \int_0^\tau d\tau ' 
\{ A^0_\ext(\tau') - A^0_\FS(\tau')\} } \psi(\tau, \bx)
\, ,
\end{eqnarray}
where $A^0_\ext(\tau') $ and $A^0_\FS(\tau') $ give 
the same external field. 
Accordingly, the pair of the covariant derivative and 
the boundary condition is transformed to 
\begin{eqnarray}
\label{eq:BC-modified}
i D_\tau =  \pd_\tau - q_f A^0_{FS} \  \ {\rm with} \ \ 
\psi(\tau =0 , \bx)= - e^{ q_f \int_0^\beta d\tau ' 
\{ A^0_\ext(\tau') - A^0_\FS(\tau')\} } \psi(\beta, \bx)
\, .
\end{eqnarray}
The modification of the boundary condition (\ref{eq:BC-modified}) 
manifests itself in the shift of the Matsubara frequency 
\begin{eqnarray}
\label{eq:Matsubara-replacement}
p^0  = i \omega_k^{B/F} +  q_f T \int_0^\beta d\tau ' 
\{ A^0_\ext(\tau') - A^0_\FS(\tau')\}
\, ,
\end{eqnarray}
where $\omega_k^{B/F} = 2k \pi T, \ (2k+1)\pi T $ 
are the Matsubara frequency 
for a boson and fermion field, respectively. 
With the gauge parameter $ A^0_\ext(\tau') - A^0_\FS(\tau')
= i q_f^{-1} \pd_\tau \a$, 
the integral over the temporal period is invariant 
under the periodic gauge transformation 
$\a(\tau=0) = \a(\tau=\beta) $. 
A non-Abelian analog of this quantity is the Polyakov loop discussed in Sec.~\ref{sec:Polyakov}.  
Since only the FS gauge component is left in the covariant derivative (\ref{eq:BC-modified}) 
after the gauge transformation, 
the temporal component does not contribute to 
the Schwinger phase (\ref{eq:Phi_S}). 
Instead, one should include the shift 
in Eq.~(\ref{eq:Matsubara-replacement}). 
This is just similar to the shift by 
a chemical potential: $ p^0 \to i \omega_k + \mu $ 
(see, e.g., Ref.~\cite{Kapusta:2006pm, Bellac:2011kqa}). 
This is also analogous to the energy shift 
due to the Aharonov-Bohm phase on a circle; 
The energy degeneracy between the clockwise and counterclockwise 
propagation is lifted due to the AB phase (see, e.g., Ref.~\cite{griffiths2003consistent}). 
One could interpret the above gauge-invariant integral 
(or a chemical potential) in a similar way 
(though the Matsubara frequency is 
not an energy spectrum in the imaginary world). 
Different gauge choices are equivalent with each other 
when the boundary condition (\ref{eq:BC-modified}) 
is invariant, i.e., when the following condition is satisfied 
$q_f \int_0^\beta d\tau '
\{ \, A^0_\ext(\tau',\bx) - A^0_\FS(\tau',\bx) \, \}
= i [\a(\beta) - \a(0)]
= 2  \pi i \lambda $ with an integer $\lambda $, 
which can be absorbed into the Matsubara frequency 
in Eq.~(\ref{eq:Matsubara-replacement}).

}

Bearing these points in mind, we compute the thermal contribution to the effective action 
in Eqs.~(\ref{HE_parallel}) and (\ref{HE_general1}) by adopting the imaginary-time formalism 
with the replacement rules $\int dp^0 \to iT\sum_{k=-\infty}^\infty$ and of Eq.~(\ref{eq:Matsubara-replacement}). 
First, we focus on the effective action in the presence of 
the parallel electric and magnetic fields in the medium rest frame. 
The matrix element (\ref{eq:K-fermion}) in the imaginary-time formalism is now given as 
\beq
\!\! {\rm tr}\, \langle x | {\rm e}^{-i\hat H s}|x\rangle 
=4 
i T\!\! \sum_{k=-\infty}^{\infty} \int\! \frac{ d^{3}p }{ (2\pi)^{3} } 
\left. \exp\left(
i\frac{p_\parallel^2}{q_f E}\tanh(q_f E s) + i\frac{ p_\perp^2}{q_f B }\tan(q_f B s) 
\right)\right|_{p^{0} = i (2k+1)\pi T +  q_f \vphi }\!
\label{eq:amplitude-T}
,
\eeq
where $\hat H= D^2 + \frac{q_f}{2}F^{\mu\nu}\sigma_{\mu \nu}$ as defined in Eq.~(\ref{Ham_fermion}). 
The Schwinger phase $ \Phi_A (x,x^\prime) $ 
after the gauge transformation (\ref{eq:gauge-transf-T}) 
vanishes in the coincidence limit, $ x \to x^\prime $.  
Performing the three-dimensional momentum integral and reorganizing the 
Matsubara sum by using the Poisson summation formula,\footnote{The Poisson summation formula for a generic function reads 
$$
\sum_{k=-\infty}^\infty f(x+k)=\sum_{\bar k=-\infty}^\infty {\rm e}^{2\pi i \bar kx}\int_{-\infty}^\infty dx' f(x'){\rm e}^{-2\pi i \bar k x'}.
$$
If one takes a Gaussian $f(x)={\rm e}^{-\sigma x^2}$, one finds a formula useful in the present calculation \cite{Dittrich:1979ux, Cangemi:1996tp,Gies:1998vt, Dittrich:2000zu}:
$$
\sum_{k=-\infty}^\infty {\rm e}^{-\sigma (k-z)^2}
=\sum_{\bar{k}=-\infty}^\infty \sqrt{\frac{\pi}{\sigma}}\, {\rm e}^{-\frac{\pi^2}{\sigma}\bar{k}^2 -2\pi i z\bar{k}}\, .
$$
}
we obtain the effective Lagrangian 
\beq 
{\cal L}^{(1)} 
&=& \sum_{f=1}^{N_{f}}
 \frac{ 1 }{8\pi^{2}} \int_{0}^{\infty} \frac{ds}{s} \, 
\e^{-i(m_{f}^{2}-i\epsilon)s} \frac{ (q_f  E) (q_f B) } { \tanh ( q_f E s)  \tan (q_f B s) }
\nonumber 
\\
&&\hspace{2cm} \times 
\left[ 1+ 2 \sum_{\bar k=1}^{\infty} (-1)^{\bar k} 
\e^{i q_f E \coth( q_f E s) \frac{ \bar k^2}{4T^{2}}  }  
\cosh(   q_f \vphi \beta ) \right]
\label{action_fermion-parallel}
\, .
\eeq
The first term in the square brackets, which is the term with $ \bar k = 0 $ in the Poisson summation, 
agrees with the zero-temperature contribution in Eq.~(\ref{HE_parallel}). 
Therefore, we find that the first and second terms in the square brackets account for 
the zero- and finite-temperature contributions, respectively.\footnote{The index $\bar{k}$ should not be confused with the one in the Matsubara summation. The Poisson summation formula is applied in order to separate the vacuum part from the temperature-dependent part.} 
The alternating sign, $ (-1)^{\bar k} $, originates from the antiperiodic boundary condition 
[cf. Eq.~(\ref{eq:gg-thermal}) below for the periodic boundary condition, where the alternating sign is absent]. 
The HE effective Lagrangian (\ref{action_fermion-parallel}) at finite temperature 
reproduces that in a magnetic field~\cite{Dittrich:1979ux} and in electric and magnetic fields~\cite{Elmfors:1994fw}.

At this stage, we should note that the system is assumed to be in a thermal equilibrium by construction of the imaginary-time formalism. 
For a system to reach such an equilibrium state, 
an external electric field, which is pumping an energy to the system, 
should be screened by the thermal particles and damped out 
\cite{Grozdanov:2016tdf, Hattori:2017usa, Hattori:2022hyo}. 
Therefore, it is not legitimate to address effects of an external electric field within the imaginary-time formalism in the rigorous sense, 
so that one should carefully check the validity of the framework for a phenomenological application to each physical situation. 
For example, there is a controversy on the medium modification of the Schwinger mechanism \cite{Cox:1984vf, Loewe:1991mn,Elmfors:1994fw, Gies:1998vt, Gies:1999vb, Gavrilov:2007hq, Kim:2007ra, Kim:2008em, Medina:2015qzc, Gould:2017fve}.\footnote{
Seemingly, there is no medium modification from the one-loop effective action~\cite{Cox:1984vf, Elmfors:1994fw, Gies:1998vt}, 
while medium modifications were put forwarded in some literature (see discussions and references in another review article~\cite{Gies:2000dw}). 
On the other hand, the interaction between a thermal photon and a fermion pair appears in the two-loop level, 
and absorption of the thermal photon by the fermion pair may ``assist'' the virtual fermion-antifermion pair to penetrate the potential barrier for the pair creation. 
This point was discussed with an explicit calculation in the low-temperature limit, $ T \ll m $~\cite{Gies:1999vb, Dittrich:2000zu}, where the population of thermal photons is sizeable, 
but that of the thermal fermions is suppressed by the Boltzmann factor. 
(Thus, possible effects of the thermal fermions, e.g., Pauli blocking, were not included.) 
Further investigations may be needed not only for technical details 
but also for finding a reasonable setup of the problem depending on phenomenological situations. 
}
Nevertheless, a finite-volume system in an electric 
field may reach a steady state, and one could measure 
relevant physical quantities (see Ref.~\cite{Endrodi:2022wym} 
for a recent study on an electric susceptibility).

The generalization to the expression in an arbitrary Lorentz frame was addressed in Ref.~\cite{Gies:1998vt} 
with a detailed account (see also Ref.~\cite{Dittrich:2000zu}). 
Therefore, we only quote the result: 
\beq 
{\cal L} ^{(1)}
&=& \sum_{f=1}^{N_{f}} \frac{ 1}{8\pi^{2}} \int_{0}^{\infty} \frac{ds}{s} \, 
\e^{-i(m_{f}^{2}-i\epsilon)s} \frac{ (q_f a) (q_f b )}{ \tanh ( q_f a s) \tan ( q_f b s) }
\nonumber \\
&&\hspace{2cm} \times 
\left[ 1+ 2 \sum_{\bar k=1}^{\infty} 
(-1)^{\bar k} \e^{ i  \frac{ q_f h (s) }{4T^{2}}  \bar k^{2} } \cosh(  q_f  \vphi \beta  \bar k)  \right]
\label{action_fermion}
\, ,
\eeq 
where the explicit temporal component $ A^0 $ in $\vphi $ 
should be also replaced by the projection $ A_u \equiv  A^\mu_\ext u_\mu$ along  the medium flow vector $u^\mu$. 
As in Eq.~(\ref{action_fermion-parallel}), the first term in the square brackets 
corresponds to the vacuum contribution, so that the parallel electromagnetic fields, $E , \, B  $, 
are simply replaced by the invariants, $ a, \, b$, outside the brackets. 
However, as mentioned above, this simple Lorentz transform does not 
work for the thermal contribution (the second term in the brackets), 
and the explicit dependence on the flow vector $u^\mu$ appears. 
In addition to $ A_u $ just mentioned above, $h(s)$ in the exponent also depends on $u^\mu$: 
\beq
 h(s) \equiv   a \frac{ b^2 + e_u^2  }{ a^2 + b^2 } \,  {\rm{coth}}(q_f a s) 
+ b  \frac{ a^2 - e_u^2  }{ a^2 + b^2 } \, {\rm{cot}}( q_f b s)
\label{eq:h-QED}
\, ,
\eeq
where the power of the electric field $ e_u^2 = - E^\mu E_\mu$ is defined 
with the covariant form of the electric field, $ E^\mu \equiv F^{\mu \nu} u_{\nu} $. 
In the medium rest frame, we have $u^{\mu} = (1,0,0,0)$ and $ E^{\mu} = (0 , \bE)  $, 
so that $e_u^2   =|\bE|^2 $, which reproduces 
Eq.~(\ref{action_fermion-parallel}) obtained 
in an earlier work~\cite{Elmfors:1994fw}.

It would be also instructive to reproduce the familiar result in the absence of the external fields. 
We focus on the thermal part. 
Taking the vanishing field-strength limit 
and maintaining only $\vphi$, we have 
\beq 
{\cal L}^\one_T
= \sum_{f=1}^{N_{f}} \frac{1}{4\pi^{2}} \sum_{\bar k=1}^{\infty} (-1)^{\bar k}   \cosh( q_f \vphi \beta  \bar k) 
\int_{0}^{\infty}  \frac{ds}{s^3} \,  \e^{-i(m_{f}^{2}-i\epsilon)s}  \e^{\frac{i}{4T^{2} s} \bar k^{2} }  
\label{eq:fermion-T-0}
\, .
\eeq 
After the rotation of the integral contour to the negative imaginary axis, 
the proper-time integral can be identified with the modified Bessel function of the second kind $ K_n(z) $ 
with the help of the integral representations~\cite{ARFKEN2013643, ModifiedBessel2nd}\footnote{
The integral in Eq.~(\ref{eq:fermion-T-0}) can be identified with the first representation (\ref{eq:Bessel1}) 
according to a property $ K_n(z) = K_{-n}(z) $ which follows from the change of the integral variable $ t \to 1/t $. 
The second representation is obtained by integrating Eq.~(7) in Ref.~\cite{ModifiedBessel2nd} by parts.
} 
\begin{subequations}
\begin{eqnarray}
K_n(z) &=& \frac12 \int_0^\infty \frac{dt}{t^{1 - n}} \, e^{- \frac{z}{2} ( t + \frac{1}{t} ) }
\label{eq:Bessel1}
\\
&=& \frac{ \sqrt{\pi}} { \Gamma(n - \frac{1}{2}) } \left( \frac{z}{2} \right)^{n-1} 
\int_{1}^\infty dy \, y (y^2-1)^{n-\frac{3}{2}} \e^{-yz}
\label{eq:Bessel}
\, ,
\end{eqnarray}
\end{subequations}
for $ {\rm Re}[z]>0 $ and $ n > 1/2  $. 
It is found that 
\begin{eqnarray}
{\cal L}^\one_T
&=& - 2 \frac{(m_f  T)^2}{\pi^2} \sum_{f=1}^{N_{f}}  \sum_{\bar k=1}^{\infty} \frac{ (-1)^{\bar k} }{ \bar k^2}  
\cosh( q_f \vphi \beta  \bar k)    K_2 \left( \frac{m_f}{T} \bar k \right)
\, .
\end{eqnarray}
By changing the integral variable as $ p = m_f (y^2-1)^{1/2} $ 
in the second integral representation (\ref{eq:Bessel}), 
one can retrieve the familiar momentum-integral form as 
\begin{eqnarray}
K_2 \big( \frac{m_f}{T} \bar k\big) 
&=& 2 \pi^2 \Big( \frac{ \bar k  }{m_f^2 T} \Big) 
\int_0^\infty \frac{d^3p}{(2\pi)^3} \e^{- \beta \sqrt{ p^2 + m_f^2} \, \bar k }
\, ,
\end{eqnarray}
where $ \beta = 1/T $ and $  \Gamma( \frac{3}{2})  = \sqrt{\pi}/2 $. 
Plugging this expression back to the effective Lagrangian and identifying the Poisson summation 
with the Taylor series $  \ln (1+x) = - \sum_{\bar k =1}^\infty \frac{ (-x)^{\bar k}}{\bar k} $, 
we find the familiar form (see, e.g., a standard textbook~\cite{Kapusta:2006pm}) 
\begin{eqnarray}
{\cal L}^\one_T=  2T   \sum_{f=1}^{N_{f}} \int \frac{d^3p}{ (2\pi)^3} \left[ \, 
\ln \Big( 1 + \e^{-  \beta( \epsilon_\bp -  q_f \vphi) } \Big) + \ln \Big( 1 + \e^{- \beta(\epsilon_\bp +  q_f  \vphi)} \Big) \, \right]
\label{eq:Therm-pot_fermion}
\, ,
\end{eqnarray}
where the one-particle energy is given by $  \epsilon_\bp = \sqrt{ \bp^2 + m_f^2}$. 
Notice that $\vphi $ appears in the positions of the chemical potential as anticipated. 
In the vanishing $\vphi$ and massless limits, 
the Stefan-Boltzmann law is reproduced 
\begin{eqnarray}
\label{eq:SB-limit}
\lim_{q_f \vphi,m_f\to 0}{\cal L}^\one_T= 
  (2 \cdot 2\cdot N_f ) \times \frac{7}{8} \cdot \frac{\pi^2 }{90} T^4
\, ,
\end{eqnarray}
where the degeneracy factor comes from two spin states, and particle and antiparticle contributions.

\subsection{Spontaneous chiral symmetry breaking: A preview of the magnetic catalysis}

\label{sec:potential-MC}

In this section, we examine the effective potential $[ V^\one \equiv  -  S_{\rm eff}^{(1)} / \int \!\ d^{4}x ] $ 
in strong QED magnetic fields at zero and finite temperature. 
Focusing on a single-flavor case, we drop the flavor sum for simplicity. 
Then, taking the vanishing $ E $ limit in Eq.~(\ref{action_fermion}), 
we have the effective potential in magnetic fields 
\begin{eqnarray}
\label{eq:V_eff-B}
V^\one = \frac{ |q_f B| }{8\pi^2}  \int_{1/\Lambda^2}^\infty \frac{ds}{s^2} \e^{- m^2 s }  \coth( q_f B s) 
\left[ \,  1 + 2 \sum_{\bar k =1}^\infty (-1)^{\bar k} \e^{- \frac{1}{4T^2 s} \bar k^2 }
\,\right]
\, ,
\end{eqnarray}
where we work in the fluid rest frame. 
While we take $\vphi=0$ and do not introduce a chemical potential, 
density effects can be studied with the same framework as an extension. 
Note also that we rotated the integral contour to the negative imaginary axis 
and inserted a UV cutoff to regularize the divergence in the vacuum part.


\subsubsection{Gap equation at zero temperature}

Specifically, we discuss the strong-field limit such that $ |eB| \gg m^2 $, 
where the LLL contribution plays the dominant role. 
Remembering the Landau-level decomposition of the HE effective action discussed in Sec.~\ref{sec:Sch-LL}, 
we find the vacuum part of the effective potential: 
\begin{eqnarray}
V^\one_\vac \sim
\rho_B \left[ \,  \frac{1}{4\pi}  \int_{1/\Lambda^2}^\infty \frac{ds}{s^2} \e^{-m^2s } \, \right]
=
\rho_B \frac{m^2}{4\pi} \left[ \, 
\frac{\Lambda^2}{m^2}  -  \ln \frac{\Lambda^2}{m^2} + \gam_{\rm E} -1 \, \right]
\label{eq:pot-B-strong}
\, ,
\end{eqnarray}
where $ \rho_B = |q_f B|/(2\pi) $ is the density of states introduced in Sec~\ref{sec:Q} 
and $\gamma_{\rm E}$ is the Euler-Mascheroni constant $\gamma_{\rm E}=0.577\cdots$. 
The proper-time integral has been reduced to the (1+1)-dimensional form 
when the cutoff is taken to be smaller than the field strength, $ \Lambda^2 \lesssim |e B| $, 
meaning that our cutoff is less than the energy gap of the higher LL.

We demonstrate an interesting property of the fermion effective potential (\ref{eq:pot-B-strong}) 
by the use of the Nambu--Jona-Lasinio model (see, e.g., Ref.~\cite{Hatsuda:1994pi}). 
Since details of the model are not important for the present discussion,\footnote{
We will discuss the roles of the gauge bosons in the next section.} 
let us assume the simplest Lagrangian: 
\begin{eqnarray}
\Lag_{\rm NJL} 
= i \bar \psi \sla \partial \psi + \frac{\lambda_{\rm NJL}}{2} (\bar \psi \psi) ( \bar \psi \psi)
\sim \bar \psi ( \sla \partial - m_\dyn) \psi - \frac{m_\dyn^2}{2 \lambda_{\rm NJL}} 
\, .
\end{eqnarray}
In the rightmost expression, we applied the mean-field approximation for 
the fermion bilinear field $ \bar \psi \psi $ 
which may fluctuate around the expectation value $ \langle \bar \psi \psi \rangle $. 
The associated dynamical mass is defined as 
$ m_\dyn = -\lambda_{\rm NJL} \langle \bar \psi \psi \rangle $ 
and the higher order fluctuations are dropped. 
Hereafter, the mass parameter contained in the potential (\ref{eq:pot-B-strong}) is 
also regarded as the dynamical mass $  m_\dyn$. 
Combining the mean-field term and the one-loop contribution, 
we get the effective potential 
\begin{eqnarray}
V^\one_\vac   =
\frac{m_\dyn^2}{2 \lambda_{\rm NJL}}
+ 
\rho_B  \frac{m_\dyn^2}{4\pi} \left[ \, 
\frac{\Lambda^2}{m_\dyn^2}  -  \ln \frac{\Lambda^2}{m_\dyn^2} + \gam_{\rm E} -1 \, \right]
\label{eq:pot-B-NJL}
\, .
\end{eqnarray}
Equivalent results were shown in Refs.~\cite{Gusynin:1994xp, Gusynin:1995nb} 
and Ref.~\cite{Fukushima:2012xw}, which however have slightly different forms from the above result. 
The differences may be attributed to the contributions of the higher Landau levels~\cite{Gusynin:1994xp, Gusynin:1995nb} and slightly different cutoff schemes introduced in 
the momentum integral \cite{Fukushima:2012xw} 
and in the proper-time integral in Eq.~(\ref{eq:pot-B-strong}), respectively.

We now find an interesting consequence of the dimensional reduction 
in the effective potential (\ref{eq:pot-B-NJL}). 
The dynamical mass is determined by the gap equation 
from the stationary condition $\partial V_\q^\vac /\partial m_\dyn =0 $, i.e., 
\begin{eqnarray}
\frac{\pi}{ \rho_B \lambda_{\rm NJL} }  
+ \frac{\gam_{\rm E}}{2} -     \ln \frac{\Lambda}{m_\dyn}  
= 0
\label{eq:pot-gap}
\, .
\end{eqnarray}
This gap equation has a nontrivial solution ($ m_\dyn \not =0 $) regardless of  the coupling strength, 
which we can immediately obtain as 
\begin{eqnarray}
m_\dyn (T=0) = \Lambda^\prime 
\exp \left( 
- \frac{\pi}{ \rho_B \, \lambda_{ \rm NJL} } 
\right)
\label{m_dyn-potential}
\, .
\end{eqnarray}
We absorbed a purely numerical order-one factor 
into the definition of the cutoff scale, $ \Lambda^\prime = \Lambda \e^{ - \gam_{\rm E}/2 } $, 
which does not have any dependence on the parameters in the model. 
Notice that the above solution has non-analytic dependences 
on the coupling constants $ \lambda_{\rm NJL} $ and $ q_f $, indicating that the dynamical mass gap 
emerges as a result of the nonperturbative effect which is captured 
by the resummation of the infinite number of diagrams. 
The overall coefficient is determined by the cutoff scale. 
This scale should be small enough $ \Lambda^2 \lesssim |q_fB| $, as noted above. 
We should also ensure that we work within the cutoff scale of the NJL model, $ \Lambda^2 \lesssim \Lambda_{\rm NJL}^2 $. 
Therefore, the smaller value, $ \Lambda^\prime \sim {\rm min}[\sqrt{q_fB}, \Lambda_{\rm NJL}] $, 
determines the size of dynamically generated mass gap (\ref{m_dyn-potential}). 

Since the gap equation (\ref{eq:pot-gap}) always has the nontrivial solution, 
the chiral symmetry breaking takes place in strong magnetic fields 
without resorting to a large value of the coupling strength. 
Remember that the NJL model in the absence of magnetic fields 
exhibits the chiral symmetry breaking only when the coupling strength is 
larger than a critical value (see, e.g., Ref.~\cite{Hatsuda:1994pi}). 
These observations suggest that the mechanism of the chiral symmetry breaking in the strong magnetic field 
should be qualitatively different from that in the absence of the magnetic field.

It is the growth of the logarithm, $ \ln \, m_\dyn$, that forces 
the effective potential to be convex upward at 
the origin and the trivial solution ($ m_\dyn = 0 $) 
to be unstable. 
The emergence of the logarithm is tracked back to the dimensionality of the integral in Eq.~(\ref{eq:pot-B-strong}), 
of which the log term goes like $ \int (ds/s^2) \cdot (-m^2_\dyn s) \sim m^2_\dyn \ln m_\dyn$ in the dimensionally reduced case, 
while $ \int (ds/s^2) \cdot (-m_\dyn^2 s)^2/s \sim m_\dyn^4 \ln m_\dyn$ 
in the usual four dimensional case at $ q_f B=0 $. 
In the latter case with the quartic prefactor, this term could be subdominant 
against the quadratic term $ m^2/\lambda_{\rm NJL} $ near the origin $ m_\dyn = 0 $, 
so that the fate of the vacuum depends on the competition between those terms, 
yielding the critical coupling strength. 
In contrast, the effective dimensional reduction gives rise to the logarithm appearing 
with the quadratic prefactor, which overwhelms the other terms 
for any value of $ \lambda_{\rm NJL} $ \cite{Gusynin:1994xp,Fukushima:2012xw}.

Therefore, the essential mechanism of the chiral symmetry breaking 
is identified with the dimensional reduction to the (1+1)-dimensions in the strong magnetic field. 
Shown above is a quick view of the so-called ``magnetic catalysis'' of the chiral symmetry breaking \cite{Gusynin:1994xp, Gusynin:1995gt, Gusynin:1995nb}, 
of which the various aspects have been investigated in the last two decades. 
We will discuss this phenomenon more intensively in Sec.~\ref{sec:dmr}.

\subsubsection{Chiral phase transition temperature}

\label{sec:chiral-T}

Let us include the finite-temperature contribution to determine the chiral phase transition temperature. 
As in the vacuum part, we take the strong-field limit 
\begin{eqnarray}
V_\q^T \sim \rho_B 
\frac{ 1 }{4\pi}  \int_{0}^\infty \frac{ds}{s^2} e^{-m_\dyn^2s }   \sum_{\bar k =1}^\infty 2 (-1)^{\bar k} e^{- \frac{1}{4T^2 s} \bar k^2 }
=
\rho_B \frac{ m_\dyn^2 }{4\pi}  \sum_{\bar k =1}^\infty  (-1)^{\bar k} \frac{8T}{m_\dyn \bar k}
 K_1 (\frac{m_\dyn}{T} \bar k )
\label{eq:pote-finiteT}
\, .
\end{eqnarray}
It is easy to check again the equivalence between the representations 
in terms of the proper-time integral and of the momentum integral. 
Applying the formula (\ref{eq:Bessel}) to Eq.~(\ref{eq:pote-finiteT}), 
we find the (1+1)-dimensional form of the potential as 
\begin{eqnarray}
V^\one_T= - \rho_B \times 2T \int_{-\infty}^\infty \frac{dp_z}{2\pi} \ln (\, 1+\e^{-\epsilon_{p_z}/T} \,)
\, ,
\end{eqnarray}
with $ \epsilon_{p_z} = \sqrt{p_z^2 + m_\dyn^2} $. 
The UV contribution is suppressed by the exponential factor 
and is cutoff by the temperature scale as expected.

Since we are interested in the behavior of the effective potential 
near the phase transition temperature, we may take the small mass limit, $ m_\dyn(T) \ll T $ assuming the second-order phase transition. 
Expanding the Bessel function in Eq.~(\ref{eq:pote-finiteT}) 
and performing the summation with respect to $ \bar k $,\footnote{
The summation can be performed with the help of the Riemann Zeta function as \cite{RiemannZeta}
$$
\sum_{\bar k =1}^\infty \frac{(-1)^{\bar k} }{ \bar k^s } = (2^{1-s}-1 ) \zeta_R(s) 
\, .
$$
Using this formula, one can get another useful formula with the logarithm: 
$$
\sum_{\bar k =1}^\infty \frac{(-1)^{\bar k} }{ \bar k^s } \ln \bar k
= (2^{1-s}-1 ) \zeta_R^\prime (s) - (2^{1-s} \ln 2 ) \zeta_R(s)
\, ,
$$
where the prime denotes the first derivative. 
}
we obtain the analytic form of the thermal part as 
\begin{eqnarray}
V^\one_T \sim - \rho_B \left[ \,  \frac{\pi T^2}{6} 
+ \frac{m_\dyn^2}{2\pi} \left( \ln \frac{m_\dyn}{\pi T} + \gam_{\rm E} - \frac{1}{2} \right) \, \right]
\, .
\end{eqnarray}
The first term corresponds to the Stefan-Boltzmann limit in the (1+1) dimensions. 
Including the vacuum part, the total potential is obtained as 
\begin{eqnarray}
V^\one&\equiv& V^\one_\vac + V^\one_T
\nnb
& \sim &
\rho_B  \left[ \, \frac{\Lambda^2}{4\pi} -  \frac{\pi T^2}{6}  
+ \frac{m_\dyn^2}{2\pi} \left(  \frac{ \pi }{ \rho_B \lambda_{\rm NJL}} +  \frac{\gamma_{\rm E} }{2}
 - \ln \frac{\Lambda}{\pi T}  \right) \, \right]
\, .
\end{eqnarray}
There is no longer the logarithm that played the crucial role 
in the chiral symmetry breaking at zero temperature. 
The chiral symmetry can be restored at a certain temperature. 
The critical temperature is read off from the coefficient of the quadratic term in $  m_\dyn^2$ 
that changes its sign across the phase transition. 
The critical temperature is now found to be  
\begin{eqnarray}
T_c = \frac{\Lambda^\prime}{\pi} \exp\left( -\frac{\pi}{\rho_B \lambda_{\rm NJL}} \right)
= \frac{1}{\pi} m_\dyn(T=0)
\label{eq:Tc-m}
\, .
\end{eqnarray}
Notice that the critical temperature agrees with the dynamical mass at $ T=0 $ 
up to the overall factor of $ \pi^{-1} $. 
Such relations are established for QED and the NJL model 
\cite{Ebert:1997um, Gusynin:1997kj, Lee:1997zj, Lee:1997uh, Fukushima:2012xw}.

According to Eq.~(\ref{eq:Tc-m}), the critical temperature of the chiral symmetry restoration is 
naturally interpreted as the point at which 
the thermal energy becomes large enough to overcome 
the dynamical mass gap and to fill the positive-energy states with thermal excitations. 
Therefore, we notice a natural behavior that the critical temperature increases 
as the dynamical mass gap increases at zero temperature. 
In the history of superconductivity, the BCS theory predicted a similar relation 
with a universal factor independent of material \cite{tinkham2004introduction}. 
In Sec.~\ref{sec:MC-strong}, we will see, however, that the lattice QCD simulation 
do not support this relation to hold in QCD. 
This is an intriguing implication that the strong-coupling dynamics in the low-energy QCD 
do not simply follow the results from the NJL model and weak-coupling QED.

\cout{\com{The following is moved to the later section.}

Therefore, the critical temperature of the chiral symmetry restoration is 
naturally interpreted as the point at which 
the thermal energy becomes large enough to overcome 
the mass gap and to fill the fermion states with the thermal excitations. 
In a word, this just means that, without significant thermal excitations, 
the chiral symmetry would not be restored. 
It should be also noted that the screening effect becomes significant 
in the gluon/photon propagator only when the temperature becomes larger than 
the quark mass (see the last part of Appendix~\ref{sec:VP_therm}). 
This is also because there must be significant thermal quark excitations (on the quark loop 
of the gluon/photon self-energy) above the quark mass gap 
so that the screening effect works. 
Once the screening effect starts cutting off the long-range interactions necessary for 
the realization of the ordered phase, 
the system will be subject to the (second-order) phase transition to the disordered phase. 
All these simple observations consistently support the relation between 
the critical temperature and the quark mass gap in Eq.~(\ref{eq:Tc-m}). 
In Sec.~\ref{sec:MC-strong}, we will see, however, that results from the lattice QCD simulation 
do not support this relation to hold in QCD. 

}

\subsection{Extension to non-Abelian theories}

Here, we invoke on another extension of the HE effective action to non-Abelian theories. 
QCD is the main target of non-Abelian theories. 
Before going into details, we briefly comment on the main features of non-Abelian problems that are not seen in the calculation of the HE action in QED. We will perform a one-loop calculation with respect to quantum fluctuations around the background gauge fields assuming that the QCD coupling constant $g$ is small enough. Then we will encounter the following differences. 

\begin{itemize}
\item In contrast to the electromagnetic fields in QED, the field strengths in non-Abelian gauge theories change covariantly under gauge transformations. Related to this, instead of imposing a gauge-variant condition $\del_\mu \nonF^a_{\nu\rho}=0$,  we rather impose the ``covariantly constant" condition $D_\lambda^{ab} \nonF^b_{\mu\nu}=0$ as introduced in Eq.~(\ref{eq:CCF}).  

\item Under the decomposition of the original gauge field into the background and the (fluctuating) dynamical fields, three-point and four-point interactions, characteristic of the non-Abelian gauge theories, induce a coupling between the background and dynamical fields. 
Since the decomposition produces quadratic terms in the dynamical gauge fields, 
we need to treat them as the gluon loop contribution (cf. Fig.~\ref{Fig:oneloopdiagrams}), 
in contrast to the QED case where the dynamical photon field only appears at higher loop diagrams.  

\item Since we have the dynamical non-Abelian gauge fields at the one-loop level, we need to include ghost fields 
when fixing the gauge of the dynamical gauge field. 
The ghost field is treated as a pure fluctuation (without background ghost fields), 
but is coupled to the background chromo-fields. 

\end{itemize}

We shall set up the classical Lagrangian as the starting point. 
Recall that we have already introduced the QCD action introduced in Eq.~(\ref{QCDaction}) 
and performed the decomposition of the non-Abelian gauge field into the dynamical fluctuation field $ a^a_\mu $ 
and classical field $ \nonA_\ext^{a \mu} $ as explained in Sec.~\ref{sec:QCD_prop} and Appendix~\ref{sec:bgd}. 
After this procedure, we have obtained the action that describes the dynamics of 
the quark field $ \psi $, the dynamical gauge field $a^a_\mu $, and the ghost field $c^a  $: 
\begin{eqnarray}
\label{eq:S_QCD}
S
= \int \!\! d^4x \left[ \,  \Lag_{\rm kin}  + \Lag_{\rm int} \, \right]
\, .
\end{eqnarray} 
The kinetic term $ \Lag_{\rm kin} $ as defined in Eq.~(\ref{eq:Lkg_d}) is bilinear with respect to 
the dynamical fields, $ \psi $, $a^a_\mu $, and $c^a $, while the interaction term $\Lag_{\rm int}$ 
given in Eq.~(\ref{eq:QCD_int}) in Appendix~\ref{sec:bgd} contains the coupling terms among more than three fields.

In addition to the external chromo-EM field $ \nonA_\ext^{a \mu} $, 
we here maintain an external Abelian field $ A_\ext^\mu $, which is included in the HE effective action discussed previously, 
and investigate interplay between QED and QCD fields in the following sections. 
Such a coexisting field configuration is realized in the ultrarelativistic heavy-ion collisions and lattice QCD simulations. 
Since the external Abelian field plays a role in the QCD dynamics only through the interaction with quarks, 
one can immediately include the Abelian field into the action (\ref{eq:S_QCD}) through 
a modification of the covariant derivative in the fundamental representation 
\begin{eqnarray}
\mathbb{D}^\mu \equiv \partial^\mu - i g \nonA_\ext^{a \mu} t^a + i q_f A_\ext^\mu
= D^\mu +  i q_f A_\ext^\mu
\, .
\end{eqnarray}
In the classical action (\ref{eq:Lkg_d}), this replaces the covariant derivative $ D^\mu $ in the quark part, 
while the covariant derivative acting on the gluon and ghost fields are intact 
because those fields do not directly interact with the external Abelian field. 
Here and below, we do not consider electrically charged gauge bosons such as the weak bosons. 
One can omit the Maxwell part $ -\frac14 F^{\mu\nu}F_{\mu\nu} $ for the constant field 
and do not consider the dynamical photon field as well. 
The dynamical photon field needs to be included when we consider the photon dynamics as in Sec.~\ref{sec:photons}, 
and radiative corrections which, however, will be subdominant to gluon radiative corrections in most of physical situations.

Now, we introduce the covariantly constant chromo-EM field, in which the field strength tensor $\nonF^a_{\mu\nu} $ associated with $ \nonA_\ext^{a \mu} $ is factorized as shown in Eq.~(\ref{eq:fact}). Accordingly, one can diagonalize the color structure in the covariant derivative, of which the diagonal component is given by  (see Sec.~\ref{sec:QCD_prop} \& Appendix~\ref{sec:bgd})
\begin{eqnarray}
\mathbb{D}^\mu_{i,f} = \partial^\mu - i w^i \nonA_\ext^{\mu} + i q_f A_\ext^\mu
\label{eq:D_QCD}
\, ,
\end{eqnarray}
where $\nonA_\ext^{\mu}$ is the Abelian-like strength of the external chromo-EM field [cf., Eq.~(\ref{eq:fund_ccf})], and $ w^i \, (i=1,2,3) $ is the effective charge introduced in Eq.~(\ref{eq:angle-fund-rep}) that linearly depends on the original strong coupling constant $g$. Notice that a mixing between the  Abelian and non-Abelian fields occurs at this point. We define the ``mixed'' field strength tensor\footnote{Compared to the usual definition of the field strength (such as ${\mathcal F}^a_{\mu\nu}t^a=\frac{i}{g}[D^\mu,D^\nu]$), we have defined $\mathbb{F}_{i,f}^{\mu \nu}$ so that it absorbs the two different coupling constants $w^i$ and $q_f$.} 
$\mathbb{F}_{i,f}^{\mu \nu} 
\equiv i [ \mathbb{D}_{i,f}^\mu, \mathbb{D}_{i,f}^\nu ]$ through the covariant derivative $\mathbb{D}_{i,f}^\mu$. It is decomposed as 
\beq
\mathbb{F}_{i,f}^{\mu \nu} = w^i {\cal F}^{\mu \nu} -  q_{f} F^{\mu \nu} 
\label{eq:F_QCD}
\, . 
\eeq 
$ {\cal F}^{\mu \nu}$ is for the magnitude of the external chromo-EM field 
introduced in Eq.~(\ref{eq:fact}), which are composed of $\vec{\mathcal{E}}$ and $\vec{\mathcal B}$, 
while $  F^{\mu \nu} $ is for the external EM field composed of $\vec{E}$ and $\vec{B}$ as usual. 
We will use notations $\aaa_{i,f} , \, \bbb_{i,f}$ for the invariants defined 
in Eqs.~(\ref{eq:a}) and (\ref{eq:b}) associated with the field strength tensor $ \mathbb{F}_{i,f}^{\mu \nu}  $ 
and also $\aaa \, ,\bbb$ associated with $ \nonF^{\mu \nu}  $.\footnote{The definitions of $\aaa$ and $\bbb$ are interchanged in some literature including Ref.~\cite{Ozaki:2015yja}.} 
The former and latter appear in the effective actions of the quark part and the Yang-Mills (gluon+ghost) part, respectively. 

\begin{figure}[t]
\begin{center}
\includegraphics[width=0.9 \textwidth]{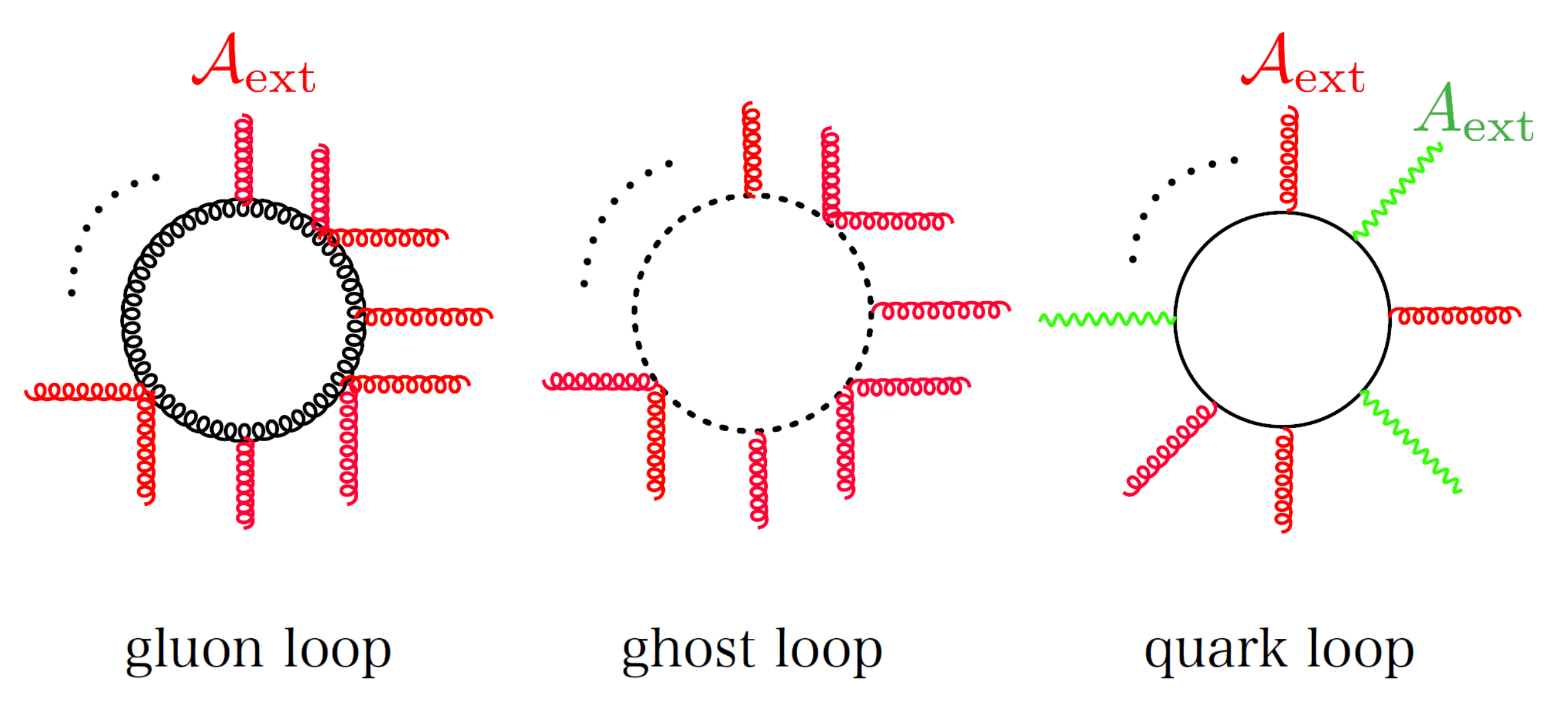}
\vspace{-0.5cm}
\end{center}
\caption{
Typical loop diagrams contributing to the resummed effective action. 
}
\label{Fig:oneloopdiagrams}
\end{figure}

By integrating out the dynamical fields, we can formally obtain the effective action in the presence of the coexistent external fields ${\cal A}^\mu_\ext$ and $A_{\rm{ext}}^{\mu}$ as 
\beq
{\rm{exp}} \Big( i S_{\rm eff}[{\cal A}^\mu_\ext, A^\mu_{\rm{ext}}] \Big)
\equiv \int {\mathscr D} a^\mu {\mathscr D}c {\mathscr D} \bar{c} {\mathscr D}\psi {\mathscr D} \bar{\psi} \ \ {\rm{exp}} \left( i S[{\cal A}^\mu_\ext, A^\mu_{\rm{ext}}, a^\mu, c, \bar{c}, \psi, \bar{\psi} ] \right)
\, .
\eeq
The effective action $ S_\eff  $ has two different types of dependences on the coupling constants $g$ and $q_f$: 
One is the type where the coupling constants are accompanied by the field strengths (i.e., $g  \mathcal{F}^{\mu\nu}, \, q_f  F^{\mu\nu}$), 
and the other is the type where the coupling constants appear alone without the field strengths. 
As in the Abelian case discussed in Sec.~\ref{sec:HE}, we consider strong external fields, 
and sum up the first-type of the dependences to the infinite order. 
The second-type is treated perturbatively. 
Then, the leading contribution comes from the bilinear terms in the dynamical (fluctuating) fields 
which constitute the one-loop diagrams shown in Fig.~\ref{Fig:oneloopdiagrams}. 
Performing the path integration over the bilinear terms, we have 
\begin{eqnarray}
 S_\eff =S^{(0)} + S^{(1)}_\quark + S^{(1)}_\gluon + S^{(1)}_\ghost 
\, ,
\label{eq:S_eff}
\end{eqnarray}
where $S^{(0)}= \int d^4x {\mathcal L}^{(0)}$ is the pure Yang-Mills action ${\mathcal L}^{(0)}=-\frac14 \nonF_{\mu\nu}^a \nonF^{a\mu\nu} $ 
and the leading-order (one-loop) contributions are obtained as 
\begin{subequations}
\beq
S^{(1)}_{\rm quark} \!\!&=&\! \!  \int \!\! d^4x {\cal L}^{(1)}_{\rm quark} =
- i  \, {\rm Tr} \, \ln \left[ i  \slashed {\mathbb D} - m \right]
= -i \, {\rm Tr} \, \ln \left[ (-1) (  \slashed {\mathbb D} \slashed {\mathbb D}  + m^2 \right) ]^{\frac12}
\label{quark-part}
\, ,
\\
S^{(1)}_{\rm gluon}\!\! &= &\!\!  \int \!\! d^4x {\cal L}^{(1)}_{\rm gluon} =
-i  \, {\rm Tr} \, \ln 
\left[ - ( D^{2})^{ab} g^{\mu \nu} + \delta^{ab} v^a (\nonF_{\alpha\beta} \J^{\alpha\beta})^{\mu\nu} 
\right]^{-\frac12} 
\, , 
\label{gluon-part}
\\
S^{(1)}_{\rm ghost}\!\! &= &\!\!  \int \!\! d^4x {\cal L}^{(1)}_{\rm ghost} =
-i  \, {\rm Tr} \, \ln \left[ ( D^{2})^{ab}  \right]^{+1} 
\, .
\label{ghost-part}
\eeq
\end{subequations}
We again used the trick on the fermion determinant 
mentioned below (\ref{Lag_original}) 
and here maintain the determinant of the sign factor 
$ \det(-1)^\frac12 $ that depends on the number of dimensions.
This factor is important when we apply the dimensional regularization in Sec.~\ref{sec:gluon-condensate}. 
In the gluon contribution, we have taken the Feynman gauge, $\xi_{g} = 1,$ for the dynamical gluon fields. 
This effective action allows one to explore the dynamics in the strong fields regime $g\mathcal{F}^{\mu\nu}, \, q_f  F^{\mu\nu}\gtrsim \Lambda^2$, 
in which the field strength exceeds the typical energy scale of the system $ \Lambda $ such as the fermion mass, temperature, and density.

The trace ``Tr"  
will be taken for the operator expectation value with the coordinate 
basis, $x$, as well as the color, Lorentz, spinor, flavor indices when exist. 
The color structures in Eqs.~(\ref{quark-part})--(\ref{ghost-part}) are all diagonalized in the covariantly constant field as discussed in Sec.~\ref{sec:QCD_prop}. 
The explicit forms of the covariant derivatives and the spin-interaction term in the gluon contribution 
are shown in Eqs.~(\ref{eq:fund_ccf}), (\ref{eq:adj_ccf}), and (\ref{eq:spin-gluon}). 
The summation with respect to the flavor degrees, as well as the color degrees 
within the covariantly constant field, only appears as the overall sum at the one-loop level. 
We will introduce these sums just below.

\subsubsection{Ghost- and gluon-loop contributions}

\label{sec:L_YM}

As far as the one-loop diagrams are concerned, the ghost and gluon contributions do not have any dependence on the quark degrees of freedom, 
and thus the ordinary (QED) electromagnetic fields are not coupled to those fields. 
Therefore, computation of the ghost and gluon contributions in QCD is equivalent to that of the pure Yang-Mills (YM) theory. 
Besides, except for some differences coming from statistics, one can compute the ghost- and gluon-loop contributions 
in a similar way as in the fermion-loop contribution discussed in Sec.~\ref{sec:finiteT}. 
We have already obtained all the necessary ingredients in the preceding sections.

Notice again that the logarithms in Eqs.~(\ref{gluon-part}) and (\ref{ghost-part}) 
can be both rewritten in the form of the proper-time integral 
\begin{subequations}
\label{effectiveaction-YM}
\beq
{\cal L}_{\ghost}^{(1)} &=& i \int_0^\infty \frac{ds}{s}\, {\rm e}^{-\epsilon s} 
\, {\rm tr}_{\rm c} \, \langle x | {\rm e}^{-i \hat H_\ghost s}|x\rangle 
\, ,
\label{effectiveaction-ghost}
\\
{\cal L}_{ \gluon}^{(1)} &=& - \frac{i}{2} \int_0^\infty \frac{ds}{s}\, {\rm e}^{-\epsilon s} 
\, {\rm tr}_{\rm c, L} \, \langle x | {\rm e}^{-i \hat H_\gluon s}|x\rangle 
\, ,
\label{effectiveaction-gluon}
\eeq
\end{subequations}
where the ``Hamiltonians'' in the ghost and gluon contributions are, respectively, given by 
\begin{subequations}\label{QCDHam}
\beq
\hat H_\ghost  &=&  ( D^{2})^{ab} 
\label{QCDHam_ghost}
\, ,
\\
\hat H_\gluon &=&  - ( D^{2})^{ab} g^{\mu \nu} + \delta^{ab} v^a (\nonF_{\alpha\beta} \J^{\alpha\beta})^{\mu\nu} 
\label{QCDHam_gluon}
\, .
\eeq
\end{subequations}
In Eqs.~(\ref{effectiveaction-ghost}) and (\ref{effectiveaction-gluon}), 
the different overall factors come from statistics of ghosts and gluons. 
The traces are taken over the color (c) index in the ghost contribution 
and the color (c) and Lorentz (L) indices in the gluon contribution. 
For the notational simplicity in these expressions and below, 
we do not subtract the contributions independent of background fields that are given 
by the free Hamiltonians $\hat H_0=\delta^{ab}\del^2$ or $-\delta^{ab}g^{\mu\nu}\del^2$. 
However, one can easily identify these contributions whenever required 
for subtraction of the divergences in the proper-time integrals.

The computation of the effective action has been reduced to 
the computation of the transition matrix element $ \langle x | {\rm e}^{-i \hat H s}|x\rangle $. 
Similar to the computation in Sec.~\ref{sec:resummed-action-constant}, 
we shall define the transition matrix elements $ K(x,x';s|A_{\rm FS}) $ as in Eq.~(\ref{eq000}) 
with the Hamiltonians (\ref{QCDHam_ghost}) and (\ref{QCDHam_gluon}) 
for the ghost and gluon contributions, respectively. 
Then, one can again establish the relation between 
$ K(x,x';s|A_{\rm FS}) $ and $\Delta (p|A)$ 
through Eqs.~(\ref{eq111}) and (\ref{eq222}). 
More specifically, $ \Delta (p|A)$ in Eq.~(\ref{eq222}) is replaced by $  \Delta_{\rm scalar} (p|A)$ in Eq.~(\ref{eq:GDp_s}) 
with a vanishing mass $ m = 0 $ for the ghost contribution, 
and by the $ \Delta_{\mu \nu}^{(a)} (p|\nonA) $ in Eq.~(\ref{eq:prop-gluon}) for the gluon contribution. 
The electric charge $ q_f $ in the $  \Delta_{\rm scalar} (p|A)$ is understood to be 
replaced by the effective charge $ v^a $ as explained in Sec.~\ref{sec:ghost-prop}. 
Note also that the statistics of the ghost field is already taken into account 
as the difference between the overall factors in Eqs.~(\ref{effectiveaction-ghost}) and (\ref{effectiveaction-gluon}), 
so that the remaining part of the ghost contribution can be treated as the scalar-particle contribution in Sec.~\ref{sec:ghost-prop}. 
With these observations, we are ready to write down the transition matrix elements (see also Ref.~\cite{Hattori:2020guh} 
for a deviation following Schwiner's paper \cite{Schwinger:1951nm}) 
\begin{subequations}
\beq
\hspace*{-9mm}
\langle x | {\rm e}^{-i\hat H_\ghost s}|x\rangle
\=
\frac{ 1 } {\cosh( v^a \aaa s)\cos( v^a \bbb s)}
\int\!\! \frac{d^4p}{(2\pi)^4} 
\exp \left( \, 
 i \frac{ p_\parallel^2}{v^a \aaa} \tanh( v^a \aaa s) + i \frac{ p_\perp^2}{v^a \bbb} \tan( v^a \bbb  s) \, \right) , \label{eq:K-ghost} \\
\hspace*{-9mm}
{\rm tr}_{\rm L} \, \langle x | {\rm e}^{-i\hat H_\gluon s}|x\rangle 
\=
2 \frac{ \cosh ( 2 v^a \aaa s )  + \cos ( 2 v^a \bbb s )} {\cosh( v^a \aaa s)\cos( v^a \bbb s)}
\nn\\
&& 
\times
\int\!\! \frac{d^4p}{(2\pi)^4} 
\exp \left( \, 
 - i \frac{ p_\parallel^2}{v^a \aaa} \tanh( v^a \aaa s) - i \frac{ p_\perp^2}{v^a \bbb} \tan( v^a \bbb  s) \, \right)
\, , 
\label{eq:K-gluon}
\eeq
\end{subequations}
where the invariants $\aaa$ and $\bbb$ are defined below Eq.~(\ref{eq:F_QCD}). 
A difference between the integrands in the ghost and gluon contributions appears 
in the cosine and hyperbolic cosine in the numerator in Eq.~(\ref{eq:K-gluon}) [and also in Eq.~(\ref{action_gluon_T}) below]. 
These factors originate from the spin interaction term in Eq.~(\ref{QCDHam_gluon}), 
which was arranged in Eq.~(\ref{eq:moment}). 
However, such a spin interaction is absent for the ghost field. 
The difference between the signs in the exponents just comes from 
the relative sign of the ghost and gluon kinetic terms in Eqs.~(\ref{eq:Lkg_d}) and (\ref{QCDHam}).


At zero temperature, one can immediately perform the Gaussian integrals 
for the four-dimensional momentum. 
At finite temperature, one needs to work on the imaginary-time formalism as in the fermion contribution, 
but with the periodic boundary condition, which leads to 
the replacements, $ \int dp^0 \to i T \sum_{n=-\infty}^\infty $ and $ p^0 \to  i 2n \pi  T + 
v^a \vphi^a$, where $\vphi^a $ is defined with $ {\mathcal A}^{a0}_\ext$ instead of $ A^{0}_\ext$ and the corresponding field 
in the FS gauge in Eq.~(\ref{eq:vphi}).\footnote{
Note that the bosonic boundary condition is taken not only for the gluon field 
but also for the ghost field, in spite of the fermionic behavior of the ghost field. 
In the present case, the difference between the periodic and antiperiodic boundary conditions 
explicitly appears as the alternating signs seen in Eq.~(\ref{action_fermion}). 
Therefore, those fields are expected to take the same boundary conditions 
so that the ghost contribution cancels the unphysical gluon contributions, 
as indeed supported by general justifications (see, e.g., Refs.~\cite{Kraemmer:2003gd, Fukushima:2017csk} and references therein). 
}
The Matsubara sum is reorganized with the help of the Poisson resummation formula, which we have examined to 
get the fermion contribution in Eq.~(\ref{action_fermion}). Then, we obtain each contribution to the effective action for chromo-EM fields at finite temperature \cite{Ozaki:2015yja}: 
\begin{subequations}
\label{eq:gg-thermal}
\beq
{\cal L}_{\rm ghost}^{(1)} 
\!\!&=&\!\!   \sum_{a=1}^{N_{c}^{2}-1} \frac{ (v^{a}\aaa) ( v^{a}\bbb) }{ 16 \pi^{2} } \!\!
\int^{\infty}_{0} \frac{ds}{s}\, \e^{-\epsilon s} 
\frac{1 }{ \sinh(v^{a}\aaa s)  \sin (v^{a}\bbb s)} 
\left[1 + 2\sum_{\bar k=1}^{\infty} \e^{ i \frac{v^a \mathfrak{h}(s)}{4T^{2}} \bar k^{2}}  
\cosh( \beta v^a \vphi^a \bar k)   \right]
\label{action_ghost_T}
\, ,
\\
{\cal L}_{\rm gluon}^{(1)} 
\!\!&=&\!\! -   \sum_{a=1}^{N_{c}^{2}-1} \frac{ (v^{a}\aaa) ( v^{a}\bbb) }{ 16 \pi^{2} }\!\!
\int^{\infty}_{0} \frac{ds}{s} \, \e^{-\epsilon s} 
\frac{\cosh(2 v^{a}\aaa s)  + \cos( 2v^{a}\bbb s) }{ \sinh(v^{a}\aaa s)  \sin (v^{a}\bbb s)} 
\left[1 + 2\sum_{\bar k=1}^{\infty} \e^{ i \frac{v^a \mathfrak{h}(s)}{4T^{2}} \bar k^{2}}  
\cosh( \beta v^a \vphi^a \bar k)  \right]
\, ,
\nn\\
\label{action_gluon_T}
\eeq
\end{subequations}
where the overall summation is taken over the color index 
and the field in $ \vphi^a$ is replaced with 
$ {\mathcal A}^{a}_u \equiv {\mathcal A}^{a\mu}_\ext u_\mu$. 
As we have identified in the fermion-loop contribution (\ref{action_fermion-parallel}), 
the vacuum and finite-temperature contributions are separated 
into the first and second terms in the square brackets, respectively. 
The $\mathfrak{h}(s)$ in the exponent is defined as in Eq.~(\ref{eq:h-QED}), but with the replacements of 
($ q_f a , \, q_f b , \, q_f e_u) $ by ($v^a \aaa , \, v^a \bbb , \,  v^a \eee_u  $), respectively, 
where the power of the electric field is defined with the field strength tensor of the chromo-field, 
${\eee}^{2}_u = - \nonF^{\mu \alpha } \nonF_{\mu \beta} u_{\alpha} u^{\beta}$. 
Namely, we have 
\beq
v^a \mathfrak{h} (s)
=   v^a \aaa \frac{ \bbb^{2} + {\eee}^{2}_u }{ \aaa^{2} + \bbb^{2}} \, {\rm{coth}}(v^a \aaa s)
+ v^a \bbb \frac{ \aaa^{2} - {\eee}^{2} _u}{\aaa^{2} + \bbb^{2}} \,  {\rm{cot}}(v^a \bbb s) 
\label{eq:h-YM}
\, .
\eeq
The sum of the gluon and ghost parts provides the Yang-Mills part of the effective action 
${\cal L}_{\rm YM}^{(1)}={\cal L}_\gluon^{(1)} + {\cal L}_\ghost^{(1)} $
in the external chromo-EM fields at finite temperature \cite{Ozaki:2015yja}.

Similar to the case of the fermion contribution in Eq.~(\ref{eq:Therm-pot_fermion}), 
we can reproduce the familiar thermodynamic potential and the Stefan-Boltzmann law 
for a free gluon gas in the vanishing limit of the external chromo-EM fields. 
Sending the field strengths to zero 
and maintaining only $\vphi^a $, 
the temperature-dependent part reads 
\begin{eqnarray}
 {\cal L}_{{\rm YM}, T} ^{(1)}(\nonE ,\nonB=0) = 
 -   \sum_{a=1}^{N_{c}^{2}-1} \frac{ 1 }{ 8\pi^{2} }
 \sum_{\bar k=1}^{\infty}\cosh(v^a \vphi^a\beta  \bar k)   
  \int^{\infty}_{0} \! \frac{ds}{s^3} \, \e^{-\epsilon s}  \e^{ \frac{i}{4T^{2} s}\bar k^{2}} 
\label{action_YM_free}
\, .
\end{eqnarray}
After the rotation of the integral contour to the negative imaginary axis, 
the proper-time integral and the summation can performed as 
\begin{eqnarray}
 {\cal L}_{{\rm YM}, T }^{(1)}  
 &=&   \sum_{a=1}^{N_{c}^{2}-1} \frac{ 1 }{ 8 \pi^{2} } 
\sum_{\bar k=1}^{\infty}\cosh(  v^a \vphi^a\beta  \bar k)   
\int^{\infty}_{0} \! \frac{ds}{s^3}   \e^{ - \frac{1}{4T^{2} s}\bar k^{2}} 
\nnb
 &=&   \frac{  T^4 }{ \pi^{2} }  \sum_{a=1}^{N_{c}^{2}-1} 
[ \, {\rm Li}_4(e^{ v^a \vphi^a \beta })+ {\rm Li}_4(e^{ -v^a \vphi^a \beta }) \, ]
\, ,
\end{eqnarray}
where the summation is identified with a series representation of the polylogarithmic function 
$ {\rm Li}_s(z) = \sum_{\bar k=1}^\infty z^{\bar k} \, / \, \bar k^s  $ \cite{Polylogarithm}. 
The momentum-integral form can be retrieved with the integral representation 
$   {\rm Li}_{s+1} ( \e^\mu) \Gamma(s+1) = -  \int_0^\infty dp \, p^s /( \e^{p-\mu} -1) $ as 
\begin{eqnarray}
 {\cal L}_{{\rm YM}, T} ^{(1)}   =   -  T\sum_{a=1}^{N_{c}^{2}-1} 
 \int \frac{ d^3 p}{(2\pi)^3} \left[ \,  \ln \big( 1- e^{ - \beta (p - v^a \vphi^a ) }\big) 
 +  \ln \big( 1- e^{ - \beta (p + v^a \vphi^a ) }  \big) \, \right]
 \, .
\end{eqnarray}
When $ \vphi^a =0  $, we have ${ \rm Li}_4( 1) = \zeta_R(4) = \pi^4/90$ and thus 
\begin{eqnarray}
\lim_{ {\mathcal A}^{a0} \to 0  }  {\cal L}_{{\rm YM}, T} ^{(1)}   =     ( N_{c}^{2}-1 ) \cdot 2  \cdot \frac{  \pi^2 }{90}  T^4
\, .
\end{eqnarray}
The Stefan-Boltzmann limit is reproduced 
with the number of color degrees of freedom $  ( N_{c}^{2}-1 )$ 
and a factor of 2. This factor should come from 
the two transverse polarization modes 
after the cancellation of the longitudinal mode 
by the ghost contribution. 
Below, we examine how this cancellation occurs 
in the presence of external chromo-fields in detail.

\subsubsection*{Cancellation of the longitudinal gluon}

When we consider a purely chromo-{\it electric} background (${\cal E}=|\vec {\cal E}|\neq 0,\ {\cal B}=0$) in the medium rest frame $u^\mu = (1,0,0,0)$, we have $\aaa \to \mathcal{E}$, $\bbb \to 0$. 
Inserting these values into Eqs.~(\ref{action_ghost_T}) and (\ref{action_gluon_T}), one can reproduce Gies' effective action\footnote{More precisely, Gies computed the effective action for the Polyakov loop and the chromo-electric fields, both of which are given by the (imaginary-)time component of the field ${\mathcal A}^0_{\rm ext}$ (see Sec.~\ref{sec:Polyakov}). 
}
at finite temperature~\cite{Gies:2000dw} 
\begin{eqnarray}
 {\cal L}_{\rm YM}^{(1)}(\nonE \neq 0,\nonB=0) = 
 -   \sum_{a=1}^{N_{c}^{2}-1} \frac{ v^{a}\nonE}{ 16 \pi^{2} }
\int^{\infty}_{0} \! \frac{ds}{s^2} \, \e^{-\epsilon s} 
\frac{\cosh(2 v^{a} \nonE s)  }{ \sinh(v^{a}\nonE s)} 
\left[1 + 2\sum_{\bar k=1}^{\infty} \e^{ i \frac{v^a \mathfrak{h}(s)}{4T^{2}}\bar k^{2}}
\cosh( \beta v^a \vphi^a \bar k)   \right]
\label{action_YM_pure_E}
\, ,
\end{eqnarray}
where $v^a \mathfrak{h}(s)=v^a {\cal E} \coth (v^a {\cal E}s)$. 
Notice that an exact cancellation occurs between the whole ghost contribution 
and the term proportional to $ \cos( 2v^{a}\bbb s)  $ in the gluon contribution in the limit, $ \bbb \to 0 $. 
Such a cancellation is, by construction, 
expected to occur for the ghost field to 
eliminate the unphysical gluonic degrees of freedom 
and is very important when gluons become on-shell particles. 
The cancellation of the longitudinal mode 
has been also discussed with the canonical quantization 
in Refs.~\cite{Ambjorn:1982nd, Ambjorn:1982bp, Ambjorn:1983ne, Tanji:2011di}. 
This can occur by the Schwinger mechanism of gluons 
in chromo-electric fields. 
While the zero-temperature part in Eq.~(\ref{action_YM_pure_E}) 
looks like a real-valued integral, 
there is a singularity at infinity $ s \to \infty $ 
as long as there is a finite electric field $ (v^a {\cal E} > \epsilon) $. 
Therefore, the result of the integral could acquire an imaginary part in the analytic continuation $ \epsilon \to 0 $, 
leading to the gluon Schwinger mechanism in a chromo-electric field. 


In case of a purely chromo-{\it magnetic} background (${\cal E}=0,\ {\cal B}\neq 0$), 
we have $\aaa \to 0$, $\bbb \to {\cal B}$, 
and reproduce the results in Refs.~\cite{Dittrich:1980nh}. 
In this case, an exact cancellation occurs 
between the whole ghost contribution and the term proportional to 
$\cosh( 2v^{a}\aaa s)$ in the gluon contribution.

It is more tempting to explicitly see cancellation of the longitudinal gluon contribution in the Landau-level representation. 
We speculated this cancellation below Eq.~(\ref{eq:g-prop-B}) 
when we identified the Zeeman splitting for gluon spin. 
Since the following arrangement does not depend on the Poison summation inside the square brackets in Eq.~(\ref{eq:gg-thermal}), 
one can focus on the zero-temperature expression for the notational simplicity. 
Decomposing the transition matrix elements into the Landau levels 
in a purely magnetic field ($ \aaa \to 0 $) [see the computations 
leading to Eqs.~(\ref{eq:G_B_scalar-LL}) and (\ref{eq:g-prop-B})], we obtain 
\begin{subequations}
\label{eq:K-YM-LL}
\beq
\langle x | {\rm e}^{-i\hat H_\ghost s}|x\rangle &=&
2\int \frac{d^4p}{(2\pi)^4}  
\, \e^{ \frac{ p_\perp^2}{\vert v^a \nonB \vert} } \sum_{n=0}^\infty  
 (-1)^n   e^{ - \epsilon s}    L_n (u_\perp)  e^{ - is (p_\para^2-\varepsilon_{\perp n}^2) }   
\label{eq:K-ghost-LL}
\, ,
\\
{\rm tr}_{\rm L} \, \langle x | {\rm e}^{-i\hat H_\gluon s}|x\rangle &=&
 2 \int \frac{d^4p}{(2\pi)^4} 
\, \e^{ \frac{ p_\perp^2}{\vert v^a \nonB \vert} } \sum_{n=0}^\infty
(-1)^{n}    e^{ - \epsilon s}   \left[ \, 
2  L_{n} (u_\perp ) e^{ - is (p_\para^2- \varepsilon_{\perp n}^2 ) }  
\right.
\label{eq:K-gluon-LL}
\\
&& \hspace{4cm}
\left.
+
 \{ L_n  (u_\perp ) + L_{n-2}  (u_\perp ) \} 
  e^{ - is (p_\para^2-  \varepsilon_{\perp n-1} ^2) }  
\, \right]
\nn
\, ,
\eeq
\end{subequations}
where $ L_{-2}(u_\perp) = L_{-1}(u_\perp) =0 $ 
and we defined $ \varepsilon_{\perp n}^2 = (2n+1) \vert v^a \nonB\vert $. 
Performing the momentum integrals in the similar manner 
as in Eq.~(\ref{eq:g-transverse}), 
we find the Landau-level representation of the effective Lagrangians\footnote{
Here, the transverse-momentum integral results in an $ n $-independent form 
\begin{eqnarray}
2 (-1)^n \int \frac{d^2p_\perp}{(2\pi)^2} e^{- \frac{u_\perp}{2}} L_n(u_\perp) = \left| \frac{  v^{a}\nonB }{ 2 \pi } \right|
\label{eq:YM-p_perp}
\, .
\end{eqnarray} 
}
\begin{subequations}
\beq
{\cal L}_{\rm ghost}^{(1)} &=&
 \frac{i}{4\pi}  \sum_{a=1}^{N_{c}^{2}-1} \left| \frac{  v^{a}\nonB }{ 2 \pi } \right|
\int^{\infty}_{0} \frac{ds}{s^2} \, \e^{-\epsilon s} 
 \sum_{n=0}^\infty   e^{ - is \varepsilon_{\perp n}^2 }
\label{action_ghost-LL}
\, ,
\\
{\cal L}_{\rm gluon}^{(1)} 
&=& -  \frac{i}{4\pi}  \sum_{a=1}^{N_{c}^{2}-1}  \left| \frac{  v^{a}\nonB }{ 2 \pi } \right|
\int^{\infty}_{0} \frac{ds}{s^2} \, \e^{-\epsilon s} 
\sum_{n=0}^\infty \left[ \, 
 e^{ - is  \varepsilon_{\perp n}^2  }  
+ \frac12
(2 - \delta_{n0} - \delta_{n1})   e^{ - is  \varepsilon_{\perp  n-1}^2 }   
\, \right]
\, .
\label{action_gluon-LL}
\eeq
\end{subequations}
In the gluon contribution, 
the first and second terms between the square brackets are identified with the contributions of the longitudinal and transverse gluons, respectively. 
This is understood from the Zeeman splitting of spin-1 particles shown in Fig.~\ref{fig:Zeeman}. 
The transverse gluon modes in each energy level are 
counted with the factor of $(2 - \delta_{n0} - \delta_{n1})  $; 
There are two-fold spin degeneracy except for 
the lowest two levels. 
Remember also that we explicitly identified 
the gluon polarization tensors $\Q^{\mu\nu}_\pm $ 
below Eq.~(\ref{eq:g-prop-B}).

Now, observe an exact cancellation between the longitudinal-gluon and ghost contributions in the all-order Landau levels. 
After the cancellation, we find the Landau-level representation of the Yang-Mills part: 
\begin{eqnarray}
 {\cal L}_{\rm YM}^{(1)} &=& 
  -  \frac{i}{8\pi}  \sum_{a=1}^{N_{c}^{2}-1} \left[ \frac{  v^{a}\nonB }{ 2 \pi } \right]
\int^{\infty}_{0} \frac{ds}{s^2} \, \e^{-\epsilon s} 
\sum_{n=0}^\infty  (2 - \delta_{n0} - \delta_{n1})    e^{ - is  (2n-1) \vert v^a \nonB\vert }  
\label{action_YM_pure_B-LLL}
\, .
\end{eqnarray}
Incidentally, similar to the discussion below Eq.~(\ref{HE_parallel-Landau}), 
we can obtain Eq.~(\ref{action_YM_pure_B-LLL}) directly from 
the effective action ${\cal L}_{\rm YM}^{(1)} $ which has been already obtained in 
Eqs.~(\ref{action_ghost_T}) and (\ref{action_gluon_T}). 
To see this, take a purely magnetic configuration (${\cal E}=0$) and use an identity 
\beq
\frac{\cos 2x}{\sin x}=i\left[\e^{ix}+\e^{-ix}+2 \sum_{n=2}^\infty \e^{-i(2n-1)x}\right]
\, .
\eeq
The cancellation holds in a coexistent chromo-electric 
and -magnetic fields. 
The cancellation of the longitudinal mode 
in the Schwinger pair production in the coexistent chromo-fields 
were also explicitly shown on the basis of the HE effective action~\cite{Hattori:2020guh} as well as 
the canonical quantization \cite{Ambjorn:1982nd, Ambjorn:1982bp, Ambjorn:1983ne, Tanji:2011di}.


\subsubsection{Quark-loop contribution}

\label{sec:L_quark}

Next, we compute the quark contribution (\ref{quark-part}). Within the covariantly constant chromo-field, 
this is a straightforward extension of the fermion-loop contributions in the Abelian field already 
given in Eqs.~(\ref{action_fermion-parallel}) and (\ref{action_fermion}). 
The ``Hamiltonian" in Eq.~(\ref{eq:amplitude-T}) is replaced by the one associated 
with the covariant derivative (\ref{eq:D_QCD}): 
\beq
\hat H_{\rm  quark} = \mathbb{D}_{i,f}^2 + \frac{1}{2} \mathbb{F}_{i,f}^{\mu\nu}\sigma_{\mu \nu}
\label{QCDHam_fermion}
\, .
\eeq
Accordingly, all the associated variables, such as $ E, \, B, \, a, \, b  $ defined 
by the covariant derivative and the external gauge field, are replaced as 
\begin{eqnarray}
(  q_f a, q_f b,  q_f \vphi ) \to ( \aaa_{i,f} ,\bbb_{i,f} , \vphi _{i,f}) 
\, ,
\end{eqnarray}
where $\varphi_{i,f} := T \int_0^\beta d\tau ' 
\{ (q_f A^\m_\ext + w_i  {\mathcal A}^\m_\ext ) 
- (q_f A^\m_\FS + w_i  {\mathcal A}^\m_\FS )\} u_\m$
in line with the previous definition (\ref{eq:vphi}) and 
$  \aaa_{i,f} ,\bbb_{i,f} $ are introduced below Eq.~(\ref{eq:F_QCD}) with $ \mathbb{F}_{i,f}^{\mu\nu} $.\footnote{
Note that the charges are included in tho definitions of 
$\varphi_{i,f} $, $ \aaa_{i,f}$, and $\bbb_{i,f} $ 
in the same way as the definition of the field strength tensor (\ref{eq:F_QCD}).
} 
Applying those replacements to Eq.~(\ref{action_fermion}), 
we find the quark contribution to the effective action (\ref{quark-part}) as \cite{Ozaki:2015sya} 
\beq 
{\cal L}_{\rm quark}^{(1)}
&=&  \sum_{i=1}^{N_{c}} \sum_{f=1}^{N_{f}} \frac{1}{8\pi^{2}} \int_{0}^{\infty} \frac{ds}{s} \, 
\e^{-i(m_{f}^{2}-i\epsilon)s} \frac{ \aaa_{i,f} \bbb_{i,f}  }{ \tanh(\aaa_{i,f}s) \tan(\bbb_{i,f}s) }
\nonumber \\
&&\hspace{2cm} \times 
\left[ 1+ 2 \sum_{\bar k=1}^{\infty} 
(-1)^{\bar k} \e^{\frac{i}{4T^{2}} \mathfrak{h}_{i,f}(s) \bar k^{2} } \cosh( \beta \varphi_{i,f} \bar k)  \right]
\label{action_quark}
\, ,
\eeq 
with 
\beq
\mathfrak{h}_{i,f}(s)
=   \aaa_{i,f} \frac{ \bbb_{i,f}^{2} + {\eee}_{i,f}^{2} }{ \aaa_{i,f}^{2} + \bbb_{i,f}^{2}} \, {\rm{coth}}(\aaa_{i,f}s)
+ \bbb_{i,f} \frac{ \aaa_{i,f}^{2} - {\eee}_{i,f}^{2} }{\aaa_{i,f}^{2} + \bbb_{i,f}^{2}} \,  {\rm{cot}}(\bbb_{i,f}s) 
\label{eq:h}
\, .
\eeq
The power of the electric field ${\eee}_{i,f}^{2} = - \mathcal{E}_{i,f}^{\mu} \mathcal{E}_{i,f \, \mu}$ 
is defined with $\mathcal{E}_{i,f}^{\mu} \equiv \mathbb{F}_{i,f}^{\mu \nu} u_{\nu} $. 
In the medium rest frame, we have $u^{\mu} = (1,0,0,0)$, $\mathcal{E}_{i,f}^{\mu} = (0,\vec{ \mathcal{E} }_{i,f} )  $, 
and thus a reduced form ${\eee}_{i,f}^{2}  = \vec{ \mathcal{E} }_{i,f}^{2} 
= (  w_{i} \vec{\mathcal{E}} - q_{f} \vec{E} )^{2} = {\aaa}_{i,f}^{2}$. 

One can again perform the Landau-level decomposition of the effective action (\ref{action_quark}) 
and also confirm the Stefan-Boltzmann law in the vanishing-field limit. 
Following the same procedure that led to Eq.~(\ref{HE_parallel-Landau}), one finds the Landau-level representation 
\beq 
{\cal L}_{\rm quark}^{(1)\vac}
=  \sum_{i=1}^{N_{c}} \sum_{f=1}^{N_{f}} \sum_{n=0}^\infty 
\left[\kappa_n \frac{|{\bbb}_{i,f}|}{2\pi}\right] \frac{i}{4\pi} 
\int_{0}^{\infty} \frac{ds}{s}\, \e^{-i(m_n^{2}-i\epsilon)s}  
\frac{{\aaa}_{i,f}}{{\rm{tanh}}({\aaa}_{i,f}s)} 
\, ,
\label{action_quark-vac-Landau-level}
\eeq 
where $\kappa_n = 2-\delta_{n 0}$. 
Here, the effective mass $m_n^2=m_f^2 + 2n|{\bbb}_{i,f}|$ 
and the Landau degeneracy factor $|{\bbb}_{i,f}|/2\pi$ depend on $i$ and $f$ through ${\bbb}_{i,f}$. 
In the absence of external fields and the massless quark limit, the Stefan-Boltzmann law reads  
\begin{eqnarray}
\lim_{{\mathcal A}^0_{i,f},m_f\to 0}{\cal L}_{\rm quark}^{(1)T} = - (2 \cdot 2\cdot N_f \cdot N_c) \times 
\frac{7}{8} \cdot \frac{\pi^2 }{90} T^4
\, ,
\end{eqnarray}
where we now have a color factor $ N_c $ as compared to Eq.~(\ref{eq:SB-limit}).

%


\subsection{Schwinger mechanism in chromo-fields}

As mentioned in the beginning of this section, one of the motivations that drove the investigations on the effective action of QCD and QED fields is the interest in the early-time dynamics in high-energy heavy-ion collisions. The effective action can be applied to the problem of particle productions from chromo-EM fields created in the early stages of the heavy-ion collisions. In particular, as we can easily understand from the diagrams  relevant for the one-loop effective action (see Fig.~\ref{Fig:oneloopdiagrams}), the chromo-electric fields allow for {\it gluon pair} production as well as quark-antiquark pair production, which is a unique feature of the non-Abelian fields. Since these processes are thought to be the essential initial step towards the formation of quark-gluon plasmas, they have been investigated in many papers. By using the effective action in the Yang-Mills theory and QCD, one can immediately find extensions of the Schwinger formula to the problems of gluon/quark production \cite{Batalin:1976uv,Yildiz:1979vv,Claudson:1980yz,Ambjorn:1982bp,Gyulassy:1986jq} (see also Refs.~\cite{Nayak:2005pf, Nayak:2005yv, Cooper:2005rk} for recent works). Also important is the application to the configurations which could be realized in collisions \cite{Casher:1978wy,Casher:1979gw}, and the evaluation of dissipative effects of the chromo-EM fields due to particle productions \cite{Kajantie:1985jh,Gatoff:1987uf,Tanji:2010eu}. 
Refined calculation of the quark and gluon production from chromo-electric fields was performed in Ref.~\cite{Taya:2016ovo}, where the effects of expanding geometry was also taken into account. In addition to the Schwinger mechanism in chromo-electric fields, the roles of chromo-magnetic fields were discussed in Refs.~\cite{Fujii:2008dd, Fujii:2009kb}, where the Nielsen-Olesen (NO) instability is found to take place in a modified way in an expanding geometry.

Below, we briefly discuss the imaginary part of the effective action that provides the vacuum persistent probability (VPP) 
for the gluon and quark productions in the presence of QED and QCD fields (cf. Sec.~\ref{sec:pair-production}). 
Furthermore, it is also found that, in the presence of both chromo-electric and chromo-magnetic fields, the gluon production rate is significantly enhanced due to the NO instability \cite{Tanji:2011di}. 
There is, however, a subtlety in this issue. 
We examine how the quark production rate is affected by the interplay between the chromo-fields and electromagnetic fields \cite{Ozaki:2015yja}.

\subsubsection{Gluon production in chromo-electromagnetic fields}

We here discuss gluon production from chromo-EM fields. 
Recall that the Yang-Mills part of the effective action is not affected by photon fields at the one-loop level.

Let us first consider the case with purely chromo-electric fields at zero temperature. 
The Yang-Mills part (\ref{action_YM_pure_E}) in chromo-electric fields has a similar proper-time integral as 
that for the HE effective action 
in electric fields (\ref{HE_pureE}). 
In a similar way as the contour integral performed there, 
one can compute the imaginary part for 
the Yang-Mills part (\ref{action_YM_pure_E}) 
by picking up the residues on the imaginary axis as 
\cite{Ambjorn:1982bp,Gyulassy:1986jq,Nayak:2005yv} 
\begin{eqnarray} 
\gamma^{}_{\rm gluon}({\mathcal E}\neq 0, {\mathcal B}=0 )
=2\Im m {\mathcal L}^{(1)}_{\rm YM}
=\sum_{a=1}^{N_c^2-1}  \frac{1}{8\pi^3} ( v^a {\mathcal E})^2 
\sum_{n=1}^\infty \frac{(-1)^{n-1}}{n^2}
\, .
\end{eqnarray}
The longitudinal gluon component has been canceled by the ghost contribution as discussed below Eq.~(\ref{action_YM_pure_E}). 
One finds that this result is consistent with that of the scalar QED (\ref{eq:SM-scalar}) except for the following three points: 
There is the color sum, an overall factor of $ 2$  that counts the number of transverse gluon modes, and no suppression factor such as ${\rm e}^{-(E_c/E)n\pi}=\e^{-m^2 n \pi/(q_fE)}$ 
for the massless nature of gluons. 
The alternating series yields a finite positive number 
$\eta(2)=\frac{\pi^2}{12}$ 
from $\eta(s)=\sum_{n=1}^\infty {(-1)^{n-1}}/{n^s}$. 
Then, the final result is obtained in a quite compact form 
\begin{eqnarray}
\gamma^{}_{\rm gluon}({\mathcal E}\neq 0, {\mathcal B}=0 )
=\frac{N_c}{96\pi}(g {\mathcal E})^2
\, ,
\end{eqnarray}
where the color sum is performed 
by the use of the identity (\ref{eq:veq}). 
The VPP quadratically decreases as a function of ${\mathcal E}$ 
without an exponential suppression. 
That is, the threshold energy to create massless gluons 
is vanishing.  
See also recent stability analyses against 
the Schwinger mechanism in non-Abelian electric fields 
(in the absence of a non-Abelian magnetic field) 
\cite{Cardona:2021ovn, Pereira:2022lbl, 
Vachaspati:2022gco, Vachaspati:2023tpt}.

When the chromo-electric and chromo-magnetic fields coexist, 
we can still discuss the gluon pair creation 
with the effective Lagrangian (\ref{eq:gg-thermal}). 
However, there is one subtle point to which 
we should pay a special attention. As we already saw, 
the dispersion relation in the lowest energy level has 
a negative energy squared, known as the NO instability. The gluon pair creation may be significantly modified by the presence of the instability. 
It is even not clear whether the unstable modes 
are well-defined on-shell excitation 
if they are created by chromo-electric fields. 
Below, we consider a chromo-magnetic field 
applied in parallel to a chromo-electric field, 
where the unstable mode can be explicitly identified 
with the Landau-level decomposition 
even in the presence of the chromo-electric field.

The effects of the NO instability was investigated in Ref.~\cite{Tanji:2011di}. 
The transition amplitude from the in-vacuum to the out-vacuum was computed in a more direct way than the method of effective action. 
That is, by using explicit solutions for the equations of motion 
of gluons and ghosts, the transition amplitude was computed as an overlap between the obtained wave functions. 
Then, the result was obtained as \cite{Tanji:2011di}
\begin{equation}
\label{eq:canonical-EB}
\gamma^{}_{\rm gluon}({\mathcal E}, {\mathcal B} )
= \frac{N_c}{8\pi^2} g^2
{\mathcal E}{\mathcal B}\sum_{\sigma=\pm}\sum_{n=0}^\infty 
\ln \left[1+ \e^{-\pi (2n+1-2\sigma)\frac{{\mathcal B}}{{\mathcal E}}}\right]\, ,
\end{equation}
where $\sigma$ and $n$ denote the gluon polarization ($\sigma=\pm 1$) and the Landau level, respectively. Notice that the lowest level ($n=0$ and $\sigma=+1$) has an exponential enhancement factor $e^{+\pi {\mathcal B}/{\mathcal E}}$, while all the other levels have exponential suppression factors. 
The difference between the enhancement and suppression factors 
stems from the sign of the squared energy that 
has a minus sign in the unstable mode. 
The above result implies that the gluon production 
is significantly enhanced by the NO instability. 
It was argued that the unstable modes can become 
well-defined excitation after the acceleration 
by the chromo-electric field that pushes 
the squared energy to a positive value. 


Now, we discuss whether the same result is reproduced 
from the effective Lagrangian (\ref{eq:gg-thermal}). 
We apply the Landau-level decomposition discussed below 
Eq.~(\ref{eq:K-YM-LL}) to the sum of the ghost and gluon contributions $ {\cal L}_{\rm YM}^{(1)}
=  {\cal L}_{\rm ghost}^{(1)}+{\cal L}_{\rm gluon}^{(1)}$, 
but now with a chromo-electric field. 
Sorting the decomposed result into the longitudinal 
and transverse gluon contributions as 
${\cal L}_{\rm YM}^{(1)} = {\cal L}_L^{(1)}+{\cal L}_T^{(1)} $, 
we find that 
\begin{subequations}
\label{eq:HE-Landau-YM}
\begin{eqnarray}
&&
{\cal L}_L^{(1)}  =\sum_{a=1}^{N_c^2-1}
\left[  \frac{| v^a \bbb|} {2\pi} \right] \sum_{n=0}^\infty
\left[ - \frac{i}{2\pi}\int_0^\infty \frac{ds}{s} 
e^{-i \epsilon_{\perp n}^2 s}   (v^a \aaa) \sinh(v^a \aaa s)  \right]
\, , 
\label{HE-Landau-L}
\\
&&
{\cal L}_T^{(1)}  = \sum_{a=1}^{N_c^2-1}
\left[  \frac{| v^a\bbb|} {2\pi} \right] \sum_{n=0}^\infty 
\left[ - \frac{i}{8\pi}\int_0^\infty \frac{ds}{s} 
\left( \e^{-i \epsilon_{\perp n-1}^2  s} 
+ \e^{-i  \epsilon_{\perp n+1}^2 s}  \right)  \frac{ v^a \aaa }{ \sinh(v^a \aaa s) } \right]
\, .
\label{HE-Landau-T}
\end{eqnarray}
\end{subequations}
We dropped the infinitesimal imaginary parameter 
because the proper-time integral in each Landau level 
does not have singularities on the real axis, 
except for the divergence at the origin which is common to 
the free theory. 
Notice that the integrand for the longitudinal mode 
$ {\cal L}_L^{(1)} $ is regular everywhere 
in the complex $ s$ plane, 
indicating that the longitudinal gluon mode is not produced. 
The NO mode sitting in the ground state of ${\cal L}_T^{(1)}$ 
is a tachyonic mode of which the spectrum is 
given as $ \epsilon_{\perp -1}^2 =  - |v^a \aaa|$. 
The proper-time integrals for the higher levels are 
well-defined by themselves and can be performed separately. 
The imaginary part from each higher-level contribution 
is just the same as that for the scalar QED 
(\ref{eq:scalarQED-Landau}) with the replacement of 
the energy spectrum $ \epsilon_{\perp n}^2 $. 
Performing the summation in Eq.~(\ref{eq:scalarQED-Landau}), 
one can reproduce the result in Eq.~(\ref{eq:canonical-EB}) 
from the different method.

Let us focus on the NO mode. 
The NO mode in Eq.~(\ref{HE-Landau-T}) is extracted as 
\begin{eqnarray}
{\cal L}_{\rm NO}^{(1)}  =
\left[  \frac{| v^a \bbb|} {2\pi} \right]
\left[ - \frac{i}{8\pi}\int_0^\infty \frac{ds}{s}  \e^{ i |v \bbb| s}  \frac{ v^a \aaa }{ \sinh(v^a \aaa s) } \right]
\, .
\label{HE-Landau-g-NO}
\end{eqnarray}
The issue is that the proper-time integral 
does not converge in the lower-half plane 
due to the tachyonic dispersion relation. 
One may not naively close the contour in the lower-half plane 
as we have done repeatedly (cf. Fig.~\ref{fig:contour_electric}).

One way to make the integral well-behaving is to include 
a gluon mass $ m_g $ so that the tachyonic region is excluded \cite{Marinov:1972nx, Karabali:2019ucc}. 
Namely, it is assumed that $ m_g^2 - |v^a\bbb| >0 $. 
The massless limit $ ( m_g \to 0)$ can be taken only after performing the proper-time integral. 
This looks like a somewhat {\it ad hoc} way. 
Yet, adopting this prescription, one can close the contour 
in the lower-half plane and then take the residues 
on the {\it negative} imaginary axis as in the other stable modes. 
After taking the massless limit, we find an exponential enhancement factor as \cite{Karabali:2019ucc, Hattori:2020guh} 
\begin{eqnarray}
\label{eq:NO-imag-enh}
2 \Im m {\cal L}_{\rm NO}^{(1)}  =
\lim_{m_g \to 0}
\left[  \frac{|v^a \bbb|} {2\pi} \right]
 \sum_{k=1}^\infty (-1)^{k-1} \frac{ |v^a \aaa| }{4\pi k} 
 e^{ - \frac{ m_g^2 - |v^a \bbb| }{ |v^a \aaa| } \pi k } 
 = 
\frac{ N_c}{8\pi^2 }  g^2 |\aaa\bbb|
\ln\big( 1 + e^{ \left| \frac{ \bbb }{ \aaa} \right| \pi } \big)
\, ,
\end{eqnarray}
where the color sum is again performed with 
the identity (\ref{eq:veq}). 
The sum of exponentials results in a finite logarithmic function 
even in the massless limit $ ( m_g \to 0)$ 
thanks to the alternating signs in the residues for bosons. 
Without the alternating signs, the sum would be 
divergent, or become a complex-valued logarithm.  
In the strong electric field limit $|\bbb/\aaa| \to 0 $, 
the exponentially enhanced imaginary part (\ref{eq:NO-imag-enh}) converges to a finite value.


\cout{

We note that, in the above result (\ref{eq:NO-imag-enh}), 
the ad hoc mass parameter $ m_g $ does not automatically 
go away in the end of calculation and the result depends on it 
unless we take the massless limit $(m_g\to 0)$. 
Although the regulated result has a well-behaving analytic continuation to the massless limit, 
the gauge invariance may not be guaranteed in this treatment. 
This contrasts to the proofs for the infrared-divergence cancellation where the mass parameter 
of gauge bosons automatically goes away 
after appropriate summation of diagrams (see, e.g., Ref.~\cite{Peskin:1995ev}). 
\com{Need to check the textbook.}

}

There is another possible way to 
perform the integral in Eq.~(\ref{HE-Landau-g-NO}). 
There is no mathematical obstacle 
to close the integral contour in the upper-half plane. 
In fact, there is a good reason to close the integral contour 
in the upper-half plane in the absence of chromo-electric fields. 
In Sec.~\ref{sec:gluon-condensate}, 
we will see that the QCD beta function is reproduced 
from the coefficients in the real part of the effective Lagrangian. 
The correct coefficient is obtained as the sum of coefficients 
computed with the contours closed in the upper-half plane for the NO mode 
and in the lower-half plane for the other modes \cite{Batalin:1976uv, Matinyan:1976mp}. 
Collecting the residues on the {\it positive} imaginary axis, 
we obtain 
\begin{eqnarray}
2 \Im m {\cal L}_{\rm NO}^{(1)} 
= \left[  \frac{|v \bbb|} {2\pi} \right]
 \sum_{k=1}^\infty (-1)^{k-1} \frac{ |v^a \aaa| }{4\pi k} 
 e^{-  \left| \frac{  \bbb }{ \aaa} \right| \pi k } 
\, .
 \label{imaginary_E-Landau-g-NO}
\end{eqnarray}
This result has an exponential suppression factor 
and is the same as the scalar QED result (\ref{eq:scalarQED-Landau}) up to the replacement of the effective mass by 
the Landau spacing $|v^a\bbb|  $, as opposed to 
the previous result (\ref{eq:NO-imag-enh}). 
One might wonder how the results could be the same 
as the scalar QED in spite of the big difference 
between the normal and tachyonic dispersion relations. 
One can compare the lowest-lying NO mode 
and the second lowest normal mode of which 
the dispersion relations are different only in the signs. 
In general, the imaginary part of the effective Lagrangian 
can be written as 
\begin{eqnarray}
2i \Im m {\cal L}^{(1)} 
\= \left[  \frac{| v\bbb|} {2\pi} \right]   
\frac{ - 1 }{8\pi} \sum_{n=0}^\infty \left
 \{ \,  F( \epsilon_{\perp n}^2 ) - [F(\epsilon_{\perp n}^2)]^\ast \, \right\}
\label{eq:L-imaga}
\, .
\end{eqnarray}
In each energy level, we have 
\begin{eqnarray}
F( \epsilon_{\perp n}^2 ) 
\= i \int_0^\infty \frac{ds}{s}  \e^{ - \epsilon s} 
 \e^{  - i \epsilon_{\perp n}^2 s }   \left[ \frac{ v \aaa }{ \sinh(v \aaa s) }  \right]
\, .
\end{eqnarray}
It is only a function of the squared form 
$\epsilon_{\perp n}^2 $ and not of a potentially 
complex-valued quantity $\epsilon_{\perp -1} $. 
The integral is convergent 
at the upper boundary $ (s \to \infty) $, 
and one can drop the infinitesimal parameter $  \epsilon$. 
Note that, as long as $ \epsilon_{\perp n}^2$ 
is a real-valued number, we have a simple relation 
\begin{eqnarray}
\label{eq:FFstar}
 F(\epsilon_{\perp n}^2 ) - [F( \epsilon_{\perp n}^2)]^\ast 
=  F( \epsilon_{\perp n}^2 ) + F( - \epsilon_{\perp n}^2 )
 \, .
\end{eqnarray}
The right-hand side is symmetric 
in the sign of $\epsilon_{\perp n}^2 $, 
and the imaginary part (\ref{eq:L-imaga}) may not depend on 
the sign of $ \epsilon_{\perp n}^2 $. 
This especially implies that the imaginary parts 
from the contributions of the lowest-lying NO mode 
and the second-lowest normal mode have the same exponential 
suppression factors. 
The exponential suppression of the NO mode 
contradicts the exponential enhancement 
in Eq.~(\ref{eq:NO-imag-enh}) from 
the gluon-mass prescription, which apparently looks puzzling. 
As mentioned above, there is, however, a good reason to 
close the integral contour in the upper-half plane, 
at least, in the absence of chromo-electric fields. 
This is a nontrivial suggestion for the choice of contours 
as elaborated in Sec.~\ref{sec:gluon-condensate}.

The choice of the contour seems to be yet controversial, 
while there are some other suggestions for the choice of 
the integral contours and consequent results 
\cite{Schanbacher:1980vq, Dittrich:1983ej, Elizalde:1984zv}.

\cout{

\com{\cite{Schanbacher:1980vq} claims the absence of an imaginary part in color B. 
\cite{Dittrich:1983ej} shows the same result below (37), but avoids discussing this issue in detail.
\cite{Elizalde:1984zv} claims an ambiguity in the sign of the imaginary part.}

To be more specific, one can compare the two low-lying modes at $ n = 0,1 $ 
that have the dispersion relations $ \zeta_{0,1} =  \mp |v\bbb|  $, respectively. 
The upper and lower signs are for $n=0  $ and $n=1  $, respectively. 
Then, one can compute the imaginary parts in those modes in parallel as 
\begin{eqnarray}
2i\Im m [{\cal L}_T^{(1^\prime)}  ]
\= \left[  \frac{| v\bbb|} {2\pi} \right]
\left[  \frac{ (- i)  }{8\pi} \int_0^\infty \frac{ds}{s}  e^{ \pm i |v \bbb|  s}   \frac{ v  \aaa  }{ \sinh(v \aaa s) } 
-   \frac{ (+ i)  }{8\pi}  \int_0^\infty \frac{ds}{s}   e^{ \mp i | v \bbb|  s}   \frac{ v  \aaa  }{ \sinh(v  \aaa  s) } 
 \right]
\nn
\\
\= \left[  \frac{| v\bbb|} {2\pi} \right]
 \frac{ (- i)  }{8\pi}  \int_0^\infty \frac{ds}{s} 
\left(  e^{ \pm i |v \bbb | s}  +  e^{ \mp i |v \bbb|  s} \right)  \frac{ v  \aaa  }{ \sinh(v  \aaa  s) } 
\, .
\end{eqnarray}
In the second line, the low-lying two modes have the same imaginary parts. 
This observation is consistent with the result in Eq.~(\ref{imaginary_E-Landau-g-NO}). 
It might still look somewhat peculiar to choose the different contour only in the Nielsen-Olesen mode. 
Nevertheless, in the absence of chromo-electric fields, 
one can confirm that the QCD beta function is reproduced 
from the coefficients in the real part of the effective Lagrangian. 
The correct coefficient is obtained as the sum of coefficients 
computed with the contours closed in the upper-half plane for the Nielsen-Olesen mode 
and in the lower-half plane for the other modes \cite{Batalin:1976uv, Matinyan:1976mp} 
(see Appendix A in Ref.~\cite{Ozaki:2013sfa}). 
The imaginary part in chromo-magnetic fields, in the absence of chromo-electric fields, 
is also obtained from the contour closed in the upper-half plane. 
Those facts may serve as nontrivial checks for the choice of contours.

}

\subsubsection{Quark pair production in QCD+QED fields}

We can study the interplay between the chromo-fields 
and electromagnetic fields in the quark-antiquark pair creation 
by the use of the quark-loop contribution (\ref{action_quark}) 
computed in the presence of both the QED+QCD fields \cite{Ozaki:2015yja}. 
The imaginary part reads 
\begin{eqnarray}
\gamma^{}_{q\bar{q}} = 2 \, {\Im}m\, \mathcal{L}_{\rm quark}^{(1)}
= \sum_{i=1}^{N_{c}} \sum_{f=1}^{N_{f}} \frac{\aaa_{i,f} \bbb_{i,f}}{4 \pi^{2}}
\sum_{n=1}^{\infty} \frac{1}{n}\, \e^{ - \frac{ m_{f}^{2} }{\aaa_{i,f}} n\pi }
 {\rm{coth}} \left( \frac{\bbb_{i,f}}{\aaa_{i,f}} n \pi \right).
\label{ImLq_full}
\end{eqnarray}
When we only have a purely chromo-electric field, 
the above result reduces to the formula obtained in Refs.~\cite{Claudson:1980yz,Gyulassy:1986jq} (see also Ref.~\cite{Nayak:2005pf}).

The result (\ref{ImLq_full}) allows one to study arbitrary combinations of the chromo-fields and electromagnetic fields 
including the dependences on the relative angles between the fields and the strengths of the fields. Several cases are investigated in Ref.~\cite{Ozaki:2015yja}. Of special interest is the ``mixed" configuration naturally realized in heavy-ion collisions. For example, in non-central heavy-ion collisions, strong magnetic fields are created in the perpendicular direction to the reaction plane in addition to the ``glasma'' fields (collinear chromo-EM fields with the same magnitude $|{\mathcal E}|=|{\mathcal B}|$ oriented to the beam direction). 
In Fig.~\ref{Fig:glasma}, we show the dependence of the light-quark production rate on the relative angle between the magnetic field and the glasma field. The production ratio is enhanced when the magnetic field exists in parallel/antiparallel to the glasma fields, and is decreased when the magnetic field is perpendicular to the glasma fields. 
Remember that we observed similar behaviors 
in Sec.~\ref{sec:Effects_of_magnetic_field} 
as the effects of the magnetic field
on the pair production rate in the Abelian fields. 
There, we found that the enhanced production is 
catalyzed by the Landau degeneracy. 
One can apply the Landau-level decomposition to 
the formula (\ref{ImLq_full}) as 
discussed in Sec.~\ref{sec:Sch-LL}.


\begin{figure}[t]
\begin{center}
\includegraphics[width=0.6 \textwidth]{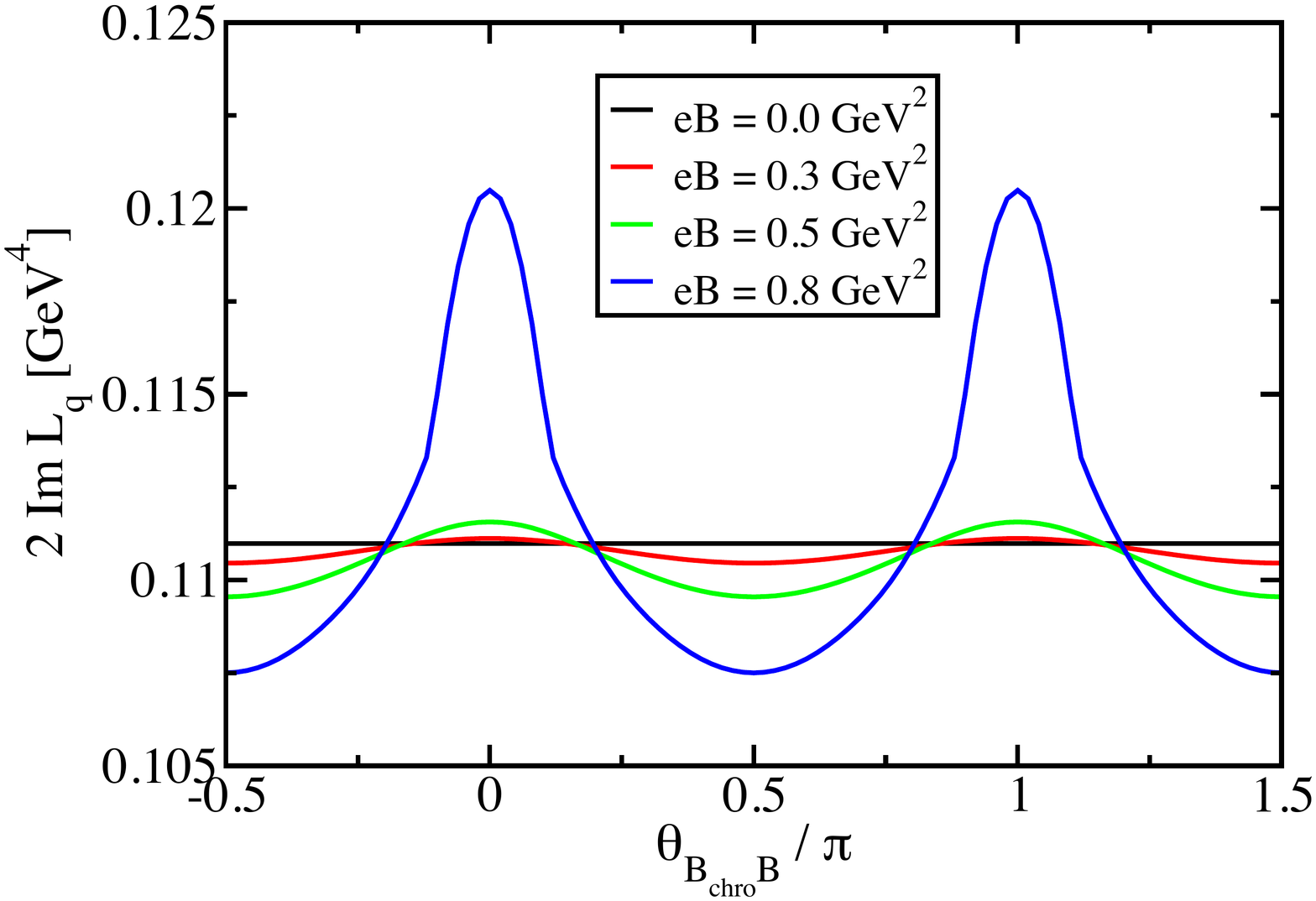}
\vspace{-1cm}
\end{center}
\caption{Light-quark production rate $2 {\Im}m \mathcal{L}_{\rm quark}^{(1)}$ in the collinear glasma fields and a QED magnetic field, 
plotted as a function of the angle between those fields. We take up quarks mass $m_q=0.5$~MeV and electric charge $q_u=+2/3 | e | $. 
The strengths of the glasma fields are taken as $g{\mathcal B}=g{\mathcal E}=1~{\rm GeV}^2$. 
}
\label{Fig:glasma}
\end{figure}

\subsection{Effective potential in QCD+QED fields: Magnetic-gluon condensation} 

\label{sec:gluon-condensate}

As we have discussed the chiral symmetry breaking in early this section, 
the effective potential provides us with insight into the vacuum state realized in a system. 
In this subsection, we discuss the QCD vacuum with the non-Abelian extended effective potential, 
and find that the QCD vacuum favors spontaneous generation of a nonzero chromo-magnetic field, 
which is often called the Savvidy vacuum~\cite{Batalin:1976uv, Savvidy:1977as, Matinyan:1976mp}. 
One finds a logarithm that induces a minimum of the effective potential at a non-zero chromo-magnetic condensation 
and see that the same logarithm gives rise to a negative beta function leading to the asymptotic freedom in QCD. 
About a half of the logarithm originates from the lowest Landau level of spin-polarized gluons in a chromo-magnetic field. 
The contribution of this lowest-lying mode also gives rise to an imaginary part of the effective action 
called the Nielsen-Olesen instability~\cite{Nielsen:1978rm}. 
We will explicitly observe those points below. 
Moreover, the asymptotic freedom is discussed as a consequence of magnetism in vacuum 
rather than the anti-screening of color charges. 

We may also discuss effects of an Abelian magnetic field on the Savvidy vacuum. 
This is related to recent lattice QCD simulations in the presence of strong magnetic fields~\cite{Ilgenfritz:2012fw, Ilgenfritz:2013ara, Bali:2013esa, Bonati:2014ksa, DElia:2015eey}, where the authors studied magnetic-field dependences of gluonic observables such as the gluon condensate and the Polyakov loop.
It has been found that the gluon condensate acquires an anisotropy induced by the external magnetic field 
and that the magnitude of the gluon condensate monotonically increases with an increasing magnetic field. 
The effective potential allows us to capture the qualitative features of those findings.

\subsubsection{Singularities in proper-time integrals}

We focus on the chromo-magnetic component of the gluon condensate. 
By taking the limit $\aaa \to 0$ in Eqs.~(\ref{eq:gg-thermal}) and (\ref{action_quark}), 
the contributions of the Yang-Mills part $ S_{\rm YM}^{(1)}  $ and the quark part $ S_{\rm quark}^{(1)} $ 
are obtained as 
\begin{subequations}
\label{action_pure_B}
\beq
S_{\rm YM}^{(1)} 
&=& -  \int \!\! d^{4}x \sum_{a=1}^{N_{c}^{2}-1} \frac{\vert v^{a}\mathcal{B} \vert}{ 16 \pi^{2} }
\int^{\infty}_{0} \frac{ds}{s^2} \e^{-\epsilon s} 
\frac{   \cos( 2 \vert v^{a}\mathcal{B} \vert s ) }{ \sin (\vert v^{a}\mathcal{B} \vert s)} 
\label{action_YM_pure_B}  
\, ,
\\
S_{\rm quark}^{(1)}
&=& \int \!\ d^{4}x \sum_{i=1}^{N_{c}} \sum_{f=1}^{N_{f}} \frac{\bbb_{i,f}}{8\pi^{2}} \int_{0}^{\infty} \frac{ds}{s^2} 
\e^{-i(m_{f}^{2}-i\epsilon)s}  {\rm{cot}}(\bbb_{i,f}s)  
\label{action_quark_pure_B}
\, . 
\eeq
\end{subequations}
Quarks are coupled to both  the chromo-magnetic field $\vec{\mathcal{B}}$ and the QED magnetic field $\vec{B}$, 
so that the invariant $\bbb_{i, f}$ is expressed with their vector sum 
\beq
\label{eq:bbb}
\bbb_{i,f}
= \sqrt{ |  w_{i} \vec{\mathcal{B}} -  q_{f} \vec{B} |^{2} } 
= \sqrt{ (w_{i} \mathcal{B} )^{2} + ( q_{f} B)^{2} - 2  w_{i}  q_{f} \mathcal{B} B \cos \theta_{\mathcal{B}B} }
\, ,
\eeq
where $\theta_{\mathcal{B} B}$ stands for the angle between $\vec{\mathcal{B}}$ and $\vec{B}$ in the coordinate space. 
As we have done in the QED case (see Sec.~\ref{sec:photons_HE-2}), 
we would possibly perform the proper-time integrals in Eq.~(\ref{action_pure_B}) 
by rotating the integral contour to the negative imaginary axis via an infinite arc in the lower-half plane. 
The effective potential is, then, obtained as $V_{\rm{eff}} = -  S_{\rm{eff}} / \int d^{4} x  $ for constant fields. 
This is still our basic strategy to get an analytic expression of the effective potential. 
However, one should note that both the integrals contain ultraviolet (UV) divergences, 
and that the integrand in the YM part contains a component 
that is not damped out on the arc in the lower-half plane. 
We shall first briefly look over these issues below.

The proper-time integrals both in Eqs.~(\ref{action_YM_pure_B}) and (\ref{action_quark_pure_B}) 
exhibit UV divergences when the lower boundary of the integral approaches the origin, $  s\to 0$. 
Expanding the integrands in the small $  s$ region, one can identify quartic and logarithmic divergences 
arising from the integrands of the forms $1/s^3  $ and $\bbb_{i,f}^2 /s $, respectively. 
The quartic divergences are a vacuum energy independent of the magnetic fields, so that we will not discuss them. 
On the other hand, the logarithmically divergent terms are proportional to 
$(v^{a}\mathcal{B})^2 \ln \Lambda$ and $ \bbb_{i,f}^2 \ln \Lambda$ (summation over $i$ and $f$ is understood) with $\Lambda$ being a UV cutoff of the proper-time integral at $ s = 1/\Lambda^2 $. 
This is a gauge-invariant regularization, sometimes called 
the proper-time regularization. 
Those logarithmic divergences induce renormalization of the coupling constants $e$ and $g  $. 
We will reproduce the beta functions in QED and QCD. 



The other issue only appears in the YM contribution (\ref{action_YM_pure_B}) 
in the asymptotic IR region, $ |s| \to \infty $. 
The asymptotic form of the integrand contains an exponentially growing factor in the lower half of the complex $s$ plane, 
and one may not naively rotate the integral contour to the negative imaginary axis. 
The origin of this singular contribution can be identified in the Landau-level decomposition of the effective Lagrangian (\ref{action_YM_pure_B-LLL}). 
The singular piece in Eq.~(\ref{action_YM_pure_B}) is nothing but the LLL contribution 
\begin{eqnarray}
 {\cal L}_{\rm YM}^{(1)\, n=0} = 
  -  \frac{i}{8\pi}  \sum_{a=1}^{N_{c}^{2}-1} \left[ \frac{ | v^{a}\nonB| }{ 2 \pi } \right]
\int_0^\infty \frac{ds}{s^2} \, \e^{ -\epsilon s + is\vert v^a \nonB \vert } 
\label{action_YM_pure_B-n0}
\, .
\end{eqnarray} 
The other all Landau levels have phases with the opposite sign 
that provide damping factors on the negative imaginary axis. 
In the next subsection, we will fully perform the proper-time integral (\ref{action_YM_pure_B}) 
with an appropriate choice of the integral contour and of the regularization for the UV divergence ($ s\to 0 $). 
In prior, it is instructive to find an alternative expression of the integral 
and to see emergence of an imaginary part. 
To this end, we go back to the Landau-level representation (\ref{eq:K-YM-LL}), 
and perform the proper-time integral first, leaving the longitudinal-momentum integral. 
After introducing a UV regularization (see Appendix~\ref{sec:energy-shift}), we obtain 
\begin{eqnarray}
\label{eq:energy-shift}
 {\cal L}_{ \YM}^{(1)} =  \sum_{a=1}^{N_{c}^{2}-1}
 \frac{ | v^a  \nonB |}{2\pi} \sum_{n=0}^\infty   \frac12 (2 -\delta_{n0} - \delta_{n1} ) 
\int \frac{dp_z}{2\pi}  \sqrt{ p_z^2 +  (2n-1) \vert v^a \nonB\vert  + i \epsilon}
\, .
\end{eqnarray} 
This expression is nothing but the (divergent) vacuum energy in the Landau levels 
and is anticipated by construction of the effective action. 
While the integrands are real-valued for $ n \geq 1 $, it is evident that 
the lowest-lying mode ($ n=0 $) has an imaginary contribution when $ p_z^2 <\vert v^a \nonB|  $. 
The imaginary part of the integral can be straightforwardly computed as\footnote{
Here, the square root is a multi-valued function on the Riemann sheet. 
However, we get a unique value without the ambiguity in Eq.~(\ref{eq:NO-imag3}) 
if we focus on the relevant IR regime from the beginning (see Appendix~\ref{sec:energy-shift}). 
} 
\begin{eqnarray}
\label{eq:NO-imag2}
 \Im m [ {\cal L}_{ \YM}^{(1) n=0} ]
=  \sum_{a=1}^{N_{c}^{2}-1} \frac{ | v^a  \nonB |}{2\pi} \cdot \frac12  
 \int_{-\sqrt{ | v^a \nonB| }} ^{\sqrt{|v^a \nonB| }}  
\frac{dp_z}{2\pi}  \sqrt{ |v^a \nonB| - p_z^2 }  
=  \sum_{a=1}^{N_{c}^{2}-1}  \frac{ | v^a  \nonB |}{2\pi}  \cdot  \frac{ 1}{ 8} |v^a \nonB|
\, .
\end{eqnarray} 
The emergence of this imaginary part 
in a chromo-magnetic field was originally 
shown by Nielsen and Olesen \cite{Nielsen:1978rm}. 
The negative value of the energy square originates from the Zeeman shift for spin-1 particles 
which is induced by the self-interactions among gauge bosons inherent in non-Abelian gauge theories. 
The NO instability does not occur in QED.\footnote{ 
Considering composite particles interacting with an Abelian magnetic field, 
``$ \rho $-meson condensation'' was put forward according to 
an analogous spin-1 Zeeman shift of the charged $  \rho$ meson spectrum. 
This observation is based on effective models of QCD \cite{Chernodub:2010qx, Chernodub:2011mc} 
and a quenched two-color lattice QCD simulation~\cite{Braguta:2011hq}. 
It has been pointed out, however, that the $ \rho $-meson condensation conflicts with the Vafa-Witten theorem 
and that the QCD inequality provides a lower bound of $ \rho $-meson mass 
by the ``connected'' neutral pion mass~\cite{Hidaka:2012mz}. 
An indication of the $  \rho$-meson condensation has not been seen 
in recent quenched lattice QCD simulations~\cite{Hidaka:2012mz,Luschevskaya:2015bea, Luschevskaya:2016epp, Bali:2017ian} (see more discussions in Sec.~\ref{sec:MC-lattice}).  
}

Bearing those singularities in the UV and IR regions in mind, we now evaluate the proper-time integrals 
for the Yang-Mills part (\ref{action_YM_pure_B}) 
and then for the quark part (\ref{action_quark_pure_B}) in order.

\subsubsection{Computation of effective Lagrangians}

We first need to regularize the UV divergences in the YM part (\ref{action_YM_pure_B}). 
To focus on the relevant integral, we define a dimensionless integral as 
\begin{eqnarray}
\Lag_{\rm YM}^{(1)} 
&=& -  \sum_{a=1}^{N_{c}^{2}-1} \frac{\vert v^{a}\mathcal{B} \vert}{ 16 \pi^{2} } I_{\rm YM}^{(1)}  
\, ,
\nnb
I_{\rm YM}^{(1)}  
&:= & \vert v^{a}\mathcal{B} \vert^{-\delta}
\mu^{2\delta} \left(  \frac{ -  i }{ 4\pi } \right)^{-\delta}
\int^{\infty}_{0} \frac{d s' }{s^{\prime \, 2-\delta} } \e^{-\epsilon s'} 
\frac{   \cos( 2  s' ) }{ \sin  s' }  
\label{eq:I-reg}
\, ,
\end{eqnarray}
where the integral variable has been scaled as $ s \to s' = \vert v^{a}\mathcal{B} \vert s $. 
Here, we have already introduced the dimensional regularization, 
and the integral $ I_{\rm YM}^{(1)}   $ is convergent on the both boundaries 
thanks to the displacement $ \delta $ as well as $  \epsilon$. 
The overall factor of $  \mu^{2 \delta}  $ is inserted to compensate
the mass dimension as before (cf. Sec.~\ref{sec:photons_HE-2}). 
We only applied the displacement to the longitudinal dimension $ d_\para = 2 - 2 \delta $, 
because the transverse momentum does not appear in the dispersion relation and the denominator of the integrand; 
The transverse-momentum integral just results in the Landau degeneracy factor. 
The UV divergences will be isolated as singular terms in the limit $\delta \to 0  $. 
As explained in Appendix~\ref{sec:integral-YM}, it is important to keep track of the overall factors, 
especially the imaginary unit $ (-i)^\delta $. 
Such a factor provides a finite contribution of order $ \delta^0 $ when combined with the divergent integral. 

The second crucial step is isolation of the NO mode that diverges on the negative imaginary axis when $ s \to - i \infty $. 
We decompose the integral as 
\begin{eqnarray}
I_{\rm YM}^{(1)} = I_{\rm NO}^{(1)} + I_{\rm res}^{(1)} 
\, ,
\end{eqnarray}
where the integrals for the NO mode and the residual all modes are given by [see Eq.~(\ref{action_YM_pure_B-n0})] 
\begin{subequations}
\begin{eqnarray}
\label{eq:INT-NO}
I_{\rm NO}^{(1)} 
&=&
\left( \frac{ 4\pi i \mu^2 }{ \vert v^{a}\mathcal{B} \vert } \right)^{\delta} 
\int^{\infty}_{0} \frac{ds}{s^{2-\delta} } \e^{-\epsilon s}  i \, e^{ is } 
\, ,
\\
\label{eq:INT-res}
I_{\rm res}^{(1)} 
&=& \left( \frac{ 4\pi i \mu^2 }{ \vert v^{a}\mathcal{B} \vert } \right)^{\delta} 
\int^{\infty}_{0} \frac{ds}{s^{2-\delta} } \e^{-\epsilon s} 
\left[ \, \frac{   \cos( 2  s ) }{ \sin  s }  - i \, e^{ is } \right]
\, .
\end{eqnarray}
\end{subequations}
Note again that both integrals are well-defined ones without divergences, 
and also that the integral for the NO mode $ I_{\rm NO}^{(1)} $ is convergent on the positive imaginary axis. 
Then, one can perform each integral straightforwardly as described in Appendix~\ref{sec:integral-YM}. 
It should be emphasized that an imaginary part only appears from the integral $ I_{\rm NO}^{(1)}  $ for the NO mode 
when we carefully include the imaginary unit $ (-i)^\delta $ in Eq.~(\ref{eq:I-reg}). 
Combining the contributions from the NO mode and the residual all modes, 
we obtain the effective Lagrangian for the Yang-Mills part~\cite{
Batalin:1976uv, Savvidy:1977as, Leutwyler:1980ma, Dittrich:1983ej, Elizalde:1984zv, Ozaki:2013sfa} 
(see also Ref.~\cite{Ozaki:2015yja} for an extension to finite temperature)
\begin{eqnarray}
\label{eq:L-YM-div}
\Lag_{\rm YM}^{(1)} 
=   \frac{1}{ 96 \pi^{2} } \sum_{a=1}^{N_c^2-1} \vert v^a \mathcal{B} \vert^2 
 \left[ \,  11 \kappa  \Big( \frac{\mu^2}{ 2 \vert v^a \nonB \vert  }  \Big)+ \ln 2  +  12 ( 1 -   \ln G ) \, \right]
+ i N_c  \frac{ \vert g \mathcal{B} \vert^2 }{ 16 \pi }
\, ,
\end{eqnarray}
where $ G $ is the Glaisher constant $ G = 1.2824 \dots $ \cite{RiemannZeta}. 
The imaginary part agrees with the previous result foreseen in the explicit vacuum-energy computation (\ref{eq:NO-imag2}). 
We defined a UV-divergent quantity 
\begin{eqnarray}
\label{eq:kappa-UV}
\kappa(x) \equiv \left( \,  \frac{1}{\delta} - \gam_E + \ln (4\pi) \,  \right) + \ln x
\, ,
\end{eqnarray}
with $  \gam_E$ being the Euler-Mascheroni constant. 
The first three terms between the brackets are the familiar combination to be subtracted in the $\MSbar$ scheme. 
Also, we find a logarithmic dependence on the chromo-magnetic field normalized by an arbitrary energy scale $ \m $.  
Both the integrals $ I_{\rm NO}^{(1)} $ and $ I_{\rm res}^{(1)} $ yield 
the terms proportional to  $ \kappa $ accompanied by numerical factors of $ 1 $ and $5/6  $, respectively. 
Namely, about a half of the total divergence and logarithm 
stem from the single lowest-lying mode.

The quark contribution (\ref{action_quark_pure_B}) contains interactions 
with both the chromo- and Abelian magnetic fields. 
In the absence of a chromo-magnetic field ($g\mathcal{B} =0$), 
it reduces to the HE Lagrangian examined in Sec.~\ref{sec:photons_HE-2} 
(up to a color factor)~\cite{Salam:1974xe, Dittrich:1975au, Dittrich:1978fc}. 
An extension to the effective Lagrangian in a chromo-magnetic field 
was investigated in Refs.~\cite{Dittrich:1983ej, Elizalde:1984zv}. 
The complete form in the coexistent QCD$ + $QED fields was obtained in Ref.~\cite{Ozaki:2013sfa} 
and was further extended to finite temperature in Ref.~\cite{Ozaki:2015yja}. 
Here, we recapitulate the calculation starting with Eq.~(\ref{action_quark_pure_B}) 
instead of the previous expression~(\ref{eq:X0-reg}) in which the UV divergences had been subtracted. 
Similar to the YM part, we use the dimensional regularization, and the regularized form of the proper-time integral reads 
\begin{eqnarray}
\Lag_{\rm quark}^\one 
\=  \sum_{i=1}^{N_{c}} \sum_{f=1}^{N_{f}} \frac{\bbb_{i,f}^{2}}{8\pi^2}  I_{\rm quark}^{(1)} 
\, ,
\nn
\\
\label{eq:I-Q}
 I_{\rm quark}^{(1)} &:=& \frac{ \mu^{2\delta} }{\bbb_{i,f}^\delta} (-1)^\delta  \left(  \frac{  i }{ 4\pi } \right)^{-\delta}
\int^{\infty}_{0} \frac{ds}{s^{2-\delta} } \e^{- i ( m^2 - i\epsilon)s} \cot( \bbb_{i,f} s)
\, .
\end{eqnarray}
To get the correct result with the dimensional regularization, 
it is again important to keep track of the overall factors (see Appendix~\ref{sec:integral-Q}). 
We can perform the remaining integral straightforwardly as done in Sec.~\ref{sec:photons_HE-2}). 
As one can explicitly confirm in Eq.~(\ref{eq:I-Q}) of Appendix~\ref{sec:integral-Q}, 
the imaginary units are completely canceled out in the quark part. 
Then, we arrive at a real-valued result 
\begin{eqnarray}
\label{eq:L-quark-div}
\Lag_{\rm quark}^{(1)} 
\=
 \sum_{i=1}^{N_{c}} \sum_{f=1}^{N_{f}} \frac{\bbb_{i,f}^{2}}{8\pi^2} 
\left[
- \big( \frac13 + 2  \bar \bbb_{i,f}^{-2} \big) \Big\{  \kappa \Big( \frac{ \mu^2 }{ m_f^2 } \Big) + 1 \Big\}
 +4  \zeta (-1, \bar \bbb_{i,f}^{-1}  ) \ln \bar \bbb_{i,f}^{-1}  + 4 \zeta'(-1, \bar \bbb_{i,f}^{-1}  )
 \right]
\, ,
\end{eqnarray}
where $  \bar \bbb_{i,f} \equiv 2  \bbb_{i,f} /m_f^2 $, 
and  $ \zeta'(z,a) \equiv \partial  \zeta(z,a)/\partial z  $ and $ \zeta(-1,a) = - ( 2a^2 - 2a + 1/3) /4 $ 
as mentioned below Eq.~(\ref{eq:X0-result}).

Note that the finite piece in Eq.~(\ref{eq:L-quark-div}) does not vanish in the vanishing field limit $ \bar \bbb_{i,f} \to 0  $, 
whereas the previous result (\ref{eq:X0-result}) vanishes in the vanishing (Abelian) field limit. 
This is just because a finite integral was previously defined with the subtraction of the free Lagrangian. 
After subtraction of the finite component in the vanishing field limit $ \bar \bbb_{i,f} \to 0  $, 
the finite piece in Eq.~(\ref{eq:L-quark-div}) agrees with the previous result (\ref{eq:X0-result}) 
up to differences in the color and flavor indices and the associate summation [see Eq.~(\ref{eq:V-quark}) below].

\subsubsection{Vacuum paramagnetism and the asymptotic freedom}

\label{sec:vacuum_magnetism}

We have obtained the one-loop quantum corrections to the effective Lagrangian 
in Eqs.~(\ref{eq:L-YM-div}) and (\ref{eq:L-quark-div}) with the UV divergences isolated in $ \kappa $. 
In this subsection, we perform the renormalization of the QED and QCD coupling constants 
and discuss an interpretation of the asymptotic freedom in terms of the vacuum magnetism.

To perform the renormalization, we shall go back to the original QCD action (\ref{QCDaction}) 
which we understand is written with the bare fields and coupling constants. 
As usual, we split the bare fields into the renormalized finite parts and residual divergent parts. 
Accordingly, the classical Lagrangian is split into the same form of the classical Lagrangian, 
which is however written with the renormalized fields and coupling constants, 
and the residual divergent terms, i.e., the counterterms (see, e.g.,  Ref.~\cite{Peskin:1995ev}). 
Now, it is understood that the divergent one-loop quantum corrections (\ref{eq:L-YM-div}) and (\ref{eq:L-quark-div}) 
have been computed by starting from the renormalized classical Lagrangian. 
Adding the one-loop quantum corrections to the classical Maxwell and Yang-Mills parts 
(at vanishing electric components), we find 
\begin{eqnarray}
\Re e[\Lag^\one_\eff (B, \nonB)  ]  \= \Lag_{\rm classical} + \Re e[  \Lag_{\rm YM}^{(1)} ] + \Lag_{\rm quark}^{(1)} 
\nnb
\=   -\frac12 (1+\delta_\nonA)  \mathcal{B}^2  - \frac12 (1+\delta_A) B^2
\label{eq:Lag-div}
\\
&&
+  \frac{11}{ 96 \pi^{2} } \sum_{a=1}^{N_c^2-1} \vert v^a \mathcal{B} \vert^2 
 \kappa  \Big (\frac{\mu^2}{ 2 \vert v^a \nonB \vert  }  \Big) 
- \frac{ 1 }{24\pi^2}  \sum_{i=1}^{N_{c}} \sum_{f=1}^{N_{f}} \bbb_{i,f}^{2} 
\kappa  \Big( \frac{\mu^2}{ m_f^2  }  \Big)
- ( \Re e [ U^\one_\YM  ] + U^\one_\quark )
\, .
\nn
\end{eqnarray}
We do not discuss the imaginary part of $  \Lag_{\rm YM}^{(1)} $ 
since it takes a finite value and does not have any UV divergence. 
All the fields and coupling constants in the above expression are renormalized finite ones 
and the counterterms are proportional to divergent parameters $ \delta_A , \, \delta_\nonA $. 
The divergent terms proportional to either $ |v^a \nonB|^2 $ or $\bbb_{i,f}^2  $ 
are absorbed by those counterterms equipped in the above expression. 
Since the one-loop effective action does not contain radiative corrections to 
the fermion mass and field strength or the vertex form factor, 
we have suppressed irrelevant counterterms for renormalization of those quantities. 
We have discarded the term $\propto m_f^4  \kappa( \frac{\mu^2}{ m_f^2 })$ 
that originates from the one-loop diagram without any external-field insertion. 
The residual finite parts are given as 
\begin{subequations}
\begin{eqnarray}
&&
U^\one_\YM (B, \nonB) =
-  \frac{1}{ 96 \pi^{2} } \sum_{a=1}^{N_c^2-1} \vert v^a \mathcal{B} \vert^2 
 \left[ \,  \ln 2  +  12 ( 1 -   \ln G ) \, \right]
 \, ,
\\
&&
U^\one_\quark  (B, \nonB) = 
- \sum_{i=1}^{N_{c}} \sum_{f=1}^{N_{f}} \frac{\bbb_{i,f}^{2}}{8\pi^2} 
\left[    - \big( \frac13 + 2  \bar \bbb_{i,f}^{-2} \big) 
 +4  \zeta (-1, \bar \bbb_{i,f}^{-1}  ) \ln \bar \bbb_{i,f}^{-1} 
 + 4 \zeta'(-1, \bar \bbb_{i,f}^{-1}  ) \right]
 \, .
\end{eqnarray}
\end{subequations}
Notice that the YM part $ U^\one_\YM (B, \nonB) $ has only a quadratic dependence on $  \nonB$ 
in spite of the all-order resummation for the external-field insertion. 
A simple reason is the massless nature of the YM part which enabled us to factorize 
all the $  \nonB$ dependences out of the proper-time integral (\ref{eq:I-reg}). 
In contrast, the magnetic-field dependences in the quark part $  U^\one_\quark  (B, \nonB)$ 
cannot be factorized due to the exponential mass dependence in the proper-time integral (\ref{eq:I-Q}), 
resulting in an intricate dependence seen in the above.

Explicit forms of the counterterms in the $ \MSbar $ scheme 
should be given as 
\begin{subequations}
\label{eq:counterterms}
\begin{eqnarray}
\delta_\nonA \= 2 \left[ \, -  \frac{11}{ 96 \pi^{2} } \sum_{a=1}^{N_c^2-1} \vert v^a \vert^2 
 \kappa  \Big (\frac{\Lambda^2}{ 2 \vert v^a \mathcal{B} \vert  }  \Big)  
 +  \frac{ g^2 }{48\pi^2}   \sum_{f=1}^{N_{f}} \kappa  \Big( \frac{\Lambda^2}{ m_f^2  }  \Big) \, \right]
 \, ,
 \\
 \delta _A \=  2  \left[ \,  \frac{ N_c }{24\pi^2}   \sum_{f=1}^{N_{f}} q_f^2
\kappa  \Big( \frac{\Lambda^2}{ m_f^2  }  \Big) \, \right]
\, ,
\end{eqnarray}
\end{subequations}
where the summation over the color indices is performed by the use of the identities (\ref{eq:w-ids}) and (\ref{eq:v-ids}). 
The UV divergences have split into the terms proportional to $ \nonB^2 $ and $B^2  $ 
since the cross terms in $\bbb_{i,f}^2  $ vanish in the color  summation. 
With those counterterms, the effective Lagrangian (\ref{eq:Lag-div}) reads 
\begin{eqnarray}
\Re e[\Lag^\one_\eff (B'/e, \nonB'/g)  ]  
\=   - \left[ \, \frac{1}{g^2} + \frac{ 1 }{24\pi^2} \left( \,   \frac{11}{ 2 } N_c  -  N_f  \, \right)  \ln  \Big( \frac{ \Lambda^2  }{\mu^2}  \Big)
 \, \right] \mathcal{B'}^2  
\nnb
&& -   \left[ \, \frac{1}{e^2} - \frac{ 1 }{12\pi^2} N_c \ln  \Big( \frac{ \Lambda^2  }{\mu^2}  \Big)  \sum_{f=1}^{N_{f}} \bar  q_f^{2}  \, \right] 
B^{\prime 2}
\label{eq:Lag-renom}
- ( \Re e [ U^\one_\YM  ] + U^\one_\quark )
\, ,
\end{eqnarray}
where $ \bar q_f : = q_f/e $. 
We introduced an arbitrary energy scale $  \Lambda$. 
This floating energy scale divides the logarithmic terms into two pieces 
included in the counterterm and the rest of the effective Lagrangian (\ref{eq:Lag-renom}). 
We rescaled the magnetic-field strengths as $ B \to B' = eB $ and $ \nonB \to \nonB' = g \nonB  $. 
Thanks to this manipulation, one can isolate the coupling constants into the classical part of the effective Lagrangian. 
The quantum corrections then do not have an explicit dependence on the coupling constants, 
and can be absorbed into effective coupling constants as 
\begin{subequations}
\begin{eqnarray}
\label{eq:running-QCD}
\frac{1}{ g^2(\Lambda)} &:=&  \frac{1}{g^2} 
+  \frac{ 1 }{24\pi^2} \left( \,   \frac{11}{ 2 } N_c  -  N_f  \, \right)  \ln  \Big( \frac{ \Lambda^2  }{\mu^2}  \Big)
\, ,
  \\
  \frac{1}{ e^2(\Lambda)} &:=&  \frac{1}{e^2} 
  -   \frac{ 1 }{12\pi^2} N_c \ln  \Big( \frac{ \Lambda^2  } {\mu^2} \Big) \sum_{f=1}^{N_f}  \bar q_f^2
  \, ,
\end{eqnarray}
\end{subequations}
where the magnitudes of the quantum corrections depend on the energy scale $ \Lambda $. 
The coupling constants $ g $ and $ e $ on the right-hand sides should be 
understood as those at the scale $ \Lambda = \mu $. 
Then, the beta functions are found to be\footnote{
We take a positive value for unit electric charge $ e>0 $, and the flavor-dependent signs 
of electric charges are included in a numerical coefficient $ \bar q_f  $ below. 
} 
\begin{subequations}
\begin{eqnarray}
\beta^\one_\QCD &=&
\Lambda \frac{\pd g(\Lambda)}{\pd\Lambda} =   -    \frac{ g^3 }{24\pi^2} \left( \,   \frac{11}{ 2 } N_c  -  N_f  \, \right) 
\, ,
  \\
\beta^\one_\QED &=&
  \Lambda \frac{\pd e(\Lambda)}{\pd\Lambda} =      \frac{   e^3 }{12\pi^2} N_c  \sum_{f=1}^{N_f}  \bar q_f^2
  \, .
\end{eqnarray}
\end{subequations}
As clear in the above derivation, 
the one-loop beta functions are independent of 
the finite terms in the counterterms (\ref{eq:counterterms}), i.e., independent of  the renormalization scheme. 
While the QED beta function $\beta^\one_\QED  $ is positive definite, 
the QCD beta function $ \beta^\one_\QCD $ can take a negative value, leading to the asymptotic freedom 
when $ N_f /N_c < 11/2 $.

\cout{

While the QED beta function $\beta^\one_\QED  $ is positive definite, 
the QCD beta function $ \beta^\one_\QCD $ can take a negative value, leading to the asymptotic freedom 
when $ N_f /N_c < 11/2 $.  
The one-loop effective coupling constant diverges when 
the right-hand side in Eq.~(\ref{eq:running-QCD}) approaches 
the QCD scale 
\begin{eqnarray}
\label{eq:QCD_scale}
\Lambda_\QCD^2 \= \mu^2  e^{ - \frac{4\pi }{ \beta_0 \alpha_s(\mu) }}
\, ,
\end{eqnarray}
where $ \beta_0 := -  (4\pi)^2 \beta^\one_\QCD/g^3 = (  11 N_c  - 2 N_f  )/3$ and $ \alpha_s = g^2/(4\pi) $. 
By the use of the QCD scale, the effective coupling constant (\ref{eq:running-QCD}) can be rewritten 
in a compact form 
\begin{eqnarray}
 \alpha_s(\Lambda) \= \frac{ 4\pi  }{ \beta_0  \ln (  \Lambda^2 /  \Lambda_\QCD^2  )  }
 \, .
\end{eqnarray}
The effective coupling constant, and the QCD scale, depends on the renormalizaton scheme 
although its derivative, i.e., the one-loop beta function, 
does not. 

}

We shall discuss an interpretation of the asymptotic freedom 
as a consequence of the paramagnetism in the Yang-Mills theory. 
As mentioned in the beginning of this section, 
the charge antiscreening effect $(\epsilon < 1  )$ and the paramagnetism $ (\mu_m>1) $ implies each other 
under the constraint of the Lorentz symmetry $ \epsilon \mu_m = 1 $, 
and one may attribute the asymptotic freedom to a consequence of 
the vacuum paramagnetism~\cite{
Hughes:1980ms, Nielsen:1980sx, Hughes:1981nw, RevModPhys.77.837} 
(see also Ref.~\cite{Grozin:2008yd} for a concise review on the asymptotic freedom in QCD). 
They are the susceptibilities at the vanishing field limit, 
and the Lorentz symmetry can be assumed. 
For free particles, the magnetic susceptibility has two 
contributions, which are spin polarization effect 
known as the Pauli paramagnetism 
and the orbital magnetic moment induced by the circular current, i.e., the Landau diamagnetism~\cite{Landau1930}. 
Therefore, the one-loop beta functions are determined 
as a consequence of competition between the vacuum energy shifts by these two effects.


\begin{itemize}

\item[]
{\it Scalar QED}.---Without a spin contribution, one can anticipate to have the Landau diamagnetism. 
There are no photon-polarization contributions to the magnetism in QED 
without self-interactions among photons. 
The vacuum diamagnetism $ (\mu_m <1) $ implies a charge screening ($ \epsilon >1 $) and a positive beta function 
as it is indeed the case in scalar QED.

\item[]
{\it Spinor QED}.---For particles with nonzero spin, one finds another contribution from the Pauli paramagnetism. 
As well-known, the magnitude of the Pauli paramagnetism is three times larger 
than that of the Landau diamagnetism in spinor QED. 
Then, one may wonder if the total contribution induces a vacuum paramagnetism (as opposed to what we know about QED). 
However, the fermion contributions to the {\it vacuum} energy should in general have 
an overall minus sign originating from the fermionic statistics. 
The energy gap between the positive- and negative-energy states 
increases due to the negative energy shift of the filled states 
in the Dirac sea. 
With this sign flip, the vacuum diamagnetism still holds in spinor QED \cite{Weisskopf:1939zz,Weisskopf:1996bu}. 


\item[]
{\it Yang-Mills theory}.---Spin-1 gauge bosons have an even larger magnetic moment 
and are coupled to the chromo-magnetic field due to the non-Abelian nature. 
The spin-polarization effect dominates over the Landau diamagnetism 
{\it without} an overall negative sign for the bosonic statistics. 
Therefore, the spin-polarization effect gives rise to the vacuum paramagnetism $ \mu_m >1 $. 
The vacuum paramagnetism and the asymptotic freedom is a salient and inherent feature of the Yang-Mills theory 
arising from the polarization effect of self-interacting spin-1 gauge bosons. 

\end{itemize}

This simple interpretation may suggest that, 
once an interacting spin-1 boson is allowed to exist in 
a theory (presumably as an elementary excitation), 
ubiquitous spin interaction generally leads to 
the asymptotic freedom. 
An early implication of the charge antiscreening effect was indeed observed 
with a (hypothetical) spin-1 vector boson interacting with an Abelian electromagnetic field~\cite{Vanyashin:1965ple}. 
Our effective Lagrangian is computed in the covariantly constant chromo-fields 
which also only have Abelian-like diagonal components in the color space. 
However, without a non-Abelian gauge symmetry, 
there is no asymptotic freedom in {\it renormalizable theories} in the four dimensions \cite{Coleman:1973sx}.


Based on the above discussion, one can divide each beta function into 
the current-induced and spin-polarization terms: 
\begin{subequations}
\begin{eqnarray}
\beta_{\rm QED} ^\one&=&  \frac{ e^3 }{8 \pi^2} (-1) \left( \frac{n_p}{6}
 - \left(\frac{g}{2} \right)^2 \right) N_c  \sum_{f=1}^{N_{f}} \bar q_f ^2 
\, ,
\\
\beta_{\rm QCD}^\one &=&   \frac{g^3} { 8\pi^2} 
\left[  (-1) \frac12 \left(\frac{n_p}{6}  - \left(\frac{g}{2} \right)^2 \right)  N_f 
+   \frac{N_c}{ 2 } \left( \frac{n_p}{6}  - (g\cdot 1)^2 \right)  \right]
\, ,
\end{eqnarray}
\end{subequations}
where $ n_p  $ is the number of polarization modes for quark and transverse gluons, i.e., $ n_p=2 $ 
and $ g=2  $ is the leading-order g-factor. 
The overall negative sign in the quark contribution originates from the negative-energy states as mentioned above. 
The quark contribution to the QCD beta function is a half of that in QED because of the color trace.\footnote{
Remember that the logarithmic divergences arise from the diagrams with two external legs, 
which yield the color trace $ \tr[t^a t^a] =1/2 , \ N_c$ for the fundamental and adjoint representations, respectively. 
} 
Compared to the quark contribution, the YM contribution is multiplied by $  N_c$ instead of $  1/2$, 
but a factor of $ 1/2 $ appears for another reason. 
Namely, the quark contribution includes contributions of particle and antiparticle pairs, 
and the eight gluons should be also paired into particles and antiparticles, resulting in the factor of $ 1/2 $; 
Recall that the effective color charges $  v^a$ appear in four pairs 
of positive and negative values (see Eq.~(\ref{eq:angle-adj-rep}) and Appendix~\ref{sec:adjoint}). 
Most importantly, the spin-1 polarization term gives rise to the negative YM contribution 
to the QCD beta function, i.e., the vacuum paramagnetism, as discussed above. 
One can also presume the beta function in scalar QED (without the color group): 
\begin{eqnarray}
\label{eq:beta-scalar}
\beta_{\rm scalar \ QED} ^\one&=&  \frac{e^3}{8 \pi^2} \frac{1 }{6} \sum_{f=1}^{N_{f}} \bar q_f ^2 
\, ,
\end{eqnarray}
with $ n_p=1 $ and without the spin-polarization term. Also, the overall sign is plus for bosons. 
An explicit calculation show that this is indeed the correct beta function in scalar QED (see Appendix~\ref{sec:beta-scalar}).


\subsubsection{Dislocated minimum of the effective potential}

\label{eq:chromo-B-condensate}

Having performed the appropriate renormalization in the above, the renormalized pieces provide an effective potential. 
The real part of the renormalized effective potential reads 
\beq
 V_{\rm{eff}} (B,\nonB)  \equiv \frac{\mathcal{B}^{2}}{2} + \frac{B^{2}}{2} 
+  V_{\rm{YM}}^{(1)}  (B,\nonB) +   V_{\quark}^{(1)}  (B,\nonB)
\, ,
\label{final_re_potential}
\eeq
with the one-loop corrections 
\begin{subequations}
\label{eq:V-one-loop}
\beq
V_{\rm{YM}}^{(1)}  (B,\nonB)
\=  - \frac{11}{ 96 \pi^{2} } \sum_{a=1}^{N_c^2-1} \vert v^a \mathcal{B} \vert^2  \ln \frac{\mu^2}{ 2 \vert v^a \nonB \vert  }
+  U^\one_\YM  (B, \nonB)
\nnb
\=
N_{c}   \frac{ 11   }{ 96 \pi^{2} } (g\mathcal{B})^{2} 
\bigg[ \,  \log \frac{ g\mathcal{B} }{ \mu^{2} } 
+ \frac{1}{N_{c}} \sum_{a=1}^{N_{c}^{2}-1} | \lambda_{\rm ad}^{a }|^2  \log| \lambda_{\rm ad}^{a}| 
- \frac{1}{11} ( 12 + \ln 4 - 12  \ln G ) \, \bigg]
\label{eq:V-YM}
, 
\\
V_\quark^{(1)}  (B,\nonB)
\= 
 \frac{ 1 }{24\pi^2}  \sum_{i=1}^{N_{c}} \sum_{f=1}^{N_{f}} \bbb_{i,f}^{2} \ln  \frac{\mu^2}{ m_f^2  }
+ \left[ \,  U^\one_\quark  (B, \nonB) - U^\one_\quark  (0,0)  \right]
\, ,
\nnb
\=
 -  \sum_{i=1}^{N_{c}} \sum_{f=1}^{N_{f}}  \frac{\bbb_{i,f}^{2}}{8\pi^2} \bigg[ \, 
 \frac{ 1 }{ 3}  \log\frac{ m_f^2 }{ \mu^{2} }  
  +  \bar \bbb_{i,f} ^{-2} - \frac13
+ 4   \zeta( -1, \bar \bbb_{i,f} ^{-1} )  \log \bar \bbb_{i,f}  ^{-1}
+ 4   \zeta^{\prime} ( -1, \bar \bbb_{i,f} ^{-1} ) 
\, \bigg]
\label{eq:V-quark}
\, .
\eeq
\end{subequations}
The logarithms from $ \kappa $ in Eq.~(\ref{eq:Lag-div}) are entirely included 
in the above effective potential $  V_{\rm{eff}} (B,\nonB)  $. 
One should thus note that the renormalization scheme here is different from 
that in Eq.~(\ref{eq:counterterms}) by finite values of the logarithms. 
We have used the identity (\ref{eq:veq}) to perform the summation in the YM part 
after decomposing the logarithm and expressed the result with $\lambda_{\rm ad}^a ( = v^a / g)$ defined there. 
Note also that the original forms of the one-loop corrections behave in the vanishing-field limit as 
\begin{subequations}
\begin{eqnarray}
&&
U^\one_\YM  (0,0)  =0 \, ,
\\
&&
U^\one_\quark  (0,0) = 
- \sum_{i=1}^{N_{c}} \sum_{f=1}^{N_{f}} \frac{\bbb_{i,f}^{2}}{8\pi^2} ( - 3\bar \bbb_{i,f}^{-2} )
=  \frac{ 3N_c }{ 32 \pi^2}  \sum_{f=1}^{N_{f}}    m_f^4
\, .
\end{eqnarray}
\end{subequations}
Therefore, we set the origin of the quark contribution so that 
it vanishes in this limit, i.e., $  V_\quark^{(1)} (0,0) =0 $ and $ V_{\rm eff} (0,0) =0 $. 
It is this subtracted form that agrees with the previous result in the Abelian magnetic field (\ref{eq:X0-result}). 
By construction, the effective potential (\ref{final_re_potential}) is invariant 
under the shift of energy scale $ \mu \to \mu + \delta \mu $, 
if the implicit energy-scale dependences in the running coupling constants are included.

We now discuss the minimum of the effective potential (\ref{final_re_potential}). 
First, we focus on the QCD part with a vanishing QED magnetic field, i.e., $V (\nonB, B=0 )$. 
At the classical level, the parabolic effective potential has a minimum at $ \nonB = 0 $. 
Importantly, the one-loop quantum correction contains 
the logarithmic term $ \propto (g \nonB)^2 \log(g \nonB) $ in the YM part $ V_{\rm{YM}}^{(1)}  $. 
This logarithm is convex upward at the origin $ g\nonB=0 $, 
and tends to cause an instability of the effective potential. 
On the other hand, the same logarithm tends to bound the effective potential at an asymptotically large magnitude of $ g\nonB $. 
Thus, the quantum correction in the YM part could induce a minimum of the effective potential at a finite strength of $ g{\mathcal B} $, 
implying that the QCD vacuum favors a spontaneous generation of the chromo-magnetic condensation. 
This point was first realized in Refs.~\cite{Batalin:1976uv, Savvidy:1977as, Matinyan:1976mp} 
and was tied to the negative beta function of the Yang-Mills theory as we have seen above. 
Such a vacuum state with the chromo-magnetic condensation is often called the Savvidy vacuum. 
There are various extensions, e.g., in Refs.~\cite{Pagels:1978dd, Leutwyler:1980ma, Adler:1981as, Adler:1982rk, Kapusta:1981nf, Dittrich:1983ej, Elizalde:1984zv, 
Cho:2002iv, Kondo:2004dg, Kondo:2006ih, Kondo:2013cka}. 
Contrary, the quark part does not have such a logarithmic dependence on $ g\nonB $, 
and tends to make the minimum shallower as confirmed with numerical plots below.

\begin{figure}[t]
\begin{minipage}{0.48\hsize}
\begin{center}
\includegraphics[width= \textwidth]{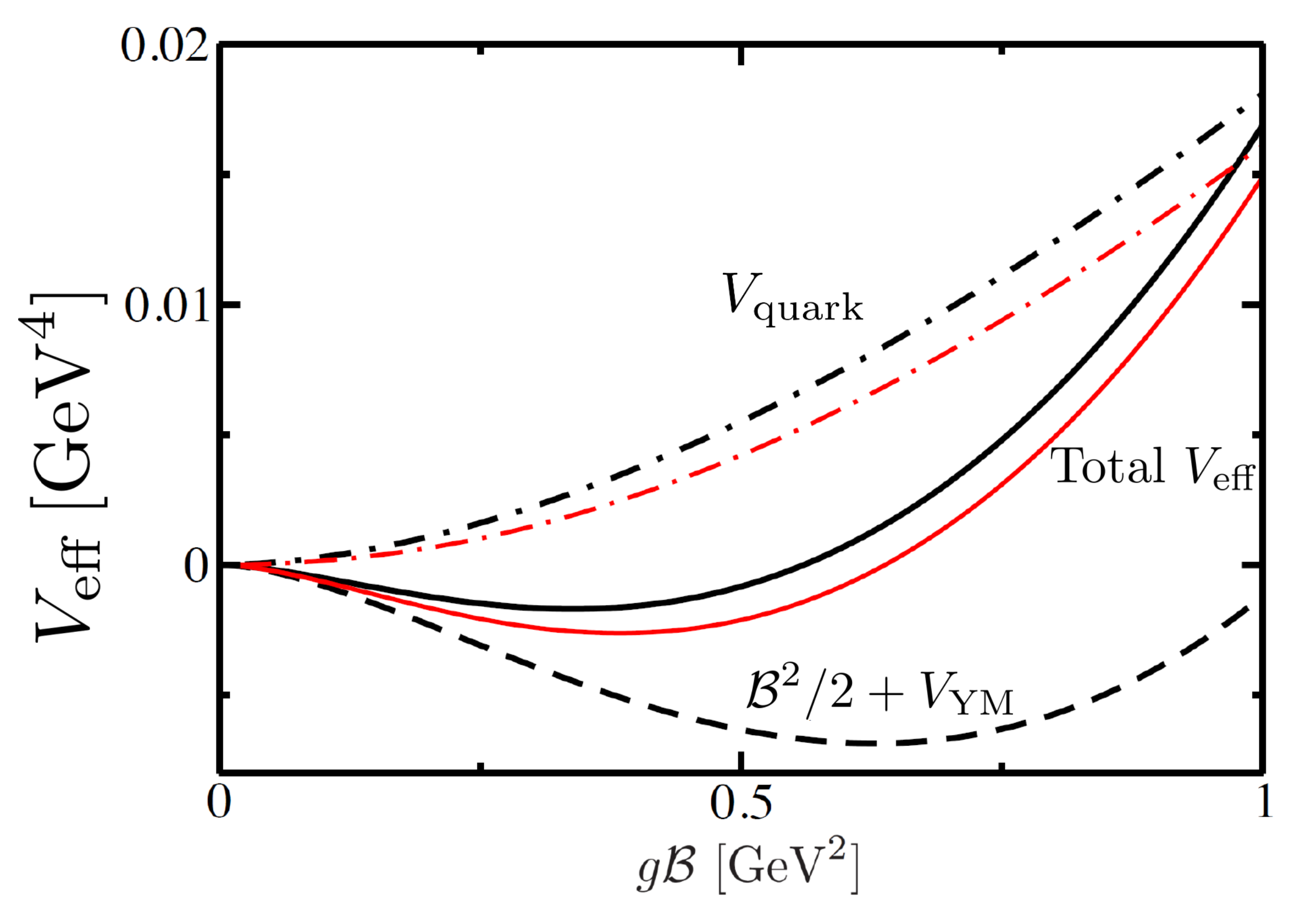}
\vspace{-0.5cm}
\caption{
YM and quark parts of the effective potential at $eB = 0 \  (0.4)$ GeV$^{2}$ 
shown with a black (red) line, as well as their sum, in the parallel configuration ($\theta_{\mathcal{B}B}$=0)~\cite{Ozaki:2013sfa}.
}
\label{Fig:breakdown-B}
\end{center}
\end{minipage}
\hspace{0.2cm}
\begin{minipage}{0.48\hsize}
\begin{center}
\vspace{-0.5cm}
\includegraphics[width=1 \textwidth]{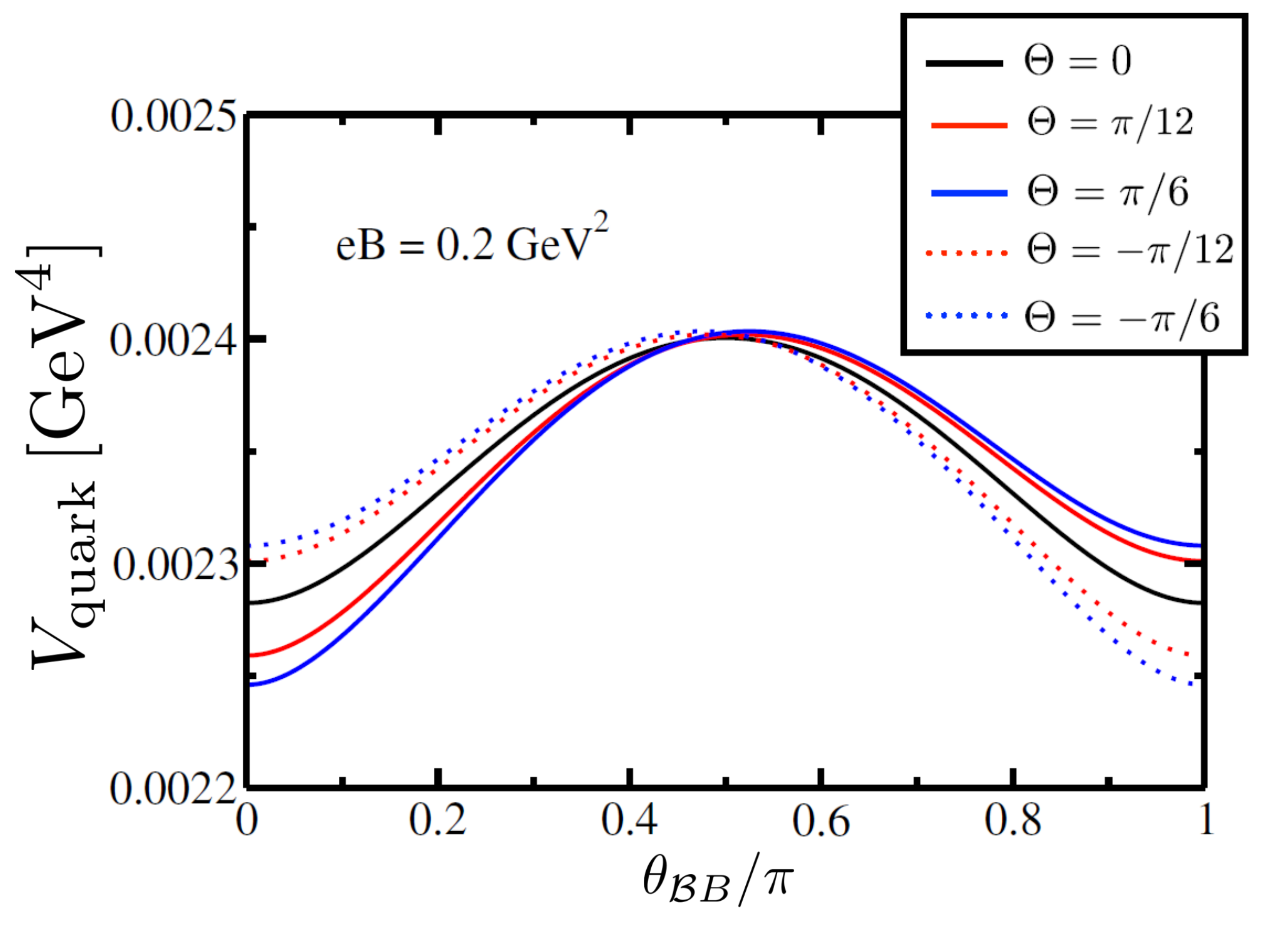}
\vspace{-1cm}
\caption{
Quark part of the effective potential $V_{\rm{quark}}$ as a function of $\theta_{\mathcal{B}B}$ 
at $g \mathcal{B} = eB = 0.2$ GeV$^{2}$~\cite{Ozaki:2013sfa}. 
Note that $ \theta = 2\pi/3 - \Theta $ in the current convention. 
}
\label{Fig:angle-dep}
\end{center}
\end{minipage}
\end{figure}

We shall confirm basic behaviors of the effective potential with numerical plots. 
We need to specify the color configuration to evaluate the remaining color summation in Eq.~(\ref{eq:V-one-loop}). 
According to Appendix~\ref{sec:cc}, we take the eigenvalues $\lambda^{i} =  \{ \pm 1/2 g ,  0\}$ 
and $\lambda_{\rm ad}^{a} =  \{\pm 1, \pm 1/2, \pm 1/2, 0, 0\}$ for $ N_c =3 $. 
The YM part (\ref{eq:V-YM}), however, depends on the color configuration only in the constant term, 
which does not change qualitative behavior of the effective potential. 
We include the three light quarks with the electric charges $q_{u} = +\frac{2}{3} |e|$ and $q_{d} = q_{s} = -\frac{1}{3} |e|$ 
and the masses $m_{u} = m_{d} = 5$ MeV and $m_{s} = 140$ MeV. 
Also, we take the strong and the electromagnetic coupling constants in such a way 
that $ g^2/(4\pi) = 1 $ and $ e^2/(4\pi) = 1 /137$ at the renormalization point $ \mu =1 $ GeV as a typical hadronic scale.

In Fig.~\ref{Fig:breakdown-B}, we plot the breakdown of the effective potential as a function of 
$ g {\mathcal B} $ in the absence of an external QED magnetic field with black lines. 
The quark part is a monotonically increasing function of $ g\nonB $, 
and tends to make the minimum of the effective potential shallower, 
competing with the logarithmic behavior in the pure YM part. 
Nevertheless, the YM part overwhelms the quark part 
and gives rise to a minimum of the effective potential at a nonzero value of $ g\nonB $.

On top of the Savvidy vacuum, one can investigate effects of an Abelian magnetic field~\cite{Ozaki:2013sfa}. 
At the one-loop order, effects of the Abelian magnetic field only enters through the quark loop (\ref{eq:V-quark}), 
and can be summarized as a replacement of $g \nonB  $ by an effective strength $ \bbb_{i,f} $ in Eq.~(\ref{eq:V-quark}). 
Therefore, the chromo-magnetic condensation would be enhanced (suppressed) 
when $ g \nonB > \bbb_{i,f} $ ($g \nonB < \bbb_{i,f} $). 
One should, however, remember that the color index $ i $ needs to be summed with a set of $ w^i$ defined in Eq.~(\ref{eq:angle-fund-rep}). 
As defined in Eq.~(\ref{eq:bbb}), the effective strength $ \bbb_{i,f} $ depends on the spatial angle $\theta_{\mathcal{B}B}$ 
and the angle $ \theta $ in the color space through $ w^i $. 
In Fig.~\ref{Fig:angle-dep}, the quark part $V_\quark$ is plotted 
as a function of $\theta_{\mathcal{B}B}$ for $g \mathcal{B} =  eB = 0.2$ GeV$^{2}$ 
and various values of the color angle $\theta  $. 
The quark part $V_\quark$ has an invariance under simultaneous discrete transformations 
$\theta_{\mathcal{B}B} \to \pi - \theta_{\mathcal{B}B}$ and $ \theta \to - \theta  $, 
because the whole set of $ \{  \bbb_{1,f}, \bbb_{2,f}, \bbb_{3,f} \} $ is invariant under this transformation. 
The angle dependences in the coordinate and color spaces are entangled with each other. 
Nevertheless, the parallel spatial configuration ($ \theta_{\nonB B} = 0 $), 
and the antiparallel one ($  \theta_{\nonB B} = \pi $) according to the above invariance, 
is favored for all the values of the color angle $ \theta $, 
suggesting that the QCD vacuum favors the parallel/antiparallel configuration 
between the chromo- and QED magnetic fields (see Ref.~\cite{Ozaki:2015yja} for more discussions). 
In other words, an anisotropy of the chromo-magnetic field 
is induced by the QED magnetic field through the interaction with the virtual quark excitations. 


\begin{figure}[t] 
\begin{minipage}{0.5\hsize}
\includegraphics[width=0.95 \textwidth]{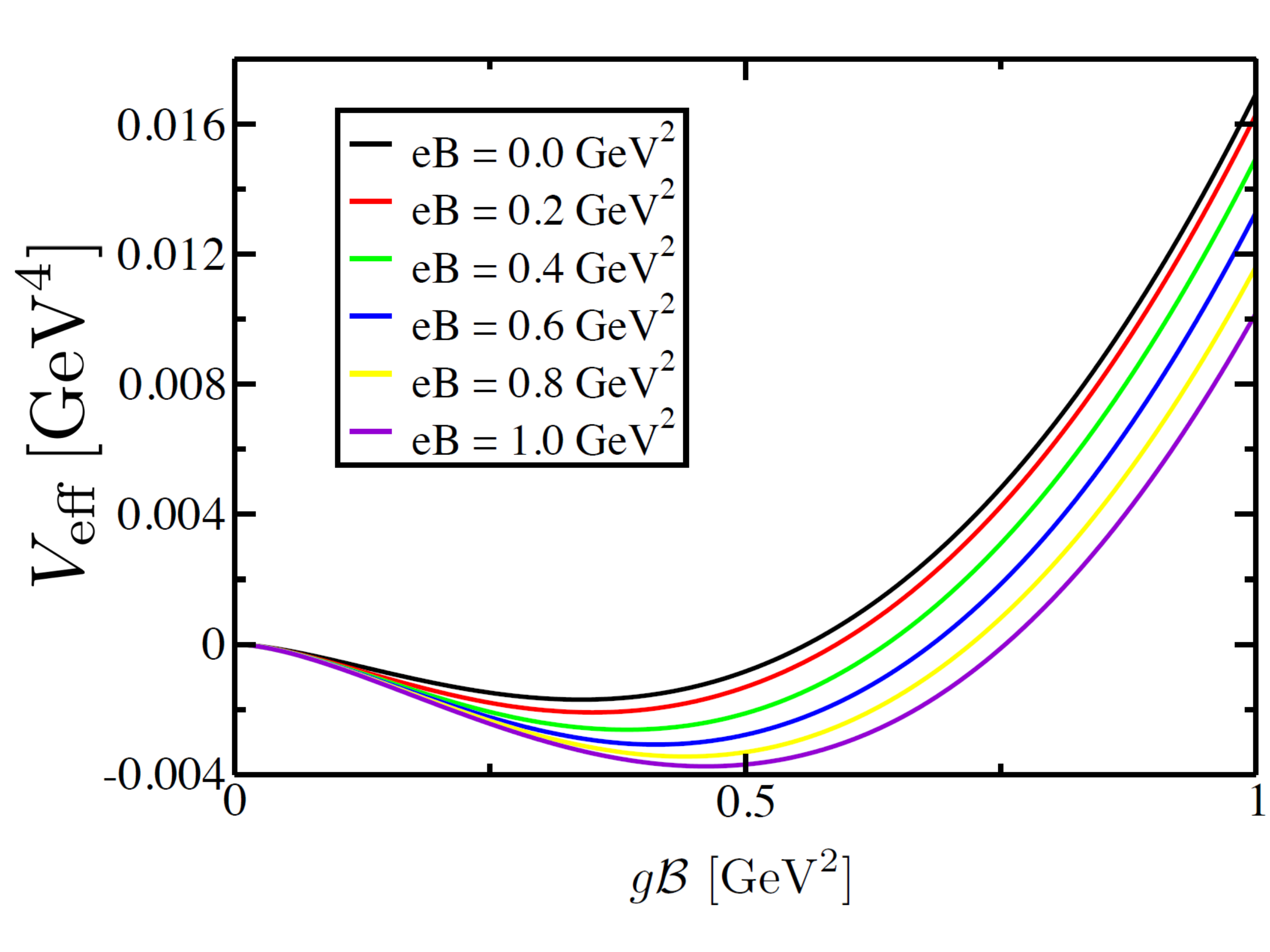}
\end{minipage}
\begin{minipage}{0.5\hsize}
\includegraphics[width=1 \textwidth]{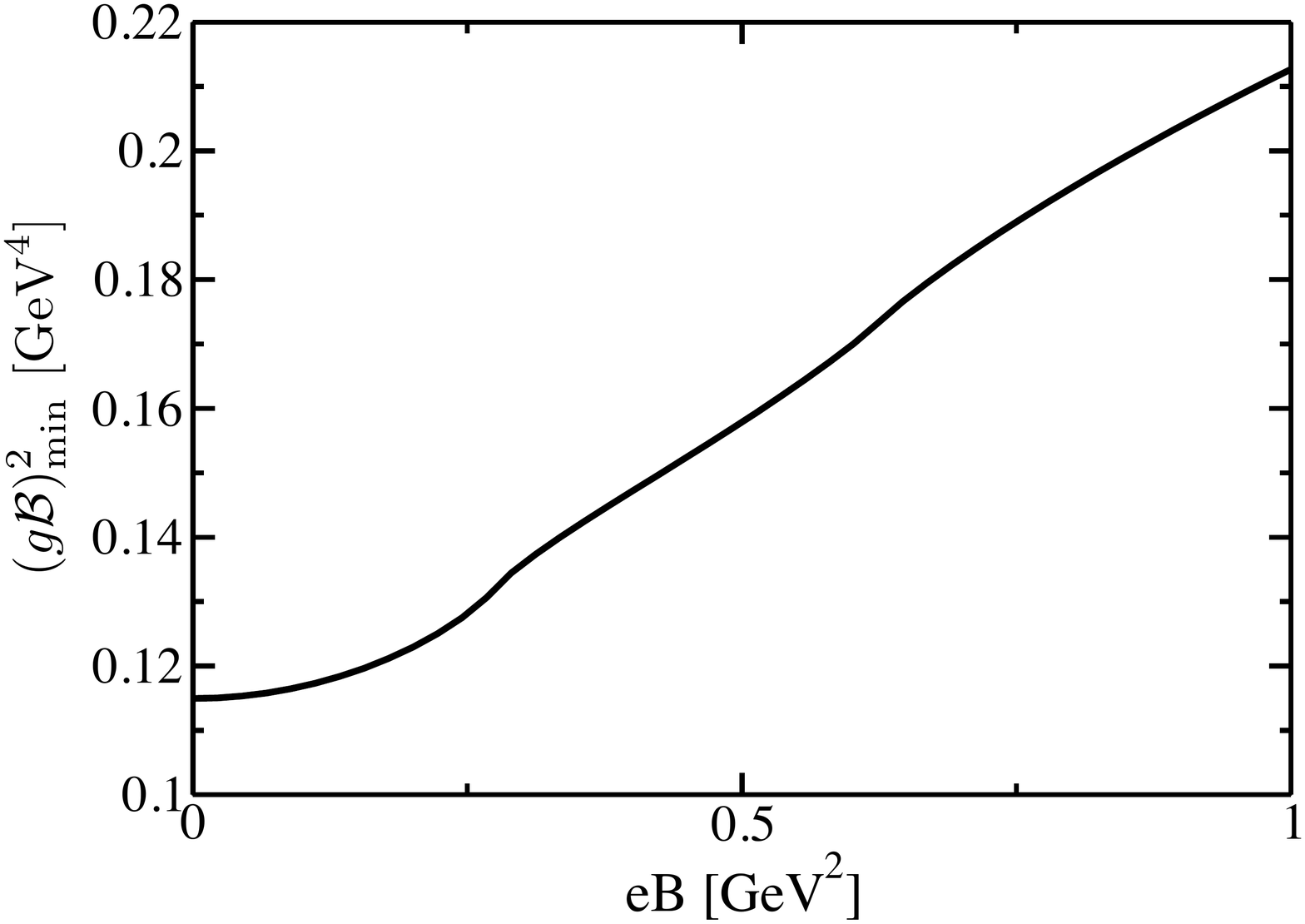}
\end{minipage} 
\caption{ Magnetic-field dependence of the QCD effective potential (left) 
and of the chromo-magnetic condensate (right)~\cite{Ozaki:2013sfa}. 
}
\label{Fig:condensate_in_B}
\end{figure}

Below, we focus on the parallel configuration $\theta_{\mathcal{B}B}$=0 with the set of $ \lambda^i $ given above. 
Going back to Fig.~\ref{Fig:breakdown-B}, we show effects of a finite $ B $ with red lines at $eB = 0.4$ GeV$^{2}$. 
The quark part $V_{\rm{quark}}$ is diminished in the presence of the QED magnetic field 
since it favors a coexistent chromo- and QED magnetic fields as we have seen above. 
Therefore, adding it to the YM part (which is intact under the application of a finite $ B $), 
one finds a deeper minimum of the total effective potential induced by the QED magnetic field. 
Namely, a finite QED magnetic field enhances 
the magnitude of the chromo-magnetic condensation,

The left panel in Fig.~\ref{Fig:condensate_in_B} shows the total effective potential $V_{\rm eff} (\mathcal{B}, B)$ 
for various strengths of the QED magnetic field in the same configurations of $\theta_{\mathcal{B}B}$ and $ w^i $ as above. 
The effective potential has a local minimum for all the shown magnetic-field strengths, 
and the location of the minimum shifts to the right as we increase the strength. 
This result suggests that the chromo-magnetic condensate $(g\mathcal{B})_{\rm{min}}^{2}$, 
which is determined from the location of the minimum,  
monotonically increases with the QED magnetic field. 
This behavior is explicitly shown in the right panel of Fig.~\ref{Fig:condensate_in_B}. 
The quark-antiquark polarizations in the vacuum play an important role in this dynamics, 
since the gluon field is not directly coupled to the external QED field. 
These behaviors qualitatively agree with the lattice QCD result \cite{Bali:2013esa}, 
where the gluonic action density is enhanced 
in the presence of the magnetic field at zero temperature.

\subsection{Polyakov-loop potential in QCD+QED fields at finite temperature}

\label{sec:Polyakov}

In the previous subsection, we discussed QCD at zero temperature. 
We here discuss an application of the effective Lagrangians 
(\ref{action_quark}), (\ref{action_ghost_T}), and (\ref{action_gluon_T}) extended to finite temperature. 
One of important quantities in finite-temperature QCD is the Polyakov loop that serves as an order parameter of 
the deconfinement phase transition in pure YM theory associated with the breaking of the center symmetry 
(see a very brief overview below and Ref.~\cite{Fukushima:2017csk} for a recent review). 
A quark field explicitly breaks the center symmetry, 
and the Polyakov loop should then be understood as an approximate order parameter in the presence of the light-quark fields. 
The QCD effective action with the background Polyakov loop 
was first computed at the one-loop level by Gross, Pisarski, and Yaffe (GPY)~\cite{Gross:1980br} 
and also by Weiss~\cite{Weiss:1980rj, Weiss:1981ev}, and is called the GPY-Weiss potential named after those authors. 
We will see that the GPY-Weiss potential is reproduced 
from the extended HE effective action obtained in this section. 
Moreover, we have further included external QED+QCD fields into the effective Lagrangian, 
and can use it to discuss how the Polyakov-loop effective potential is modified by those external fields~\cite{Gies:2000dw, Ozaki:2015yja}. 
It is remarkable that the Polyakov loop, as well as the gluon condensate discussed in the previous subsection, 
has been measured with recent lattice QCD simulations in magnetic fields~\cite{Ilgenfritz:2012fw, Ilgenfritz:2013ara, Bali:2013esa, Bonati:2014ksa, DElia:2015eey} (see Sec.~\ref{sec:MC-lattice} 
for more recent developments). 
These facts motivate us to study the deconfinement phase transition especially in the presence of external electromagnetic fields.

The Polyakov loop is defined by the closed Wilson line in the temporal direction (or imaginary-time direction denoted as the fourth component):
\beq
\Phi (\vec{x})
&=& \frac{1}{N_{c}} 
{\rm{Tr}} \ \mathcal{P} \ {\rm{exp}} \left\{ ig \int^{\beta}_{0} d\tau {A}_{4}^{a} (\tau, \vec{x}) t^{a} \right\}
\, ,
\label{defPolyakovLoop}
\eeq
where $\beta = 1/T$, and $\mathcal{P}$ is the path-ordered product. 
The thermal average is taken with the QCD action. 
The confinement phase corresponds to the vanishing value of 
the Polyakov loop $ \langle \Phi \rangle  \to 0$, 
yielding a diverging free energy $ F_q $ of an infinitely heavy quark embedded 
in the medium, $ F_q = - T \ln \langle \Phi \rangle \to \infty $.
On the other hand, a deconfinement phase corresponds to a finite value $\langle \Phi \rangle \neq 0$ 
which breaks the center symmetry of the color SU($  N$) group.

We first briefly summarize the basic points, taking the SU($2$) case as a simple example.  
Working in the Polyakov gauge for a time-independent field $A_4^a(\vec x)=\phi(\vec x)\delta^{a3}$, 
we can perform a functional integral with respect to fluctuations around the field $\phi(\vec x)$. 
In this way, one obtains the effective action for $\phi(\vec x)$. 
This procedure is nothing but what we have performed in this section: 
We divided the gluon field $A_\mu^a$ into a background field ${\cal A}_\mu^a$ 
and a fluctuation field $a^{a}_{\mu}$, and then integrated the latter. 
Therefore, we expect that the aforementioned well-known results are reproduced by the proper-time method.

We divide the background field into the constant part 
and the coordinate-dependent part as $\mathcal{A}_{\mu}^{a}(x) = (\bar{\A}_{\mu} + \hat{\A}_{\mu}(x)) n^{a}$. 
The zeroth component of the first constant term $\bar{\A}_{\mu}$, 
after the Wick rotation, gives the Polyakov loop defined in Eq.~(\ref{defPolyakovLoop}), 
and the second term generates the external electromagnetic fields, $  \F_{\mu \nu}^{a} $. 
We work in the Polyakov gauge for $\bar{\A}_{4}^{a}$ and the static limit~\cite{Weiss:1980rj}:
\beq
\bar{\A}_{4}^{a} = \bar{\A}_{4}\, \delta^{3 a}, \ \ \ \partial_{4} \bar{\A}_{4} = 0
\, .
\eeq
This gauge condition is compatible with the covariantly constant condition in Eq.~(\ref{eq:CCF}), 
so that one can include both of those fields \cite{Gies:2000dw}. 
Note that we employ the Polyakov gauge with the color direction $\delta^{a3}$ 
even for the SU($N_c$) case,\footnote{
In the literature, the temporal component of the gauge field $\bar{\A}^{a}_{4}$ in the Polyakov gauge is often expressed by $N_{c}-1$ real scalar fields. In our formalism, these color degrees of freedom are encoded in the color eigenvalues $w^{i} \ (i=1, \ldots, N_{c})$ and $v^{a} \ (a=1, \ldots, N_{c}^{2} -1)$. 
Choosing the third direction of the color unit vector, $n^{a} = \delta^{a 3}$, 
we pick up one particular field $\bar{\A}_{4}$ (but with the color eigenvalues attached). 
This convention results in a simple expression for the Polyakov loop shown 
in Eq.~(\ref{simple_Polyakov_loop}) after taking the trace over the eigenvalues $ v^{a} $. 
}
and the color unit vector $n^a$ should be here understood as $n^a=\delta^{3a}$. 
Introducing a dimensionless field 
\beq
C = \frac{g {\bar{\A}_{4} } }{ 2 \pi T }
\, , 
\eeq
we can express the Polyakov loop as
\begin{subequations}
\label{simple_Polyakov_loop}
\begin{eqnarray}
\Phi
&=& {\rm{cos}} (\pi C) \qquad \qquad \quad \ \ {\rm{for \ SU(2) }}
\, ,  \nonumber \\
\Phi
&=& \frac{1}{3} \Big\{ 1 + 2{\rm{cos}}( \pi C) \Big\} \quad \ {\rm{for\ SU(3)}}
\,  . 
\end{eqnarray}
\end{subequations}

\subsubsection{GPY-Weiss potential at finite temperature}

Let us proceed to computing the effective potential for the Polyakov loop. 
First, we only maintain the $ \bar{\A}_{4}^{a}  $ for the Polyakov loop without external electromagnetic fields 
and confirm that the current formalism correctly reproduces 
the GPY-Weiss potential \cite{Gross:1980br, Weiss:1980rj, Weiss:1981ev}. 
In the beginning of the section, we already obtained the general form of the effective action 
in the presence of the temporal component of the gauge field. 
Therefore, identifying the $\bar{\A}^{a}_{4}$ for 
the Polyakov loop 
as $\A^{a}_{0} = i \bar{\A}^{a}_{4}  $ in Eqs.~(\ref{action_YM_pure_E}) and (\ref{action_quark}), 
we obtain the YM and quark contributions to the effective potential as 
\begin{subequations}
\beq
V_\YM 
&=& - \frac{(4-2) }{ 16 \pi^{2} } \sum_{a=1}^{N_{c}^{2}-1} 
\sum_{\bar k=1}^{\infty} \cos \left( \frac{ gv^{a}  \bar{\A}_{4}  \bar k} {T} \right) \frac{16 T^4}{\bar k^4} 
\, ,
\\
V_\q
&=& \frac{N_f }{4\pi^{2}} \sum_{i=1}^{N_{c}} 
\sum_{\bar k=1}^{\infty} (-1)^{\bar k}  {\rm{cos}}\left( \frac{ gw^{i} \bar{\A}_{4} \bar k}{T} \right) 
\frac{16 T^4}{ \bar k^4} 
\, .
\eeq
\end{subequations}
We discarded the vacuum contributions (zero-point energies) which are independent of $ \bar{\A}_{4} $, 
and took the massless limit for the quark part, assuming 
that $T \gg  m_f  $. 
The last factor in each contribution originates from the proper-time integral 
which is now finite both in the UV and IR regimes. 
The cancellation between the unphysical degrees of freedom in the gluon and ghost sectors 
results in the number of physical degrees of freedom, 
$  4-2 =2$ attached in front of the YM part. 
The summation over $ \bar k $ can be analytically performed as\footnote{
This summation can be regarded as a Fourier series, 
of which the original function is known as~\cite{BernoulliPolynomial}
$$
\sum_{\bar k =1}^\infty \frac{\cos(x \bar k)}{\bar k^4} 
= - \frac{\pi^4}{3} B_4 \left( \frac{x}{2\pi}\right)
= - \frac{\pi^4}{3} \left[ \,  \left( \frac{x}{2\pi} \right)^2 
\left( \frac{x}{2\pi} -1 \right)^2 - \frac{1}{30}  \, \right]
\, ,
$$
where $ B_n(z) $ is the Bernoulli polynomial, 
and its argument is here understood to be $ z \ {\rm modulo} \ 1 $ for the first equality to hold for the periodic function. 
Likewise, one can get the formula for the fermionic part by shifting the argument as 
$$
\sum_{\bar k =1}^\infty (-1)^{\bar k} \frac{\cos(x \bar k)}{\bar k^4} 
= - \frac{\pi^4}{3} B_4 \left( \frac{x+\pi}{2\pi}\right)
\, .
$$
A similar formula is easily obtained in the (1+1)-dimensional case, 
which we will use in the strong magnetic field limit below. 
}
\begin{subequations}
\beq
V_\YM 
&=& - \pi^2 T^4 \sum_{a=1}^{N_{c}^{2}-1}  
\left[ \, \frac{1}{45} - \frac{2}{3} (C^{a})^2 \left( C^a-1 \right)^2
\, \right]
\label{eq:Weiss-YM-SUN}
\, ,
\\
V_\q
&=&  2\pi^2 T^4 N_f \sum_{i=1}^{N_{c}}  
\left[ \, \frac{1}{45} - \frac{2}{3}  (C^{i})^2 (  C^i - 1 )^2 \, \right]
\label{eq:Weiss-q-SUN}
\, ,
\eeq
\end{subequations}
where $  C^a = g v^a \bar{\A}_{4} / ( 2 \pi T)$ and 
$C^i = (g w_i \bar{\A}_{4} / T  + \pi)/(2\pi) $ modulo $  1$. 
The shift in $ C^i  $ by a constant term originates from the alternating signs 
in the Poisson resummation, which, therefore, reflects the fermionic statistics. 


\begin{figure}[t]
\begin{minipage}{0.5\hsize}
\includegraphics[width=1 \textwidth
]{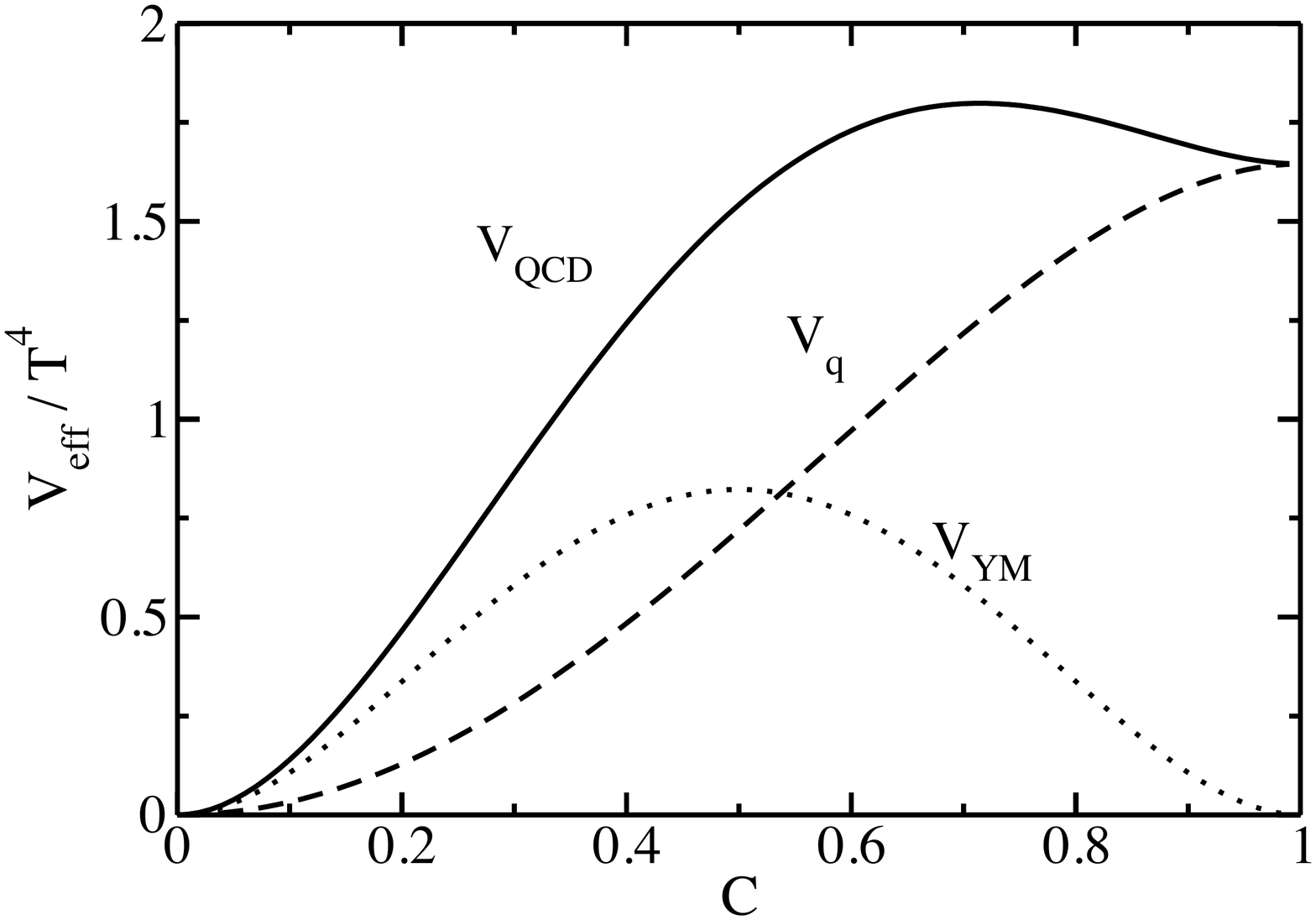}
\end{minipage}
\begin{minipage}{0.5\hsize}
\includegraphics[width=1 \textwidth
]{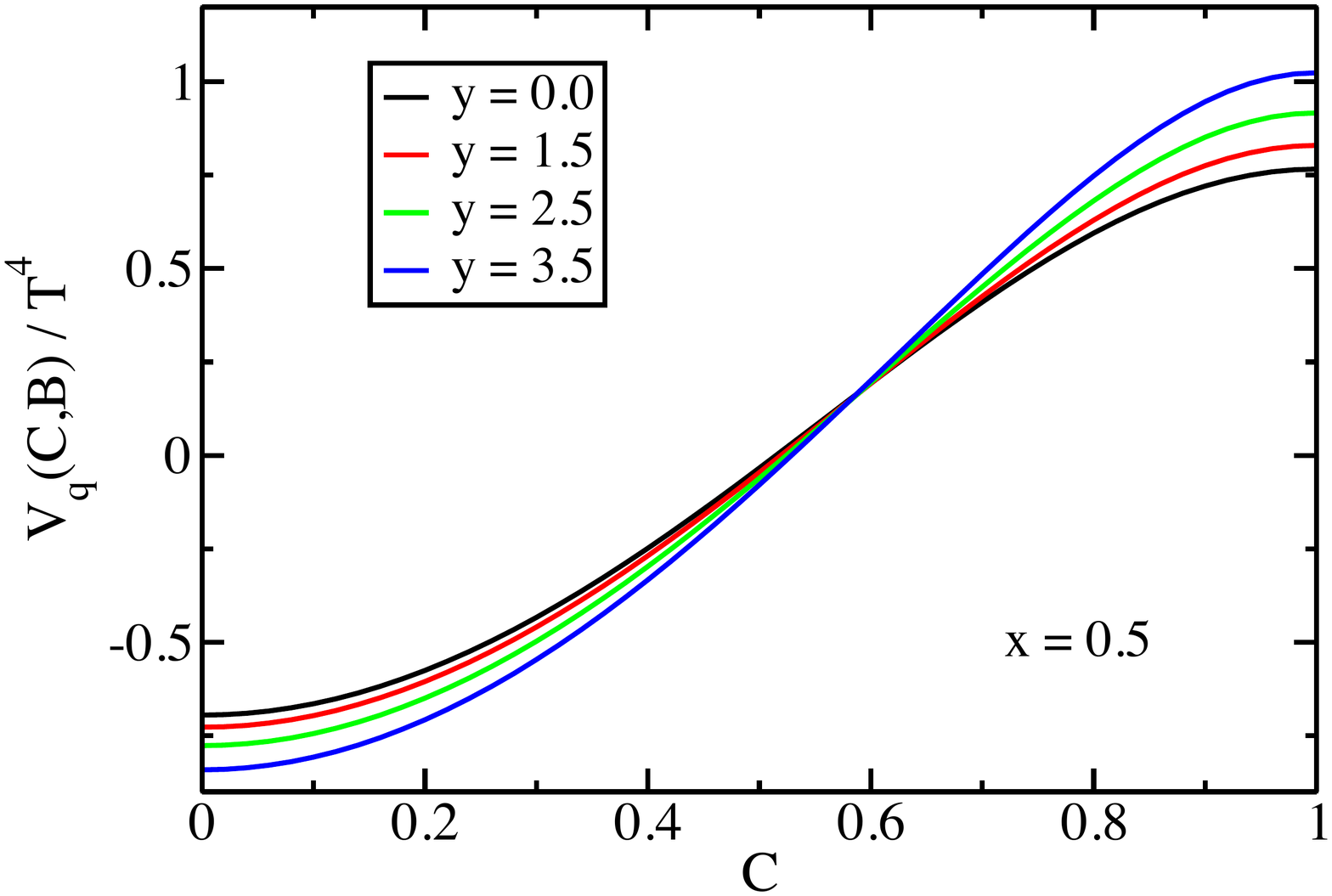}
\end{minipage}
\vspace{-0.5cm}
\caption{
One-loop Polyakov-loop potential as a function of 
$C= g \bar{\A}_{4} /( 2 \pi T )$ 
without external fields, $ E,B , {\cal E, B} = 0 $ (left), 
and the quark part for several values of magnetic fields (right). 
We defined dimensionless parameters as $x = m^{2}/T^{2}$ and $y = eB/T^{2}$, 
and subtracted the constant terms which are independent of $C$. 
}
\label{Fig:WeissPotential}
\end{figure}

To study the essence of the Polyakov loop potential, 
we shall focus on the case of the color SU(2) group and a single-flavor fermion. 
Then, using $w^{i} = \pm 1/2$ and $v^{a} = 0, \pm 1$,\footnote{
These eigenvalues can be found by following the description in Appendix~\ref{sec:cc}. 
} 
one can reproduce the well-known form of the potential~\cite{Gross:1980br, Weiss:1980rj,Weiss:1981ev}
\beq
V_{\rm{eff}}[C]=V_{\rm YM}[C]+V_\q [C]\, ,
\eeq
where the YM and quark parts are, respectively, given by 
\begin{subequations}
\beq
V_{\rm YM}[C]&=&- \frac{ 3 }{ 45 } \pi^{2} T^{4} + \frac{4}{3} \pi^{2} T^{4} C^{2} (1-C)^{2}
\, ,
\label{Weiss_YM}
\\
V_{\rm q}[C]&=&- \frac{7}{90} \pi^{2} T^{4} + \frac{1}{6} \pi^{2} T^{4} C^{2} ( 2 - C^{2} )
\, .
\label{Weiss_quark}
\eeq
\end{subequations}
We normalized the field as $  C = g  \bar{\A}_{4} / ( 2 \pi T)$, 
and these expressions are valid in a cycle of the periodic YM part, $ 0 \leq C \leq 1 $. 
While the quark part is also periodic as clearly seen in Eq.~(\ref{eq:Weiss-q-SUN}), 
its period is now twice as large as that of the YM part because of the difference between 
the eigenvalues $  w^i$ and $v^a  $ that reflect the group representations. 
The first term in each contribution again corresponds to the Stefan-Boltzmann limit, 
which obviously holds the center symmetry.

In the left panel of Fig.~\ref{Fig:WeissPotential}, 
we show the effective potential $V_{\rm{eff}}[C]$ and its breakdown. 
In the YM part, degenerated minima appear at $C=0$ and $C = 1$, 
reflecting the center symmetry $C\to C+1$ in SU(2). 
Therefore, selecting one of the two minima spontaneously breaks the center symmetry. 
This result is thought to be natural, because our one-loop computation is reliable 
in the high-temperature regime where the system is expected to be in the deconfined phase. 
The quark part explicitly breaks the center symmetry (due to the doubled period), 
so that the potential at $C=0$ and $C=1$ are no longer degenerated. 
In the presence of the quark part, the origin $C=0$ is favored, which corresponds to the deconfined phase.

This picture of the deconfined phase is intimately related to the color-electric screening 
of the external field $  \bar{\A}_{4} $ \cite{Weiss:1981ev, Fukushima:2017csk}. 
Since the above potential was obtained from the effective action for the $  \bar{\A}_{4} $, 
one can immediately read off the thermal mass $ m_{\rm th} $ from 
the quadratic term $(m_{\rm th}^2/2)   \bar{\A}_{4} $ or the Debye screening mass 
\begin{eqnarray}
m^2_{\rm sc} = 2 m_{\rm th}^2
=  \frac{1}{3}  \left( N_c + \frac{ N_f}{2} \right) (gT)^2
\, ,
\end{eqnarray}
where $ N_c =2 $ in the present case. 
This expression agrees with the Debye screening mass obtained 
from the one-loop gluon self-energy in the static limit \cite{Gross:1980br, Bellac:2011kqa}. 
The larger the screening mass is, 
the more the minimum at $ C=0 $ is stabilized with a larger potential curvature 
as expected from the screening of the confining force.


\subsubsection{GPY-Weiss potential in a magnetic field}

\label{sec:Weiss}

Now we come back to our most general effective Lagrangians (\ref{action_quark}), (\ref{action_ghost_T}), and (\ref{action_gluon_T}). 
By using those formal results, one can study modifications of the Polyakov-loop effective action 
in an electric field \cite{Gies:2000dw, Ozaki:2015yja} and in a magnetic field \cite{Ozaki:2015yja}. 
Here, we look into the effects of a magnetic field in some more detail 
as they are related to lattice QCD simulations \cite{Bruckmann:2013oba, Endrodi:2015oba}.

We take the vanishing limit of the electric field and the chromo-EM fields, $E, \mathcal{E}, \mathcal{B} \to 0$, whereas keep the Polyakov loop $\bar{\A}_{4}$ and the magnetic field $B$ nonzero in the medium rest frame. 
The YM part is not coupled to the magnetic field at the one-loop order, 
which is thus unchanged from Eq.~(\ref{Weiss_YM}). 
Therefore, we may write the effective potential as 
\beq
V_{\rm eff} [C, B] = V_{\rm YM}[C] + V_{\rm q} [C, B]
\label{eq:V_eff-Polyakov}
\, ,
\eeq
where the quark part is given as~\cite{Ozaki:2015yja, Bruckmann:2013oba} 
\beq
V_{\rm q} [C, B] =  \sum_{i=1}^{N_{c}} 
 \sum_{\bar k=1}^{\infty} 2 (-1)^{\bar k}  {\rm{cos}} \left( \frac{ g w_{i}\bar{\A}_{4}  \bar k }{ T}  \right) 
\frac{\rho_{B}}{4\pi}
\int^{\infty}_{0} \frac{ds}{s^{2}} \e^{-m_{f}^{2}s - \frac{\bar k^{2}}{4T^{2}s} }
\, {\rm{coth}}( |q_{f} B|s ) 
\label{PolyakovLoopwithB}
\, .
\eeq
We again focus on the case of $N_{c}=2$ and $ N_f =1 $ with $ q_f B = eB $ and $ m_f =m $. 
Qualitative features of the following results do not depend on the sign of the quark electrical charge 
or the direction of the magnetic field, and the fractional factor in the quark electrical charge is neglected for simplicity. 
In this case, we have 
\beq
V_{\rm q}[ C, B ]
\=   \sum_{\bar k=1}^{\infty} 2 (-1)^{\bar k}  {\rm{cos}} \left( C\pi \bar k \right)
\frac{\rho_B}{4\pi} 
\int^{\infty}_{0} \frac{ds }{s^{2}} \e^{-m^{2} s - \frac{\bar k^{2}}{4T^{2}s}} 
\,  {\rm{coth}}( |eB| s )
\label{eq:polyakov-B}
\, .  
\eeq
For a general strength of the magnetic field, we need to resort to numerical evaluation. 
The integral is well convergent. 
In the right panel in Fig.~\ref{Fig:WeissPotential}, 
we show the magnetic-field dependence of the quark part 
with $x=m^2/T^2=0.5$ and various values of $ y = |eB|/T^2 $. 
We find an enhancement of the explicit center-symmetry breaking 
as we increase the magnetic-field strength.

%
%
%

One can confirm this tendency with an analytic form of the potential in the strong magnetic field. 
As we saw in Sec.~\ref{sec:potential-MC}, the effective potential is subject to 
the effective dimensional reduction in the presence of the strong magnetic field. 
In the same way, the proper-time integral in Eq.~(\ref{eq:polyakov-B}) 
is factorized as a product of the density of states $  \rho_B$ 
and the (1+1)-dimensional form of the potential, when $ \coth(|eB|s ) \to 1 $ in the strong field limit 
(see Sections~\ref{sec:Sch-LL} and \ref{sec:L_YM} for the Landau-level decompositions). 
We can perform the proper-time integral and the summation 
in parallel to the (3+1)-dimensional case demonstrated in Eq.~(\ref{Weiss_quark}). 
Then, we find the asymptotic form of the effective potential \cite{Ozaki:2015yja}
\beq
V_{\rm q}[C, B] \to
\rho_B \times \pi T^2 \left( -  \frac{1}{3} +  C^2  \right)
\label{Weiss_quark-strongB}
\, .
\eeq
This means that the quark contribution, which breaks the center symmetry, 
is proportional to the magnetic-field strength, confirming the tendency seen in Fig.~\ref{Fig:WeissPotential} (right).

As we saw above, the potential curvature is proportional 
to the electric screening mass or the thermal mass of the $ \bar{\A}_{4} $ field. 
The screening mass is read off from the asymptotic form of the quark part (\ref{Weiss_quark-strongB}) as 
\begin{eqnarray}
\label{eq:screening-A0}
m^2_{\rm sc} =  \rho_B \frac{ g^2}{4\pi}  + \frac{N_c}{6}  (gT)^2
\, .
\end{eqnarray}
The first term exactly reproduces the screening mass from the LLL quark loop 
discussed in Sec.~\ref{sec:photons} and Appendix~\ref{sec:VP_vac} (with a factor of $  1/2$ from the color trace). 
This term grows with $ \rho_B $, while the second term from the YM part is independent of the magnetic field. 
The minimum at $ C=0 $ is more stabilized by the growing quark-loop contribution to the screening mass. 
This observation is again consistent with the tendency seen in Fig.~\ref{Fig:WeissPotential}, 
and implies that the deconfinement phase-transition temperature decreases 
as we apply a stronger external magnetic field.

Strictly speaking, the above perturbative treatment may not be justified in such a phase-transition region. 
Nevertheless, from the above analytic results and intuitive discussions, 
we may learn a qualitative tendency that the center symmetry is more strongly broken when the screening effect is stronger. 
One should, however, note that the above result is obtained in the vanishing quark mass limit. 
The screening effect would be suppressed with a finite (constituent) quark mass 
due to the competition between the temperature scale and the mass gap. 
A striking question is how strong the screening effect is near the phase transition temperature 
when the (constituent) quark mass is generated from the inherent strong-coupling nature in QCD 
and is possibly modified by a strong magnetic field. 
This poses an interesting question since, in early this section, we have glanced at the fact 
that the strong magnetic field ``catalyses'' the chiral symmetry breaking. 
One might then expect that a strong magnetic field enlarges the mass gap 
and the phase transition temperature increases due to suppression of the screening effect. 
However, the recent lattice simulations suggest an opposite behavior, i.e., 
a decreasing behavior of the transition temperature as we increase the magnetic-field strength~\cite{Ilgenfritz:2012fw, Ilgenfritz:2013ara, Bali:2013esa, Bonati:2014ksa, DElia:2015eey}. 
A key issue is the non-trivial magnetic-field dependence of the mass gap in the strongly coupled QCD 
that distinguishes it from weak-coupling QED and possibly naive effective models of QCD. 
In the next section, we focus on the low-energy dynamics in strong magnetic fields 
and will come back to the recent results from lattice QCD simulations.




\section{Low-energy dynamics in strong magnetic fields}

\label{sec:dmr} 
As discussed in Sec.~\ref{sec:Q}, the charged-particle spectrum 
is subject to the Landau quantization in magnetic fields. 
Since the Landau-level spacing increases with the magnetic-field strength $ \sqrt{q_f B} $, 
the higher Landau levels are decoupled from the lowest-Landau level (LLL) 
when one focuses on the low-energy excitations (cf. \fref{fig:Zeeman}). 
For massless fermions, the low-energy spectrum is, therefore, identified with the dispersion relation in the LLL [see \eref{eq:fermion-rela}] 
\begin{eqnarray}
\ep^{}_{\rm LLL} = \pm p_z
\, ,
\label{eq:e_LLL}
\end{eqnarray}
where the magnetic field is again assumed to be applied in the $z$ direction. 
This dispersion relation is identical to that of a free particle in the (1+1) dimensions, giving rise to 
an effective dimensional reduction to the longitudinal space along the magnetic field. 
In this section, we discuss consequences of this dimensional reduction occurring in the low-energy regime. 
The residual two-dimensional phase space is degenerated 
since there is no preferred position of a cyclotron motion 
in the transverse plane. 
The density of the degenerate states is given by $ |q_f B|/(2\pi) $. 
The (1+1)-dimensional kinetic term for the LLL state reads 
\begin{eqnarray}
S_{\rm LLL}^{\rm kin} = \int  dt \int  \frac{dp_z}{2\pi} \bar \psi^{}_{\rm LLL} (p_z)
 \left( i\partial_t \gam^0 - p_z \gam^3 \right) \psi^{}_{\rm LLL} (p_z)
\label{eq:L_LLL}
\, .
\end{eqnarray} 
Here, we have suppressed the degeneracy label 
for notational simplicity (see Sec.~\ref{sec:Q} for more details).

\subsection{Analogy with the dense system, effective dimensional reduction, and infrared scaling dimensions}

\label{sec:dense-mag}

\begin{figure}
     \begin{center}
              \includegraphics[width=0.8\hsize]{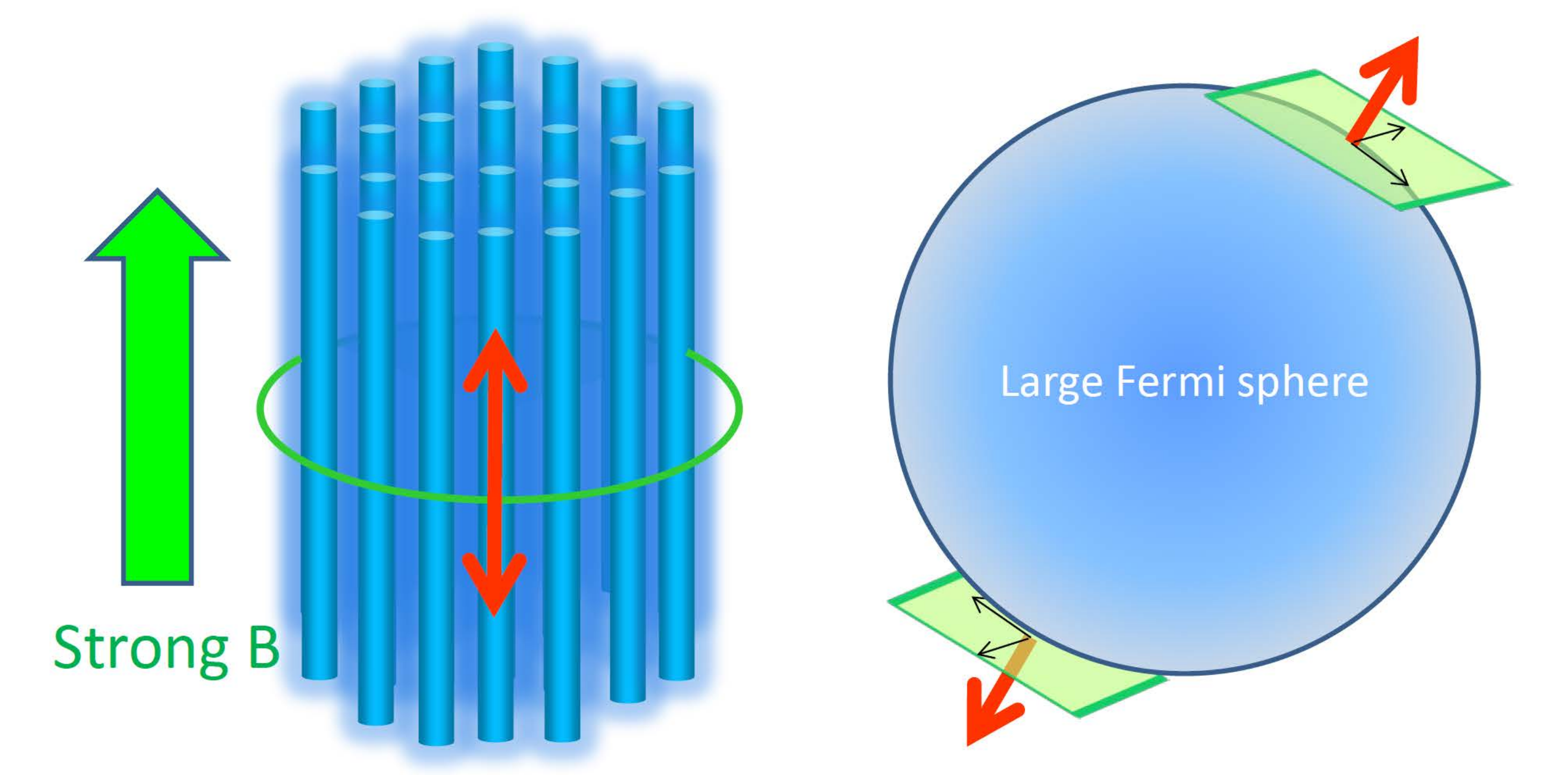}
     \end{center}
\vspace{-1cm}
\caption{Dimensional reduction in a strong magnetic field and a dense system: 
Shown are squeezed cyclotron orbits in a strong magnetic field (left) and 
low-energy excitations near a large Fermi sphere (right). 
Low-energy excitations only cost energy in the one-dimensional directions 
along the magnetic field and normal to the Fermi surface (red arrows). 
There are large degeneracies in the two-dimensional phase spaces transverse to the magnetic field 
and tangential to the Fermi surface (green planes).}
\label{fig:dim}
\end{figure}

To get clear physical insights into consequences of the dimensional reduction, 
we begin with pointing out an analogy between the systems in the strong magnetic field and at high fermion density. 
In the presence of a large Fermi sphere, the low-energy excitations near the Fermi surface 
also exhibits a dimensional reduction (see \fref{fig:dim}). 
This is because, in the small curvature limit, the excitation energy in the vicinity of the Fermi surface 
is degenerated.  
Namely, the excitation energy, measured from the Fermi surface, 
only depends on the one-dimensional momentum normal to the Fermi surface (red arrows). 
The energy difference among the states on a patch of the tangential plane (green plane) can be infinitesimally small as compared to the Fermi energy as we focus on a lower and lower energy scale.

In condensed matter physics, the dimensional reduction near the Fermi surface has been known to 
bring about rich physical consequences, 
i.e., superconductivity \cite{Bardeen:1957mv} and the Kondo effect \cite{kondo1964resistance}.\footnote{
The Kondo effect was proposed to explain 
an anomalous behavior in a temperature dependence of electrical resistance. 
Existence of a local minimum at a certain temperature 
had been a long-time mystery since experimentally observed in particular kinds of alloys. 
}
They are sometimes called the Fermi surface effects. 
One may understand these phenomena on the basis of the instabilities induced by the (1+1)-dimensional infrared (IR) dynamics. 
In high-energy physics under the strong magnetic field, 
the counterparts will be the chiral symmetry breaking 
accompanied by the formation of the chiral condensates 
in QED and QCD 
as clearly stated by Gusynin et al.~\cite{Gusynin:1994xp} 
and the magnetically induced QCD Kondo effect 
proposed by Ozaki et al.~\cite{Ozaki:2015sya}. 
According to the analogy, the former implies that the chiral symmetry breaking occurs even in QED 
in spite of the inherent small coupling constant. 
This is called the ``magnetic catalysis'' of the chiral symmetry breaking (see Ref.~\cite{Miransky:2015ava} for a large list of references). 
In Sec.~\ref{sec:potential-MC}, 
we have already previewed the magnetic catalysis 
in terms of the effective potential in the strong magnetic field. 
It is worth mentioning here that, while the $S$-wave Cooper pairs in superconductors are 
unstable in magnetic fields due to the spin-flip effect, 
the chiral condensate, where the fermion and antifermion  have the magnetic moments with different signs, 
is a stable configuration against magnetic fields~\cite{Klevansky:1989vi, Klevansky:1991ey, Klevansky:1992qe, 
Suganuma:1991dw, Suganuma:1990nn, Schramm:1991ex}. 
The QCD Kondo effect is a newly proposed phenomenon that leads to 
a strong correlation between light quarks forming a bulk matter 
and a heavy quark embedded in it as an impurity \cite{Hattori:2015hka, Ozaki:2015sya}.

In the modern language, the emergences of the superconductivity and the Kondo effect 
are informed from the renormalization-group (RG) flow \cite{Polchinski:1992ed, Stone, hewson1997kondo}. 
Namely, if an interaction term in a low-energy effective theory is found to be marginal or relevant 
as the energy scale is reduced to toward the Fermi energy, 
the effects of the interaction become important and potentially induce an infrared strong-coupling regime. 
On the other hand, if an interaction term is found to be irrelevant, 
the system will evolve into a nearly non-interacting Fermi gas. 
In case of the superconductivity, the four-Fermi operator for conduction electrons turns out to be marginal 
when a pair of electrons has the opposite momentum, i.e., the BCS configuration \cite{Polchinski:1992ed}. 
In case of the Kondo effect, the four-Fermi operator for 
the coupling between a conduction electron and an impurity becomes marginal \cite{anderson1970poor, Wilson:1974mb}. 
One can extract these insights simply by examining the scaling dimensions of the fermion fields. 
It is thus instructive to first compare the scaling dimensions 
in the strong magnetic field and in the dense system.

We shall determine the scaling dimension of the LLL fermion field 
when the excitation energy scale is reduced as 
$ \ep^{}_{\rm LLL} \to s \ep^{}_{\rm LLL} $ ($t \to s^{-1}t$) with $  s<1$.
Here, we discuss the massless dispersion relation (\ref{eq:e_LLL}) to demonstrate 
the essential mechanism of the instabilities. 
Since the LLL fermion has the (1+1)-dimensional dispersion relation (\ref{eq:e_LLL}), 
the longitudinal momentum $  p_z$ is also transformed as $ p_z \to s p_z $. 
On the other hand, the transverse momentum, which serves as the label of the degenerate states 
and does not appear in the dispersion relation (\ref{eq:e_LLL}), is not transformed. 
Therefore, when the kinetic term (\ref{eq:L_LLL}) is invariant 
under the scale transformation, the LLL fermion field scales as $ s^{-1/2} $ in the low-energy dynamics.\footnote{
We count the scaling dimensions in the mixed representations $ \psi^{}_\LLL (t, \bp) $ and so on, 
since the dimensional reduction occurs in the momentum space.
}

As for the dense system, one can explicitly identify the relevant excitations by 
performing an expansion with respect to a large chemical potential, $1/\mu  $. 
To organize this expansion, we decompose the momentum $p^\mu$ into 
the Fermi momentum $  \mu \bv^{}_{\rm F}$ and the small residual momentum $ \ell^\mu = (\ell^0 , \bl)$ as 
\begin{eqnarray}
\label{eq:mom_decomp}
p^0 = \ell^0
\, , \quad 
p^i = \mu v_{\rm F}^i + \ell^i
\, ,
\end{eqnarray}
where the energy $ p^0$ is measured from the Fermi surface 
and the magnitude of the Fermi velocity is unity $ |\bv^{}_{\rm F}|=1 $. 
For a given Fermi velocity, the corresponding plane wave is factorized as 
\begin{eqnarray}
\psi (x) = \sum_{\bv_{\rm F}} \e^{i \mu \bv^{}_{\rm F} \cdot \bx} \psi(x; \bv^{}_{\rm F})
\, .
\end{eqnarray}
The high-density effective field theory (HDEFT) is an expansion in a phase space 
around a fixed direction of the Fermi velocity $ \bv^{}_{\rm F} $. 
The entire phase space is spanned by the patches of those subspaces 
as represented by the summation over $ \bv_{\rm F} $ \cite{Hong:1998tn, Hong:1999ru, Hong:2004qf, 
Casalbuoni:2000na, Beane:2000ms, Nardulli:2002ma, Schafer:2003jn, Hattori:2019zig}. 
One can perform a mode expansion with respect to the residual momentum $ \ell^\mu$ as 
\begin{eqnarray}
\psi(x; \bv^{}_{\rm F}) = \int_{\ell \ll \mu} \!\! \frac{d^4\ell}{(2\pi)^4} \e^{-i\ell^\mu x_\mu} \psi(\bl ; \bv^{}_{\rm F})
\, .
\end{eqnarray}
Projecting out the states above the Dirac sea 
$\psi_+ (x;\bv^{}_{\rm F}) \equiv \frac{1}{2} (1+ \gam^0 \bv^{}_{\rm F} \cdot \bgam) \psi(x;\bv^{}_{\rm F}) $, 
one gets the leading-order effective Lagrangian 
\begin{eqnarray}
\label{eq:HDEF}
\mathcal{S}_{\rm HD}^{\rm{kin}} = \int \!\! d^4 x \, \bar \psi(x) (i\sla \partial + \mu \gam^0) \psi(x)
= \int \!\! dt \sum_{\bv^{}_{\rm F}} \int\!\! \frac{d^2\bl_\perp d\ell_\para} {(2\pi)^3} 
\bar \psi_+ (\bl;\bv^{}_{\rm F}) ( i\partial_t - \ell_\para) \gam^0  \psi_+ (\bl;\bv^{}_{\rm F}) + \mathcal O (1/\mu)
\, ,
\end{eqnarray}
where we defined $\ell_\para \equiv \bv^{}_{\rm F} \cdot \bl$ and $ \bl_\perp \equiv \bl - \ell_\para \bv^{}_{\rm F} $. 
The excitations of the fermion field $\psi_+ $, i.e., particle and hole states, are thus found to 
satisfy the dispersion relation 
\begin{eqnarray}
\ell^0 =  \bv^{}_{\rm F} \cdot \bl 
\label{eq:disp_mu}
 \, .
\end{eqnarray} 
Clearly, this is a linear dispersion relation in the (1+1) dimension with 
the momentum component normal to the Fermi surface as already discussed in an intuitive way (see Fig.~\ref{fig:dim}). 
The remaining two-dimensional momentum $ \bl_\perp $ labels the degenerate states on a patch. 
Therefore, when the energy scale is reduced to the Fermi energy ($  \ell^0 =0$) as 
$\ell^{0} \to s \ell^{0}$ ($t \to s^{-1}t$) with $  s<1$, 
only the $ \ell_\para $ scales as $\ell_\para \to s \ell_\para$, 
and the tangential momentum $ \bl_\perp  $ is intact. 
Following from the invariance of the kinetic term under the scale transformation, 
one finds that the fermion field $\psi_+(\bl; \bv^{}_{\rm F})$ scales as $s^{-1/2}$.

In the above, we have found that the fermion fields have the same scaling dimensions 
in the strong magnetic field and at high density. 
We now discuss the scaling dimensions of the interaction terms relevant 
for the superconductivity/magnetic catalysis and then the Kondo effect.

\subsubsection*{Superconductivity and magnetic catalysis}

An effective four-Fermi interaction among conduction electrons 
or light quarks in dense quark matter can be written as 
\begin{eqnarray}
\mathcal{S}_{\rm HD}^{\rm{int}}\!\! &=&\!\! 
\int \!\! dt \prod_{i=1,2,3,4} \sum_{\bv_{\rm F}^{(i)}} \int \frac{d^2\bl_\perp^{(i)} d\ell_\para^{(i)}} {(2\pi)^3} 
\, G \left[\bar \psi_+ (\bl^{(4)} ;\bv_{\rm F}^{(4)} )  \hat  \gam^\mu_\para  \psi_+ (\bl^{(2)};\bv_{\rm F}^{(2)}) \right]
\left[ \bar \psi_+ (\bl^{(3)};\bv_{\rm F}^{(3)}) \hat  \gam_\mu^\para  \psi_+ (\bl^{(1)} ;\bv_{\rm F}^{(1)} ) \right]
\nn
\\
&&  \hspace{3cm} \times
\delta^{(3)}( \bp^{(1)}+\bp^{(2)}-\bp^{(3)}-\bp^{(4)} )
\label{eq:fermi_SC}
\, ,
\end{eqnarray}
where $\hat \gam_\para^\mu = (\gam^0, (\bg \cdot \bv^{}_{\rm F}) \bv^{}_{\rm F})  $ and $ \bp $ is given in \eref{eq:mom_decomp}. 
$  G$ is an effective coupling constant. 
Since the Fermi momentum $ \sim \mu $ is much larger than the fluctuation near the Fermi surface $ \bl $, 
one may in general neglect the $ \bl $ in the momentum conservation. 
Therefore, the delta function does not scale in this case. 
By using the scaling dimensions discussed below \eref{eq:disp_mu}, 
the four-Fermi operator turns out to scale as $ s^{+1} $, 
meaning that the interaction is in general irrelevant in the low-energy dynamics.

However, there is an exception. 
When the scattering fermions have the opposite Fermi momenta, 
$ \mu\bv_{\rm F}^{(1)}+\mu \bv_{\rm F}^{(2)} =0 $, which is called the (S-wave) BCS configuration, 
the momentum conservation in the order of $ \mu $ is automatically satisfied. 
Since the delta function only has the order-$ \ell $ quantities, 
the delta function scales as $ s^{-1} $ when $ \ell_\para \to s \ell_\para $. 
Therefore, in the BCS configuration, the four-Fermi operator scales as $ s^0 $ and thus is lifted to a marginal operator. 
This observation suggests that a logarithmic correction emerges in the one-loop scattering diagram shown in \fref{fig:loop}, 
and that the effective coupling constant $ G $ grows in attractive channels. 
The emergence of the Landau pole, where the coupling constant diverges, has been shown 
for the color superconductivity in the dense quark matter \cite{Evans:1998ek, Evans:1998nf, Schafer:1998na}. 
Furthermore, it was shown that the effect of the unscreened magnetic gluon 
significantly enhances the growth of the coupling constant, 
so that the Landau pole appears in a larger scale than the would-be scale found in the absence of the long-range interaction \cite{Son:1998uk, Hsu:1999mp}. 
We will also discuss dependences on interaction types 
with several examples in the following subsections.

Bearing the above discussion in mind, we now proceed to the magnetic catalysis. 
The effective four-Fermi interaction in the LLL reads  
\beq
\mathcal{S}_{\rm LLL}^{\rm{int}}
&=& \int dt \, \prod_{i=1,2,3,4} \int \!\! dp_{z}^{(i)} 
\, G \left[ \bar{\psi}_{\rm{LLL}}(p^{(4)}_{z}) \gamma^{\mu}_{\parallel} \psi_{\rm{LLL}}(p^{(2)}_{z})\right]
\left[ \bar{\psi}_{\rm{LLL}}(p_{z}^{(3)}) \gamma_{\parallel \mu} \psi_{\rm{LLL}} (p^{(1)}_{z}) \right]
\nn
\\
&&  \hspace{3cm} \times
\delta( p_{z}^{(1)} + p_{z}^{(2)} - p_{z}^{(3)} - p_{z}^{(4)} )
\label{four_int_B} 
\, ,
\eeq
where $ G$ is an effective coupling constant and $ \gam_\para^\mu =(\gam^0,0,0,\gam^3) $. 
Again, the transverse momenta are not written explicitly. 
Because of the (1+1)-dimensional dispersion relation (\ref{eq:e_LLL}), the delta function scales as $  s^{-1}$. 
Therefore, when the LLL fermion field scales as $ s^{-1/2} $, this interaction term scales as $s^{0}$, 
meaning that the four-Fermi operator is marginal in the strong magnetic field \cite{Hong:1996pv, Hong:1997uw}. 
Thus, the scattering amplitude is expected to acquire the logarithmic correction, 
and can be depicted with the same diagram as in the case of superconductivity (cf. \fref{fig:loop}). 
The chiral symmetry will be broken even in weak-coupling theories like QED. 
This is an analog of the well-known fact that superconductivity is induced by any weak attractive interaction. 
We readily understand those facts for the dimensional reason. 
The dimensional reduction occurs in strong magnetic fields enough to compensate the smallness of coupling strengths in underlying theories. 
Therefore, the magnitudes of the emergent scale 
can depend on the coupling strengths and are not universal, 
while the emergence of the Landau pole itself 
is independent of the coupling strengths.

We are now confident about the occurrence of the logarithmic corrections 
and are in position to explicitly construct the RG equations. 
In Sec.~\ref{sec:RG_MC}, we will see that the effective coupling constant $  G$ indeed 
runs into the Landau pole and induces the magnetic catalysis. 

\begin{figure}
	\begin{center} 
		\includegraphics[width=\hsize]{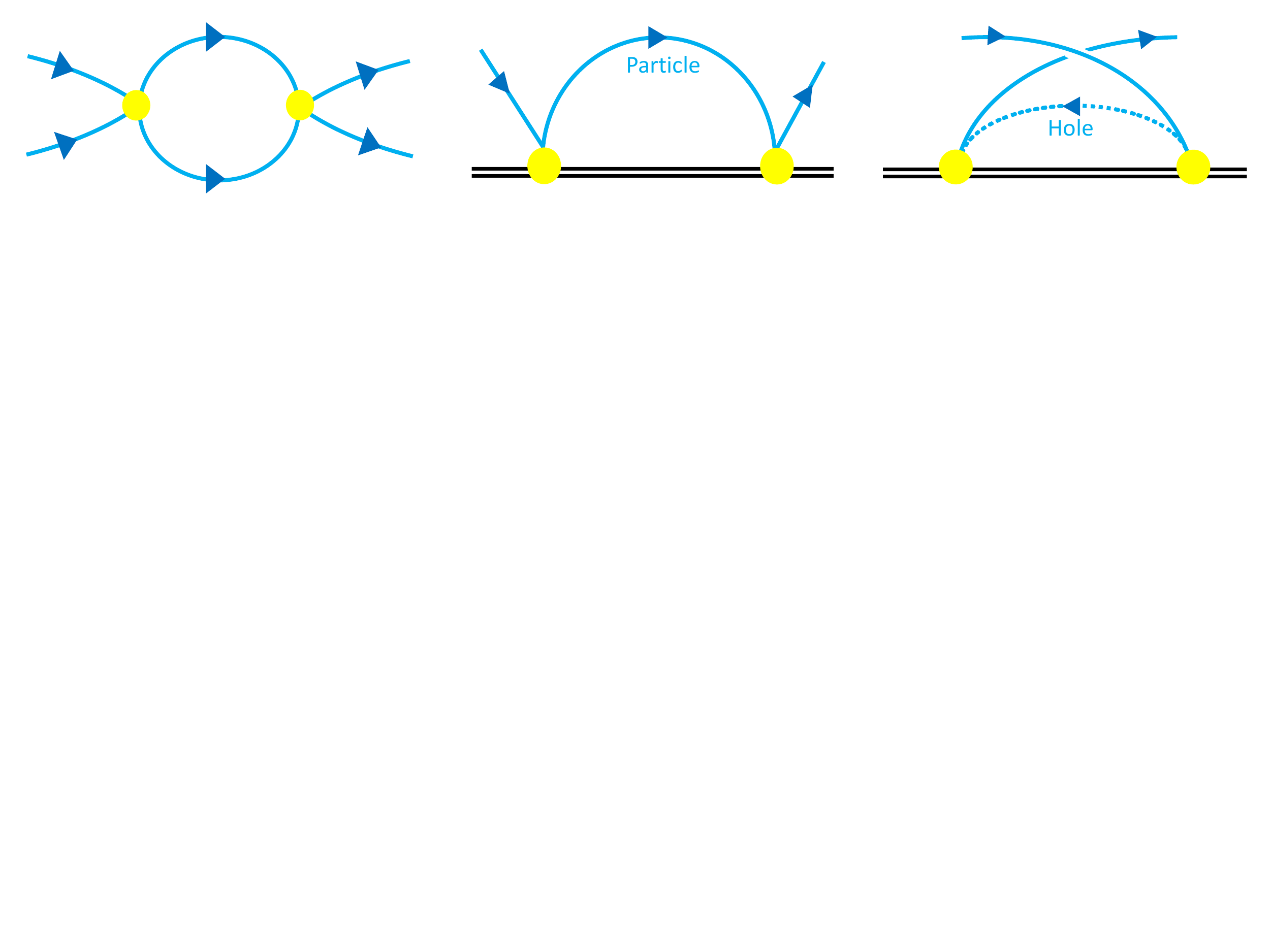}
	\end{center}
\caption{The leading quantum corrections to the four-Fermi interactions 
in superconductor (left) and in the Kondo effect (right two). 
In the Kondo diagrams, the double lines denote impurities (heavy quarks for QCD), 
and the vertices have generators of a non-Abelian symmetry.}
\label{fig:loop}
\end{figure}

\subsubsection*{The Kondo effect in dense quark matter and in strong magnetic fields}

We proceed to the QCD Kondo effect that occurs in the scattering 
between the light and heavy quarks \cite{Hattori:2015hka, Ozaki:2015sya}. 
Here, we introduce heavy quarks as impurities. 
While the bulk properties of a quark matter are predominantly determined 
by dynamics of light quarks such as up, down, and strange quarks, 
a small amount of heavy quarks (charm and bottom quarks) can affect the transport properties via impurity scatterings. 
We shall consider a spatially localized impurity which does not have a spatial velocity. Such a situation is also 
realized in condensed matter systems as well. 
In other words, we consider the limit of an infinite heavy-quark mass. 
In such cases, the spatial translational invariance, 
and thus the spatial momentum conservation, 
are absent due to the localized impurities, while the energy remains conserved with the temporal translational invariance.

A systematic expansion with respect to a large heavy-quark mass $1/m^{}_{\rm H}$ 
is formulated with the heavy-quark effective field theory (HQEFT) \cite{Manohar:2000dt}. 
In the HQEFT, a heavy-quark momentum $p^\mu$ is divided into hard and residual parts, $p^\mu = m^{}_{\rm H}v^\mu + k^\mu$ with $v^\mu=(1,0,0,0)$ for a static impurity. 
We factorize the hard-momentum part $\e^{-im^{}_{\rm H}v\cdot x}$ from the heavy-quark field $\Psi$ 
and define the reduced field  $\Psi_{+} = \frac12( 1 + \sla v) \e^{im^{}_{\rm H}v\cdot x} \Psi$ 
with the positive-energy projection. 
Then, the kinetic term in the leading order of the HQEFT is given by \cite{Manohar:2000dt} 
\beq
S_{\rm H}^{\rm{kin}}
= \int  dt \int \frac{ d^{3} \bk}{(2\pi)^3} \, \Psi_{+}^\dagger (\bk) \, i \partial_{t} \Psi_{+}(\bk) + {\mathcal O}(1/m^{}_{\rm H})
\label{kin_Q}
\, ,
\eeq
Note that the mass term is absent because we have already removed the hard-scale dynamics in defining the $\Psi_+$. 
When the spatial velocity is vanishing, the spatial derivative does not appear in the kinetic term, 
and the residual momentum $ \bk $ does not scale. 
This kinetic term indicates that the heavy-quark field $\Psi_{+}$ does not scale ($s^{0}$) when $t \to s^{-1}t$.

Now that we have determined the scaling dimensions of both the light- and heavy-quark fields, 
we can investigate their interactions. 
In the dense light-quark matter, the four-Fermi interaction between the light and heavy quarks is given by 
\beq
S^{\rm{int}}_{\rm HD}
&=& \int dt 
\sum_{\bv_{\rm F}^{(1)},\, \bv_{\rm F}^{(3)}} 
\int \frac{d^2\bl_\perp^{(1)} d\ell_\para^{(1)}} {(2\pi)^3} \frac{d^2\bl_\perp^{(3)} d\ell_\para^{(3)}} {(2\pi)^3} 
\int \frac{d^3\bk^{(2)}}{(2\pi)^3}  \frac{d^3\bk^{(4)}}{ (2\pi)^3}
\nn
\\
&& \hspace{2cm} \times
G \Big[ \bar \psi_+ (\bl^{(3)};\bv_{\rm F}^{(3)}) t^a \psi_+ (\bl^{(1)};\bv_{\rm F}^{(1)}) \Big]
\Big[ \bar \Psi_{+} (\bk^{(4)}) t^{a} \Psi_{+} (\bk^{(2)}) \Big]
\, .
\label{Effective_Int_QCDKondo}
\eeq
Note again that the recoil effect is negligible in the heavy-quark limit, 
or, in other words, the spatial translational invariance is broken by the localized impurity. 
Accordingly, the above interaction term does not have a delta function for the spatial momentum conservation. 
Plugging the scaling dimensions of the fields and of the momenta discussed above, 
we find that the light-heavy four-Fermi operator is marginal.

From the above simple analysis, one expect that the four-Fermi interaction 
acquires logarithmic quantum corrections from the next-to-leading order scattering diagrams shown in Fig.~\ref{fig:loop} . 
However, we will see in Sec.~\ref{sec:QCD-Kondo} that, without noncommutative matrices on the interaction vertices, 
the logarithmic corrections from the two diagrams exactly cancel each other. 
Therefore, in the Kondo systems, a non-Abelian property of the interaction is essential 
to make the logarithmic contributions physical. 
In the QCD Kondo effect, the color matrices naturally play a role so that 
such a complete cancellation does not occur.\footnote{
The original Kondo effect would be the first observation of asymptotic freedom with a non-Abelian interaction. 
However, the mechanism of the Kondo effect by the effective dimensional reduction 
should be distinguished from that of the asymptotic freedom in the (3+1) dimensional QCD. 
The latter is induced by nonlinear gluon dynamics arising as a consequence of the non-Abelian gauge symmetry, 
which manifests itself in the negative contribution to the (leading-order) beta function~\cite{Gross:1973id, Politzer:1973fx}. 
}

Let us discuss the magnetically induced QCD Kondo effect 
that is caused by 
the scatterings between the LLL fermions and heavy quark. 
In the heavy-quark limit $ m_{\rm H}^2 \gg eB $, effects of the magnetic field on the heavy quark, such as the Lorentz force, 
are suppressed, so that we do not consider those effects directly acting on the heavy quark. 
Then, the role of the magnetic field is to confine the light quarks in the LLL, 
and the four-Fermi operator is given by
\beq
\mathcal{S}_{\rm LLL}^{\rm{int}}
= \int dt \int  \frac{dp_{z}^{(1)}} {2\pi} \frac{dp_{z}^{(3)}} {2\pi} 
\int  \frac{d^3\bk^{(2)}}{(2\pi)^3}  \frac{d^3\bk^{(4)}}{ (2\pi)^3}
\, G \left[ {\psi}_{\rm{LLL}} ^\dagger (p^{(3)}_{z}) t^a \psi_{\rm{LLL}}(p^{(1)}_{z})\right]
\left[ \Psi_+^\dagger (\bk^{(4)}) t^a \Psi_+ (\bk^{(2)}) \right]
\label{int_term_MQCDKondo}
\, .
\eeq
We find that the four-Fermi operator (\ref{int_term_MQCDKondo}) in the magnetic field 
has a scaling dimension $s^{0}$, and thus is marginal as in the Kondo effect in the dense system discussed above. 
Also, the non-Abelian properties of QCD are again essential 
for the incomplete cancellation of logarithmic effects, 
leading to the magnetically induced Kondo effect \cite{Ozaki:2015sya}. 
In Sec.~\ref{sec:QCD-Kondo}, we will explicitly see how the logarithm survives with non-Abelian interactions.

\subsection{Magnetic catalysis of the chiral symmetry breaking}

\label{subsec:MC-weak}

The very early studies of the chiral symmetry breaking/restoration in external electric and magnetic fields 
were carried out based on the effective models of QCD in Refs.~\cite{Klevansky:1989vi, Klevansky:1991ey, Klevansky:1992qe}, 
Refs.~\cite{Suganuma:1991dw, Suganuma:1990nn}, and Ref.~\cite{Schramm:1991ex}.\footnote{
Large portions of these studies by the first two groups 
were aimed at the investigation of the chiral symmetry breaking/restoration 
in chromo-electromagnetic fields, as a mimic of color flux tubes, rather than in ordinary electromagnetic fields. 
A recent lattice QCD simulation supports the partial restoration of the chiral symmetry 
in the color flux tube spanned between a quark and an antiquark \cite{Iritani:2015zwa}.
} 
The authors of these papers observed tendencies 
toward chiral symmetry restoration in electric fields 
and toward stronger dynamical breaking in magnetic fields, 
compared to the case without the external fields. 
The basic and intuitive interpretations were based on the facts that 
the electric fields act to separate the quark and antiquark both in the coordinate and momentum spaces 
and that the magnetic fields stabilize their spin-singlet configuration. 
This picture is in contrast to the effects of the external fields on the Cooper pairs in a superconductor, 
because the Cooper pairs are composed of like-sign charges. 
Also, the authors of Ref.~\cite{Schramm:1991ex} discussed a possibility that 
the spatial localization due to the formation of the cyclotron orbits may enhance the $ q \bar q $ pairing in a strong magnetic field.

It was Gusynin, Miransky, and Shovkovy who clearly pointed out a deep analogy 
between the low-energy dynamics near the Fermi surface 
and in the strong magnetic fields (see Ref.~\cite{Gusynin:1994xp} for a concise discussion). 
They initiated the paradigm of the dimensional reduction in the strong magnetic fields 
first in the (2+1)-dimensional systems \cite{Gusynin:1994re, Gusynin:1994va} 
and soon later in the (3+1)-dimensional systems \cite{Gusynin:1994xp, Gusynin:1995gt, Gusynin:1995nb} 
(see also brief summary in the proceedings \cite{Miransky:1995rb}). 
The concept of the dynamical symmetry breaking due to the dimensional reduction, {\it the magnetic catalysis}, 
was established by focusing on the Nambu--Jona-Lasinio (NJL) model and QED at weak coupling. 
Namely, the dynamical chiral symmetry breaking was shown to 
occur no matter how small the coupling strengths are,  
without the help of any inherent nonperturbative interaction in the model/theory.\footnote{
Without a magnetic field, the dynamical symmetry breaking in the NJL model only occurs for 
a large enough coupling strength beyond the critical value (see below).} 
In the previous subsection, we have already seen that the four-Fermi operator for the LLL fermions 
is marginal only for the dimensional reason irrespective of the coupling constants in the underlying theories.

The mechanism of the magnetic catalysis was confirmed by various methods. 
Presumably, the most established method is solving the gap equations derived from 
the Schwinger-Dyson (SD) equation with appropriate approximations 
or from the stationary condition of the effective potential.  
Those studies are summarized already in comprehensive review articles in 
Refs.~\cite{Shovkovy:2012zn, Andersen:2014xxa, Miransky:2015ava} (see also Sec.~\ref{sec:potential-MC}).  
The spontaneous chiral symmetry breaking is also characterized by 
a nontrivial solution of the Bethe-Salpeter equation 
for the Nambu-Goldstone (NG) mode \cite{Gusynin:1995gt, Gusynin:1995nb}, 
in which the constituent mass is identified with the dynamical mass gap induced by the magnetic catalysis. 
It was also discussed that existence of the NG modes indicates avoidance of the Coleman-Mermin-Wagner theorem 
that states the absence of spontaneous breaking of continuous symmetries 
in the spacetime dimensions equal to or lower than (1+1) dimensions. 
The point is that, while the charged fermion in the LLL is confined in $(1+1)$ dimensions, 
the neutral composite particles, say $ \pi^0 $, 
do not form the Landau levels and are not subject to the dimensional reduction.\footnote{
Charged pions are no longer NG modes in a magnetic field due to the explicit breaking of the isospin symmetry. 
Note also that a strong enough magnetic field can resolve the inner structure of the composite particles. 
One needs to investigate the dispersion relation of a neutral pion 
on the basis of the fundamental degrees of freedom in QCD in such a strong field (see Sec.~\ref{sec:MC-strong}). 
}
Their fluctuations are, therefore, not large enough to prevent the system from rolling into the broken phase.

In the following, we shed light on the method of renormalization group (RG) 
\cite{Hong:1996pv, Hong:1997uw, 
Fukushima:2012xw, Scherer:2012nn, Hattori:2017qio}. 
In line with the scaling argument given in the previous subsection, 
we explicitly obtain the size of the dynamical mass gap which agrees with the result from the aforementioned methods. 
We first focus on the (3+1)-dimensional systems with the NJL model and weak-coupling theories, or more specifically QED. 
The magnetic catalysis in a planar system shares the same idea of the dimensional reduction, 
but has different features due to the dimensional reasons. 
The interested reader is referred to review articles \cite{Shovkovy:2012zn, Miransky:2015ava}. 
Then, we proceed to the magnetic catalysis in QCD in Sec.~\ref{sec:MC-strong} 
with detailed summary of stimulating results from recent lattice QCD simulations.

\subsubsection{Renormalization-group analysis}

\label{sec:RG_MC}

We derive the RG equation for the effective four-Fermi interaction. 
More specifically, we shall compute the scattering amplitudes $ \M(\Lm) $ at a given energy scale $ \Lam $ 
for the fermion and antifermion pair that forms the chiral condensate. 
Since the four-Fermi operator in the LLL has a marginal scaling dimension, 
we anticipate that the effective coupling constant acquires a logarithmic quantum correction  
which drives the system toward a strong-coupling regime. 
When the energy scale is reduced from $ \Lm $ to $ \Lm - \delta \Lm $, 
the energy-scale dependence of $ \M(\Lm) $ is captured 
by the integration of the loop momentum over this infinitesimal energy band. 
Since the LLL excitation has the linear dispersion relation, 
the integral can be schematically written as 
\begin{eqnarray}
\delta \mathcal{M} (\Lm)= \# \int_{\Lm - \delta \Lm}^\Lm \!\! \frac{dk_z}{2\pi} \frac{1}{k_z}
\, .
\end{eqnarray}
The logarithmic IR correction will be then absorbed by the renormalization of the effective coupling constant, 
and we are naturally led to the RG equation. 
By solving the RG equation with a given initial condition, we will find a Landau pole emerging at a certain IR scale. 
The size of the dynamically generated mass gap should be associated with this dynamical IR scale. 

\begin{figure}[t]
	\begin{center}
		\includegraphics[width=0.5\hsize]{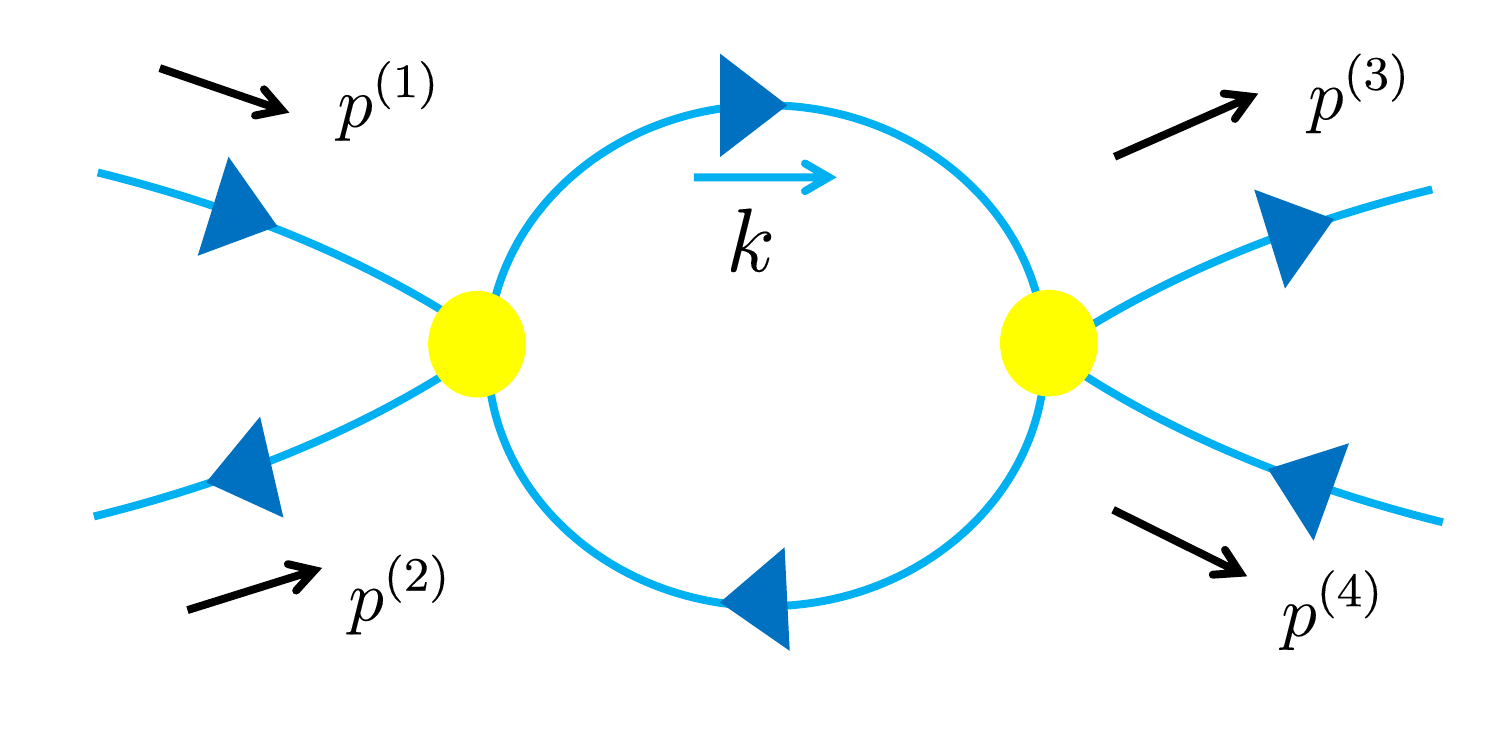}
	\end{center}
	\vspace{-1cm}
\caption{Scattering diagram for the chiral condensate.}
\label{fig:box}
\end{figure}

The relevant scattering channels for the formation of the chiral condensate 
are those between the fermion-antifermion pairs in the opposite chiralities, 
so that the two relevant spinor structures are given by 
$[ \, \bar{u}^{}_{\rm{R/L}} (p^{(3)}_{\parallel})  \gamma_{\parallel \mu} u^{}_{\rm{R/L}} (p^{(1)}_{\parallel}) \, ]\, 
[ \, {v}^{}_{\rm{L/R}} (p^{(4)}_{\parallel} ) \gamma^{\mu}_{\parallel}  \bar{v}^{}_{\rm{L/R}} (p^{(2)}_{ \parallel})  \, ]  
$ with the spinors of the fermion $ u $ and the antifermion $ v $ in the LLL. 
The chirality is denoted as ${\rm R/L} $. 
The scattering amplitudes are the same for the both channels, 
so that we suppress these trivial spinor structures below for notational simplicity.

First, we take the NJL model as one of the simplest examples. 
The coupling constant $ G_{\rm NJL} $ does not have any energy-scale dependence in the classical Lagrangian. 
Then, the leading-order scattering amplitude is just given by 
the four-Fermi coupling constant as 
\beq
\mathcal{M}_0= G_{\rm NJL}
\label{tree_amp_magneticfields}
\, .
\eeq
Next, by using the propagator in the LLL (\ref{eq:prop-LLL}), 
the one-loop amplitude shown in Figs.~\ref{fig:box} is written down as 
\beq
\label{Amp_1-loop}
i \mathcal{M}_1
=  G^{2}_{\rm NJL} \rho^{}_B \int \frac{ d^{2} k_{\parallel} }{ (2\pi)^{2} }
\left[ \bar{u} (p^{(3)}_{\parallel}) \gamma_{\parallel}^ {\mu}
S_\para ( k_{\parallel}  ) 
\gamma_{\parallel}^ {\nu} u(p^{(1)}_{ \parallel}) 
\right] 
\left[ {v}(p^{(4)}_{ \parallel})  \gamma_{\parallel \mu} 
S_\para (k_{\parallel} - P_\parallel) 
\gamma_{\parallel \nu} \bar{v} (p^{(2)}_{ \parallel})\right] 
\, ,
\eeq 
where $ P_\para = p^{(1)}_{\parallel} + p^{(2)}_{ \parallel}  $ is the total momentum of the pair 
and the longitudinal part of the propagator is given by 
$ S_\para (k_\para) =  i \sla k_\para/(k_\para^2 + i \epsilon) \prj_+ $. 
The Landau degeneracy factor $ \rho^{}_B = |eB|/(2\pi)$ is reproduced from the transverse-momentum integral, 
so that we are left with the two-dimensional loop integral as a natural consequence of the dimensional reduction. 
After performing the elementary $k^{0}$-integral (see an appendix in Ref.~\cite{Hattori:2017qio}), 
one finds the origin of the magnetic catalysis: 
\beq
\label{eq:kz}
\delta \mathcal{M}_1 (\Lm) = \rho^{}_B
G^{2}_{\rm NJL} \int^{\Lambda }_{\Lambda - \delta \Lm} 
\frac{ d k_{z} }{ 2\pi } \frac{ 1 }{ | k_{z} | } 
\, .
\eeq
Performing the integral, the logarithmic increment from the one-loop correction is found to be 
\beq
\delta \mathcal{M}_1 (\Lm) 
= \rho^{}_B 
\frac{ G^{2} _{\rm NJL} (\Lm)  }{\pi}\log \frac{\Lambda }{\Lambda - \delta \Lm} 
\label{eq:MC-one-loop}
\, ,
\eeq
where we have introduced the effective coupling $G_{\rm NJL}(\Lambda )$ at the energy scale $\Lambda$.

From the scattering amplitudes in Eqs.~(\ref{tree_amp_magneticfields}) and (\ref{eq:MC-one-loop}), 
the RG equation is obtained as 
\beq
\Lambda \frac{d}{d\Lambda} [\rho^{}_B G_{\rm NJL}(\Lambda)]=-\frac{1}{\pi} [\rho^{}_B G_{\rm NJL}(\Lambda)]^2
\label{RGeq_MC}
\, , 
\eeq
where $ \rho^{}_B G_{\rm NJL}(\Lambda) $ is a dimensionless combination. 
One can immediately read off the beta function from Eq.~(\ref{RGeq_MC}), 
which is found to be negative. 
When the interaction is attractive ($ G_{\rm NJL}>0 $), 
the magnitude of the interaction strength increases in the low-energy domain. 
Namely, we find that the four-Fermi interaction is a marginally relevant operator. 
The effect of a magnetic field on the beta function 
can be clearly summarized in Fig.~\ref{fig:beta_B} \cite{Fukushima:2012xw} (cf. also Fig.~2 in Ref.~\cite{Scherer:2012nn}). 
When $ B = 0 $, the chiral symmetry breaking occurs only when the initial value of 
the coupling constant in the RG evolution is larger than the critical value located at the UV fixed point. 
Namely, the chiral symmetry breaking cannot occur without the help of the strong-coupling nature in the low-energy QCD. 
While an analogy between QCD and superconductor has provided fruitful insights 
into high-energy physics in the history \cite{Nambu:1961tp, Nambu:1961fr,Hatsuda:1994pi}, 
the microscopic mechanisms of the symmetry breaking are actually quite different in this regard. 
The effect of the magnetic field pushes the beta function 
down to the negative region as we increase the magnetic field, 
and the beta function finally becomes negative in the entire region. 
Then, the broken phase is favored irrespective of the initial value of the coupling constant. 
We have emphasized the dimensional mechanism 
through the evaluation of the (1+1)-dimensional integral 
and the scaling argument in the previous subsection; 
In this regard, the magnetic catalysis of 
the chiral symmetry breaking 
provides a closer analogy to superconductivity. 
We also discussed the dimensional reduction in terms of the effective potential in Sec.~\ref{sec:potential-MC}. 


\begin{figure}[t]
     \begin{center}
              \includegraphics[width=0.5\hsize]{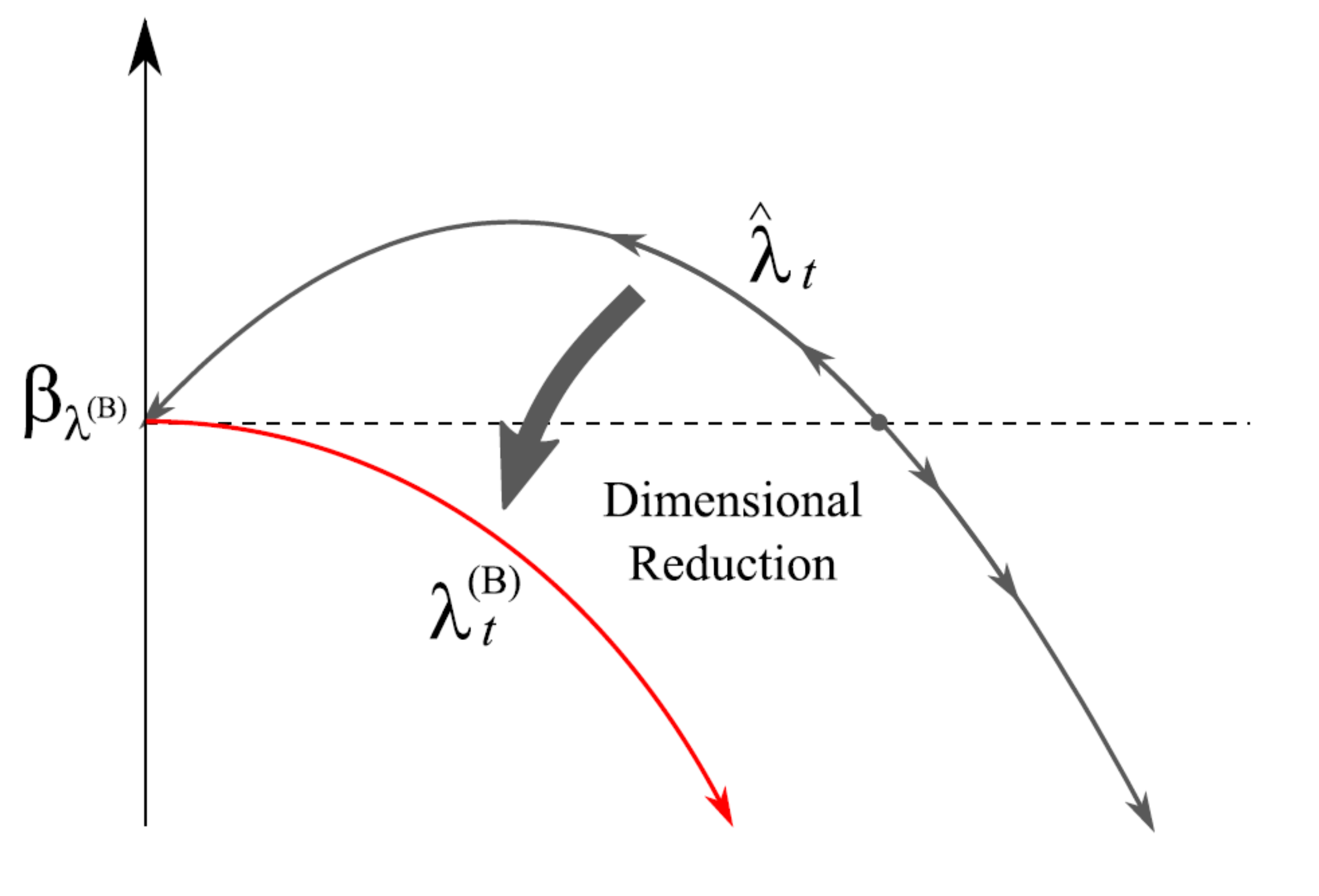}
     \end{center}
\vspace{-1cm}
\caption{The beta function for the four-Fermi interaction in vacuum (black) 
and in a strong magnetic field (red) \cite{Fukushima:2012xw} (See also Fig.~2 in Ref.~\cite{Scherer:2012nn}).
The horizontal axis shows the coupling constant. 
When the effective spacetime dimension is reduced to the (1+1) dimensions,  
the beta function is negative for any value of the coupling constant. 
}
\label{fig:beta_B}
\end{figure}

It is now easy to solve the RG equation (\ref{RGeq_MC}). 
Starting from an initial UV scale $\Lambda_0  $, 
the solution at the energy scale $\Lambda (<\Lambda_0)$ is obtained as 
\begin{eqnarray}
G_{\rm NJL}(\Lam) = \frac{  G_{\rm NJL}(\Lambda_0)  }{ 1 + \pi^{-1} \rho^{}_B G_{\rm NJL}(\Lambda_0 ) \ln(\Lambda/\Lambda_0)}
\, .
\end{eqnarray}
The effective coupling $G_{\rm NJL}(\Lambda)$ grows with a decreasing $\Lambda$, and diverges at a Landau pole. 
This implies a breakdown of the perturbative picture and an emergence of nonperturbative dynamics. 
The presence of the Landau pole gives rise to 
a dynamical IR scale 
\begin{eqnarray}
\label{eq:L_IR-NJL}
\Lambda_{\rm IR}  = \Lambda_0  \exp \left( -\, \frac{\pi}{\rho^{}_B G_{\rm NJL}(\Lambda_0)} \right)
\, .
\end{eqnarray}
One may take an initial UV scale $ \Lambda_0 = \sqrt{eB} $, 
below which the dynamics is dominated by the LLL fermions. 
The size of the dynamical mass should be associated with the emergent scale, so that $m_\dyn \sim \Lm_{\rm IR}  $. 
Therefore, the order of the dynamical mass gap is found to be\footnote{
Hereafter, a quantity $ eB $ is understood as a positive value. 
We omit the symbols of absolute value for notational simplicity. } 
\cite{Fukushima:2012xw, Hattori:2017qio}
\begin{eqnarray}
m^{}_\dyn  \sim \sqrt{eB}\,  \exp \left( -\, \frac{\pi}{ \rho^{}_B G_{\rm NJL}(\sqrt{eB})} \right)
\label{eq:m_dyn_NJL}
\, .
\end{eqnarray}
This result agrees with that obtained from the solution 
of the gap equation \cite{Gusynin:1994xp, Fukushima:2012xw} (see Sec.~\ref{sec:potential-MC}). 
Because of the dimensional reduction, 
each fermion-antifermion pair in the chiral condensate $ \langle  \bar f f \rangle $ 
is squeezed within the size of the cyclotron motion $\sim 1/\sqrt{eB}  $, 
and the size of the chiral condensate is given by that of the squeezed condensate 
multiplied by the Landau degeneracy factor, $\langle \bar f f \rangle \sim  \Lm_{\rm IR} \cdot |e B|/(2\pi) $.

Since we have examined the scaling argument for superconductivity in the previous section, 
it is instructive to revisit the analogy between superconductivity and the magnetic catalysis.  
The above mass gap indeed has a quite similar form 
as that of the energy gap 
emerging in superconductivity \cite{Gusynin:1994xp}: 
$\Delta \sim \omega_{\rm{D}} \, {\rm{exp}}( - c^{\prime} / \rho_{\rm{F}} G_{\rm{S}} )$, 
where $ G_{\rm{S}} $ and $c^{\prime}$ are the coupling constant and a numerical constant of order one, respectively. 
The Debye frequency $\omega^{}_{\rm{D}}$ specifies the band width near the Fermi surface 
where an attractive phonon interaction is effective, while $\rho^{}_{\rm{F}}$ is the density of states at the Fermi surface. 
They are counterparts of the UV scale $  \Lambda_0 $ 
and the Landau degeneracy factor $ \rho^{}_B $ in the magnetic field, respectively. 
The physics behind this similarity is of course the analogous dimensional reductions 
occurring in the IR dynamics (cf. Fig.~\ref{fig:dim}).

\subsubsection{Magnetic catalysis in weak-coupling gauge theories}

\label{sec:RG-gauge}

In the above, we explicitly identified the origin of the logarithmic quantum correction to the four-Fermi interaction 
and discussed the similarity to superconductivity. 
It should be noticed that, in both the magnetic catalysis and superconductivity, 
the exponent depends on the effective coupling constant multiplied 
by the density of states (see Eq.~(\ref{eq:m_dyn_NJL}) and discussions below it). 
This dependence arises because the coupling constant of the four-Fermi interaction model $ G_{\rm NJL} $ 
has a mass dimension that needs to be compensated by another dimensionful quantity in the exponent. 
Putting it differently, this exponent is not a universal quantity and depends on details of the interactions. 
Therefore, the next natural issue is investigating the magnetic catalysis with different types of interactions. 
Especially, it is interesting to apply the RG analysis to gauge theories 
where the four-Fermi interaction is generated by gauge-boson exchanges with a dimensionless coupling constant. 
Below, we examine the magnetic catalysis in the weak-coupling QED 
which serves as a theoretically well-controlled and nontrivial example of 
the chiral symmetry breaking in gauge theories.

To build a bridge to the scaling argument, 
we first construct an effective four-Fermi interaction from a photon exchange interaction (cf. Fig.~\ref{fig:MC-diagrams}). 
An important point in this treatment is that the effective four-Fermi coupling 
potentially acquires an energy-scale dependence arising from the photon propagator. 
This is indeed the case because the photon propagator has 
an obvious energy-scale dependence in the UV region which behaves as the inverse square momentum. 
Moreover, this dependence will be modified in the IR region if there is a screening effect. 
Therefore, the effective coupling could have multiple origins of the energy-scale dependences 
in addition to the quantum correction stemming from the dimensional reduction, 
reflecting properties of the interactions in underlying theories. 
We establish a simple method to include such multiple energy-scale dependences into the RG equations. 
This enables us to keep track of the modifications 
and clarify how they are finally reflected in the mass gap \cite{Hattori:2017qio}. 
This effective theory should be distinguished from the 
na\"ive four-Fermi interactions, like the NJL model discussed above, 
which do not respect intrinsic energy-scale dependences in underlying theories such as QCD.

\begin{figure}
     \begin{center}
              \includegraphics[width=0.8\hsize]{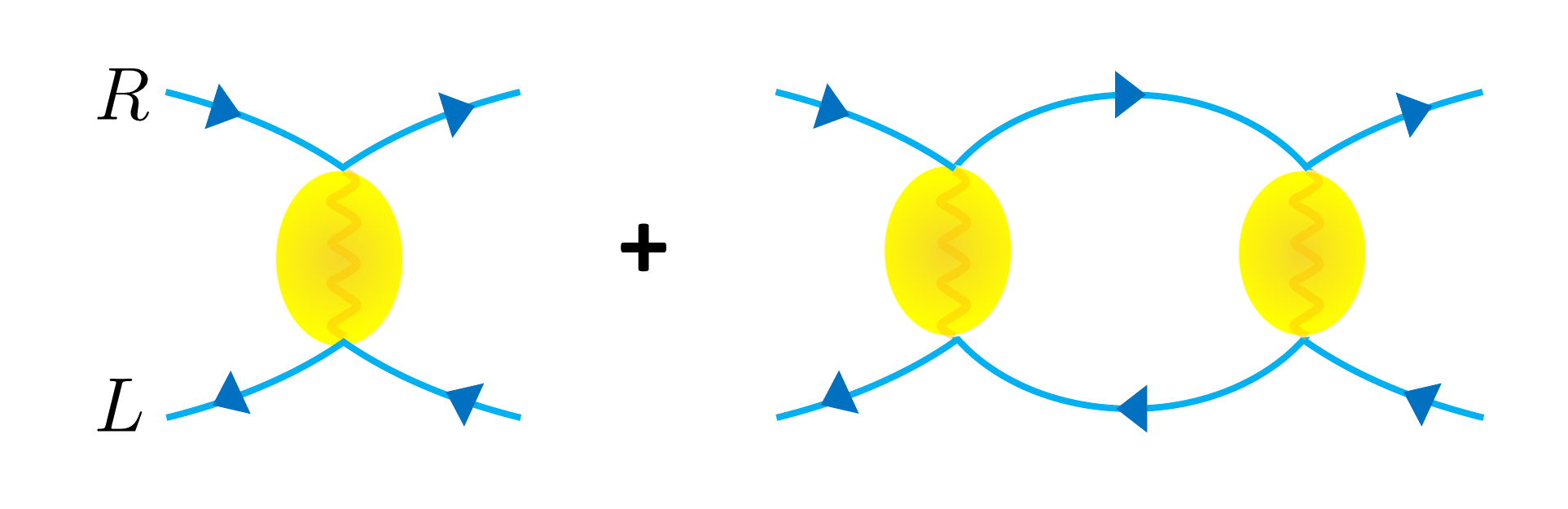}
     \end{center}
\vspace{-0.5cm}
\caption{Scattering diagrams contributing to the RG flow in the magnetic catalysis.  
Yellow blobs denote the effective coupling which are defined 
by integrating the photon propagator in Eq.~(\ref{effecG_magMC}). }
\label{fig:MC-diagrams}
\end{figure}

We introduce the photon propagator in the strong magnetic field
{\it a la} Gusynin et al.~\cite{Gusynin:1998zq, Gusynin:1999pq}: 
\beq
i D_{\mu \nu} (q)
&=& \frac{ g_{\parallel}^{\mu \nu} }{ q^{2} - m_{\gamma}^{2} }
+ \frac{ g_{\perp}^{\mu \nu} }{ q^{2} }  
- \frac{ q_{\perp}^{\mu} q_{\perp}^{\nu} + q_{\perp}^{\mu} q_{\parallel}^{\nu} + q_{\parallel}^{\mu} q_{\perp}^{\nu} }{ (q^{2})^{2} }
\, .
\label{photon_prop_inp}
\eeq 
The polarization of the LLL fermion and antifermion pairs induces the static screening mass 
(see Sec.~\ref{sec:vp} and Appendix~\ref{sec:screening}) 
\begin{eqnarray}
\label{eq:m_gam}
m_{\gamma}^{2}  = N_f \rho_B \cdot \frac{e^2}{\pi}
\, ,
\end{eqnarray}
where $ N_f $ is the number of fermion flavors. 
This screening mass is a gauge-invariant quantity 
and is a close analog of the Schwinger mass~\cite{Schwinger:1962tn, Schwinger:1962tp}. 
Properties of this screening effect have been discussed in the literature~\cite{
Gusynin:1998zq, Gusynin:1999pq, Fukushima:2011nu, Hattori:2012je, Fukushima:2015wck, Hattori:2017xoo}. 
The second and third terms are coupled to neither the LLL fermion loop nor the scattering LLL fermions 
because the LLL fermion current, $ j^\mu_\LLL = \bar \psi_\LLL \gam^\mu_\para \psi_\LLL$, 
is longitudinal to the magnetic field. 
Thus, those terms are irrelevant in the present discussion. 
Now, we note that the photons live in the ordinary four dimensions while the LLL fermion has the (1+1)-dimensional dispersion relation (\ref{eq:e_LLL}). 
Therefore, we define the effective coupling constant $ G $ by integrating out 
the transverse momentum in the photon propagator as \cite{Hattori:2015aki, Ozaki:2015sya}  
\beq
G (\sqrt{-q_\para^2}) \equiv 
(-ie)^{2} \int \frac{ d^2 \bq_{\perp}} { (2\pi)^{2} }
\frac{ 1 }{ q_{\parallel}^{2} - \bq_{\perp}^{2} - m_\gam^2 } \, \e^{ - \frac{ \bq_{\perp}^{2} }{ 2 e B } } 
e^{ - i \frac{ P_y q_x}{eB} }
\label{effecG_magMC}
\, .
\eeq
In this way, we obtain an effective (1+1)-dimensional interaction in the form of Eq.~(\ref{four_int_B}). 
The Gaussian factor comes from the transverse part of 
the fermion wave functions 
as discussed in Sec.~\ref{sec:Ritus-Feynman}. 
This factor cuts off the UV region of the integral. 
We note that the Schwinger phase is not written explicitly. 
In general, the Schwinger phases from the vertices remain in Eq.~(\ref{effecG_magMC}) \cite{Hattori:2015aki}. 
The resultant Schwinger phase factor $ \exp( - i P_y q_x/|eB|) $ with $ {\bm P} \equiv \bp^{(1)} +\bp^{(2)} $ 
can be ignored in the region of our interest with a small $q_\perp $ (see below).

\cout{ 
In the present case, the resultant Schwinger phase is $ \exp( - i P_y q_x/|eB|) $ with $ {\bm P} \equiv \bp^{(1)} +\bp^{(2)} $. 
However, this phase factor reduces to one when we consider a homogeneous condensate 
with a vanishing total momentum ($  {\bm P}=0 $). 
}
 
When the energy and momentum scales of the fermions and antifermions are of the order of $ \Lm $, 
the momentum transfer is $ - \Lm^2 \lesssim q_\para^2 \lesssim 0 $, 
where we are only interested in the space-like region contributing to the fermion-antifermion scatterings. 
If the momentum transfer is much larger than the screening mass scale $(\Lambda \gg \Lambda_{\rm sc} = m_\gam) $, 
the photon propagator reduces to the free propagator, 
and the effective coupling constant depends on the energy scale $G (\sqrt{-q_\para^2}) \sim G ( \Lambda)  $. 
In the opposite limit, if the momentum scale $ \Lambda $ is much smaller than $ m_\gam $, 
the photon exchange can be well approximated by the four-Fermi contact interaction 
that is independent of the momentum transfer $ q_\para^2 $. 
In this case, the effective theory reduces to the (1+1)-dimensional four-Fermi interaction discussed in the previous section. 
Therefore, it is important to specify the hierarchy of the scales in the problem. 
The relevant scales are the initial UV scale $ \Lambda_\UV $, the screening mass $ \Lambda_{\rm sc} = m_\gam $, 
and a possible emergent IR scale $ \Lambda_\IR $ from the Landau pole. 
The initial UV scale $ \Lambda_\UV $ can be taken at $  \sqrt{eB} $ so that 
the higher Landau levels are decoupled from the LLL dynamics. 
One can consider three cases: 
\begin{subequations} \label{eq:QED_hierarchy}
\begin{eqnarray}
 \Lambda_{\rm sc} \gg  \Lambda_\UV \gg \Lambda_\IR 
 \label{eq:QED1}
 \, ,
\\
\Lambda_\UV \gg  \Lambda_{\rm sc} \gg  \Lambda_\IR 
 \label{eq:QED2}
\, ,
\\
\Lambda_\UV \gg  \Lambda_\IR  \gg  \Lambda_{\rm sc} 
 \label{eq:QED3}
\, .
\end{eqnarray}
\end{subequations}
In the first case, the classical effective coupling is a constant independent of the momentum transfer $q_\para^2  $ 
all the way through the RG evolution from $ \Lambda_\UV $ to $ \Lambda_\IR  $. 
In the second case, the effective coupling has an energy-scale dependence even at the classical level 
in the regime between $ \Lambda_\UV $ and $  \Lambda_{\rm sc}  $, 
while it again reduces to a constant below $ \Lambda_{\rm sc} $. 
Figure~\ref{fig:RG-MC} shows the hierarchy in this case. 
In the third case, or simply when the screening effect is neglected, 
the classical effective coupling has an energy-scale dependence arising from the free photon propagator 
all the way through the RG evolution. 
We shall examine how those differences in the photon exchanges 
manifest themselves in the magnitude of the emergent IR scale 
resulting from the RG evolution driven by the quantum logarithmic corrections.

\begin{figure}
     \begin{center}
              \includegraphics[width=0.6\hsize]{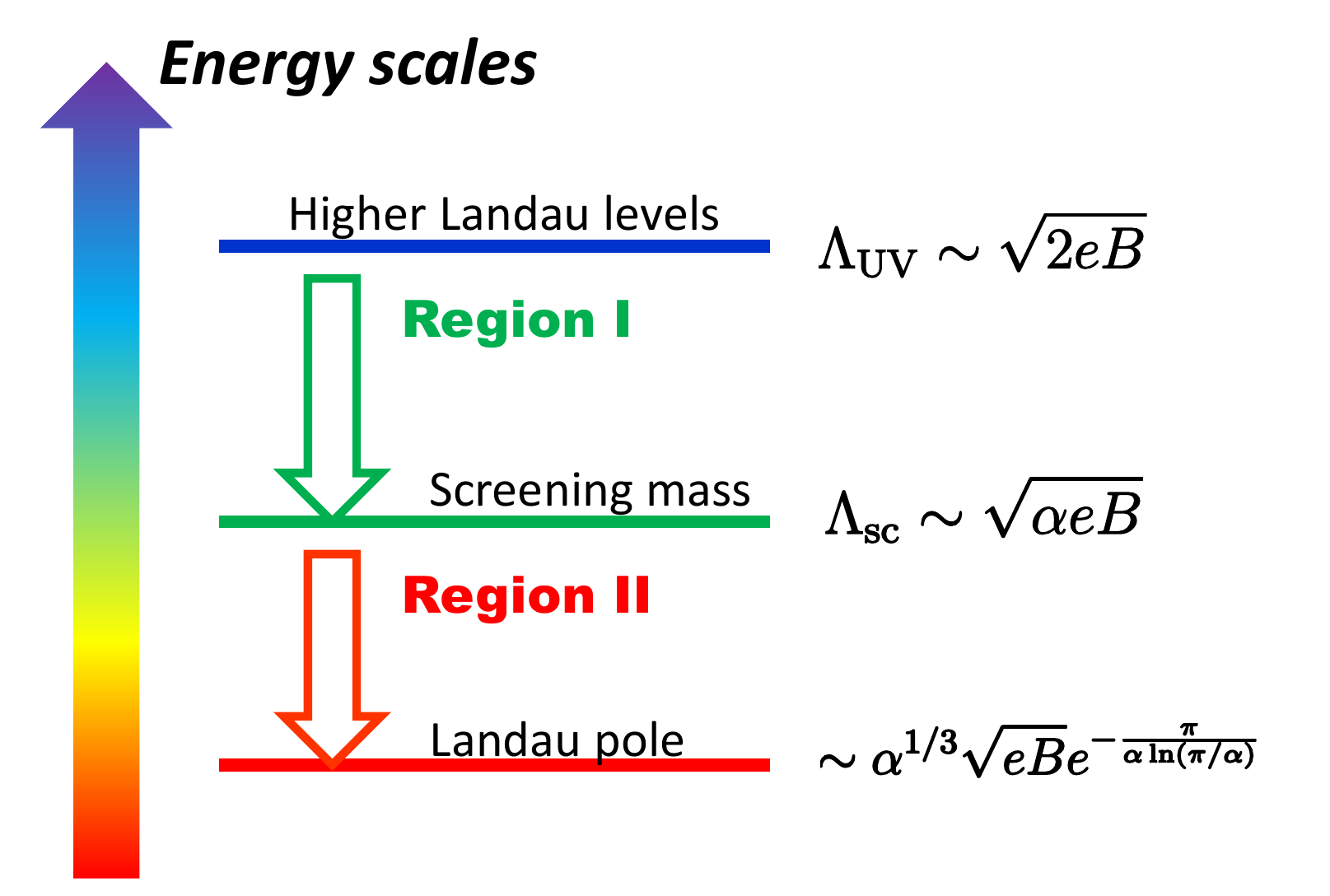}
     \end{center}
\vspace{-0.5cm}
\caption{Hierarchy of the energy scales in the RG evolution. 
The relevant region from the UV cutoff down to the Landau pole 
is divided into two regions by the scale of the screening mass~\cite{Hattori:2017qio}.
} 
\label{fig:RG-MC}
\end{figure}

\subsubsection*{Contact interaction in ``many-flavor QED''}

We first consider the case of a strong screening effect (\ref{eq:QED1}). 
The screening mass scale $\Lambda_{\rm sc} $ becomes much larger 
than the initial UV scale at $ \Lambda_\UV  $ 
if the number of fermion flavors $ N_f $ is sufficiently large~\cite{Gusynin:2003dz}. 
This theory is realized when $ N_f \alpha \gg1 $ according to Eq.~(\ref{eq:m_gam}) and may be called the ``many-flavor QED.'' 
Though this is a sort of hypothetical theories, we can use the many-flavor QED 
as a control test for our understanding of interaction effects. 
In this case, explicit integration in Eq.~(\ref{effecG_magMC}) provides 
\begin{eqnarray}
\label{eq:contact}
G = \rho_B \cdot \frac{ e^2}{m_\gam^2 } = \frac{  \pi }{N_f}
\, ,
\end{eqnarray} 
which is a constant up to possible quantum corrections. 
The mass gap is simply given by the previous result in Eq.~(\ref{eq:m_dyn_NJL}) by identifying  $\rho^{}_B  G_{\rm NJL} \to G$.
Therefore, we get the mass gap 
\begin{eqnarray}
m^{}_\dyn \sim \sqrt{eB} \, e^{ -  N_f }
\, .
\end{eqnarray}
This agrees with the result from the SD equation \cite{Gusynin:2003dz}.

\subsubsection*{Unscreened QED as an example of non-contact interactions}

We consider the third case (\ref{eq:QED3}). 
When there is no screening mass $ ( m_\gam=0) $, 
the photon propagator has an obvious energy dependence $ \sim q_\para^{-2} $. 
Here, we call this model ``unscreened QED.'' 
When the energy scale of the fermionic degrees of freedom is reduced from $ \Lm  $ to $ \Lm-\delta\Lm  $, 
the lower boundary of the photon momentum scale changes from $ - \Lm^2 \lesssim q_\para^2  $ 
to $ - (\Lm-\delta \Lam)^2 \lesssim q_\para^2  $. 
As a consequence, even the tree-level amplitude acquires an increment, 
which contrasts to the constant (classical) coupling constant discussed 
in the many-flavor QED and the NJL model in Eq.~(\ref{tree_amp_magneticfields}). 
In Eq.~(\ref{effecG_magMC}), the energy-scale dependence only appears 
from a small $ |\bq_\perp| $ regime such that $ |\bq_\perp|^2 \lesssim | q_\para^2| \lesssim \Lambda_\UV$, 
where the Gaussian factor and the Schwinger phase factor reduce to unity and the $ \Lambda $ specifies the lower cut-off of the integral. 
Thus, the increment is immediately obtained as 
\begin{eqnarray}
\delta \M_0(\Lm ) 
= G( \Lm - \delta \Lm  ) - G(  \Lm ) 
\sim 
\alpha \int ^{\Lm^2}_{(\Lm-\delta\Lm)^2} \frac{d (q^2_\perp)}{q^2_\perp}
\, .
\end{eqnarray}
Importantly, this increment gives rise to a logarithm 
\begin{eqnarray}
\delta \M_0(\Lm ) = 2 \alpha \log \frac{\Lambda}{\Lam - \delta \Lam}
\label{eq:tree-unscreen}
\, ,
\end{eqnarray}
which partly drives the RG flow and will be finally reflected in the mass gap.

This situation is somewhat, though not exactly, similar to that in color superconductivity in dense quark matter. 
There is only a dynamical screening effect in the color-magnetic interaction in the dense QCD, 
so that the color-magnetic interaction has a rather long-range nature 
as compared to the color-electric interaction which is cut off by a static Debye screening effect. 
In Ref.~\cite{Son:1998uk}, Son showed that the energy-scale dependence of the color-magnetic interaction 
plays an important role in obtaining the correct RG equation. 
Similar to Eq.~(\ref{eq:tree-unscreen}), there is an additional logarithm from the one-gluon exchange diagram 
that finally modifies the exponent of the emergent IR scale from $ \exp(-  c_1/g^2) $ to $ \exp(- c_2/g) $ 
with $ c_{1,2} $ being numerical constants. 
This is a significant effect when the QCD coupling constant $ g $ is a small number. 
We should learn a lesson from this result that the color-magnetic interaction, 
or any kind of energy-scale dependent interactions, 
cannot be naively approximated by the NJL-type contact interaction.

From the one-loop contribution $ \M_1$, we again obtain the logarithmic quantum correction. 
This contribution is the same as that of the NJL model up to possible differences in subleading corrections. 
Therefore, combining the logarithms from the tree-level contribution (\ref{eq:tree-unscreen}) 
and the quantum correction (\ref{eq:MC-one-loop}), the RG equation for the unscreened QED is obtained as 
\begin{equation}
\Lambda \frac{d}{d\Lambda}G(\Lambda)=-{2\alpha } -\frac{1}{\pi}  G^2(\Lambda)
\label{RGeq-unsc_MC}
\, .
\end{equation}
Note that the first term on the right-hand side was absent in Eq.~(\ref{RGeq_MC}) 
and that a similar inhomogeneous RG equation was obtained 
in the aforementioned color superconductivity~\cite{Son:1998uk,Hsu:1999mp}. 
The solution of the above RG equation is obtained as 
\beq
G(\Lambda)
&=&  \sqrt{ 2 \alpha \pi } \,\, {\rm{tan}} 
\left[ - \sqrt{ \frac{  \alpha }{ 2 \pi } }\, \log \frac{ \Lambda^2 }{ \Lm_\UV ^2} 
+  {\rm{arctan}} \left( \frac{ G(\Lm_\UV )} { \sqrt{ 2\pi\alpha } } \right)  \right]
\, .
\eeq
Taking the initial scale $\Lambda_\UV = \sqrt{eB}$, 
the initial value of the effective coupling (\ref{effecG_magMC}) is evaluated as 
\begin{eqnarray}
 G (\Lm_\UV) 
 \simeq \alpha \int_{\Lambda_\UV^2}^{2eB} \frac{ d(q_\perp^2)}{q_\perp^2} 
 = \alpha \,  \log (2eB/ \Lm_{\UV}^{2}) \ll1
 \, .
 \end{eqnarray} 
Neglecting the higher-order terms in $  G (\Lm_\UV)  \ll 1 $, the above solution of the RG equation reads 
\beq
G(\Lambda) \simeq \sqrt{ 2 \alpha \pi } \,\, {\rm{tan}} 
\left( - \sqrt{ \frac{  \alpha }{ 2 \pi } }\,  \log \frac{ \Lambda^2 }{ 2eB}  \right)
\label{GI_MC}
\, .
\eeq 
As we reduce the energy scale $\Lambda$, the solution (\ref{GI_MC}) hits the Landau pole. 
The emergent IR scale can be determined from the location of the Landau pole 
\beq
- \sqrt{ \frac{ \alpha }{ 2 \pi } }  \log\frac{ \Lambda_{\rm{IR}}^{2} }{ 2 eB } 
= \frac{ \pi }{2} 
\, ,
\eeq
which immediately yields 
\beq
m^{}_{\rm{dyn}}  \sim \Lm_\IR \sim \sqrt{2eB} \, \exp\left( - \frac{ \pi }{2} \sqrt{ \frac{ \pi }{ 2 \alpha } } \right)
\label{eq:mdyn-unsc}
\, .
\eeq
Notice that the exponent of this mass gap is different from that in the NJL model (\ref{eq:m_dyn_NJL}), 
because of the long-range photon interaction. 
There is no explicit dependence on the magnetic field in the exponent, so that 
the dependence of $ m_{\rm{dyn}} $ on the magnetic field is milder than that in the NJL model. 

As mentioned earlier, the dynamical mass (\ref{eq:mdyn-unsc}) was also obtained as the constituent mass 
of the NG mode which emerges when the chiral symmetry is spontaneously broken in a magnetic field \cite{Gusynin:1995gt, Gusynin:1995nb}. 
Namely, it was shown that there exists a nontrivial solution of the Bethe-Salpeter equation for the NG boson 
of which the spectrum is composed of the constituent mass of the same form as in Eq.~(\ref{eq:mdyn-unsc}). 
Consistent results were obtained with the Schwinger-Dyson equation 
in the rainbow approximation \cite{Leung:1996qy, Lee:1997zj} 
and the RG analysis \cite{Hattori:2017qio} up to differences in order-one constants. 
This is not an accidental coincidence. 
From the diagrammatic point of view, the RG equation corresponds to 
the resummation of the ladder diagrams with the multiple photon exchanges (cf. Fig.~\ref{fig:MC-diagrams}). 
This is nothing but the Bethe-Salpeter equation in the ladder approximation. 
Also, one can notice a similarity between the rainbow and ladder diagrams 
since the rainbow approximation of the SD equation only holds the planar diagrams of the fermion self-energy. 
Cutting the intermediate fermion propagator in the SD equation provides the same ladder diagrams.

Taking the unscreened QED as an example, 
we have discussed how the energy-scale dependences 
from the underlying theory, 
together with the quantum corrections, 
can be consistently taken into account 
in the effective theory in terms of the RG evolution.

\subsubsection*{RG analysis with the static screening effect}

Lastly, we consider the second case (\ref{eq:QED2}), where 
the screening mass sets an intermediate scale $  \Lm_{\rm sc} \equiv m_\gam$ 
in between the UV and IR scales (see Fig.~\ref{fig:RG-MC}). 
A successful effective theory should correctly take into account 
the intrinsic energy-scale hierarchy in the underlying theories. 
This is an important issue which 
is shared with various systems involving multiple scales. 
We will find a prescription to reproduce the correct form of 
the dynamical mass gap in the ``screened QED'' \cite{Hattori:2017qio} 
that has been obtained with other methods \cite{Gusynin:1998zq, Gusynin:1999pq}.

The basic strategy is summarized in Fig.~\ref{fig:RG-MC}, where the RG evolution 
from the UV to IR scales are divided into two stages by the screening mass scale $ \Lm_{\rm sc}  $. 
When the scale of interest $  \Lm$ is larger than $ \Lm_{\rm sc} $, we call it Region I ($ \Lm > \Lm_{\rm sc} $), 
while the deeper IR region is called Region II ($ \Lm < \Lm_{\rm sc} $). 
An important difference in these regions is found in the integral for the effective coupling (\ref{effecG_magMC}). 
The screening mass is negligible in Region I, while it is important in Region II as the IR cutoff of the integral. 
As we have learned with the unscreened QED just above, 
one should include the energy-scale dependence from the tree-level amplitude in Region I. 
On the other hand, the effective coupling does not have such an energy-scale dependence in Region II 
since the integral region is bounded by 
the constant IR cutoff $  \Lm_{\rm sc}$, 
instead of the floating scale of interest $ -q_\para^2 \sim \Lm^2 $ [see a discussion below Eq.~(\ref{effecG_magMC})]. 
As a consequence, the RG evolutions in these regions are governed by different RG equations.

We construct a set of RG equations, one for each region. 
When solving the RG equations, we switch over from 
the RG equation for Region I 
to that for Region II by smoothly connecting the two solutions 
at the intermediate scale $ \Lm_{\rm sc}  $. 
This strategy was proposed in Ref.~\cite{Ozaki:2015sya} for the RG analysis of the Kondo effect 
in a magnetic field (see Sec.~\ref{sec:QCD-Kondo}) and then applied to the magnetic catalysis \cite{Hattori:2017qio}. 
Performing the integral in Eq.~(\ref{effecG_magMC}), we obtain the effective coupling in each region 
\beq
G(\Lm)
&\simeq& \left\{  \begin{array}{ll}
{\alpha} \,  \log\left( 2 eB / \Lambda^{2}  \right) & 
\quad {\rm Region \ I}
\\
{\alpha} \,  \log \left( 2 eB / m_{\gamma}^{2}  \right) &  
\quad  {\rm Region \ II}
\end{array} 
\right. .
\label{effeG_MC}
\eeq
To emphasize the energy-scale dependence, it would be useful to put the results in different forms 
\beq
G(\Lm - \delta \Lm) - G(\Lm)
&\simeq& \left\{  \begin{array}{ll}
2 {\alpha} \, \log \left( \frac{ \Lambda }{ \Lambda - \delta \Lm } \right) & 
\quad {\rm Region \ I}
\\
0 &  
\quad {\rm Region \ II}
\end{array} 
\right. .
\label{effeG_MC2}
\eeq
As discussed above, there is the logarithmic dependence in Region I, while no such dependence in Region II. 
We have already learned all machinery to 
construct RG equations with the previous models. 
The RG equation in each region reads 
\begin{subequations}\label{RGeq-I-II_MC}
\beq
&
\Lambda \frac{d}{d\Lambda}G(\Lambda)=-{2\alpha } -\frac{1}{\pi}  G^2(\Lambda)
\ \ \ \ \  {\rm Region \ I}
\, , 
\label{RGeq-I_MC}
\\
&
\Lambda \frac{d}{d\Lambda}G(\Lambda)=-\frac{1}{\pi} G^2(\Lambda)
\hspace{1.7cm}  {\rm Region \ II}
\label{RGeq-II_MC}
\, .
\eeq
\end{subequations}

Now, we solve the RG equations and connect the two solutions at the intermediate scale $ \Lm_{\rm sc}$. 
From the solution for the RG equation (\ref{RGeq-I_MC}), 
one finds the running coupling constant at the lower boundary of Region I ($ \Lm = \Lm_{\rm sc}$) 
\beq
G(\Lm^{}_{\rm sc}) \simeq   
\alpha \, \log \frac{ 2eB }{ m_{\gamma}^{2} }
+ \frac{ \alpha^2 }{ 6\pi} \left( \log \frac{ 2eB }{ m_{\gamma}^{2} }\right)^3
\, ,
\label{solution_mgamma}
\eeq 
where we performed an expansion with respect to a small quantity 
$\alpha \, \log( 2eB / m_{\gamma}^2) = \alpha \,  \log( \pi / \alpha) \ll 1$. 
The leading-order term in $\alpha  $ corresponds to the tree-level coupling constant (\ref{effecG_magMC}), 
while the subsequent terms explain the growth of the coupling constant driven by the quantum correction in Region I.

When the scale $\Lambda$ enters Region II, we use the RG equation (\ref{RGeq-II_MC}) 
with the initial condition at the upper boundary of Region II, i.e., $ \Lm^{}_{\rm sc}$, which has been obtained 
as the result of the RG evolution in Region I 
in Eq.~(\ref{solution_mgamma}). 
Then, we find the solution in Region II  
\beq
G( \Lambda) = \frac{ G (\Lm^{}_{\rm sc}) }{ 1 + \pi ^{-1} G(\Lm^{}_{\rm sc})  
\log ( \Lambda/ \Lm_{\rm sc}  )} 
\, .
\label{Solution_of_RegionII}
\eeq
Clearly, this solution has a Landau pole and indicates the emergence of the dynamical IR scale at 
\begin{eqnarray}
\Lm_{\rm IR} = 
\Lm_{\rm sc} \, {\rm e}^{-{\pi}/{G(\Lm_{\rm sc})} }
\label{IRScale_MC}
\, .
\end{eqnarray}
Therefore, plugging Eq.~(\ref{solution_mgamma}) into $G(\Lambda_{\rm sc})$, 
we obtain the magnitude of the dynamical mass gap \cite{Hattori:2017qio}
\beq
m^{}_{\rm{dyn}}\sim \Lambda_{\rm IR} 
\sim m_{\gamma}\, \exp \left\{ - \frac{ \pi }{ \alpha \,  \log(\pi / \alpha ) } + \log \left( \frac{ \pi }{ \alpha } \right)^{\frac{1}{6}} \right\} 
=
 \sqrt{ 2 eB } \left( \frac{ \alpha }{\pi} \right)^{\frac{1}{3}} \! \exp \left( - \frac{ \pi }{ \alpha \, \log (\pi / \alpha ) } \right)
\label{DynamicalMass}
\, ,
\eeq
where the first and second terms in the exponent correspond to those in Eq.~(\ref{solution_mgamma}), respectively. 

Remarkably, the result in Eq.~(\ref{DynamicalMass}) agrees 
with that from the SD equation~\cite{Gusynin:1998zq, Gusynin:1999pq}. 
One can identify the origins of the overall factors of $ \sqrt{eB} $, $ \alpha^{1/3} $, 
and the exponent in Eq.~(\ref{DynamicalMass}) in the language of the RG method \cite{Hattori:2017qio}. 
Especially, in the passing from Eq.~(\ref{IRScale_MC}) to Eq.~(\ref{DynamicalMass}), 
the factor of $ \alpha^{1/3} $ is obtained as a product of 
the $  \alpha$ dependences from different origins 
in $ \Lm_{\rm sc} = m_{\gamma}$ and $ G(\Lm_{\rm sc}) $. 
Reproducing the correct fractional power of $ \alpha^{1/3} $ 
serves as a consistency check of 
the hierarchy scheme implemented above.

\subsubsection*{Short summary}

Here, we summarize the results of the RG analyses on the magnetic catalysis. 
As suggested by the scaling argument, 
the logarithmic quantum correction gives rise to the Landau pole 
in the RG evolution of the effective four-Fermi coupling. 
The energy-scale dependence of the photon-exchange interaction, which is intrinsic in QED, 
can be included in the RG analysis with the effective four-Fermi coupling (\ref{effecG_magMC}) 
constructed from the photon propagator. 
Depending on the hierarchy shown in Eq.~(\ref{eq:QED_hierarchy}), 
we constructed the appropriate RG equations and obtained the correct forms of the dynamical mass gap, 
which are itemized as 
\begin{itemize}
\item[$ \bullet $] $ m_\gam \gg \Lm_\UV$ (``Many-flavor QED'' or equivalent to the NJL model)

\begin{itemize} 
\item[] $  m^{}_\dyn  \sim \sqrt{eB}  \exp \left( - N_f \right) \ll  \sqrt{eB}  \exp \left( - \frac{1}{\alpha}\right)$ 
\end{itemize}


\item[$ \bullet $] $ \Lm_\UV \gg m_\gam \gg \Lm_\IR $ (``Screened QED'')

\begin{itemize}
\item[] $  m^{}_{\rm{dyn}} \sim  \sqrt{ 2 eB } \, \alpha^{1/3} \exp \left( - \frac{ \pi }{ \alpha \, \log(\pi / \alpha ) } \right) $
\end{itemize}


\item[$ \bullet $] $ \Lm_\IR \gg m_\gam  $ or $ m_\gam =0  $ (``Unscreened QED'')

\begin{itemize}
\item[] $ m^{}_{\rm{dyn}}  \sim \sqrt{2eB} \, \exp \left( - \frac{ \pi }{2} \sqrt{ \frac{ \pi }{ 2 \alpha } } \right)$
\end{itemize}

\end{itemize}
In the many-flavor QED, an inequality $ N_f \gg \alpha^{-1}$ as well as $ \alpha \ll1 $ 
need to be satisfied although there is no explicit coupling dependence in the exponent. 
In case of the second hierarchy ($ \Lm_\UV \gg m_\gam \gg \Lm_\IR $), 
the photon exchange is screened below the scale $ m_\gam $, but is not screened above this scale. 
We constructed the two separate RG equations for these regions, 
and connected their solutions at 
the intermediate scale $ m_\gam $. 
For $  \alpha \sim 1/137$, the mass gap is suppressed in the screened QED 
as compared to that in the unscreened QED as expected. 
All these results agree with those from other equivalent methods as discussed above.

\subsection{The QCD Kondo effect}

\label{sec:QCD-Kondo}

\begin{figure}
     \begin{center}
              \includegraphics[width=0.8\hsize]{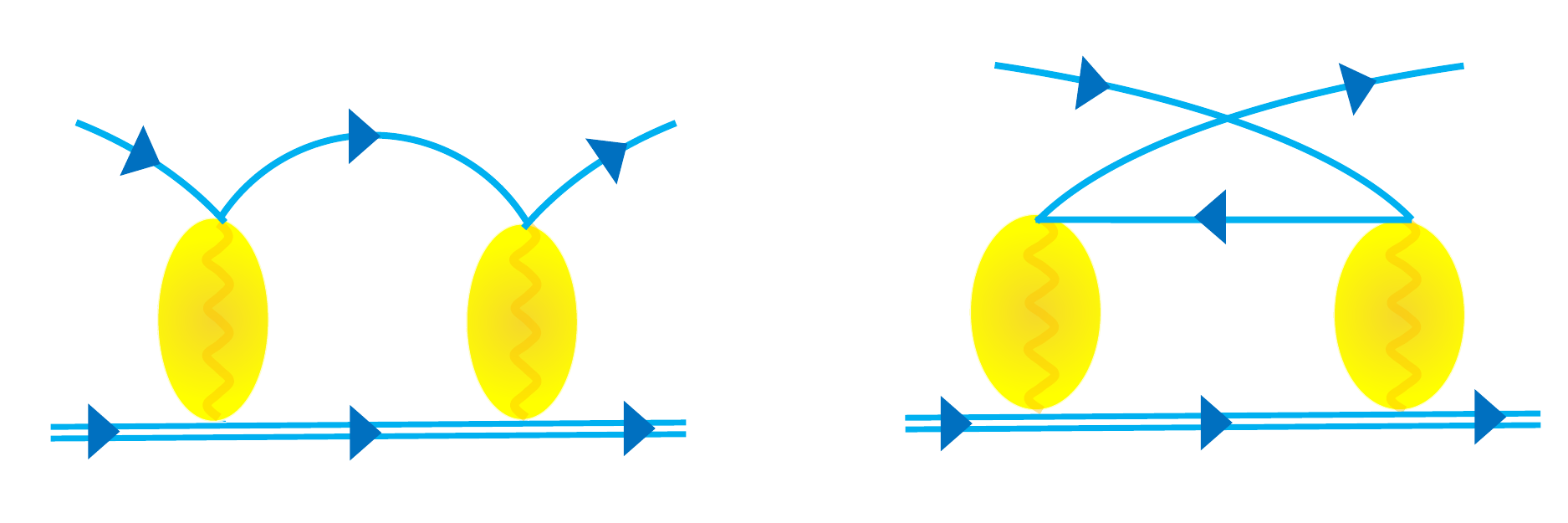}
     \end{center}
\vspace{-1cm}
\caption{Kondo diagrams. 
If there were not a non-Abelian interaction, 
the logarithmic enhancements from those one-loop diagrams completely cancel each other. 
}
\label{fig:KondoDiagrams}
\end{figure}

In this subsection, we proceed to the ``QCD Kondo effect'' occurring in dense quark matter and in a strong magnetic field.
In condensed matter physics, an anomalous temperature dependence of the electric resistivity, which is now called the Kondo effect, 
has been experimentally discovered a long time before J. Kondo pointed out the essential mechanism~\cite{Kondo:1964nea}. 
The existence of a minimum at a certain temperature was a puzzle 
since the electrical resistivity arising from impurity scatterings had been thought to monotonically decrease 
with a decreasing temperature and converge to a constant value 
at zero temperature. 
Then, J. Kondo showed a logarithmic enhancement of 
the resistivity arising from the next-to-leading order scattering diagrams. 
This quantum correction overwhelms classical impurity effects 
at low temperature, and thus gives rise to a minimum in the temperature dependence of the resistivity. 
We can anticipate this result from our scaling argument examined in the early this section. 
The resistivity would increase in the low temperature if there is an emergent strong-coupling regime 
in the interaction between the conduction electrons and an impurity. 
We will see how the strong-coupling regime arises on the basis of the RG method which we are already familiar with.

There are three important conditions for the occurrence of the Kondo effect. 
They are (i) the existence of the Fermi surface, (ii) quantum fluctuations (loop effects), 
and (iii) a non-Abelian property of the interaction 
between charge carriers and an impurity. 
As we have seen in the scaling argument in Sec.~\ref{sec:dense-mag}, 
the interplay between Conditions (i) and (ii) induces logarithmic enhancements 
of the scattering amplitudes between light and heavy particles, e.g., a conduction electron and an impurity. 
There are two relevant diagrams in the next-to-leading order as shown in Fig.~\ref{fig:KondoDiagrams} (see also Fig.~\ref{fig:loop}). 
The intermediate state in the left diagram is a particle-like state, 
while that in the right diagram is a hole-like state with an annihilation on the vertex. 
As we will see explicitly, the logarithms from those diagrams have opposite signs. 
This is because the boundaries of the loop integrals 
approach the Fermi surface from above or below 
depending on whether a diagram is dominated by 
a particle or hole contribution. 
Then, one needs to satisfy Condition (iii) 
so that the logarithms from the two diagrams do not completely cancel each other with noncommutative matrix structures on the vertices. 
The third condition is satisfied with the (pseudo)spin interaction in condensed matter physics. 
Once the necessary conditions and their roles are identified, 
one can look for analogous systems which also satisfy those conditions.

Recently, it is proposed that the Kondo effect occurs in dense quark matter 
with heavy-quark impurities~\cite{Hattori:2015hka} (see also an earlier work \cite{Yasui:2013xr} 
and a short review \cite{Yasui:2018ntm}). 
The non-Abelian interaction is naturally provided by the color-exchange interaction in QCD, 
and we do not need a spin interaction as 
a possible origin of non-Abelian interactions.\footnote{
In fact, the Bohr magneton and thus the spin interaction are suppressed by the heavy-quark mass. 
}
This is called the QCD Kondo effect.  
Based on the analogy between the dimensional reductions at high density and in a strong magnetic field, 
the QCD Kondo effect was shown to occur 
in a strong magnetic field even in the absence of 
the Fermi surface \cite{Ozaki:2015sya}. 
It should be noticed that a strong magnetic field spoils 
the spin flip interaction since the spin direction is frozen 
in a strong magnetic field due to the Zeeman effect: 
It may be difficult to expect a spin interaction 
as an origin of non-Abelian interactions. 
The ``magnetically induced QCD Kondo effect'' can still occur 
as a result of the interplay between 
the light quarks strongly interacting with a magnetic field 
and gluons indifferent to a magnetic field.


Below, we discuss an essence of the QCD Kondo effect in dense quark matter, 
and then proceed to the magnetically induced QCD Kondo effect, 
focusing on the role of the dimensional reductions.

\subsubsection{The QCD Kondo effect in dense quark matter}

Here, we briefly review the RG approach to the QCD Kondo effect in dense quark matter. 
We consider the amplitude of the light-quark excitations near the Fermi surface 
scattering off a heavy-quark impurity at a certain energy scale $\Lambda$. 
This amplitude defines the effective coupling between the light and heavy quarks. 
Then, we derive an RG equation that governs the evolution of the effective coupling when we reduce the energy scale $\Lambda$. 
In this analysis, we assume a sufficiently large chemical potential $\mu \gg \Lambda_{\rm{QCD}}$, 
so that a perturbative analysis with respect to 
a small value of the QCD coupling constant $g$ is justified. 
We also assume that the heavy-quark mass is much larger 
than typical scales of the system such as the chemical potential, i.e., $m^{}_{\rm H} \gg \mu$.

Based on the perturbation theory for the effective interaction at high density (\ref{Effective_Int_QCDKondo}), 
we can evaluate the scattering amplitude between a light quark and a heavy-quark impurity.
In the high-density QCD, the effective coupling $G$ in Eq.~(\ref{Effective_Int_QCDKondo}) 
can be obtained from the one-gluon exchange.  
When a heavy-quark mass is much larger than the other scales in the system, 
one may organize an expansion with respect to $1/ m^{}_{\rm H}$. 
In the leading order of the expansion, i.e., in the heavy-quark limit $  m^{}_{\rm H} \to \infty$, 
only the temporal component of the gauge field is coupled to the heavy quark. 
Therefore, the color electric interaction becomes the dominant interaction in this system, 
and the color magnetic interaction is suppressed by $1/ m^{}_{\rm H}$ at the vertices. 
Below, we consider the $S$-wave scattering channel. 
Then, the $S$-wave projection of the color-electric interaction gives the effective coupling\footnote{
We use the hard dense loop (HDL) gluon propagator
$
D^{\mu\nu}(k) = P_L^{\mu\nu}/(k^2 - \Pi_L) + P_T^{\mu\nu}/(k^2-\Pi_T)
$
where 
$
P_T^{\mu\nu} = \delta^{\mu i} \delta^{\nu j} 
\left(\delta^{ij} - k^i k^j/\vert \bk \vert^2 \right)
$ and $
P_L^{\mu\nu} = - \left(g^{\mu\nu} - k^\mu k^\nu/k ^2 \right) - P_T^{\mu\nu}
$. 
The longitudinal component of the gluon self-energy provides 
the screening mass $  \Pi_L  \to m^{}_{\rm D} $ in the static limit. 
Note also that a similar $  S$-wave projection was performed for the Cooper pairing 
in Refs.~\cite{Hsu:1999mp, Son:1998uk, Evans:1998ek}. 
}
\beq
G\, \delta^{ab}
&=& - (ig)^{2} \frac{1}{2}  \int^{1}_{-1} d \, \cos\theta \ D_{00}^{ab} (p^{(1)} - p^{(3)} ; \mu ) 
\nonumber \\
&\simeq& \frac{g^2}{4 \mu^{2} } \delta^{ab} \int^{1}_{-1} 
 \frac{d \, {\rm{cos}} \theta }{  ( 1 - {\rm{cos}} \theta ) + m_{\rm D}^{2} /(2 \mu^{2} ) } 
 \nonumber \\
&\simeq& \frac{ g^{2} }{ 4 \mu^{2} }\,  \log\left( \frac{ 4 \mu^{2} }{ m_{\rm{D}}^{2} } \right) \delta^{ab}
\label{eq:G-Kondo}
\, .
\eeq
The Debye mass is given as $m_{\rm{D}}^{2} = \alpha_{s} \mu^{2} / \pi$. 
We put the initial and final momenta of the light quark on the Fermi surface 
since the Kondo effect occurs in such a low-energy regime. 
Then, we have integrated the angle between the Fermi velocities 
$\bm{v}_{\rm F}^{(1)}$ and $\bm{v}_{\rm F}^{(3)}$. 
Note that the effective interaction is diagonal in the color space. 
Using this effective coupling, the tree amplitude is expressed as
\beq
\mathcal{M}^{S{\rm{-wave}}}_{0}
&=&  G \sum_{r=1}^{N_c^2-1} (t^{r})_{i j} (t^{r})_{l m}
\, ,
\label{eq:M0-Kondo}
\eeq
where we have suppressed the trivial spinor structure and summed over the color matrix $ t^r $.

Next, we consider the one-loop amplitudes.
There are two relevant diagrams for the Kondo effect, 
i.e., the box and crossed diagrams shown in Fig.~\ref{fig:KondoDiagrams}. 
Based on the effective coupling $G$ and the effective Lagrangians (\ref{eq:HDEF}) and (\ref{kin_Q}), 
those amplitudes are written down as (see Ref.~\cite{Hattori:2019zig} for further details) 
\begin{subequations}
\label{amplitude_Kondo}
\beq
i \mathcal{M}_{1, \, \rm{box}}^{S\rm{-wave }}
&=& -(iG)^{2} \, \mathcal{T}^{\rm{(a)}} \sum_{\bm{v}^{}_{\rm F}}  \int \frac{ d^{4}\ell}{ (2\pi)^{4}} 
\frac{ i^{2} }{ ( -\ell^{0} + i \epsilon )( \ell^{0} - \ell_{\parallel}  + i \ell^0 \epsilon ) } 
\label{amplitude_a_Kondo}
\, ,
\\
i \mathcal{M}_{1, \, \rm{crossed}}^{S\rm{-wave }}
&=& -(iG)^{2}\, \mathcal{T}^{\rm{(b)}} \sum_{\bm{v}^{}_{\rm F}}  \int \frac{ d^{4}\ell}{ (2\pi)^{4}} 
\frac{ i^{2} }{ ( +\ell^{0} + i \epsilon )( \ell^{0} - \ell_{\parallel}  + i  \ell^0 \epsilon )} 
\label{amplitude_b_Kondo}
\, ,
\eeq
\end{subequations}
where the infinitesimal imaginary part is assigned so that 
a positive (negative) value of $ \ell^0 $ corresponds to a particle (hole) excitation near the Fermi surface. 
Again, we put the initial and final momenta of the light quark on the Fermi surface, i.e., $\ell_i^\mu = \ell_f^\mu = 0  $. 
As required by Condition (iii), there is a difference between the color-matrix structures in the two diagrams, 
of which the explicit forms are given as 
\begin{subequations} \label{eq:Tab}
\begin{eqnarray}
\mathcal{T}^{({\rm{a}})}_{i j; lm} &=& 
\sum_{s,r}^{N_c^2-1} \sum_{k, k^{\prime} }^{N_c} (t^{r})_{i k } (t^{s})_{k j} (t^{r})_{l k^{\prime}} (t^{s})_{k^{\prime} m }
\, ,
\\
\mathcal{T}^{\rm{(b)}}_{ij ; lm} &=& 
\sum_{s,r}^{N_c^2-1} \sum_{k, k^{\prime} }^{N_c}  (t^{r})_{i k } (t^{s})_{k j} (t^{s})_{ l k^{\prime}} (t^{r})_{k^{\prime} m }
\, .
\end{eqnarray}
\end{subequations}
Performing the integrals in Eq.~(\ref{amplitude_Kondo}), we have 
\begin{subequations}
\label{amplitude_Kondo2}
\beq
 \mathcal{M}_{1, \, \rm{box}}^{S\rm{-wave }}
&=& - G^{2} \, \mathcal{T}^{\rm{(a)}} \, \rho^{}_{\rm{F}}  
\int \frac{ d \ell_{\parallel} }{ \ell_{\parallel} }\, \theta(\ell_\para)
\label{amplitude_a_Kondo2}
\, ,
\\
 \mathcal{M}_{1, \, \rm{crossed}}^{S\rm{-wave }}
&=& - G^{2} \, \mathcal{T}^{\rm{(b)}}\, \rho^{}_{\rm{F}}  
\int \frac{ d \ell_{\parallel} }{ \ell_{\parallel} }\, \theta(-\ell_\para)
\, ,
\label{amplitude_b_Kondo2}
\eeq
\end{subequations}
where the spinor structures are the same as that of the tree amplitude 
and are suppressed for notational simplicity. 
The step functions stem from the locations of the poles 
with the infinitesimal imaginary parts,\footnote{
One obtains the same result in the both cases where the contour is enclosed in the upper- and lower-half planes. 
} 
and indicate that the box and crossed diagrams correspond to the particle and hole contributions. 
The density of states on the Fermi surface $\rho^{}_{\rm{F}}$ has been obtained as\footnote{
In the high-density effective theory, 
the Fermi velocity $  \bm{v}^{}_{\rm F}$ is the label of patches on the Fermi surface \cite{Schafer:2003jn}. 
Therefore, the combination of the sum over the Fermi velocity 
and the two-dimensional integral on a patch cover the whole Fermi sphere.} 
\beq
\rho^{}_{\rm{F}}
= \sum_{\bm{v}^{}_{\rm F}} \int \frac{ d^{2} \ell_{\perp} }{ (2\pi)^3 } = \frac{ 4 \pi \mu^{2} }{ ( 2\pi)^{3} }
= \frac{ \mu^{2} }{ 2\pi^{2} }
\, .
\eeq
As in the analysis of the magnetic catalysis, 
we obtain the one-loop amplitude given by a product of 
the density of states and the logarithm from 
the remaining integrals in Eq.~(\ref{amplitude_Kondo2}).

Then, the sum of the two one-loop amplitudes, integrated over a thin momentum shell, is obtained as
\beq
\mathcal{M}_{1}^{S\rm{-wave}}
&=&  \mathcal{M}_{1, \, \rm{box}}^{S\rm{-wave }} + \mathcal{M}_{1, \, \rm{crossed}}^{S\rm{-wave }} 
\nonumber 
\\
&=&  - G^{2}  \rho^{}_{\rm{F}} \,  \log\left( \frac{ \Lambda }{ \Lambda - \delta \Lambda} \right) 
\left(\mathcal{T}^{\rm{(a)}}-\mathcal{T}^{\rm{(b)}}\right) 
\, .
\eeq
The relative minus sign in the curly brackets originates from the step functions in Eq.~(\ref{amplitude_Kondo2}), 
and would lead to a cancellation between the logarithms 
if the interaction were an Abelian type. 
In case of a non-Abelian interaction, 
one may use the following decomposition 
\begin{subequations}
\beq
\mathcal{T}^{({\rm{a}})}_{ij; lm}
&=&   \frac{ N_{c}^{2} - 1 }{ 4N_{c}^{2} } \delta_{ij} \delta_{lm} - \frac{1}{N_{c}} 
\sum_r^{N_c^2-1}  (t^{r})_{ij} (t^{r})_{lm}
\, , 
\label{Ta}
\\ 
\mathcal{T}^{({\rm{b}})}_{ij; lm} 
&=&  \frac{ N_{c}^{2} -1 }{ 4 N_{c}^{2} } \delta_{ij } \delta_{lm}+\left(\frac{N_c}{2} -\frac{1}{N_c} \right) 
\sum_r^{N_c^2-1}   (t^{r})_{ij} (t^{r})_{lm} 
\, .
\label{Tb}
\eeq
\end{subequations}
To get the above expressions, one can apply an identity in Eq.~(A.38) of Ref.~\cite{Peskin:1995ev} to Eq.~(\ref{eq:Tab}). 
In this case, the cancellation is not complete, and we find a term surviving in the sum 
\beq
\mathcal{M}_{1}^{S\rm{-wave}}
&=&  G^{2} \frac{ N_{c} }{ 2 } \, \rho^{}_{\rm{F}} \, \log \left( \frac{ \Lambda }{ \Lambda - \delta \Lambda} \right) 
\sum_r^{N_c^2-1} (t^{r})_{ij} (t^{r})_{lm} 
\, .
\label{eq:M1_total}
\eeq
There remains a logarithmic quantum correction that modifies the effective coupling $  G$.

Now, combining the results in Eqs.~(\ref{eq:M0-Kondo}) and (\ref{eq:M1_total}), 
we obtain the RG equation 
\beq
\Lambda \frac{ d G }{ d \Lambda }
= - \frac{ N_{c} }{ 2 } \rho^{}_{\rm{F}} G^{2}.
\eeq
The solution of this RG equation is found to be 
\beq
G(\Lambda)
= \frac{ G (\Lambda_0) }{ 1 + 2^{-1}  N_{c} \rho^{}_{\rm{F}} G(\Lambda_0) 
 \log(\Lambda /\Lambda_0) }
\, .
\label{Solution_QCDKondo}
\eeq
Here, $\Lambda_0$ is the initial energy scale, 
and the initial condition of $G$ is given by the tree-level result in Eq.~(\ref{eq:G-Kondo}) as 
$G(\Lambda_0)=( g^{2}/4\mu^{2} )  \log ( 4\mu^{2} / m_{\rm D}^{2} )$. 
When the interaction is attractive $[ G (\Lambda_0) > 0 ]$, 
the effective coupling (\ref{Solution_QCDKondo}) is enhanced according to a negative beta function 
as the energy scale $\Lambda$ is reduced. 
This behavior is characteristic in the Kondo effect as pointed out first by J. Kondo \cite{Kondo:1964nea}. 
We know analogous patterns in the occurrence of the Landau poles 
for superconductivity and the asymptotic freedom in QCD, 
though the asymptotic freedom in QCD is induced 
by nonlinear gluon interactions instead of the dimensional reduction as mentioned in Sec.~\ref{sec:dense-mag}. 
We read off the location of the Landau pole that is called the Kondo scale: 
\beq
\Lambda_{\rm{K}}
= \Lambda_0 \, {\rm{exp}} \left( - \frac{ 2 }{ N_{c} \rho^{}_{\rm{F}} G(\Lambda_0) } \right) 
= \mu \, {\rm{exp}} \left( - \frac{ 4 \pi }{ N_{c} \alpha_{s}  \log( 4 \pi / \alpha_{s}  ) } \right)
\, ,
\eeq
where we took the initial energy scale as $ \Lambda_0 = \mu $.  
As we approach the Kondo scale, the system becomes non-perturbative 
no matter how small the initial coupling $G(\Lambda_0)$ and $\alpha_{s}$ are. 
Intuitively, the carriers are more strongly trapped around an impurity due to this strong-coupling nature. 
Therefore, the resistance is enhanced below the Kondo temperature $ T_{\rm K} \sim \Lambda_{\rm K} $, 
and there emerges a local minimum at $  T_{\rm K}$ in the temperature dependence of the resistance. 
The reader is referred to Ref.~\cite{Yasui:2017bey} for estimates of the transport coefficients.

The Kondo effect contains rich nonperturbative physics. 
One of important observations among those is that the sign of the two-loop beta function 
depends on the number of scattering channels. 
In case of the QCD Kondo effect, the number of channels $k$ corresponds to that of the flavor degrees of freedom, 
while the non-Abelian group corresponds to the color symmetry group SU($N_{c}$). 
This is called the multi-channel Kondo effect in which the sign of the beta function 
could be changed for a sufficiently large number of the flavor channels as compared to $ N_{c} $. 
This perturbative observation already suggests that it is not straightforward to 
settle the fate of the Kondo effect below the Kondo scale, 
and one needs to ultimately invoke on nonperturbative methods. 
There are seminal works in condensed matter physics which 
developed powerful nonperturbative methods 
such as the Wilson's numerical renormalization group~\cite{Wilson:1974mb} 
and applications of the Bethe ansatz~\cite{Andrei:1980fv, Wiegmann:1980} 
and of the (1+1)-dimensional conformal field theory (CFT)~\cite{Affleck:1990iv, Affleck:1995ge}. 
For the quark matter, the multi-channel Kondo effect was first investigated in Ref.~\cite{Kanazawa:2016ihl}, 
and the CFT approach was applied in Ref.~\cite{Kimura:2016zyv,Kimura:2018vxj} (see also Ref.~\cite{Kimura:2020uhq}). 
In the latter, the CFT analysis was generalized to the $k$-channel SU($N$) Kondo effect with general integers $k$ and $N$. 
Besides, there are a number of developments 
in the QCD Kondo effect including the mean-field analyses 
\cite{Yasui:2016svc,Yasui:2016yet,Yasui:2017izi,Suzuki:2017gde,Yasui:2017bey,Fariello:2019ovo,Hattori:2019zig,Suenaga:2019car,Suenaga:2019jqu,Kanazawa:2020xje,Araki:2020fox,Araki:2020rok,Suenaga:2020oeu,Ishikawa:2021bey,Suenaga:2021wio}.

\subsubsection{The magnetically induced QCD Kondo effect}


Next, we discuss the magnetically induced QCD Kondo effect~\cite{Ozaki:2015sya}. 
We will derive the RG equation for the amplitude of the light quark, 
which carries the electric charge $q_{f}$, scattering off the heavy-quark impurity in the strong magnetic field. 
Here, we take $q_{f}B \gg T^2$ so that the dynamics of light quarks is dominated by the LLL. 
Under this assumption, the gluon screening is also dominantly provided by the quark loop in the LLL. 
Effects of the magnetic field directly exerting on the heavy quark are suppressed by 
an inverse factor of the heavy quark mass $  m^{}_{\rm H}$, 
so that those effects are assumed to be small, $q_fB /m_{\rm H}^2 \ll 1  $.

The computation below goes in parallel to that for the QCD Kondo effect with a similar kinematics in the heavy-light quark scattering and is also partly similar to that for the magnetic catalysis in the LLL (cf. Sec.~\ref{sec:RG_MC}). 
Also, we take the same strategy as in the RG analysis of the magnetic catalysis in the ``screened QED.'' 
Namely, the RG evolution starts from the initial scale $\Lambda_0 \sim \sqrt{q_{f}B}$,\footnote{
As in Sec.~\ref{sec:RG_MC}, we omit the symbols of absolute value for $ |q_{f}B| $.
}
and is divided into the two stages above and below the scale of the gluon screening mass (cf. Fig.~\ref{fig:RG-MC}). 
Similar to Eq.~(\ref{effecG_magMC}), 
we introduce an effective coupling for the heavy-light system 
\beq
G
= (ig)^{2}  \int \frac{ d^{2} \bm{q}_{\perp} }{ (2\pi)^{2} }\, \frac{ g^{}_{00} }{ q_{\parallel}^{2} - \bm{q}_{\perp}^{2} -m_{g}^{2}  }\, {\rm{e}}^{- \frac{ \bm{q}_{\perp}^{2} }{ 4 q_{f}B } }
\, .
\label{effe1+1coupling_int_kondo_in_B}
\eeq
The temporal component of the metric $ g^{00} $ reflects the fact that 
only the electric gluon is relevant in the leading order of the heavy-quark expansion. 
The temporal component of the gluon propagator has 
a static screening mass $m_{g}^{2} = (\alpha_{s} / \pi) q_{f} B  $ in the strong magnetic field 
which is a half of the photon mass (\ref{eq:m_gam}) due to the color trace 
(up to the replacement of $ \alpha_s $ by $\alpha  $ and $  q_fB $ by $ eB$ for $  N_f =1 $). 
While the numerical factor in the Gaussian was one half in Eq.~(\ref{effecG_magMC}), 
it is now one fourth since only the light quarks are subject to the Landau-level discretization. 
For the same reason, the Schwinger phase has a slightly different form 
which is however negligible in the current computation \cite{Ozaki:2015sya}.

We shall perform the integral in Eq.~(\ref{effe1+1coupling_int_kondo_in_B}) 
with respect to the transverse momentum $\bm{q}_{\perp}$. 
Remember that we performed a similar integral to get the results in Eq.~(\ref{effeG_MC}). 
As discussed there, the lower boundary of the integral is given by $\Lambda$ in Region I ($\Lambda > m_{g}$), 
while it should be replaced by $m_{g}$ in Region II ($\Lambda < m_{g}$). 
In both regions, the upper boundary of the integral is given by $4 q_{f}B$ due to the Gaussian factor. 
With these integral regions, we find that 
\beq
G 
\simeq \left\{  \begin{array}{ll}
{\alpha_{s}} \,  \log \left( \frac{ {4} q_{f} B }{ \Lambda^{2} } \right) & \quad \text{Region  I} 
\\
{\alpha_{s}} \,  \log\left( \frac{ {4} q_{f} B }{ m_{g}^{2} } \right) &  \quad  \text{Region II}
  \end{array} \right. .
 \label{Effective1+1GluonExchange}
\eeq
Based on the effective interaction term in Eq.~(\ref{int_term_MQCDKondo}), 
we evaluate the scattering amplitude at the energy scale $\Lambda$.  
In terms of the effective coupling $G$, the tree-level amplitude can be expressed as 
\beq
 \mathcal{M}_{0}^{\rm {LLL}} 
= G \sum_r^{N_c^2-1} (t^{r})_{ij} (t^{r})_{lm} 
\, .  
\label{Leading_amp}
\eeq
The trivial spinor structures are again suppressed for notational simplicity. 
The logarithmic dependence of the effective coupling 
on $\Lambda $ contributes to the RG evolution in Region I.

Next, we consider the one-loop scattering amplitudes. 
The two relevant one-loop diagrams in the Kondo system (cf. Fig.~\ref{fig:KondoDiagrams}) can be written down as 
\begin{subequations}
\beq
 \mathcal{M}^{\rm {LLL}}_{1, \, {\rm{box}} }
&=& G^{2} \, \mathcal{T}^{({\rm{a}})} \  
 \int_{\Lambda - \delta \Lambda}^{\Lambda} \frac{ dk_{z} }{2\pi }\, \frac{ 1 }{ -k_{z} }
 \, ,
\label{one-loop_amp_a}
\\
  \mathcal{M}^{{\rm LLL}}_{1, \, {\rm{crossed}}}
&=& G^{2} \mathcal{T}^{({\rm{b}})}  \, 
 \int_{\Lambda - \delta \Lambda}^{\Lambda} \frac{ dk_{z} }{ 2\pi } \frac{1}{k_{z}}
 \, ,
\label{one-loop-amplitude_b}
\eeq
\end{subequations}
where the spinor part is the same as that of the tree-level amplitude 
and the color factors are already given in Eq.~(\ref{eq:Tab}). 
The total one-loop amplitude at the energy scale $\Lambda - \delta \Lambda$ is thus found to be 
\beq
\mathcal{M}^{\rm LLL}_{1}
&=&   {\mathcal{M}^{{\rm LLL}}_{1, \, {\rm{box}}}} + {\mathcal{M}^{{\rm LLL}}_{1, \, {\rm{crossed}}}} 
\nonumber \\
&=&  G^{2} \frac{N_{c}}{2}\, 
 \log\left( \frac{ \Lambda }{ \Lambda - \delta \Lambda } \right) 
\sum_r^{N_c^2-1} (t^{r})_{ij}(t^{r})_{lm} \, .
\label{total_amp1}
\eeq
The logarithmic contributions from the box and crossed diagrams 
do not cancel each other thanks to the non-Abelian property of QCD. 
The strong magnetic field does not spoil the non-Abelian property, 
and the QCD Kondo effect is compatible with the dimensional reduction induced by the Landau quantization; 
The non-Abelian property would be lost in 
a strong magnetic field if it needed a spin interaction.

Combining the scattering amplitudes at the tree and one-loop levels, 
we find the RG equations for the effective coupling $G$: 
\begin{subequations}
\beq
&&
\Lambda \frac{d}{d\Lambda}G(\Lambda)=-{2\alpha_{s} } -\frac{N_c}{4\pi}  G^2(\Lambda) 
\quad \text{Region  I} 
\label{RGeq-I_magKondo}
\, ,
\\
&&
\Lambda \frac{d}{d\Lambda}G(\Lambda)=-\frac{N_c}{4\pi} G^2(\Lambda) 
\quad \quad \quad \  \ \, \text{Region II}
\, .
\label{RGeq-II_magKondo}
\eeq
\end{subequations}
The first term in the RG equation (\ref{RGeq-I_magKondo}) originates from 
the logarithmic dependence of the tree-level amplitude. 
These RG equations are quite similar to those for the magnetic catalysis in Eq.~(\ref{RGeq-I-II_MC}).  
As in the previous analysis, we successively solve the RG equations in Region I and II, 
and connect the solutions at the screening mass~\cite{Ozaki:2015sya} (see Sec.~\ref{sec:RG_MC}). 
Explicitly, the solution at $\Lambda = m_g $ is obtained from Eq.~(\ref{RGeq-I_magKondo}) as 
\beq
G(m_{g})
&=& \alpha_{s} \, \log\frac{ 4 q_{f} B }{ m_{g}^{2} }
\left\{ 1 + \frac{1}{3} \left( \sqrt{ \frac{ N_{c} \alpha_{s} }{ 8 \pi } }  \log\frac{ 4 q_{f} B }{ m_{g}^{2} } \right)^{2} + \cdots \right\}
\, .
\eeq
Then, the final solution for $ \Lambda < m_g $ is obtained 
from Eq.~(\ref{RGeq-II_magKondo}) as  
\beq
G(\Lambda)
&=& \frac{ G(m_{g}) }{ 1 + \frac{1}{4\pi} N_{c}  G(m_{g})\, \log( \Lambda / m_{g} )}
\label{Solution_magKondo}
\, .
\eeq 
The solution (\ref{Solution_magKondo}) has a Landau pole, 
of which the location gives rise to the Kondo scale 
\beq
\Lambda_{\rm K}
&\simeq& \sqrt{q_{f}B}\, \alpha_{s}^{{ 1/3} }\, {\rm{exp}} \left\{ - \frac{ {4} \pi }{ N_{c} \alpha_{s}  \log\left( {4} \pi / \alpha_{s} \right) } \right\}.
\label{Kscale2}
\eeq 
The expression of the Kondo scale resembles 
that of the dynamical mass induced by 
the magnetic catalysis (\ref{DynamicalMass}) 
as a result of the screening effects 
included by the same hierarchy scheme.

If the Kondo scale (\ref{Kscale2}) only depends on the magnetic field through the prefactor, 
it rapidly increases with an increasing $B$. 
However, the QCD coupling may be evaluated at $\sqrt{q_{f}B}$ consistently to the fact that 
the effective coupling in the Kondo scale must be evaluated at the initial scale $\Lambda \sim \sqrt{q_{f}B}$. 
In that case, the dependence of the running QCD coupling constant $\alpha_{s}$ on the magnetic field strength 
induces a nontrivial $  B$-dependence of the Kondo scale. 
Inserting the running QCD coupling $\alpha_{s} (q_{f}B)^{-1} \simeq b_{0} \,  \log ( q_{f}B / \Lambda_{\rm{QCD}}^{2} )$ with $b_0 = ( 11 N_{c} - 2 N_{f} ) / 12 \pi $ into Eq.~(\ref{Kscale2}), we find the following $B$-dependence: 
\beq
\frac{ \Lambda_{\rm{K}}^{2} }{ \Lambda_{\rm{QCD}}^{2} }
&\simeq& \left[  b_0 \, \log  \left( \frac{ q_{f}B }{ \Lambda^{2}_{\rm{QCD}} } \right) \right]^{-2/3} 
\left( \frac{ q_{f} B}{ \Lambda_{\rm{QCD}}^{2} } \right)^{1 - 2 \gamma } 
\, ,
\label{B-dep_Kscale}
\eeq 
where $\gamma = ( 4 \pi b_0 / N_{c} ) /  \log \left\{ 4 \pi b_0 \,  \log( q_{f}B / \Lambda_{\rm{QCD}}^{2} ) \right\}$ corresponds to the anomalous dimension for $\Lambda_{K}(B)$. 
The Kondo scale (\ref{B-dep_Kscale}) increases slowly, but monotonically, as we increase the magnetic field. 
A similar behavior was discussed for the dynamical quark mass generated 
by the magnetic catalysis in the weak-coupling regime of QCD~\cite{Miransky:2002rp}.


We have seen strong statements in the presence of 
a strong magnetic field: 
No matter how small the coupling strength is, 
an interaction between light fermions and antifermions 
induces the spontaneous chiral symmetry breaking \cite{Gusynin:1994xp, Gusynin:1995gt, Gusynin:1995nb} 
and an interaction between light and heavy fermions 
induces the Kondo effect \cite{Ozaki:2015sya}. 
Then, what if there are both light-light and light-heavy interactions? Which of the magnetic catalysis and the Kondo effect is stronger? 
This is like {\it the shield and spear paradox} 
in ancient China. The competition gives rise to 
a quantum phase transition \cite{Hattori:2022lnh}.

\subsection{Magnetic catalysis in QCD}

\label{sec:MC-strong}


In Sec.~\ref{subsec:MC-weak}, we discussed the magnetic catalysis occurring 
in a weak-coupling theory, i.e., QED, as a consequence of the dimensional reduction. 
One might wonder what happens in QCD as a natural extension, 
where the chiral symmetry breaking plays a crucial role in the formation of the hadron spectrum. 
Indeed, the chiral symmetry in magnetic fields was first investigated with 
the effective models of QCD in the early works \cite{Klevansky:1989vi, Klevansky:1991ey, Klevansky:1992qe, 
Suganuma:1991dw, Suganuma:1990nn, Schramm:1991ex}. 
It then attracted even more attention after the clear recognition of the dimensional reduction mechanism \cite{Gusynin:1994xp, Gusynin:1995gt, Gusynin:1995nb}. 
An incomplete list of numerous studies on the QCD extensions 
includes those by the chiral perturbation theory~\cite{Shushpanov:1997sf, Agasian:1999sx, Agasian:2000hw,  Agasian:2001ym, Agasian:2001hv, Cohen:2007bt, Werbos:2007ym, Agasian:2008tb, Andersen:2012dz, Andersen:2012zc}, 
various types of chiral models~\cite{
Babansky:1997zh, Ebert:1999ht, Inagaki:2003yi, Fraga:2008qn, Klimenko:2008mg, Menezes:2009uc, Boomsma:2009yk, Fayazbakhsh:2010bh,Fukushima:2010fe, Gatto:2010qs, Mizher:2010zb, Gatto:2010pt, Nam:2011vn, Skokov:2011ib, Kashiwa:2011js, Chatterjee:2011ry, Avancini:2012ee, Andersen:2012jf, Ferrari:2012yw, Fraga:2012ev, Fraga:2012fs, Ruggieri:2013cya, Fraga:2013ova, Andersen:2013swa, Allen:2013lda, Ferreira:2013oda, Kamikado:2013pya, Ferreira:2014kpa, Fayazbakhsh:2014mca, Andersen:2014oaa}, 
and holographic models \cite{Preis:2010cq, Erdmenger:2011bw, Mamo:2015dea}.

The most interesting part of studying the magnetic catalysis in QCD is the interplay between 
the intrinsic strong-coupling nature in the asymptotic-free gauge theory 
and that induced by the quantum many-body effect via the dimensional reduction irrespective of 
the coupling strength in underlying microscopic theories. 
Such interplay has been investigated with the first-principle lattice QCD simulations over the last decade. 
The Monte-Carlo simulation is feasible since external magnetic fields do not cause 
the notorious sign problem (as long as the action is free of the sign problem 
in the absence of the magnetic field). 
This is opposed to the cases of electric fields. 
A lot of studies after around 2010 is driven by 
outcome of the lattice QCD simulations 
which we summarize below in detail. 
The reader is also referred to Ref.~\cite{DElia:2012ems} for a review on the lattice formulation, 
Refs.~\cite{Shovkovy:2012zn, Preis:2012fh, Fraga:2012rr, Gatto:2012sp} for complementary review articles 
written in the early stages just after the lattice QCD results came out, 
and Refs.~\cite{Andersen:2014xxa, Andersen:2021lnk} 
for a large volume of references and summary of technical details 
that may be missed in the discussions below.

\subsubsection{Results from lattice QCD simulations}

\label{sec:MC-lattice}

Here, we summarize the major results from the lattice QCD simulations at zero baryon density. 
In most cases, we do not mention details of each simulation set-up, for which the reader is referred to original papers in the reference list.

\subsubsection*{Summary of summary}

First, let us summarize what we summarize. 
The main body of the summary will be somewhat long 
as we cite as many simulation results as possible 
in a chronicle way to clarify the issues resolved 
and not yet resolved as of the end of 2022. 
We also would like to give some comments and interpretations. 
Before doing so, we summarize minimum materials.

\begin{itemize}

\item Chiral condensate at zero temperature: 
The lattice QCD simulations have shown the enhancement of 
the chiral condensate at zero temperature as expected 
from the magnetic catalysis. 
However, there appeared a deviation between the results from 
the lattice QCD simulation and the chiral perturbation theory 
in the regime of strong magnetic fields.

\item Chiral symmetry restoration at finite temperature: 
The simulation results at finite temperature were thought to be 
puzzling. The magnitude of the chiral condensate 
decreases as we increase the magnetic-field strength 
in spite of the increase at zero temperature. 
This phenomenon was named the {\it inverse magnetic catalysis}. 
It also appeared that the pseudo-critical temperature 
for the crossover chiral phase transition {\it decreases} 
as we increase the magnetic-field strength 
in contrast to the increase in QED and the NJL model 
(cf. Sec.~\ref{sec:chiral-T}).

\item Deconfinement phase transition: 
The deconfinement phase transition temperatures 
read off from the Polyakov loop and the strange-quark susceptibility decrease as we increase 
the magnetic-field strength in a similar manner 
as the chiral phase transition temperature. 
Also, the heavy-quark potential has been measured. 
The string tension along the magnetic-field direction 
decreases as we increase the magnetic-field strength.

\item Diagnosis of the inverse magnetic catalysis I - III: 
A magnetic field couples to quarks in the QCD action 
and observable operators. 
To identify the mechanism leading to the inverse magnetic catalysis, simulations were performed by turning off either of these couplings. 
It appeared that the chiral condensate is suppressed by 
the coupling to the QCD action near the transition temperature, 
while the coupling to the QCD action at low temperature 
as well as the coupling to the observable operators 
at any temperature act to enhance the chiral condensate.

One can also tune the quark mass as a part of simulation set-up 
and confirm that the inverse magnetic catalysis only occurs with 
a nearly physical-point quark mass where 
the dynamical quarks are well excited. 
Nevertheless, even with a large quark mass, 
the chiral phase transition temperature was 
found to decrease irrespective of 
the presence/absence of the inverse magnetic catalysis. 
In this respect, it is more useful to regard the inverse magnetic catalysis and the decrease of the chiral phase transition temperature as separate phenomena, 
and rather discuss the decreasing behaviors of 
all the aforementioned pseudocritical temperatures 
in parallel as a consequence of the magnetic enhancement 
of the dynamical-quark effects, that is, 
an enhancement of the screening effect.


\item Exploring the further strong-field regime: 
When we increase the magnetic-field strength, 
the simulation results exhibit a systematic tendency 
for an enhancement of 
the strength of the chiral phase transition. 
Very recent simulation results suggested that 
the chiral phase transition is still a crossover transition 
at $eB = 4 \ \GeV^2$ but turns into a first-order transition 
at $eB = 9 \ \GeV^2$. This can be the first indication 
of a first-order phase transition emerging 
directly from the QCD action once it is confirmed. 
There are also discussions about the existence 
of a first-order phase transition 
and an associated critical point in the deconfinement phase transition.

\item Hadron spectrum: Simulation results for 
neutral and charged $\pi $, $\rho $, $ K$, $K^\star $ 
are available so far at zero temperature. 
Very recent results from dynamical simulations 
suggest drastic modifications of 
the results from the quenched simulations. 
The Gell-Mann--Oaks--Renner relation in magnetic fields 
was confirmed by the lattice QCD simulation for the first time. 
The measurement of the heavy-quark potential was extended 
to the ever strongest magnetic field $eB = 9 \ \GeV^2$. 
At finite temperature, the meson correlation length, 
or the screening mass, was measured very recently.

\item Magnetization: The magnetic susceptibility of 
the QCD matter was measured in the temperature range 
$100 \lesssim T \lesssim 300 \ \MeV$, indicating 
diamagnetism at $T \lesssim 150 \ \MeV $ 
and paramagnetism in the higher temperature region. 
The quark spin contribution was measured separately 
by an independent method. 
The quark spin contribution is found to be diamagnetic 
within the available data set up to $T \lesssim 200 \ \MeV$. 
However, the quark spin and orbital contributions should 
converge to the Pauli paramagnetism and the Landau diamagnetism, respectively, at asymptotically high temperature 
where quarks and gluons are expected to behave as free particles.  
The convergence could be slow with the logarithmic running 
of the QCD coupling constant. 

\end{itemize}

\subsubsection*{Chiral condensate at zero temperature}

\begin{figure}
     \begin{center}
              \includegraphics[width=\hsize]{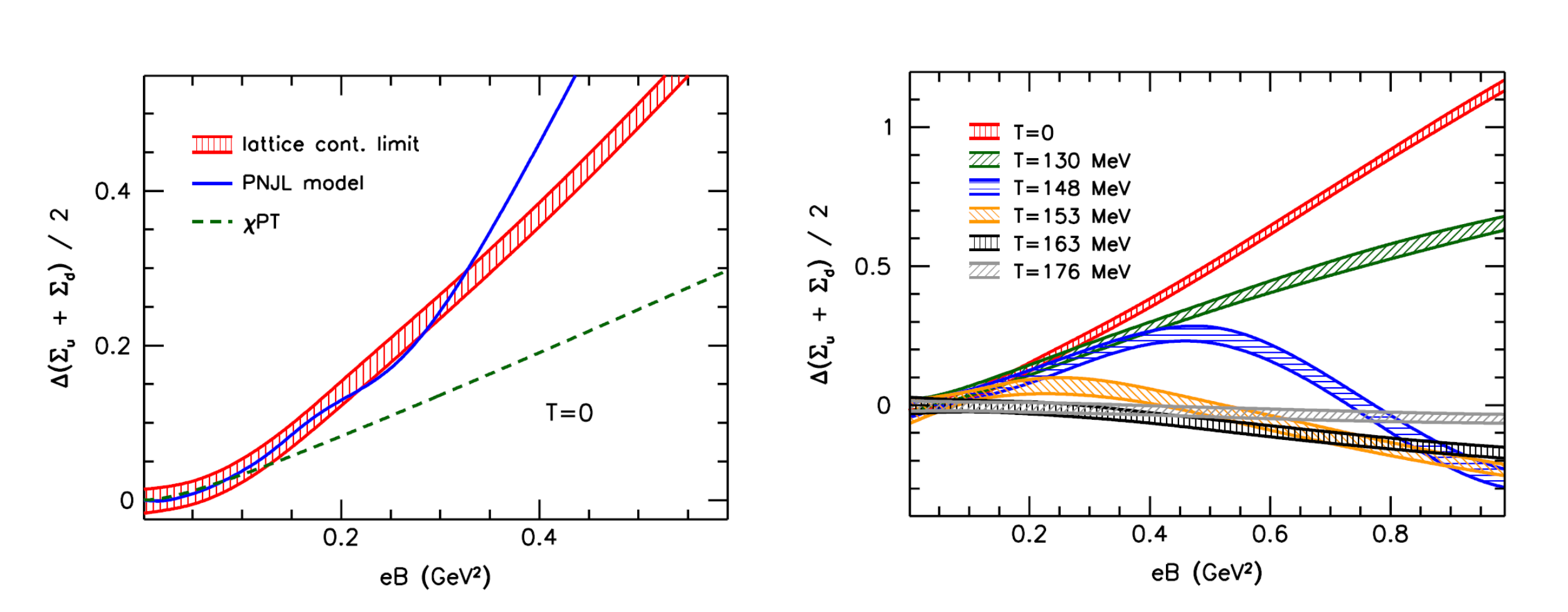}
     \end{center}
\vspace{-0.7cm}
\caption{Lattice QCD simulation of the chiral condensate at $ T=0 $ (left) and $ T \not = 0 $ (right) 
with the physical quark mass \cite{Bali:2012zg}. 
The chiral condensate at nonzero $ eB $ is shown as the difference, 
$ \Delta \Sigma = \Sigma(eB) - \Sigma(0) $, 
and is averaged over the contributions of $  u$ and $d  $ quarks. 
The lattice simulation result is compared with the results from the NJL model \cite{Gatto:2010pt} 
and the two-loop chiral perturbation theory \cite{Andersen:2012dz, Andersen:2012zc}. 
}
\label{fig:lattice-chiral}
\end{figure}

As expected from the magnetic catalysis, the lattice QCD  simulations, performed at zero temperature, have shown an enhancement of the chiral condensate with an increasing magnetic-field strength. 
A numerical simulation in a background magnetic field 
was first performed for the two-color and quenched QCD \cite{Buividovich:2008wf}\footnote{
We add that the earliest numerical simulations were 
performed in an abelianized chromo-magnetic field \cite{Cea:2005td, Cea:2007yv}. 
} 
and then the three-color and quenched QCD \cite{Braguta:2010ej}. 
The simulation was refined with the inclusion of 
the dynamical quarks \cite{DElia:2010abb, DElia:2011koc}. 

\cout{
As we have seen in the preceding sections, 
effects of the magnetic field on the formation of the chiral condensate is two fold. 
The magnetic field induces the dimensional reduction leading to the magnetic catalysis 
and also modifies the screening effects due to the quark loops inserted in the gauge-boson propagators. 
The latter effect is absent in the infinitely heavy quark limit, i.e., the quenched simulations, 
and becomes larger as the quark mass is reduced to the physical point. 
Therefore, the quark mass may play an important role 
(both with and without the magnetic fields). 
The authors of Ref.~\cite{DElia:2011koc} analyzed those two effects separately 
and proposed a further analysis with tuning of the quark mass (see below for comments on the mass dependence). 
\com{There was overlap between the above and the comment below. 
The above may be removed.}
}

A lot of numerical improvements, such as simulations with the physical quark mass, 
larger system volumes, and finer lattice spacings, were achieved in 
Ref.~\cite{Bali:2012zg}. 
In the left panel of Fig.~\ref{fig:lattice-chiral}, 
the lattice QCD simulation result is compared with 
the two-loop chiral perturbation theory \cite{Andersen:2012dz, Andersen:2012zc} 
and Polyakov-loop extended NJL model  \cite{Gatto:2010pt}. 
The chiral condensate exhibits a quadratic 
increase in the small $ eB $ region. 
This is a natural behavior since the chiral condensate is a charge-conjugation even quantity. 
The chiral perturbation theory indeed reproduces the quadratic increase $ \sim \order( eB^2/ (f_\pi^2 m_\pi^2) ) $ 
with $ f_\pi $ and $ m_\pi $ being the pion decay constant 
and pion mass, respectively \cite{Cohen:2007bt, DElia:2011koc}.\footnote{
The symmetry argument for the quadratic dependence assumes 
an analytic dependence of the chiral condensate on $ |eB| $ at $ eB=0 $; 
Otherwise, the leading correction could be $ |eB|^\nu $ with $ 0<  \nu < 2$. 
If there is any non-analytic dependence, 
it should be studied with a finite pion mass 
so that one has a dimensionless expansion parameter $eB/m_\pi^2$ \cite{Cohen:2007bt, DElia:2011koc}. 
}
However, it should be noticed that the simulation result 
turns into a linear increase 
in the large $  eB$ region.\footnote{
The earlier result with a heavier quark mass had shown a quadratic increase \cite{DElia:2011koc} (see below for comments on the quark mass dependence.) 
}
In this region, the chiral perturbation theory 
deviates from the simulation result. 
A very recent work suggests that this linear increase 
continues to the ever largest magnetic-field strength at 
$  eB =  9 \ \GeV^2$ \cite{DElia:2021tfb}.

The chiral perturbation theory is a low-energy effective theory of QCD with 
the low-lying degrees of freedom, 
pions as the most important contribution~\cite{Shushpanov:1997sf, Agasian:1999sx, Agasian:2000hw,  Agasian:2001ym, Agasian:2001hv, Cohen:2007bt, Werbos:2007ym, Agasian:2008tb, Andersen:2012dz, Andersen:2012zc}. 
Thus, its results should agree with the first-principle calculations for QCD in relatively weak magnetic fields 
$  eB / \Lambda_\QCD^2 \ll 1 $. 
The computation at small values of $ m_\pi^2, \, eB \ll \Lambda_\QCD^2 $, 
but for an arbitrary ratio $ m_\pi^2/eB $, was performed in Ref.~\cite{Cohen:2007bt, Werbos:2007ym} 
as an extension of the computation at the chiral limit $ (m_\pi^2/eB =0) $ \cite{Shushpanov:1997sf}. 
The chiral perturbation theory, however, would not work as we increase the magnetic field strength, which appears as 
the deviation observed in 
the left panel of Fig.~\ref{fig:lattice-chiral}. 
This is because the low-lying degrees of freedom 
can be strongly modified by strong magnetic fields. 
For example, due to the explicit isospin-symmetry breaking in magnetic fields, 
charged pions are no longer the Nambu-Goldstone (NG) bosons; 
This is consistent with 
the observation that the ground-state energy levels of spinless charged particles 
are gapped out by the Landau quantization $ m_{\pi^\pm}^2 (B) =  m_{\pi^\pm}^2(B=0) + |eB| $. 
Also, even neutral pions could be modified 
when internal motion of quarks and antiquarks are 
strongly constrained by strong magnetic fields. 
Eventually, naive hadronic effective theories break down when a large number of magnetic lines 
penetrates through the composite particles with the flux density larger 
than the QCD scale $ eB \sim \Lambda_{\rm QCD}^2 \sim 0.04 \ {\rm GeV}^2 $. 
One needs to reconstruct hadrons with quarks and gluons in strong magnetic fields.

\subsubsection*{Chiral symmetry restoration at finite temperature}

The next question would be how the chiral symmetry restoration occurs at finite temperature. 
One may think that the transition temperature $T_c $ increases as we increase the magnetic field. 
We have indeed found that $ T_c \sim \sqrt{| eB|} \sim m_\dyn$ in Sec.~\ref{sec:potential-MC} 
for the NJL model \cite{Fukushima:2012xw} and for QED \cite{Gusynin:1997kj, Lee:1997uh}. 
One can expect such results in these theory and model 
since the long-range correlation can be screened 
by the thermal excitations only when the excitations are activated over the mass gap of the order of $ m_\dyn \sim \sqrt{| eB|} $. 
In other words, the emergent scale is only one available scale 
that characterizes both $ m_\dyn $ and $ T_c  $ 
associated with the magnetic catalysis. 
In QCD, we, however, have an intrinsic scale, 
i.e., the QCD scale $\Lambda_\QCD $, 
which could change the story.

Temperature dependence of the chiral condensate 
and associated chiral susceptibility has been studied 
with the lattice QCD simulations. 
The simulation result at the physical point, i.e., 
at the physical value of the current quark mass, has shown that 
the transition temperature {\it decreases} with an increasing magnetic field \cite{Bali:2011qj}. 
We summarize this surprising outcome 
and follow-up results in the followings. 

In the right panel of Fig.~\ref{fig:lattice-chiral}, 
one finds that, while the chiral condensate is an increasing function of $ eB $ at zero and low temperatures, 
it turns into a decreasing function at higher temperatures. 
This turnover is more clearly seen in the left panel in Fig.~\ref{fig:lattice-phase}; 
While the magnitudes of the chiral condensate is 
in an ascending order in the low temperature region, 
it becomes a descending order in the high temperature region.  
Furthermore, the transition temperature, read off from the chiral condensate, decreases with an increasing $eB $. 
The suppression of the chiral condensate 
observed near the transition temperature is 
dubbed the ``inverse magnetic catalysis.'' 
To be specific, we shall call the suppression of 
the chiral condensate the inverse magnetic catalysis 
rather than the decrease of the transition temperature. 
It appeared that those two phenomena 
are not necessarily equivalent with each other 
as observed in more recent simulation results 
(see Diagnosis II below).

In the right panel of Fig.~\ref{fig:lattice-phase}, 
the decreasing behavior of the transition temperature 
is shown by a red band with error estimates \cite{Endrodi:2015oba}. 
Interpolation to a newer data point at $ eB = 3.25 \ \GeV^2$, which appeared after the original work \cite{Bali:2011qj}, 
suggests a monotonic decrease up to $ eB = 3.25 \ \GeV^2$. 
The other data for the transition temperatures, 
read off from different observables, are discussed just below. 
Summary of the early numerical studies is available in Ref.~\cite{Bali:2013cf} 
(see also Refs.~\cite{Ilgenfritz:2012fw, Ilgenfritz:2013ara} for the two-color QCD).  

\begin{figure}
     \begin{center}
              \includegraphics[width=\hsize]{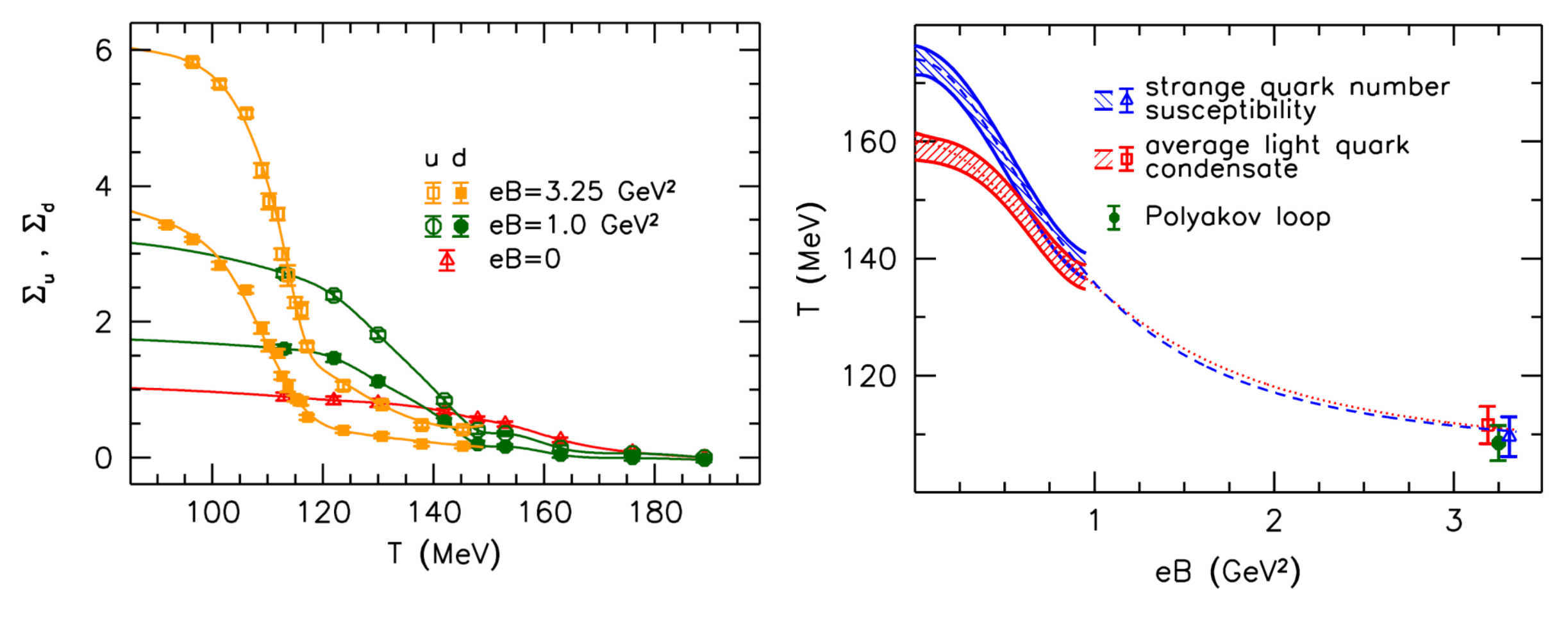}
     \end{center}
\vspace{-0.7cm}
\caption{ 
The temperature dependence of 
the $ u \,(d)$-quark condensate $ \Sigma_u \, (\Sigma_d)$ (left) and 
the decreasing behavior of the transition temperatures $ T_c $ 
read off from the strange-quark susceptibility, 
the light-quark condensate, and the Polyakov loop (right), 
both taken from \cite{Endrodi:2015oba}.  
}
\label{fig:lattice-phase}
\end{figure}

\subsubsection*{Deconfinement phase transition}

Phases of QCD in high and low temperature regions 
are distinguished by the color confinement property, 
as well as by the magnitude of 
the chiral condensate discussed above. 
The Polyakov loop serves as an order parameter for the deconfinement phase transition in a pure gluon theory. 
The strange-quark susceptibility is also studied 
as an alternative indicator for the deconfinement phase transition with quarks (see, e.g., Refs.~\cite{Aoki:2006br, Aoki:2009sc, Borsanyi:2010bp, Bazavov:2011nk}) 
since the Polyakov loop is not a strict order parameter when the center symmetry is explicitly broken by inclusion of quarks (cf. Sec.~\ref{sec:Polyakov} and references therein).\footnote{
While sensitivity of the Polyakov loop to the deconfinement phase  transition should increase as the quark mass is increased to approach infinity, the sensitivity is yet questionable 
with light quarks near the physical point. 
The strange-quark susceptibility 
is not a strict order parameter either, 
but is expected to capture a difference 
between the confinement and deconfinement phases. 
That is, the strange-quark density fluctuation 
is enhanced in the deconfinement phase 
where the expected temperature scale $\gtrsim 150 \MeV $ 
is larger than the current strange-quark mass. 
In the confinement phase, the fluctuation is suppressed 
since strangeness is confined in hadrons heavier 
than the temperature scale. 
Accordingly, the strange-quark susceptibility can work better as an indicator 
of the confinement property than light-quark observables. 
}
For the sake of reference, we note that 
the transition temperatures at a vanishing magnetic field, which are read off from the Polyakov loop and the strange-quark susceptibility, 
are $170 \ \MeV $ and $169 \ \MeV $ 
up to error estimates, respectively \cite{Aoki:2009sc}; 
The difference between those transition temperatures 
are much smaller than the magnetic-field effects shown 
in the right panel of Fig.~\ref{fig:lattice-phase}.

In the right panel of Fig.~\ref{fig:lattice-phase}, 
the phase transition temperature read off from 
the strange-quark susceptibility is shown 
by the blue line \cite{Bali:2013cf} with a new data point 
at $ eB = 3.25 \ \GeV^2$ \cite{Endrodi:2015oba}. 
This transition temperature decreases 
as we increase the magnetic-field strength.

The Polyakov loop was also measured in Refs.~\cite{DElia:2010abb, Bruckmann:2013oba, Bornyakov:2013eya, Endrodi:2015oba, DElia:2018xwo, Endrodi:2019zrl, Ding:2020inp}. 
In the right panel of Fig.~\ref{fig:lattice-phase}, 
the transition temperature read off from the Polyakov loop 
is shown by a marker at $ eB = 3.25 \ \GeV^2$ \cite{Endrodi:2015oba}. 
The transition temperature at a vanishing magnetic field is 
about $170 \ \MeV $ as mentioned above \cite{Aoki:2009sc}. 
Interpolation between those two points suggests 
a decrease of the transition temperature read off from 
the Polyakov loop as well 
(see also Ref.~\cite{DElia:2018xwo} for the decreasing behavior 
observed at a larger pion mass $m_\pi  = 440 \ \MeV$).

Those observations suggest that 
the deconfinement phase transition 
occurs significantly at lower temperatures 
in strong magnetic fields than at a vanishing magnetic field 
{\it albeit} in a crossover manner. 
Those two transition temperatures 
as well as that from the chiral condensate 
exhibit the similar decreasing behaviors. 
Also, the three data points at $ eB = 3.25 \ \GeV^2$ indicates  degeneracy among the three transition temperatures, 
while, at a vanishing magnetic field, 
there is splitting between the transition temperatures 
from the chiral condensate and strange-quark susceptibility  
by about $ 10-15 \ \MeV$. 

Measurement of the heavy-quark potential provides 
complementary information of the confinement properties and 
a more direct access to modification of 
the confinement potential profile 
such as the spatial anisotropy induced 
by magnetic fields \cite{Bonati:2014ksa, Bonati:2016kxj, Bonati:2017uvz, DElia:2021tfb, DElia:2021yvk} (see the summary of hadron spectrum given below for more discussions). 
It is worth adding that a heavy-quark potential 
in a strong {\it electric} field was also investigated with lattice simulation 
{\it albeit} for a sign-problem free theory 
that contains two fermions having the same magnitudes and opposite signs of electric charges \cite{Yamamoto:2012bd}. 
A strong electric field gives rise to a tendency towards 
deconfinement because the linear confinement potential is 
diminished by a voltage that is also 
a linear function of the distance. 
The critical electric field, which cancels the confinement potential, is given by the string tension.

\subsubsection*{Diagnosis of the inverse magnetic catalysis I: Sea-quark effect}

\begin{figure}[t]
\begin{center}
   \includegraphics[width=0.9\hsize]{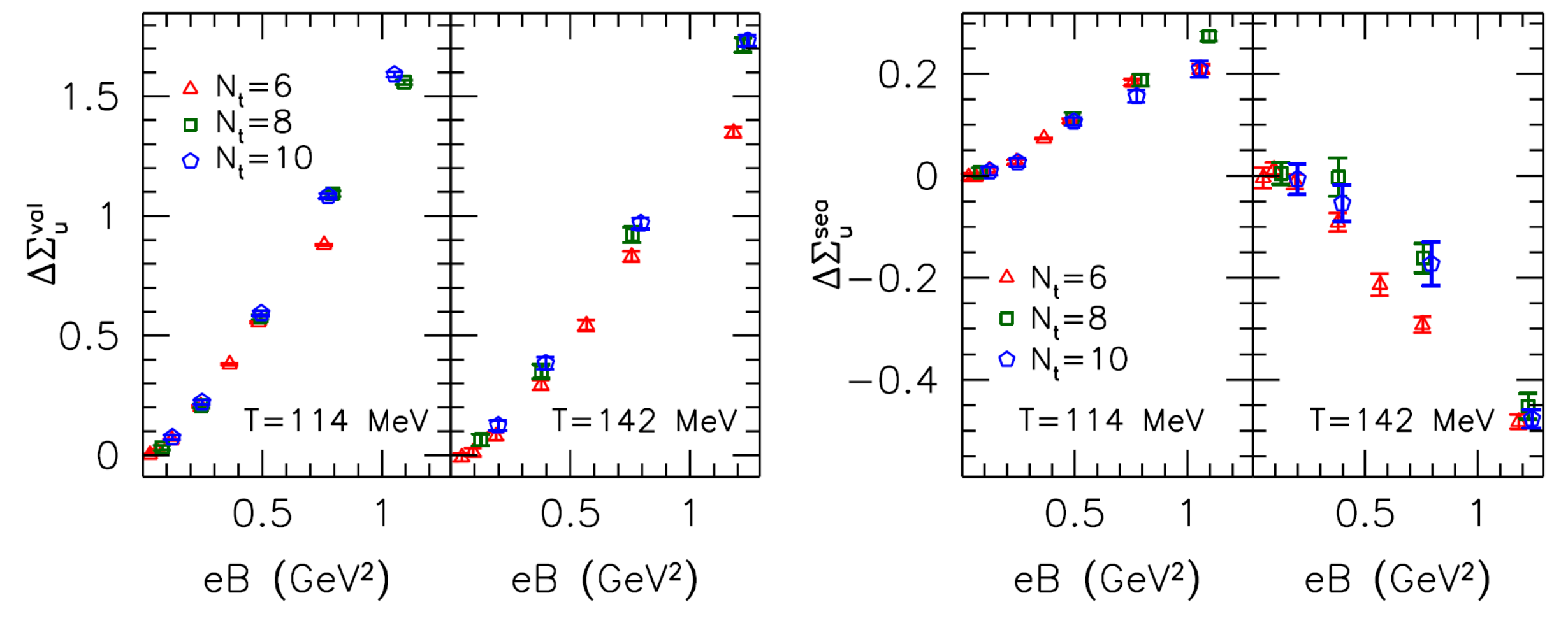}
\end{center}
\vspace{-0.5cm}
\caption{Magnitudes of the chiral condensate 
with the valence-quark effect (left) 
and with the sea-quark effect (right), 
taken from Ref.~\cite{Bruckmann:2013oba}. 
The chiral condensate at a vanishing magnetic field is subtracted 
to isolate the magnetic-field effects. 
}
  \label{fig:valence_vs_sea}
\end{figure}

The inverse magnetic catalysis, 
which refers to the suppression of the chiral condensate, 
and the decrease of the chiral transition temperature 
shed light on distinctive properties of QCD 
not seen in QED and the chiral models. 
It should be a consequence of nonperturbative aspects of QCD probed by strong magnetic fields. 
The deconfinement transition temperatures, read off from 
the Polyakov loop and the strange-quark susceptibility,  
also show the decreasing behaviors similar to 
that of the chiral transition temperature.

To identify the mechanism leading to the inverse magnetic catalysis, it is useful to investigate the breakdown of 
the magnetic-field effects as originally proposed in Ref.~\cite{DElia:2011koc}. 
In the lattice QCD simulations, magnetic fields couple 
to the quark fields in the observable operator $  \bar \psi \psi$ and the QCD action; We shall call them the ``valence-quark effect'' and ``sea-quark effect,'' respectively, 
according to Ref.~\cite{Bruckmann:2013oba} 
where improved numerical simulations were reported. 
The simulation results have shown that 
the ``valence-quark effect'' from the former coupling 
enhances the chiral condensate at any temperatures 
as expected from the magnetic catalysis \cite{Bruckmann:2013oba}. 
Along with the Banks-Casher relation (in the massless limit), 
this effect may be rephrased as an enhancement of 
quark zero modes by the Landau degeneracy. 
Similar interpretation holds with 
the low-energy spectral density in massive cases. 
On top of this picture for non-interacting quarks, 
QCD interaction effects can change 
the dependence on the magnetic-field strength 
through modification of the spectral density 
at the quantitative level (see Ref.~\cite{Bruckmann:2017pft} for 
a distribution of the Dirac eigenvalues 
for interacting quarks on the lattice).

In contrast to the valence-quark effect, 
the sea-quark effect exhibits a turnover 
from an enhancement of the chiral condensate in low temperature 
to a suppression near the transition temperature \cite{Bruckmann:2013oba}. 
The breakdown is shown in Fig.~\ref{fig:valence_vs_sea}.\footnote{
Note that the full result for the chiral condensate 
is not simply reconstructed as a sum of 
the valence-quark and sea-quark contributions 
beyond the linear order in the magnetic-field strength. 
}
The plot origin is located with subtraction of 
the chiral condensate at a vanishing magnetic field. 
The left panel shows the valence-quark effect on 
the magnitude of the chiral condensate, 
where the magnetic field in the fermion determinant, 
resulting from the fermionic Gaussian path integral of 
the QCD action, is switched off by hand. 
Plots at two different temperatures exhibit 
similar increasing behaviors. 
The right panel shows the sea-quark effect, 
where the magnetic field acting on the observable operator 
is switched off by hand. 
The sea-quark effect acts to enhance the chiral condensate 
at the lower temperature (see also Ref.~\cite{DElia:2011koc}
for a consistent result with a larger quark mass). 
However, the sea-quark effect turns to 
suppressing the chiral condensate 
near the transition temperature. 
The inverse magnetic catalysis occurs 
when this turnover occurs in the sea-quark effect 
and the suppression due to the sea-quark effect is strong enough 
to overwhelm the enhancement due to the valence-quark effect. 
Therefore, the sea quarks, which are also often called the dynamical quarks, are identified as the main player in the inverse magnetic catalysis.  
More recent simulations have shown the valence- and sea-quark effects on the light-quark condensate at various temperatures \cite{Ding:2022tqn}, 
which confirms the results in Fig.~\ref{fig:valence_vs_sea}. 

The sea-quark effect is also identified by 
measuring the gluon condensate. 
There is no counterpart of the valence-quark effect 
in the gluon condensate simply because magnetic fields do not directly couple to gluons. 
The simulation results have shown that 
the gluon condensate gives rise to a turnover similar to 
that of the sea-quark effect on the chiral condensate. 
That is, the gluon condensate is enhanced 
in low temperature but is suppressed 
near the transition temperature \cite{Bali:2013esa}, 
which may be interpreted as a consequence of 
the sea-quark effect on the gluon condensate.
Notice also that the Polyakov loop discussed above 
is another purely gluonic observable, 
and thus the decreasing transition temperature 
read off from the Polyakov loop should originate 
from the sea-quark effect.

Also, one should note that the sea-quark effect is blind to 
the charges of observable operators by definition, 
while the valence-quark effect can depend on 
electric charges of quark fields 
if the quark flavor sum is not taken in an observable operator. 
Note also that the flavor sum in the QCD action is always taken 
for the sea-quark effect. 
Numerical simulation results 
have shown that there is no significant difference 
between the transition temperatures read off from 
the $u $-quark condensate and the $ d$-quark condensate \cite{Endrodi:2015oba}; 
Both of the transition temperatures exhibit 
almost the same decrease in spite of the charge difference. 
Such a charge-blind effect should originate from 
the see-quark effect. 
On the other hand, 
a charge-dependent effect is found in the absolute magnitudes of 
the $u $-quark condensate and the $ d$-quark condensate 
shown in the left panel of Fig.~\ref{fig:lattice-phase}; 
The magnitude for $d$ quark is almost a half of 
that for $ u$ quark (see also Ref.~\cite{DElia:2011koc} 
for an earlier simulation result). 
This difference may imply the existence of 
an overall degeneracy factor $\sim |q_f B|/(2\pi)$ 
with a quark electric charge $q_f $ stemming from 
the valence-quark spectrum distribution 
as mentioned above as well. 
See below for a dependence on the valence-quark mass.

\cout{\com{Previous version}
Importance of the sea-quark effect is also implied by 
another observation that there is no significant splitting 
between the transition temperatures read off from 
the $u $-quark condensate and the $ d$-quark condensate \cite{Endrodi:2015oba}; 
Both the transition temperatures exhibit almost the same decrease 
in spite of the difference between their electric charges 
in the observable operators $\bar u u $ and $\bar d d $. 
The difference between their electric charges instead 
manifests itself in a difference between the absolute magnitudes of 
the $u $-quark condensate and the $ d$-quark condensate 
shown in the left panel of Fig.~\ref{fig:lattice-phase}; 
The magnitude of the $d$-quark is almost 
half of the $ u$-quark condensate (see also Ref.~\cite{DElia:2011koc}). 
This difference may imply the existence of 
an overall degeneracy factor $\sim |q_f B|/(2\pi)$ 
with an electric charge $q_f $ stemming from 
the valence-quark spectrum. 
However, the slope shape in each curve, 
and thus the transition temperature, 
is found to be governed by work of the sea quarks. 
\com{Tc does not necessarily depend on charges?}
}

\subsubsection*{Diagnosis of the inverse magnetic catalysis II: Quark mass dependences}

While the sea-quark effect was identified as 
the driving force for the inverse magnetic catalysis, 
the magnitude of the sea-quark effect depends on 
a value of the current quark mass 
in the numerical simulation set-up. 
The sea-quark effect is absent in the infinitely heavy quark limit, i.e., the quenched simulations, 
and becomes larger as the quark mass is 
reduced to the physical point. 
Thus, the quark mass can be an important control parameter 
among other simulation parameters \cite{DElia:2011koc, Bali:2012zg}. 
Besides, there could be a dependence on the valence-quark mass as well because the effective coupling strengths 
between quark fields in observable operators and gauge fields 
is suppressed by a quark mass. 
This dependence can be studied by comparing the light-quark condensate 
and the strange-quark condensate (with the same sea-quark mass) \cite{Ding:2022tqn}.

In the early studies, neither the suppression of the chiral condensate nor the decrease of the transition temperature 
was observed in quenched simulations \cite{Buividovich:2008wf, Braguta:2010ej} and simulations with a large current quark mass \cite{DElia:2010abb, DElia:2011koc} where the sea-quark effect is suppressed.  
Comparing the simulation set-ups, 
a large current quark mass and/or 
the unimproved staggered fermion 
used in the early works \cite{DElia:2010abb, DElia:2011koc} 
were thought to be possible reasons why 
those phenomena were not observed \cite{DElia:2018xwo}.

The quark-mass dependence was elaborated 
in recent works with improved staggered fermions \cite{DElia:2018xwo, 
Endrodi:2019zrl} 
and with the unimproved staggered fermion \cite{Ding:2020jui}. 
With improved staggered fermions, 
the simulation results exhibit the inverse magnetic catalysis 
near the transition temperature 
when the quark mass is small, 
but the magnetic catalysis at any temperature 
when the quark mass is large \cite{DElia:2018xwo, Endrodi:2019zrl}. 
Namely, it was found that, 
as the current quark mass is increased at a fixed temperature, 
a turnover from the inverse magnetic catalysis 
to the magnetic catalysis occurs somewhere 
between the corresponding pion masses $ m_\pi=343 \ \MeV $ 
and $m_\pi=664 \ \MeV   $ 
when $ eB=0.425 , \ 0.85 \ \GeV^2 $ \cite{DElia:2018xwo}. 
An updated work provided a consistent estimate of the turnover point to be $ m_\pi \sim 500 \ \MeV$ when $ eB = 0.6 \ \GeV^2 $ (see Ref.~\cite{Endrodi:2019zrl} for the exact value, 
the issue of scale setting, and the simulation setup).

However, form the above simulations with the {\it improved} staggered fermions, it appeared that the transition temperature still {\it decreases} 
even when the inverse magnetic catalysis, the suppression of the chiral condensate, does not occur 
with a large quark mass \cite{DElia:2018xwo, Endrodi:2019zrl}.
To disentangle possible lattice artifacts due to 
the discretization methods, 
numerical simulations were performed with the {\it unimproved} staggered fermion in Refs.~\cite{Tomiya:2017cey, Ding:2020jui} 
as a mimic of the early works \cite{DElia:2010abb, DElia:2011koc}. 
In this set-up, the transition temperature {\it increases} 
with an increasing magnetic-field strength 
and the inverse magnetic catalysis does not occur 
at any temperature. 
Therefore, it was concluded that 
the increasing transition temperature observed 
in the early works \cite{DElia:2010abb, DElia:2011koc} 
is not ascribed to a large quark mass 
but to the use of unimproved staggered fermion \cite{Ding:2020inp}.

One can also study a dependence on the valence-quark mass. 
This serves as not only a control test but also a physical set-up 
that provides us with an insight into 
possible differences between the light-quark and strange-quark condensates. 
The strange-quark condensate as well as the light-quark 
and light-strange mixed condensates were computed 
very recently with dynamical simulations \cite{Ding:2022tqn}. 
The results for the strange-quark condensate suggests that 
the valence-quark effect is stronger than the sea-quark effect at any temperatures and the inverse magnetic catalysis does not occur 
when the valence-quark mass is as large as the physical strange-quark mass. 
Nevertheless, the transition temperature, read off from the strange-quark condensate, decreases with an increasing magnetic-field strength. 
Namely, the balance between the valence-quark and sea-quark effects is affected by the valence-quark mass. 
In this simulation, the strange-quark and light-quark masses are fixed at the physical value and one tenth of the strange-quark mass, respectively, which corresponds to 
kaon mass $ \sim 507 \ \MeV $ and pion mass $\sim 220 \ \MeV $.



It appeared in the last decade that the decreases of the transition temperatures are commonly seen in 
various observables of QCD thermodynamics, i.e., the $u $-quark condensate, the $ d$-quark condensate, the $s $-quark condensate, the gluon condensate, the Polyakov loop, and the strange-quark susceptibility, in spite of the distinctive properties of 
the observable operators. 
The common main player is the sea-quark effect, of which 
the temperature dependence was found to be particularly important. 
In contrast, we have seen just above that 
whether or not the inverse magnetic catalysis occurs 
depends on the observable operators, 
i.e., the valence-quark mass, due to the competition with 
the valence-quark and sea-quark effects. 
Therefore, one can get a better insight 
if we regard the inverse magnetic catalysis of 
the chiral condensate as a separate issue 
from the common decreases of the transition temperatures.



It is worth adding that, to investigate connections 
between the chiral restoration and meson properties at finite temperature, 
a Ward-Takahashi identity was derived for the chiral transformation in Ref.~\cite{Ding:2020hxw} (see also Ref.~\cite{Bali:2017ian}) 
and was confirmed with the aforementioned dynamical simulations at finite temperature \cite{Ding:2022tqn}. 
The Ward-Takahashi identity relates the chiral condensate 
to the spacetime integral of the meson-meson correlator 
in the pseudoscalar channel. 
The same authors investigated possible correlations between 
the chiral condensate and the meson screening mass 
that is read off from the exponential decay of 
the meson-meson correlator 
at the long {\it spatial} distance limit 
as originally computed in Ref.~\cite{Detar:1987kae, Detar:1987hib} (see the section of ``hadron spectrum'' below for a difference between the screening mass and the pole mass). 
The dynamical simulation results have shown that 
the valence-quark and sea-quark effects act on the screening mass 
with opposite tendencies. 
That is, the screening masses in the pseudoscalar channels are 
enhanced by the sea-quark effect as we increase 
the magnetic-field strength, 
while they are suppressed by the valence-quark effect \cite{Ding:2022tqn}. 
Here, the valence-quark effect refers to a direct coupling 
of a magnetic field to the meson operator in the correlator.

Also, comparison between the light-quark and strange-quark channels suggested a correlation that 
the sea-quark effect overwhelms the valence-quark effect 
both on the meson screening mass and the chiral condensate 
when the inverse magnetic catalysis occurs \cite{Ding:2022tqn}. 
However, a direct connection between the screening mass 
and the chiral restoration are yet elusive.  
This may be partly due to lack of clear understanding 
of what the temperature dependence of the screening mass 
tells us in the confinement phase and near the transition temperature even in the absence of magnetic fields; 
Note that this screening mass is not of gluons 
or the confinement force. 
It may be interesting, for instance, to check 
degeneracy between the scalar and pseudoscalar channels 
as well as between the vector and axial ones 
as a signal of chiral symmetry restoration rather than temperature dependences in single channels; 
There are such studies in the absence of magnetic fields (see Refs.~\cite{Bazavov:2019www, DallaBrida:2021ddx} and references therein).

\subsubsection*{Diagnosis of the inverse magnetic catalysis III: Interpreting the sea-quark effect}

We have seen that the temperature dependence of 
the sea-quark effect is particularly important 
as observed in Fig.~\ref{fig:valence_vs_sea} (see also Ref.~\cite{Ding:2022tqn}). 
Now, we shall speculate how the sea-quark effect can 
enhance the chiral condensate in low temperature and 
suppress it near the transition temperature. 
In Sec.~\ref{subsec:MC-weak},  
we have identified two-fold effects of magnetic fields on the formation of the chiral condensate 
{\it albeit} with perturbative calculations. 
The modification of the screening effect, 
through the fermion loops inserted in the gauge-boson propagators, corresponds to the sea-quark effect, 
while the dimensional reduction on the scattering fermions, 
leading to the magnetic catalysis, 
corresponds to the valence-quark effect. 
In general, screening effects diminish 
the magnitude of particle pairing or condensation. 


According to the above correspondence, 
the overall tendency in the temperature dependence 
can be understood as follows. 
The sea-quark effect leads to the {\it enhancement} of the chiral condensate in low temperature 
because of a {\it suppression} of the screening effect 
by a dynamical quark mass in strong magnetic fields 
that is larger than that at a vanishing magnetic field. 
The dynamical quarks can be, nevertheless, excited 
by thermal energy at finite temperature. 
Once the minimum energy cost, the dynamical quark mass, 
is overcome by thermal energy, 
the screening effect is {\it enhanced} by abundant 
low-energy excitations in the Landau degeneracy 
as opposed to the low temperature case. 
Then, the sea-quark effect leads to the {\it suppression} 
of the chiral condensate 
because of the {\it enhancement} of the screening effect. 

An implicit condition assumed above is 
that the  quarks in strong magnetic fields 
have a small mass enough to develop the screening effect 
near the transition temperature. 
Once the mass threshold is well overcome by thermal energy, 
the screening effect is enhanced as mentioned just above. 
However, the quark mass could grow as we increase 
the magnetic-field strength. 
In such a case, the suppression of the screening effect, 
and associated enhancement of the chiral condensate 
by the sea-quark effect, 
would continue all the way from zero to transition temperatures;  
We would not see the inverse magnetic catalysis. 
This is the case of QED and the NJL model mentioned earlier 
due to the monotonic growth of the dynamical quark mass 
and may also be what is going on in the lattice QCD simulations 
with a large current quark mass \cite{DElia:2018xwo, Endrodi:2019zrl} 
(where generation of a large dynamical mass is also 
expected due to a suppression of the screening effect 
by the large current mass).

One should investigate whether the total quark mass 
from the current and dynamical contributions 
remains small enough in strong magnetic fields. 
This question was posed and investigated 
in Refs.~\cite{Kojo:2012js, Kojo:2013uua}. 
As discussed in Sec.~\ref{subsec:MC-weak}, 
the magnitude of the dynamical mass is 
sensitive to interaction details. 
We will discuss this point later this section 
with a QCD-motivated interaction. 
This is an interesting and basic issue for 
understanding QCD in strong magnetic fields.


\subsubsection*{Exploring the further strong-field regime}


 
We have seen that the transition temperatures 
decrease up to $eB = 3.25 \GeV^2$. 
Here, we pay attention to the order of phase transition as well. 
In the left panel of Fig.~\ref{fig:lattice-phase}, 
the chiral phase transition remains a crossover transition 
without a singularity or discontinuity in the curves. 
Nevertheless, we notice that the slope in the transition region 
becomes steeper as we increase the magnetic-field strength, 
which has been pointed out in the very early studies \cite{DElia:2010abb, Bali:2011qj}. 
One might expect that the phase transition could turn into 
a singular one as we further increase 
the magnetic-field strength.

More careful numerical analyses by the finite-size scaling method 
yet confirm the crossover transition up to $eB = 3.25 \ \GeV^2$ 
and, at the same time, the tendency toward the singular phase transition \cite{Endrodi:2015oba}. 
Those results motivate us to explore the phase structures 
in further strong magnetic fields. 
A first-order phase transition was indeed observed 
with the unimproved staggered fermion \cite{Ding:2020inp},\footnote{
Without improved staggered fermion, 
the transition temperature increases 
with an increasing magnetic field as mentioned 
in the above diagnosis. 
The first-order phase transition is observed 
in the measurement of the Polyakov loop as well \cite{Ding:2020inp}. 
} 
while determining the critical temperature and magnetic field 
was left to improved simulations. 
Very recent improved simulation results 
suggested that the chiral phase transition 
remains a crossover transition at $eB = 4 \ \GeV^2 $ 
but is a first-order transition at $eB = 9 \ \GeV^2 $ \cite{DElia:2021yvk}. 
The first-order phase transition temperature was 
estimated to be around $63 \ \MeV $, 
while the pseudo-critical temperature at $eB = 4 \ \GeV^2 $ 
is around $98 \ \MeV $. 
The transition temperature still keeps decreasing from 
$113 \ \MeV$ at $eB=3.25 \ \GeV^2$ shown in Fig.~\ref{fig:lattice-phase}, though 
one needs to be careful about comparison among 
the data obtained with different simulation set-ups. 
Those are the very first reports of the indication for 
a first-order phase transition in QCD, 
including the case in the absence of magnetic fields, 
and certainly deserves further confirmation and extensions, 
which requires improvements in simulation set-ups (see discussions in Ref.~\cite{DElia:2021yvk}). 
Scanning the phase structures 
at various magnetic-field strengths 
is a necessary extension to draw the first-order transition line 
and locate  an associated critical point. 
In this perspective, it is interesting to investigate 
the critical phenomena near the possible critical point.

In Ref.~\cite{DElia:2021yvk}, the same authors also investigated 
the quark-antiquark potential obtained from the Wilson loop 
at the above magnetic-field strengths  
$eB = 4 , \  9 \ \GeV^2 $. 
Temperature is fixed at a single point $T \sim 86 \ \MeV $ 
that is in between the transition temperatures 
at $eB = 4 \ \GeV^2 $ and $eB = 9 \ \GeV^2 $ mentioned above. 
It was shown that the linear confinement potential disappears at $eB = 9 \ \GeV^2 $, 
while there remains a linear potential at $eB = 4 \ \GeV^2 $ though the potential becomes anisotropic. 
Provided that those results suggest the deconfinement transition, 
the system is in a confinement and chirally broken phase 
at $eB = 4 \ \GeV^2 $ and $T \sim 86 \ \MeV $ 
and in a deconfinement and chirally symmetric phase 
at $eB = 9 \ \GeV^2 $ and $T \sim 86 \ \MeV $. 
Those data are still consistent with 
the degeneracy between the chiral and deconfinement 
phase transition temperatures, which is explicitly 
observed in case of $eB=3.25 \ \GeV^2$ 
(see Fig.~\ref{fig:lattice-phase}).

It is interesting to further investigate 
the confinement properties in strong magnetic fields. 
It was suggested that a critical point 
exists in the {\it deconfinement phase transition} 
in strong magnetic fields \cite{Cohen:2013zja}. 
This argument is based on an assumption that 
the dynamical quark mass keeps increasing 
by the magnetic catalysis 
in the asymptotic magnetic-field strength. 
In such a case, all the ``heavy'' quarks are decoupled 
from the low-energy dynamics, which leaves 
a pure gluonic system realized in the low-energy regime. 
The order of a deconfinement phase transition in a pure gluonic system may be first order \cite{Yaffe:1982qf}. 
On the other hand, the deconfinement phase transition 
is known to be crossover at a vanishing magnetic field 
due to the existence of quarks. 
Therefore, one can expect the existence of a critical point 
somewhere in between the vanishing and asymptotic magnetic-field strengths.

There are two issues to be clarified. 
Assuming the ``infinitely heavy'' quarks 
means that the chiral symmetry is broken 
at the deconfinement phase transition temperature 
in the effective pure gluonic system. 
On the other hand, we have seen the degeneracy 
of the transition temperatures 
read off from the chiral condensate, the strange-quark susceptibility, and the Polyakov loop 
up to the magnetic-field strength mentioned above. 
Especially, the deconfinement phase transition 
for the pure gluonic system should be well captured by 
the Polyakov loop. 
Thus, it should be clarified whether or not there is room for splitting between the transition temperatures 
from the chiral condensate and the Polyakov loop 
in further strong magnetic fields. 
The other issue is related to an anisotropic nature of 
the possible pure gluonic system with respect to 
the direction of the strong magnetic field. 
While magnetic fields do not directly couple to gluons, 
the pure gluonic action may contain 
an anisotropic dielectric constant 
due to an anisotropic screening effect 
stemming from the quark loops in magnetic fields \cite{Miransky:2002rp} 
(see also Sec.~\ref{sec:photons} for a QED analog). 
The issue is whether the deconfinement phase transition 
is first order in such an anisotropic pure gluonic system \cite{ Cohen:2013zja}. 
The lattice simulation has been performed 
for the anisotropic gluonic system by treating 
the anisotropic dielectric constant as a free parameter \cite{Endrodi:2015oba}. 
A first-order deconfinement phase transition 
in the anisotropic pure gluonic system was concluded 
with the finite-size scaling method. 
Determining the phase transition temperature 
in physical units is left as an open question. 

\cout{
If the deconfinement phase transition occurs 
at a lower temperature than the chiral phase transition, 
there can be an interesting phase structure 
in the deconfinement phase transition as well. 
In Ref.~\cite{Cohen:2013zja}, the existence of a critical point 
was suggested in the deconfinement phase transition 
if the magnetic catalysis acts to generate 
an ``infinitely'' large dynamical quark mass 
at the asymptotic magnetic-field strength. 
In such case, all the quarks are decoupled 
from the low-energy dynamics, and 
a pure gluonic system is left with 
an anisotropic dielectric constant due to 
the anisotropic screening effect by the LLL quark loops. 
A pure gluonic system realized there may exhibits 
a first-order deconfinement phase transition \cite{}, 
so that there can be a critical point 
at a finite magnetic-field strength 
since the deconfinement phase transition is known to be 
crossover without an external magnetic field.

There are two issues to be clarified. 
The appearance of infinitely heavy quarks seem 
to have a tension to the tendency of the merging phase transition temperatures read off from the chiral condensate and the Polyakov loop up to the aforementioned numerical data. 
Whether or not splitting occurs is an interesting question. 
The other issue is whether the deconfinement phase transition 
is first order in the anisotropic pure gluon theory \cite{Miransky:2002rp, Cohen:2013zja}. 
The result is interpreted as a first-order phase transition 
from the singular behavior of the susceptibility and 
the finite-size scaling method \cite{Endrodi:2015oba}. 
Due to the absence of a priori known dimensionful scales 
in the asymptotic large B system, determination of 
the phase transition temperature in physical units 
is still left as an open question. 

}

\cout{
From estimates of the lattice scale available at this moment, 
the transition temperature may be a decreasing function of $ eB $, 
which is consistent with the prediction in \cite{Miransky:2002rp}. 
}

\subsubsection*{Hadron spectrum}

As mentioned earlier, strong magnetic fields can modify 
internal structures of composite particles 
and resulting spectra 
when dense magnetic lines penetrate through the bodies. 
Besides an interest in the composite structures on its own, 
studying QCD thermodynamics requires information 
of how pions and other low-lying hadron spectra get modified 
in strong magnetic fields 
since hadrons, rather than quark quasiparticles, 
serve as elementary excitations in the confinement phase. 
Lattice simulations have been providing 
light-meson spectrum \cite{Bali:2011qj, Hidaka:2012mz, 
Luschevskaya:2014lga, Luschevskaya:2015bea,
Luschevskaya:2016epp, Bali:2017ian, Ding:2020hxw} 
as well as baryon spectrum \cite{Endrodi:2019whh} 
(see also Ref.~\cite{Buividovich:2010tn} for the two-color QCD). 
All of those measurements were performed with 
the quenched simulations except for a charged pion in Ref.~\cite{Bali:2011qj}, a neutral pion in Ref.~\cite{Bali:2017ian},\footnote{A neutral pion mass was measured by the use of the gauge configuration generated in Ref.~\cite{Bali:2011qj} 
for the sake of comparison to the quenched and Wilson-fermion result in Ref.~\cite{Bali:2017ian}.} 
and the recent paper \cite{Ding:2020hxw}. 
The dynamical simulation in the last paper 
was extended to measurements of neutral pseudoscalar mesons at finite temperature for the study of QCD thermodynamics \cite{Ding:2022tqn}. 
One can also measure other properties such as 
the decay constants\footnote{
Due to a preferred spatial orientation provided by a magnetic field, there appear two decay constants that parametrize 
the hadronic matrix element of the weak current which has one Lorentz index \cite{Fayazbakhsh:2013cha}.} \cite{Bali:2018sey, Ding:2020hxw} 
and the anisotropic wave functions of light mesons \cite{Hattori:2019ijy}.

\begin{figure}[t]
\begin{center}
   \includegraphics[width=0.6\hsize]{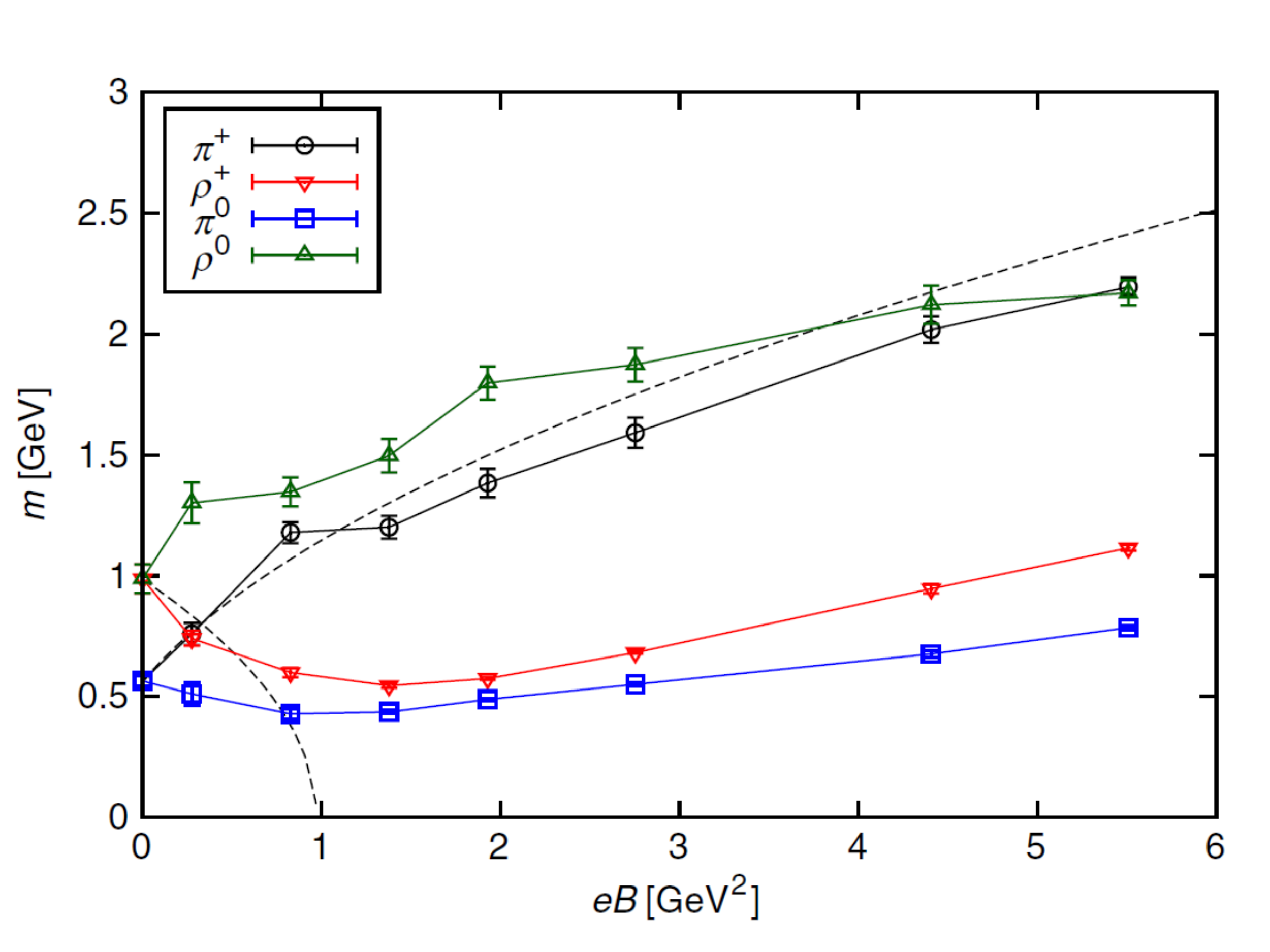}
\end{center}
\vspace{-0.5cm}
\caption{Light-meson spectra in magnetic fields \cite{Hidaka:2012mz}. The dashed lines show the Landau quantization of a point-like charged pion and rho-meson in a naive hadronic picture. 
}
  \label{fig:light_mesons}
\end{figure}

Figure \ref{fig:light_mesons} shows the simulation results 
on the light-meson spectra \cite{Hidaka:2012mz}. 
The dashed lines show the Landau quantization of a point-like charged pion and rho-meson with the lowest Zeeman energy. 
In the charged-pion spectrum, one finds that 
the simulation result increases in a similar way to 
the dashed line. 
Other simulation results also showed increasing behaviors \cite{Bali:2011qj, Luschevskaya:2015bea, Bali:2017ian}. 
However, a recent dynamical simulation suggested 
a turnover to a decreasing behavior 
with a peak near $eB \sim 0.6 \ \GeV^2 $ \cite{Ding:2020hxw}. 
No such peak structure has been observed with the quenched simulations in Fig.~\ref{fig:light_mesons} and Refs.~\cite{Luschevskaya:2015bea, Bali:2017ian} 
or the dynamical simulation up to $eB \sim 0.4 \ \GeV^2 $ \cite{Bali:2011qj}. 
A charged kaon from the dynamical simulation also exhibits 
a similar peak structure at almost the same position (see Fig.~9 in Ref.~\cite{Ding:2020hxw}). 
Notice also that, even though no such drastic behavior is seen 
in the quenched simulations, there is still a systematic tendency for a deviation from the point-like spectrum in the strong-field regime in Fig.~\ref{fig:light_mesons} 
and Refs.~\cite{Bali:2011qj, Luschevskaya:2015bea, Bali:2017ian}, 
where the point-like spectrum for a charged pion 
overshoots the quenched simulation results. 
Those findings may suggest modification of composite structures 
and deserve further study in the strong-field regime.

As for the charged rho-meson states,\footnote{
Note that, in quenched simulations, rho mesons are stable 
against the dominant decay channel to two pions 
unlike in reality.} 
while the dashed line for the lowest Zeeman energy 
follows the simulation result 
up to an intermediate magnetic-field strength, 
there is a significant deviation in the strong-field regime $ eB \gtrsim 0.5 \ \GeV^2 $. 
This is not very surprising since the magnetic-field strength is 
already much larger than $\Lambda_\QCD^2 $ 
and becomes comparable to the decreasing rho-meson mass. 
Whereas a vanishing rho-meson mass was 
thought to be an indication for a rho-meson condensation in 
strong magnetic field \cite{Chernodub:2010qx, Chernodub:2011mc},\footnote{
A negative Zeeman energy dominates over 
the zero-point energy in the Landau quantization 
when charged particles have spin larger one, 
leading to a vanishing mass at a certain critical magnetic field. 
This applies to charged rho-mesons granted that 
they are point-like particles. 
}   
the simulation result exhibits a slow down of the decreasing behavior (red curve), and a discussion with the QCD inequality does not support this scenario either \cite{Hidaka:2012mz}. 
No indication for a rho-meson condensate, at least that conjectured with the point-like picture, is observed 
in later simulations \cite{
Luschevskaya:2015bea, Luschevskaya:2016epp, Bali:2017ian}. 
There is a discrepancy among those simulation results 
in the asymptotic behavior: 
The later results exhibit a saturating behavior rather than the increasing behavior seen in the strong-field regime of  Fig.~\ref{fig:light_mesons}. 
This discrepancy needs to be revolved in the future studies. 
Increasing behaviors in the other two spin states (not discussed in Fig.~\ref{fig:light_mesons}) 
were observed in Ref.~\cite{Bali:2017ian}. 
One of those spin states needs investigation for 
the mixing with a charged pion,\footnote{
Due to the breaking of a spatial rotational symmetry in magnetic fields, only the spin component along a magnetic field 
remains a good quantum number. Therefore, a spinless particle 
and the spin-zero state of spin-1 particle can 
have the same quantum numbers and can be mixed with each other. 
} which was, however, concluded to be small \cite{Bali:2017ian}.

In Fig.~\ref{fig:light_mesons}, the neutral pion and rho-meson spectra also vary as we increase the magnetic-field strength, 
though they are just constants in a naive hadronic picture. 
While one finds a non-monotonic behavior in the neutral pion 
in Fig.~\ref{fig:light_mesons} \cite{Hidaka:2012mz}, 
a monotonic decreasing behavior was observed in a later simulation ~\cite{Luschevskaya:2014lga}. 
The origin of the discrepancy was pointed out 
in Ref.~\cite{Bali:2017ian} as a lattice artifact 
inherent in the Wilson fermions used in Ref.~\cite{Hidaka:2012mz}. 
The Wilson term, added to the Dirac operator to kill the doublers, generates an artificial magnetic-field dependence 
of the Dirac eigenvalue at finite lattice spacing, 
which acts to increase the quark mass 
as we increase the magnetic-field strength.\footnote{
The Wilson term is proportional to the Klein-Gordon operator. 
Recall that the lowest eigenvalue of the (non-interacting) 
Klein-Gordon operator increases linearly with an increasing magnetic field due to the Landau quantization (see Appendix A in Ref.~\cite{Bali:2017ian}). 
}
Renormalizing the quark mass to 
subtract the magnetic-field dependence, it was shown that 
the neutral pion spectrum monotonically decreases 
and then almost saturates with an increasing magnetic field, 
which resolves the discrepancy mentioned above \cite{Bali:2017ian}. 
A recent dynamical simulation confirmed the monotonic decrease 
and also suggested that 
the effect of dynamical quarks on a neutral pion 
is negligible at {\it zero temperature} 
(see below for a finite temperature case) \cite{Ding:2020hxw}. 
Note that they are ``connected pions'' where a quark-antiquark pair annihilation is not taken into account. 
The mass of neutral kaon was also measured in Ref.~\cite{Ding:2020hxw}, exhibiting a similar, but weaker, decreasing behavior 
due to a physical strange-quark mass heavier than those of up and down quarks. 

As for neutral rho mesons, one finds an increasing behavior 
in Fig.~\ref{fig:light_mesons} for the nonzero spin states 
that do not suffer from the mixing with a neutral pion. 
Such an increasing behavior was observed in later simulations as well \cite{Luschevskaya:2014lga, Bali:2017ian}. 
Neglecting the mixing effect, 
the spin-zero state was measured in Ref.~\cite{Luschevskaya:2014lga}, exhibiting a decreasing behavior. 
The effects of the mixing and the disconnected diagrams 
for neutral pion and rho-meson yet await final conclusions 
(see discussions in Ref.~\cite{Ding:2020hxw}). 
Once the mass splitting between the spin-zero and -nonzero states 
is confirmed, this tendency may be understood from spin configurations favored in magnetic fields. 
Spin configurations inside neutral mesons, 
where a valence quark and antiquark pair has the opposite electric charges, 
favor the spin-zero states (including neutral pions) 
and disfavor the nonzero spin states 
due to the Zeeman energy cost.\footnote{ 
In the Landau quantization picture, this means that 
the spin-zero states can be configured 
with a quark and antiquark both in the LLL, 
while the spin-nonzero states have to include at least 
a hLL which costs energy (see, e.g, an appendix in Ref.~\cite{Hattori:2015aki}).} 

Fitting the measured mass shifts 
by a polynomial of the magnetic-field strength, 
one can quantifies the effect of composite structures 
by the polynomial coefficients. 
The $ g$-factor, which appears as the linear-order coefficient, 
was read off for rho mesons \cite{Luschevskaya:2015bea, Luschevskaya:2016epp, Bali:2017ian} as well as vector kaons $K^\ast $  \cite{Luschevskaya:2016epp}. 
The dipole magnetic polarizability, which appears as the quadratic-order coefficient, 
was read off for vector mesons \cite{Luschevskaya:2014lga, Luschevskaya:2015bea} and pseudoscalar mesons \cite{Luschevskaya:2014lga, Bali:2017ian, Ding:2020inp}.

It was shown for the first time by lattie QCD simulation 
that the Gell-Mann--Oaks--Renner (GOR) relation  
holds in magnetic fields by separately computing the chiral condensate, neutral meson mass, and decay constant 
at zero temperature \cite{Ding:2020hxw} (see also Ref.~\cite{Agasian:2001ym} for chiral perturbation theory). 
The corrections to the GOR relation was found to be 
small for light quarks and mild for strange quark 
that actually become smaller than the corrections 
at a vanishing magnetic field. 
Recall that neutral meson mass decreases with 
an increasing magnetic field though the chiral condensate 
increases at zero temperature. 
Those behaviors are compatible with one another 
in the GOR relation with the aid of an increasing decay constant. 
The decreasing neutral meson masses may imply 
a decreasing chiral transition temperature 
as thermal fluctuations of the (NG) bosons drive 
chiral restoration.

At finite temperature, the meson correlation length was 
measured in Ref.~\cite{Ding:2022tqn} 
as we mentioned earlier in the discussions about 
the sea-quark effect on the chiral condensate. 
The correlation length is defined by the long-distance 
behavior of the meson-meson correlator \cite{Detar:1987kae, Detar:1987hib}, where the correlator is expected to decay exponentially, characterized by the correlation length on the shoulder. 
The spatial direction is taken along a magnetic field. 
The dynamical simulations were performed 
in the pseudoscalar channels not only for light quarks 
but also strange quarks. 
As an overall tendency, the correlation lengths take almost constant values below $100\  \MeV$ and rapidly decrease as we further increase temperature across the transition temperature. 
Understanding those results is left as an open question. 
It will be important to first understand 
the high-temperature behavior 
with the dimensionally reduced effective theory 
as discussed in the absence of a magnetic field \cite{Hansson:1991kb, Koch:1992nx, Ishii:1994ku,Hansson:1994nb,Laine:2003bd, Brandt:2014uda, Shuryak:2022qzi}.\footnote{
One can construct an effective theory at high-temperature 
by utilizing a property in the imaginary-time formalism 
that the compact temporal coordinate 
shrinks in the high-temperature limit. 
Regarding the remaining spatial three dimensions as 
an analytic continuation from a fictitious (2+1) dimensional Minkowski spacetime, the screening mass can be formulated as 
a ``quarkonium spectrum'' in this effective theory with a large fermion mass given by the lowest Matsubara frequency \cite{Hansson:1991kb, Koch:1992nx, Ishii:1994ku,Hansson:1994nb,Laine:2003bd, Brandt:2014uda, Shuryak:2022qzi} (see Ref.~\cite{DallaBrida:2021ddx} for recent simulation 
in the (very) high-temperature region). 
} 
The inverse of the correlation length is 
called the static screening mass that is different from 
the pole mass, i.e., an energy gap in a dispersion relation (cf. Appendix~\ref{sec:photon_mass} for the difference 
between these two masses seen in a perturbative analog).\footnote{  
The energy gap of the lowest-lying state is 
measured with the exponential decay 
in the long (Euclidean) time correlation, 
while the screening mass is measured 
with the long spatial correlation. 
By construction of the imaginary-time formalism, 
the length of temporal coordinate is fixed by an inverse temperature, and energy gaps 
cannot be directly measured in this static formulation. 
There is not an energy selection in the measurement of the screening mass that should, therefore, include a contribution of the whole spectral function up to the temperature scale. 
The screening mass approaches the energy gap at zero temperature 
as anticipated from the Lorentz symmetry 
and confirmed by the recent simulation \cite{Bazavov:2019www} and approaches twice the lowest quark Matsubara frequency in the high-temperature limit where quarks are free of the confinement.  
}
The dispersion relations of the NG bosons may be, nevertheless,  
determined by a combination of the screening mass, 
decay constant, and axial isospin susceptibility 
that are all {\it static} quantities as pointed out in Refs~ \cite{Son:2001ff, Son:2002ci}; 
The axial isospin susceptibility could be measured in the future studies. 
It is interesting to explicitly see correlations between 
QCD thermodynamics and the NG-boson spectrum at finite temperature.

\cout{

\pp{

a direct connection between the screening mass 
and the phase transition are yet elusive. 
This may be largely due to lack of clear understanding 
of what the temperature dependence of the screening mass 
tells us in the confinement phase and near the transition temperature even in the absence of magnetic fields; 
Note that this screening mass is not of gluons 
or the confinement force. 
For instance, in the studies without magnetic fields, 
degeneracy between the scalar and pseudoscalar masses 
as well as between the vector and axial ones 
is taken as a signal of chiral symmetry restoration rather than temperature dependences in single channels (see Refs.~\cite{Bazavov:2019www, DallaBrida:2021ddx} and references therein).

}

\cgd{

The dynamical simulation results have showed that 
the screening masses in the pseudoscalar channels are 
enhanced by the sea-quark effect as we increase 
the magnetic-field strength, 
while they are suppressed by the valence-quark effect. 
Here, the valence-quark effect refers to a direct coupling 
of a magnetic field to the meson operator in the correlator.  
The measurements were performed not only for 
the light-quark channel but also strange-quark channels. 
The above tendencies in the valence-quark and sea-quark effects 
were observed both in the light-quark and strange-quark channels, 
while the magnitudes of the effects are suppressed by a strange-quark mass. 
In the strange-quark channel, the sea-quark effect is not 
strong enough to induce an enhancement of 
the screening mass and the inverse magnetic catalysis of the strange-quark condensate \cite{Ding:2022tqn}. 
} 
\pp{
While the correlated behaviors between 
the meson screening mass and the chiral condensate were observed 
in the valence-quark and sea-quark effects, 
a direct connection between the screening mass 
and the phase transition are yet elusive. 
This may be largely due to lack of clear understanding 
of what the temperature dependence of the screening mass 
tells us in the confinement phase and near the transition temperature even in the absence of magnetic fields; 
Note that this screening mass is not of gluons 
or the confinement force. 
For instance, in the studies without magnetic fields, 
degeneracy between the scalar and pseudoscalar masses 
as well as between the vector and axial ones 
is taken as a signal of chiral symmetry restoration rather than temperature dependences in single channels (see Refs.~\cite{Bazavov:2019www, DallaBrida:2021ddx} and references therein). 
It is also interesting to investigate the high-temperature region 
with the dimensionally reduced effective theory \cite{Hansson:1991kb, Koch:1992nx, Ishii:1994ku,Hansson:1994nb,Laine:2003bd, Brandt:2014uda, Shuryak:2022qzi} (see the above footnote) and 
the real-time propagation of the NG bosons 
determined by a combination of the screening mass, 
decay constant, and axial isospin susceptibility 
that are all {\it static} quantities below the transition temperature as pointed out in Refs~ \cite{Son:2001ff, Son:2002ci}. 

}

To investigate further connections between meson properties 
and the chiral restoration at finite temperature, 
a Ward-Takahashi identity was derived for the chiral transformation in the same paper \cite{Ding:2020hxw} (see also Ref.~\cite{Bali:2017ian}) 
and confirmed with dynamical simulations in magnetic fields 
and at finite temperature \cite{Ding:2022tqn}. 
The Ward-Takahashi identity relates the chiral condensate 
to \cgd{the spacetime integral of} the meson-meson correlator 
in the pseudoscalar channel. 
\cgd{
A meson property encoded in the correlator can be 
extracted by taking the long-distance limit \cite{Detar:1987kae, Detar:1987hib}, where the correlator is expected to decay exponentially, characterized by the correlation length on the shoulder. 
The inverse of the correlation length may be 
called the static screening mass.\footnote{
\cgd{
The static screening mass is different from 
an energy gap in a dispersion relation (cf. Appendix~\ref{sec:photon_mass} for the difference 
between these two masses seen in a perturbative analog). 
By construction of the imaginary-time formalism, 
the length of temporal coordinate is fixed by an inverse temperature, and energy gaps, which are measured in vacuum 
from the temporal correlation, 
cannot be directly measured in this static formulation. 
\pp{
There is not an energy selection to the lowest-lying state like the hadron mass measurement in vacuum, and the screening mass should include the contribution of the whole spectral function up to the temperature scale. 
}
The screening mass approaches the energy gap at zero temperature 
as anticipated from the Lorentz symmetry 
and confirmed by the recent simulation \cite{Bazavov:2019www} and approaches twice the lowest quark Matsubara frequency in the high-temperature limit where quarks are free of the confinement 
as observed in the original numerical simulations \cite{Detar:1987kae, Detar:1987hib}. 
Corrections at high-temperature can be obtained with the aid of 
the dimensional reduction in the temporal direction. 
Regarding the remaining spatial three dimensions as 
an analytic continuation from a fictitious (2+1) dimensional Minkowski spacetime, the screening mass can be formulated as 
a ``quarkonium spectrum'' in this effective theory with a large fermion mass given by the lowest Matsubara frequency \cite{Hansson:1991kb, Koch:1992nx, Ishii:1994ku,Hansson:1994nb,Laine:2003bd, Brandt:2014uda, Shuryak:2022qzi} (see Ref.~\cite{DallaBrida:2021ddx} for recent simulation in the high-temperature region). 
} 
}
The dynamical simulation results have showed that 
the screening masses in the pseudoscalar channels are 
enhanced by the sea-quark effect as we increase 
the magnetic-field strength, 
while they are suppressed by the valence-quark effect. 
Here, the valence-quark effect refers to a direct coupling 
of a magnetic field to the meson operator in the correlator.  
The measurements were performed not only for 
the light-quark channel but also strange-quark channels. 
The above tendencies in the valence-quark and sea-quark effects 
were observed both in the light-quark and strange-quark channels, 
while the magnitudes of the effects are suppressed by a strange-quark mass. 
In the strange-quark channel, the sea-quark effect is not 
strong enough to induce an enhancement of 
the screening mass and the inverse magnetic catalysis of the strange-quark condensate \cite{Ding:2022tqn}. 
} 
\pp{
While the correlated behaviors between 
the meson screening mass and the chiral condensate were observed 
in the valence-quark and sea-quark effects, 
a direct connection between the screening mass 
and the phase transition are yet elusive. 
This may be largely due to lack of clear understanding 
of what the temperature dependence of the screening mass 
tells us in the confinement phase and near the transition temperature even in the absence of magnetic fields; 
Note that this screening mass is not of gluons 
or the confinement force. 
For instance, in the studies without magnetic fields, 
degeneracy between the scalar and pseudoscalar masses 
as well as between the vector and axial ones 
is taken as a signal of chiral symmetry restoration rather than temperature dependences in single channels (see Refs.~\cite{Bazavov:2019www, DallaBrida:2021ddx} and references therein). 
It is also interesting to investigate the high-temperature region 
with the dimensionally reduced effective theory \cite{Hansson:1991kb, Koch:1992nx, Ishii:1994ku,Hansson:1994nb,Laine:2003bd, Brandt:2014uda, Shuryak:2022qzi} (see the above footnote) and 
the real-time propagation of the NG bosons 
determined by a combination of the screening mass, 
decay constant, and axial isospin susceptibility 
that are all {\it static} quantities below the transition temperature as pointed out in Refs~ \cite{Son:2001ff, Son:2002ci}. 

}

} 

\cout{

To investigate further connections between meson properties 
and the chiral restoration at finite temperature, 
a Ward-Takahashi identity was derived for the chiral transformation in the same paper \cite{Ding:2020hxw} (see also Ref.~\cite{Bali:2017ian}) 
and confirmed with dynamical simulations in magnetic fields 
and at finite temperature \cite{Ding:2022tqn}. 
The Ward-Takahashi identity relates the chiral condensate 
to \cgd{the spacetime integral of} the meson-meson correlator 
in the pseudoscalar channel. 
A meson property encoded in the correlator can be 
extracted by taking the long-distance limit. 
In this limit, the correlator is expected 
to decay exponentially with the correlation length on the shoulder that is dominated by the contribution 
of the lowest-lying mode, i.e., the NG boson; 
The inverse of the correlation length may be 
called the static screening mass of the NG boson.\footnote{
\cgd{
The static screening mass is different from 
an energy gap in dispersion relations (cf. Appendix~\ref{sec:photon_mass} for the difference 
between these two masses seen in a perturbative analog). 
By construction of the imaginary-time formalism, 
the length of temporal coordinate is fixed by an inverse temperature, and energy gaps, which are measured in vacuum 
from the temporal correlation (see above references), 
may not be measured directly in this formulation; 
They could possibly be related to one another 
by an analytic continuation \cite{Detar:1987kae, Detar:1987hib}. 
} 
}
The dynamical simulation results showed that 
the screening masses of the NG bosons are 
enhanced by the sea-quark effect as we increase 
the magnetic-field strength, 
while they are suppressed by the valence-quark effect. 
Here, the valence-quark effect refers to a direct coupling 
of a magnetic field to the meson operator in the correlator. 
This means that the screening effect on the long-range interaction is stronger in magnetic fields, 
driving chiral restoration. 
The measurements were performed not only for 
the light-quark channel but also strange-quark channels. 
The above tendencies of the valence-quark and sea-quark effects 
were observed both in the light-quark and strange-quark channels, 
while the magnitudes of the effects are suppressed by a strange-quark mass. 
In the strange-quark channel, the sea-quark effect is not 
strong enough to induce an enhancement of 
the screening mass and the inverse magnetic catalysis of the strange-quark condensate \cite{Ding:2022tqn}. 

}

\begin{figure}[t]
\begin{center}
   \includegraphics[width=0.9\hsize]{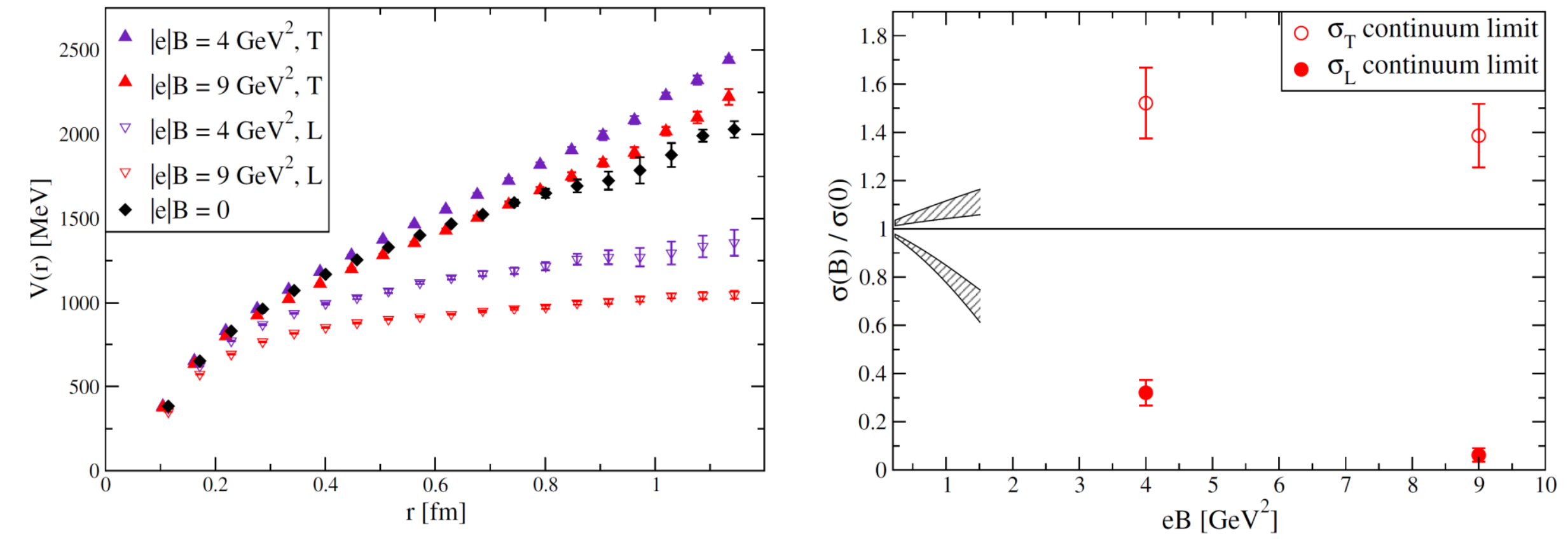}
\end{center}
\vspace{-0.5cm}
\caption{
The heavy-quark potential in strong magnetic fields (left) and 
the ratio of the string tension $\sigma $ in a magnetic field to that at a vanishing magnetic field (right). 
Both are taken from Ref.~\cite{DElia:2021tfb}. 
The subscripts ``L'' and ``T'' denote the cases where 
the quark and antiquark pair is located 
along the longitudinal and transverse 
directions to the magnetic field, respectively. 
The dashed gray regions show the continuum extrapolations 
of the earlier results in Ref.~\cite{Bonati:2016kxj}.
}
  \label{fig:cornell}
\end{figure}

Besides the light-quark observables above, 
the heavy-quark potential was also measured at zero and finite temperatures and fitted by the Cornell potential \cite{Bonati:2014ksa, Bonati:2016kxj, DElia:2021tfb, DElia:2021yvk}. 
The left panel in Fig.~\ref{fig:cornell} shows 
the latest results in Ref.~\cite{DElia:2021tfb} 
for the heavy-quark potential $V(r) $ 
in the ever strongest magnetic fields as a function of 
the distance $ r$ between a quark and antiquark pair. 
The subscripts ``L'' and ``T'' denote the cases where 
the pair is located along the longitudinal and transverse 
directions to the magnetic field, respectively. 
One can clearly see an anisotropy in the heavy-quark potential 
especially in a long-distance part beyond $r = 0.4 \ \fm $. 
This anisotropy is captured as the magnetic-field dependence of the string tension $ \sigma(B)$ in the Cornell potential. 
The right panel of Fig.~\ref{fig:cornell} shows the string tension normalized by that at a vanishing magnetic field. 
In the longitudinal case, the string tension is suppressed 
and almost vanishes at the strongest magnetic field, 
though it still has a finite value. 
In the transverse case, the string tension is enhanced 
to seemingly saturate in strong magnetic fields. 
By plugging the measured heavy-quark potential 
into the Schr\"odinger equation, one can study quarkonium spectra as done with the earlier data \cite{Bonati:2015dka} (see also Refs.~\cite{Alford:2013jva,Cho:2014exa,  Cho:2014loa, Hattori:2016emy, Suzuki:2016kcs, Yoshida:2016xgm, Iwasaki:2018pby,Iwasaki:2018czv, Iwasaki:2021nrz} for related works). 
Medium effect at finite temperature generally suppresses 
the string tension \cite{Bonati:2016kxj}. 
It is interesting to pursue a signal of the deconfinement 
transition at finite temperature in line with 
the aforementioned work \cite{DElia:2021yvk}.

As a complementary measurement, 
one can measure a strength of the chromo-electric field 
along the color flux tube spanned between a heavy-quark pair, 
which is the dominant source of the confinement force 
(especially in the heavy-quark limit) \cite{Bonati:2018uwh, DElia:2021tfb}.  
This quantity can be defined in a gauge-invariant way. 
The simulation results confirm that 
the chromo-electric field and the string tension exhibit 
the same tendency in the spatial anisotropy. 
The screening mass in the deconfinement phase was also measured 
for both the electric and magnetic components \cite{Bonati:2017uvz}.

The above tendency in the confinement force 
can be understood from the screening effect on 
the chromo-electric field by quark excitations in the LLL 
that can only excite along magnetic fields in the strong-field limit. 
In the longitudinal case, the LLL quarks can couple to 
the chromo-electric field extending along the magnetic field. 
The screening effect is enhanced by the Landau degeneracy, 
which {\it suppresses} the string tension as compared to the case 
at a vanishing magnetic field (see the right panel of Fig.~\ref{fig:cornell}). 
In the transverse case, the LLL quarks do not couple to 
the chromo-electric field extending in perpendicular 
to the magnetic field, 
and the chromo-electric field is less screened. 
Therefore, as we increase the magnetic field, 
the transverse string tension is {\it enhanced} 
since quarks fall into the LLL from the hLL. 
Eventually, the transverse string tension {\it saturates} 
after quarks are completely decoupled. 
The saturation value is determined by 
the residual screening effect by gluon excitations, 
which explains the saturation of 
the transverse string tension in the right panel of 
Fig.~\ref{fig:cornell}.

\subsubsection*{Magnetization}

Response of vacuum and/or medium to an external magnetic field 
induces magnetization. 
In Sec.~\ref{sec:vacuum_magnetism}, we have discussed 
the vacuum responses that lead to renormalization of divergent bare quantities. 
Here, we summarize finite temperature results from lattice QCD simulations.  
Magnetization manifests itself in 
an energy shift in thermodynamic systems, 
and can be defined as a derivative of 
the partition function by a magnetic-field strength. 
However, one cannot simply compute 
this derivative on periodic lattices, 
or a compact surface in general, 
due to the quantization of a magnetic field 
required by uniqueness of the phase.\footnote{
An arbitrary closed path on a compact surface is 
a common perimeter of an area $S $ 
and of the rest of the surface, which requires uniqueness 
of the phase, i.e., 
$\exp( i q_f \oint A_\mu dx^\mu) =
\exp( i q_f S B) = \exp( \, - i q_f ( S_0 - S)B \, ) $ 
with the total area $S_0 $, 
and thus the quantization $q_f B = 2 \pi n/S_0 $ (see, e.g., Refs.~\cite{AlHashimi:2008hr,DElia:2012ifm, Bonati:2013lca, Bonati:2013vba} for more details). 
}
Magnetization has been nevertheless investigated 
with lattice QCD simulations by bypassing this issue. 
We also note that the magnetization studied with lattice QCD simulation so far is in the linear response regime 
with respect to a magnetic field. 
Namely, the magnetization is regarded as a linear function 
of a magnetic field and the magnetic susceptibility corresponds to the linear slope parameter independent of a magnetic field. 
Investigating the nonlinear effects, which become important in strong magnetic fields, are left as open issues (cf. Sec.~\ref{sec:photon-vp} for nonlinear magnetic permeability).

\begin{figure}[t]
\begin{center}
   \includegraphics[width=\hsize]{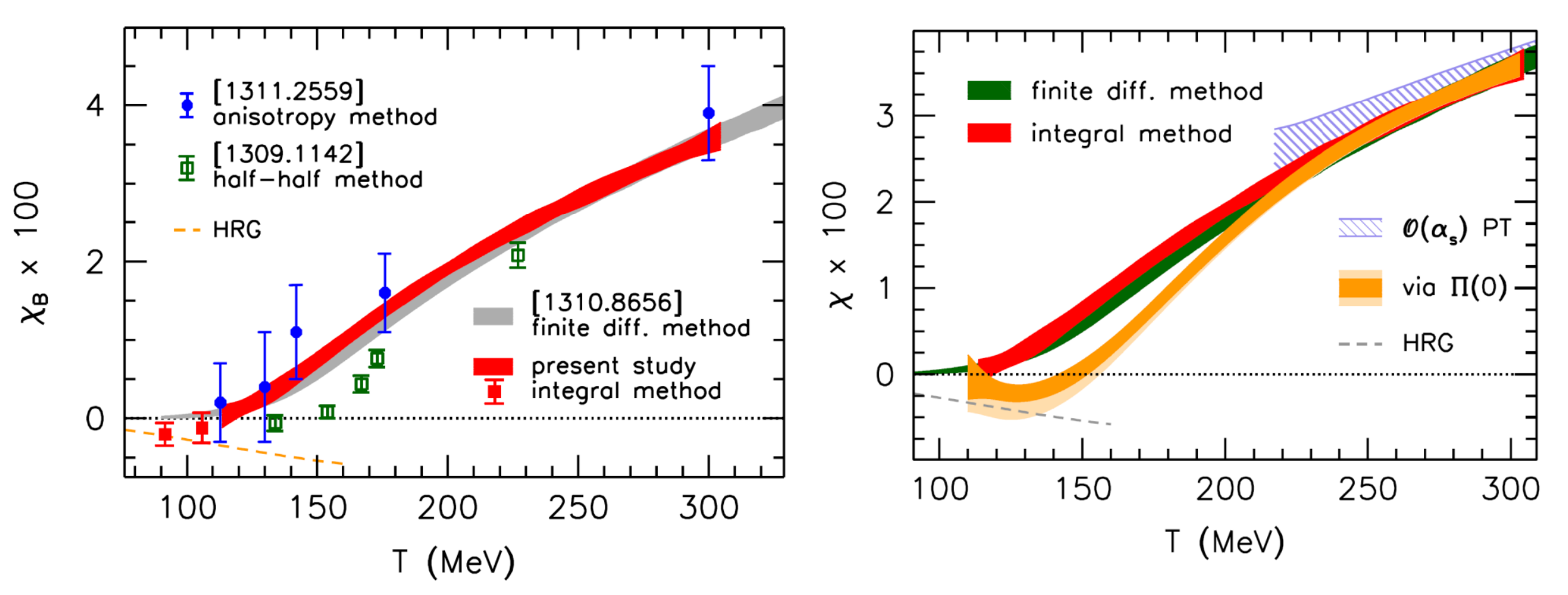}
\end{center}
\vspace{-1cm}
\caption{Magnetic susceptibility $ \chi_B$ (left), taken from Ref.~\cite{Bali:2014kia}. 
Shown data are from Refs.~\cite{Bali:2013owa, Bonati:2013vba, Levkova:2013qda, Bali:2014kia}. 
The latest update is shown with the orange band (right) \cite{Bali:2020bcn} 
together with the results from hadron resonance gas (HRG) model 
and perturbation theory (PT). 
}
  \label{fig:magnetization}
\end{figure}

Various approaches to the issue of 
the magnetic-field quantization 
have been proposed as indicated in the left panel of Fig.\ref{fig:magnetization}. 
The first result was obtained with a physics-motivated approach 
denoted as the ``anisotropy method'' 
where the magnetization is read off from the pressure anisotropy between the parallel and perpendicular components with respect to the magnetic-field direction \cite{Bali:2013esa, Bali:2013owa} (see a review paper~\cite{Hattori:2022hyo} 
for magnetization captured by the energy-momentum tensor). 
Subsequent results were obtained with 
the ``finite difference method'' 
where an interval between the quantized magnetic fields 
is interpolated by artificial systems with the Dirac strings \cite{Bonati:2013lca, Bonati:2013vba}, 
the ``half-and-half'' method where 
each half of the system volume is exposed to magnetic fields 
in the opposite directions so that the total magnetic flux is maintained vanishing \cite{Levkova:2013qda}, 
and the ``integral method'' 
where a simple mathematical trick relates 
the pressure anisotropy at the physical point 
to that at the infinite quark mass that can be analyzed by a large-mass expansion \cite{Bali:2014kia}. 
Overall, those approaches provided consistent results 
for the magnetic susceptibility $ \chi$ as 
shown in the left panel of Fig.\ref{fig:magnetization}. 
Some of those results are compared with the latest result 
in the right panel \cite{Bali:2020bcn} 
where the magnetic susceptibility 
was tied to the current-current correlator $\Pi $ 
of the Kubo formula type with a suitably chosen perturbation. 
As long as one focuses on the linear response regime, 
numerical simulations at a vanishing magnetic field 
should suffice as in this method, reducing numerical costs. 
See also a recent work \cite{Buividovich:2021fsa} 
for magnetization at finite density within the two-color QCD.

Now, we discuss the results in Fig.~\ref{fig:magnetization} at finite temperature. 
Those results are renormalized in such a way that 
the susceptibility vanishes at zero temperature 
with subtraction of the (divergent) vacuum contribution. 
In the high-temperature limit, 
the simulation results approach the free-quark contribution 
denoted as the perturbation theory (PT) 
and exhibit paramagnetism $ \chi > 0 $.  
Remember that, according to the discussions in Sec.~\ref{sec:vacuum_magnetism}, 
free fermions with spin-1/2 give rise to paramagnetism 
as a difference between the Pauli paramagnetism from a spin polarization and the Landau diamagnetism from eddy currents; 
The absolute magnitude of the former is three times larger 
than that of the latter. 
This ratio holds at finite temperature as well 
since the magnetic-field dependence in the effective action (\ref{action_quark}) is factorized as an overall factor 
in front of the sum of the vacuum and finite-temperature contributions when an electric field is absent in the medium rest frame (see Refs.~\cite{Bali:2014kia, Bali:2020bcn} 
for details in finite temperature calculations).

A significant update was made by carefully taking 
a continuum limit as shown by the orange band \cite{Bali:2020bcn}. 
One finds a diamagnetic region below 
the chiral phase transition temperature $ \sim 155 \ \MeV$ where 
the magnetic susceptibility takes negative values.\footnote{
A nice global parameterization over the whole temperature region 
is available in Ref.~\cite{Bali:2020bcn}, 
covering both the paramagnetic and diamagnetic regions. 
} 
The diamagnetic region was observed in the preceding result 
in lower temperature $\lesssim 100 \ \MeV $ \cite{Bali:2014kia} 
(see the left panel in Fig.~\ref{fig:magnetization}), 
but appeared to extend 
to a higher temperature as shown in the right panel. 
The existence of the diamagnetic region may be 
attributed to the contribution 
of abundant charged pions in the confinement phase 
that only provide diamagnetic contributions 
without spin (cf. Sec.~\ref{sec:vacuum_magnetism}). 
The simulation result indicates a rough agreement 
with that from the hadron resonance gas (HRG) model 
shown by the dashed line that includes contributions of 
other hadrons as well as pions up to around $ 1 \ \GeV$ masses (see Ref.~\cite{Endrodi:2013cs} for a list of included hadrons). 
Also, the reader is referred to results from 
functional renormalization group \cite{Kamikado:2014bua} 
and chiral perturbation theory \cite{Hofmann:2021bac}.

\subsubsection*{Spin and orbital contributions to magnetic susceptibility}

\begin{figure}[t]
\begin{center}
   \includegraphics[width=0.6\hsize]{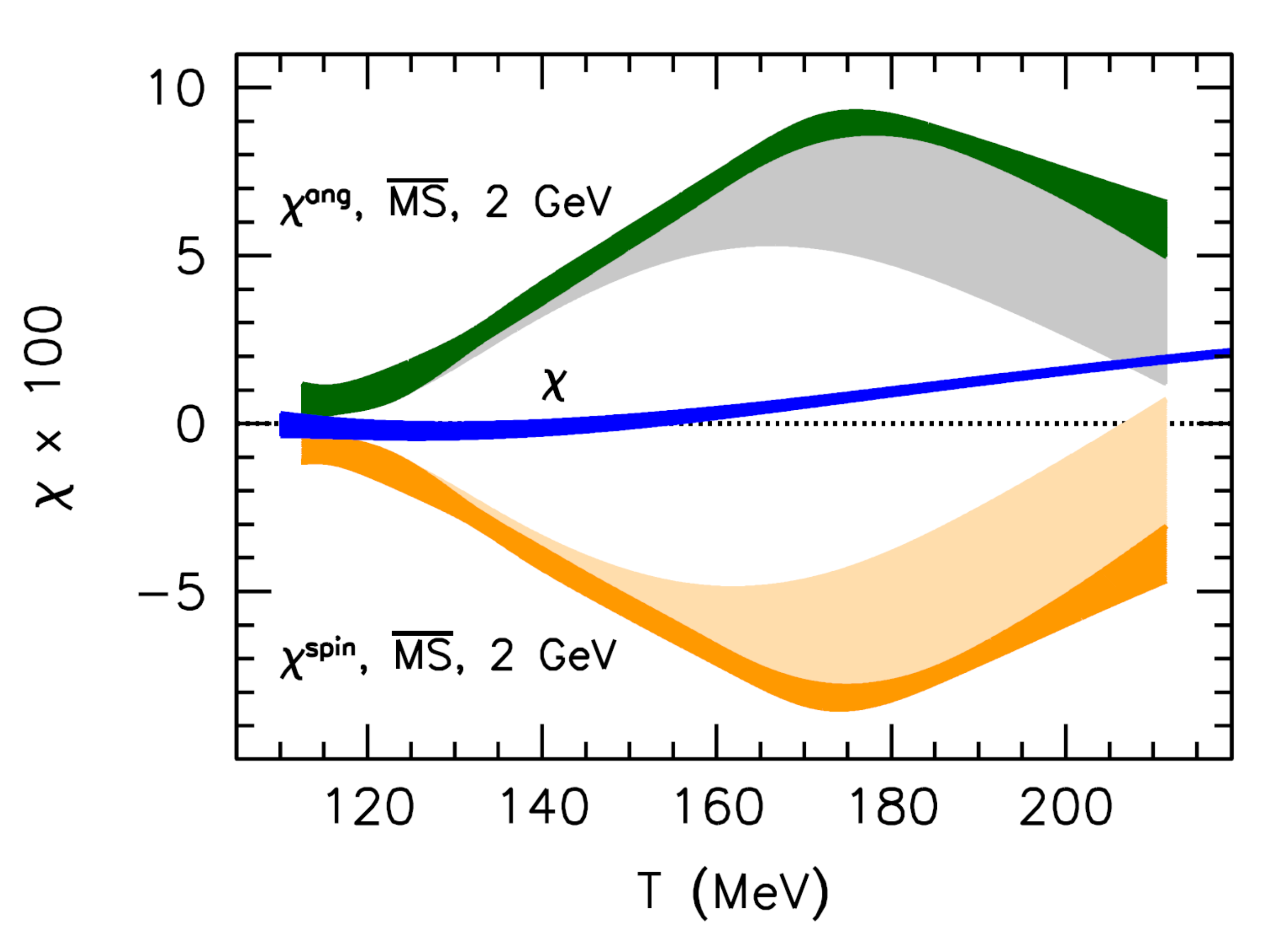}
\end{center}
\vspace{-0.7cm}
\caption{The total and quark-spin magnetic susceptibility 
shown by the blue and orange bands, respectively \cite{Bali:2012jv}. 
Each of them are renormalized to vanish at zero temperature. 
The green line shows the difference between 
the total amount and quark-spin contribution. 
}
  \label{fig:magnetization-breakdown}
\end{figure}

The simulation results in Fig.~\ref{fig:magnetization} 
suggest that, as we decrease temperature starting from 
the asymptotically high temperature, 
the paramagnetic region is smoothly connected to 
the diamagnetic region in the chirally broken phase. 
Since the asymptotically high temperature region 
is governed by free particles, 
this transition should stem from the QCD interaction effects. 
To investigate interaction effects, 
it is useful to separate spin and orbital contributions. 
The simulation results showed that the quark-spin contribution 
is diamagnetic below $T \lesssim 200 \ \MeV $ 
which is still far from the Pauli paramagnetism 
due to the free quarks at the asymptotically high temperature.

To investigate the breakdown, one should decompose 
the fermion determinant in the partition function (cf. Eq.~(\ref{Lag_original}) for the version without gluon fields). 
Similar to the decomposition in Eq.~(\ref{eq:K-GandZeeman}), 
one can decompose the product of the Dirac operators 
as $\sla D \sla D = D_\mu D^\mu 
+ \frac{g}{2} G^{\mu\nu} \sigma_{\mu\nu}
+ \frac{q_f}{2} F^{\mu\nu} \sigma_{\mu\nu} $ 
with $\sigma^{\mu\nu} =i/2 [\gam^\mu, \gam^\nu]$, 
where $F^{\mu\nu} $ is for a constant magnetic field 
and $  G^{\mu\nu} = G^{a \mu\nu} t^a $ 
is the gluon field strength tensor. 
The covariant derivative contains the gluon field 
as well as the constant magnetic field. 
The last term gives rise to the Zeeman effect 
in the fermion dispersion relation (see Sec.~\ref{sec:relativistic-fermion}). 
Therefore, it is reasonable to define 
a quark-spin contribution to magnetization 
by the contribution from the Zeeman term.\footnote{
More precisely speaking, this is a polarization of magnetic moment instead of spin itself 
and thus has additive contributions from 
antiquarks and quark flavors, 
while the spin angular momentum is subject to cancellation 
among particles with opposite electric charges 
when the magnetic moment is polarized in a magnetic field. 
}
Accordingly, an orbital contribution 
is defined as the contribution from 
the rest of the above decomposition. 
Note that the orbital contribution defined here 
contains not only quark contributions but also gluon contributions.

A nonzero gluon contribution to magnetization 
was indeed observed by measuring the anisotropic pressure, 
that is more specifically, the gluon field strength tensor 
squared and traced over the color index \cite{Bali:2013esa, DElia:2015eey}. 
The chromo-magnetic field has a larger magnitude 
along the magnetic field than in the perpendicular direction, 
while the chromo-electric field has 
a smaller magnitude along the magnetic field. 
This tendency is also seen in a perturbative expansion 
of the Heisenberg-Euler effective action 
to the second orders in the gluon fields 
and the magnetic field, i.e., the quantum correction 
from the box diagrams \cite{Bali:2013esa}. 
A full-order analysis both in the magnetic field and chromo-magnetic field, but without a chromo-electric field, 
is performed in Ref.~\cite{Ozaki:2015yja}; 
See Sec.~\ref{eq:chromo-B-condensate} 
and the discussions about 
Fig.~\ref{Fig:angle-dep} there. 
The magnitude of anisotropy in the gluon contribution 
is found to be smaller than that from the quark contribution, 
and was neglected in the ``anisotropy method'' \cite{Bali:2013esa, Bali:2013txa}. 
However, the gluon contributions are included 
in the other methods in Fig.~\ref{fig:magnetization} where 
any decomposition is not performed.

Since the quark-spin contribution is defined 
by the Zeeman term proportional to $\sigma^{\mu\nu} $, 
one can make a connection to 
measurement of the tensor condensate $\LR{\bar \psi \sigma^{\mu\nu} \psi} $ \cite{Bali:2012jv, Bali:2020bcn}. 
The tensor condensate can be nonzero 
because of a preferred orientation provided 
by the magnetic field, and thus is proportional to 
the magnetic-field strength in the linear response regime.\footnote{
Also, note that, 
since $ \sigma^{\mu\nu}$ and $ \gam^5$ commute with each other, 
the tensor condensate is nonzero only when 
there is chirality mixing on the quark propagator 
induced by a nonzero current quark mass 
and/or the chiral symmetry breaking. 
}
The magnetic susceptibility is accordingly read off from 
the linear slope parameter of the tensor condensate 
or is tied to a correlator at a vanishing magnetic field 
similarly to the aforementioned method 
for the total susceptibility \cite{Bali:2020bcn}; 
It was concluded that the former method with the simulation 
at nonzero magnetic fields works better 
with a smaller statistical error 
for the measurement of the tensor condensate. 
We add that the tensor condensate can be probed 
in terms of the operator product expansion as well \cite{Ioffe:1983ju, Balitsky:1983xk, Ball:2002ps, Vainshtein:2002nv, Gubler:2015qok}; 
Some of those results were compared with the simulation results  \cite{Bali:2012jv, Bali:2020bcn}.

The earlier simulation results suggested a diamagnetic quark-spin contribution up to $ T \lesssim 170 \ \MeV$ \cite{Bali:2012jv}. 
This result was confirmed in Ref.~\cite{Bali:2020bcn} where precision of the renormalization procedure was improved and temperature dependence was explicitly studied as shown in Fig.~\ref{fig:magnetization-breakdown}. 
The blue and orange bands show 
the total magnetic susceptibility 
from the correlator method (cf. Fig.~\ref{fig:magnetization}) 
and the quark-spin contribution from the tensor condensate, respectively.\footnote{
For clarity, one should note that 
the orange band in Fig.~\ref{fig:magnetization-breakdown} 
is for the quark-spin contribution. 
The total magnetic susceptibility shown by 
the orange band in Fig.~\ref{fig:magnetization} is 
shown by the blue band in Fig.~\ref{fig:magnetization-breakdown}. 
} 
They are the results of two independent measurements 
and are both renormalized to vanish at zero temperature. 
The green band shows the contribution other than the quark-spin contribution, i.e., the difference between 
the total and the quark-spin contribution 
(and is not from an independent measurement). 
The lighter orange band shows systematic uncertainties 
originating from an approximation (see Ref.~\cite{Bali:2020bcn} for details). 
One finds that the {\it diamagnetic quark-spin} contribution, 
which corresponds to negative values, 
extends from a low temperature to 
above the chiral and deconfinement phase transition temperatures. 
Accordingly, the rest part, shown by the green band, 
appeared to be positive, indicating a paramagnetic contribution. 
As compared to the Pauli paramagnetism for free fermions, 
the QCD interaction effects even induce a sign change 
in the spin magnetic susceptibility, 
exhibiting a significant impact on magnetism. 

The quark-spin contribution should eventually 
approaches the Pauli paramagnetism in higher temperature 
in consistent with the asymptotic freedom. 
Approaching the free-quark limit appears to require 
a sign change in the quark-spin contribution 
at a certain temperature $T \gtrsim 200 \ \MeV $. 
This behavior has not been directly confirmed by simulations; 
Though, it may be natural to expect that the orange band 
in Fig.~\ref{fig:magnetization-breakdown} keeps increasing 
above the phase transition temperatures 
to eventually reach the free-quark limit. 
Recall also that the total susceptibility is 
confirmed to approach the free-quark limit 
above $T \gtrsim 250 \ \MeV $ in Fig.~\ref{fig:magnetization}. 
In the lower temperature, 
the QCD interaction effects should be responsible for 
the deviation between the simulation result and 
the free-quark limit. 
One such effect is the color confinement 
where quark spin is forced to configure hadron spin. 
Inside charged pions, not both of the spin magnetic moments 
can be aligned in parallel to an external magnetic field, 
which diminishes the paramagnetic response of quark spin. 
One can see within the rather large uncertainties 
that the spin magnetic susceptibility 
takes a minimum value near the phrase transition temperatures.

The rest part, shown by the green band, 
is interpreted as a sum of 
the quark and gluon orbital contributions 
since gluon spin does not directly respond to a magnetic field. 
A nonzero gluon orbital contribution has been 
observed in the form of an anisotropic gluon pressure 
as mentioned above, 
but should become negligible in the high-temperature limit 
where free gluons do not respond to a magnetic field. 
Therefore, the green band is expected to approach 
the orbital contribution from free quarks 
that is the Landau diamagnetism with a negative susceptibility. 
This behavior has not been directly confirmed 
by simulations either; 
Within the plot range, the green band yet exhibits 
a positive sign and is required 
to change its sign above $T \gtrsim 200 \ \MeV $ 
to approach the free-quark limit.

\subsubsection*{Phases of magnetism}

More studies are demanded to understand 
the QCD interaction effects on magnetization 
near the phase transition region. 
It is useful to explicitly confirm whether and how the spin and orbital contributions approach their high-temperature limits. 
Approach to the free-quark limit could be slow 
since the run of the QCD coupling constant is logarithmic in nature; Lattice QCD simulations have shown 
a slow approach of the energy density and pressure 
of the quark-gluon plasma to the Stefan-Boltzmann limit 
(see, e.g., Refs.~\cite{Borsanyi:2013bia, HotQCD:2014kol}). 
Also, it is useful to pin down the temperature 
at which the spin magnetic susceptibility 
takes its minimum value. 
This temperature can be correlated 
with the chiral/deconfinement phase transition temperature. 
Along this line, it is interesting to pursue the correlation 
between magnetism and the chiral/deconfinement phase transition 
in stronger magnetic fields. 
The simulation results in Fig.~\ref{fig:magnetization} shows 
a crossover transition in consistent with 
the chiral and deconfinement crossover transitions 
at a vanishing magnetic field. 
However, the strengths of the chiral phase transition 
is enhanced with an increasing magnetic-field strength 
as we discussed above. 
This tendency may manifest itself in the magnetic susceptibility 
in the form of a sharper transition than the smooth behaviors 
seen in Figs.~\ref{fig:magnetization} and 
\ref{fig:magnetization-breakdown}. 
Magnetism in strong magnetic fields is left as an open question.

\subsubsection{Quark excitations in strong magnetic fields}

We saw three main challenges regarding 
the chiral symmetry breaking and restoration in QCD. 
(i) The magnitude of the chiral condensate at zero temperature 
is not fully understood at the quantitative level; 
Effective models of QCD and the chiral perturbation theory 
do not explain the linear dependence on the magnetic-field strength observed by the lattice simulations 
in the strong magnetic fields, though the increasing behavior 
is in a qualitative agreement with 
the picture of the magnetic catalysis. 
(ii) In spite of the increase of the chiral condensate 
in low temperature, the chiral condensate decreases 
as we increase the magnetic-field strength in higher temperature, 
which is known as the inverse magnetic catalysis. 
(iii) The transition temperature decreases 
as we increase the magnetic-field strength. 
Those issues can be related to one another 
since the transition temperature should be basically correlated with the dynamical quark mass as mentioned earlier.

What would be a possible mechanism 
that suppresses the magnetic catalysis of the chiral symmetry breaking? 
As mentioned in the beginning of Sec.~\ref{subsec:MC-weak}, 
one crucial point in the realization of the magnetic catalysis was that 
the NG bosons remain (3+1) dimensional excitations, avoiding 
the Coleman-Mermin-Wager theorem that inhibits the spontaneous symmetry breaking 
in (1+1) dimensions~\cite{Gusynin:1994xp, Gusynin:1994re, Gusynin:1994va}. 
This is simply because neutral mesons 
do not feel the Lorentz force and, on the other hand, 
charged mesons are no longer the NG bosons 
in magnetic fields. 
Neutral pion fluctuations are parametrically smaller than 
the quark fluctuations that are enhanced by the logarithmic factor and the Landau degeneracy. 


However, the composite structures of mesons need more investigation 
when the magnetic-field strength becomes larger than their typical scales $ \sim  \Lambda_\QCD $. 
Such strong magnetic fields can cause 
strong modifications of their inner structures 
because the quarks and antiquarks are subject to the dimensional reduction inside mesons. 
Then, the NG bosons could be also subject to the dimensional reduction 
and acquire the Landau degeneracy. 
If the dimensional reduction occurs 
at the level of neutral mesons as well, 
the magnitude of the chiral symmetry breaking 
should be diminished by the mesonic fluctuations as stated 
by the Coleman-Mermin-Wager theorem~\cite{Fukushima:2012kc}. 
In the language of the effective potential (see Sec.~\ref{sec:potential-MC}), 
the dimensionally reduced mesonic fluctuations provide the logarithmic singularity 
multiplied by the Landau degeneracy factor. 
In that case, the magnitude of the mesonic fluctuations 
is comparable in magnitude to the logarithmic singularity from the quark fluctuations that is the origin of the magnetic catalysis. 
Those two logarithmic fluctuations compete with each other 
due to opposite overall signs originating from the statistics.

The above argument suggests importance of investigating the meson spectra in strong magnetic fields. 
It was shown within the resummation of ring diagrams that 
the dispersion relation of the neutral pion indeed exhibits 
the dimensional reduction in the strong magnetic field, 
and the resulting logarithmic singularity 
diminishes the magnitude of the chiral symmetry breaking \cite{Fukushima:2012kc}. 
This mechanism was named the {\it magnetic inhibition}. 
The enhancement of the mesonic fluctuations is in favor of 
the milder slope of the chiral condensate at zero temperature 
and the inverse magnetic catalysis at finite temperature.

Yet, there are more to be clarified at the fundamental level. 
The transverse size of the mesonic fluctuations should depend on 
the interaction range between the quark and antiquark pair. 
One may thus wonder what aspects of the QCD interaction 
could give rise to the dimensionally reduced mesonic fluctuations as compared to QED and typical effective models. 
The mesonic excitations were more systematically 
studied by explicit construction 
of the Bethe-Salpeter equations not only for the NG bosons 
but also light charged mesons \cite{Kojo:2012js, Hattori:2015aki}. 
Below, we discuss roles of QCD-motivated interactions 
in the spectra of mesonic fluctuations. 
However, even before constructing mesons, 
one needs to determine the constituent quark mass 
in strong magnetic fields 
because the low-lying meson spectrum is predominantly 
determined by the constituent quark mass. 
One should discuss roles of QCD-motivated interactions 
on the determination of the constituent quark mass.

\subsubsection*{Saturation of the dynamical quark mass}

Remember that we obtained the phase transition temperature 
for the NJL model in Eq.~(\ref{m_dyn-potential}) 
that is proportional to the dynamical mass at zero temperature \cite{Ebert:1997um} 
\begin{eqnarray}
T_c  \sim m_\dyn(T=0)
\label{eq:Tc-m-2}
\, .
\end{eqnarray}
A similar relation is established in QED \cite{Gusynin:1997kj, Lee:1997zj, Lee:1997uh} 
(see also, e.g., Ref.~\cite{tinkham2004introduction} for an analogous relation in the BCS theory). 
Therefore, the critical temperature of the chiral symmetry restoration is 
naturally interpreted as the point at which the thermal energy becomes large enough to overcome 
the mass gap and to fill the fermion states with thermal excitations. 
This just means that the chiral symmetry would not be restored 
without a significant amount of thermal excitations. 
This simple observation is expected to hold on general grounds. 
Thus, the magnitude of the dynamical mass, rather than of the chiral condensate, 
may provide a direct clue to understand the results from the lattice QCD simulations. 

The mass gap, the excitation energy of quasiparticle and antiparticles, 
can be determined by the Schwinger-Dyson (SD) equation in magnetic fields. 
The SD equation has been used for QED and the NJL model 
at zero temperature \cite{Leung:1996qy, Shushpanov:1997sf}
and finite temperature \cite{Gusynin:1997kj, Lee:1997zj, Lee:1997uh}. 
It may be worth mentioning that 
the temperature dependence of the superconducting gap, 
the counterpart of the dynamical mass, 
has been also often discussed on the basis of 
the BCS theory in the weak-coupling limit (see, e.g., Ref.~\cite{tinkham2004introduction}). 
However, in the nonperturbative QCD, 
the excitation energy could be quite different from 
that in weak-coupling theories.

We shall try to grab a key aspect of 
the nonperturbative gluon interactions. 
Such effects on the SD equation were first discussed by Kojo and Su \cite{Kojo:2012js, Kojo:2013uua} 
and further in Refs.~\cite{Watson:2013ghq, Mueller:2014tea, Mueller:2015fka, Hattori:2015aki}. 
Below, we discuss the SD equation in the rainbow approximation. 
All the fermions on the external and internal lines 
are assumed to stay in the LLLs since 
the inter-Landau-level transition is suppressed 
by a large Landau spacing in the strong-field limit.

\begin{figure}[t]
\begin{center}
   \includegraphics[width=0.6\hsize]{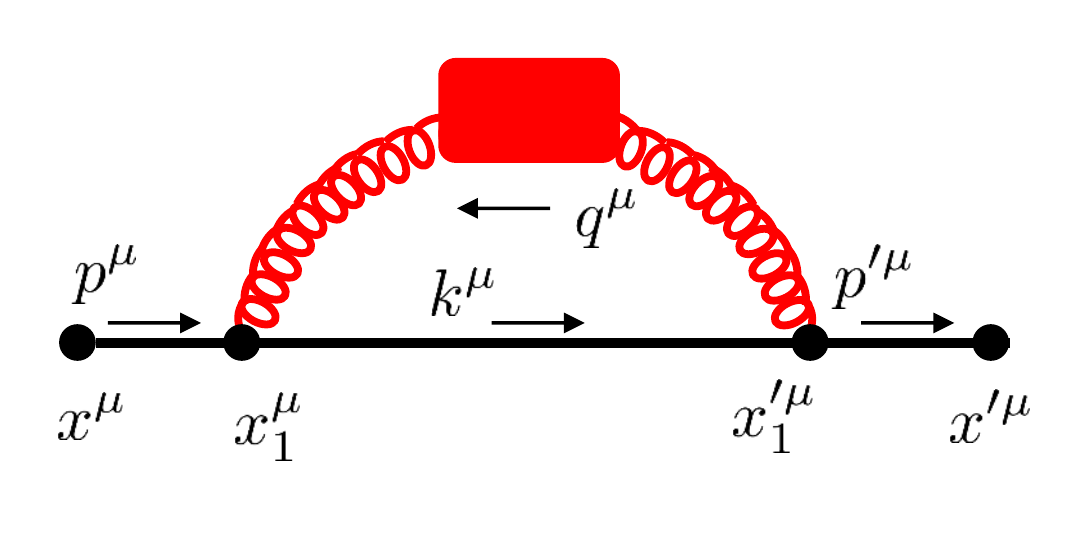}
\end{center}
\vspace{-1cm}
\caption{Quark self-energy in the coordinate space. 
The quark lines, denoted by the black lines, 
are dressed by a magnetic field. 
}
  \label{fig:SD-eq}
\end{figure}

\cout{

\cgd{

We would like to get Green's function $ S(x^\mu ,x ^{\prime\mu} ) $ of the Dirac operator such that 
\begin{eqnarray}
\label{eq:Dirac-S}
(i \slashed D _x -m )  S(x^\mu ,x ^{\prime\mu}) = i \delta^{(4)}(x^\mu -x ^{\prime\mu})
 \, .
\end{eqnarray}
Notice that the Dirac operator can be rewritten as 
\begin{eqnarray}
i \slashed D  -m 
\= ( i \slashed \partial _\parallel -m) 
- \sqrt{2|q_f B|}  \gamma^1\left(\hat a \prj_+ + \hat a^\dagger \prj_- \right) 
 \, ,
 \label{Dirac_Ritus}
\end{eqnarray}
with the spin-projection operator $ \prj _\pm 
=  \left(1 \pm  i s_f  \gamma^1 \gamma^2  \right)/2 $. 
We now project the propagator onto the LLL wave function 
\begin{eqnarray}
S(x^\mu ,x ^{\prime\mu}) \= \int \frac{d^2p_\para}{(2\pi)^2} \int \frac{dp_y}{2\pi} 
e^{-ip_\para \cdot (x_\para-x'_\para) + i p_y (y-y')}
\nnb
&&
\times
\phi_\LLL (x - \frac{ p_y}{q_fB} )  {\mathcal S}_\LLL(p_\para) 
\phi^\ast_\LLL  (x - \frac{ p_y}{q_fB} )  \prj_+
\, .
\end{eqnarray} 
We denote the first and second components of the four vector $  x^\mu$ 
as $ x $ and $y  $, respectively. 
We have chosen the Landau gauge $ A^\mu = (0, 0, Bx ,0) $ for the external magnetic field. 
Then, the second component of the fermion momentum $ p_y := p^2$ is still a good quantum number, 
and the plane wave $ e^{ i p_y y} $ serves as the eigenfunction of the Dirac operator. 
This also suggests that the LLL states can be labelled by $ p_y  $ as $   | 0, p_y \rangle  $ 
with the lowest principal quantum number $  \hat a^\dagger \hat a | 0, p_y \rangle = 0  $ 
and the normalization $ \langle 0, p_y  | 0, p'_y \rangle  = \delta(p_y - p_y') $. 
The rest part of the LLL wave function is given by the Hermite polynomial 
(see, e.g., Appendix in Ref.~\cite{Hattori:2020htm}) 
\begin{eqnarray}
e^{ i p_y y}  \phi_\LLL  (x - \frac{ p_y}{q_fB} )  := \langle x^\mu_\perp | 0, p_y \rangle 
= e^{ i p_y y}  \Big( \frac{ |q_f B| }{ \pi } \Big)^{\frac14} e^{ -  \frac{ |q_fB| }{2} (x - \frac{ p_y}{q_fB} )^2  }
\, .
\end{eqnarray} 
We will shortly clarify how the gauge dependence of the wave function is made tractable. 
The propagator takes a simple form in this basis called 
the Ritus basis \cite{Ritus:1972ky, Ritus:1978cj}, 
just because it is spanned by the eigenfunctions of the Dirac operator. 
After the LLL projection, we find 
\begin{eqnarray}
{\mathcal S}_\LLL(p_\para)  = \frac{i}{ \sla p_\para - m}
\, .
\end{eqnarray}
This serves as Green's function of Eq.~(\ref{eq:Dirac-S}), 
assuming that the completeness is here saturated by the (degenerate) ground states 
\begin{eqnarray}
\delta^{(2)} (x_\perp^\mu - x^{\prime\mu}_\perp) = \sum_{n=0}^\infty \int \frac{dp_y}{2\pi} 
 \langle x^\mu_\perp | n, p_y \rangle \langle n, p_y | x^{\prime\mu}_\perp \rangle
\sim   \int \frac{dp_y}{2\pi}  \langle x^\mu_\perp | 0, p_y \rangle \langle 0, p_y | x^{\prime\mu}_\perp \rangle
\, .
\end{eqnarray}
If one includes all the higher Landau levels (hLLs), 
they provide additive contributions to the fermion propagator 
that is given by summation over the Landau levels (see, e.g., Ref.~\cite{Gusynin:1995nb}). 

}

\pp{

\begin{eqnarray}
\Sigma(x^\mu, x^{\prime\mu}) 
\= S(x^\mu, x^{\prime\mu}) + \sum_{n=1}^\infty
\int d^4x_1 \cdots d^4x_n
S(x_n^\mu, x^{\prime\mu}) \Sigma(x_{n-1}^\mu, x_n^{\mu}) 
 S(x_{n-2}^\mu, x_{n-1}^{\mu}) \Sigma(x_{n-3}^\mu, x_{n-2}^{\mu}) 
 \cdots
  S(x^\mu, x_1^{\mu})  
\end{eqnarray}

}

} 

As discussed in Sec.~\ref{sec:resum},  
the quark propagator is factorized into the Schwinger phase 
and the translation-invariant part as $ S (x^\mu,x^{\prime\mu}) 
=e^{   i \Phi_A(x^\mu,x^{\prime\mu}) } \bar S(x^\mu-x^{\prime\mu})$. 
The Schwinger phase is given as $\Phi(x^\mu,x^{\prime\mu}) 
= \frac{q_fB}{2} (x+x')(y-y') $ 
in the Landau gauge specified in Eq.~(\ref{eq:Landau-g}).\footnote{Note that $x $ and $y $ ($x'$ and $y'$) are 
the first and second spatial components of 
$ x^\mu$ ($x^{\prime\mu} $).}  
Plugging those ingredients, the coordinate representation of 
the tree-level quark propagator can be arranged as 
\begin{eqnarray}
S^\zero_\LLL (x,x') &:=& e^{ i \Phi_A(x,x')} 
\int \frac{d^4 p}{(2\pi)^4} e^{ -i p \cdot (x'-x)} S_\LLL (p)
\nnb
\= \prj_+ \int \frac{d^2 p_\para}{(2\pi)^2}  
\int \frac{d p_y}{2\pi}   \frac{i}{\sla p_\para -m}
 \nnb
 &&\times
e^{- i p_\para \cdot(x'_\para-x_\para)} e^{  i p_y(y'-y)}  
\phi_\LLL \big( \frac{ x - s_f \ell_f^2 p_y }{\ell_f}  \big)
\phi_\LLL \big( \frac{ x' - s_f \ell_f^2 p_y }{\ell_f}  \big)
\, ,
\end{eqnarray}
where $  S_\LLL (k)$ is the translation-invariant part of the fermion propagator (\ref{eq:prop-LLL}) in the momentum space. 
We explicitly extracted the wave function in the Landau gauge 
(\ref{eq:WF_Landau}) which is here denoted as 
$\phi_\LLL ( \xi )= e^{- \xi^2/2}/ \sqrt{\pi^{1/2} \ell_f}$ 
without the plane-wave factor for the $y $ direction. 
The above arrangement is essentially a basis change 
from the Fourier basis to the Ritus basis 
(see Appendix~\ref{sec:relation}). 
Then, the one-loop quark propagator with 
the gluon dress can be written as (cf. Fig.~\ref{fig:SD-eq})
\begin{eqnarray}
S^\one_\LLL (x,x')
&:=& \int d^4x_1 \int d^4 x_1^{\prime}
S(x^\mu,x_1^{\mu})\gam^{\mu_1} S(x_1^\mu,x_1^{\prime\mu})
D_{\mu_1\nu_1} (x_1^\mu,x_1^{\prime\mu})
\gam^{\nu_1} S(x_1^{\prime\mu},x^{\prime\mu})
\nnb 
\= \prj_+ 
\int \frac{d^2 p_\para}{(2\pi)^2}  \int \frac{d p_y}{2\pi} 
 e^{ i p_\para \cdot x_\para - i p_y y} 
\phi_\LLL \big( \frac{ x - s_f \ell_f^2 p_y }{\ell_f}  \big)
\nnb
&& \times
\int \frac{d^2 p'_\para}{(2\pi)^2}  \int \frac{d p'_y}{2\pi} 
 e^{ - i p_\para \cdot x'_{\para} + i p'_yy'} 
\phi_\LLL \big( \frac{ x' - s_f \ell_f^2 p'_y }{\ell_f}  \big)  
\nnb
&& \times
\frac{i}{\sla p_\para -m}
\Big[ - i \Sigma_\LLL^\one( p_\para^\mu,p_y )
(2\pi)^4 \delta^{(2)}(p_\para^\mu-p_\para^\mu) \delta(p_y-p_y')
\rho_B \Big] 
\frac{i}{\sla p'_\para -m}
\, .
\end{eqnarray}
The external quark lines are factorized as the three-dimensional integral measures and the quark wave functions, 
and we are left with the amputated quark self-energy 
$\Sigma_\LLL^\one( p_\para^\mu,p_y ) $. 
The three-dimensional delta functions and 
the Landau degeneracy factor are also factorized 
as we know that the integrals of the wave functions should 
provide them. 
The explicit form of the amputated quark self-energy reads 
\begin{eqnarray}
&& \hspace{-1cm}
-i \Sigma_{\LLL}^\one (p_\para^\mu, p_y) 
(2\pi)^4 \delta^{(2)}(p_\para^\mu-p_\para^\mu) \delta(p_y-p_y')
\rho_B 
\nnb
\=  (ig)^2 C_2 \prj_+  \int d^4x  \int d^4x' 
 e^{ - i (p_\para^\mu x_\mu - i p_y y)
+ i (p_\para^{\prime\mu} x'_\mu - i p'_y y') } 
\phi_\LLL( \frac{ x - s_f \ell_f^2 p_y}{\ell_f} )
\phi_\LLL^\ast  ( \frac{ x' - s_f \ell_f^2 p_y'}{\ell_f} )
\nnb
 &&\times
 e^{   i \Phi_A(x^\mu,x^{\prime\mu}) } 
 \int \frac{d^4 k}{(2\pi)^4}  e^{-i k \cdot (x'-x) } 
 \int \frac{d^4 q}{(2\pi)^4}  e^{-i q \cdot (x-x') } 
\gam^\mu_\para  S_\LLL (k) \gam^\nu_\para D_{\mu\nu} (q)
\nn
\, ,
\end{eqnarray}
where $ D_{\mu\nu} (q)$ is 
the gluon propagator in the momentum space. 
Note that the gluon propagator is not necessarily 
a free propagator but is a resummed one that includes 
nonperturbative effects. 
We assumed that the gluon propagator is diagonal 
in the color space, so that 
the self-energy is also diagonal in the color space 
with the factor of $C_2 = (N_c^2-1)/(2N_c) $.  
All the fermion lines in the LLLs come with 
the spin projection operators $ \prj_+$, 
and the gamma matrices sandwiched in between them can 
take only the parallel components, i.e., 
$\prj_+ \gamma^\mu  \prj_+ = \prj_+ \gamma_\para^\mu$. 
We suppress the spinor and diagonal color indices 
of the self-energy.

One can extract the three-dimensional momentum conservation 
by changing the integral variables to 
$ X^\mu=(x^\mu+x^{\prime\mu})/2$ 
and $\tilde x^\mu = x^\mu-x^{\prime\mu} $. 
After this arrangement, we find that  
\begin{eqnarray}
\Sigma_{\LLL}^\one (p_\para^\mu, p_y)
\= i (ig)^2 C_2 \prj_+  
\rho_B^{-1} \int dX  \int d  \tilde x \ 
\phi_\LLL \big( \frac{ X+\frac{\tilde x}{2} - s_f \ell_f^2 p_y}{\ell_f} \big)
\phi_\LLL ^\ast \big( \frac{ X-\frac{\tilde x}{2} - s_f \ell_f^2 p_y}{\ell_f} \big)
 \nnb
&&\times
 \int \frac{d^4 k}{(2\pi)^4}   \int \frac{d^4 q}{(2\pi)^4}  
 e^{ - i (k_x - q_x) \tilde x }
 (2\pi)^3  \delta^{(2)} ( p_\para - k_\para + q_\para) 
 \delta( p_y - k_y + q_y - q_f B X )
  \nnb
&&\times
\gam^\mu_\para  S_\LLL (k) \gam^\nu_\para D_{\mu\nu} (q)
\, ,
\end{eqnarray}
where the remaining integral variables $ X$ and $ \tilde x$ 
are the first spatial components of $ X^\mu$ and $\tilde x^\mu $. 
Inserting the explicit forms of the LLL wave function 
and the LLL propagator, we arrive at  
\begin{eqnarray}
\label{eq:quark-sf-one-loop}
\Sigma_\LLL^\one (p_\para^\mu, p_y)  
= 
\prj_+   \int \frac{d^2 k_\para}{(2\pi)^2}  
\gam^\mu_\para \frac{ \tilde D_{\mu\nu} (k_\para - p_\para)}
{\sla k_\para} \gam^\nu_\para   
\, ,
\end{eqnarray} 
where we defined 
\begin{eqnarray}
\label{eq:reduced-gluon}
\tilde D_{\mu\nu} (k_\para)
= g^2 C_2 \int \frac{d^2 q_\perp}{(2\pi)^2} 
 e^{ - \frac{|\bq_\perp|^2}{2|q_fB| } } 
 D_{\mu\nu} (k_\para; \bq_\perp)
 \, .
\end{eqnarray}
The three components of the external momentum, 
$p_\para^\mu $ and $ p_y$, are conserved in the Landau gauge. 
$p_y $ labels the degenerate fermion states 
in homogeneous magnetic fields. 
The self-energy does not depend on $p_y $ 
(except for the delta function), meaning that 
all the degenerate states acquire 
the same self-energy correction as expected. 
The rotational symmetry with respect to 
the magnetic-field direction 
has been restored despite of the use of the Landau gauge. 
This implies that the self-energy does not depend on 
the gauge choice for the external magnetic field.


Now, we proceed to the SD equation 
in the rainbow approximation, following 
familiar procedures in the absence of a magnetic field. 
We resum the quark self-energy linked by 
the tree-level quark propagators $S^\zero_\LLL (x,x')$ 
in the form of the geometrical series. 
Further, nesting the fermion self-energy, we obtain 
the SD equation in the rainbow approximation 
\begin{eqnarray}
\label{eq:SD-LLL}
 \Sigma_\LLL (p_\para^\mu)  
\=  \prj_+  
\int \frac{d^2 k_\para}{(2\pi)^2}  
\gam^\mu_\para \frac{\tilde D_{\mu\nu} (k_\para - p_\para)}
{ \sla k_\para - \Sigma_\LLL (k_\para) } \gam^\nu_\para   
\, .
\end{eqnarray}
We denote the solution for the SD equation as 
$ \Sigma_\LLL  $ that contains 
higher loop effects within the rainbow approximation. 
Notice that this equation has a (1+1)-dimensional form 
after the transverse gluon momentum is integrated out 
in the reduced propagator (\ref{eq:reduced-gluon}). 
The magnetic-field dependence is not explicit in this (1+1)-dimensional form, and is solely encoded in the Gaussian factor 
in the reduced gluon propagator (\ref{eq:reduced-gluon}). 
Below, we discuss how the reduced gluon propagator controls 
the density of intermediate virtual states 
that contribute to the self-energy correction. 
This is a crucial point since the magnitude of 
the self-energy correction depends on 
how much of the degenerate LLLs, 
spread over the transverse plane, 
participates in the intermediate state.

The semi-classical picture helps us 
with understanding the basic physics involved 
in the SD equation (\ref{eq:SD-LLL}). 
Initially, a quark on the external line is confined in a cyclotron orbit 
and can propagate only in the longitudinal direction. 
By emitting a gluon in the intermediate virtual state, 
the quark can hop from the initial orbit to any one of the degenerate orbits aligned in the transverse plane. 
Then, it comes back to the original position by absorbing the gluon back. 
Here, we understand that 
all the degenerate LLL states contribute to the self-energy 
if this transverse hopping distance can be infinite, 
while a limited number of the degenerate states contributes 
if the hopping distance is finite.



More specifically, in the Landau gauge (\ref{eq:Landau-g}), 
one of the canonical transverse momentum 
$\xc =  p_y/(q_f B) $ specifies 
the location of the cyclotron orbit. 
A quark at $ \xc $ hops to another position $\xc^\prime =  k_y/(q_f B) $ in the intermediate virtual state. 
According to the canonical-momentum conservation, 
the hopping distance is specified by the gluon momentum $ q_y$ 
as $ \Delta x :=  \xc'-\xc = q_y/(q_f B)$. 
It should be emphasized that 
the {\it soft} gluon momentum corresponds to 
a {\it short} distance in the transverse plane 
instead of a long distance in usual Fourier analyses. 
This difference stems from the specific property, 
i.e., $\xc =  p_y/(q_f B) $, 
of the basis in the presence of magnetic fields. 
At the quantum level, the hopping rate should be captured by 
the overlap among the two quark wave functions and the plane-wave basis for the gluon. 
This convolution results in the Gaussian factor 
in the reduced gluon propagator (\ref{eq:reduced-gluon}) 
where the integrand depends on the ratio of the hopping distance 
to the magnetic length $ \Delta x^2/\ell_f^2  $ 
as expected from the above observation. 


Now, we focus on the gluon spectrum since it controls the transverse size of the virtual fluctuations. 
Namely, a typical hopping distance is determined by the momentum dependence of the gluon propagator. 
One can examine two extreme cases.

\begin{itemize}

\item[•] {\it Momentum-independent interactions.---}
If the gluon propagator is constant or extremely flat in the transverse-momentum dependence, 
the hopping can occur at any spatial scale. 
This means that all the degenerate states participate in 
the fluctuations, 
enhancing the quark self-energy by the Landau degeneracy factor. 
Indeed, the Gaussian integral in Eq.~(\ref{eq:reduced-gluon}) 
results in the Landau degeneracy factor 
when the gluon propagator does not depend on 
the transverse momentum.

\item[•] {\it IR-dominant interactions.---}
In the opposite extreme case where the gluon propagator 
has a dominant support in the soft momentum region 
$ \Delta x^2/\ell_f^2  = |\bq_\perp|^2 / |q_f B| \ll 1$, the Gaussian factor is approximately a constant in the integral. 
What is remarkable is that the integral no longer depends on the magnetic-field strength, 
and neither does the quark self-energy \cite{Kojo:2012js}. 
The dimensionful factor may be provided by a typical momentum scale 
characterizing the gluon propagator instead of the Landau degeneracy factor. 
This scale works as an effective cutoff of the hopping distance. 

\end{itemize}

\cout{
The latter extreme case occurs when the typical hopping distance becomes smaller than the cyclotron radius.\footnote{ 
For a fixed gluon momentum, the scaling of the coordinate $ \sim 1/ q_fB $ is faster 
than the shrinking of the cyclotron radius $ \sim 1/ \sqrt{ q_fB} $ when we increase the magnetic-field strength. 
}
Under such a condition, the degenerate states outside 
the initial cyclotron orbit do not contribute to the integral. 
Therefore, the effective density of states 
that can participate in the self-energy correction 
is insensitive to how large the density of degenerate orbits 
is in the entire transverse plane. 

}

What should we expect to happen in the low-energy QCD? 
One should expect that the momentum dependence 
of the gluon propagator has a larger weight 
in the IR regime $|\bq_\perp|^2  \ll \Lambda_\QCD^2 $ 
than the UV regime, where $ \Lambda_\QCD $ is the QCD scale. 
The existence of this emergent IR scale 
is a definite property of QCD that 
divides the perturbative and nonperturbative regimes 
and distinguishes QCD from QED. 
This situation is close to the latter of 
the above two extreme cases. 
Therefore, the dynamical quark mass approaches 
a value $ \sim \Lambda_\QCD $ 
as we increase the magnetic-field strength to the regime 
$ \Delta x^2/\ell_f^2  = \Lambda_\QCD^2 / |q_f B| \ll 1$. 
This offers a saturation mechanism of the dynamical quark mass 
when we increase the magnetic-field strength beyond $\Lambda_\QCD^2 $. 
The saturation of the dynamical quark mass 
was first illustrated with 
a confinement-potential model \cite{Kojo:2012js} 
and is supported by other approaches \cite{Kojo:2013uua, Watson:2013ghq, Mueller:2014tea, Mueller:2015fka, Hattori:2015aki}.

One can interpret the reduced gluon propagator 
(\ref{eq:reduced-gluon}) as a quasi-(1+1)-dimensional interaction 
realized by the cooperative effects of the IR-dominant 
gluon propagator and the Landau quantization 
in the quark and antiquark eigenstates. 
Accordingly, both quarks and gluon interactions are subject 
to effective dimensional reductions 
even though magnetic fields do not directly couple to gluons. 
Then, one can expect that bound states or particle pairing, 
such as the chiral condensate and mesons, also reduce to 
(1+1) dimensional forms as speculated earlier. 


It would be also interesting to investigate the saturation mechanism of the dynamical quark mass 
with the RG method (see Refs.~\cite{Skokov:2011ib, Fukushima:2012xw, Kamikado:2013pya, 
Andersen:2014oaa, Braun:2014fua, Mueller:2015fka, Aoki:2015mqa, Hattori:2017qio} for RG analyses). 
In Sec.~\ref{sec:RG-gauge}, we discussed 
the equivalence between the analyses by 
the RG method and the Schwinger-Dyson equation 
and stressed the importance of the scale hierarchy 
in the RG evolution (cf. Fig.~\ref{fig:RG-MC}). 
The IR-dominant gluon interactions may modify 
the RG evolution of the four-Fermi operator 
below the QCD scale from that in QED. 


\cout{

What should we expect to happen in the low-energy QCD? 
The former extreme case corresponds to typical 
contact-interaction models that fail to consistently explain 
the results from the lattice QCD simulations. 
However, the gluon propagator should certainly 
has a momentum dependence. 
While it is not simple to obtain 
a nonperturbative gluon propagator in the low-energy QCD, 
the existence of an IR scale, i.e., the QCD scale $ \Lambda_\QCD $, 
is a definite property of QCD, dividing the perturbative and nonperturbative regimes. 
The existence of this emergent scale and the associated 
strong-coupling regime below $  \Lambda_\QCD $ distinguishes QCD from QED. 
One may then expect that the gluon propagator has a larger weight 
in the IR regime $|\bq_\perp|^2  \ll \Lambda_\QCD^2 $ than the UV regime. 
This is an approximate realization of the latter extreme case, 
and the dynamical quark mass approaches a saturation value $ \sim \Lambda_\QCD $ 
when the hopping distance becomes smaller than the cyclotron radius 
$ \Delta x^2/\ell_f^2  = \Lambda^2 / |q_f B| \ll 1$. 
This point was first illustrated with 
a confinement-potential model \cite{Kojo:2012js} 
and is supported by other approaches in Refs.~\cite{Kojo:2013uua, Watson:2013ghq, Mueller:2014tea, Mueller:2015fka, Hattori:2015aki}. 
\com{Some references to the QCD papers.} 
\cgd{
In short, the cooperative effects of the IR dominance 
and the strong magnetic field induce 
an additional ``dimensional reduction'' in 
the interaction range even though the magnetic fields 
do not directly couple to gluons. 
Then, one can expect that bound states, such as 
the chiral condensate and mesons, also reduce to 
(1+1) dimensional forms as speculated earlier. 

}
} 

\subsubsection*{Parametric estimate of the chiral condensate}

One can construct a relation between the dynamical quark mass 
and the chiral condensate 
in the spirit of the constituent quark model. 
With insertion of the self-energy $\Sigma_\LLL $ 
into the quark propagator (\ref{eq:prop-LLL}), 
the chiral condensate is given by a loop of 
the resummed quark propagator as 
\cite{Shushpanov:1997sf, Kojo:2012js}\footnote{
The Schwinger phase cancels for the one-point function, 
indicating that the chiral condensate is 
independent of the gauge choice 
for the external magnetic fields.} 
\begin{eqnarray}
\lan \bar \psi \psi \ran 
= 2i \int \frac{d^4p}{(2\pi)^4} 
e^{ - \frac{|\bp_\perp|^2}{|q_fB|}}
\tr \Big[  \frac{ 1} { \sla p_\para -\Sigma_\LLL } \Big]
\sim    \frac{|q_fB|}{2\pi} m_\dyn
\, .
\end{eqnarray} 
The transverse momentum integral results 
in the Landau degeneracy factor. 
The remaining integral is dominated by 
the infrared contribution, and is proportional to 
$m_\dyn = \Sigma_\LLL(0) $ for the dimensional reason 
in a qualitative estimate. 
The dynamical quark mass quantifies the magnitude of the (1+1) dimensional pairing between a quark and an antiquark, 
while the Landau degeneracy factor counts 
the density of the (1+1) dimensional pairing 
aligned in the transverse plane.

The above factorization is a universal consequence of the dimensional reduction, 
so that the microscopic details of the low-energy QCD 
is solely captured by the magnitude of $ m_\dyn $. 
When the saturation mechanism works for 
the constituent quark mass $ m_\dyn \sim \Lambda_\QCD $ in the strong magnetic fields, 
the magnitude of the chiral condensate is estimated to be \cite{Kojo:2012js, Kojo:2013uua}
 \begin{eqnarray}
 \label{eq:chiral-condensate}
\lan \bar \psi \psi \ran &\sim&   \frac{|q_fB|}{2\pi} \Lambda_\QCD
\, .
\end{eqnarray}
This estimate is in favor of the linear dependence of the chiral condensate on the magnetic-field strength 
that has been observed in the lattice QCD simulations at zero and low temperature (cf. Fig.~\ref{fig:lattice-chiral}).

Putting it in the other way around, the linear increase of the chiral condensate 
does not necessarily imply an increase of the dynamical quark mass. 
As long as the dynamical quark mass remains as small as the QCD scale $ \Lambda_\QCD$, 
the chiral restoration temperature can also remain as small as the QCD scale 
or should even decrease because the Landau degeneracy factor 
enhances the thermal fluctuations \cite{Kojo:2012js}. 
Therefore, the above scenario can offer a consistent 
explanation of the magnetic-field dependence of the magnetic catalysis in the low-temperature region 
and the decrease of the transition temperature; 
Those two issues are intimately related to one another 
via the quark excitation spectrum of the strongly coupled QCD 
in the strong magnetic fields. 


In the above estimate, one can see that 
the estimate of the chiral condensate 
(\ref{eq:chiral-condensate}) overshoots 
the lattice QCD result if the dynamical quark mass grows 
with the magnetic field strength $m_\dyn \sim q_f B $ 
as seen in Fig.~\ref{fig:lattice-chiral}. 
Such a problem occurs when the whole degenerate states 
contribute to the quark self-energy with short-range interactions 
as discussed below Eq.~(\ref{eq:SD-LLL}).


%
%

\subsubsection{Debye screening effect and the deconfinement phase transition}

\begin{figure}[t]
     \begin{center}
              \includegraphics[width=0.5\hsize]{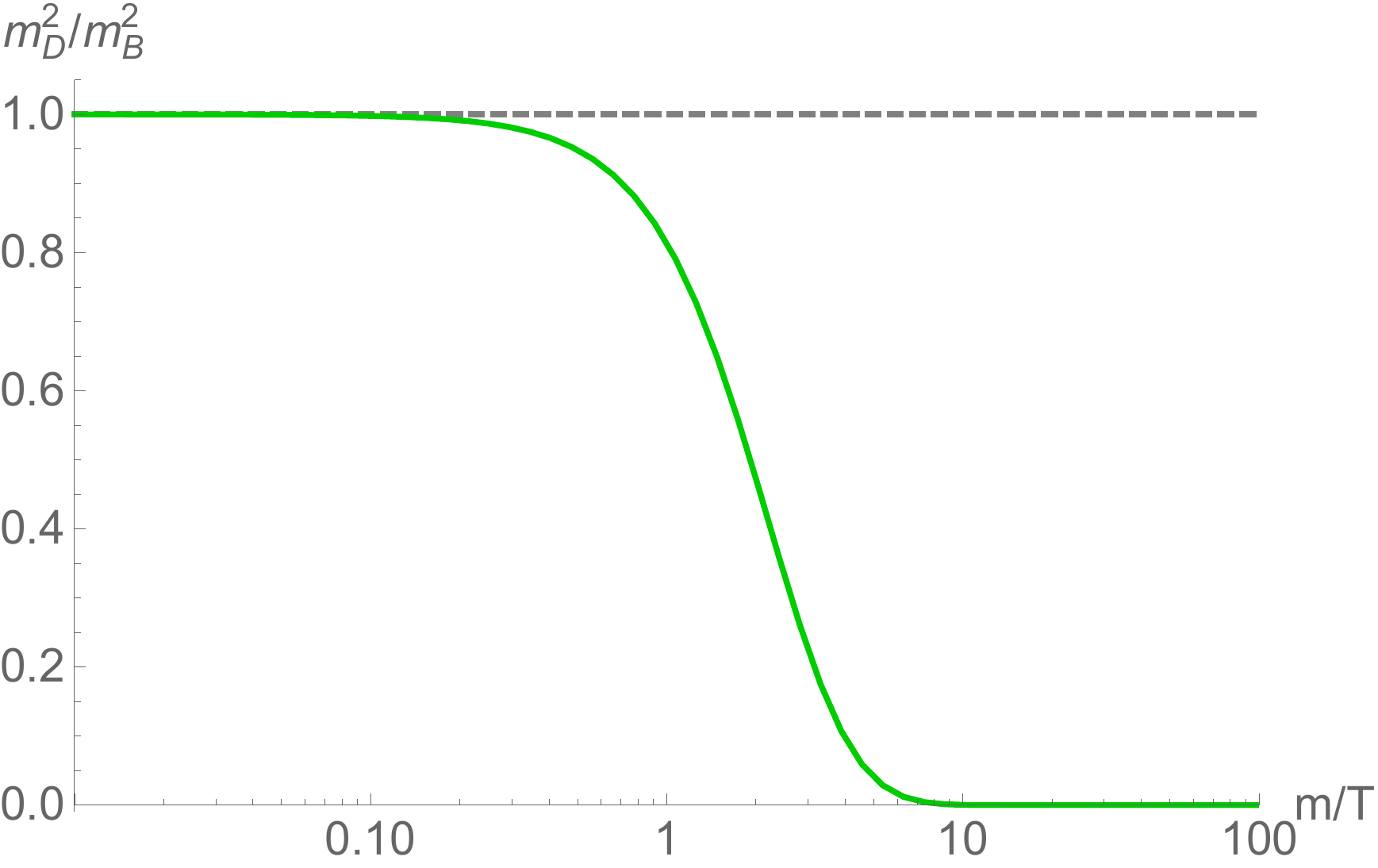}
     \end{center}
\vspace{-0.5cm}
\caption{Debye screening mass from the massive quark loop in the LLL (\ref{eq:Debye-sc-2}). }
\label{fig:vp_finiteT}
\end{figure}

Based on the estimate of the dynamical quark mass, one can also discuss the deconfinement phase transition in QCD. 
Once the thermal excitations are activated, the quark-antiquark pair excitations develop the Debye screening effect 
on long-range gluon interactions, 
weakening the confinement force. 
We have discussed the interpretation for the suppression 
of the longitudinal string tension observed 
in the lattice QCD simulations (See Fig.~\ref{fig:cornell}).

The effective potential for 
the Polyakov loop, the approximate order parameter of 
the deconfinement phase transition, explicitly 
indicates that the deconfined phase is favored 
when the Debye screening effect becomes larger 
(see Sec.~\ref{sec:Polyakov}). 
As we have discussed around Eq.~(\ref{eq:screening-A0}), 
the Debye screening mass is enhanced by the Landau degeneracy factor in the strong magnetic field, 
giving rise to the tendency of lowering phase transition temperature. 
This computation was done with massless quarks. 
Here, we revisit the Debye screening effect with a finite quark mass 
that is identified with the dynamical quark mass discussed above.

The screening effect is captured by the gluon self-energy $  \Pi_\para^{\temp }  (q) $ 
with $ q^\mu $ being the gluon momentum. 
In QCD, there are both quark- and gluon-loop contributions. 
We focus on the quark-loop contribution 
since it is enhanced by the Landau degeneracy factor 
and dominates over the gluon-loop contribution in the strong magnetic fields. 
Computational details of the gluon self-energy at finite temperature 
are summarized in Appendix~\ref{sec:screening} (see also Ref.~\cite{Hattori:2022uzp}). 
Using the one-loop gluon self-energy 
with an arbitrary fermion mass on the loop, 
the Debye screening mass $m_D $ is defined as\footnote{
We first take the static limit $  q^0  \to 0$ and then the long wavelength limit $ q_z \to 0 $. 
We also take the homogeneous limit $ |q_\perp|\to 0 $ in the transverse plane.} 
[see Eq.~(\ref{eq:Debye-sc})]: 
\begin{eqnarray}
m_D^2  
\=  -  m_B^2   \lim_{q_z \to 0}   \frac{m^2}{  q_z} 
 \prj \int_{-\infty}^\infty \frac{d  p_z}{ \epsilon_p} \frac{ n( \epsilon_p)  } {  p_z -  q_z }
\label{eq:Debye-sc-2}
\, .
\end{eqnarray} 
The integral is defined with the Cauchy principal value denoted by $  \prj  $. 
The LLL dispersion relation and the thermal distribution function are given 
as $ \epsilon_p = \sqrt{p_z^2 + m^2} $ and $n(\epsilon) = 1/(e^{\epsilon_p/T} + 1) $, respectively. 
In the massless limit ($m=0  $), the Debye screening effect is given by 
the Schwinger mass $ m_B^2 = g^2 |q_fB|/(2\pi)^2  $ multiplied by the Landau degeneracy factor 
and a color group factor $  1/2$. 
While the integral in Eq.~(\ref{eq:Debye-sc-2}) looks vanishing when $ m/T \to 0 $, 
the singularity at the origin $ (p_z =0)$ reproduces the nonzero value in the massless limit $ m_D = m_B $ [see Eq.~(\ref{eq:Debye-sc-infinite})]. 
After the factorization of the Landau degeneracy factor, 
the result of integration can only be a function of the dimensionless combination $ m/T $. 
It is clear that the result of integration is sizable only when the Fermi distribution function has 
a support in the low-energy region $ p_z \sim q_z \sim 0 $; 
The Debye screening mass can be developed only when 
the quark mass is small enough to have thermal excitations, 
i.e., $m/T \lesssim 1  $. 
In Fig.~\ref{fig:vp_finiteT}, 
one can confirm those behaviors from 
the numerical integration of Eq.~(\ref{eq:Debye-sc-2}).


For a qualitative discussion, we identify the fermion mass $ m $ 
with the dynamical quark mass $m_\dyn $. 
This provides a good approximation when the quark self-energy 
is inserted in the quark propagator 
and the loop integral is dominated by the infrared contribution  
(and this should be the case in the dimensionally reduced system). 
When the saturation mechanism of the dynamical quark mass 
gives $m \sim \Lambda_\QCD $, 
the Debye screening mass is developed 
when $ T \sim  \Lambda_\QCD $ or even lower 
with the help of the Landau degeneracy factor. 
This observation suggests a decrease of the deconfinement phase transition temperature in favor of the observation by the lattice QCD simulation \cite{Bruckmann:2013oba, Endrodi:2015oba}. 
The decrease of the phase transition temperatures in 
the deconfinement transition, as well as 
the chiral restoration discussed above, 
may not be explained if the dynamical quark mass 
increases with an increasing magnetic-field strength.



\begin{figure}
\begin{center}
\includegraphics[width=0.65 \textwidth]{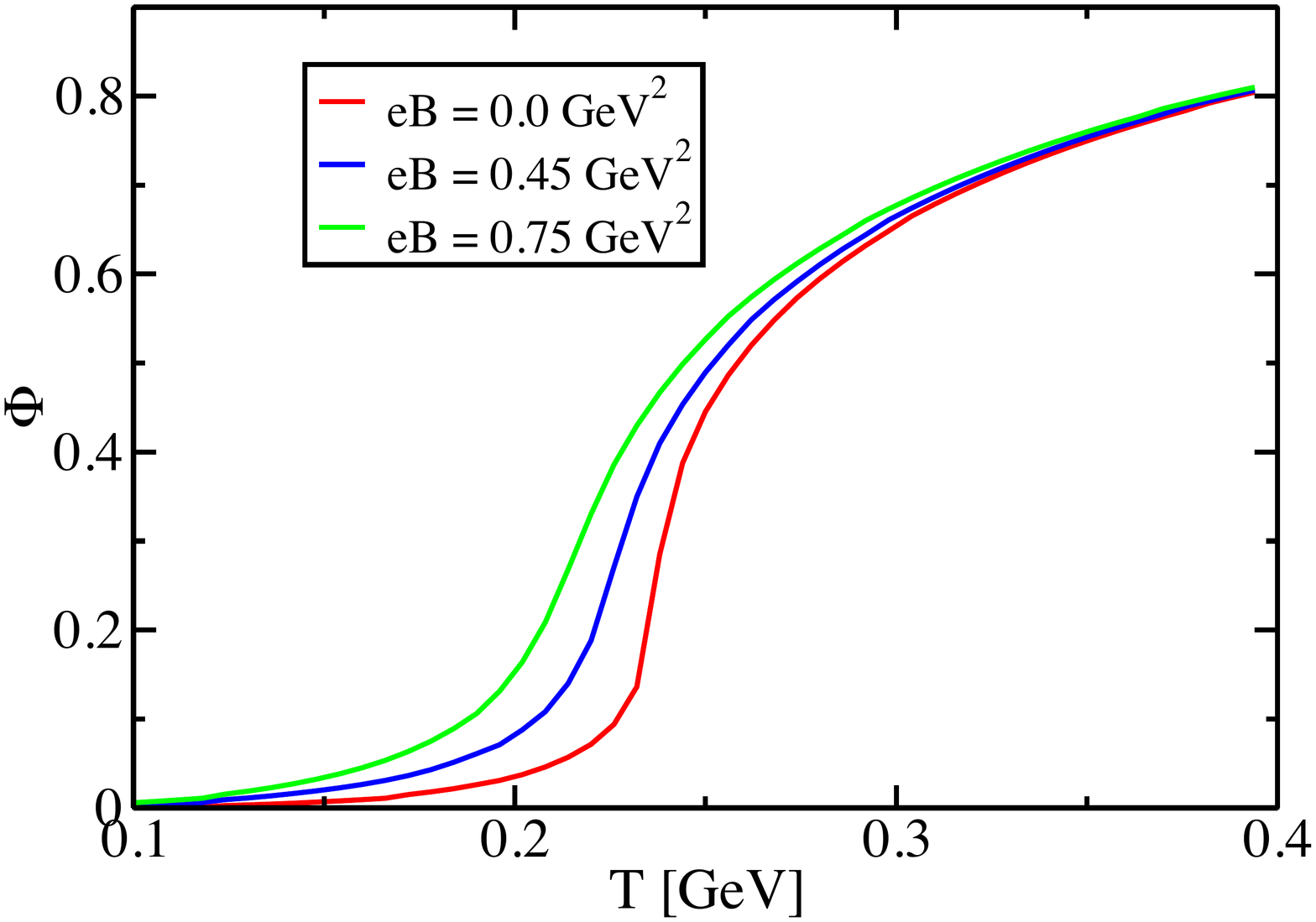}
\end{center}
\vspace{-1cm}
\caption{
Temperature dependence of the Polyakov loop $ \Phi $ determined from 
the effective potential at different values of the magnetic-field strength \cite{Ozaki:2015yja}.
}
\label{Fig:B-dep_PolyakovLoop}
\end{figure}

To quantify the above discussions, we discuss 
the Polyakov loop with a finite quark mass \cite{Ozaki:2015yja, Lo:2020ptj}. 
We parametrize the nonperturbative gluonic potential 
from the quenched lattice QCD results as \cite{Roessner:2006xn} 
\begin{eqnarray}
\mathcal{U}[C]
&=& -\frac{1}{2}a(T) \Phi^{2} + b(T)\, {\rm{ln}} \left[ 1 - 6 \Phi^{2} + 8 \Phi^{3} - 3 \Phi^{4} \right]
\label{phenomenological_potential}
\, ,
\end{eqnarray}
with
\begin{eqnarray}
a(T) = a_{0} + a_{1}(T_{0}/T) + a_{2}(T_{0} / T )^{2}, \ \ \ \ b(T) = b_{3}(T_{0}/T)^{3}
\, .
\end{eqnarray}
The $  C$ variable is related to the Polyakov loop $ \Phi $ via Eq.~(\ref{simple_Polyakov_loop}). 
Here, the parameters are given as 
$a_{0} = 3.51,\, a_{1} = -2.47,\, a_{2} = 15.2,\, b_{3} = -1.75$, and $T_{0} = 270$ MeV for $N_{c}=3$. 
This fitting result replaces the perturbative one $V_{\rm YM}[C] $ in Eq.~(\ref{eq:V_eff-Polyakov}). 
On top of this, we include the quark contribution to examine the tendency of the magnetic-field effects. 
The quark contribution is included at the one-loop level as in Sec.~\ref{sec:Weiss}, 
so that the strong-coupling effects in QCD are approximately included through 
the magnitude of the dynamical quark mass. 
For simplicity, we take the color eigenvalues $ w _i = \pm 1/2, \, 0$. 
Then, the quark part $V_{\rm quark}[C,B]$ in Eq.~(\ref{PolyakovLoopwithB}) 
takes the same form for $ N_c = 2 $ and $  3$ 
because the quark with the eigenvalue $w_{3}=0$ does not contribute to the potential. 
Therefore, we obtain the quark contribution to the effective potential by 
performing the numerical integration (\ref{eq:polyakov-B}) as previously.

In Fig.~\ref{Fig:B-dep_PolyakovLoop}, we show the temperature dependence of 
the Polyakov loop $ \Phi $ determined from the stationary point of the effective potential \cite{Ozaki:2015yja}.  
In this analysis, we fix the dynamical quark mass at $ 350 $ MeV 
that is independent of the magnetic field and is of the order of $\Lambda_{\rm{QCD}}$.
The quark contribution is enhanced as we increase the magnetic-field strength, 
leading to an enhancement of explicit breaking of the center symmetry. 
In turn, this gives rise to a tendency of decreasing transition temperature $T_c(B)<T_c(B=0)$ 
in the strong magnetic field $ |eB| > \Lambda_\QCD^2 $.



\subsubsection{Mesonic excitations toward the phase transitions}

In the low-energy QCD, hadronic states should serve as a good basis 
rather than quarks and gluons because of the color confinement. 
Therefore, it is important to identify the hadron spectra 
in strong magnetic fields. 
Knowing the hadron spectra enables us to 
compute the pressure function (or the partition function) 
in the so-called hadron resonance gas model. 
It is a tradition to discuss the deconfinement phase transition 
with the hadron resonance gas model. 
Also, the chiral condensate can be in principle 
computed by taking a derivative of 
the pressure function (or the partition function) 
with respect to the current quark mass.

Magnetic-field effects have been included into 
the hadron resonance gas model 
as the Landau quantization and the Zeeman effect 
on point-like charged hadrons \cite{Endrodi:2013cs}. 
Such point-like picture works in the weak magnetic field regime, 
but may break down in the strong magnetic field regime. 
Indeed, when the Zeeman energy becomes larger than 
a hadron mass at the zero magnetic field, 
the squared energy falls into a negative value. 
However, we have seen in the above that lattice QCD simulations 
do not support such collapsing behaviors 
(see Fig.~\ref{fig:light_mesons}). 
The internal structures need to be reconstructed 
when dense magnetic lines penetrate through hadrons, 
which can cure problems in the naive treatment. 
It is particularly important to investigate neutral pions 
that remain the NG boson and govern the thermodynamics. 
Those issues are also related to one of the original issues 
posed in the beginning of this subsection; 
The conventional chiral perturbation theory 
does not work in the strong magnetic field regime 
$q_f B \gtrsim \Lambda_\QCD^2 $.

\cout{

Light mesons-----------------

\cite{Klevansky:1989vi, Klevansky:1991ey}

BS amplitude discussed for NG boson in catalysis in QED \cite{Gusynin:1995nb, Gusynin:1995gt}

\cite{Fukushima:2012kc}

chiral perturbation \cite{Andersen:2012dz, Andersen:2012zc}

\cite{Kojo:2012js} \cite{Hattori:2015aki}

\cite{Fayazbakhsh:2013cha}

\cite{Andreichikov:2013zba}

\cite{Colucci:2013zoa}

Taya \cite{Taya:2014nha}

Hidden local symmetry \cite{Kawaguchi:2015gpt}

Mei Huang \cite{Liu:2014uwa} 

Indian group \cite{Ghosh:2016evc}

\cite{Andreichikov:2016ayj}

\cite{Wang:2017vtn} \cite{GomezDumm:2017jij}

NJL \cite{Liu:2018zag, Mao:2018dqe}

String tension \cite{Cao:2019azh}

Nucleon \cite{Andreichikov:2013pga, Yakhshiev:2019gvb}
Lattice and NJL \cite{Endrodi:2019whh}

}

Magnetic-field effects on meson properties were studied 
with various frameworks \cite{Klevansky:1989vi, Klevansky:1991ey,
Gusynin:1995nb, Gusynin:1995gt, Chernodub:2010qx, 
Chernodub:2011mc, Callebaut:2011ab, Hidaka:2012mz, 
Fukushima:2012kc, Andersen:2012dz, Andersen:2012zc, Kojo:2012js,
Callebaut:2013wba, Cai:2013pda, Bu:2012mq, Cai:2013kaa, 
Fayazbakhsh:2013cha, Andreichikov:2013zba, 
Colucci:2013zoa, Taya:2014nha, Liu:2014uwa, Hattori:2015aki, 
Kawaguchi:2015gpt, Andreichikov:2016ayj, Ghosh:2016evc, 
Wang:2017vtn, GomezDumm:2017jij, 
Liu:2018zag, Mao:2018dqe} (see also Refs.~\cite{Andreichikov:2013pga, Yakhshiev:2019gvb, Endrodi:2019whh} for baryons). 
Among others, meson states can be constructed from 
the Bethe-Salpeter equation in a traditional way. 
The quark propagator should be dressed by the self-energy that gives the dynamical quark mass. 
Another ingredient is the gluon-exchange interactions 
between a quark and an antiquark. 
One can start with the simplest set-up, for example, 
in the LLL approximation and the ladder approximation \cite{Hattori:2015aki}. 
Similar to the case of the SD equation 
discussed above, one will find a convolution among 
the quark and antiquark wave functions and 
the gluon propagator on the ladder. 
As a result, one finds a reduced gluon propagator 
analogous to that in Eq.~(\ref{eq:reduced-gluon}), 
though they are not exactly the same due to different kinematics 
(see Ref.~\cite{Hattori:2015aki} for an explicit form). 
The effective interaction range in 
the transverse plane is again determined by the ratio of 
the cyclotron radius to $1/\Lambda_\QCD $ 
when the gluon interaction is dominated by 
the infrared region below $ \Lambda_\QCD $.

When there is an IR dominance in the gluon propagator, 
soft gluons only mediate the interactions among 
nearby quark and antiquark states in the transverse plane. 
Therefore, the meson wave functions are squeezed 
in (1+1)-dimensional forms. 
Indeed, prolate shapes of mesonic wave functions 
have been observed 
with lattice QCD simulations \cite{Hattori:2019ijy}. 
The formation of such squeezed mesons saves the energy cost 
due to the string extension in the transverse direction. 
The meson spectrum is dominantly determined by 
the internal eigenmodes along the magnetic field, 
and the spectrum of ``light'' mesons will not strongly 
depends on the magnetic-field strength once we 
are in the strong-field regime \cite{Hattori:2015aki}.\footnote{
Good quantum numbers for meson states are different from 
those in the absence of magnetic fields 
(see an appendix in Ref.~\cite{Hattori:2015aki}). 
Light mesons here mean the low lying modes 
in the strong magnetic field. 
} 
Therefore, a typical light meson mass should remain 
of the order of the saturation value 
of the dynamical quark mass $\sim \Lambda_\QCD$. 
Moreover, the meson spectrum should succeed the Landau degeneracy 
of its constituent quark and antiquark unless 
the gluon interaction breaks the translation invariance 
in the transverse plane.


When the low-energy dynamics is governed by such 
light and degenerate meson excitations, 
the system is more easily boiled up at finite temperature 
with the aid of the large degeneracy, 
as compared to the case without magnetic fields 
when the same amounts of energy are injected to the systems. 
Namely, near the phase transition region $T\sim \Lambda_\QCD $ 
and in the strong-field regime $ eB \gtrsim \Lambda_\QCD^2 $, 
the thermodynamic pressure grows as $ eB \times T^2 $ 
with an increasing magnetic field, 
and is enhanced as compared to the Stefan-Boltzmann law 
$ T^4 \sim \Lambda_\QCD^4 $ in the (3+1) dimensions. 
The magnetically enhanced pressure will lead to 
a decrease of the phase transition temperatures 
\cite{Kojo:2012js, Hattori:2015aki}. 
Numerical analyses are yet left as an open problem 
(see Ref.~\cite{Kojo:2021gvm} for recent demonstration 
within a nonrelativistic model). 
Further studies are desired in cooperation with 
the progress in lattice QCD simulations on the hadron spectrum. 
It is also very interesting to discuss 
the magnetization and the possible critical point 
suggested by the lattice QCD simulations. 
Those simulation results are summarized earlier in this section.

\cout{

\com{Below may be removed.}
We finally discuss internal eigenmodes of 
the meson states based on an intuitive illustration 
in Fig.~\ref{fig:Hall_drift}, 
instead of looking into technical details of 
the Bethe-Salpeter equation. 
\cgd{
The red arrows show a binding force stemming from 
gluon interactions in the transverse plane. 
Without magnetic fields, the internal eigenmodes consist of 
string oscillations along the binding force. 
When an external magnetic field is applied, 
the motion along the binding force is bent by the Lorentz force, 
inducing drift motion in perpendicular to the binding force 
as indicated with the blue thick arrows.\footnote{
More precisely, each of quark and antiquark form 
small cyclotron orbits in the LLLs, and the center coordinates 
of the cyclotron orbits, indicated with the orange circles, 
drift in perpendicular to the binding force. 
} 
An unlike-sign charge pair feels the Lorentz force in the same direction, 
while a like-sign charge pair in the opposite directions.\footnote{
To understand the directions of the Lorentz force, it should be noticed that 
the binding force is blind to signs of electric charges 
as it originates from the QCD interactions. 
} 
Therefore, the quark and the antiquark inside neutral mesons 
drift in the same direction, 
generating translational modes of the meson states. 
On the other hand, the quark and the antiquark inside 
charged mesons drift in the opposite directions, 
generating relative rotational modes. 
This picture can be confirmed with 
analytic structures of the Bethe-Salpeter equation\footnote{
In case of charged mesons, 
the difference between absolute values of fractional charges 
results in a difference between magnitudes of the Lorentz force, 
distorting the relative rotational motion. 
To find eigenmodes precisely, 
one needs to analyze the Bethe-Salpeter equation in detail. 
} 
\cite{Hattori:2015aki}. 
}
Numerical analyses are yet left as an open problem. 
Further studies are desired in cooperation with 
the progress in lattice QCD simulations mentioned earlier. 

}



\cout{ 

\begin{eqnarray}
\lan \bar \psi \psi \ran &=&
\frac{1}{Z_\QCD} \int \!\! {\mathscr D} A_{\rm g}  \int  \!\! {\mathscr D} \bar \psi \int  \!\! {\mathscr D}\psi  \ 
\frac{\partial}{\partial m} e^{ i S_{\rm QCD}(B)} 
\nn
\\ 
&=& \frac{1}{Z_\QCD} \int  \!\! {\mathscr D} A_{\rm g} \ e^{ i S_{\rm g}} 
\det[ \sla D(B,A_{\rm g}) + m] \, \tr[ \frac{1}{\sla D (B,A_{\rm g}) + m} ] 
\end{eqnarray}
where the covariant derivative $  D (B,A_{\rm g})$ contains the gluon field $ A_g $ and the external magnetic field $ B $. 
The gauge fixing and color indices are suppressed for notational simplicity. 

\begin{eqnarray}
\frac{ \partial }{\partial m} \det( \sla D + m) 
&= & \frac{ \partial }{\partial m} e^{ \tr [ \ln( \sla D + m)] }
\nn
\\
&= & \det( \sla D + m)  \frac{ \partial }{\partial m} \tr [ \ln( \sla D + m)] 
\nn
\\
&= & \det( \sla D + m)  \tr[ \frac{1}{\sla D + m} ] 
\end{eqnarray}

}


%
%
%
%

\cout{
 
\subsubsection*{Further developments}


In the presence of a preferred direction along an external field, 
the tensor condensate $ \langle \bar \psi \sigma^{\mu\nu} \psi \rangle$ can take a finite value \cite{Ioffe:1983ju}. 
This quantity was identified with the ``spin'' contribution to the magnetization $ \propto  \partial \log Z/\partial B $ 
with $ Z$ being the QCD partition function in a magnetic field, 
while the other part comes from the ``orbital angular momentum'' \cite{Bali:2013esa}, {\it albeit} gauge dependent. 
The magnetization and associated pressure anisotropy has been measure 
with the lattice QCD simulations \cite{Bali:2012jv, Bonati:2013lca, Bonati:2013vba, 
Endrodi:2013cs, Bali:2013esa, Bali:2013owa, Levkova:2013qda}. 
This tensor condensate also appears in the operator product expansion \cite{Ioffe:1983ju}, 
and contributes to the hadron spectroscopy by the QCD sum rule \cite{Machado:2013yaa, Gubler:2015qok}.

\gray{
{\it Electric field effect.---} Effects of (chromo-)electric fields on the condensates have been studied in the literature \cite{Klevansky:1989vi, Klevansky:1991ey, Klevansky:1992qe,  Suganuma:1991dw, Suganuma:1990nn, Cohen:2007bt, Cao:2015dya, Iritani:2015zwa}. 
The systems under electric fields are inherently dynamical (though they could reach a steady state), 
and thus one may not draw an equilibrium phase diagram 
in an electric field. 
Nevertheless, it may be a challenging issue to consider the real-time dynamics in an electric field \cite{Copinger:2018ftr}. 
 }

}


%

\section{Quantum anomaly in transport phenomena}

\label{sec:transport} 
In the last section, we discuss the transport phenomena in magnetic fields. 
Computing the vector and axial-vector currents, 
we find the {\it chiral magnetic effect} and {\it chiral separation effect} 
that belong to a family of the anomaly-induced transport phenomena 
or, in short, {\it anomalous transport phenomena} named after their relations to quantum anomaly. 
We then discuss the dynamics of helicities driven by the chiral magnetic effect.

\subsection{Chiral magnetic/separation effect}

\label{sec:Ritus-CME}

\begin{figure}[t]
     \begin{center}
              \includegraphics[width=0.25\hsize]{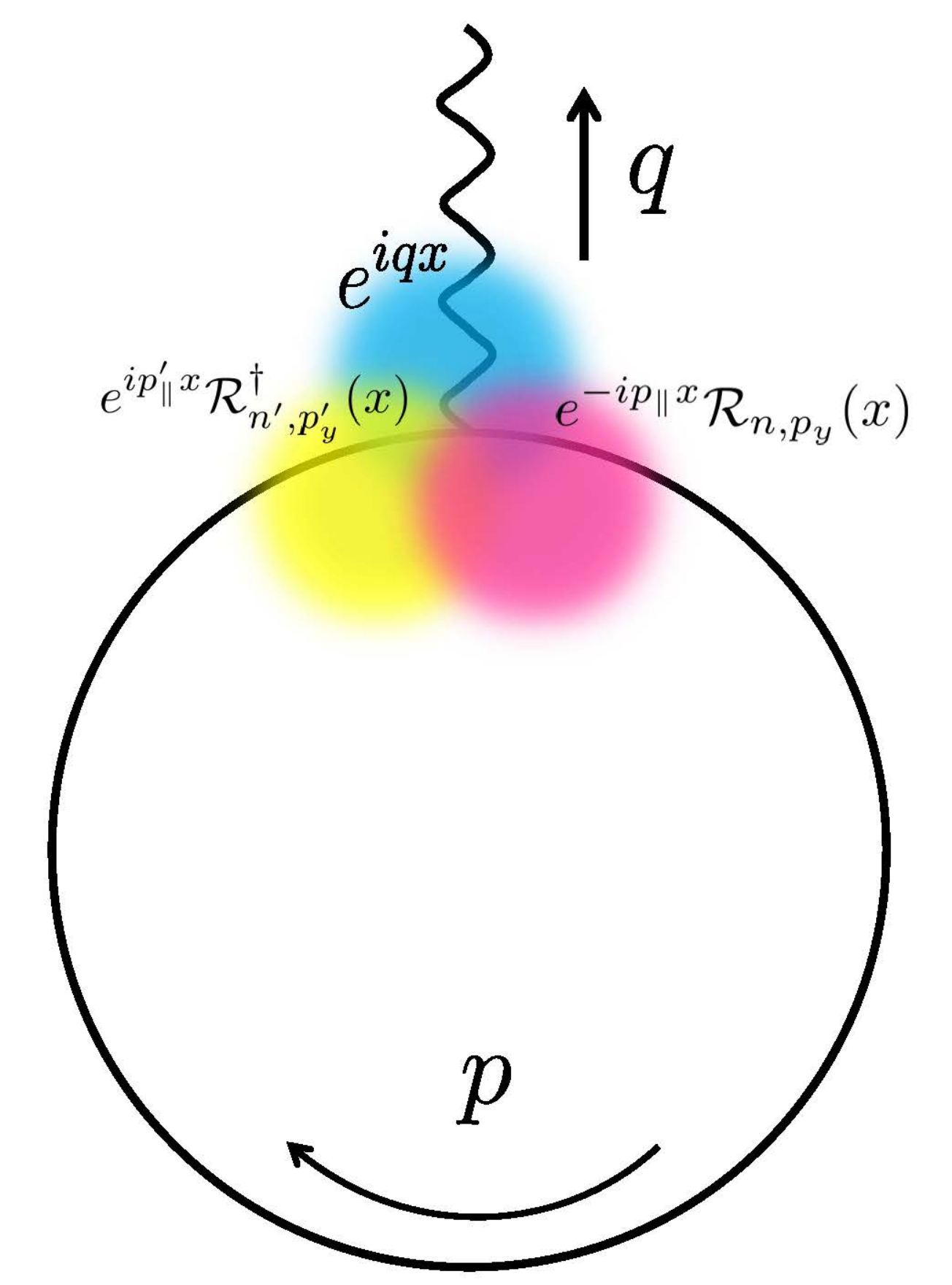}
     \end{center}
\vspace{-0.5cm}
\caption{Current in an external magnetic field with the Ritus basis.
}
\label{fig:tadpole}
\end{figure}

We focus on the currents transported by massless fermion carriers. 
Mass effects are discussed in the subsequent section and Appendix~\ref{sec:massive}. 
Since the chirality is a good quantum number for massless fermions, 
we compute the current in the chirality basis 
\begin{eqnarray}
\label{eq:current-definition}
 j^\mu_{R/L} (x) =  \langle \bar \psi(x) \gam^\mu \prj_{R/L} \psi(x) \rangle
 \, ,
 \end{eqnarray} 
where the chirality projection operator is defined by $ \prj_{R/L} = ( 1\pm \gam^5)/2 $ 
with $ \gam^5 $ given below Eq.~(\ref{eq:metrics}). 
In this subsection, the upper and lower signs are 
understood as those for the right- and left-handed chirality, respectively. 
The brackets denote the thermal average at finite temperature and density 
with a finite chemical potential $ \mu_{R/L} $ in the chirality basis.

Recall that we examined the Landau levels of the relativistic fermions in Sec.~\ref{sec:relativistic-fermion}. 
The Ritus basis (\ref{eq:Ritus}) provides the eigenspinor in a magnetic field, 
so that the spatially homogeneous currents may be decomposed into contributions from the Landau levels 
according to the Ritus-basis mode expansion (\ref{eq:Ritus-mode}) in the Landau gauge. 
Assuming that the magnetic field is applied in the third direction without loss of generality, 
the currents along the magnetic field are given as 
\begin{eqnarray}
j^3_{R/L} (q=0) \= \frac{1}{L_x L_y} \int d^3 x \, e^{ - i 0 \cdot \bx}  j^3_{R/L} (x) 
 \nnb
 \=   \frac{1}{L_x L_y}   \int d^3 x  \sum_{\kappa=\pm}  \sum_{\kappa'=\pm}
  \sum_{n=0}^{\infty} \sum_{n'=0}^{\infty}
  \int \frac{d p_z}{2\pi}   \int  \frac{dp_y dp'_y}{(2\pi)^2}  \frac{1}{\sqrt{4 \epsilon_n \epsilon_{n'}}}
\label{eq:current-RL-0}
\\
&&
\times \Big[ \  \langle a_{p_{n'}, p'_y}^{\kappa' \dagger}  a_{p_n, p_y}^{\kappa} \rangle  
  \bar u ^{\kappa'} (p_{n'})   \Ritus_{n',p_y'}^\dagger(x_\perp)  
  \gam^3 \prj_{R/L} \Ritus_{n,p_y} (x_\perp)   u ^\kappa (p_n)
   e^{  i (\epsilon_{n'} - \epsilon_{n}) t} 
  \nnb
&& 
  +  \langle b_{\bar{p}_{n'}, p'_y}^{\kappa'}  b_{\bar{p}_{n}, p_y}^{\kappa \dagger}  \rangle 
 \bar v ^{\kappa'} (\bar p_{n'})    \Ritus_{n',p_y'}^\dagger(x_\perp)
   \gam^3 \prj_{R/L} \Ritus_{n,p_y} (x_\perp)  v ^\kappa (\bar p_n)   e^{ - i (\epsilon_{n'} - \epsilon_{n}) t} 
\ \Big]
\nn
 \, ,
 \end{eqnarray}  
with $L_{x,y}  $ being the system lengths in the transverse directions. 
The Landau levels are given as $\epsilon_n = \sqrt{ p_z^{ 2} + 2 n |q_f B| }  $ and $ \epsilon_0 = |p_z| $ is understood.\footnote{
$\epsilon_0 $ is the positive-energy part of 
the massless dispersion relation (\ref{eq:gappless}). 
Here, the positive- and negative-energy solutions 
are already separated explicitly in Eq.~(\ref{eq:current-RL-0}). 
} 
We use the fact that  $  \prj_\pm$ commutes with $ \gam^3 $ and $ \prj_{R/L} $ 
and then the orthogonal relation (\ref{eq:orthogonal-p}) to find that 
\begin{eqnarray}
\label{eq:current-RL}
j^3_{R/L} (q=0) \=  \frac{|q_fB|}{2\pi}
 \sum_{\kappa=\pm}  \sum_{\kappa'=\pm} \sum_{n=0}^{\infty}  
  \int \frac{d p_z}{2\pi}  \frac{1}{2 \epsilon_n }
\\
&&
\times \Big[  \langle a_{p_{n}, p_y}^{\kappa' \dagger}  a_{p_n, p_y}^{\kappa} \rangle  
  \bar u ^{\kappa'} (p_{n})  \gam^3  I_n  \prj_{R/L} u ^\kappa (p_n) \
 +  \langle  b_{\bar{p}_n, p_y}^{\kappa'} b_{\bar{p}_{n}, p_y}^{\kappa \dagger}  \rangle 
 \bar v ^{\kappa'} (\bar p_{n})   \gam^3  I_n \prj_{R/L}  v ^\kappa (\bar p_n)  
 \Big]
 \nn
 \, .
 \end{eqnarray} 
The Landau degeneracy factor is reproduced from the $ y $ integral in the Landau gauge (see the last part of Sec.~\ref{sec:Landau-g}). 
The thermal expectation value is assumed to be diagonal 
in the spin basis, i.e., $ \langle a_{p_{n}, p_y}^{\kappa' \dagger}  a_{p_n, p_y}^{\kappa} \rangle 
= \delta_{\kappa\kappa'} f(\epsilon_n - \mu_{R/L}) $, 
where $ f(\epsilon) = 1/[\exp(\epsilon /T) +1] $ is the Fermi-Dirac distribution function 
Likewise, we have $ \langle  b_{\bar{p}_n, p_y}^{\kappa'} b_{\bar{p}_{n}, p_y}^{\kappa \dagger}  \rangle 
= - \delta_{\kappa\kappa'} f(\epsilon_n + \mu_{R/L}) $, where the contribution from the Dirac sea is dropped.

Taking the summation with Eq.~(\ref{eq:spin-sum}), we have 
\begin{eqnarray}
\label{eq:current-RL-2}
j^3_{R/L} (q=0) 
 \=  \frac{|q_fB|}{2\pi} \sum_{n=0}^{\infty}    \int \frac{d p_z}{2\pi}  \frac{1}{2 \epsilon_n }
\{  f(\epsilon_n - \mu_{R/L})    -   f(\epsilon_n + \mu_{R/L}) \}  \tr[ \sla p_n   \gam^3  I_n  \prj_{R/L} ]
 \, .
 \end{eqnarray}   
Because of the ``unit matrix $  I_n $'' defined in Eq.~(\ref{eq:id-magnetic}), 
the result of trace depends on the Landau levels. 
In case of the hLLs $ (n\geq 1) $, one finds that 
the chirality projection operator does not play any role: 
 \begin{eqnarray}
  \tr[ \sla p_n   \gam^3  I_n  \prj_{R/L} ] =  \frac12 \tr[ \sla p_n   \gam^3 ] = 2 p_z
\, .
 \end{eqnarray}
Remarkably, the $ p_z $ integral in Eq.~(\ref{eq:current-RL}) exactly vanishes for all the hLLs 
due to the linear dependence of the trace on $ p_z $, and there is no contribution to the current. 
This is because there are an equal number of the carriers 
moving in parallel and antiparallel to the magnetic field 
due to the spin degeneracy, which leads to the vanishing net currents.\footnote{ 
This contrasts to the case of the Ohmic current 
induced by an electric field, 
which has a finite net current. 
The difference can be understood with the parity property: 
An electric field is a parity-odd quantity, 
while a magnetic field is a parity-even quantity. 
Accordingly, acceleration by the electric field induces an asymmetric momentum distribution of carriers with respect to 
the sign flip $  p_z \to - p_z$, resulting in a nonzero Ohmic current \cite{Hattori:2016lqx, Hattori:2016cnt, 
Fukushima:2017lvb, Fukushima:2019ugr}. 
}

In the LLL, because of the spin projection operator from $ I_0 $, 
the chirality projection operator plays a crucial role. 
The trace result depends on the chirality as 
  \begin{eqnarray}
  \tr[ \sla p_0   \gam^3  I_0  \prj_{R/L} ] 
  =  \frac14 \left(  \tr[ \sla p_n   \gam^3 ] 
  +   \tr[ \sla p_0   \gam^3 (is_f \gam^1\gam^2) (\pm \gam^5) ] \right)
= p_z \pm   s_f \epsilon_0
  \, .
 \end{eqnarray}
After cancellation of all the hLLs, 
the LLL contribution is left as 
\begin{eqnarray}
j^3_{R/L} (q=0) 
 \= \pm s_f  \frac{|q_fB|}{2\pi}  \int_{0}^\infty \frac{d p_z}{2\pi}  
\{  f(\epsilon_0 - \mu_{R/L})    -   f(\epsilon_0 + \mu_{R/L}) \}  
\label{eq:currents-1}
 \, .
 \end{eqnarray}         
The overall sign depends on the chirality since 
the chirality and the momentum flow are locked 
with each other in the LLL (cf. Figs.~\ref{fig:LL} and \ref{fig:chirality}). 
Notice that the currents are odd functions of $ \mu_{R/L} $ 
as also expected from the charge-conjugation properties. 
The linear term should not depend on temperature 
for the mass dimensions to work on the both sides. 
In Eq.~(\ref{eq:currents-1}), the Landau degeneracy factor 
provides mass-dimension two instead. 
The integrals can be exactly performed as 
\begin{eqnarray}
 \int_{0}^\infty \frac{d p_z}{2\pi} 
\{  f(\epsilon_0 - \mu_{R/L})- f(\epsilon_0 + \mu_{R/L}) \}  
= \frac{\mu_{R/L}}{2\pi} 
\label{eq:current-final}
\, .
\end{eqnarray}
We have found that the currents do not explicitly 
depend on temperature. 
Although each integral in Eq.~(\ref{eq:current-final}) yields temperature-dependent terms, 
they are canceled out when we take the difference between the two integrals.


Wrapping up the above computation, we have found that the hLL contributions exactly vanish 
and that the LLL contributions do not depend on temperature in the massless limit. 
Since the LLL fermions can only move along the magnetic field, 
the current in the LLL only has the longitudinal component 
as easily proven with the identity $ \prj_\pm \gam \prj_\pm = \gam_\para \prj_\pm  $. 
Therefore, one can restore the vector forms of the currents 
\begin{subequations}
\label{eq:CME-CSE}
\begin{eqnarray}
{\bm j}_V = {\bm j}_R + {\bm j}_L = q_f \frac{\mu_A}{2\pi^2} {\bm B}
\, ,
\\
{\bm j}_A = {\bm j}_R - {\bm j}_L  = q_f \frac{\mu_V}{2\pi^2} {\bm B}
\, ,
\end{eqnarray}
\end{subequations}
where the chemical potentials in the $ V/A $ basis are 
related to those in the $ R/L $ basis as $ \mu_{V/A} = ( \mu_R \pm \mu_L)/2 $ \cite{Alekseev:1998ds, Fukushima:2008xe}. 
The vector and axial-vector currents induced by the magnetic field are now called the chiral magnetic effect (CME) 
and chiral separation effect (CSE), respectively \cite{Vilenkin:1980fu, Nielsen:1983rb, Alekseev:1998ds, 
Son:2004tq, Metlitski:2005pr, Newman:2005as, Kharzeev:2007jp, Fukushima:2008xe}. 
It is instructive to compare the CME with the familiar Ohmic current $ q_f \bj_V =\sigma_{\rm Ohm} \bE $. 
Since the electric and magnetic fields have opposite parity, so do the CME and Ohmic conductivities. 
This means that the CME only occurs in a parity-odd environment 
that is characterized by the axial chemical potential $ \mu_A $ 
as clear from the asymmetric dependence on the R and L chiralities in the definition of $ \mu_A $. 
The vector chemical potential in the CSE is also necessary 
for satisfying the charge-conjugation and parity properties. 
Notice that the CME/CSE will induce a vector/axial charge separation, 
which acts as a source of the subsequent CSE/CME. 
This mutual induction gives rise to a longitudinal wave propagating along the magnetic field, 
which is called the {\it chiral magnetic wave} \cite{Newman:2005as, Kharzeev:2010gd, Burnier:2011bf, Stephanov:2014dma}.

\subsection{Manifestation of the chiral anomaly}

The above results on the currents (\ref{eq:CME-CSE}) 
have been confirmed with various methodologies 
in terms of quantum field theory 
\cite{Vilenkin:1980fu, Alekseev:1998ds, Newman:2005as, Fukushima:2008xe, 
Kharzeev:2009pj, Hong:2010hi, Landsteiner:2011cp, Hou:2011ze, Warringa:2012bq, Hongo:2019rbd}, 
hydrodynamics \cite{Son:2009tf, Neiman:2010zi, Sadofyev:2010pr, Kharzeev:2011ds, 
Lin:2011aa, Jensen:2012jy, Banerjee:2012iz, Hattori:2017usa}, 
the AdS/CFT correspondence \cite{
Newman:2005as, Lifschytz:2009si, Yee:2009vw, Rebhan:2009vc, Gynther:2010ed, Kalaydzhyan:2011vx}, 
quantum kinetic theories \cite{Pu:2010as, Gao:2012ix, Son:2012wh, Son:2012zy, Stephanov:2012ki, Chen:2012ca, 
Hidaka:2016yjf, Mueller:2017lzw, Mueller:2017arw, Carignano:2018gqt, Lin:2019ytz, Chen:2021azy}, 
and lattice gauge theories \cite{Buividovich:2009zzb, Buividovich:2009wi, Buividovich:2010tn, 
Yamamoto:2011gk, Yamamoto:2011ks, Bali:2014vja}.\footnote{
These studies have also shown an analogy/difference between 
the effects of a magnetic field and a vorticity in a chiral fluid  (see also other early studies 
in Refs.~\cite{Vilenkin:1978hb, Vilenkin:1979ui, Vilenkin:1980zv, Erdmenger:2008rm}). 
Induction of the currents by a vorticity is called the chiral vortical effect. 
More recent studies include analyses of the interplay between 
a magnetic field and a vorticity/rotation \cite{Mameda:2015ria, Chen:2015hfc, 
Hattori:2016njk, Ebihara:2016fwa, 
Chernodub:2016kxh, Chernodub:2017ref, Lin:2021sjw, Yamamoto:2021gts} (see Ref.~\cite{Fukushima:2018grm} for a review). 
} 
The central issue addressed in these studies was the relation 
of the CME/CSE currents to the chiral anomaly. 
It has been established by those studies that 
the factors of $ 1/(2\pi^2)$ in Eq.~(\ref{eq:CME-CSE}) 
stem from the ``anomaly coefficient $ C_A$'' involved in 
the anomalous axial Ward identity \cite{Adler:1969gk, Bell:1969ts} 
\begin{eqnarray}
\partial_\mu j^\mu_A = q_f^2 C_A \bE \cdot \bB
\label{eq:anomaly}
\, ,
\end{eqnarray}
where we have $ C_A = 1/(2\pi^2) $ for a single-flavor (colorless) Dirac fermion coupled to the electromagnetic field.

\cout{
Indeed, the CME/CSE can be written as 
$ \bj_{V/A} =\sigma_{\rm CME/CSE}  \bB $ 
with the conductivities 
\begin{eqnarray}
\label{eq:CME-conductivity}
\sigma_{\rm CME/CSE} = q_f^2 C_A \mu_{A/V} 
\, ,
\end{eqnarray}
with the ``anomaly coefficient $ C_A$'' involved in 
the anomalous axial Ward identity \cite{Adler:1969gk, Bell:1969ts} 
\begin{eqnarray}
\partial_\mu j^\mu_A = q_f^2 C_A \bE \cdot \bB
\label{eq:anomaly}
\, .
\end{eqnarray}
We have $ C_A = 1/(2\pi^2) $ for a single-flavor (colorless) Dirac fermion coupled to the electromagnetic field. 
\cgd{
Here, we have multiplied the currents (\ref{eq:CME-CSE}) 
by the electric charge $ q_f $, so that the CME conductivity $ \sigma_{\rm CME} $ 
is for an electric current $ q_f \bj_V $ and $ \sigma_{\rm CSE} $ is given by 
a chemical potential $ q_f \mu_V $ conjugate to the electric charge density. 
}
}

\begin{figure}
     \begin{center}
              \includegraphics[width=0.9\hsize]{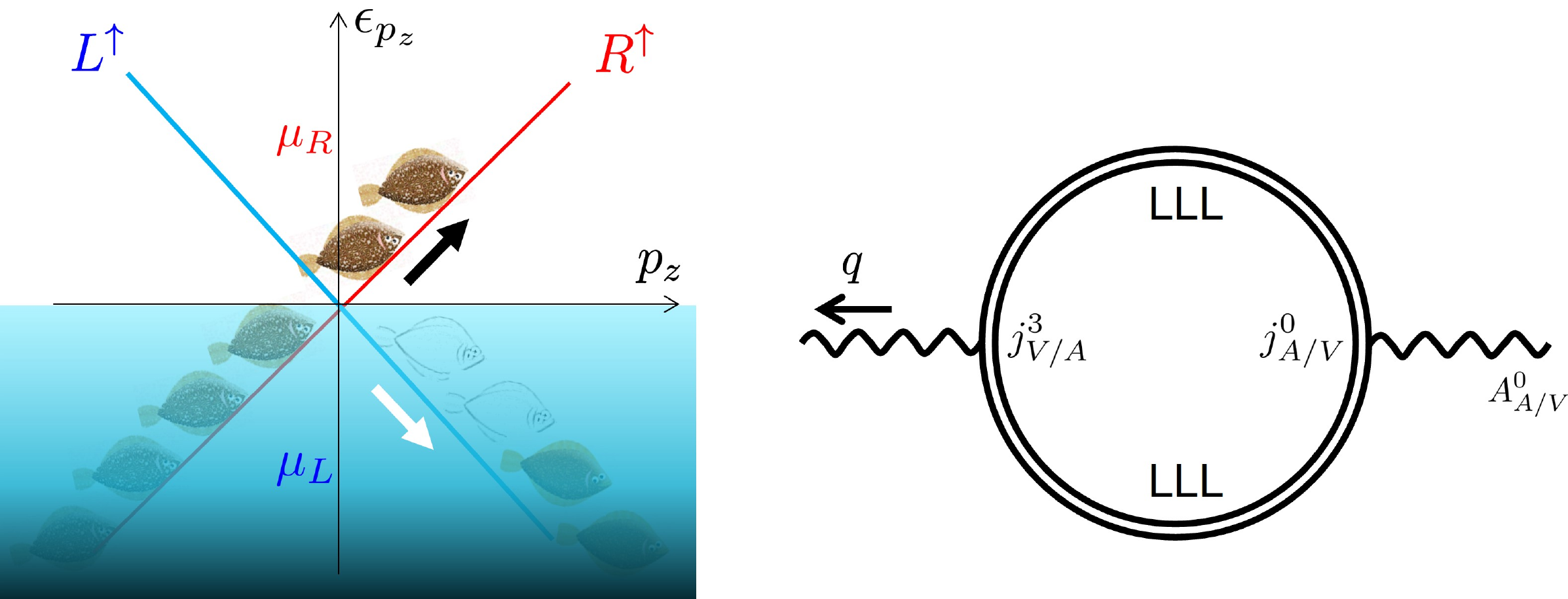}
     \end{center}
\vspace{-0.7cm}
\caption{Quick views of the chiral anomaly. 
Perturbed by an electric field $  (q_f E>0)$, right-handed (left-handed) flounders appear (disappear) 
across the surface of the bottomless Dirac sea (left panel, see also Fig.~\ref{fig:LL}). 
Fishermen can only catch the right-handed ones. 
Reinterpretation of the tadpole diagram in Fig.~\ref{fig:tadpole} by getting rid of all vanishing pieces (right panel). 
Internal double lines show the LLL fermion propagators coupled to the vector/axial chemical potential $ A^0_{V/A} $. 
}
\label{fig:ChiralAnomaly}
\end{figure}

Figure~\ref{fig:ChiralAnomaly} shows a quick view of the intimate relation of the chiral magnetic effect to the chiral anomaly. 
The left panel shows the ``spectral flow'' across the level-crossing point at the origin when an electric field is applied in parallel to the magnetic field. 
We discussed the spectral flow in Sec.~\ref{sec:massless}. 
After the spectral flow occurred, one finds currents 
due to acceleration of (negative-energy) particles 
by an electric field\footnote{
The acceleration of negative-energy particles, 
as an analog of valence electrons, would not induce currents 
if the dispersion relation were not directly connected 
to the positive-energy states or the conduction band. 
The linear dispersion relation is 
a special property of the massless LLL 
where the particle and antiparticle states, 
as an analog of the conduction and valence bands, 
are directly connected with each other. 
See Appendix~\ref{sec:massive} for a massive case. 
} 
as well as a nonzero chirality imbalance ($ \mu_R = - \mu_L $) 
from the bottomless Dirac. 
Since the currents and the chirality imbalance are created 
by the same mechanism, their coefficients should be related to one another.

\cout{ 

Figure~\ref{fig:ChiralAnomaly} shows a quick view of the intimate relation of the chiral magnetic effect to the chiral anomaly. 
The left panel shows creation of the chirality imbalance ($ \mu_A \not = 0 $) 
in the LLL state from the bottomless Dirac sea when an electric field is applied in parallel to the magnetic field. 
As discussed in Sec.~\ref{sec:massless}, acceleration of particles and antiparticles 
in the LLL induces the ``spectral flow'' across the level-crossing point at the origin, 
leading to the chirality imbalance, $ \mu_R = - \mu_L $~\cite{Nielsen:1983rb, Ambjorn:1983hp}. 
The imbalanced Fermi surfaces immediately indicate occurrence of the CME current 
as a consequence of the one-way motion in the LLL. 
Note that the chirality imbalance is necessary for a nonzero CME current 
to avoid a cancellation between the right- and left-handed contributions. 

}


In the right panel of Fig.~\ref{fig:ChiralAnomaly}, 
we more explicitly identify the relation of the currents 
to the chiral anomaly in a diagrammatic language. 
This two-point correlator composed of the LLL loop is nothing but 
the chiral anomaly diagram in the (1+1) dimensions (see Appendix~\ref{sec:anomaly-LLL}). 
On the other hand, this diagram identifies 
the nonzero contribution among the tadpole diagrams in Fig.~\ref{fig:tadpole} which we used to compute the currents. 
We have found that the currents only have 
the contributions from the LLL and 
are linear in the chemical potential, 
whereas the tadpole diagrams originally include 
contributions from all the Landau levels 
and all orders in the chemical potential. 
Therefore, what we computed previously was nothing but 
the chiral anomaly diagram.


In order to explicitly see the currents reproduced 
from the anomaly diagram in Fig.~\ref{fig:ChiralAnomaly}, 
one can construct the Kubo formulas, that is, relations between the currents and the thermal correlators: 
\begin{subequations}
\label{eq:current1}
\begin{eqnarray}
j_{V}^3 = \lim_{\omega,\bq\to0} \Pi ^{30}_{(VA)} (\omega,\bq) A^0_{A}
\, ,
\\
j_{A}^3 = \lim_{\omega,\bq\to0} \Pi ^{30}_{(AV)} (\omega,\bq) A^0_{V}
\, .
\end{eqnarray}
\end{subequations}
The axial/vector chemical potential is introduced as a temporal component 
of a constant axial/vector gauge field $ A^0_{A/V} = \mu_{A/V} $, 
which are coupled to the axial/vector current as $ (j_{A/V})_{\mu} A_{A/V}^{ \mu} $. 
Thus, the retarded correlator is given by $ \Pi ^{\mu\nu}_{(VA)/(AV)} (x) = \langle[ j_{V/A}^\mu ,j_{A/V}^\nu] \rangle \theta(x^0) $ 
where $ j_{V/A}^\mu $ is the current in the LLL. 
Similar to the purely (1+1) dimensional system, the vector and axial-vector currents are related to each other 
via a simple identity $ \gam_\para^\mu \gam^5 \prj_\pm = \mp s_f \epsilon _\para^{\mu \nu} \gam_\nu \prj_\pm$ 
with the antisymmetric tensor $ \epsilon_\para^{03} = -\epsilon_\para^{30} =1 $, 
i.e., $ j_A^\m = \mp s_f \epsilon _\para^{\mu \nu} j_{V\nu} $. 
Therefore, all we need to do is calculating the vector-vector correlator in the LLL 
which is an analog of that in the Schwinger model~\cite{Schwinger:1962tn, Schwinger:1962tp}. 
This is a simple exercise given in Appendix~\ref{sec:VP_vac}, 
and the result is shown in Eq.~(\ref{eq:massless-vac}) for the massless case.\footnote{
One should just remove the trace of color matrices and set $ g=1 $ 
to match the result in Eq.~(\ref{eq:massless-vac}) to the present case. 
} 
Plugging the correlator into Eq.~(\ref{eq:current1}), we have 
\begin{eqnarray}
\label{eq:CME-correlator}
j_{V/A}^3
& =&
 -  \frac{ q_f B }{2\pi}  \cdot \frac{ \mu_{A/V} }{\pi} 
\lim_{\omega,\bq\to0}    e^{ - \frac{ |\bq_\perp|^2 }{2 |q_f B| } }
 \frac{1}{q_\para^2}  \epsilon _{\para \, \rho}^{3} ( q_\parallel^2 g_\parallel^{\rho0} -  q_\parallel^\rho q_\parallel^0)
\nn
\\
& =&
\frac{ q_f B }{2\pi}  \cdot  \frac{ \mu_{A/V} }{\pi} 
\left[ \, \lim_{\omega,q_z\to0}   \frac{ - q_z^2 }{ \omega^2 - q_z^2 } \, \right]
\, .
\end{eqnarray}
The two limits do not commute with each other (see discussions below). 
Taking the zero-frequency limit first, one can reproduce the CME/CSE currents (\ref{eq:CME-CSE}).

According to the above Kubo-formula results, one can interpret those currents as the anomalous currents in the (1+1) dimensions 
multiplied by the Landau degeneracy factor. 
As shown in Appendix~\ref{sec:VP_therm}, there is no temperature correction to the anomalous two-point correlator 
in the massless case. This is consistent with the previous observation in Eq.~(\ref{eq:current-final}). 
The anomalous currents succeed the salient feature of the chiral anomaly, that is, nonrenormalizability 
in the hydrodynamic limit $ (\omega,\bq\to0) $, 
meaning that the functional forms of the currents do not depend on interactions and energy scales. 
Indeed, all the theories mentioned in the beginning, 
form ultraviolet to infrared (effective) theories, 
confirmed the same form of the anomalous current (\ref{eq:CME-CSE}). 
In Appendix~\ref{sec:anomaly-massive}, 
we also discuss a relation of the chiral anomaly 
from the LLL two-point correlator in the effective (1+1) dimensions to the chiral anomaly from the familiar triangle diagrams in (3+1) dimensions. 
They agree with one another in the homogeneous limit 
such that $ |\bq_\perp|/|q_fB| \to 0$. 
This implies that, in the hydrodynamic limit $ (\omega,\bq\to0) $, 
a perturbative computation 
linear in the magnetic field as well as the chemical potential 
provides the same currents as the above LLL result  
because one can construct an anomalous triangle diagram with 
a perturbative magnetic field, axial current (vector current), 
and vector chemical potential (axial chemical potential).

We now briefly return to the issue of the noncommutative limits in Eq.~(\ref{eq:CME-correlator}). 
In general, the vanishing frequency and momentum limits 
of a function may not agree with each other 
without the Lorentz symmetry 
which can be broken by, e.g., the presence of a medium. 
Nevertheless, it was pointed out in Ref.~\cite{Satow:2014lva} 
that the hydrodynamic limit for the anomalous currents 
can be taken irrespective of the ordering of the two limits 
after appropriate resummation of interaction effects. 
This could be one of special, and thus specific, properties of 
the anomalous currents, 
but can be understood in the following intuitive way. 
Notice that the hydrodynamic limit is a long spacetime limit, 
while the one-loop calculation corresponds to 
a non-interacting limit. 
To achieve the hydrodynamic limit, 
one needs to resum interaction effects even when the coupling constant is small 
since the mean-free time and path are finite and are smaller than 
the hydrodynamic spacetime scale, which is taken to be infinite. 
Interestingly, the authors also showed that the interaction effects 
cancel out in the final expression of the CME current if one takes the hydrodynamic limit. 
This result supports the nonrenormalizability of the CME current in the hydrodynamic limit. 
Similar issues of noncommutative limits had been known in the one-loop diagram calculation 
of the currents in a weak magnetic field \cite{Kharzeev:2009pj, Hou:2011ze}.

%

Finally, we add that the effect of the color confinement, 
i.e., nonzero contributions of hadronic operators 
to the CME current, was shown 
on the basis of the Wess-Zumino-Witten action \cite{Imaki:2019rlm}, while a vanishing contribution was found in 
an earlier work \cite{Fukushima:2012fg}. 
The discrepancy originates from the difference 
between the Wess-Zumino-Witten actions employed there 
(see Ref.~\cite{Imaki:2019rlm} for discussions 
about the discrepancy 
and evaluation of the hadronic operators).

\subsubsection*{Experimental realization of the chiral magnetic effect}

The CME discussed in relativistic heavy-ion collisions \cite{Kharzeev:2007jp, Fukushima:2008xe} 
have triggered diverse research activities in many other systems including astrophysics, cosmology, and condensed matter physics (see 
Refs.~\cite{Kharzeev:2013jha, Kharzeev:2013ffa, Liao:2014ava, Kharzeev:2015znc, 
Miransky:2015ava, Huang:2015oca, Kharzeev:2015kna, Hattori:2016emy, Landsteiner:2016led} for reviews), 
{\it albeit} there had been various proposals equivalent to the CME 
in the chronicle [see the references above Eq.~(\ref{eq:anomaly})].

Among the recent progress, we first briefly mention 
a remarkable experiment in condensed matter physics. 
In early times, the anomalous current, which is now called the CME, 
was proposed in a lattice system \cite{Nielsen:1983rb}. 
However, experimental measurement was achieved only recently thanks to development of the Dirac/Weyl semimetals \cite{murakami2007phase,wan2011topological}. 
In such materials, the axial charge is dynamically created with a parallel electric and magnetic field, 
so that one will get a chiral imbalance $ \mu_A \propto E B $. 
This imbalance induces the CME contribution to 
the longitudinal current $ j_\para = ( \sigma_{\rm Ohm} + c B^2) E $ together with the Ohmic current. 
A constant $ c $ depends on competition between the anomaly effect 
and relaxation effects due to chirality-flipping processes which may depend on details of materials. 
The CME gives rise to a suppression of the resistance 
as we increase the applied magnetic-field strength. 
This quadratic suppression signals the occurrence of the CME in the Weyl/Dirac semimetals 
and is referred to as the (anomalous) negative magnetoresistance \cite{Son:2012bg, Zyuzin:2012tv, 
PhysRevLett.113.247203, Gorbar:2013dha,PhysRevB.91.245157}. 
In experiments, one can study the dependence of the current on the relative angle between the electric and magnetic fields. 
The negative magnetoresistance has been observed only when 
an electric and magnetic field are applied nearly in parallel to each other (see Fig.~\ref{fig:semimetal}) \cite{PhysRevLett.111.246603, 
Li:2014bha, xiong2015evidence, Huang:2015eia}, 
where the chiral anomaly plays a dominant role. 
The reader is referred to Refs.~\cite{Burkov:2015hba, Armitage:2017cjs, yan2017topological, felser2022topology} 
for recent reviews and lists of other experimental papers.

In case of relativistic heavy-ion collisions, one cannot directly measure the current induced in the quark-gluon plasma (QGP) 
and has less control of electromagnetic-field configurations. 
Nevertheless, the former issue could be solved by measuring the two-particle angle correlations 
that have different tendencies for the like-sign and unlike-sign pairs 
when a charge separation in QGP is induced by the CME current (see Refs.~\cite{Kharzeev:2013ffa, 
Liao:2014ava, Kharzeev:2015znc, Kharzeev:2015kna, Skokov:2016yrj, Hattori:2016emy} for reviews).

In early measurements, nonzero signals were observed 
by the STAR collaboration at Relativistic Heavy Ion Collider (RHIC)\cite{STAR:2009wot,STAR:2009tro,STAR:2013ksd} 
and the ALICE collaboration at the Large Hadron Collider (LHC) \cite{ALICE:2012nhw} at the top collision energies. 
The collision energy dependence was studied in 
the Beam Energy Scan program I by the STAR collaboration 
where the signal decreases as the collision energy 
is decreased \cite{STAR:2014uiw}. 
However, the CMS collaboration at the LHC pointed out that 
the same magnitudes of signals are observed in high-multiplicity 
proton-nucleus collisions as well as nucleus-nucleus collisions \cite{CMS:2016wfo, CMS:2017lrw}. 
This implies contamination of the signal due to non-CME effects 
stemming from the momentum correlations 
driven by collective motion and/or resonance decay 
since the CME signal is not expected to seen 
in proton-nucleus collisions 
where the induced magnetic fields have less spatial homogeneity. 
The STAR collaboration confirmed the presence of 
such background effects with proton-nucleus 
and deuteron-nucleus collisions at RHIC \cite{STAR:2019xzd}. 
Since then, there have been lots of efforts 
to disentangle the CME signals from 
the background effects by those collaborations 
as well as the ALICE collaboration \cite{ALICE:2017sss, ALICE:2020siw, STAR:2020gky, STAR:2021pwb, STAR:2022ahj} 
(see also Ref.~\cite{Zhao:2019hta} for an experimental review).

The magnetic-field effects could be disentangled 
from the background effects if one can compare 
different collision events with appropriately chosen nuclides. 
The comparison between the proton-nucleus 
and nucleus-nucleus collisions posed an important challenge, 
but one has to deal with the backgrounds and the magnetic fields 
both in different magnitudes. 
We need to change either of the magnitudes while 
keeping the other magnitude intact. 
A way to achieve distinct magnitudes of the flow backgrounds, 
without changing the magnitude of the magnetic field, 
is proposed with the use of deformed uranium nuclei 
\cite{Voloshin:2010ut}.  
The tip-tip and body-body collisions generate 
different magnitudes of the collective flows with 
similar magnitudes of the magnetic fields. 
On the other hand, distinct magnitudes of the magnetic fields 
can be achieved by the ``isobar collisions'' 
\cite{Voloshin:2010ut, Deng:2016knn}. 
Isobars have different atomic numbers 
but have the same mass numbers, 
so that one can create different magnitudes of magnetic fields 
without changing the QCD background effects that are blind to electrical charges carried by protons. 
Results of the isobar data analysis had been 
awaited for a few years and 
were released after careful blind analyses 
in Ref.~\cite{STAR:2021mii}. 
However, the results did not satisfy 
the predefined criteria for the CME signature 
and rather exhibited different magnitudes of 
the background effects for different isobars. 
Seemingly, the background effects are blind to electric charges 
but have relevant sensitivities to 
the shapes and nuclear structures of isobars. 
One can find a nice summary of the experimental progress 
in the proceedings \cite{Wang:2022eoo}. 
Those effects are still under investigation as of the end of 2022.



\begin{figure}
     \begin{center}
              \includegraphics[width=0.9\hsize]{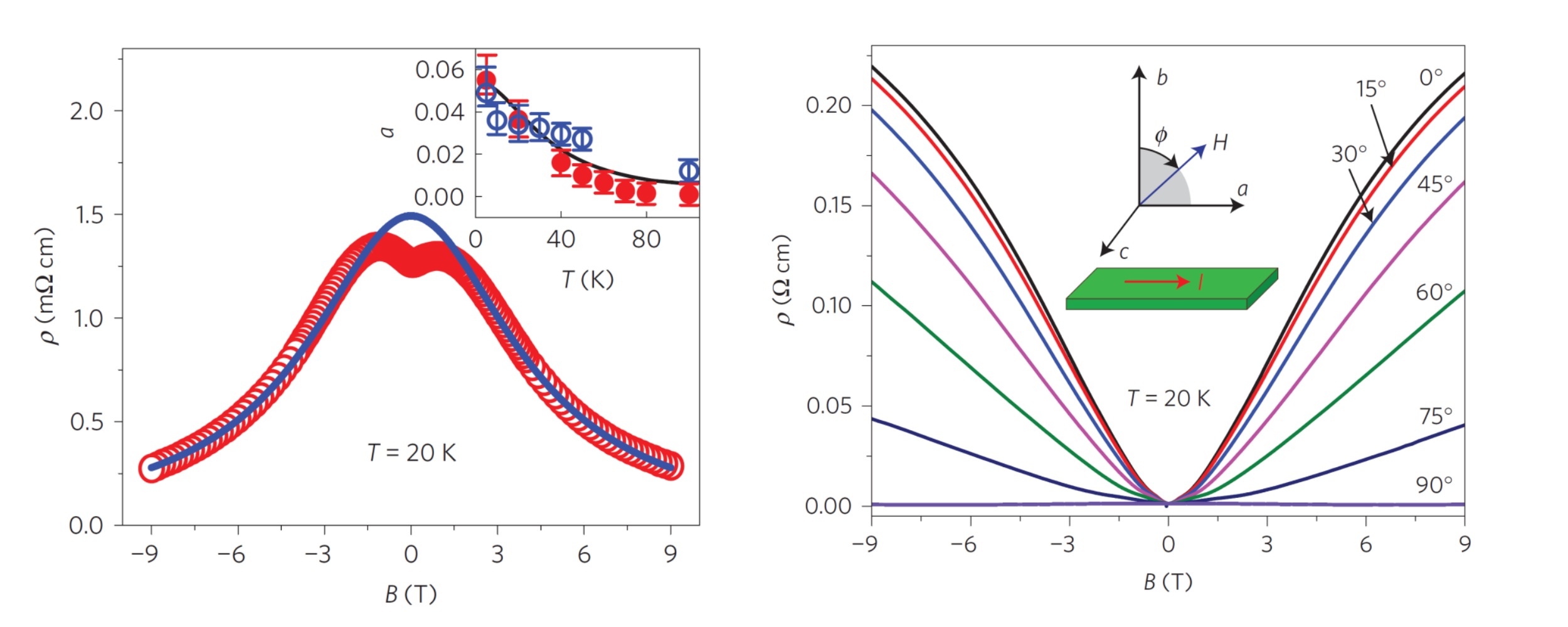}
     \end{center}
\vspace{-1cm}
\caption{``Negative magnetoresistance'' measured with the Weyl semimetal~\cite{Li:2014bha}. 
The resistance $ \rho $ decreases with an increasing magnetic field strength (left), which is observed 
only when the electric and magnetic fields are applied nearly in parallel ($ \phi = 90^\circ $ in the right panel). 
Note that the scales of the vertical axes in 
the left and right figures are different by three orders. 
}
\label{fig:semimetal}
\end{figure}

\subsection{Axial-charge dynamics: Helicity conversion and the chiral plasma instability}

\label{sec:CPI}

The CME conductivity is invariant under the time-reversal transformation 
since both the vector current and the magnetic field are odd under this transformation, 
implying the non-dissipative nature of the CME current \cite{Kharzeev:2011ds}. 
However, this is true only if the axial charge was a conserved quantity. 
One should note that the axial charge is only a conserved quantity 
in the classical and massless-fermion theory. 
Namely, the chiral symmetry is explicitly broken by the chiral anomaly at the quantum level 
and/or by a finite fermion mass. 
In such situations, the axial chemical potential, which works 
as the source of the CME, is no longer a static or conserved quantity 
protected by a symmetry. The chirality imbalance dissipates in real-time dynamics. 
Therefore, the dynamics of the axial charge can break the time-reversal symmetry, 
and a consistent description of the coupled dynamics of the axial charge and the CME is 
important for the long-time off-equilibrium evolution. 
Prohibition of the {\it equilibrium} CME current in an external magnetic field 
has been shown with the generalized Bloch theorem \cite{Yamamoto:2015fxa} 
(see also Refs.~\cite{Landsteiner:2012kd, Landsteiner:2016led, vazifeh2013electromagnetic, Zubkov:2016tcp}). 
In contrast, the vector charge is a conserved quantity protected by the $ U(1)_V $ symmetry, 
and the associated vector chemical potential is a well-defined equilibrium quantity. 
This means that one can investigate mass corrections to the CSE in 
a thermal equilibrium state specified by a vector chemical potential 
\cite{Metlitski:2005pr, Gorbar:2013upa, Lin:2018aon} (see Appendix.~\ref{sec:massive}).

The triangle anomaly diagrams give rise to the nonzero divergence of 
the axial-vector current (\ref{eq:anomaly}) in the massless case. 
Computing the triangle diagrams but with massive fermion loops, 
one finds that the triangle diagrams split into two terms and 
result in two terms in the divergence of 
axial-vector current as \cite{Bell:1969ts, Adler:1969gk} 
\begin{eqnarray}
\partial_\mu j^\mu_A = q_f^2 C_A \bE \cdot \bB + 2i m_f \langle \bar \psi \gamma^5 \psi \rangle
\label{eq:anomaly-mass}
\, .
\end{eqnarray}
In Appendix~\ref{sec:anomaly-massive}, 
we reproduce this computation and provide 
an explicit form of $\langle \bar \psi \gamma^5 \psi \rangle $  in Eq.~(\ref{eq:anomaly-massive}). 
The first term on the right-hand side is the same as that in the massless case (\ref{eq:anomaly}). 
In spite of the mass dependences of the fermion propagators, 
the mass dependences in this term go away in the end of the day. 
Mass dependences of the triangle diagrams yield 
the other term in the form of the pseudoscalar condensate $ \langle \bar \psi \gamma^5 \psi \rangle $ 
at the leading nonvanishing order in insertion of external fields. 
More generally, the pseudoscalar condensate $ \langle \bar \psi \gamma^5 \psi \rangle $ may have 
higher-order corrections as well as medium corrections \cite{Hattori:2022wao}. 
The presence of this term is suggested 
by the massive Dirac equation as well. 
The massive triangle diagrams, i.e., 
the sum of two terms on the right-hand side as a whole, 
are suppressed as the fermion mass becomes large 
and eventually vanish in the infinite-mass limit 
as shown in Eq.~(\ref{eq:offset}). 
The absence of the axial-charge generation is 
also understood from the pattern of spectral flow on 
the parabolic dispersion curves for massive fermions \cite{Ambjorn:1983hp} (see also Ref.~\cite{Copinger:2018ftr}).

Integrating the anomaly equation (\ref{eq:anomaly-mass}) for a massive fermion, we find that 
\begin{eqnarray}
\label{eq:helicity-conservation}
\frac{d }{dt} [ Q_A + q_f^2 \frac{C_A }{2} Q_M] = \Gamma_m
\, ,
\end{eqnarray}
where the fermion chirality, the magnetic helicity, and the chirality mixing rate are, respectively, given by 
\begin{subequations}
\begin{eqnarray}
Q_A &=& 
\int d^3 \bx \ j_A^0
\, ,
\\
Q_M &=&  
\int d^3 \bx \ {\bm A} \cdot \bB
\, ,
\\
\Gamma_m &=& 
2i m_f \int d^3 \bx \ \langle \bar \psi \gamma^5 \psi \rangle
\, .
\end{eqnarray}
\end{subequations}
The magnetic helicity $ Q_M $, which originates from the Chern-Simons current, 
is a gauge-invariant quantity if the surface term vanishes. 
In the massless case ($ \Gam_m=0 $), we may define the conserved total helicity 
\begin{eqnarray}
 Q_{\rm tot} = Q_A +  q_f^2 \frac{ C_A}{ 2} Q_M 
 \end{eqnarray} 
such that $d  Q_{\rm tot} /dt =0  $. 
However, the two helicities are not conserved separately. 
Namely, the chiral-anomaly effect allows for the mutual conversion 
between the fermionic chirality and magnetic helicity when the electromagnetic field is dynamical. 
As we will see shortly, the CME current actually serves as a physical mechanism for this mutual conversion. 
On the other hand, the mass effect explicitly breaks 
the conservation law due to the chirality mixing as shown 
in Eq.~(\ref{eq:helicity-conservation}). 
If one invokes on QCD effects, the QCD sphaleron transition is another mechanism of the helicity change  \cite{McLerran:1990de, Arnold:1998cy, Moore:2010jd}, 
though its transition rate is suppressed in the weak-coupling regime.

We shall examine the dynamics of helicities with simple equations. 
The time evolution of the electromagnetic field is governed by the Maxwell equation 
\begin{eqnarray}
\label{eq:CSM}
\nabla \times \bB - \frac{ \partial \bE}{\partial t}  = \bj = \sigma_\CME (t) \bB + \sigma_{\Ohm} \bE
\, .
\end{eqnarray}
For simplicity, we assume that the Ohmic conductivity $  \sigma_{\Ohm}  $  is a constant quantity 
and that the CME current is given by $ \bj_\CME (t,\bx) = \sigma_ \CME(t) \bB(t,\bx) $ 
with $ \sigma_\CME (t) = q_f^2 C_A \mu_A(t)$. 
Remember that the relation of the CME conductivity $ \sigma_\CME $ to $  C_A $ and $ \mu_A $ 
is only proven in the hydrodynamic limit in the presence of a constant axial chemical potential, 
and the CME conductivity at a finite frequency can be subject to interaction effects 
\cite{Jimenez-Alba:2015bia, Kharzeev:2016sut} (see also Ref.~\cite{Kharzeev:2009pj}). 
We here intend to exemplify the dynamics of the axial charge in the slow-variation limit.

An important observation is that Eq.~(\ref{eq:CSM}) can be regarded as 
an eigenvalue equation for the magnetic field, i.e., $ \nabla \times \bB = \sigma_\CME (t) \bB $, 
when the electric field is absent. 
Such eigenvectors are given by the circular-polarization vectors 
in case of the plane-wave solutions such as $ ( 1, \pm i , 0) e^{ik_z z} $ 
and by the Chandrasekhar-Kendall (CK) states $  \bW_{\ell,m}^\pm (\bx,k) $ 
as the most general solution \cite{Chandrasekhar1957}. 
The CK states have desired good properties 
\begin{subequations}
\begin{eqnarray}
&&
\nabla \times \bW_{\ell,m}^\pm (\bx,k) = \pm k \bW_{\ell,m}^\pm (\bx,k) \, ,
\\
&&
\int d^3\bx \bW_{\ell,m}^a (\bx,k) \cdot \bW_{\ell^\prime,m^\prime}^b (\bx,k^\prime)  
= \delta_{ab}  \frac{\pi}{k^2} \delta(k-k^\prime) \delta_{\ell,\ell^\prime} \delta_{mm^\prime} 
\, ,
\end{eqnarray}
\end{subequations}
where $ k $ is the modulus of the momentum vector and $ \ell,m $ are angular-momentum indices.\footnote{
The CK states, or the Beltrami fields, are defined as (see Refs.~\cite{Hirono:2015rla, Xia:2016any, Qiu:2016hzd} 
and references therein for more details)
\begin{eqnarray}
\bW_{\ell m}^\pm (\bx , k) = \bT_{\ell , m}^\pm (\bx , k)  \mp \bP_{\ell , m}^\pm (\bx , k) 
\end{eqnarray}
where 
\begin{eqnarray}
 \bT_{\ell , m}^\pm (\bx , k)  =
- \frac{i}{\sqrt{\ell(\ell+1)}}  j_\ell(kr)   (\bx \times \nabla)  Y_\ell^m (\theta,\phi)  
\, ,
\quad
 \bP_{\ell , m}^\pm (\bx , k) = \frac{i}{k} \nabla \times \bT_{\ell , m}^\pm (\bx , k) 
\nn
\, ,
\end{eqnarray}
with the spherical Bessel function $ j_\ell(kr) $ and 
the spherical harmonic function $ Y_\ell^m (\theta,\phi)  = e^{im\phi} P_\ell^m (\cos \theta) $. 
} 
The first equation implies the transversality, $ \nabla \cdot \bW_{\ell,m}^\pm (\bx,k) = 0 $. 
By using the CK state, we can decompose the gauge potential $ \bA $, magnetic field $  \bB$, and electric field $ \bE $ as 
\begin{subequations}
\label{eq:CK-decomp}
\begin{eqnarray}
\bA (t,\bx) &=& \sum_{\ell,m} \int_0^\infty \frac{dk}{\pi} 
k [ \calA_{\ell,m}^+(t,k) \bW_{\ell,m}^+ (\bx,k) +  \calA_{\ell,m}^-(t,k) \bW_{\ell,m}^-  (\bx,k)  ]
\, ,
\\
\bB (t,\bx) &=& \sum_{\ell,m} \int_0^\infty \frac{dk}{\pi} 
k^2 [ \calA_{\ell,m}^+(t,k) \bW_{\ell,m}^+ (\bx,k) - \calA_{\ell,m}^-(t,k) \bW_{\ell,m}^-  (\bx,k)  ]
\, ,
\\
\bE (t,\bx) &=&- \sum_{\ell,m} \int_0^\infty \frac{dk}{\pi} 
k [ \dot{\calA}_{\ell,m}^+(t,k) \bW_{\ell,m}^+ (\bx,k) + \dot{ \calA}_{\ell,m}^-(t,k) \bW_{\ell,m}^-  (\bx,k)  ]
\, ,
\end{eqnarray}
\end{subequations} 
where $ \calA^\pm_{\ell,m} $ is the power spectrum of each helicity mode. 
Accordingly, the magnetic helicity and the magnetic and electric energies are decomposed as 
\begin{subequations}
\label{eq:B-decomp}
\begin{eqnarray}
Q_M(t) &=&  \sum_{\ell,m}   \int_0^\infty \frac{dk}{\pi} k 
[ \, (  \calA_{\ell,m} ^+ (t,k) )^2 - ( \calA_{\ell,m} ^- (t,k) ) ^2  \, ]
\, ,
\\
\calE_M(t) &=&  \frac12 \int d^3\bx \, |\bB (t,\bx)|^2
=  \frac{1}{2} \sum_{\ell,m}  \int_0^\infty \frac{dk}{\pi} k^2 
[  \, (  \calA_{\ell,m} ^+ (t,k) )^2 + ( \calA_{\ell,m} ^- (t,k) ) ^2  \, ]
\, ,
\\
\calE_E (t) &=&   \frac12 \int d^3\bx\,  |\bE (t,\bx)|^2
= \frac{1}{2} \sum_{\ell,m}   \int_0^\infty \frac{dk}{\pi}  
[ \, ( \dot{ \calA}_{\ell,m} ^+ (t,k) )^2 + ( \dot{\calA}_{\ell,m} ^- (t,k)) ^2  \, ]
\, .
\end{eqnarray}
\end{subequations}

Applying the CK decomposition (\ref{eq:CK-decomp}) to the Maxwell-Chern-Simons equation (\ref{eq:CSM}), 
we obtain the equation of motion for the helicity components of the gauge field 
\begin{eqnarray}
\label{eq:EoM-A} 
\ddot {  \calA}_{\ell,m} ^\pm (t,k) = - \{ k^2 \mp  k \sigma_\CME(t)  \}  \calA_{\ell,m} ^\pm (t,k) 
-  \sigma_\Ohm \dot {  \calA}_{\ell,m} ^\pm  (t,k)  
\, .
\end{eqnarray}
Note that the $ k $ modes are still coupled 
through $ \sigma_\CME(t)  $ that depends on the $ k $ spectrum. 
Without the CME term, the first term on the right-hand side serves as a ``restoring force'' 
that induces an oscillation mode near the center position at $  \calA_{\ell,m} ^\pm (t,k) =0 $. 
However, the CME term pushes the motion away from the origin if the sign of the coefficient, $ k^2 \mp  k \sigma_\CME(t) $, is flipped. 
Then, the magnetic spectrum $ \calA_{\ell,m} ^\pm (t,k) $ 
could grow as long as the constrains from the total energy and helicity conservations are satisfied. 
The Ohmic current works as a ``friction'' that provides a damping effect.

One can visualize the above observations 
by constructing an effective potential. 
Without the dissipative effects ($\sigma_\Ohm  = 0$ and $ m_f =0  $), 
we find an integral of motion [$ d \calE_{\ell,m} ^\pm (k)/dt = 0 $]: 
\begin{eqnarray} 
\calE_{\ell,m} ^\pm (k)  = \frac{1}{2} (\dot { \calA}_{\ell,m} ^\pm  (t,k)  )^2 
+ V_k^\pm [  \calA_{\ell,m} ^\pm  (t,k)   ]
\, ,
\end{eqnarray}
where the effective potential is given by 
\begin{eqnarray} 
  V_k^\pm [  \calA_{\ell,m} ^\pm  (t,k) ] =  \frac{1}{2}k^2 ( \calA _{\ell,m} ^\pm  (t,k) )^2 
 \mp \frac{1}{2} k \int_0^t dt^\prime \sigma_\CME(t^\prime)
  \frac{d \ }{dt^\prime} ( \calA_{\ell,m} ^\pm  (t',k) )^2
 \, .
\end{eqnarray}
We dropped the constant of integration which does not play a role in the following. 
The total energy is found to be 
\begin{eqnarray} 
\calE_\tot =
\sum_{\ell,m} \sum_{h=\pm}\int \frac{ dk }{ \pi } \calE_{\ell,m} ^h (k)   = \calE_E (t) + \calE_M (t)
+ \int_0^t dt^\prime \int d^3 \bx \, \bj_\CME (t^\prime, \bx ) \cdot \bE(t^\prime, \bx)
\, .
\end{eqnarray}
As mentioned above, this total energy is a conserved quantity 
in the absence of the dissipative effects ($\sigma_\Ohm  = 0$ and $ m_f=0  $). 
The third term shows an energy conservation 
between the fermionic and electromagnetic sectors 
due to the CME current. 
Here, we only consider a dynamical electric field 
without an external one.


We are now interested in how the magnetic spectrum evolves, starting with 
a vanishing initial spectrum $ \calA _{\ell,m} ^\pm  (0,k) = 0 $ and magnetic helicity $ Q_M(0)=0 $, 
but with a finite CME current $ \sigma_\CME(0) \not = 0  $. 
To examine the stability of the effective potential within the present initial conditions, 
we shall focus on the quadratic terms near the origin ($ \calA _{\ell,m} ^\pm  (t,k) = 0 $). 
Then, we can replace $  \sigma_\CME(t) $ by an initial value $  \sigma_\CME(0) $. 
Performing the integral, we get \cite{Kaplan:2016drz} 
\begin{eqnarray}
 V_k [ \calA _{\ell,m} ^\pm  (t,k) ] &\sim& 
    \frac{1}{2} \{ (k\mp k_\ast)^2 - k_\ast^2  \} ( \calA _{\ell,m} ^\pm  (t,k) )^2 
    + \order \left(  ( \calA _{\ell,m} ^\pm  (t,k))^4 \right)
 \, ,
\end{eqnarray}
where $  k_\ast \equiv  \sigma_\CME (0)/2 =   q_f^2 C_A \mu_A(0) /2$.  
Either of the helicity modes have a semipositive-definite 
coefficient in front of the quadratic term, 
depending on the sign of $  \mu_A(0) $. 
This means a stability at the origin in that helicity mode. 
On the other hand, the coefficient in the other mode takes a negative value when $ 0\leq k \leq |k_\ast| $, 
indicating an upward convexity near the origin (cf. Fig.~\ref{fig:CPI} for each $ \ell, \, m, \, k $). 
Therefore, in this momentum regime, 
the latter helicity mode is selectively excited 
in response to an infinitesimal perturbation. 
As a consequence, a helical magnetic field is generated 
by the CME current. 
The presence of such an instability was recognized in early works \cite{Carroll:1989vb, Garretson:1992vt} 
and was investigated as a possible mechanism for generation of the primordial magnetic field 
in the inflation era \cite{Garretson:1992vt} and the electroweak phase transition \cite{Joyce:1997uy, Field:1998hi, Semikoz:2004rr, Laine:2005bt}\footnote{
A similar idea with the chiral vortical effect was considered in Ref.~\cite{Vilenkin:1982pn}. 
Note also that, when the Lagrangian has an axion term or a topological term $ \Lag_a \propto  \theta F^{\mu\nu} \tilde F_{\mu\nu} $, 
we get a current $ j^\mu  \propto ( \partial_\nu \theta)  \tilde F^{\mu\nu}  $. 
This current can be identified with the CME current with $ \mu_A \propto \partial_t \theta $ in Eq.~(\ref{eq:CSM}). 
Therefore, the axion electrodynamics \cite{Wilczek:1987mv} contains the same type of instability as implied in Ref.~\cite{Turner:1987bw}. 
} 
or, in the opposite way, for the baryogenesis from the magnetic helicity \cite{Giovannini:1997eg, Giovannini:1999wv, Giovannini:1999by}. 
Now, it is often referred to as the chiral plasma instability (CPI) \cite{Akamatsu:2013pjd}. 
As shown in Fig.~\ref{fig:CPI}, the CPI can be understood as a cycle of positive feedbacks as follows~\cite{Akamatsu:2014yza}. 
First, when there is an initial $ \mu_A $ and a seed magnetic field $  B_z$, they induce the CME current $ J_z $ 
and subsequently a magnetic field $ B_\theta $ according to the Amp\`ere's law. 
Then, the CME current $ J_\theta $ induced by $ B_\theta $ provides a {\it positive} feedback to the seed magnetic field $ B_z $.

More recently, the instability has been reinvestigated in cosmology \cite{Boyarsky:2011uy, Tashiro:2012mf, Kamada:2016cnb, Kamada:2016eeb, Fujita:2019pmi, Domcke:2019mnd, Schober:2020ogz, 
Mukaida:2021sgv, Domcke:2022kfs, Fujita:2022fwc}, 
neutron-star/supernova physics \cite{Akamatsu:2013pjd, Ohnishi:2014uea, Dvornikov:2014uza, Grabowska:2014efa, Sigl:2015xva, Yamamoto:2015gzz, Kaplan:2016drz, Masada:2018swb}, 
and relativistic heavy-ion collisions \cite{Chernodub:2010ye, Tuchin:2014iua,  Manuel:2015zpa, Hirono:2015rla, Xia:2016any}. 
The presence of the instability was also shown 
with the chiral magnetohydrodynamics \cite{Hattori:2017usa} 
(see also Ref.~\cite{Hattori:2022hyo} and references therein 
for a review and detailed discussions about the recent reformulation of magnetohydrodynamics).


 \begin{figure}
\begin{minipage}{0.6\hsize} 
	\begin{center} 
		\includegraphics[width=0.9\hsize]{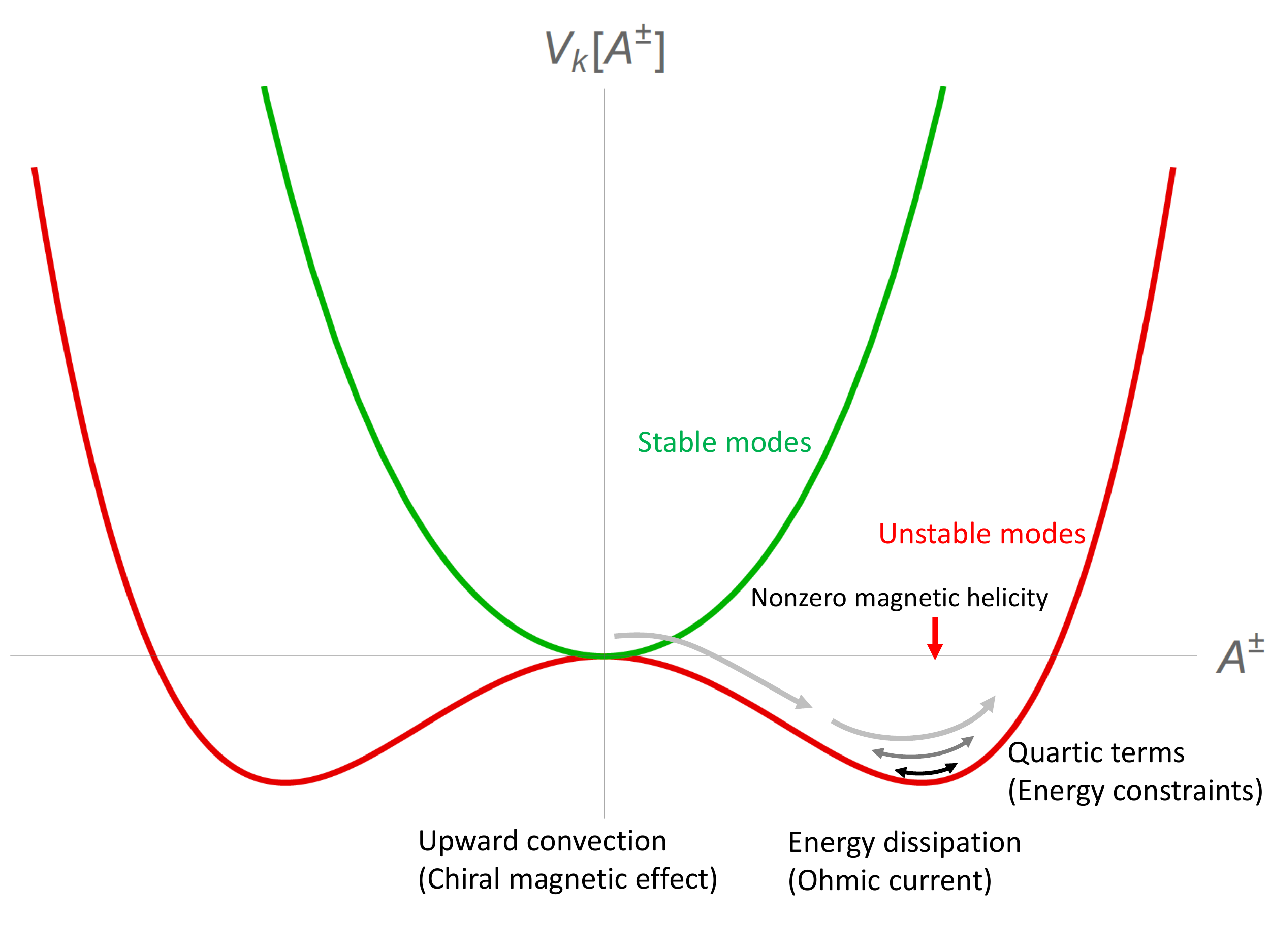} 
	\end{center}
\end{minipage}
\begin{minipage}{0.4\hsize}
	\begin{center}
\includegraphics[width=\hsize]{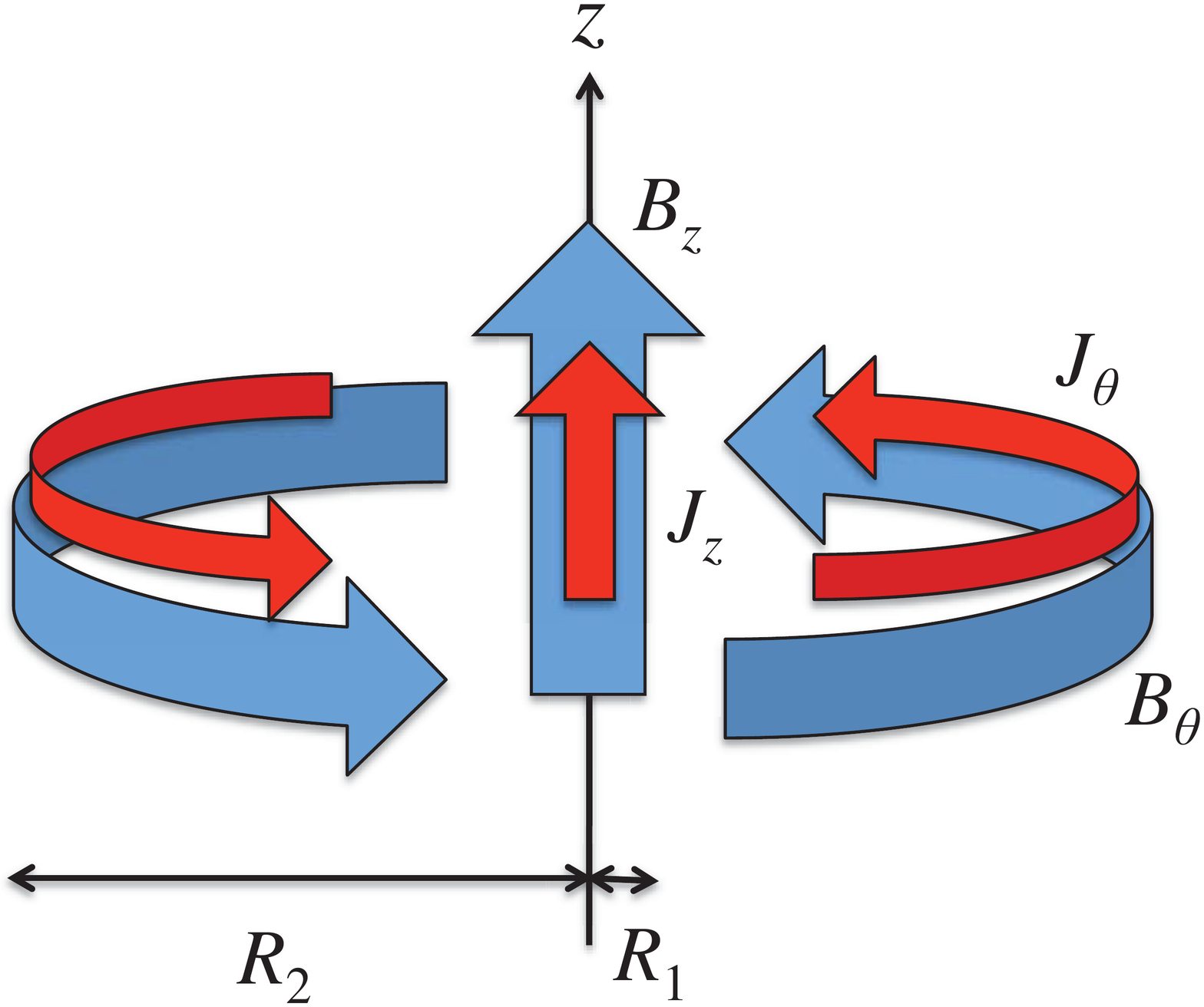}
	\end{center}
\end{minipage}
\caption{
An effective potential for the magnetic spectrum with an unstable convexity at the origin (left). 
A schematic picture of the positive feedback leading to the chiral plasma instability (right) \cite{Akamatsu:2014yza}. 
}
\label{fig:CPI}
\end{figure}

What would be the fate of the instability leading to the helical magnetic field generation? 
The effective potential $ V_k[ \calA _{\ell,m} ^\pm  (t,k) ] $ at a large $\calA _{\ell,m} ^\pm  (t,k) $ should be 
bounded so that the total energy of the system is conserved. 
The presence of such a quartic term was explicitly constructed for a non-interacting Fermi gas \cite{Kaplan:2016drz}. 
Then, the unstable mode will oscillate in the vicinity of the local minimum at a non-vanishing helical spectrum. 
It depends on the energy constraint how large the helical spectrum is at the position of the local minimum. 
On top of this, if one maintains the Ohmic current in Eq.~(\ref{eq:EoM-A}), 
the oscillatory mode will be damped out due to the energy dissipation 
and the system will fall in the local minimum. 
Since the total helicity $ Q_\tot$ is conserved, the gain in the magnetic helicity $  Q_M$ 
needs to be supplied from the fermion chirality $ Q_A $. 
Therefore, the CME, and thus the instability, 
will be diminished as $ Q_A $ is consumed.

What about effects of the mode coupling through the conservation of the total energy and helicity? 
In Eq.~(\ref{eq:B-decomp}), we notice that a unit magnetic helicity stores an energy $ k $. 
Therefore, a larger spatial structure is energetically favored for the magnetic field to carry a nonzero magnetic helicity. 
Under the constraint of the total-helicity conservation, this implies a tendency that short-scale structures 
of the magnetic-field lines are transformed to large-scale structures. 
This phenomenon was pointed out as the {\it inverse energy cascade} of the magnetic helicity 
in the dynamo theory with the magnetohydrodynamic turbulence (without the CME current) 
\cite{frisch_pouquet_leorat_mazure_1975, pouquet_frisch_leorat_1976, Christensson:2000sp} 
and has been applied as the mechanism for extending the correlation length of 
the primordial magnetic field in the universe \cite{ Baym:1995fk, Cornwall:1997ms, Son:1998my, Field:1998hi, Vachaspati:2001nb} (see Refs.~\cite{Durrer:2013pga, Subramanian:2015lua} for reviews). 
We point out that the CME term has the same form as the ``$ \alpha $-effect'' 
in the dynamo theory, where the pseudo-scalar coefficient 
is provided by the fluid kinetic helicity associated with 
the rotational motion instead of the fermion chirality (see, e.g., textbooks~\cite{biskamp1997nonlinear, davidson2002introduction}). 
Subsequently, the inverse energy cascade with the chiral anomaly was intensively investigated 
\cite{Boyarsky:2011uy, Tashiro:2012mf, Manuel:2015zpa, Sigl:2015xva, Hirono:2015rla, 
Yamamoto:2016xtu, Xia:2016any, Kamada:2016cnb, Kamada:2016eeb, Brandenburg:2017rcb, Domcke:2019mnd} 
and the effects of an inhomogeneous $ \mu_A (t,\bx) $ were also examined \cite{Boyarsky:2015faa, Gorbar:2016klv}. 
Numerical simulation of the chiral MHD evolution in supernovae 
was performed in Ref.~\cite{Masada:2018swb}.

However, the total helicity also dissipates due to 
the chirality mixing by the mass term in Eq.~(\ref{eq:helicity-conservation}). 
In the presence of the explicit chiral symmetry breaking, the chirality is not a conserved quantity 
and should vanish once a system reaches an equilibrium state: 
The CME current vanishes in the equilibrium state. 
A question is if the unstable mode grows fast enough in a transient state before the chirality imbalance is damped out. 
An answer depends on details of individual systems. 
The helicity-flipping effects had been mostly neglected in the aforementioned works for the high-temperature epoch in cosmology, 
but are investigated in Ref.~\cite{Boyarsky:2011uy} for the electroweak theory 
and more recently in Refs.~\cite{Kamada:2016cnb, Boyarsky:2020cyk, Boyarsky:2020ani} in detail. 
In case of neutron-star/supernovae physics, there is an estimate suggesting 
that the Rutherford scatterings of electrons off ambient protons 
have a significant impact on the dissipation of the electron helicity \cite{Grabowska:2014efa, Kaplan:2016drz}. 
Also, an estimate of the maximal field strength from the total energy conservation was performed in Ref.~\cite{Sigl:2015xva}. 
Currently, it is still a challenging issue to create magnetic fields over the cosmological/astrophysical spatial scales 
in the universe and in the neutron stars/magnetars on the basis of the CPI. 


There have been also a number of studies on 
the axial-charge dynamics in the context of the quark-gluon plasma. 
The helicity-flip rate by the gluon Compton scattering was estimated in Ref.~\cite{Manuel:2015zpa}. 
Such a perturbative process gives a much longer relaxation time than the lifetime of the quark-gluon plasma. 
However, it has been known that the relativistic heavy-ion collisions create 
a ``strongly-coupled'' quark-gluon plasma, 
which may demand further estimates beyond the perturbation theory as well. 
More recently, the kinetic theories for massive fermions are being developed 
to describe the coupled off-equilibrium dynamics among the vector and axial charges 
\cite{Weickgenannt:2019dks, Gao:2019znl, Hattori:2019ahi, Wang:2019moi, Sheng:2020oqs, Lin:2021mvw}, 
which may be called the ``axial kinetic theory'' (see Ref.~\cite{Hidaka:2022dmn} for a review). 
Compared to the ``chiral kinetic theory'' for massless fermions~\cite{Pu:2010as, Gao:2012ix, Son:2012wh, 
Son:2012zy, Stephanov:2012ki, Chen:2012ca, Hidaka:2016yjf,Mueller:2017lzw, Mueller:2017arw, Carignano:2018gqt, Lin:2019ytz, Chen:2021azy}, 
the axial kinetic theory contains a larger number of dynamical degrees of freedom 
since fermion spin is no longer aligned to the momentum in the massive theories. 
A smooth connection between the massive and massless theories are shown in Ref.~\cite{Hattori:2019ahi} (see also Ref.~\cite{Sheng:2020oqs}). 
It is the next indispensable step toward description of the relaxation dynamics 
to implement the collisional effects in the quantum kinetic equations \cite{Li:2019qkf, Yang:2020hri}. 
Also, there has been progress in the real-time numerical simulation for the axial-charge generation, 
the CME, and the CPI~\cite{Tanji:2016dka, Mueller:2016ven, Mace:2016shq, Tanji:2018qws, Mace:2019cqo, Schlichting:2022fjc}.

In the context of QCD thermodynamics, 
the theta angle in the presence of a magnetic field 
and an Euclidean electric field was measured 
by the Monte Carlo lattice QCD simulation \cite{DElia:2012ifm}. 
Another simulation for the topological susceptibility 
in a magnetic field (without an electric field) showed that 
the magnitude of the spatial correction of the topological charge fluctuation is enhanced, but an anisotropy with respect to 
the parallel and perpendicular directions to the magnetic field is small at $eB=1.1 \ \GeV $ \cite{Bali:2013esa}.



\section{Summary}

\label{sec:summary}

In this review article, we presented fundamentals and applications of the quantum dynamics under strong fields. 
In the first half, we summarized the fundamental points 
in a pedagogical way so that the reader can get the techniques 
required in the strong-field physics. We discussed the Landau quantization by a gauge-invariant formulation, 
the Ritus basis, and the proper-time method. 
In the proper-time resummation, the external fields are 
extended to non-Abelian fields within 
the covariantly constant fields.

All of these concepts and techniques were then 
applied to computation of physical quantities in the second half. 
We discussed the nonlinear QED effects such as 
the Schwinger mechanism and the vacuum birefringence. 
The interplay between QED and QCD was discussed on 
the bases of the Heisenberg-Euler effective action 
in the coexistent QED and QCD background fields, 
including the discussion about the Polyakov loop 
in strong magnetic fields at finite temperature. 
Then, starting with the basic concept of the dimensional 
reduction in strong magnetic fields, 
we discussed the magnetic catalysis of the chiral condensate 
and the magnetically induced Kondo effect. 
These nonperturbative phenomena are induced by 
quantum many-body effects irrespective of the strength of 
the coupling constant in the underlying theories. 
We emphasized this point based on the analogy between the systems 
in the strong magnetic fields and in the dense matter. 
What is even more interesting is the interplay between 
such nonperturbative many-body effects 
and the intrinsic nonperturbative interaction 
in the low-energy QCD. 
We provided detailed summary of the novel results from 
the lattice QCD simulations in the last decade 
and interpretations putting an emphasis on the importance of 
the infrared-dominant interaction in QCD. 
Finally, we discussed the anomaly-induced transport phenomena, 
called the chiral magnetic effect. 
As we discussed on the basis of the effective potential, 
the vector current in the chiral magnetic effect 
induces the chiral plasma instability 
and the associated conversion of the fermion helicity 
to the magnetic helicity as a consequence of chiral anomaly; 
This instability is expected to be a potential mechanism of 
amplifying coherent magnetic fields over the astronomical scales.

The strong-field physics is still a growing research field 
with the advent of experiments using high-intensity laser field, 
condensed matter materials, and heavy-ion collisions 
as well as astrophysical observations. 
The interested reader is referred to the references 
and the topical review papers cited in the previous sections. 
We hope that this review article provides 
useful introduction for many readers who would like 
to understand the basic aspects of the strong-field physics.

\section*{Acknowledgement}

The authors thank Gergely Endr\H{o}di, 
Kenji Fukushima, Yoshimasa Hidaka, 
Masaru Hongo, Xu-Guang Huang, Toru Kojo, Gergely Mark\'o, 
Daisuke Satow, 
Igor Shovkovy, Hidetoshi Taya, Naoki Yamamoto, 
and Di-Lun Yang for useful discussions and comments. 
KH thanks Norihiro Hizawa, Hiroki Ohata, and 
Kotaro Uzawa, brilliant students in Kyoto University, 
for carefully reading the manuscript and providing useful feedbacks. 
K.H. and S.O. thank KEK for hospitalities and financial supports where a part of this work was achieved. 
This work is partially supported by JSPS KAKENHI under grant Nos.~20K03948 and 22H02316. 
The research of S.O. is supported by MEXT-Supported Program for the Strategic Foundation at Private Universities, ``Topological Science" under Grant No.~S1511006.

\appendix

\section{Wave functions in magnetic fields}

\subsection{Landau gauge}

\label{sec:wf_Landau}

The wave function of the ground state can be obtained by solving an equation 
\begin{eqnarray}
\langle x \vert \hat a \vert 0 , \xc \rangle = 0
\, .
\end{eqnarray}
Now, in the Landau gauge, the coordinate representation of the annihilation operator is given by 
\begin{eqnarray}
\hat a = - i \frac{\ell_f}{\sqrt{2} } \{ \, 
\frac{ \partial \ }{\partial x} +  \ell_f^{-2} (\hat x - x_c)
\, \} 
= - i \frac{ 1 }{\sqrt{2}  } e^{- \frac{ \xi^2 }{2} }  \frac{\partial \ }{\partial \xi} e^{ \frac{ \xi^2 } {2} }
\, ,
\end{eqnarray}
where $ \xi = (x - x_c)/\ell_f$. 
The derivative is assumed to act on what follows on the right as well as on the Gaussian. 
The explicit form of the above condition reads 
\begin{eqnarray}
\frac{\partial \ }{\partial \xi} \left( \, e^{ \frac{\xi^2}{2} } \tilde \phi_0  \, \right) = 0
\, ,
\end{eqnarray}
where the wave function at $ n=0 $ is denoted as $ \tilde \phi_0 =  \langle x \vert 0 , \xc \rangle$. 
Therefore, we find 
\begin{eqnarray} 
\tilde \phi_0(\xi) = C_L  e^{ - \frac{\xi^2}{2} }
\, ,
\end{eqnarray}
with a factor of $ C_L=( \ell_f \pi^{\frac{1}{2}} )^{-1/2}$ that comes from the normalization 
\begin{eqnarray}
\int dx \vert \tilde \phi(x) \vert^2 = 1
\, .
\end{eqnarray}

The wave function of the higher Landau levels can be obtained by multiplying the creation operators as 
\begin{eqnarray}
\tilde \phi_n (\xi)   = \langle x \vert \frac{(a^\dagger)^n}{\sqrt{n!}} \vert 0 , \xc \rangle
\, .
\end{eqnarray}
The coordinate representation of the creation operator is  
\begin{eqnarray}
\hat a^\dagger = - i \frac{\ell_f}{\sqrt{2} } \{ \, 
\frac{ \partial \ }{\partial x} -  \ell_f^{-2} (\hat x - x_c)
\, \}
= - i \frac{1}{\sqrt{2} } e^{ \frac{\xi^2}{2} }  \frac{\partial \ }{\partial \xi} e^{ -\frac{ \xi^2 } {2}}
\, .
\end{eqnarray}
Therefore, the wave function in the general Landau level is obtained as\footnote{ 
Since the overall phase $ i^n = \exp(i n \pi/2) $ depends on the Landau level $ n $, 
it is important to maintain it when considering the overlap between the wave functions of different Landau levels. 
Here are two of examples: Only when we maintain this factor, 
the fermion propagator from the Ritus method 
agrees with that from the proper-time method 
[cf. Eq.~(\ref{eq:prop-n})] 
and the Ward identity is satisfied (see an appendix in Ref.~\cite{Hattori:2020htm}). 
This phase does not appear if one takes an alternative choice of the Landau gauge, $A_x = - By $ 
or interchange $\pi_x $ and $ \pi_y$ in the definition of the canonical pair in Eq.~(\ref{eq:CA}). 
} 
\begin{eqnarray}
\tilde \phi_n (\xi) = \frac{1}{\sqrt{n!}} \left( - \frac{i}{\sqrt{2}  }\right)^n e^{ \frac{\xi^2}{2} } 
\frac{\partial^n \ }{\partial \xi^n } \left( \, e^{ - \frac{\xi^2}{2} } \tilde \phi_0 (\xi)  \, \right)
= C_L \frac{ i^n }{\sqrt{ 2^n n!}  } e^{ - \frac{\xi^2}{2} } H_n(\xi)
\, ,
\end{eqnarray}
where the Hermite polynomial is defined in Eq.~(\ref{eq:H_poly}). 
Including the plane-wave part, we get the wave function shown in Eq.~(\ref{eq:WF_Landau}).

\subsection{Symmetric gauge}

\label{sec:wf_symm}

As in the case of the Landau gauge, 
one can get the explicit form of the wave function 
by using the coordinate representation of the creation and annihilation operators. 
In the symmetric gauge, we, however, have the two sets of the operators, 
of which the coordinate representation are given by 
\begin{eqnarray}
\hat a \, , \hat a^\dagger &=&
-i \frac{\ell_f}{\sqrt{2}} [ \, 
( \,  \frac{\partial \ }{\partial x} \pm i s_f \frac{\partial \ }{\partial y} \, )
\pm \frac{1}{2\ell_f^2} ( x \pm is_f y) \, ]
\, ,
\\
\hat b \, , \hat b^\dagger &=&
\pm \frac{\ell_f}{\sqrt{2}} [ \, 
( \,  \frac{\partial \ }{\partial x} \mp i s_f \frac{\partial \ }{\partial y} \, )
\pm \frac{1}{2\ell_f^2} ( x \mp is_f y) \, ]
\, ,
\end{eqnarray}
where the upper (lower) signs on the right-hand side  
are for $ \hat a $ and $ \hat b$ ($ \hat a^\dagger$ and $ \hat b^\dagger$).\footnote{
While the above explicit forms are specific to 
the symmetric gauge, 
the algebra of $ \hat b $,  $ \hat b^\dagger $, 
as well as $ \hat a $,  $ \hat a^\dagger $, 
holds independently of the gauge choice 
since $ \hat b $,  $ \hat b^\dagger $ 
are originally defined with the center coordinates in Eq.~(\ref{eq:b-operators})
} 
Introducing a complex coordinate 
\begin{eqnarray}
\zeta = \frac{1}{\ell_f} (x + i s_f y) \, , \quad \bar \zeta = \frac{1}{\ell_f} ( x - i s_f y)
\, ,
\end{eqnarray}
the coordinate representation of the annihilation operators are given by 
\begin{eqnarray}
\hat a = - i \sqrt{2} e^{- \frac{\vert \zeta \vert^2}{4}} 
\frac{\partial \ }{\partial \bar \zeta} e^{ \frac{\vert \zeta \vert^2}{4}} 
\, , \quad
\hat b =  \sqrt{2} e^{- \frac{\vert \zeta \vert^2}{4}} 
\frac{\partial \ }{\partial  \zeta} e^{ \frac{\vert \zeta \vert^2}{4}} 
\, .
\end{eqnarray}
Then, \eref{eq:S-00} can be written as 
\begin{eqnarray}
&&
\frac{\partial \ }{\partial \bar \zeta} \left( \, e^{ \frac{\vert \zeta \vert^2}{4}} 
\phi_{00} (\zeta, \bar \zeta) \, \right)
= \frac{\partial \ }{\partial \zeta } \left( \, e^{ \frac{\vert \zeta \vert^2}{4}} 
\phi_{00} (\zeta, \bar \zeta)  \, \right) = 0
\, .
\end{eqnarray}
Therefore, we obtain the LLL wave function as 
\begin{eqnarray}
\phi_{00} (\zeta, \bar \zeta) = C_S \, e^{ - \frac{\vert \zeta \vert^2}{4}}
\, ,
\end{eqnarray}
where the normalization constant $ C_S = (2\pi\ell_f^2)^{-\frac{1}{2}}$ is determined from an integral 
\begin{eqnarray}
\int \!\! dx \int \!\! dy \, \vert  \phi_{00} \vert^2 = 1
\, .
\end{eqnarray}

The wave functions of the other states $ \phi_{nm} = \langle \bx \vert  n , m \rangle$ can be obtained 
by operating the creation operators 
\begin{eqnarray}
\hat a^\dagger = - i  \sqrt{2} e^{ \frac{\vert \zeta \vert^2}{4}} 
\frac{\partial \ }{\partial \zeta} e^{ - \frac{\vert \zeta \vert^2}{4}} 
\, , \quad
\hat b^\dagger = - \sqrt{2} e^{ \frac{\vert \zeta \vert^2}{4}} 
\frac{\partial \ }{\partial \bar  \zeta} e^{ - \frac{\vert \zeta \vert^2}{4}} 
\, .
\end{eqnarray}
Noting that $\hat a^\dagger + i \hat b = i \bar \zeta/\sqrt{2} $, we find 
\begin{eqnarray}
\phi_{nm} (\zeta , \bar \zeta) = \frac{ 1}{\sqrt{n! m!}}
\langle \bx \vert ( \hat b^\dagger )^m (  - i \hat b + i \frac{\bar \zeta}{\sqrt{2}} )^n \vert 0,0\rangle
= \frac{ i^n}{\sqrt{ 2^n n! m!}}
\langle \bx \vert ( \hat b^\dagger )^m  \bar \zeta^n \vert 0,0\rangle
\, .
\end{eqnarray}
To make a connection to a special function, we further arrange the derivative operators as 
\begin{eqnarray}
\phi_{nm} (\zeta , \bar \zeta) 
&=&  \frac{ i^n (-1)^m 2^{\frac{\ell}{2}}}{\sqrt{ n! m!}}
e^{ \frac{\vert \zeta\vert^2}{4} } \frac{ \partial^m \ }{\partial \bar \zeta^m } 
\left( \,  \bar \zeta^n  e^{ - \frac{\vert \zeta\vert^2}{4} } \phi_{00} (\zeta , \bar \zeta)  \, \right)
\nonumber
\\
&=&   C_S \frac{ i^n (-1)^m }{ 2^{\frac{\ell}{2}} \sqrt{ n! m!}}
e^{ \frac{\rho}{2} } \zeta^\ell 
\frac{ \partial^m \ }{\partial \rho^m } \left( \, \rho^n  e^{ -\rho } \, \right)
\, ,
\end{eqnarray}
where we have put $ \ell = m-n $ and $  \rho = |\zeta|^2/2$. 
Then, one can immediately apply the definition of the associated Laguerre polynomial (\ref{eq:Laguerre}). 
When the angular momentum $ \ell$ is a negative integer, 
the upper index of the Laguerre polynomial is positive ($ \alpha = - \ell $). 
To maintain the upper index positive for a positive $\ell  $ as well, 
one may use a formula 
\begin{eqnarray}
L_m^{-\ell} (\rho) = \frac{(m-\ell)!}{m!} (-\rho)^\ell L_{m-\ell}^{\ell} (\rho)
\label{eq:Laguerre-negative}
\, .
\end{eqnarray}
The final result is shown in Eq.~(\ref{eq:wf-sym}) with $ r = \ell_f  \sqrt{ \vert \zeta\vert^2}$.

\cout{

\section{Notes on the Ritus-basis method}

\subsection{Relation to the proper-time method}

\label{sec:relation}

\com{Notations of the RItus basis changed on Dec. 3, 2020. Please check throughout. }

In Sec.~\ref{sec:Feynman_rules}, we have obtained the fermion propagator (\ref{eq:prop-Ritus}). 
There, the particularly simple form is realized thanks to the use of the Ritus basis that 
provides the eigenspinor of the Dirac operator in the external constant magnetic field. 
In general, one can, nevertheless, choose any other basis. 
Here, we convert the fermion propagator (\ref{eq:prop-Ritus}) obtained in the Ritus basis 
to that in the Fourier basis, i.e., the plane-wave basis.  
The result agrees with the propagator obtained by the use of the proper-time method in Sec.~\ref{sec:prop-time}.

The wave function in the Landau gauge (\ref{eq:WF_Landau}) depends on 
the coordinate and momentum in a particular combination, i.e., $ x - s_f \ell_f^2 p_y $. 
Therefore, by shifting the integral variable $ p_y \to p'_y = p_y - s_f (x+x')/(2 \ell_f^{2})$ 
in Eq.~(\ref{eq:prop-Ritus-x}), one finds 
\begin{eqnarray} 
S(x, x') &=& e^{   i \frac{q_fB}{2} (x+x')(y-y') } \bar S(x-x')
\nn
\, ,
\\
\label{eq:S(x)}
\bar S(x-x') &\equiv&
\sum_{n=0}^{\infty} \int \frac{d^2 p_\para}{(2\pi)^2} \int  \frac{dp'_y}{2\pi} 
e^{ -  i p_\para \cdot (x _\para - x'_\para) }   
\\
&&
\times
 \Ritus_{n, p_y^\prime} \left( \frac{ (x-x')/2 - s_f \ell_f^2 p'_y  }{\ell_f}  \right)  
 S(p_n) 
 \Ritus_{n,p_y^\prime}^\dagger  \left( \frac{ - (x-x')/2 - s_f \ell_f^2 p'_y  }{\ell_f}  \right)   
 \nn
 \, .
\end{eqnarray}
Notice that the phase factor, arising from the shift of the plane-wave part in the wave function, 
is factorized from the residual part $ \bar S(x-x')  $ that has a manifest translational invariance. 
This phase is nothing but the Schwinger phase introduced in Sec.~\ref{sec:FS}. 
This can be confirmed as follows by taking the Landau gauge (\ref{eq:Landau-g}) in Eq.~(\ref{eq:Phi_line}): 
\begin{eqnarray}
\label{eq:SF-Landau}
\Phi_A = - q_f \int_0 ^1 A_2 \frac{ d \xi^2 }{d w} dw
=  q_f B \int_0 ^1 \{x' + ( x - x') w \} (y-y') dw
=   q_f B  \frac{x+x'}{2} (y-y')
\, ,
\end{eqnarray}
where the straight path starts from $  x^{\prime \mu}$ toward $ x^{\mu} $, 
and is parametrized as $  \xi^\mu (w) = x^{\prime \mu} + ( x^\mu - x^{\prime\mu}) w, \ w \in [0,1]$. 
The rightmost side agrees with the Schwinger phase in Eq.~(\ref{eq:S(x)}) 
that has an explicit dependence on the gauge field and breaks the translational invariance in the transverse plane. 
In case of the Landau gauge, this factor also explicitly breaks the rotational invariance along the magnetic field.

Below, we examine the Fourier component of the translation-invariant part: 
\begin{eqnarray}
\label{eq:S(x)-Fourier}
\bar S(p) &=& \int d^4 x \, e^{ i p\cdot x} \, \bar S(x) 
\, .
\end{eqnarray}
Inserting the explicit form of the Ritus basis (\ref{eq:Ritus}) in the Landau gauge (\ref{eq:WF_Landau}), we find 
\begin{eqnarray}
\bar S(p)
%
=  \sum_{n=0}^\infty \frac{ i } { p_n^2 - m^2 } 
\left[  ( \sla p_\para +m )  \left (  \bar \Ham_{n , n} \prj_+ + \bar \Ham_{n-1 , n-1}  \prj_- \right) 
+ i \sqrt{ 2n |q_f B| }
 \gam^1 \left(  \bar \Ham_{n-1 , n}  \prj_+ -  \bar \Ham_{n , n-1}  \prj_- \right)
  \right] 
   \, .
  \nn
  \\
 \label{eq:prop-n}
\end{eqnarray}
The imaginary unit in front of $ \gam^1 $ is a remnant of a factor of $ i^n $ in the wave function (\ref{eq:WF_Landau}). 
We have split $ \sla p_n $ in Eq.~(\ref{eq:S(x)}) into 
the longitudinal part $ \sla p_\para $ and the transverse part $ \gam^1 p_n^1 $, 
motivated by the properties of the spin projection operators: 
$ \prj_\pm \gam_\para^\mu  \prj_\pm  = \prj_\pm \gam_\para^\mu $, 
$ \prj_\pm \gam_\para^\mu  \prj_\mp = 0 $, 
$ \prj_\pm \gam_\perp^\mu  \prj_\pm  =0$, 
and $ \prj_\pm \gam_\perp^\mu  \prj_\mp  = \prj_\pm \gam_\perp^\mu $. 
Remember that each Landau level has the two-fold spin degeneracy. 
In Eq.~(\ref{eq:prop-n}), the first and second terms correspond to the spin-up and -down states (depending on $ s_f $), 
while the remaining two terms the mixing between the degenerate spin states. 
The overlap between two wave functions is encoded in the coefficient function 
\begin{eqnarray}
 \bar \Ham_{n , n'} (p_x, p_y) &\equiv& \int dx \, e^{ i p_x x}
\Ham_n \left (\frac{ x/2 - s_f \ell_f^2 p_y  }{\ell_f} \right) 
\Ham_\np \left (\frac{ - x/2 - s_f \ell_f^2 p_y  }{\ell_f} \right) 
\, .
\end{eqnarray}
By using an identity $ \Ham_{n} (-x)= (-1)^n \Ham_{n} (x)$ and inserting Eq.~(\ref{eq:Hermite-f-p}) into the above, 
one can further arrange it as 
\begin{eqnarray}
 \bar \Ham_{n , n'} (p_x, p_y) 
 &=& (-1)^\np  \int dx \, e^{ i p_x x}
\Ham_n \left (\frac{ x/2 - s_f \ell_f^2 p_y  }{\ell_f} \right) 
\Ham_\np \left (\frac{  x/2 + s_f \ell_f^2 p_y  }{\ell_f} \right) 
\nn
\\
&=&  (-1)^\np c_n c_\np e^{ - \ell_f^2 |\bp_\perp|^2  }
\times 2 \ell_f  \int dx \, e^{ -  \frac{1}{4 \ell_f^{2}} x^2  }
H_n \left ( x + i \ell_f  p_\perp  \right) H_\np \left ( x + i \ell_f  p_\perp^\ast  \right) 
\, ,
\end{eqnarray}
where $c_n = 1/ (  2^n n! \pi^{\frac{1}{2}} \ell_f )^{1/2}   $. 
We also defined a complex variable $ p_\perp := p_x + i s_f p_y $ with its complex conjugate $p_\perp^\ast $ 
and norm $ |p_\perp|^2 =  p_x^2 + p_y^2 =  |\bp_\perp|^2$. 
By the use of the formula (\ref{eq:HH-L}), we find 
\begin{subequations}
\begin{eqnarray}
 \bar \Ham_{n , n} ( |\bp_\perp| ) &=& 
2 (-1)^n e^{ - \ell_f^2 |\bp_\perp|^2  } L_n\left( 2 \ell_f^2 |\bp_\perp|^2 \right)
\, ,
 \\
 \bar \Ham_{n-1 , n}  ( |\bp_\perp| )  &=& \bar \Ham^\ast_{n , n-1} ( |\bp_\perp| )
 =
2 i  (-1)^n e^{ - \ell_f^2 |\bp_\perp|^2  }\sqrt{\frac{2}{n}} \,  \ell_f \,
p_\perp L^1_{n-1} \left( 2 \ell_f^2 |\bp_\perp|^2 \right)
\, .
\end{eqnarray}
\end{subequations}
It turns out that those coefficients depend on $ p_\perp^\mu $ only in the form of the norm $ |\bp_\perp|  $. 
This indicates restoration of the rotational invariance in the transverse plane, 
though it is not quite obvious before combining the two wave functions in the Landau gauge.

Inserting those expressions back to Eq.~(\ref{eq:prop-n}), 
we obtain the fermion propagator in the Fourier basis: 
\begin{eqnarray}
\bar S(p) =
2 i e^{ - \ell_f^2 |\bp_\perp|^2  } \sum_{n=0}^\infty (-1)^n  
 \frac{( \sla p_\para +m )  ( \, L_n \prj_+ - L_{n-1} \prj_-)  - 2  \sla p_\perp L_{n-1}^1 } 
 { p_n^2 - m^2 } 
 \, ,
\end{eqnarray}
where we omitted the arguments of the Laguerre polynomials, $ 2 \ell_f^2 |\bp_\perp|^2= 2|\bp_\perp|^2/|q_fB| $. 
This expression agrees with the propagator obtained from the proper-time method in Eq.~(\ref{eq:fermion_prop-LL}). 
$  \bar S(p)$ has a manifest rotational invariance around the direction of the magnetic field 
in spite of the use of the Landau gauge that breaks the rotational invariance. 
This implies that the gauge dependence has been completely factorized in the Schwinger phase $ \Phi_A $, 
and that the other part in Eq.~(\ref{eq:S(x)}) is a gauge-invariant quantity. 
The spin-mixing terms mentioned below Eq.~(\ref{eq:prop-n}) are nonzero 
only when the transverse momentum $ p_\perp^\mu $ is finite. 
Recall that $ p_\perp^\mu $ is introduced as the Fourier mode in Eq.~(\ref{eq:S(x)-Fourier}) 
and is different from the components in $p_n^\mu  $ defined below Eq.~(\ref{eq:free}). 
The square of this momentum is given by $ p_n^2 = (p^0)^2 - p_z^2 - 2n |q_f B|$.

\subsection{The Ward identity for each pair of the Landau levels}

\label{sec:Ward_id}

Here, we confirm that the vertex function (\ref{eq:Ritus-vertex2}) satisfies the Ward identity 
when the fermion lines are put on-shell. 
We shall start with a simple check and see how one can generally show 
the conservation of the vector current $ \bar \psi \gam^\mu \psi $ 
when the fermion field is coupled to the external electromagnetic field. 
The fermion field obeys the Dirac equation: 
\begin{eqnarray}
( i \sla D - m) \psi = 0
\, , \quad
\bar \psi ( i \overleftarrow{  \sla D}^\ast + m)  = 0
\, ,
\end{eqnarray}
where the covariant derivative contains the external gauge field 
and the asterisk denotes the complex conjugate. 
Then, one can simply use the Dirac equation to find 
\begin{eqnarray}
i \partial_\mu ( \bar \psi \gam^\mu \psi) 
&=& 
\bar \psi  i\overleftarrow{\sla \partial} \psi + \bar \psi  i\sla \partial \psi 
\nn
\\
&=& 
\bar \psi ( - q_f  \sla A^\ast - m) \psi + \bar \psi (  q_f \sla A + m) \psi 
\nn
\\
&=& 0
\, ,
\end{eqnarray}
for the real field $ A^\ast_\mu  (x) =A_\mu (x)$. 
Therefore, the current is conserved as it should be.

Now, it is more tempting to see how the vertex function (\ref{eq:Ritus-vertex2}) satisfies 
the Ward identity after the Landau level is introduced with the Ritus basis, 
which serves as a nontrivial check of the correctness of the vertex function. 
The contraction of the vertex function with the photon momentum reads 
\begin{eqnarray}
&&
\hspace{-1.5cm}
q_\mu \bar u(p'_\np)  \Gam^\mu_{n^\prime, n} (q_{x,y})   u(p_n)  
\nn
\\
&=& 
\bar u(p'_\np) ( \sla p_\para - \sla p'_\para) \left( \prj_+ \bar \Gam_{n^\prime, n} + \prj_- \bar \Gam_{n^\prime-1,n-1} \right) u(p_n)
+ \bar u(p'_\np)  \sla q_\perp \left( \prj_+ \bar \Gam_{n^\prime-1,n} + \prj_- \bar \Gam_{n^\prime,n-1} \right)  u(p_n)
\nn
\\
&=& 
\sqrt{2|q_f B|} \, \bar u(p'_\np) \gam^1 \Big[ 
\prj_+ \left(  \sqrt{n} \bar \Gam_{n^\prime-1,n-1} - \sqrt{\np} \bar \Gam_{n^\prime, n}  
- \bar q_\perp \bar \Gam_{n^\prime-1,n}  \right)
\nn
\\
&&
 + \prj_- \left( \sqrt{n}  \bar \Gam_{n^\prime, n} -\sqrt{\np} \bar \Gam_{n^\prime-1,n-1}   
 - \bar  q_\perp^\ast \bar \Gam_{n^\prime,n-1}  \right)
 \Big] u(p_n)
 \label{eq:Ward-id0}
\end{eqnarray} 
where the complex variables $ q_\perp = q_x + is_f q_y $ and 
$ \bar q_\perp  = q_\perp/ \sqrt{ |2 q_f B| } $ are defined below Eq.~(\ref{eq:Gamma-L}) 
and should not be confused with the four vector $ q_\perp^\mu $. 
To reach the second line, we used the momentum conservation of the parallel components, 
the ``Dirac equation''~(\ref{eq:free}), and the spin eigenequation, $ i \gam^1 \gam^2 \prj_\pm = \pm s_f \prj_\pm $. 
Below, we prove two identities 
\begin{subequations}
\begin{eqnarray}
\label{eq:id1}
&& 
\sqrt{n} \bar \Gam_{n', n} - \sqrt{n'} \bar \Gam_{n'-1,n-1} - \bar q_\perp^\ast \bar \Gam_{\np, n-1} = 0
\, ,
\\
&&
\label{eq:id2} 
\sqrt{n} \bar  \Gam_{n'-1,n-1} - \sqrt{n'}  \bar  \Gam_{n',n} - \bar q_\perp \bar  \Gam_{n'-1,n} = 0
\, .
\end{eqnarray}
\end{subequations}

We provide a gauge-invariant proof of the identities (\ref{eq:id1}) and (\ref{eq:id2}) 
by the use of gauge-invariant operators~(\ref{eq:CA-rela}). 
According to the definition of the scalar vertex function (\ref{eq:Ritus-form}), we have 
\begin{eqnarray}
\bar q_\perp^{\, \ast} \bar \Gam_{n^\prime, n-1}(q_\perp) 
&=&
 \frac{i}{ \sqrt{2 |q_f B|}} \int \!\! d^2x_\perp \, e^{i q_\perp \cdot x_\perp}
[\, \phi^\ast _{n^\prime,\qn^\prime} (x_\perp) 
(  \overleftrightarrow{ \partial }^{1} -  i  s_f  \overleftrightarrow{ \partial}^{2} ) 
\phi _{n-1,\qn} (x_\perp)  \,]
\nn
\\
&=&
 \frac{1}{ \sqrt{2 |q_f B|}} \int \!\! d^2x_\perp \, e^{i q_\perp \cdot x_\perp}
 \phi^\ast _{n^\prime,\qn^\prime} (x_\perp) 
\big\{   ( i \overrightarrow{ D}^{1}   +  s_f  \overrightarrow{ D}^{2} ) 
- ( i \overleftarrow{ D}^{1}  +  s_f  \overleftarrow{ D}^{2})^\ast 
 \big\}
 \phi _{n-1,\qn} (x_\perp) 
 \, .
 \nn
 \\
\end{eqnarray}
In the first line, we performed the partial integral, 
and then the derivatives no longer act on the exponential factor. 
In the second line, we used $ A^\ast_\mu  (x) =A_\mu (x)$ for the real field. 
According to Eq.~(\ref{eq:CA-rela}), the above equation immediately indicates 
that the scalar vertex function satisfies the identity (\ref{eq:id1}). 
The other identity (\ref{eq:id2}) can be proven in the same manner: 
\begin{eqnarray}
\bar q_\perp \bar \Gam_{n^\prime-1, n}(q_{x,y}) 
&=&
 \frac{i}{ \sqrt{2 |q_f B|}} \int \!\! d^2x_\perp \, e^{i q_\perp \cdot x_\perp}
[\,  \phi^\ast _{n^\prime-1, \qn^\prime} (x_\perp) 
(  \overleftrightarrow{ \partial }^{1}  + i  s_f  \overleftrightarrow{ \partial}^{2} ) 
\phi _{n,\qn} (x_\perp)  \, ]
\nn
\\
&=&
 \frac{1}{ \sqrt{2 |q_f B|}} \int \!\! d^2x_\perp \, e^{i q_\perp \cdot x_\perp}
 \phi^\ast _{n^\prime-1, \qn^\prime} (x_\perp) 
\big\{   ( i \overrightarrow{ D}^{1}   -  s_f  \overrightarrow{ D}^{2} ) 
- ( i \overleftarrow{ D}^{1}  -  s_f  \overleftarrow{ D}^{2})^\ast 
 \big\}
\phi _{n,\qn} (x_\perp) 
\, .
\nn
\\
\end{eqnarray}
Again, this equation immediately indicates 
that the scalar vertex function satisfies the identity (\ref{eq:id2}). 
Plugging those identities into Eq.~(\ref{eq:Ward-id0}), 
we have proven that the vertex function satisfies the Ward identity, i.e., 
\begin{eqnarray}
\label{eq:Ward-id}
q_\mu \bar u(p'_\np)   \Gam^\mu_{n^\prime n}(q_{x,y})  u(p_n)  = 0 
\, .
\end{eqnarray} 
This identity is satisfied for each pair of the Landau levels $ n $ and $  n'$ 
and the continuous momenta $ p_z, \, p_z' $.

Incidentally, we prove the identities (\ref{eq:id1}) and (\ref{eq:id2}) with the Landau gauge 
in which we have obtained an explicit form of the scalar vertex function (\ref{eq:Gamma-L}). 
Since the vertex function in the Landau gauge is expressed with the associated Laguerre polynomial, 
we will use recursive relations \cite{AssociatedLaguerrePolynomial}: 
\begin{subequations}
\begin{eqnarray}
&&
\label{eq:Laguerre1}
z L_{n-1}^{k+1} (z) = - \left[ n L_n^k (z)  - (n+k) L_{n-1}^k (z)  \right]
\, ,
\\
&&
\label{eq:Laguerre2}
L^k_n (z) = L^{k+1}_n (z) - L^{k+1}_{n-1} (z) 
\, .
\end{eqnarray}
\end{subequations}
We collectively denote $ n $-independent irrelevant factors in $\bar  \Gam_{n', n}  ^{\rm (L)}$ as 
\begin{eqnarray}
g = 2\pi \delta(p_y - p^\prime _y -q_y)  \; e^{ - i q_x \frac{p_y + p^\prime _y}{2 q_f B} } 
e^{- \frac{1}{2} |\bar q_\perp|^2}  
\, ,
\nn
\end{eqnarray}
which is common to all three terms in the identities (\ref{eq:id1}) and (\ref{eq:id2}). 
Following the straightforward arrangements, the last term in Eq.~(\ref{eq:id1}) reads 
\begin{eqnarray} 
\bar q_\perp^{\, \ast}   \bar \Gam_{n',n-1}^{\rm (L)} (q_{x,y}) 
&=&
g (-1)^{\Delta n+1} e^{  i s_f \Delta n \theta_q}
\sqrt{ \frac{ (n-1)! }{n^\prime! } }  |\bar q_\perp|^{\Delta n} 
\times |\bar q_\perp|^2 L_{n-1}^{\Delta n+1} ( |\bar q_\perp|^2 )
\nn
\\
&=& g (-1)^{\Delta n} \bar q_\perp^{\, \Delta n} 
\Big [ \sqrt{n}  \sqrt{ \frac{ n! }{\np! } } L_{n}^{\Delta n} ( |\bar q_\perp|^2 )
-  \sqrt{\np}  \sqrt{ \frac{ (n-1)! }{ (\np-1) ! } } L_{n-1}^{\Delta n}  ( |\bar q_\perp|^2 ) \Big]
\, ,
\end{eqnarray} 
where $ \theta_q $ and $ |\bar q_\perp| $ are defined below Eq.~(\ref{eq:Gamma-L}). 
We used the recursive relation (\ref{eq:Laguerre1}). 
The right-hand side is nothing but the first two terms in Eq.~(\ref{eq:id1}). 
On the other hand, the last term in Eq.~(\ref{eq:id2}) is arranged as 
\begin{eqnarray} 
\bar q_\perp  \bar \Gam_{n'-1,n}^{\rm (L)} (q_{x,y}) 
&=&
g (-1)^{\Delta n-1} e^{  i s_f \Delta n \theta_q}
\sqrt{ \frac{ n! }{ (\np-1)! } }  |\bar q_\perp|^{\Delta n} 
 L_{n}^{\Delta n-1} ( |\bar q_\perp|^2 )
\nn
\\
&=&
g (-1)^{\Delta n} \bar q_\perp^{\, \Delta n}  
\Big [   \sqrt{n}  \sqrt{ \frac{ (n-1)! }{ (\np-1) ! } } L_{n-1}^{\Delta n}  ( |\bar q_\perp|^2 )
- \sqrt{\np}  \sqrt{ \frac{ n! }{\np! } } L_{n}^{\Delta n}  ( |\bar q_\perp|^2 )
\Big]
\, ,
\end{eqnarray} 
where we used the recursive relation (\ref{eq:Laguerre2}). 
The right-hand side is nothing but the first two terms in Eq.~(\ref{eq:id2}). 
Therefore, one can again conclude the validity of 
the identities (\ref{eq:id1}) and (\ref{eq:id2}), 
and thus the Ward identity (\ref{eq:Ward-id}) in the Landau gauge. 

}


\section{Solutions for the resummed propagators}

\subsection{Schwinger phases on polygons}

\label{sec:polygon}

\begin{figure}
     \begin{center}
              \includegraphics[width=0.5\hsize]{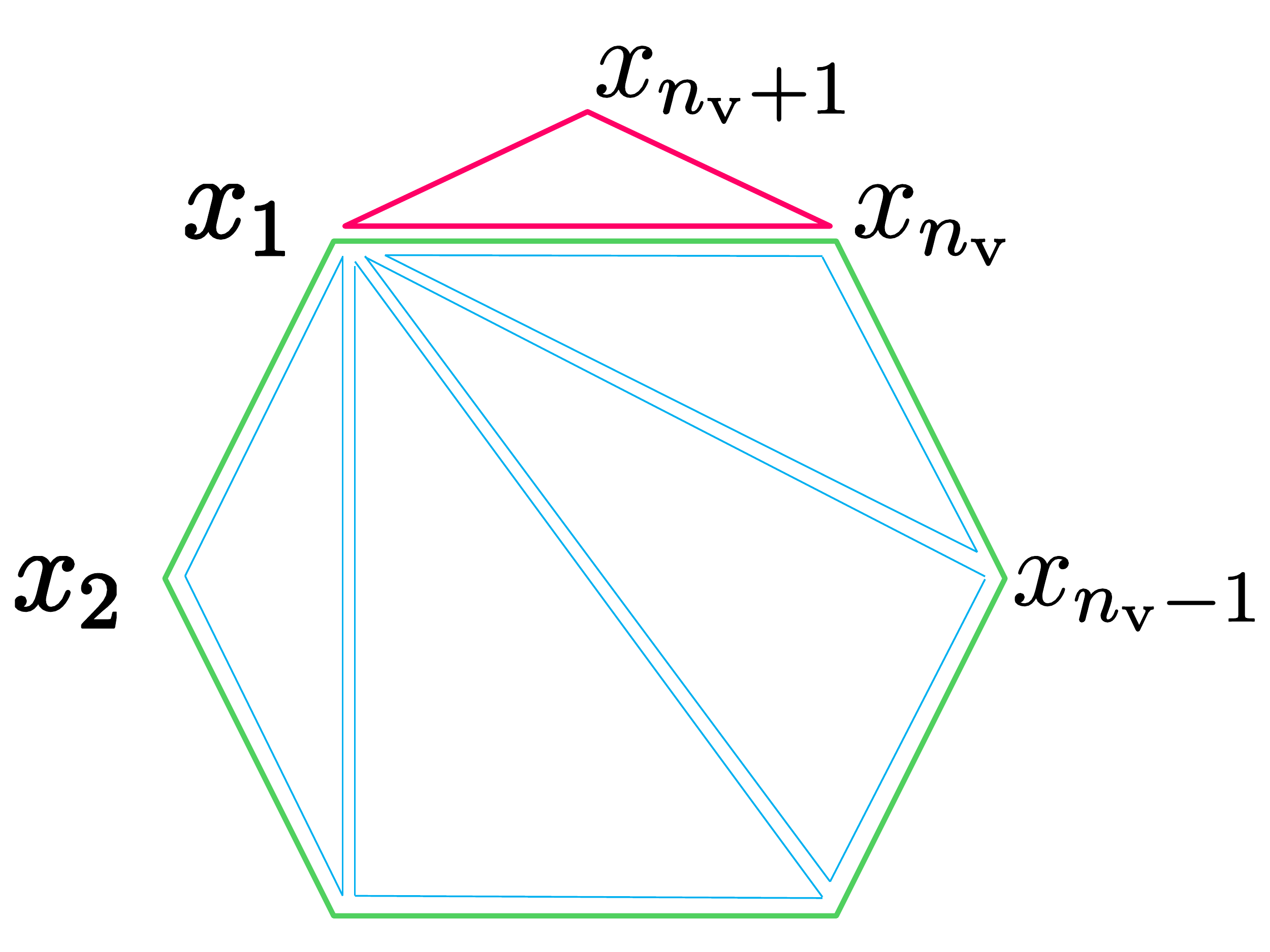}
     \end{center}
\caption{Schwinger phases on the $  (n_{\rm v}+1)$-gon, which can be obtained 
as the sum of those on the $  n_{\rm v}$-gon (green) and the triangle (magenta).}
\label{fig:polygon}
\end{figure}

In Fig.~\ref{fig:Sphase}, we show closed loops with different numbers of vertices, $ n_{\rm v} = 1$, $2 $, and $ 3$. 
When $ n_{\rm v} = 1$ which includes the case of the Heisenberg-Euler effective action discussed in Sec.~\ref{sec:HE}, 
one finds a vanishing phase as in Eq.~({\ref{eq:vanish}). 
However, when $ n_{\rm v} \geq 2$, 
the integrand differs on each segment by the initial coordinate $ y$ in the argument of the propagator (\ref{eq:transf}). 
in the gauge choice (\ref{eq:FS-condition}). 
Therefore, one finds in general that 
\begin{eqnarray}
 \Phi_A^{(n_{\rm v}) } :=  \Phi_A(x_2, x_1) + \Phi_A(x_3,x_2) \cdots +\Phi_A(x_1, x_{n_{\rm v}}) 
\not = \Phi_A(x_1, x_1) 
\, .
\end{eqnarray}
This statement can be checked 
by choosing a particular path in Eq.~(\ref{eq:Phi_S}), 
that is, a straight path parametrized as $ \xi^\mu (w) = y^\mu + (x^\mu-y^\mu)w$, $ w \in [0,1]$. 
This straight line is shown by blue lines in Fig.~\ref{fig:Sphase}. 
In this case, the translation-breaking term in Eq.~(\ref{eq:Phi_S}) vanishes, and we find that 
\begin{eqnarray}
\Phi_A(x,y) =   - q_f  \int_0^1  A_\mu(\xi (w))  \frac{d \xi^ \mu}{dw} dw
\, .
\label{eq:Phi_line}
\end{eqnarray} 
The integrand explicitly depends on the line element $ \xi(w)$ on which one performs the integral. 
For $ n_{\rm v} = 3$, we immediately notice the fact 
that the phase $ \Phi_A(z,x) $ on the segment from $ x$ to $z $ 
is different from the sum of phases $\Phi_A (y, x) + \Phi_A(z,y) $ when one takes a detour via the vertex $ y$. 
Therefore, the Schwinger phase is nonvanishing for $ n_{\rm v} \geq 3$. 
We have an exceptional case, that is, $ n_{\rm v} = 2$, which includes the vacuum polarization diagrams. 
In this case, the Jacobians on the two lines only differ by an overall sign, 
so that the Schwinger phase identically vanishes thanks to the cancellation. 
An explicit form of this integral is computed with the Landau gauge in Eq.~(\ref{eq:SF-Landau}).

\cout{

\begin{figure}[t]
     \begin{center}
           \includegraphics[width=0.9\hsize]{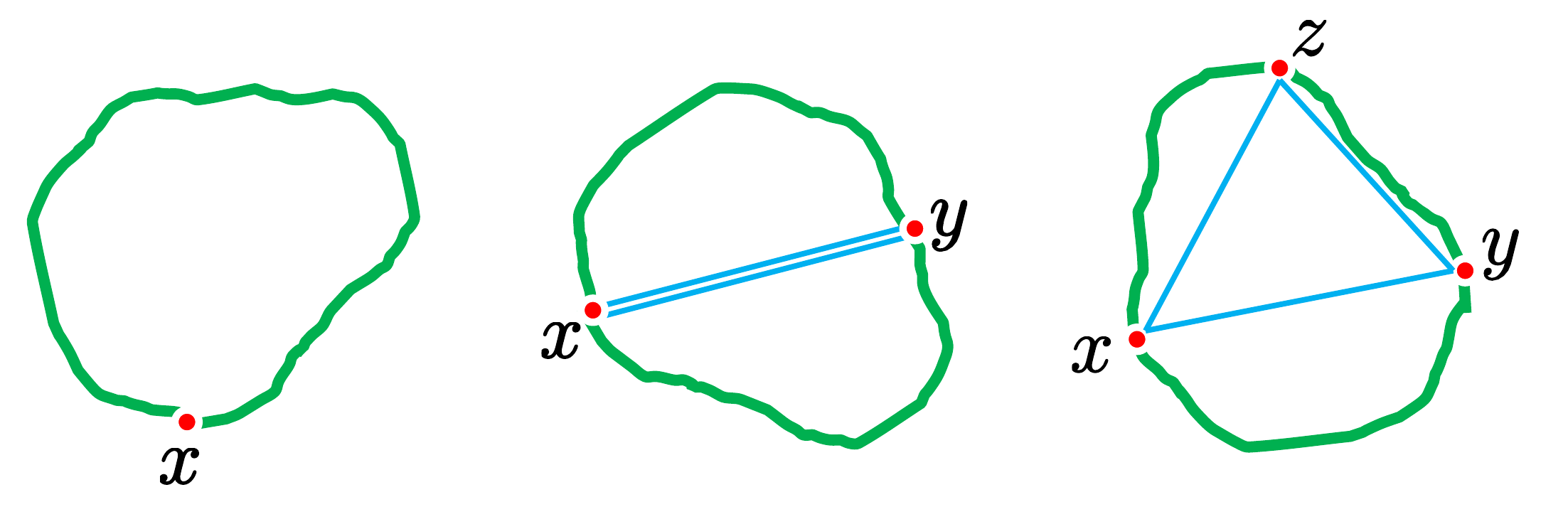}
     \end{center}
\caption{Schwinger phases on closed paths. 
}
\label{fig:Sphase}
\end{figure}

}

While the Schwinger phase is not a gauge- or translation-invariant quantity on each segment, 
one can show that the sum of the Schwinger phases associated with a closed fermion loop is 
a gauge-invariant quantity and can be cast into a translation-invariant form. 
To see this, notice first that the sum of the first two terms in \eref{eq:Phi_S} is by itself a curl-free quantity 
and does not depend on a specific coordinate $ y^\nu $. 
Therefore, they do not contribute to an integral along a closed path. 
This readily means invariance of closed fermion loops 
with respect to the gauge of external fields. 
The only one nonzero contribution comes from 
the third term that depends on $ y^\nu $. 
Assembling this contribution from all the segments, 
one finds the total Schwinger phase on a closed path with $ n_{\rm v} $ vertices:\footnote{
One can confirm the total Schwinger phase (\ref{eq:Phi-n}) 
starting from the integral along the straight path (\ref{eq:Phi_line}). 
The Schwinger phase on a segment can be arrange as 
\begin{eqnarray}
\Phi_A(x,y) &=&   q_f  \int_0^1  [\partial_{\xi^\nu}A_\mu(\xi) ]  \frac{d \xi^ \nu}{dw}   \xi^\mu  dw
\nn
\\
&=&   \frac{ q_f}{2} F_{\mu\nu} x^\mu y^\nu
+  \frac{ q_f}{2}   \int_y^x [\, \partial_{\xi^\mu}A_\nu(\xi) +  \partial_{\xi^\nu}A_\mu(\xi) \,]  \xi^\mu  d \xi^ \nu
\nn
\, .
\end{eqnarray} 
One can show that the integrand in the second term is a curl-free quantity. 
Therefore, the contribution of this term vanishes when integrated along a closed path, 
and the sum of the first term results in Eq.~(\ref{eq:Phi-n}). 
}
\begin{eqnarray}
\label{eq:Phi-n}
\Phi_A^{(n_{\rm v})}  =  \frac{q_f}{2} F_{\mu\nu} \sum_{m=1}^{n_{\rm v}} x_{m+1}^\mu x_m^\nu    
\, ,
\end{eqnarray}
where $  x_{n_{\rm v}+1}^\mu \equiv x_1^\mu$ is understood for the closed path. 
One can arrange this total phase in a manifestly translation-invariant form as follows. 
Explicitly, the first few cases read 
\begin{subequations}
\begin{eqnarray}
\Phi_A^{(2)} &=&  \frac{q_f}{2} F_{\mu\nu} ( x_2^\mu x_1^\nu +  x_1^\mu x_2^\nu ) =0
\, ,
\\
\Phi_A^{(3)} &=& \frac{q_f}{2} F_{\mu\nu} (x_{3}^\mu - x_2^\mu) (x_2^\nu - x_1^\nu)
\label{eq:Phi_3}
\, ,
\\
\Phi_A^{(4)} &=& \frac{q_f}{2} F_{\mu\nu} \{\,  (x_{4}^\mu - x_3^\mu) (x_3^\nu - x_1^\nu)
+ (x_{3}^\mu - x_2^\mu) (x_2^\nu - x_1^\nu) \, \}
\label{eq:Phi_4}
\, .
\end{eqnarray}
\end{subequations}
As shown just below, 
the Schwinger phase on an $  n_{\rm v}$-gon can be arranged as 
\begin{eqnarray}
\Phi_A^{(n_{\rm v})} = \frac{q_f}{2}  \sum_{m=1}^{n_{\rm v}-1} F_{\mu\nu} (x_{m+2}^\mu - x_{m+1}^\mu) (x_{m+1}^\nu - x_1^\nu)
\label{eq:Schwinger_n}
\, , 
\end{eqnarray}
where again $  x_{n_{\rm v}+1}^\mu \equiv x_1^\mu$. 
This expression only depends on positions in the form of differences. 
Therefore, one can conclude that the total Schwinger phase on a closed fermion loop is 
a gauge- and translation-invariant quantity.

One can prove the general relation (\ref{eq:Schwinger_n}) 
by the use of mathematical induction. 
Assuming that this statement is true for the Schwinger phase $ \Phi_A^{(n_{\rm v})} $ for a $  n_{\rm v}$-gon, 
we construct a ($  n_{\rm v}+1$)-gon by inserting a vertex between $ x_1 $ and $ x_{n_{\rm v}} $ (see Fig.~\ref{fig:polygon}). 
This suggests to consider a quantity 
\begin{eqnarray}
\phi \equiv \Phi_A^{(n_{\rm v})} + \frac{q_f}{2} F_{\mu\nu} (x_{n_{\rm v}+1}^\mu - x_{n_{\rm v}}^\mu) (x_{n_{\rm v}}^\nu - x_1^\nu)
\, ,
\end{eqnarray}
where the second term is the Schwinger phase for a triangle attached on the $  n_{\rm v}$-gon. 
Since the second term has a manifest translational invariance, 
the $  \phi$ also has the translational invariance according to the assumption for $ \Phi_A^{(n_{\rm v})} $. 
Expanding the $  \phi$, we find a cancellation of the Schwinger phases associated with the diagonal line 
between the vertices $ x_1 $ and $ x_{n_{\rm v}} $ as 
\begin{eqnarray}
\phi = \frac{q_f}{2} F_{\mu\nu} \left[  \, 
(\,  \sum_{m=1}^{n_{\rm v}-1} x_{m+1}^\mu x_m^\nu + x_1^\mu x_{n_{\rm v}}^\nu \, )
+ (\, x_{n+1}^\mu x_{n_{\rm v}}^\nu + x_{1}^\mu x_{n_{\rm v}+1}^\nu - x_1^\mu x_{n_{\rm v}}^\nu \, )
\, \right]
=  \frac{q_f}{2} F_{\mu\nu} \sum_{m=1}^{n_{\rm v}+1} x_{m+1}^\mu x_m^\nu 
\, ,
\end{eqnarray}
where $  x_{n_{\rm v}+2}^\mu \equiv x_1^\mu$. 
The rightmost side is nothing but $ \Phi_A^{(n_{\rm v}+1)} $, 
indicating that $ \Phi_A^{(n_{\rm v}+1)} $ has a translational invariance when $ \Phi_A^{(n_{\rm v})} $ does. 
In Eqs.~(\ref{eq:Phi_3}) and (\ref{eq:Phi_4}), 
we have already explicitly seen that the first two nonzero cases $  \Phi_A^{(3)}$ and $  \Phi_A^{(4)}$ have translational invariances. 
Therefore, we conclude that the Schwinger phase $  \Phi_A^{(n_{\rm v})}$ 
for an arbitrary $n_{\rm v}  $ $ (\geq 3) $ has a translational invariance. 
In other words, an explicit expression for $ n_{\rm v}\geq 3 $ is simply obtained by summing 
the translation-invariant Schwinger phases on the triangle sub-diagrams [cf. \fref{fig:polygon} and \eref{eq:Schwinger_n}]. 
Since the phases on the diagonal lines cancel each other in the sum, 
we get the sum of the phases on the sides as demanded.



One can take a derivative of the Schwinger phase 
by using the fact that the integral is path-independent 
once the initial and terminal points are fixed, i.e., 
$ \left[ \int_y^x + \int_x^{x+\delta x} +  \int_{x+\delta x}^y \right] d\xi^\alpha f_\alpha (\xi; y) = 0 $, where 
we put the integrand of the Schwinger phase (\ref{eq:Phi_S}) 
as $ f_{\mu}(\xi; y) = A_\mu(\xi) 
+\frac{1}{2}F_{\mu \nu} (\xi^\nu - y^\nu )$. 
Following the definition of partial derivative, one finds that 
\begin{eqnarray}
\label{eq:D}
\frac{\partial \ }{\partial x^\mu} \Phi_A (x,y) 
&=& - q_f \lim_{\Delta \to 0} \frac{1}{\Delta } 
\left[  \int_y^{x +\delta x} \!\! - \int_y^x  \right]  d\xi^\alpha f_\alpha (\xi; y) 
\nonumber
\\
&=& - q_f  \lim_{\Delta \to 0} \frac{1}{\Delta } \int_x^{x+\delta x}  \!\! d\xi^\alpha f_\alpha (\xi; y) 
\nonumber
\\
&=& - q_f  f_\mu (x; y) 
\, ,
\end{eqnarray}
where only the $ \mu$-th component is displaced as $ \delta x^\alpha = \Delta \delta^{ \mu \alpha}  $ by definition. 
The above result readily means that 
\begin{eqnarray}
D_x^\mu {\rm e}^{i\Phi_A(x,y)} = {\rm e}^{i\Phi_A(x,y)} \{\, \partial_x^\mu -  \frac{i}{2} q_f F^{\mu\nu} (x_\nu - y_\nu) \, \}
\, ,
\nn
\end{eqnarray}
which is shown in Eq.~(\ref{eq:D-Phi}) in the main text. 
After the derivative went through the Schwinger phase, 
the terms between the braces are given in a manifestly gauge- and translation-invariant form 
and are actually the covariant derivative expressed in the Fock-Schwinger gauge (\ref{eq:FS}) with $ x_0^\mu = y^\mu $ 
as expected from the covariance of the covariant derivative.

\cout{
This result is expected from the covariance of the derivative operator 
which can be evaluated in the Fock-Schwinger gauge and 
then transformed to a general gauge by the Schwinger phase. 
When the Dirac operator $ (i \sla D_x -m) $ is operated on the both sides of Eq.~(\ref{eq:transf}), 
the derivative relation (\ref{eq:D-Phi}) also implies that 
\begin{eqnarray}
 (i \sla D_x -m) S(x,y|A) 
 =  e^{ i\Phi_A(x,y) } i \delta^{(4)} ( x-y)
 \, ,
\end{eqnarray}
where we used Eq.~(\ref{eq:Gx}). Recalling that $ \Phi_A(x,x)=0 $, we notice that 
the right-hand side reads $ i \delta^{(4)} ( x-y) $ when $ x=y $ and, otherwise, vanishes. 
Therefore, we confirm that the resummed propagator $ S(x,y|A) $ in a general gauge 
is the Green's function of the Dirac operator, 
although it does not have a translation invariance like in familiar cases.  
}

\subsection{Fermion propagator}

\label{sec:resum_prop}

We solve Eq.~(\ref{eq:DELp}) by using an ansatz \cite{Dittrich:1985yb}. 
Since the operator in Eq.~(\ref{eq:DELp}) contains the squared momentum $p^2$ 
and the second derivative $\partial_p \partial_p$, 
we assume an ansatz given by 
\begin{eqnarray}
\Delta(p|A) = \frac{1}{i} \int_0^\infty \!\! ds \ e^{-i\kappa^2 s + M(s) }
\, ,
\label{eq:ansatz_q}
\end{eqnarray}
with a momentum bilinear 
\begin{eqnarray}
M(s) = i p^{\alpha} X _{\alpha \beta}(s) p^\beta + Y(s)
\, ,
\end{eqnarray}
where $X^{\alpha\beta}(s)  = X^{ \beta \alpha}(s)$ is a symmetric tensor. 
Substituting the ansatz for Eq.~(\ref{eq:DELp}), we find that 
the equation has a structure 
\begin{eqnarray}
1 = \frac{1}{i} \int_0^\infty \!\!ds \ g(s) e^{-f(s)} 
\label{eq:ans}
\ \ ,
\end{eqnarray} 
in terms of 
\begin{subequations}
\begin{eqnarray}
f(s) &=& - i p^\alpha X_{\alpha  \beta}(s) p^{\beta} - Y(s) + i \kappa^2 s   
\, ,
\label{eq:fff}  \\
g(s) &=&  p^\alpha \left( \id_\alpha^{\ \beta} 
- q_f^2 X_{\alpha  \rho}(s) F^{ \rho \sigma} F_{\sigma \lambda} X^{\lambda \beta}(s) \right) p_{\beta} 
+ \frac{i q_f^2}{2} F_{\alpha \beta} X^{\beta \gamma} (s) F_{\gamma}^{\ \, \alpha} 
- \kappa^2 
\ \ .
 \label{eq:ggg}
\end{eqnarray}
\end{subequations}
Therefore, if these functions satisfy the following three conditions 
\begin{subequations}
\begin{eqnarray}
&& \frac{d}{ds} f(s) = \frac{1}{i} g(s) 
\, ,
\label{eq:con1} 
\\
&&\lim_{s\rightarrow \infty} {\rm Re} f(s) \rightarrow \infty 
\, ,
\label{eq:con2} 
\\
&&f(0) = 0  
\label{eq:con3}
\ \ ,
\end{eqnarray}
\end{subequations}
our ansatz (\ref{eq:ansatz_q}) works well. 
That is, the integral with respect to the proper-time variable $s$ would be simply carried out as 
\begin{eqnarray}
\frac{1}{i} \int_0^\infty \!\!ds \ g(s) e^{-f(s)} = - \left[ \  e^{-f(s)} \ \right]_{0}^{\infty} = 1
\, \ ,
\label{eq:EEEEEEEEEEEEEEE}
\end{eqnarray} 
indicating that Eq.~(\ref{eq:ans}) is satisfied.

We shall look for the explicit expressions of $X(s)$ and $Y(s)$ 
that satisfy the first condition (\ref{eq:con1}). 
Then, we check if they meet the second and third conditions. 
and a finite value $e^{-f(0) } = 1$ at the lower boundary. 
Inserting Eqs.~(\ref{eq:fff}) and (\ref{eq:ggg}) into Eq.~(\ref{eq:con1}), 
one finds that $X_\alpha^{\ \beta}(s)$ and $Y(s)$ should satisfy a set of equations 
\begin{subequations}
\begin{eqnarray}
&&\frac{d}{ds} X_\alpha^{\ \, \beta}(s) \ = \ \1_\alpha^{\ \beta} 
- q_f^2 X_{\alpha  \rho}(s) F^{ \rho \sigma} F_{\sigma \lambda} X^{\lambda \beta}(s) 
\, ,
\label{eq:eqX}
\\
&&\frac{d}{ds} Y(s) \ = \  - \frac{q_f^2}{2} F_{\alpha \beta} X^{\beta \gamma} (s) F_{\gamma}^{\ \, \alpha} 
\, \ . 
\label{eq:eqY} 
\end{eqnarray}
\end{subequations}
Owing to elementary relations among the hyperbolic functions, 
the solution of the first equation (\ref{eq:eqX}) is found to be
\begin{eqnarray}
X_{\alpha \beta}(s) = \left[ (q_f F)^{-1} \tanh(q_f Fs) \right] _{\alpha \beta}
\, \ , 
\label{eq:X}
\end{eqnarray}
where $(F^{-1}) ^{\alpha \beta }$ is the inverse matrix of the field strength tensor. 
Inserting this solution into the second equation (\ref{eq:eqY}), we have 
\begin{eqnarray}
\frac{d}{ds} Y(s) \ = \  - \frac{q_f}{2} \left[ \tanh(q_fFs) \right] _{\mu \nu} F^{\nu \mu} 
\, \ .
\end{eqnarray}
Then,  the solution is found to be 
\begin{eqnarray}
Y(s) = - \frac{1}{2} {\rm Tr} \left[ \ln \left\{ \cosh(q_f F s) \right\} \right]
\, \ .  \label{eq:Y}
\end{eqnarray}

Now, we check if the conditions in Eqs.~(\ref{eq:con2}) and (\ref{eq:con3}) are satisfied. 
To see this, one should note the following two points. 
First, $X^{\alpha \beta}(s)$ is a real-valued matrix, 
so that we have $  {\rm Re}\ f(s) = - Y(s)$. 
Second, the logarithm in $Y(s) $ diverges in the limit, $s \rightarrow \infty$. 
The second point is valid regardless of the sign of any element in $ F_{\mu \nu}$ 
since the hyperbolic cosine is an even function. 
We confirm that the second condition (\ref{eq:con2}) is satisfied as 
\begin{eqnarray}
\lim_{s\rightarrow \infty} {\rm Re}\ f(s) 
& =&  \lim_{s \rightarrow \infty} \frac{1}{2} {\rm Tr} \left[ \ln \left\{ \cosh(q_f F s) \right\} \right]  +  \epsilon s 
= \infty
\, \ .
\end{eqnarray}
On the other hand, in the limit of vanishing $s$, we have an expansion 
\begin{eqnarray}
\lim_{s\rightarrow 0} f(s) &=& \lim_{s\rightarrow 0} \left[
- i p^2 s + \frac{1}{2} {\rm Tr} \left[ \ln\left ( \1 + O(s^2) \right) \right] + i \kappa^2 s 
\right]  
= 0
\, \ ,
\end{eqnarray}
which indicates that the third condition (\ref{eq:con3}) is successfully satisfied. 

This completes solving Eq.~(\ref{eq:DELp}). 
The results of this appendix is summarized in Eqs.~(\ref{eq:Dscalar})--(\ref{eq:DscalarY}).

\subsection{Equivalence with the Ritus basis method}

\label{sec:relation}


Here, we show an equivalence of the proper-time method 
discussed in Sec.~\ref{sec:prop-time} 
with the Ritus-basis method discussed 
in Sec.~\ref{sec:Ritus-Feynman}. 
They should be equivalent with one another 
since the different between those methods 
merely consists in the choice of basis 
used to define fermion excitations. 
We have seen in Sec.~\ref{sec:Ritus-Feynman} that 
the Ritus-basis is the eigenfunction of the Dirac operator, 
while the plane-wave basis used in the proper-time method, 
as in the familiar perturbation theory without external fields, 
is no longer an eigenfunction of the Dirac operator 
in external fields. 
In short, we examine the change of basis below.

According to the operation of the Dirac operator 
on the Ritus basis (\ref{Ritus_u}), 
one can find the propagator 
\begin{eqnarray}
S (x , x^{\prime }|A_{\rm L})
&=& 
\sum_{n=0}^{\infty} \int \frac{d^2 p_\para}{(2\pi)^2} \int  \frac{dp_y}{2\pi} 
e^{ -  i p_\para (x-x') }   \Ritus_{n,p_y} (x_\perp)    
\left[ \, \frac{ i }{ \sla p_n - m  } \, \right]
\Ritus_{n,p_y}^\dagger  (x'_\perp)   
\label{eq:prop-Ritus-x}
\, ,
\end{eqnarray} 
where $A_{\rm L} $ refers to the external field 
in the Landau gauge 
that we used to implement the explicit form of the Ritus basis 
in Sec.~\ref{sec:Ritus-Feynman}. 
Here, the transformation kernel is given by the Ritus basis 
instead of the familiar Fourier basis. 
By using the orthogonal relation (\ref{eq:orthogonal-x}), 
one can immediately confirm that $S (x , x^{\prime }|A_{\rm L})$ is the Green's function of the Dirac operator, i.e., 
$ ( i \sla D_x - m) S(x, x^{\prime}|A_{\rm L})
=  i \delta^{(4)} ( x - x') $. 
In the Ritus-basis space, 
the propagator has a simple form $ i/(\sla p_n - m)$ 
in contrast to the resummed propagator (\ref{eq:fermion_prop-LL}) 
in the proper-time method. 
This is simply because the propagator (\ref{eq:prop-Ritus-x}) 
is an inverse Dirac operator evaluated with its eigenfunction. 
A drawback in the Ritus-basis method is appearance 
of complex interaction vertices. 
In case of the QED interaction, 
the vertex functions are given by convolutions of 
the two Ritus basis for fermions 
and the Fourier basis for a photon, 
and are no longer the delta functions 
for the four-dimensional momentum conservation (see, e.g., Ref.~\cite{Hattori:2020htm}).


Below, we show that the propagator (\ref{eq:prop-Ritus-x}) 
agrees with that in the proper-time method (\ref{eq:fermion_prop-LL}) up to the Schwinger phase for the Landau gauge. 
The wave function in the Landau gauge (\ref{eq:WF_Landau}) depends on 
the coordinate and momentum in a particular combination, i.e., $ x - s_f \ell_f^2 p_y $. 
Therefore, by shifting the integral variable $ p_y \to p'_y = p_y - s_f (x+x')/(2 \ell_f^{2})$ 
in Eq.~(\ref{eq:prop-Ritus-x}), one finds that 
\begin{eqnarray} 
&&
\label{eq:S(x)}
S(x, x^{\prime}|A_{\rm L}) 
= e^{   i \frac{q_fB}{2} (x+x')(y-y') } \bar S(x-x')
\, ,
\\
&&
\bar S(x-x') \equiv
\sum_{n=0}^{\infty} \int \frac{d^2 p_\para}{(2\pi)^2} \int  \frac{dp'_y}{2\pi} 
e^{ -  i p_\para \cdot (x _\para - x'_\para) }   
 \Ritus_{n, p_y^\prime} \left( \frac12 (x-x') \right)  
 S(p_n) 
 \Ritus_{n,p_y^\prime}^\dagger  \left( - \frac12 (x-x') \right)
 \nn
 \, .
\end{eqnarray} 
Notice that the phase factor, arising from the shift of the plane-wave part, 
is factorized from the residual part $ \bar S(x-x')  $ that has a manifest translational invariance. 
This phase is nothing but the Schwinger phase introduced in Sec.~\ref{sec:FS}. 
This can be confirmed as follows by taking the Landau gauge (\ref{eq:Landau-g}) in Eq.~(\ref{eq:Phi_line}): 
\begin{eqnarray}
\label{eq:SF-Landau}
\Phi_A = - q_f \int_0 ^1 A_2 \frac{ d \xi^2 }{d w} dw
=  q_f B \int_0 ^1 \{x' + ( x - x') w \} (y-y') dw
=   q_f B  \frac{x+x'}{2} (y-y')
\, ,
\end{eqnarray}
where the straight path starts from $  x^{\prime \mu}$ toward $ x^{\mu} $, 
and is parametrized as $  \xi^\mu (w) = x^{\prime \mu} + ( x^\mu - x^{\prime\mu}) w, \ w \in [0,1]$. 
The rightmost side agrees with the Schwinger phase in Eq.~(\ref{eq:S(x)}) 
that has an explicit dependence on the gauge field and breaks the translational invariance in the transverse plane. 
In case of the Landau gauge, this factor also explicitly breaks the rotational invariance around the magnetic field.

Below, we focus on the Fourier transformation 
of the translation-invariant part: 
\begin{eqnarray}
\label{eq:S(x)-Fourier}
\bar S(p) &=& \int d^4 x \, e^{ i p\cdot x} \, \bar S(x) 
\, .
\end{eqnarray}
Inserting the explicit form of the Ritus basis (\ref{eq:Ritus}) in the Landau gauge (\ref{eq:WF_Landau}), we find that 
\begin{eqnarray}
\bar S(p)
%
=  \sum_{n=0}^\infty \frac{ i } { p_n^2 - m^2 } 
\left[  ( \sla p_\para +m )  \left (  \bar \Ham_{n , n} \prj_+ + \bar \Ham_{n-1 , n-1}  \prj_- \right) 
+ i \sqrt{ 2n |q_f B| }
 \gam^1 \left(  \bar \Ham_{n-1 , n}  \prj_+ -  \bar \Ham_{n , n-1}  \prj_- \right)
  \right] 
   \, .
  \nn
  \\
 \label{eq:prop-n}
\end{eqnarray}
The imaginary unit in front of $ \gam^1 $ is a remnant of a factor of $ i^n $ in the wave function (\ref{eq:WF_Landau}). 
We have split $ \sla p_n $ in Eq.~(\ref{eq:S(x)}) into 
the longitudinal part $ \sla p_\para $ and the transverse part $ \gam^1 p_n^1 $, 
motivated by the properties of the spin projection operators: 
$ \prj_\pm \gam_\para^\mu  \prj_\pm  = \prj_\pm \gam_\para^\mu $, 
$ \prj_\pm \gam_\para^\mu  \prj_\mp = 0 $, 
$ \prj_\pm \gam_\perp^\mu  \prj_\pm  =0$, 
and $ \prj_\pm \gam_\perp^\mu  \prj_\mp  = \prj_\pm \gam_\perp^\mu $. 
Remember that each Landau level has the two-fold spin degeneracy. 
In Eq.~(\ref{eq:prop-n}), the first and second terms correspond to the spin-up and -down states (depending on $ s_f $), 
while the remaining two terms the mixing between the degenerate spin states. 
The overlap between two wave functions is encoded in the coefficient function 
\begin{eqnarray}
 \bar \Ham_{n , n'} (p_x, p_y) &\equiv& \int dx \, e^{ i p_x x}
\Ham_n \left (\frac{ x/2 - s_f \ell_f^2 p_y  }{\ell_f} \right) 
\Ham_\np \left (\frac{ - x/2 - s_f \ell_f^2 p_y  }{\ell_f} \right) 
\, .
\end{eqnarray}
By using an identity $ \Ham_{n} (-x)= (-1)^n \Ham_{n} (x)$ and inserting Eq.~(\ref{eq:Hermite-f-p}) into the above, 
one can further arrange it as 
\begin{eqnarray}
 \bar \Ham_{n , n'} (p_x, p_y) 
=  (-1)^\np c_n c_\np e^{ - \ell_f^2 |\bp_\perp|^2  }
\times 2 \ell_f  \int dx \, e^{ -  \frac{1}{4 \ell_f^{2}} x^2  }
H_n \left ( x + i \ell_f  p_\perp  \right) H_\np \left ( x + i \ell_f  p_\perp^\ast  \right) 
\, ,
\end{eqnarray}
where $c_n = 1/ (  2^n n! \pi^{\frac{1}{2}} \ell_f )^{1/2}   $. 
We also defined a complex variable $ p_\perp := p_x + i s_f p_y $ with its complex conjugate $p_\perp^\ast $ 
and norm $ |p_\perp|^2 =  p_x^2 + p_y^2 =  |\bp_\perp|^2$. 
By the use of the formula 
\begin{eqnarray}
\label{eq:HH-L}
\int \!\! dx \ e^{-x^2} H_m (x+y) H_n (x+z) = 2^n \sqrt{ \pi } \, m! \, z^{n-m} L_m^{n-m} (-2 yz)
\, ,
\end{eqnarray}
for $ n -m \geq 0 $,\footnote{
When $  m - n >0 $, one can simply interchange roles of the two Hermite polynomials to 
have a positive index on the Laguerre polynomial:  
$\int \!\! dx \ e^{-x^2} H_m (x+y) H_n (x+z) = 2^m \sqrt{ \pi } \, n! \, y^{m-n}   L^{m-n}_{n} (-2 yz)  $. 
This agrees with an expression obtained by applying the formula~(\ref{eq:Laguerre-negative}) to Eq.~(\ref{eq:HH-L}) for $  m - n >0 $. 
}
we find that 
\begin{subequations}
\begin{eqnarray}
 \bar \Ham_{n , n} ( |\bp_\perp| ) &=& 
2 (-1)^n e^{ - \ell_f^2 |\bp_\perp|^2  } L_n\left( 2 \ell_f^2 |\bp_\perp|^2 \right)
\, ,
 \\
 \bar \Ham_{n-1 , n}  ( |\bp_\perp| )  &=& \bar \Ham^\ast_{n , n-1} ( |\bp_\perp| )
 =
2 i  (-1)^n e^{ - \ell_f^2 |\bp_\perp|^2  }\sqrt{\frac{2}{n}} \,  \ell_f \,
p_\perp L^1_{n-1} \left( 2 \ell_f^2 |\bp_\perp|^2 \right)
\, .
\end{eqnarray}
\end{subequations}
It turns out that those coefficients depend on $ p_\perp^\mu $ only in the form of the norm $ |\bp_\perp|  $. 
This indicates restoration of the rotational invariance in the transverse plane, 
though it is not quite obvious before combining the two wave functions in the Landau gauge.

Inserting those expressions back to Eq.~(\ref{eq:prop-n}), 
we obtain the fermion propagator in the Fourier basis: 
\begin{eqnarray}
\bar S(p) =
2 i e^{ - \ell_f^2 |\bp_\perp|^2  } \sum_{n=0}^\infty (-1)^n  
 \frac{( \sla p_\para +m )  ( \, L_n \prj_+ - L_{n-1} \prj_-)  - 2  \sla p_\perp L_{n-1}^1 } 
 { p_n^2 - m^2 } 
 \, ,
\end{eqnarray}
where we omitted the arguments of the Laguerre polynomials, $ 2 \ell_f^2 |\bp_\perp|^2= 2|\bp_\perp|^2/|q_fB| $. 
This expression agrees with the propagator (\ref{eq:fermion_prop-LL}) obtained in the proper-time method. 
$  \bar S(p)$ has a manifest rotational invariance around the direction of the magnetic field 
in spite of the use of the Landau gauge that breaks the rotational invariance. 
This implies that the gauge dependence has been completely factorized in the Schwinger phase $ \Phi_A $, 
and that the other part in Eq.~(\ref{eq:S(x)}) is a gauge-invariant quantity. 
The spin-mixing terms mentioned below Eq.~(\ref{eq:prop-n}) are nonzero 
only when the transverse momentum $ p_\perp^\mu $ is finite. 
Recall that $ p_\perp^\mu $ is introduced as the Fourier mode in Eq.~(\ref{eq:S(x)-Fourier}) 
and is different from the components in $p_n^\mu  $ defined below Eq.~(\ref{eq:free}). 
The square of the latter momentum is 
given as $ p_n^2 = (p^0)^2 - p_z^2 - 2n |q_f B|$ 
and reproduces the pole positions at the Landau levels.

\subsection{Gluon propagator}

\label{sec:resum_prop-gluon}

We solve Eq.~(\ref{eq:D-g}) for the resummed gluon propagator 
in parallel to the procedure performed in Appendix~\ref{sec:resum_prop} for fermions. 
In this subsection, we drop the color indices for notational simplicity. 
Similar to \eref{eq:ansatz_q}, we assume an ansatz  
\begin{eqnarray}
\Delta_{\mu \nu}^{(a)} (p| \nonA_\FS) = \frac{1}{i} \int_0^\infty \!\! ds \ 
e^{\bar M(s) }  [e^{-i s v^a ( \nonF_{\alpha\beta} \J^{\alpha\beta})} ]_{\mu \nu}
\, \ ,
\label{eq:ansatz_g}
\end{eqnarray}
with a momentum bilinear 
\begin{eqnarray}
\bar M (s) = - i p^{\alpha} X _{\alpha \beta} (s) p^\beta + Y(s)
\, \ ,
\end{eqnarray}
where $X^{\alpha\beta}(s)  = X^{ \beta \alpha}(s)$ is again a symmetric tensor.\footnote{ 
It is convenient to put a minus sign in front of the momentum bilinear for gluons, 
while it was a positive sign in Eq.~(\ref{eq:ansatz_q}) for the fermions. 
In the vanishing field limit, this relative sign correctly 
reproduces the overall sign of the free propagator. 
}
Inserting the ansatz (\ref{eq:ansatz_g}) into Eq.~(\ref{eq:D-g}), 
one finds an equation of the form 
\begin{eqnarray}
\delta_\mu^\sigma &=& 
\frac{1}{i} \int_0^\infty \!\! ds \ \hat {\mathsf g}_{\mu \nu} (s) e^{ - \hat {\mathsf f}^{\nu \sigma} (s) }
\, ,
\end{eqnarray}
and the problem reduces to solving the following equations for $X^{\alpha\beta}(s)   $ and $Y(s)  $: 
\begin{subequations}
\begin{eqnarray}
\hat {\mathsf f}_{\mu \nu} (s)
&=&
( i p^\alpha X_{\alpha\beta} (s)  p^\beta - Y(s) )  g_{\mu\nu}+ i s v^{a} ( \nonF_{\alpha\beta} \J^{\alpha\beta} ) _{\mu \nu}
\, ,
\\
\hat {\mathsf g}_{\mu \nu} (s)
&=&
- p^\alpha \left\{ \delta_{\alpha}^{\beta} 
- ( v^{a})^2 X_{\alpha  \rho}(s) \nonF^{ \rho \sigma} \nonF_{\sigma \lambda} X^{\lambda \beta}(s) \right\} p_\beta g_{\mu \nu}
\nonumber
\\
&&\hspace{0.5cm}
+ i  \frac{ (v^a)^2 }{2}  (\nonF_{\alpha \beta} X^{\beta \gamma} (s) \nonF_{\gamma}^{\ \, \alpha} ) g_{\mu \nu }
- v^{a} ( \nonF_{\alpha\beta} \J^{\alpha\beta} )_{\mu \nu}
\ \ .
\end{eqnarray}
\end{subequations}
These equations are very similar to (\ref{eq:ans})--(\ref{eq:ggg}). 
The major difference is just the additional Lorentz structures. 
As in Sec.~\ref{sec:resum_prop}, we notice that our ansatz 
in terms of $X^{\alpha \beta} (s)$ and $Y(s)$ works 
if the above functions $\hat {\mathsf f}_{\mu \nu} (s)$ and $\hat {\mathsf g}_{\mu \nu} (s)$ satisfy a relation 
\begin{eqnarray}
\frac{ d \,}{ds} \hat {\mathsf f}_{\mu \nu} (s) = \frac{1}{i} \hat {\mathsf g}_{\mu \nu} (s)
\ \ .
\end{eqnarray}
This leads to the same differential equations as those in Eqs.~(\ref{eq:eqX}) and (\ref{eq:eqY}) 
up to the replacement of the coupling constant $ q_f \to - v^a $. 
We already have the solutions in Eqs.~(\ref{eq:X}) and (\ref{eq:Y}). 
Therefore, by using the same $X^{\alpha \beta} (s)$ and $Y(s)$ as those for the fermion propagator, 
the resummed gluon propagator is obtained as shown in \eref{eq:prop-gluon}. 


\section{Perturbative propagators in weak external fields}

\label{sec:prop-weak}


Here, we examine a perturbative interaction between 
a charged particle and an external field 
and obtain propagators for a charged scalar particle 
and a fermion with perturbative insertion of external fields. 
We also confirm that those propagators agree with 
those obtained from a perturbative expansion of 
the resummed propagator shown in Sec.~\ref{sec:resum}.

\subsection{Scalar QED}

The scalar QED Lagrangian (\ref{eq:Lqed_s}) is expanded as 
\begin{eqnarray}
\Lag 
=
\phi^\ast (-\partial^2 - i q_f \partial^\mu A_\mu + 2i q_f A^\mu \partial_\mu + q_f^2 A^2 - m^2) \phi
\, ,
\end{eqnarray}
where the covariant derivative $D^\mu$ is defined in Eq.~(\ref{eq:covariantD-QED}). 
As in the resumed propagator, we use the Fock-Schwinger gauge (\ref{eq:FS}). 
Then, we have $  \partial_\mu A_\FS^\mu=0$ and 
$ A_\FS^\nu \partial_\nu = \frac{1}{4}  F^{\mu \nu} (x_\mu \partial_\nu - x_\nu \partial_\mu)$. 
The last factor in this expression is a generator of the Lorentz group. 
Since $\phi(x)$ is a scalar field, a Lorentz transform 
$F^{\mu \nu}\Lambda_{\mu\nu} \phi(x)$ by an ``angle'' $F^{\mu\nu}$ vanishes. 
Therefore, the three-point vertex $\phi^\ast  A^\mu \partial_\mu \phi$ does not play any role 
in the Fock-Schwinger gauge.
Removing those vanishing terms, we have 
\begin{eqnarray}
\Lag &=& \phi^\ast (-\partial^2 + q_f^2 A_\FS^2 - m^2) \phi
\, .
\end{eqnarray}
Now, an exact propagator is formally written as 
\begin{eqnarray}
G (p) &=& \frac{i}{p^2-m^2+q_f^2A_\FS^2}
\nn
\\
&=& G_0(p) \sum_{n=0}^\infty\left[ \, i q_f^2 A_\FS^2 G_0(p) \,\right]^n
\, ,
\end{eqnarray}
where the free propagator is given by $G_0(p) = i/(p^2-m^2 )$. 
We have defined the translation-invariant part by $ G_n (p_1,p_2)  = (2\pi) ^4 \delta(p_1-p_2) G_n (p_1)  $. 
This expansion provides explicit expressions of perturbative propagators on an order-by-order basis 
which we denote as $  G(p) =  \sum_{n=0}^\infty G_n(p)$. 
This series contains the free propagator $ G_0(p) $ as the first term. 
Note that the Schwinger phase, which breaks the translation symmetry, 
is not shown explicitly in the above exact propagator.

We shall see the perturbative propagator at $ n=1  $ which comes with the squared gauge field. 
Following the above expansion, we find that\footnote{
The absence of the three-point vertex can be explicitly seen as follows. 
Inserting an external field between two free propagators, we have 
$
A^\mu_\FS \partial_\mu G_0 (z-y) \propto 
 F^{\mu\nu} \int \!\! d^4k  e^{-i(z-y)k}  \frac{\partial \ }{\partial k^\mu} [ \, k_\nu G_0(k) \, ]
$. 
However, the derivative of the free propagator is symmetric in the Lorentz indices, 
and vanishes when contracted with the field strength tensor. 
This is true in all orders. 
}
\begin{eqnarray}
G_1 (p_1,p_2) &=& 
\integ dx  \integ dy   \integ dz \, e^{ i p_1 x - ip_2y } G_0(x-z) [ i q_f^2 A_{\rm FS}^2(z) ] G_0(z-y)
%
%
%
%
%
%
\nonumber
\\
&=& (2\pi) ^4 \delta(p_1-p_2) \times (-i) \frac{q_f^2}{2^2}  F^{\mu\alpha} F^\nu_{\ \, \alpha}  
G_0(p_1)  \frac{\partial^2 G_0(p_1)}{\partial p_1^\mu \partial p_1^\nu} 
\label{eq:G1}
\, .
\end{eqnarray}
It is important to take the origin of the Fock-Schwinger gauge field as $ x_0^\mu = y^\mu $. 
Otherwise, one would get an additional term given by a derivative of delta function. 
This choice is similar to the one we made in Sec.~\ref{sec:prop_half}, 
and, thus, the Schwinger phase has the same form as shown in Sec.~\ref{sec:FS}. 
The appearance of the delta function in Eq.~(\ref{eq:G1}) indicates that 
the other part has a manifest translational invariance as expected. 
Inserting the second derivative of the free propagator, 
we obtain the leading correction to the propagator 
\begin{eqnarray}
\label{eq:boson-1}
G_1 (p) =
 \frac{ 2 i q_f^2}{(p^2-m^2)^3} F_{\mu\alpha} F_\nu^{\ \alpha}  
\left( \,   \frac{ p^\mu p^\nu}{p^2-m^2} - \frac{1}{4} g^{\mu\nu} \, \right)
\, .
\end{eqnarray}

The same result is obtained from an expansion of the resummed propagator (\ref{eq:scalar}) with respect to $q_fF^{\mu\nu} $. 
In the leading order of the external field strength, we have 
\begin{subequations}
\begin{eqnarray}
&&
X^{\mu\nu}(s) \sim s - \frac{q_f^2}{3} F^{\mu\alpha} F_\alpha^{\ \, \nu} s^3
\, ,
\\
&&
Y(s) \sim -  \frac{q_f^2}{4} F_{\mu}^{\ \, \alpha} F_\alpha^{\ \, \mu} s^2
\, ,
\end{eqnarray}
\end{subequations}
Plugging these approximations into the resummed propagator (\ref{eq:scalar}), 
the proper-time integral is performed as 
\begin{eqnarray}
G(p|A_\FS) &\sim&
 \int _0^\infty \!\! ds \, e^{ i s (p^2-m^2 + i \epsilon)}
\left[ 1  + q_f^2 \left(  \frac{is^3}{3} p_\mu F^{\mu\alpha} F^\nu_{\ \, \alpha} \, p_\nu 
+  \frac{s^2}{4} F^{\mu \nu} F_{ \mu\nu}  \right)
 \right]
 \nn
 \\
&=& G_0(p) + G_1(p)
\, .
\end{eqnarray}
The leading term is the free propagator. 
The next term agrees with $G_1(p)  $ in Eq.~(\ref{eq:boson-1}). 
In this way, the higher-order propagators can be obtained straightforwardly 
with a systematic expansion of the resumed propagator. 

For example, when parallel electric and magnetic fields are applied, 
one can use Eq.~(\ref{eq:FF_mn}) to simplify the above propagator. 
Especially, when one of the field, e.g., an electric field, is absent ($ E=0 $), we have a simple form 
\begin{eqnarray}
G_1 (p) =  -  \frac{ i (q_fB)^2}{(p^2-m^2)^4} \left[ \,   (p_\parallel^2 - p_\perp^2) - m^2 \, \right]
 \, .
\end{eqnarray}

\subsection{Spinor QED}

The perturbative fermion propagators are obtained 
in the same manner as the boson propagators discussed just above. 
We show the first two terms in the perturbative expansion. 
The leading correction term is obtained by inserting one external field between free fermion propagators as  
\begin{eqnarray}
S_1(p_1,p_2) &=& (-i q_f) \integ dx \integ dy \, e^{ i p_1 x - i p_2y } S_0(x-z) \slashed A_\FS(z) S_0(z-y)
\nn
\\
&=& (2\pi)^4 \delta^4(p_2-p_1)  \left( - \frac{1}{4}  q_f F^{\mu\nu}   \right) 
\left( \, \frac{\partial \ }{\partial p_2^{\mu}} - \frac{\partial \ }{\partial p_1^{\mu}} \, \right) 
 S_0(p_1)  \gamma_\nu  S_0(p_2)  
 \, ,
\end{eqnarray}
where the free propagator is given by $ S_0 (p) = i/(\sla p + m) $. 
Carrying out the derivatives acting on the free propagator, we find that 
\begin{eqnarray}
\label{eq:fermion1}
S_1(p) = \frac{i}{4}  q_f  F_{\mu\nu}  
\frac{(\slashed p + m) \, \sigma^{\mu\nu} + \sigma^{\mu\nu} (\slashed p + m)}{ (p^2 - m ^2)^2 } 
\, .
\end{eqnarray}
Similarly, the second-order term is obtained as 
\begin{eqnarray}
\label{eq:fermion2}
S_2(p) = - \frac{i}{4} q_f^2 F_{\alpha\beta} F_{\mu\nu} 
\frac{ (\sla p +m ) ( \, f^{\alpha\beta\mu\nu} + f^{\alpha\mu\beta\nu} 
+f^{\alpha\mu\nu\beta} \,)(\sla p +m )   }{(p^2-m^2)^5}
\, ,
\end{eqnarray}
where $  f^{\alpha\beta\mu\nu} = \gam^\alpha(\sla p +m ) \gam^\beta(\sla p +m ) \gam^\mu(\sla p +m ) \gam^\nu$. 
As shown in the previous subsection for the boson propagator, 
these propagators can be obtained from an expansion of the resummed propagator (\ref{eq:electron}) 
with respect to $ q_f B $. 
It is worth mentioning that these propagators were used to compute the Wilson coefficients 
for the gluon operators in the operator product expansion~\cite{Novikov:1983gd, Nikolaev:1982rq, Reinders:1984sr} 
and in magnetic fields \cite{Machado:2013yaa, Cho:2014exa, Cho:2014loa, Gubler:2015qok}. 
In Sec.~\ref{sec:selfenergy-weak_B}, 
we will use those perturbative propagators 
to compute the photon/gluon self-energy in a weak magnetic field.


\section{QCD Lagrangian in external chromo-electromagnetic fields}

\subsection{Decomposition into external and fluctuation fields}

\label{sec:bgd}

We summarize the QCD Lagrangian in an external chromo-electromagnetic field. 
According to Eq.~(\ref{eq:aA}), the field strength tensor $\nonF^{a\mu\nu}$ is also decomposed into 
two parts, that is, the external and fluctuation fields. 
After a simple arrangement, the field strength tensor is decomposed as 
\begin{eqnarray}
\nonF^{a}_{\mu\nu}  \rightarrow
\nonF_{\mu\nu}^a  + D_\mu^{ac} a_\nu^c - D_\nu^{ac} a_\mu^c
+ g a_\mu^b a_\nu^c f^{abc} 
\,  ,
\end{eqnarray}
where the field strength tensor on the right-hand side 
is for the external field $\nonF_{\mu\nu}^a t^a = \frac{i}{g} [D_\mu,D_\nu]$ 
which follows from the covariant derivative defined in Eq.~(\ref{eq:Dext}). 
Then, the Yang-Mills Lagrangian is expanded into 
a kinetic term, three- and four-point vertices of the fluctuation field as 
\begin{eqnarray}
- \frac{1}{4} \nonF^{a\mu\nu} \nonF^a_{\mu\nu} \rightarrow \Lag_{\rm gluon}^{\rm kin} + \Lag_{\rm gluon}^{(3)} + \Lag_{\rm gluon}^{(4)}
\, \ ,
\end{eqnarray}
where 
\begin{eqnarray}
\Lag_{\rm gluon}^{\rm kin} &=& 
- \frac{1}{4} \left\{
( D_\mu^{ac} a_\nu^c - D_\nu^{ac} a_\mu^c )^2 + 2g \nonF^{a \mu\nu} a_\mu^b a_\nu^c f^{abc}
\right\}
\, ,
\\
\Lag_{\rm gluon}^{(3)} &=& - g ( D_\mu^{ad} a_\nu^d ) a^{b \mu} a^{c \nu} f^{abc}
\, ,
\\
\Lag_{\rm gluon}^{(4)} &=& - \frac{g^2}{4} ( a_\mu^b a_\nu^c f^{abc} ) ( a^{d \mu} a^{e \nu} f^{ade} )
\,  .
\end{eqnarray}
The three-point vertex $\Lag_{\rm gluon}^{(3)}$ contains the external field in the covariant derivative. 
Diagrammatically, this coupling is understood by replacing 
one of the legs in the four-point gluon vertex 
by the external field.

We shall fix the gauge of the fluctuation field $a^{a\mu}$. 
For a given external field, the gauge transformation of $a^{a\mu}$ by an angle $ \alpha^a$ reads 
\begin{eqnarray}
a^{a}_{\mu} \rightarrow 
a_\mu^a + \frac{1}{g} D_\mu \alpha^a + f^{abc} a_\mu^b \alpha^c
\,  .
\end{eqnarray}
Compared with the usual gauge transformation, 
the derivative in the second term has been replaced by the covariant derivative with the external field. 
Therefore, the derivative in the gauge-fixing term 
in terms of the Faddeev-Popov trick 
is also replaced by the covariant derivative as 
\begin{eqnarray}
\Lag_{\rm gauge}  = - \frac{1}{2 \xi_{g} } ( D^{ab \mu} a_\mu^b)^2
\, .
\end{eqnarray}
Integrating the kinetic term by parts, the gauge-fixed kinetic term is obtained as 
\begin{eqnarray}
\Lag_{\rm gluon}^{\rm kin} + \Lag_{\rm gauge} 
=
- \frac{1}{2} a_\mu^a \left\{
- (D^2)^{ac} g^{\mu\nu}  + ( 1-  \frac{1}{\xi_{g} } )  (  D^{ \mu} D^{ \nu} )^{ac}
- 2 g \nonF^{b\mu\nu} f^{abc} 
\right\} a_\nu^c
\label{eq:Lkg}
\, .
\end{eqnarray}
Also, associated with the above gauge-fixing procedure, there arises the ghost term 
\begin{eqnarray}
\Lag_{\rm ghost} = - \bar c^a (D^2)^{ac} c^c -  g \bar c^a ( D_\mu^{ab} a^{d\mu} f^{bdc} ) c^c 
\, ,
\end{eqnarray}
which also contains the external field in the form of the covariant derivative.

Including all the terms for quarks, gluons, and ghosts, 
the Lagrangian $\Lag$ in the external field is wrapped up as 
\begin{eqnarray}
\Lag  &=& \Lag_{\rm quark} + \Lag_{\rm gluon} + \Lag_{\rm ghost}
\label{eq:Lqcd}
\, ,
\end{eqnarray}
where 
\begin{subequations}
\begin{eqnarray}
\Lag_{\rm quark}  &=& \bar \psi (i \slashed D-m) \psi - ig \bar \psi \slashed a^a (t^a) \psi 
\, ,
\\
 \Lag_{\rm gluon} &=& \left( \Lag_{\rm gluon}^{\rm kin} + \Lag_{\rm gauge} \right) + \Lag_{\rm gluon}^{(3)} + \Lag_{\rm gluon}^{(4)}
 \, ,
\\
\Lag_{\rm ghost} &=&  - \bar c^a (D^2)^{ac} c^c  -  g \bar c^a ( D_\mu^{ab} a^{d\mu} f^{bdc} ) c^c 
\, .
\end{eqnarray}
\end{subequations}
The first term in each line provides the kinetic term, 
of which the inverse is the resummed propagator shown in Sec.~\ref{sec:QCD_prop}. 
The other terms provide the perturbative vertices that connect those resummed propagators, assembled as 
\begin{eqnarray}
\Lag_{\rm int}  = 
- ig \bar \psi \slashed a^a (t^a) \psi 
+ \Lag_{\rm gluon}^{(3)} + \Lag_{\rm gluon}^{(4)}
 -  g \bar c^a ( D_\mu^{ab} a^{d\mu} f^{bdc} ) c^c 
 \label{eq:QCD_int}
\, .
\end{eqnarray}


\subsection{Covariantly constant external fields}

\label{sec:cc}

As shown in \eref{eq:fact} for the covariantly constant field, 
the field strength tensor is decomposed into the color direction and the strength. 
Therefore, the covariant derivative is also decomposed as 
\begin{eqnarray}
D^\mu &=& \partial^\mu - i g ( n^a t^a) \nonA_\ext^\mu 
\label{eq:D-fact}
\, ,
\end{eqnarray}
where the Abelian-like field $  \nonA_\ext^\mu$ does not have the color index. 
The color matrix $\T \equiv n^a t^a$ in \eref{eq:D-fact} can be diagonalized 
because the $N^2 -1 $ generators of $  SU(N)$ are Hermitian matrices. 
Below, we explicitly find the diagonal forms in the fundamental and adjoint representations.

\subsubsection{Fundamental representation}

We notice that the diagonalized matrix should have $N-1$ independent components, 
because the generators are traceless. 
Since the rank of $  SU(N)$, namely the dimension of the Cartan subgroup, is $N-1$, 
we can decompose the matrix $\T_{\rm f}$ 
by using the simultaneously diagonalized $  SU(N)$ generators, 
where the subscript denotes the matrix $\T$ 
in the fundamental representation. 

We shall discuss the case of $ SU(3) $. 
In order to represent the diagonalized $\T_{\rm f}$, 
we can use the two diagonal matrices out of the eight Gell-Mann matrices. 
They are $t^3 = {\rm diag} (1/2,\ -1/2,\ 0) $ and $t^8 = {\rm diag}(1/(2\sqrt{3}),\ 1/(2\sqrt{3}),\ -1/\sqrt{3} )$. 
Thus, the diagonalized form will be  
\begin{eqnarray}
U \T_{\rm f} U^\dagger = \alpha_1 t^3 + \alpha_2 t^8
\label{eq:eq0}
\, ,
\end{eqnarray}
where $ U $ is the unitary matrix that diagonalizes $ \T_{\rm f} $. 
To get convenient forms of the coefficients $ \alpha_1 $ and $ \alpha_2 $, 
we square and cube the both sides of \eref{eq:eq0} 
and then carry out the trace of those quantities. 
Then, the left-hand side results in (one half of) 
the following quantities  
\begin{subequations}
\begin{eqnarray}
C_1 &=& n^a n^a = 1
\, ,
\\
C_2 &=& d^{abc} n^a n^b n^c
\, ,
\end{eqnarray}
\end{subequations}
where the first and second lines are for the square and cube, respectively. 
We used the normalization $n^a n^a =1$ 
and a completely symmetric constant $d^{abc} \equiv  \tr[ \{ t^a, t^b \} t^c ] $. 
The coefficients $  C_1$ and $C_2  $ are the Casimir invariants in $SU(3)$. 
Comparing the above expressions with the square and cube of the right-hand side, 
we obtain two relations 
\begin{subequations}
\begin{eqnarray}
C_1 &=& \alpha_1^2 + \alpha_2 ^2
\, ,
\\
C_2 &=& \frac{1}{2 \sqrt{3} } ( 3 \alpha_1^2 \alpha_2 - \alpha_2^3 ) 
\,  .
\end{eqnarray}
\end{subequations}
The first relation suggests to parametrize the coefficients as 
$\alpha_1 = \sqrt{C_1} \cos \theta $ and $\alpha_2 = \sqrt{C_1} \sin \theta $, respectively. 
Thus, the angle $\theta $ is specified by the gauge invariant quantities $C_1=1$ and $C_2$ as 
\begin{eqnarray}
\sin 3\theta = 2 \sqrt{3} C_2
\label{eq:angle-fnd-rep}
\, .
\end{eqnarray}
The three diagonal elements, $U \T_{\rm f} U^\dagger =  {\rm diag} ( \lambda^{1},\lambda^{2}, \lambda^{3} )$, 
are then specified as 
\begin{eqnarray}
\lambda^k 
= \frac{ 1 }{ \sqrt{3} }  \sin \left (  \frac{2}{3}k \pi - \theta\right)
\, ,\ \ \ \  (k = 1,2,3)
\, .
\end{eqnarray}
The diagram in Fig.~\ref{fig:weight} shows the magnitude of $ \lambda^k $ 
specified by the angle $ \theta $ in the color space. 
Multiplying the QCD coupling constant $  g$, we have Eq.~(\ref{eq:angle-fund-rep}). 
Following the properties of the generators, one gets useful identities: 
\begin{subequations}
\label{eq:w-ids}
\begin{eqnarray}
\sum_{k=1}^{N_{c}} \lambda^k &=& \sum_{a=1}^{N_{c}^{2}-1}   \tr[ t^a ] n^a = 0
\label{eq:w}
\, ,
\\
\sum_{k=1}^{N_{c}} (\lambda^k)^2 &=&\sum_{a=1}^{N_{c}^{2}-1} \sum_{b=1}^{N_{c}^{2}-1} \tr[ t^a t^b  ] n^a n^b = \frac{1}{2}
\label{eq:weq}
\, .
\end{eqnarray}
\end{subequations}

\begin{figure}[t]
     \begin{center}
              \includegraphics[width=0.5\hsize]{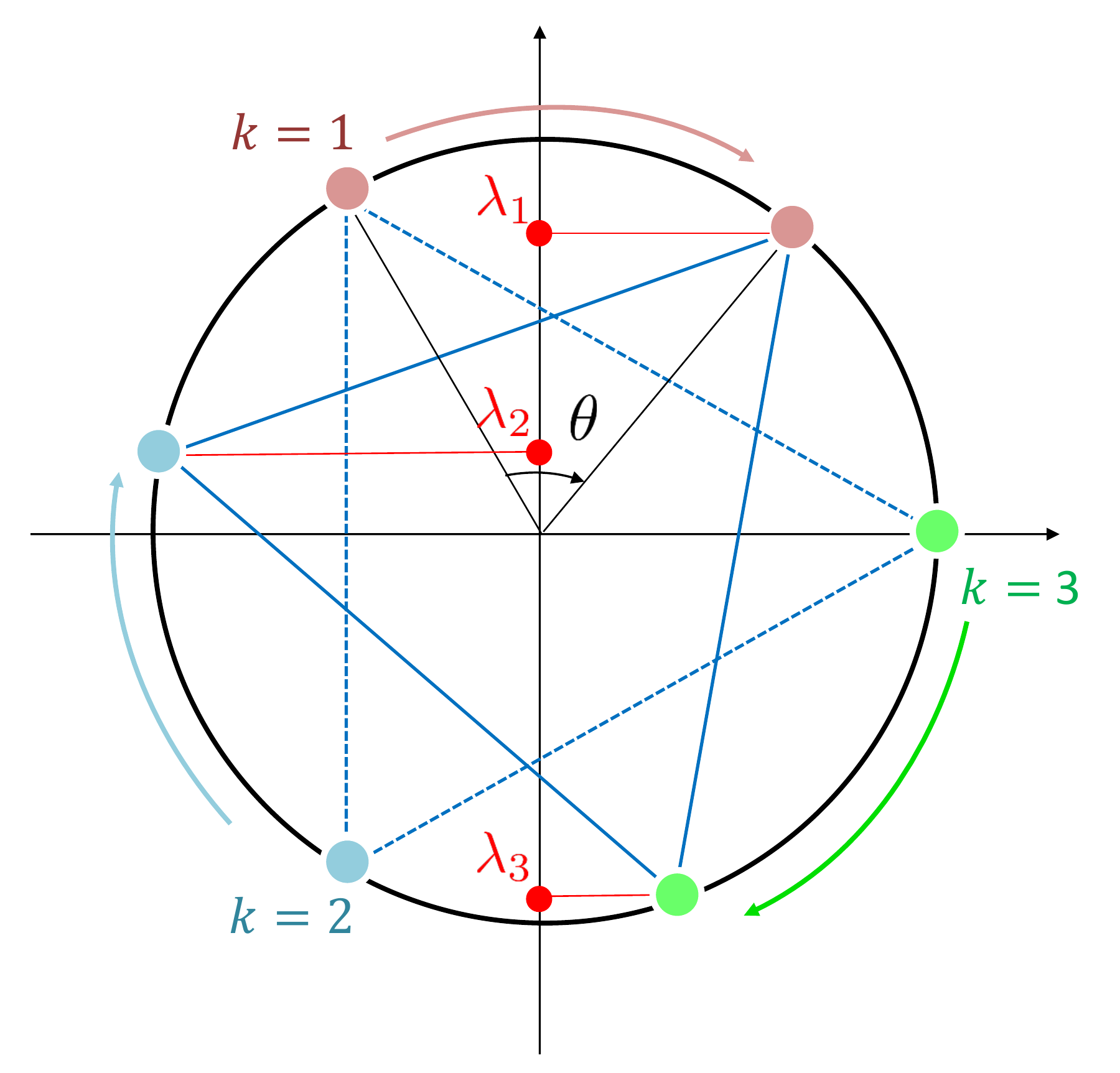}
     \end{center}
\vspace{-0.5cm}
\caption{
Diagrammatic representation of the eigenvalues $  \lambda^k$ 
that determine the effective color charges in the covariantly constant field. 
Note a difference from the definition of $ \theta $ in Ref.~\cite{Tanji:2010eu}. 
}
\label{fig:weight}
\end{figure}

\subsubsection{Adjoint representation}

\label{sec:adjoint}

Similar to the case of the fundamental representation, 
one can diagonalize the Hermitian matrix $ \T_{\rm ad}^{ac} = i f^{abc} n^b $ in the adjoint representation as well. 
In the case of $ SU(3) $, the number of independent degrees of freedom is again two. 
This can be explicitly seen as follows. 

We first count the number of independent eigenvalues. 
Since the eigenvalue $\lambda $ of the Hermitian matrix is real, 
the complex conjugate of the eigenvalue equation reads 
$ \T_{\rm ad}^{\ast ac}  u^{\ast c} = \lambda u^{\ast a} $ 
with $ u^a $ being the eigenvector corresponding to $\lambda $. 
The completely antisymmetric property of the structure constant, 
together with the Hermitian property of the generator, indicates that all components of $   f^{abc} $ are real, 
and thus $  \T_{\rm ad}^{\ast ac} = - \T_{\rm ad}^{ ac} $. 
This fact leads to $ \T_{\rm ad}^{ac}  u^{\ast c} = - \lambda u^{\ast a} $, meaning that 
the eigenvalue $ \lambda $ appears with a partner which has the opposite sign. 
The associated eigenvector is given by the complex conjugate $  u^{\ast a}$. 
Also, we find that $  n^c$ is a trivial eigenvector of $ \T_{\rm ad}^{ac} $ giving a zero eigenvalue. 
Therefore, apart from the pair of zero eigenvalues, 
the number of the independent eigenvalues in $ SU(3) $ is 
equal to $ (8-2)/2-1 =2$, 
where the minus one stems from a constraint $\T_{\rm ad}^{ac} \T_{\rm ad}^{ca} =  N \delta^{ac} n^a n^c =  N $ 
for the adjoint representation of the $ SU(N) $ group.

Based on this observation, 
the diagonalized form can be decomposed by two independent diagonal matrices as 
\begin{eqnarray}
U \T_{\rm ad} U^\dagger = \beta_1 T^3 + \beta_2 T^8
\label{eq:eq2}
\, ,
\end{eqnarray}
where we may choose $  T^3 = {\rm diag} ( t^3, 0, - t^3,0)$ and $  T^8 = {\rm diag} ( t^8, 0, - t^8,0)$ 
by using the diagonal Gell-Mann matrices and two vanishing components. 
Then, computing the square of \eref{eq:eq2}, we have 
\begin{eqnarray}
N C_1 = \beta_1^2 + \beta_2 ^2
\, ,
\end{eqnarray}
where $ C_1 = n^a n^a $ is normalized to be one. 
As in the fundamental representation, 
this relation suggests to put $\beta_1 = \sqrt{ N C_1} \cos \theta_{\rm ad} $ 
and $\beta_2 = \sqrt{  N C_1} \sin \theta_{\rm ad} $. 
Thus far, it is similar to what we have done for the fundamental representation. 
However, we find that the symmetric tensor $ d^{abc} $ identically vanishes in the adjoint representation, 
so that it does not make any sense to compute the cube. 
We, therefore, compute the sixth power of \eref{eq:eq2} to find 
\begin{eqnarray}
\label{eq:angle_adj}
\theta_{\rm ad} = \frac{1}{6} \cos^{-1} \left( \, 10 - \frac{ 144 C_2^\prime }{  C_1^3 } \right)
\, ,
\end{eqnarray}
where 
\begin{eqnarray}
C_2^\prime = \tr[ (t^a_{\rm ad} t^b_{\rm ad} t^c_{\rm ad} n^a n^b n^c)^2]
\, .
\end{eqnarray}
These equations establish the connection between the angle $ \theta_{\rm ad} $ in the adjoint representation 
and the gauge invariant quantities $ C_1 $ and $ C_2^\prime $. 
We are now able to specify the eight diagonal elements, $ \hat \T_{\rm ad} =  {\rm diag} ( \lambda^{1}_{\rm ad}, \lambda^{2}_{\rm ad}, \lambda^{3}_{\rm ad}, 0, - \lambda^{1}_{\rm ad}, - \lambda^{2}_{\rm ad}, - \lambda^{3}_{\rm ad}, 0 )$, 
in a gauge-invariant way by using the angle $\theta_{\rm ad}   $ as 
\begin{eqnarray}
\lambda^a _{\rm ad}=  \frac{ \sqrt{NC_1} }{2\sqrt{3}} \sin \left ( \frac{2}{3} a \pi -   \theta_{\rm ad}  \right)
\, ,\quad  (a = 1,2,3)
\label{eq:diagonal-adj}
\, .
\end{eqnarray}
Multiplying the QCD coupling constant $  g$, we have Eq.~(\ref{eq:angle-adj-rep}). 
The counterparts of identities (\ref{eq:w}) and (\ref{eq:weq}) are obtained as 
\begin{subequations}
\label{eq:v-ids}
\begin{eqnarray}
\sum_{a=1}^{N_{c}^{2}-1} \lambda^{a} _{\rm ad}&=& \sum_{a=1}^{N_{c}^{2}-1}  \tr[ t_{\rm ad}^a  ] n^a = 0
\label{eq:v}
\, ,
\\
\sum_{a=1}^{N_{c}^{2}-1} (\lambda^{a}_{\rm ad})^2 &=& \sum_{a=1}^{N_{c}^{2}-1} \sum_{b=1}^{N_{c}^{2}-1}  \tr[ t_{\rm ad}^a t_{\rm ad}^b  ] n^a n^b = N_c
\label{eq:veq}
\, .
\end{eqnarray}
\end{subequations}


\section{QCD effective action in a chromo-magnetic field}

We provide supplementary computational notes on the QCD effective action 
in a chromo- and Abelian magnetic field discussed in Sec.~\ref{sec:gluon-condensate}.

\subsection{Vacuum energy in the Landau levels}

\label{sec:energy-shift}

Here, we find an alternative form of the Landau-level representation (\ref{action_YM_pure_B-LLL}). 
Performing the transverse-momentum integral in Eq.~(\ref{eq:K-YM-LL}), 
there remain the longitudinal-momentum and proper-time integrals in the form 
\begin{eqnarray}
\label{eq:energy-shift-0}
 {\cal L}_{ \YM}^{(1)} = -i \frac{ | v^a  \nonB |}{2\pi} \sum_{n=0}^\infty   \frac12 (2 -\delta_{n0} - \delta_{n1} ) 
  \int \frac{d^2p_\para}{(2\pi)^2}    \int_0^\infty \frac{ds}{s}
e^{ - is \{ p_\para^2  -i \epsilon - (2n-1) \vert v^a \nonB\vert\} }   
\, .
\end{eqnarray}
The energy and proper-time integrals can be also performed as 
\begin{eqnarray}
&&
 \lim_{\delta \to 0} \mu^{2\delta}  \int \frac{d^2p_\para}{(2\pi)^2} \int_0^\infty \frac{ds}{s^{1-\delta}} 
 e^{ -i \{ p_\para^2 - (2n-1) \vert v^a \nonB\vert - i \epsilon\} s }
 \nnb
&& \quad \quad =
\frac{ (-i)^{\frac12}}{2 \sqrt{\pi}} \lim_{\delta \to 0} \mu^{2\delta}  \int \frac{dp_z}{2\pi} \int_0^\infty \frac{ds}{s^{\frac32-\delta}} 
e^{ i \{ p_z^2 +(2n-1) \vert v^a \nonB\vert + i \epsilon \} s }
\nn
\\
&& \quad \quad =
\frac{(-i)^{\frac12} }{2 \sqrt{\pi}}  \lim_{\delta \to 0} \mu^{2\delta}  \int \frac{dp_z}{2\pi} \Gamma( \delta - \frac12 )
 \big[ \, -i (p_z^2 +(2n-1) \vert v^a \nonB\vert + i \epsilon ) \, \big]^{\frac12-\delta} 
 \nnb
&& \quad \quad =
 i  \int \frac{dp_z}{2\pi}  \sqrt{ p_z^2 +   (2n-1) \vert v^a \nonB\vert + i \epsilon}
\, ,
\end{eqnarray}
where we regularized the UV divergences with a displacement $\delta  $ 
and $  \Gamma(z) = \int_0^\infty ds s^{z-1} e^{-s}$. 
Plugging this result back into Eq.~(\ref{eq:energy-shift-0}), 
we obtain the expression in Eq.~(\ref{eq:energy-shift}) 
that is given by the integration of the energy levels, 
i.e., the vacuum energy in the form of the Landau levels~\cite{Nielsen:1978rm}.

We shall focus on the region where $ p_z^2 +  (2n-1) \vert v^a \nonB\vert  <0 $. 
This only occurs in the IR regime where $ n=0 $ and $ p_- < p_z < p_+ $ with $ p_\pm = \pm \sqrt{|v^a \nonB|} $. 
Then, we can rotate the contour to the lower half plane to get 
\begin{eqnarray}
 \int \frac{d^2p_\para}{(2\pi)^2} 
\int_0^\infty \frac{ds}{s } e^{ - i ( p_\para^2 + |v^a \nonB| - i \varepsilon)s }
&=&
\frac{ (-i)^{\frac12}}{2 \sqrt{\pi}} 
 \int_{p_-}^{p_+}  \frac{dp_z}{2\pi} (-i)^{ - \frac12 } \int_0^\infty \frac{ds}{ s^{\frac32 }} e^{ - |p_z^2  -  |v^a \nonB|  | s }
\nnb
&=&
-  \int_{p_-}^{p_+}  \frac{dp_z}{2\pi}  \sqrt{  |v^a \nonB| - p_z^2  } 
\, .
\end{eqnarray}
The contribution from this IR regime gives rise to an imaginary part of the Yang-Mills effective Lagrangian 
\begin{eqnarray}
\label{eq:NO-imag3}
i \Im m[ {\cal L}_{ \YM}^{(1)} ]
= i \frac{ | v^a  \nonB |}{2\pi}   \frac12   \int_{p_-}^{p_+}  \frac{dp_z}{2\pi}   \sqrt{  |v^a \nonB| - p_z^2  } 
= i  \frac{ | v^a  \nonB |}{2\pi}  \cdot  \frac{ 1}{ 8} |v^a \nonB|
 \, .
\end{eqnarray}
A consistent result is also shown in Eq.~(\ref{eq:NO-imag2}).

\subsection{Yang-Mills part}

\label{sec:integral-YM}

We first identify the origins of the overall factors in the regularized integral (\ref{eq:I-reg}). 
To understand each factor in the regularized integral (\ref{eq:I-reg}), 
one needs to track it back to the Gaussian integral in Eq.~(\ref{eq:K-gluon}). 
In the $ d_\para  $ dimensions, the Gaussian integral is defined and performed as 
\begin{eqnarray}
\label{eq:Gaussian-gluon}
4 \int \frac{d^2p_\perp d^{d_\para} p_\para}{(2\pi)^{2+d_\para }} \exp\Big[
-  i \frac{p_\perp^2}{ |v^a \nonB|}\tan( |v^a \nonB| s) -  i p_\parallel^2 s
\Big]
&=&
\frac{ 4 }{(2\pi)^{2+d_\para} }
\cdot \frac{ |v^a \nonB| \pi}{   i \tan(|v^a \nonB|)} \cdot i \left( \frac{ \pi}{  i s  } \right)^{d_\para/2}
\nn
\\
&=& \left[\,  - \frac{i}{4\pi^2}  \frac{|v^a \nonB|}{\tan (|v^a \nonB| s)} \cdot \frac{1}{s}  \, \right] 
\left(  \frac{ -i}{ 4\pi s } \right)^{-\delta} 
\, ,
\end{eqnarray}
where $ \delta := (2  - d_\para)/2 $. 
The overall factor of $  \mu^{2 \delta}  $ is inserted to compensate
the mass dimension as before (cf. Sec.~\ref{sec:photons_HE-2}). 
The last factor on the right-hand side $[  -i / ( 4\pi s ) ]^{-\delta} $ is taken over to the regularized integral $ I_{\rm YM}^{(1)}   $.  
The first factor $ \vert v^{a}\mathcal{B} \vert^{-\delta} $ in Eq.~(\ref{eq:I-reg}) appears from 
the rescaling of the integral variable $ s \to s' = \vert v^{a}\mathcal{B} \vert s $.

The integrals in Eqs.~(\ref{eq:INT-NO}) and (\ref{eq:INT-res}) for 
the Nielsen-Olesen (NO) and residual modes are performed separately. 
The proper-time integral for the NO mode (\ref{eq:INT-NO}) is rotated to the upper-half plane, 
i.e., $ \int_0^\infty ds f(s) = i \int^{\infty}_0 ds f( + is ) $, 
since the integral diverges in the lower-half plane. 
Then, the integral for the NO mode is straightforwardly performed along the positive imaginary axis as
\begin{eqnarray}
I_{\rm NO}^{(1)} 
&=& 
\left( \frac{ 4\pi i \mu^2 }{ \vert v^{a}\mathcal{B} \vert } \right)^{\delta} 
\frac{i^2}{i^{2-\delta}} \int^{\infty}_{0} \frac{ds}{s^{2-\delta} }  e^{ - s } 
\nn
\\
&=&  
\left( \frac{ 4\pi  \mu^2 }{ \vert v^{a}\mathcal{B} \vert } \right)^{\delta} 
i^{2\delta}  \Gamma( -1 + \delta)
 \nn
\\
&=&   - \kappa \left( \frac{ \mu^2 }{ 2 \vert v^a \nonB \vert} \right) - \ln 2 - 1  - i \pi + \order (\delta^1)
\label{eq:I-NO}
\, ,
\end{eqnarray}
where we used an identity $ a^\delta = e^{ \delta \log a} $. 
This result has an imaginary part as foreseen in Eq.~(\ref{eq:NO-imag2}). 
We defined a UV-divergent quantity $ \kappa (x) := [  1/\delta- \gam_E + \ln (4\pi) ] + \ln x $ 
as in Eq.~(\ref{eq:kappa-UV}) with the Euler-Mascheroni constant $  \gam_E$.

On the other hand, the integral for the residual part $ I_{\rm res}^{(1)}  $ can be computed 
as usual by rotating the integral contour to the lower-half plane: 
\begin{eqnarray}
\label{eq:I_res2}
I_{\rm res}^{(1)} 
=
\left( \frac{ 4\pi i \mu^2 }{ \vert v^{a}\mathcal{B} \vert } \right)^{\delta} 
\frac{1}{ (-i)^{2-\delta} }  \int^{\infty}_{0} \frac{ds}{s^{2-\delta} } 
\left[ \,\frac{   \cosh(  2  s ) }{ \sinh s }  -  e^{ + s } \right]
\, .
\end{eqnarray}
Notice that the overall imaginary units cancel out in this expression in contrast to the case in the NO mode (\ref{eq:I-NO}) 
that gives rise to the imaginary part in the final result (\ref{eq:L-YM-div}). 
The regularized integral (\ref{eq:I_res2}) can be expressed with the Hurwitz zeta function introduced in Sec.~\ref{sec:photons_HE-2} 
via another formula.\footnote{
Starting from the integral representation of the Hurwitz zeta function (\ref{eq:Hurwitz-zeta}), 
one can easily show a relation 
\begin{eqnarray}
\zeta (w;  \frac{x+3}{2} )  =  
- \zeta (w;  \frac{x+1}{2} ) + \frac{2^w}{\Gamma(w)} 
\left[ \, 
- \int _0^\infty\frac{ ds}{ s^{1-w}} e^{- x s}   e^{   s } 
+\int _0^\infty\frac{ ds}{ s^{1-w}} e^{- x s}  \frac{ \cosh(2s)}{\sinh s}
 \, \right]
\nn
 \, ,
\end{eqnarray}
which, when $ x = 0 $ for massless gluons, reduces to 
\begin{eqnarray}
\label{eq:zeta-gluon}
 \int _0^\infty\frac{ ds}{ s^{1-w}} \left[ \,  \frac{ \cosh(2s)}{\sinh s} -   e^{   s } \,  \right]
= \Gamma(w) \left[ \,  2 ( 1-2^{-w}) \zeta_R (w)   -  1   \, \right]
 \, .
\end{eqnarray}
In the above, we used relations to the Riemann zeta function $ \zeta_R(w) $, i.e., 
$ \zeta( w, 1/2) = ( 2^w-1) \zeta_R(w)  $ and $  \zeta( w, 3/2) =  \zeta_R( w, 1/2) - 2^w$ (cf. Ref.~\cite{HurwitzZetaFunction}). 
An expansion of the Riemann zeta function reads 
\begin{eqnarray}
\zeta_R( -1 + \delta) = - \frac{1}{12} + \left( \frac{1}{12}  - \ln G \right) \delta + \order(\delta^2)
\nn
\, ,
\end{eqnarray}
with $ G $ being the Glaisher constant $ G = 1.2824 \dots $ \cite{RiemannZeta}. 
}
Applying the formula (\ref{eq:zeta-gluon}) to the proper-time integral~(\ref{eq:I_res2}), we get 
\begin{eqnarray}
I_{\rm res}^{(1)} 
&=& 
- \left( \frac{ 4\pi \mu^2 }{ \vert v^{a}\mathcal{B} \vert } \right)^{\delta} 
\times \Gamma(-1+\delta) \left[ \,  2 ( 1-2^{1-\delta}) \zeta_R(-1+\delta)   -  1   \, \right]
\nn
\\
&=&   - \frac{5}{6}   \kappa ( \frac{ \mu^2 }{ 2 \vert v^a \nonB \vert} )  - 1 + 2 \ln G +  \frac{11}{6}  \ln 2 + \order( \delta^1 )
\, ,
\end{eqnarray}
where $ \kappa(x)$ is defined above. 
Combining the contributions of the NO and residual modes, 
we obtain the complex-valued result in Eq.~(\ref{eq:L-YM-div}).

\subsection{Quark part}

\label{sec:integral-Q}

As in the Yang-Mills part, we first identify the origins of the overall factors in the regularized integral (\ref{eq:I-Q}). 
One should maintain an overall factor of $ (-1)^\delta $ which comes from the determinant $  \det(-1)$ 
associated with the arrangement below Eq.~(\ref{Lag_original}). 
Explicitly, one should note that 
$ \det \left[ (-1) (  \slashed {\mathbb D} \slashed {\mathbb D}  + m^2 \right) ]^{\frac12}
= (-1)^\delta  \det \left[  \slashed {\mathbb D} \slashed {\mathbb D}  +  m^2 \right]^{\frac12}$ in Eq.~(\ref{quark-part}). 
Also, similar to Eq.~(\ref{eq:Gaussian-gluon}), the Gaussian integral for fermions (\ref{eq:K-fermion}) yields 
\beq
\label{eq:Gaussian-quark}
4 \int \frac{d^2p_\perp d^{d} p_\para}{(2\pi)^{2+d }} \exp\Big(
i\frac{p_\perp^2}{\bbb_{i,f}}\tan(\bbb_{i,f} s) + i p_\parallel^2 s
\Big)
= \left[\,  - \frac{i}{4\pi^2}  \frac{\bbb_{i,f}}{\tan (\bbb_{i,f} s)} \cdot \frac{1}{s}  \, \right] 
\left(  \frac{  i }{ 4\pi s } \right)^{-\delta} 
\, ,
\eeq
where the sign of the imaginary unit is, however, opposite to that in Eq.~(\ref{eq:Gaussian-gluon}). 
Combining those factors, we arrive at the regularized integral (\ref{eq:I-Q}).

Remember that we have performed the proper-time integral for the fermionic effective action 
in Sec.~\ref{sec:photons_HE-2}, albeit it was for electrons in an Abelian magnetic field. 
In the same manner, one can rotate the integral contour to the lower-half plane and use the formula~(\ref{eq:zeta-fermion}). 
Then, the regularized integral (\ref{eq:I-Q}) reads 
\begin{eqnarray}
 I_{\rm quark}^{(1)} 
&\equiv& \frac{ (4\pi i \mu^{2})^\delta }{\bbb_{i,f}} 
\int^{\infty}_{0} \frac{ds}{s^{2-\delta} } \e^{- i ( m^2 - i\epsilon)s} \cot( \bbb_{i,f} s)
\nn
\\
&=&-   \left(  \frac{ 4\pi  \mu^{2} } {\bbb_{i,f}} \right)^{\delta} 
\int^{\infty}_{0} \frac{ds}{s^{2-\delta} } \e^{- 2 \bar \bbb_{i,f}^{-1}  s}  \coth s
\nn
\\
&=& -   \left(  \frac{ 4\pi  \mu^{2} } {\bbb_{i,f}} \right)^{\delta} 
\Gamma(-1+\delta) \left[ \, 2^{2- \delta}  \zeta ( - 1 +\delta , \bar \bbb_{i,f}^{-1} ) - (2 \bar \bbb_{i,f}^{-1} )^{1-\delta} \, \right]
\, ,
\end{eqnarray}
where $  \bar \bbb_{i,f} = 2  \bbb_{i,f} /m_f^2 $. 
The imaginary units are canceled out in the second line after the rotation 
of the integral contour, leaving a real-valued integral. 
The divergent pieces can be isolated in the expansion with respect to $ \delta $: 
\begin{eqnarray}
 I_{\rm quark}^{(1)} 
 \= - \big( \frac13 + 2  \bar \bbb_{i,f}^{-2} \big) \Big\{  \kappa \Big( \frac{ \mu^2 }{ m_f^2 } \Big) + 1 \Big\}
 +4  \zeta (-1, \bar \bbb_{i,f}^{-1}  ) \ln \bar \bbb_{i,f}^{-1}  + 4 \zeta'(-1, \bar \bbb_{i,f}^{-1}  )
\, ,
\end{eqnarray} 
where $ \zeta'(z,a) \equiv \partial  \zeta(z,a)/\partial z  $ 
and $ \zeta(-1,a) = - ( 2a^2 - 2a + 1/3) /4 $ as mentioned below Eq.~(\ref{eq:X0-result}). 
We arrive at the result in Eq.~(\ref{eq:L-quark-div}) with $ \kappa(x) $ defined below Eq.~(\ref{eq:I-NO}) 
in the previous subsection for the YM part.

\subsection{Scalar QED}

\label{sec:beta-scalar}

The Gaussian integral for scalar QED is the same as that for the quark part (\ref{eq:Gaussian-quark}). 
Therefore, the regularized integral is given as 
\begin{eqnarray}
{\cal L}_{\rm scalar }^{(1)} &=& -\frac{1}{16\pi^2}   I_{\rm scalar}^{(1)}  
\, ,
\nnb
 I_{\rm scalar}^{(1)} 
&\equiv& \frac{ (4\pi i \mu^{2})^\delta } { q_fB }
\int^{\infty}_{0} \frac{ds}{s^{2-\delta} } \e^{- i ( m^2 - i\epsilon)s} \frac{1}{ \sin(  q_fB  s) }
\, ,
\end{eqnarray}
The above integral can be identified with an integral formula that can be derived from 
the integral representation of the Hurwitz zeta function (\ref{eq:Hurwitz-zeta}):
\begin{eqnarray}
\zeta (w;  \frac{x+1}{2} ) 
&=&   \frac{2^{w-1}}{\Gamma(w)} \int _0^\infty\frac{ ds}{ s^{1-w}} e^{- x s}  \frac{1}{\sinh s}
\label{eq:zeta-scalar}
 \, .
\end{eqnarray}
Applying this formula, we have 
\begin{eqnarray}
 I_{\rm scalar}^{(1)} 
&=&-   \left(  \frac{ 4\pi  \mu^{2} } { q_fB } \right)^{\delta} 
\int^{\infty}_{0} \frac{ds}{s^{2-\delta} } \e^{ - \frac{m^2}{q_f B} s}  \frac{1}{ \sinh s }
\nn
\\
&=& -   \left(  \frac{ 4\pi  \mu^{2} } { q_fB }  \right)^{\delta} 
\Gamma(-1+\delta)  2^{2- \delta}  \zeta ( - 1 +\delta , \frac{m^2}{2 q_f B} + \frac12 )
\nn
\\
\= 4 \zeta( -1  , \frac{m^2}{2 q_f B} + \frac12 ) \left[ \, \kappa\big( \frac{\mu^2}{q_f B} \big) + 1 - \ln 2   \, \right]
+ 4 \zeta^\prime ( -1  , \frac{m^2}{2 q_f B} + \frac12 )
\, ,
\end{eqnarray}
where $ \zeta( -1, \frac{\zeta+1}{2} ) = ( 1 - 3 \zeta^2 )/24 $. 
Therefore, the one-loop correction in scalar QED also contains a logarithmic divergence with a quadratic dependence on $ B $. 
The numerical coefficient in front of $ \kappa $ is found to be $   - 1 / (16\pi^2)  \times (4/24) = - 1/(2 \cdot 8\pi^2) (1/6)$, 
which results in the beta function (\ref{eq:beta-scalar}). 
Note that a factor of 1/2 is canceled in the renormalization-group equation 
since the logarithmic dependence on the renormalization scale 
has a quadratic form in the above expressions.



\section{Gluon/photon self-energy in magnetic fields}

\label{sec:screening}

Photons/gluons are not directly coupled to an external $ U_{\rm em}(1) $ field 
since they do not carry electric charges. 
Nevertheless, their self-energies are affected by the external fields through fermion loops. 
Such effects give rise to various consequences. 
For example, photons propagating in the external fields 
exhibit the vacuum birefiringence (see Sec.~\ref{sec:photons}). 
Also, we will see below that they acquire a gauge-invariant mass in a strong magnetic field 
due to the effective dimensional reduction to the (1+1) dimensions 
as an analog of the Schwinger mass \cite{Schwinger:1962tn, Schwinger:1962tp} that is also related to chiral anomaly. 
Besides, it is widely known that the medium-induced self-energies 
are important to obtain the collective excitations 
and the screening effects (see, e.g., Refs.~\cite{Bellac:2011kqa,Blaizot:2001nr} and references therein 
in the absence of external fields). 
The screening effect is a necessary ingredient 
when computing transport coefficients and production rate of the electromagnetic probes. 
In the last part of this appendix, 
we discuss how the external magnetic fields give rise to novel properties of the photons/gluons via interplay 
with the medium effects. 
Below, we first summarize various properties of 
the gluon/photon propagator 
in the presence of both the external field and the medium. 
While we focus on the perturbative computations, it is worth mentioning 
that the screening effect has been also investigated by the lattice QCD simulations 
as a nonperturbative method \cite{Bonati:2014ksa, Bonati:2016kxj, Bonati:2017uvz}.

\subsection{General form of gluon/photon propagator}

We shall consider the tensor structure of the photon/gluon propagator 
that can be constrained by the structure of the photon/gluon self-energy. 
In Eq.~(\ref{eq:Pex}), we found the tensor structures that satisfy the Ward identity in QED. 
There are two new tensor structures in addition to the one in the ordinary vacuum 
because the external field breaks a part of the spatial rotational symmetries. 
The gluon/photon propagator with the resummation of the self-energies 
was discussed in Ref.~\cite{Hattori:2012je} at zero temperature and density. 
There is an additional tensor structure when there is both constant electric 
and magnetic fields (see, e.g., Ref.~\cite{Dittrich:2000zu}).

More recently, this result was generalized to the gluon/photon propagator 
in the presence of both a magnetic field and medium~\cite{Hattori:2017xoo}. 
In such a case, one needs to take into account splitting of the longitudinal and transverse modes 
with respect to the gluon/photon momentum because the Lorentz symmetry is broken in the medium. 
Correspondingly, the transverse projection tensor, $g^{\mu\nu}-q^\mu q^\nu/q^2$, 
for the Lorentz-symmetric system is split into two structures. 
Then, the tensor structure of the self-energy can be written as 
\begin{eqnarray}
\label{eq:selfenergy}
\Pi^{\mu \nu}_R(q)
= \sum_{i=T, L, \parallel, \perp}\Pi_i(q) P^{\mu\nu}_i(q)  
\, ,
\end{eqnarray}
where the projection operators are defined by 
\begin{eqnarray}
P^{\mu\nu}_T(q) =    g^{\mu i} g^{\nu j } \Big (\, \delta_{ij}-\frac{q_i q_j }{\bq^2} \, \Big)
\, , \ \ \ 
P^{\mu\nu}_L(q) =  \big( g^{\mu\nu} - \frac{q^\mu q^\nu}{q^2} \big)  - P^{\mu\nu}_T(q) 
\, ,
\end{eqnarray}
for the splitting of the longitudinal and transverse components in the medium rest frame, 
and 
\begin{eqnarray} 
P^{\mu\nu}_\parallel(q) = g^{\mu\nu}_\parallel-\frac{q^{\mu}_\parallel q^{\nu}_\parallel }{q^2_\parallel} 
\, , \ \ \ 
P^{\mu\nu}_\perp(q) =   g^{\mu\nu}_\perp-\frac{q^{\mu}_\perp q^{\nu}_\perp }{q^2_\perp} 
 \, ,
\end{eqnarray}
for the splitting of the parallel and perpendicular components 
with respect to the external magnetic field (applied in the $  z$ direction). 
These four transverse structures have been known to 
appear in the perturbative computation of the self-energies 
in the magnetic field and in the medium with the hard thermal loop approximation. 
Whereas they exhaust the transverse tensor structures in QED, 
there could appear non-transverse structures from general consideration in QCD 
where the color current is not necessarily conserved within the fermion sector 
due to color charges carried by gluons.  
Below, we, however, do not take into account those potential structures since 
nonzero components have not been found thus far in perturbative computation in magnetic fields. 
It is still an open problem to investigate those potential components. 
The reader is referred to Ref.~\cite{Hattori:2017xoo} for more discussions,  
where useful properties of the projection tensors are also summarized.

Including the self-energy (\ref{eq:selfenergy}), 
the retarded gluon/photon propagator $D_R^{\mu\nu}(q)$ is given by 
\begin{align}
\label{eq:D-D0Pi}
i D^R_{\mu\nu}(q) &= [( i  D_0(q))^{-1}+\Pi^R(q)]^{-1}_{\mu\nu}
\, ,
\end{align}
where we suppressed the color indices for the notational simplicity and the inverse matrix in the Minkowski space is defined as 
$D^{\mu\nu} D^{-1}_{\nu\alpha}=\delta^\mu_\alpha$. 
A relatively simple form of the resummed propagator is obtained in the covariant gauge, 
so that we use a bare propagator 
\begin{align}
\label{eq:D-free}
i  D_0^{\mu\nu}(q) = \frac{1}{q^2} \big( g^{\mu\nu} - \frac{q^\mu q^\nu}{q^2} \big)
+ \xi_g \frac{q^\mu q^\nu}{(q^2)^2}
\, ,
\end{align}
where $\xi_g$ is the gauge-fixing parameter. 
The photon/gluon energy $q^0$ contains an infinitesimal imaginary part ($q^0+i\epsilon$) for the retarded function. 
Plugging these ingredients, we find the resummed propagator 
\begin{eqnarray}
\label{eq:Dmunu-general}
i  D_R^{\mu\nu}(q)
&=&  \frac{1}{\Delta}
\left[(q^2-\Pi_\parallel-\Pi_L)P_T^{\mu\nu}(q)  +(q^2-\Pi_\parallel-\Pi_T)P_L^{\mu\nu}(q) 
+\Pi_\parallel P_\parallel^{\mu\nu}(q)  +D_\perp(q)P_\perp^{\mu\nu}(q)
\right]
\nn
\\
&&
+\xi_g \frac{q^\mu q^\nu}{(q^2)^2}
\, , 
\end{eqnarray} 
where two functions are defined by 
\begin{eqnarray}
\label{eq:Delta-general}
\Delta &\equiv& 
(q^2-\Pi_T)(q^2-\Pi_L) 
-\Pi_\parallel\left[ q^2-\Pi_T a \frac{q^2}{q^2_\parallel}-\Pi_L(1-a)\frac{(q^0)^2}{q^2_\parallel }
\right]
\, , 
\\
D_\perp(q) &\equiv& 
\frac{1}{q^2-\Pi_T-\Pi_\perp} 
\Bigl[\Pi_\parallel(\Pi_L-\Pi_T)(1-a)\frac{(q^0)^2}{q^2_\parallel} 
+\Pi_\perp(q^2-\Pi_L-\Pi_\parallel)\Bigr] 
\label{eq:Dperp-general}
\, , 
\end{eqnarray}
with $a\equiv (q_z)^2/|\bq|^2$. 
We assumed that the self-energy corrections (\ref{eq:selfenergy}) 
are diagonal in the color space. 
Once the above propagator is obtained, one can, for example, 
extract dispersion relations of the gluons/photons from pole positions 
and also use it as a building block in diagram computations. 
More information including the Coulomb-gauge propagator, 
physical meaning of each mode, collective excitations, 
and screening effects are available in Ref.~\cite{Hattori:2017xoo, Hattori:2022uzp, Hattori:2022wao}.

\subsection{Anomalous dynamics in the lowest Landau level}

One can estimate the order of each self-energy component associated 
with a magnetic field and a medium as 
$ \Pi_\para \sim \alpha_s eB  $ and  $ \Pi_{L,T} \sim \alpha_s T^2  $, respectively.\footnote{
Here, we have $  \alpha_s $ for the coupling between a dynamical gluon and the one-loop polarization, 
which should be replaced by $  \alpha_\EM $ for a photon. 
Note also that we do not consider an external chromo fields discussed in Sections~\ref{sec:QCD_prop} and \ref{sec:HE_QCD} 
which would induce an additional contribution of 
a gluon loop to $ \Pi_\para $. 
} 
The physical scales of the system are the magnetic field strength $ eB $, 
temperature $ T $, and/or chemical potential $ \mu $. 
The former dependence originates from the Landau degeneracy factor. 
When a magnetic field is so strong that $ eB \gg T^2 $, 
the magnetized fermion-loop contribution $ \Pi_\para \sim \alpha_s eB  $ is much larger than 
the thermal-loop contributions $ \Pi_{L,T} \sim \alpha_s T^2  $. 
Also, we have a vanishing component $ \Pi_\perp =0 $ in the strong-field limit as discussed in Sec.~\ref{sec:photons} 
because the LLL fermions do not have fluctuations perpendicular to the magnetic field. 
Therefore, considering the momentum regime $ p^2 \sim \Pi_\para \sim eB $ 
where the thermal contribution is negligible (see Ref.~\cite{Hattori:2017xoo} for 
discussions in other regimes), we have a simple expression of the photon/gluon propagator 
\begin{eqnarray}
\label{eq:Photon-prop-LLL}
i  D_R^{\mu\nu}(q)
=   \frac{1}{q^2} \left[ \, P_0^{\mu\nu} + \frac{ \Pi_\para }{ q^2 - \Pi_\para} P_\para^{\mu\nu} 
+   \xi_g \frac{q^\mu q^\nu}{q^2} \, \right]
\, .
\end{eqnarray} 
Especially, when the on-shell LLL current $ j^\mu_\LLL $ is coupled to 
the photon/gluon propagator, we have 
\begin{eqnarray}
j_{\LLL\, \mu}  i  D_R^{\mu\nu}(q) =   \frac{ j_\LLL^\nu }{ q^2 - \Pi_\para} 
\, ,
\end{eqnarray}
where we used the Ward identity $ q_\mu j_{\LLL}^{ \mu}   =0$ 
and the fact that the LLL current $  j_{\LLL}^{ \mu}  $ is only nonvanishing in the $ 0,3 $ components. 
Those expressions motivate us to elaborate the computation of the self-energy, 
focusing on the relevant component in the strong magnetic field 
\begin{eqnarray}
\label{eq:Pi_LLL}
\Pi_R^{\mu\nu} = \Pi_\para P_\para^{\mu\nu}  
\, .
\end{eqnarray} 
The screening effect arising from this component has been discussed 
in various contexts \cite{Gusynin:1998zq, Gusynin:1999pq, 
Fukushima:2011nu, Ozaki:2015sya, Fukushima:2015wck, Hattori:2016emy, 
Bandyopadhyay:2016fyd, Li:2016bbh, 
Hattori:2016lqx, Hattori:2016cnt, Hattori:2017xoo, Hattori:2017qio}. 
Note that only one of the photon/gluon modes is screened; 
This mode has the electric field oscillating along the strong magnetic field \cite{Hattori:2012je, Hattori:2017qio}. 
The other modes are neither screened nor play any role in the dynamics of the LLL fermions, 
simply because those modes are not coupled to the LLL fermions 
confined in the (1+1) dimensions, like electrons confined in polarizers. 
We first summarize rather technical aspects in computation of 
the one-loop self-energy in the strong magnetic field 
and then discuss physical consequences of the self-energy correction. 


\subsubsection{Vacuum contribution: Massless and massive cases}

\label{sec:VP_vac}

We first briefly summarize computation of the vacuum contribution. 
By using the LLL propagator (\ref{eq:prop-LLL}), the self-energy in the strong magnetic field is written as 
\begin{eqnarray}
\label{eq:selfenergy-gluon}
i \Pi_R^{\mu\nu} (q) =  - (ig)^2 
\int \!\! \frac{d^4p}{(2\pi)^4} 
\tr [ \gam^\mu S_\LLL(p) \gam^\nu S_\LLL(p+q) ]
\, .
\end{eqnarray}
The color factor $  \tr[t^a t^b] = 1/2\delta^{ab} $ 
should be attached for gluons, which we suppress 
for the notational simplicity. 
For notational simplicity, we do not explicitly write the flavor sum or the color sum. 
The latter simply yields an overall factor of $ N_c $ in the quark-loop contributions to the photon self-energy 
(with the replacement of the coupling constant, $ g \to q_f $) and is absent for a lepton loop. 
Also, one should take care of the difference in the electric charges $ q_f $ as well as the mass 
when carrying out the flavor sum. 
Remember that the gauge-dependent Schwinger phase goes away 
in this self-energy diagram [see discussions below Eq.~(\ref{eq:Phi_line})].

Inserting the explicit form of the LLL propagator (\ref{eq:prop-LLL}) which is 
factorized into the parallel and perpendicular parts, 
we find that the one-loop self-energy (\ref{eq:selfenergy-gluon}) is also factorized as 
\begin{eqnarray}
\label{eq:VV}
i \Pi^{\mu\nu}_{R} (q) = 
I_\perp (q_\perp^2) i \Pi_{1+1}^{\mu\nu} (q_\parallel)
\, ,
\end{eqnarray}
where
\begin{subequations}
\begin{eqnarray}
I_\perp (q_\perp^2) &=& 2^2 \int \!\! \frac{d^2p_\perp}{(2\pi)^2} 
e^{ - \frac{ 1 }{ |q_f B| } (|\bp_\perp|^2 + |\bp_\perp + \bq_\perp |^2 ) } 
\, ,
\\
i \Pi_{1+1}^{\mu\nu} (q_\parallel) &=& - (ig)^2 \int \!\! \frac{d^2p_\parallel}{(2\pi)^2} 
\frac{\tr [\,  \gam^\mu_\parallel  i (\sla p_\parallel + m) \prj_+
\gam_\parallel^\nu  i ( \, (\sla p_\parallel + \sla q_\parallel) + m\, ) \prj_+ \, ]}
{ (p_\parallel^2 -m^2)( \,  (p_\parallel + q_\parallel)^2 -m^2 \,)}
\, .
\end{eqnarray}
\end{subequations}
Here, the spinor trace is taken at the four dimensions. 
Performing the elementary integral for the transverse momentum, 
we find that 
\begin{eqnarray}
I_\perp (q_\perp^2) = \rho_B  e^{ - \frac{ |\bq_\perp|^2 }{2 |q_f B| } }
\, .
\label{eq:I_Gaussian}
\end{eqnarray}
The transverse integral results in the Landau degeneracy factor $ \rho_B =  |q_f B | /( 2\pi) $.

On the other hand, the longitudinal part is the polarization tensor in the (1+1)-dimensional QED, 
i.e., the Schwinger model \cite{Schwinger:1962tp}. 
After the standard treatment with the help of the Feynman parameter, 
the longitudinal integral is cast into the form 
\begin{eqnarray}
\label{eq:int_para0}
i \Pi_{1+1}^{\mu\nu} (q_\parallel) &=&  
- g^2 \frac{4}{2} \int_0^1 \!\! dx \left[ \, 
\int \!\! \frac{d^2\ell_\parallel}{(2\pi)^2} 
\frac{ 2\ell_\parallel^\mu \ell_\parallel^\nu - \ell_\parallel^2 g_\para^{\mu\nu}}{(\ell_\parallel^2 - \Delta_\para)^2}
+ 
\alpha^{\mu\nu}
\int \!\! \frac{d^2\ell_\parallel}{(2\pi)^2} 
\frac{ 1 }{(\ell_\parallel^2 - \Delta_\para)^2}
\, \right]
\, , 
\end{eqnarray}
where $ \ell_\para^\mu = p_\para^\mu + x(1-x) q_\para^\mu $, 
$\Delta _\para = m^2 - x(1-x)q_\para^2  $, 
and $ \alpha^{\mu\nu} =
m^2 g_\parallel^{\mu\nu} + x(1-x) ( q_\parallel^2 g_\parallel^{\mu\nu} - 2 q_\parallel^\mu q_\parallel^\nu)$. 
The factor of 4 from the four-dimensional trace is multiplied by 
the factor of $ 1/2$ from the spin projection operator, 
resulting in the factor of 2 just like the two-dimensional trace. 
Also, one needs to be careful about the gauge-invariant regularization. 
We use the dimensional regularization, and the numerator of the first integral is proportional to 
$ \epsilon_2 g_\parallel^{\mu\nu}  $ with $  \epsilon_2 \equiv  (2-d)$, 
which appears to vanish in the two dimensions. 
However, a factor of $  1/\epsilon_2$ arises from the integral that 
provides a finite contribution and is necessary for the gauge-invariant result: 
\begin{eqnarray}
\Pi_{1+1}^{\mu\nu} (q_\parallel) =  -  \frac{ g^2}{\pi}  
( q_\parallel^2 g_\parallel^{\mu\nu} -  q_\parallel^\mu q_\parallel^\nu)
\int_0^1 \!\! dx x(1-x) \Delta_\para^{-1} 
\label{eq:int_para}
\, .
\end{eqnarray}
The $ x $ integral can only be a function of $ q_\para^2/m^2 $ (up to an overall factor of $1 /q_\para^2 $) 
as a consequence of the Landau degeneracy 
and the gauge-invariance. 
Especially, without the mass scale $( m =0)$, the integral can only be a pure number. 
Below, we consider the massless and massive cases in order.

The integral (\ref{eq:int_para}) is quite simple in the massless case, 
i.e., $ \int_0^1 \!\! dx x(1-x) \Delta_\para^{-1} = - 1/q_\para^2 $. 
Plugging the expressions in Eqs.~(\ref{eq:I_Gaussian}) and (\ref{eq:int_para}), we have 
\begin{eqnarray}
\Pi_R^{\mu\nu} (q) = 
m_B^2 e^{ - \frac{ |\bq_\perp|^2 }{2 |q_f B| } }  
 \frac{ q_\parallel^2 g_\parallel^{\mu\nu} -  q_\parallel^\mu q_\parallel^\nu}{q_\para^2 + i \omega \epsilon}
\label{eq:massless-vac}
\, ,
\end{eqnarray}
where $ \omega \equiv q^0 $ and we put 
$ \omega \to \omega + i \epsilon $ for the retarded correlator. 
We defined 
\begin{eqnarray}
 m_B^2 :=  \rho_B \cdot \frac{ g^2 }{ \pi } 
\label{eq:Schwinger-mass}
 \, ,
\end{eqnarray} 
where a factor of 1/2 from the color factor 
should be attached for gluons. 
Then, the self-energy correction to the propagator (\ref{eq:Photon-prop-LLL}) is found to be 
\begin{eqnarray}
\label{eq:Pi-massless}
\Pi_\para 
= 
m_B^2 e^{ - \frac{ |\bq_\perp|^2 }{2 |q_f B| } } 
\, .
\end{eqnarray} 
This function is independent of $ \omega $ and $ q_z $. 
Therefore, 
the photon propagator (\ref{eq:Photon-prop-LLL}) acquires 
a pole mass 
in a gauge-invariant manner 
as a consequence of the $ 1/q_\para^2 $ pole in Eq.~(\ref{eq:massless-vac}) 
that is required on the dimensional ground. 
This mechanism is an analog with that in 
the Schwinger model \cite{Schwinger:1962tn, Schwinger:1962tp}. 
In the vanishing momentum limit $(|\bq_\perp| \to 0)  $, 
the gauge-invariant mass is given by $m_B $ in Eq.~(\ref{eq:Schwinger-mass}) that is 
the Schwinger mass $ g^2/\pi $ multiplied by the Landau degeneracy factor $ \rho_B$.

The imaginary part in Eq.~(\ref{eq:massless-vac}) is proportional to 
$ ( q_\para^2 g_\para^{\mu\nu} - q^\mu_\para q^\nu_\para) \delta(q_\para^2) $, 
that vanishes when contracted with the photon/gluon polarization vectors 
and summed over the polarization modes. 
This implies that the pair production in the LLL is strictly prohibited in the massless limit 
since the imaginary part provides the squared tree amplitude of the pair production 
from a single photon/gluon. 
The physical reason behind this prohibition is the absence of chirality mixing. 
Namely, fermions and antifermions, which would satisfy 
the momentum conservation, belong to 
different chirality eigenstates (see Sec.~\ref{sec:massless}), 
and are not directly coupled with each other. 
A similar prohibition mechanism is known as 
the ``helicity suppression'' in the leptonic decay of 
charged pions, where the muon channel dominates over the electron channel due to a mass suppression factor in the decay rate 
\cite{Donoghue:1992dd, Zyla:2020zbs}.


In terms of kinematics, it is rather natural that 
a vanishing energy-momentum transfer $ (q_\para^2  =0 )$ does not create a pair. 
A vanishing energy-momentum transfer implies an adiabatic process, 
and may not create a particle and an antiparticle 
that usually belong to different energy bands even in the massless limit. 
There is, however, an exception in the (1+1) dimensions, where the particle and antiparticle 
branches are directly connected with each other. 
This is the spectral flow discussed in Sec.~\ref{sec:massless}, 
where particles and antiparticles can be created 
in adiabatic shifts along single energy bands. 
Nevertheless, the net charge creation vanishes as seen above. 
Instead, the adiabatic shift creates a nonzero axial charge 
and breaks the chiral symmetry as we further discuss below. 


In the massive case, we need to perform the remaining integral in Eq.~(\ref{eq:int_para}). 
This can be done analytically as\footnote{
The correspondence between the notations in this appendix 
and in Sec.~\ref{sec:vp} is summarized as 
$P_\parallel^{\mu\nu} \leftrightarrow   P_2^{\mu\nu} / q_\parallel^2 $, 
$\Pi_\para \leftrightarrow - q_\parallel^2 \chi_1^\LLL$, 
and $I(q_\parallel^2) \leftrightarrow I^0_{0\Delta}(q_\parallel^2)/2$ 
[cf. Eqs.~(\ref{eq:Pex}) and (\ref{eq:selfenergy})]. 
} 
\begin{eqnarray}
\Pi_\para =  m_B^2 \, e^{-\frac{\vert \bq_\perp \vert^2}{2 \vert q_fB\vert}}
 \bigg[ \, 1- I \big( \frac{q_\parallel^2}{4m^2} \big) \,  \bigg] 
\label{eq:vac_massive}
\, .
\end{eqnarray}
Because of the mass scale, the expressions between the square brackets can be 
a function of a dimensionless combination $ q_\parallel^2/(4m^2)  $. 
The explicit form of the integral is obtained as 
\begin{eqnarray}
I(x + i  \omega  \epsilon) = 
\left\{
\begin{array}{ll}
\frac{1}{2} \frac{1}{ \sqrt{x(x-1) } }
\ln  \frac{ \sqrt{x(x-1) } - x }{  \sqrt{x(x-1) } + x } 
& x < 0 
\\
\frac{1}{ \sqrt{x(1 - x) } }\arctan \frac{x } { \sqrt{x(1-x) } } 
& 0 \leq x < 1
\\
\frac{1}{2} \frac{1}{ \sqrt{x(x-1) } }
\left[ \,\ln  \frac{x -  \sqrt{x(x-1) } }{x +  \sqrt{x(x-1) } } 
+ i \, \sgn(\omega) \pi \, \right]
& 1 \leq x
\end{array}
\right.
\label{eq:I}
\, .
\end{eqnarray} 
This function is plotted in Fig.~\ref{fig:I}. 
We took the retarded prescription $ \omega \to \omega + i\epsilon $ that 
yields an infinitesimal displacement $ q_\para^2 \to q_\para^2 + i \omega \epsilon $.\footnote{
We used a relation $ \arctan \frac{x } { \sqrt{x(1-x) } }  
= \frac{i}{2} \ln \frac{\sqrt{x(x-1) }  -  x}{\sqrt{x(x-1) }  +  x} $ when $ x < 0 $ or $ 1 \geq x  $. 
When $ x\geq 1 $, the argument of the logarithm takes a negative value, 
so that the logarithm becomes a complex-valued function, $ \arctan \frac{x } { \sqrt{x(1-x) } }  
= \frac{i}{2} [ \ln \frac{ |\sqrt{x(x-1) }  -  x|}{\sqrt{x(x-1) }  +  x} + i  \sgn( \omega) \pi ] $. 
The sign of the imaginary part depends on that of the infinitesimal displacement. 
When $ x < 0 $, the argument of the logarithm is semi-positive definite, i.e., 
$ \frac{ \sqrt{x(x-1) } - x }{  \sqrt{x(x-1) } + x }  = \frac{ - x }{ ( \sqrt{x(x-1) } + x)^2 }  \geq0  $. 
} 
The advanced and causal correlators can be computed in the same manner. 
As discussed in Sec.~\ref{sec:photons}, the polarization tensor acquires an imaginary part 
in the regime $ q_\para^2 \geq 4m^2 $, indicating decay of photon/gluon into a fermion and antifermion pair. 
One can consider the light- and heavy-fermion limits relative to the photon energy, i.e., 
$ m^2 \ll |q_\para^2| $ and $ m^2 \gg |q_\para^2| $, respectively. 
In those limits, one finds that 
\begin{eqnarray}
\label{eq:I-limits}
\lim_{x \to \infty} I(x) = 0 \, , \quad \lim_{x \to 0} I(x) = 1
\end{eqnarray}
In the light-fermion limit, one finds the finite photon mass $ \sim m_B $ 
that is induced by the polarization of the nearly massless fermions. 
In contrast, in the heavy-fermion limit, one finds that $ \Pi_\para \to 0 $, 
meaning that the polarization effects are suppressed. 
This is simply because the fermions have too large a mass gap 
to be excited by a soft photon momentum.\footnote{
The author (K.H.) thanks Toru Kojo for useful discussions about those limiting behaviors 
and differences/similarities in finite temperature/density.}


The massless limit of the imaginary part is more subtle. 
Let us carefully examine how the massless limit (\ref{eq:massless-vac}) is reproduced. 
Notice that the integral (\ref{eq:I}) vanishes everywhere in the massless limit except 
the pair production threshold at $ q_\para^2 = 4m^2 $, 
where the imaginary part diverges as $ q_\para^2 $ approaches $ 4m^2 $ from above. 
Thus, the functional form of the imaginary part approaches a delta function in the massless limit: 
\begin{eqnarray}
\label{eq:massless-limit-vac}
\lim_{m \to 0} \frac{  1} {q_\para^2}   \Im m \Pi_\para( q_\para^2+ i\omega\epsilon)
\= - \lim_{m \to 0} \frac{1} {q_\para^2}   \Im m I( \frac{q_\para^2}{4m^2} + i\omega\epsilon)
m_B^2 e^{ - \frac{ |\bq_\perp|^2 }{2 |q_f B| } }   
\nnb
&\to&  - C_0 \delta(q_\parallel^2) \{ \theta(\omega) - \theta(-\omega) \} m_B^2 e^{ - \frac{ |\bq_\perp|^2 }{2 |q_f B| } }  
\, .
\end{eqnarray}
The coefficient $ C_0 $ can be fixed by the fact that the imaginary part satisfies a sum rule
\begin{eqnarray}
\int_{- \infty}^\infty \!\! dq_\para^2  \frac{1} {q_\para^2}   \Im m  I( \frac{q_\para^2}{4m^2} + i\omega\epsilon)
&=& \pi\{ \theta(\omega) - \theta(-\omega) \} \int_{1}^\infty \!\! dx \frac{ 1 }{2x\sqrt{x(x-1)} }
\nn
\\
&=& \pi \{ \theta(\omega) - \theta(-\omega) \}
\label{eq:sum-rule}
\, .
\end{eqnarray}
Remarkably, the result of the integral is 
independent of the mass parameter \cite{Smilga:1991xa, Baier:1991gg, Adam:1992hs}, 
so that this sum rule should hold in the massless limit. 
Such a sum rule was shown in the four dimensions even earlier~\cite{Dolgov:1971ri, Horejsi:1985qu}. 
Therefore, we get $  C_0 = \pi $, and can reproduce the correct massless limit for the imaginary part 
that agrees with Eq.~(\ref{eq:massless-vac}). 
The integral in Eq.~(\ref{eq:sum-rule}) is dominated by 
the contribution of an integrable singularity 
at the threshold $ q_\para^2 = 4m^2 $ (cf. Fig.~\ref{fig:I}). 
The threshold behavior depends on the number of 
phase-space dimensions in a factor of $ (q^2 - E_{\rm th}^2)^{(d-2)/2} $ with $ d $ and $E_{\rm th}  $ 
being the number of dimensions and the threshold energy. 
Here, we have $d=1 $ because of 
the effective dimensional reduction. 




\subsubsection{Chiral anomaly: Massless and massive cases} 

\label{sec:anomaly-LLL}
 
Now, consider the currents carried by the LLL fermions 
\begin{subequations}
\begin{eqnarray}
\label{eq:jV-def}
j_V^\mu \= q_f \bar \psi_\LLL \gam^\mu \psi_\LLL
\, ,
\\
\label{eq:jA-def}
j_A^\mu \= \bar \psi_\LLL  \gam^\mu \gam^5 \psi_\LLL
\, ,
\end{eqnarray}
\end{subequations}
where $ \psi_\LLL $ is the fermion spinor in the LLL that is an eigenstate of 
the spin projection operator $  \prj_+$ (cf. Sec.~\ref{sec:Ritus-Feynman}). 
Those currents can be induced in response to an external $U(1)$ gauge field $ A^\mu $.\footnote{
This is an additional external gauge field superimposed on the one inducing the external magnetic field. 
Note also that the color trace in Eq.~(\ref{eq:massless-vac}) should be removed 
as we consider a coupling to the $ U(1) $ gauge field 
and that the coupling constant should be replaced as $ g \to q_f $. 
There is no mixing between the photon and gluon (without an external color field). 
}  
By the use of a relation among the gamma matrices $ 
\gam_\para^\mu \gam^5 \prj_\pm = \mp s_f \epsilon _\para^{\mu \nu} \gam_{\para \nu} \prj_\pm$, 
we find a relation between the LLL contributions to 
the vector and axial-vector currents along the magnetic field\footnote{
Here, the axial-vector current (\ref{eq:jA-def}) does not carry an electric charge, 
which is, however, included in the vector current (\ref{eq:jV-def}) 
and the electromagnetic coupling $j_{V}^\mu A_\mu  $ in Eq~(\ref{eq:anomaly-LLL-1}). 
} 
\begin{eqnarray}
q_f j_A^\mu = - s_f  \epsilon _\para^{\mu \nu} j_{V \nu}
\label{eq:V-A}
\, ,
\end{eqnarray}
where we introduced an antisymmetric tensor $  \epsilon _\para^{\mu \nu}  $ 
which only has two non-vanishing components, $ \epsilon_\para^{03} = -\epsilon_\para^{30} =1 $. 
This relation is an analog of the well-known relation in the (1+1) dimensional QED 
and is understood as a consequence of the effective dimensional reduction 
and the spin polarization in the LLL.  
Equation (\ref{eq:V-A}) relates the vector--vector correlator $ \Pi^{\mu\nu}_R $, which we just computed, 
to the vector--axial-vector correlator 
\begin{eqnarray}
\langle j_A^\mu j_V^\nu \rangle_R = - s_f  \epsilon _{\para \, \rho}^{\mu} q_f^{-1} \Pi^{\rho\nu}_R  
\, .
\end{eqnarray}
We shall consider a linear response to the gauge field $ A_\mu $. 
According to the transverse form of the polarization tensor (\ref{eq:int_para}), 
the divergences of the currents read \cite{Hattori:2022wao}
\begin{subequations}
\label{eq:Ward-ids}
\begin{eqnarray}
&&
{\rm F.T.} \ \pd_\mu j_V^\mu =  -i q_\mu  \Pi^{\mu\nu}_R   A_\nu =0
\label{eq:Ward-id1}
\, ,
\\
&&
{\rm F.T.} \ \pd_\mu j_A^\mu = -i q_\mu \langle j_A^\mu j_V^\nu \rangle_R A_\nu 
=  s_f  q_f^{-1} \Pi_\para \tilde E  _\para
\label{eq:anomaly-LLL-1}
\, ,
\end{eqnarray}
\end{subequations}
where F.T. stands for the Fourier transform 
and $ \tilde E_\para(q) = - i \epsilon _\para^{\mu \nu} q_\mu A_\nu $ is the parallel (or antiparallel) 
component of an electric field along the magnetic field. 
Note that $  \tilde E_\para(q) $ is the Fourier spectrum, while $ B $ is a constant magnitude. 
In case of the quark-loop contributions, one should count the number of color degrees of freedom 
circulating on the loop, that comes from the simple color trace $\tr[\id_c] =N_c  $. 
The Ward identity (\ref{eq:Ward-id1}) is a consequence of the gauge invariance in QED. 
However, the axial current may not be conserved in Eq.~(\ref{eq:anomaly-LLL-1}) in general.

We first consider the massless case. 
Inserting the polarization tensor (\ref{eq:Pi-massless}) into Eq.~(\ref{eq:anomaly-LLL-1}), 
one finds the axial Ward identity (AWI) 
\begin{eqnarray}
{\rm F.T.} \ \pd_\mu j_A^\mu 
=  s_f   \rho_B  \frac{q_f}{\pi} \tilde E_\para \, e^{-\frac{|\bq_\perp|^2}{2|q_fB|}} 
=  \frac{ q_f^2}{2\pi^2} \tilde \bE \cdot \bB  \, e^{-\frac{|\bq_\perp|^2}{2|q_fB|}} 
\label{eq:anomaly-LLL-2}
\, .
\end{eqnarray}
The expression in the middle of Eq.~(\ref{eq:anomaly-LLL-2}) 
is given by the Landau degeneracy factor $\rho_B $ multiplied by 
the chiral anomaly in purely (1+1) dimensions 
(without any other external line), i.e., 
$- i  q_\mu \langle j_A^\mu j_V^\nu \rangle_{(1+1)} A_\nu =  (q_f/\pi) \tilde E$ 
(see, e.g., Ref.~\cite{Peskin:1995ev}). 
This is a consequence of the dimensional reduction 
in the strong magnetic field \cite{Nielsen:1983rb}. 
At the same time, the rightmost side in Eq.~(\ref{eq:anomaly-LLL-2}) exhibits the same form as 
the chiral anomaly in the (3+1) dimensions from the triangle diagrams. 
Notice that the correlator $ \Pi^{\mu\nu}_R $ indeed 
contains the triangle diagrams as a part of the resummed series 
with respect to the external magnetic field. 
However, due to the Gaussian factor that originates from the transverse wave functions, 
the contribution of the LLL fermions only saturates the (3+1)-dimensional chiral anomaly relation 
in the long-wavelength and/or strong-field limits where $ |\bq_\perp|^2/|q_fB| \to 0 $. 
In such a limit, the total current is composed of copies of 
the (1+1)-dimensional currents with the transverse density $ \rho_B $ since the axial current and the electromagnetic fields do not 
resolve the cyclotron motion of the radius $ \sim 1/\sqrt{|q_fB|} $.\footnote{
At the mathematical level, it is not obvious how the chiral anomaly 
in the (3+1) dimensions is reproduced from the LLL contribution. 
The familiar triangle diagrams are the lowest-order contribution 
in the series expansion with respect to the coupling constant. 
On the other hand, the LLL contribution is obtained only after summing 
the all-order one-loop diagrams with respect to the external-field insertion 
and then expanding it in another series in terms of the Landau levels: The LLL contribution contains the higher-order diagrams in the coupling constant that give rise to 
a non-analytic dependence 
of the Gaussian on the coupling constant. 
The leading terms in those two expansions could differ from each other in general. 
}

In the massive case, $ \Pi_\para $ in Eq.~(\ref{eq:vac_massive}) 
provides the AWI 
\begin{eqnarray}
\label{eq:AWI}
{\rm F.T.} \ \pd_\mu j_A^\mu =
 \frac{ q_f^2}{2\pi^2} \tilde \bE \cdot \bB  \Big[ \, 1- I \big( \frac{q_\parallel^2}{4m^2} \big) \,  \Big] 
  e^{-\frac{|\bq_\perp|^2}{2|q_fB|}}  
  \, .
\end{eqnarray}
We have already seen that $ \Pi_\para  $ approaches the expression in the massless case 
when $ m^2 / q_\para^2 \to 0 $. 
In the opposite limit, we have $ \Pi_\para \to 0 $ when $ m^2/ q_\para^2  \to \infty $, 
so that the axial current is effectively conserved, i.e., $ \pd_\mu j^\mu_A =0$ 
in such a heavy-mass and/or soft-photon limit. 
This is a natural result because the weak electric field, perturbatively applied 
in Eq.~(\ref{eq:anomaly-LLL-1}), cannot create any massive on-shell fermions over the mass gap. 
The mass dependence plays a crucial role in determining the magnitude of the divergence of the current.

It should be emphasized that 
the effective conservation of $ j_A^\mu $ in the above limit 
does not contradict with the chiral anomaly at all. 
Actually, the term proportional to the $ I $ function in Eq.~(\ref{eq:vac_massive}) 
is nothing but the matrix element of the pseudoscalar condensate 
$ 2i m \langle \bar \psi \gam^5 \psi \rangle $ that arises 
from the explicit chiral symmetry breaking by a finite fermion mass. 
Namely, as explicitly shown in Sec.~\ref{sec:anomaly-massive}, 
the anomaly diagrams composed of the massive fermion loops 
are split into two terms 
that are the mass-independent anomalous term 
and mass-dependent matrix element of the pseudoscalar condensate. 
The anomaly diagram as a whole vanishes when the fermion mass is sent to infinity 
just because all the internal lines go far off-shell. 
One could rephrase this vanishing result as an exact cancellation between the anomalous term 
and the pseudoscalar condensate term~\cite{Ambjorn:1983hp, Smilga:1991xa}. 
While the above result and discussion hold 
only for a perturbative electric field, 
the cancellation was further shown for 
arbitrary electric-field strength 
as long as the electromagnetic fields are taken to be constant fields \cite{Copinger:2018ftr}.\footnote{
It was also shown that a nonzero axial charge is created 
by the Schwinger mechanism in the real-time formalism. 
} 
The cancellation may not be always exact 
when there is an energy-momentum transfer from 
inhomogeneous electromagnetic fields (or dynamical photons) as discussed 
in Sec.~\ref{sec:anomaly-massive} 
(see also Refs.~\cite{Ambjorn:1983hp, Hattori:2022wao}).

\subsubsection{Chiral anomaly from dispersion integral} 

\begin{figure}[t]
     \begin{center}
              \includegraphics[width=0.5\hsize]{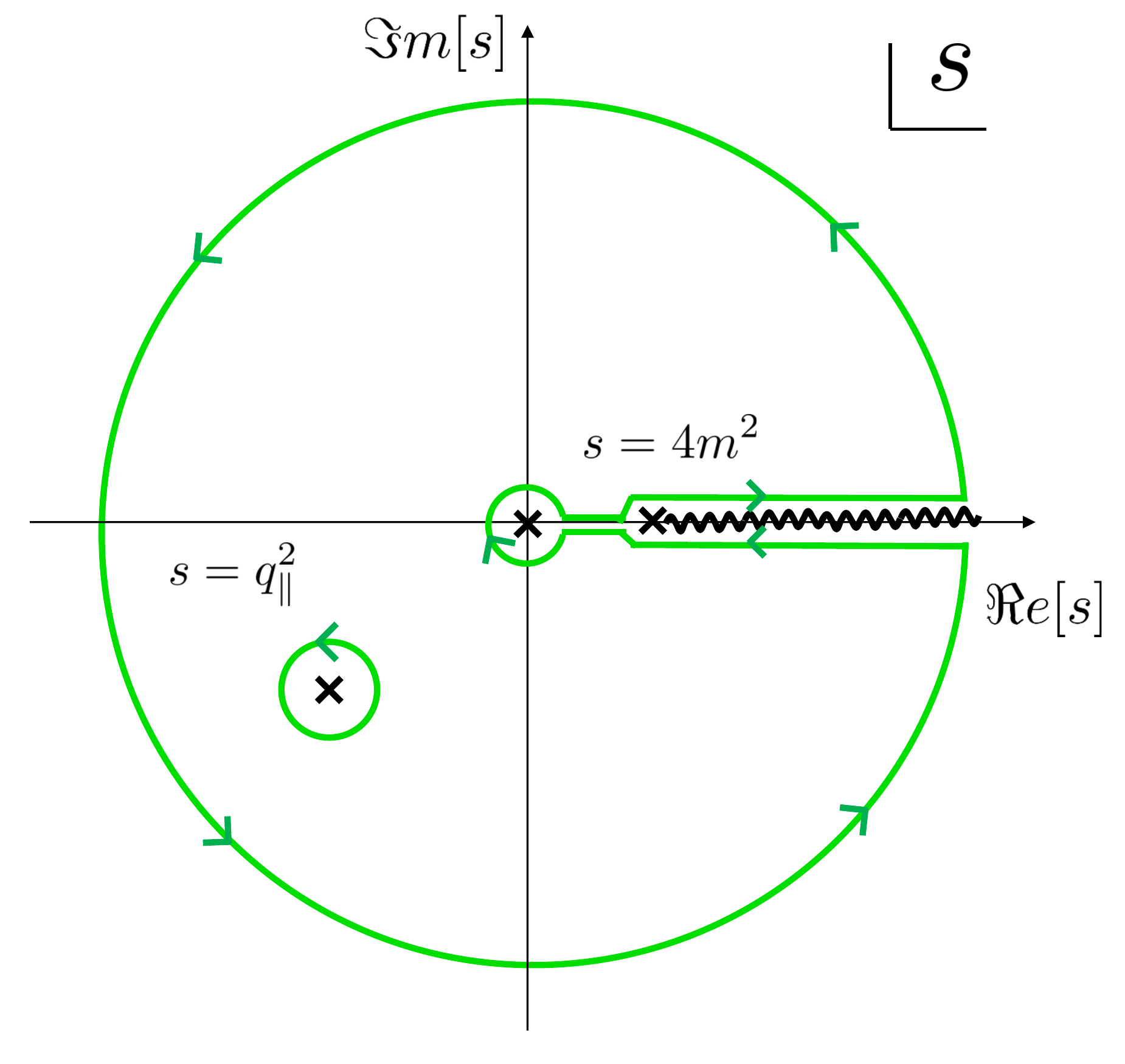}
     \end{center}
\vspace{-1cm}
\caption{Dispersion relation from analyticity in the complex $  s$ plane.}
\label{fig:dispersion-relation}
\end{figure}

Chiral anomaly is often explained, a little bit technically, as a consequence of 
the gauge-invariant regularization of the UV divergence. 
The UV divergence appears in the real part of the polarization tensor. 
On the other hand, the real part can be tied to the imaginary part via the dispersion integral. 
One may, therefore, identify the origin of the chiral anomaly with 
the physical processes captured by the imaginary part \cite{Dolgov:1971ri, Smilga:1991xa}. 
Of course, the imaginary part is free of any UV divergence since it corresponds to the squared amplitude 
of the tree diagram (at the lowest order in the coupling constant). 
In short, the chiral anomaly from the (superficially divergent) UV dynamics 
is connected to the IR dynamics.\footnote{
It is also worth emphasizing that the anomaly diagram does not yield any UV-divergent term in the end. 
Remember that, in Eq.~(\ref{eq:int_para0}), we only needed the regularization 
not to miss the terms of order $ (\epsilon_2)^0 $, 
and that there were no divergent poles like $  1/\epsilon_2$ after all. 
This ``marginal'' degree of the divergence is precisely the reason why 
the anomalous term gets independent of the fermion mass (see Sec.~\ref{sec:anomaly-massive}). 
}

We shall first confirm that the polarization tensor satisfies the dispersion integral 
that captures the analyticity in the complex plane (see Fig.~\ref{fig:dispersion-relation}). 
We begin with a simple identity stemming from a contour integral that encloses a pole at $ s = q_\para^2 $: 
\begin{eqnarray}
f (q_\para^2) = \frac{1}{2\pi i}  \oint ds \frac{ f(s) }{  s - q_\para^2 } 
\, ,
\end{eqnarray}
where $  f(s)$ is an arbitrary complex function that is analytic inside the closed contour. 
We apply the above formula to $ f (s) =  \Pi_\para (s)  /s$. 
Expanding the integral contour to an infinitely large arc, we find that 
\begin{eqnarray}
\label{eq:dis-integral}
\frac{1}{q_\para^2} \Pi_\para (q_\para^2) 
=  - \frac{\Pi_\para (0)}{ - q_\para^2} + 
  \frac{1}{\pi }  \int_{4m^2}^\infty ds \frac{ \Im m \Pi_\para(s) }{ s ( s - q_\para^2 ) }
\, ,
\end{eqnarray}
which is often called the dispersion integral and is another kind of sum rules [cf. Eq.~(\ref{eq:sum-rule})]. 
For the moment, we assume a kinematical condition $ q_\para^2 < 4m^2 $, 
which can be relaxed with the help of the analytic continuation afterwards.

In the massive case, the polarization tensor (\ref{eq:vac_massive}) 
has a branch cut along the real axis $ s \geq 4m^2 $ 
that corresponds to the pair creation from a single photon. 
This continuum contributes to the second term in the dispersion integral (\ref{eq:dis-integral}). 
On the other hand, we have $ \Pi_\para (0) = 0 $ as discussed below Eq.~(\ref{eq:I}), so that the first term vanishes. 
Inserting the imaginary part in Eq.~(\ref{eq:I}) into the dispersion relation 
leads to an elementary integral 
\begin{eqnarray}
\frac{1}{q_\para^2} \Pi_\para (q_\para^2)  
\=  \sgn(\omega) m_B^2 \, e^{-\frac{\vert \bq_\perp \vert^2}{2 \vert q_fB\vert}}
\frac{1}{2  }  \int_{ 4m^2}^\infty ds 
\frac{ -  4m^2  }{ s (s - q_\para^2 )  \sqrt{s(s-4m^2)}  }
\nnb
\= \sgn(\omega)  m_B^2 \, e^{-\frac{\vert \bq_\perp \vert^2}{2 \vert q_fB\vert}}
\frac{1}{q_\para^2} \Big[ \,1 - I ( \frac{q_\para^2}{4m^2}  ) \, \Big]
\, .
\end{eqnarray} 
This result agrees with the right-hand side of Eq.~(\ref{eq:vac_massive}), 
indicating that the polarization tensor satisfies the dispersion relation. 
Therefore, we identify the pair creation from a single photon 
or a finite-frequency electric field 
as the physical process that induces 
the non-conservation of the axial current in the massive case. 
Namely, a fermion and antifermion pair of the same helicity, 
which belong to different chirality sectors, is 
created thanks to the chirality mixing by a finite mass term. 
This 1-to-2 process is not allowed in the (3+1) dimensions 
but is allowed in the effective (1+1) dimensions 
with the Landau levels. 
The created axial charge contains the contribution 
of the explicit breaking term $2i m \langle \bar \psi \gam^5 \psi \rangle $ as discussed below Eq.~(\ref{eq:AWI}), 
and thus does not necessarily agree with 
that in the massless case (\ref{eq:anomaly-LLL-2}) 
except for in some limits.

The above 1-to-2 process is different from 
the physical process encoded in the imaginary part of 
the triangle diagrams, that is, the Compton scattering \cite{Dolgov:1971ri}. 
Interestingly, the same non-conservation equations are 
induced by the different physical processes 
in the weak and strong magnetic fields, 
up to the aforementioned difference in the Gaussian factor. 
In the subsequent appendix, 
we will discuss an explicit connection 
between the triangle diagrams 
and the polarization diagram in the LLL.


The dispersion integral (\ref{eq:dis-integral}) 
is simpler in the strictly massless case $ (m=0) $. 
In Eq.~(\ref{eq:massless-vac}), the polarization tensor 
does not have a branch cut due to the prohibition of 
the photon/gluon decay in the absence of chirality mixing 
as discussed below (\ref{eq:Schwinger-mass}). 
It is this prohibition that makes the chiral anomaly 
so simple in the massless theory 
by excluding the physical processes other than 
the spectral flow discussed in Sec.~\ref{sec:massless}. 
The spectral flow stems from the residue at the origin $\Pi_\para(0)  $. 
One can immediately find that the dispersion relation (\ref{eq:dis-integral}) is satisfied in the strictly massless case [cf. Eq.~(\ref{eq:Pi-massless})]. 
Clearly, the analytic structures in the complex $ s $ plane are different in the strictly massless and massive cases. They are not continuously related.

Summarizing the dispersion analysis, 
we found that the relevant physical processes are 
different in the strictly massless and massive cases.  
In the strictly massless case, 
the spectral flow gives rise to the chiral anomaly 
from the adiabatic shifts of the whole occupied states 
along the dispersion lines from the UV to IR regions (see Sec.~\ref{sec:massless}). 
This is the only process that can occur without the chirality mixing, which is the reason for 
the simplicity in the massless case. 
In the massive case, the 1-to-2 pair creation gives rises to 
the axial charge creation. This is a diabatic process 
dominated by the IR contribution, i.e., 
the near-threshold divergence. 
There can be contributions from the higher-order processes. 
In the massive case, there is no contribution from 
the adiabatic spectral flow in vacuum \cite{Ambjorn:1983hp}, 
while there are finite contributions at finite temperature 
and/or density \cite{Hattori:2022wao} (see Sec.~\ref{sec:AWI-medium}).

\subsection{Fermion-mass dependences of the anomaly diagrams}

\label{sec:anomaly-massive}

We here explicitly see that the anomaly diagrams with massive-fermion loops 
can be split into two terms that are nothing but the anomalous and pseudoscalar-condensate terms. 
More specifically, this means that the term proportional to $I$ function in Eq.~(\ref{eq:vac_massive}) 
is identified with the matrix element of the pseudoscalar condensate. 
Also, we show that the familiar triangle diagrams in the four dimensions 
are split into the two terms in the same manner.

It is often said that the chiral anomaly is independent of the fermion mass 
since the anomalous term originates from a UV divergence. 
We further clarify this statement and carefully discuss 
mass dependences of the anomaly diagrams. 
The nonzero divergence of the axial-vector current (\ref{eq:anomaly-LLL-1}), 
which we refer to as the axial Ward identity (AWI), 
certainly has mass dependences as a whole. 
What one can explicitly show is that the anomalous term is independent of the fermion mass. 
The reason is rather because the anomaly diagrams are {\it not} divergent after all, 
though they are superficially divergent. They are even not logarithmically divergent. 
This ``marginal'' degree of divergence gives rise to the mass independence of the anomalous term.

The remaining term is explicitly identified with the matrix element of the pseudoscalar condensate 
and depends on the fermion mass. 
This term vanishes in the massless limit, reproducing the chiral anomaly. 
In contrast, this term exactly offsets the anomalous term in the large-mass limit, 
leading to effective {\it conservation} of the axial-vector current. 
Therefore, the mass dependence plays a crucial 
to control the behaviors of the total AWI. 
The above exact cancellation is anticipated 
because both of the terms 
originate from the same diagram which is only split into the two terms {\it a posteriori}. 
The cancellation is nothing more than a simple statement that 
diagrams generically vanish when all internal lines go far off-shell due to an infinitely large mass.


The presence of mass dependences in the AWI may be further anticipated as follows. 
(1) In general, finite terms in divergent diagrams do have mass dependences; 
A mass term just does not change the degree of divergence. 
The anomaly diagrams are even {\it not} divergent after all, and should have mass dependences as a whole. 
(2) The imaginary part of the anomaly diagrams correspond to 
the scattering rates of the on-shell processes. 
In case of the LLL, it is the pair creation from a single gluon/photon as discussed above \cite{Smilga:1991xa, Hattori:2022wao}. 
In case of the triangle diagrams, it is an interference between the amplitudes of 
the Compton scattering and decay of the axial current to a fermion pair \cite{Dolgov:1971ri}. 
No one would believe that those processes are independent of the fermion mass. 
The real part, reconstructed from the imaginary part via the dispersion integral, 
should also have mass dependences as seen above for the LLL. 
No UV regularization procedure is involved. 
(3) If there were not any mass dependence in the AWI, 
heavy fermions, or even undiscovered heavy fermions if any, contribute to the anomalous term 
with the same magnitudes as the light fermions do, 
changing the chiral-anomaly coefficients by factors. 
However, we have not detected such ``anomalous'' contributions to the chiral anomaly. 
This may imply that the cancellation with the pseudoscalar-condensate term 
occurs for heavy fermions when they are too heavy to be excited 
by the axial-vector current of a given energy scale. 


Below, we follow the conventions and strategy given in the standard textbook \cite{Peskin:1995ev}, 
where the authors explain the chiral anomaly from the familiar triangle diagrams 
in the massless case with the dimensional regularization. 
We thus first examine the triangle diagrams with massive fermions 
so that the reader can directly compare the massive and massless cases. 
We then examine the anomaly diagram from the LLL massive fermions.

\subsubsection{Massive triangle diagrams 
and the limit of constant magnetic fields}

\label{sec:triangle-constant}

Following the strategy with the dimensional regularization described in Ref.~\cite{Peskin:1995ev}, 
we compute the triangle diagrams with the massive-fermion loops. 
We essentially compute the matrix element $ \M^{\mu\nu\lambda} $ such that 
\begin{eqnarray}
\label{eq:matrix-anomaly}
\int d^4 x e^{-iq x} \langle p,k |\pd _\mu j_A^\mu (x) | 0\rangle
= (2\pi)^4 \delta^{(4)}( p+k -q) i q_\mu 
\M^{\mu\nu\lambda}(q;p,k) \epsilon^\ast_\nu (p) \epsilon^\ast_\lambda (k)
\, ,
\end{eqnarray}
where $ p,k $ are the photon momenta and $  \epsilon_\nu  $ is the photon polarization vector. 
There are two triangle diagrams with the opposite fermion charge flows: 
\begin{subequations}
\label{eq:triangles}
\begin{eqnarray}
\label{eq:triangles-1}
 \M^{\mu\nu\lambda} _1 \= (-1) (-ie)^2 \int \frac{d^4 \ell}{(2\pi)^4} 
 \frac{ i^3 \tr[ \gam^\mu \gam^5 ( \sla \ell - \sla k + m ) \gam^\lambda
 ( \sla \ell + m ) \gam^\nu (\sla \ell + \sla p + m) ]}
 { \{ (\ell-k)^2 -m ^2 \} \{ \ell^2 -m^2 \} \{ (\ell+p)^2 - m ^2\} }
 \, ,
 \\
\label{eq:triangles-2}
  \M^{\mu\nu\lambda} _2 \= (-1) (-ie)^2 \int \frac{d^4 \ell}{(2\pi)^4} 
 \frac{ i^3 \tr[ \gam^\mu \gam^5 ( \sla \ell - \sla p + m) \gam^\nu 
 (\sla \ell + m ) \gam^\lambda (\sla \ell + \sla k +m) ]}
 { \{ (\ell-p)^2 -m^2\} \{ \ell^2 -m^2\} \{ (\ell+k)^2 -m^2\}  }
 \, ,
\end{eqnarray}
\end{subequations}
where $ \M^{\mu\nu\lambda} = \M^{\mu\nu\lambda}_1+\M^{\mu\nu\lambda} _2$. 
Each integral is linearly divergent, and one needs to regularize it in a gauge-invariant manner. 
To apply the dimensional regularization, 
it is important to make sure the properties of the $ \gam^5 $. 
The $ \gam^5 $ is extended in such a way that 
$ \{\gam^5, \gam^\mu\} =0 $ for $ \mu = 0,1,2,3 $ 
and $ [ \gam^5, \gam^\mu ] =0 $ for the other components. 
The extra components in the momentum is denoted with tilde like $ \tilde  \ell ^\mu$.\footnote{ 
This corresponds to $\ell_\perp ^\mu $ in the notations of Ref.~\cite{Peskin:1995ev}. 
We do not use their notations with $ \para,\perp $ 
to avoid possible confusions with the parallel and perpendicular components 
with respect to the magnetic field in this paper.} 
The external momenta $ q^\mu, p^\mu, k^\mu $ have vanishing components 
in the extra dimensions, $ \tilde q^\mu =\tilde p^\mu =\tilde  k^\mu =0$. 
The extra components of $ \gam ^\mu$ commute with $ \gam^5 $, i.e., $ [\tilde {\sla \ell}, \gam^5] = 0 $. 

Then, by the use of an identity 
\begin{eqnarray}
\label{eq:anom-id2}
q_\mu \gam^\mu \gam^5 
= ( \sla \ell + \sla p -m)  \gam^5 +  \gam^5 ( \sla \ell - \sla k -m)
-  2 \gam^5  ( \tilde{ \sla \ell}  - m) 
\, ,
\end{eqnarray}
\cout{
we can arrange the matrix element as 
\begin{eqnarray}
i q_\mu \M^{\mu\nu\lambda} _1
\= e^2  \int \frac{d^d \ell}{(2\pi)^d}  
  \Big[ \, 
  \frac{ \tr[    \gam^5 ( \sla \ell - \sla k + m) \gam^\lambda ( \sla \ell + m) \gam^\nu ]}
 { \{ (\ell-k)^2 -m^2\} \{ \ell^2 -m^2\}  }
 -
 \frac{  \tr[ \gam^5 ( \sla \ell +m) \gam^\nu (\sla \ell  + \sla p + m)\gam^\lambda ]}
 { \{  \ell^2 -m^2\} \{ (\ell+p)^2 -m^2\} }
\,  \Big]
\nnb
&& + (-1) (-ie)^2  \int \frac{d^d \ell}{(2\pi)^d} 
 \frac{ \tr[  (- 2) \gam^5 (\tilde{ \sla \ell}  -m) ( \sla \ell - \sla k + m ) \gam^\lambda
 ( \sla \ell + m ) \gam^\nu (\sla \ell + \sla p + m) ]}
 { \{ (\ell-k)^2 -m ^2 \} \{ \ell^2 -m^2 \} \{ (\ell+p)^2 - m ^2\} } 
 \, .
 \label{eq:anom-massive}
\end{eqnarray}
Since all the integrals have been regularized, 
it is legitimate to shift the integral variable in the first term as $ \ell^{'\mu} = \ell ^\mu - k^\mu $. 
After the shift, the first two terms are found to be antisymmetric 
under the simultaneous interchange $ (p;\nu) \leftrightarrow (k; \lambda) $. 
This implies that those two terms are cancelled 
with the corresponding terms from the other matrix element $q_\mu \M^{\mu\nu\lambda} _2 $ 
so that the sum of the two triangle diagrams $ q_\mu \M^{\mu\nu\lambda}  $ becomes symmetric as a whole. 
The point is that there appears the third term after the regularization 
that is proportional to the extra component of the momentum $ \tilde{ \sla \ell}  $. 
Also, the third term contains the mass term that originates from the identity (\ref{eq:anom-id2}). 
Dropping the terms to be cancelled, we denote the persistent two terms as 
}
we find the regularized matrix element 
\begin{eqnarray}
i q_\mu \M^{\mu\nu\lambda} _1 
= i q_\mu \Delta\M^{\mu\nu\lambda} _{1(a)} + i q_\mu \Delta \M^{\mu\nu\lambda} _{1(m)}
\, ,
\end{eqnarray}
where 
\begin{subequations}
\begin{eqnarray}
i q_\mu \Delta\M^{\mu\nu\lambda} _{1(a)}
\=-2 (-1) (-ie)^2  \int \frac{d^d \ell}{(2\pi)^d} 
 \frac{ \tr[   \gam^5  \tilde{ \sla \ell}  ( \sla \ell - \sla k + m ) \gam^\lambda
 ( \sla \ell + m ) \gam^\nu (\sla \ell + \sla p + m) ]}
 { \{ (\ell-k)^2 -m ^2 \} \{ \ell^2 -m^2 \} \{ (\ell+p)^2 - m ^2\} } 
 \, ,
\\
\label{eq:PS-condensate-1}
i q_\mu \Delta  \M^{\mu\nu\lambda} _{1(m)}
\= 2m (-1) (-ie)^2 \int \frac{d^d \ell}{(2\pi)^d} 
 \frac{ \tr[   \gam^5  ( \sla \ell - \sla k + m ) \gam^\lambda
 ( \sla \ell + m ) \gam^\nu (\sla \ell + \sla p + m) ]}
 { \{ (\ell-k)^2 -m ^2 \} \{ \ell^2 -m^2 \} \{ (\ell+p)^2 - m ^2\} } 
 \, .
\end{eqnarray}
\end{subequations}
We denote the corresponding terms from 
the other triangle diagram as 
$  \Delta\M^{\mu\nu\lambda} _{2(a)}$ and $  \Delta\M^{\mu\nu\lambda} _{2(m)} $, respectively, 
and have dropped the terms that cancel each other 
between $ \M^{\mu\nu\lambda} _1 $ and $\M^{\mu\nu\lambda} _2 $. 
Moreover, we will not need to compute 
those two triangle diagrams separately 
since they provide the same contributions. 
The point is that there appears 
$q_\mu \Delta\M^{\mu\nu\lambda} _{1(a)} $ 
proportional to $  \tilde{ \sla \ell}$ after the regularization. 
Below, we demonstrate how $ q_\mu \Delta\M^{\mu\nu\lambda} _{1(a)} $ results in the {\it mass-independent} anomalous term, 
though the above expression yet contains mass terms. 
Also, the second term $ q_\mu \Delta  \M^{\mu\nu\lambda} _{1(m)} $ is identified with 
the leading nonvanishing matrix element of the pseudoscalar condensate in the coupling-constant expansion.

We first compute the anomalous term 
$  q_\mu \M^{\mu\nu\lambda} _{1(a)} $. 
Introducing the Feynman parameter integral and then performing the momentum shift 
$  \ell^{\prime\mu} = \ell^\mu - xk^\mu + yp ^\mu = \ell ^\mu+ P^\mu$, 
the integral is arranged as 
\begin{eqnarray}
i q_\mu \Delta \M^{\mu\nu\lambda} _{1(a)}
\=  (- 2) 2!e^2  \int_0^1dx \int_0^{1-x}dy 
\nnb
&&\times \int \frac{d^d \ell}{(2\pi)^d}  
 \frac{ \tr[  \gam^5 \tilde{ \sla \ell} ( \sla \ell - \sla k - \sla P + m ) \gam^\lambda
 ( \sla \ell - \sla P + m ) \gam^\nu (\sla \ell + \sla p - \sla P + m) ]}
 { (\ell^2 -  \Delta )^3 } 
 \, ,
\end{eqnarray}
where $ \Delta = m^2 - \{ x (1-x) k^2  + y   (1-y) p^2 + 2 x y  \, p\cdot k \} $. 
Note that the external momenta $ p^\mu,k^\mu $ do not have nonzero components in the extra dimensions, 
so that $  \tilde \ell^\mu $ is not shifted by $ P^\mu $. 
Performing the spinor trace, one arrives at only one integral 
\begin{eqnarray}
\label{eq:mom-integral-triangle}
 \int \frac{d^d \ell}{(2\pi)^d}   \frac{ \tilde  \ell ^\mu \tilde \ell^\nu  } {  (  \ell^2 +  \Delta )^3 } 
 \=  \frac{d-4}{d}   g^{\mu\nu}   \times \frac{ i (-1)^2}{(4\pi)^{d/2}} \frac{d}{2} 
  \frac{\Gamma(-\frac{d-4}{2})}{\Gamma(3)}  \Delta^{\frac{d-4}{2} } 
  \, .
\end{eqnarray}
Notice that this integral is {\it not} divergent because of the overall dimensional factor 
that suppresses the divergent pole as $ (d-4) \Gamma(- \frac{d-4}{2}) = - 2 +  \order( d-4) $. 
This means that one can take the limit $ d \to 4 $ in all the places, 
indicating that the mass-dependent factor goes away, $\Delta^{  \frac{d-4}{2} } \to 1 $. 
If there remained a divergence, 
the mass-dependent factor $  \Delta $, 
as well as a divergent pole, would remain 
in the result of the integral. 
We have found the mass-independent anomaly coefficient 
\begin{eqnarray}
\label{eq:anomalous}
 i q_\mu [\,\Delta \M^{\mu\nu\lambda} _{1(a)} +\Delta \M^{\mu\nu\lambda} _{2(a)} \,]
   \epsilon^\ast_\nu (p) \epsilon^\ast_\lambda (k)
= -   \frac{ e^2  }{ 16 \pi^2 }  \epsilon^{\a\nu\b\lambda}
 \langle p,k |  F_{\a\nu} F_{ \b \lambda} | 0\rangle
 \, ,
\end{eqnarray} 
where the two terms on the left-hand side provide 
the same contributions. 
The antisymmetric tensor comes from 
the spinor trace with $\gamma^5 $.
This is only a part of the triangle diagrams 
(\ref{eq:matrix-anomaly}).

Now, we consider the other term $\M^{\mu\nu\lambda}_{1(m)} $ 
proportional to the fermion mass 
in Eq.~(\ref{eq:PS-condensate-1}). 
Only one difference of $\M^{\mu\nu\lambda}_{1(m)} $ 
from the original triangle (\ref{eq:triangles-1}) is 
the replacement of the axial vector vertex $ \sla q \gam^5 $ 
by the pseudoscalar vertex $ \gam^5 $ 
via the identity (\ref{eq:anom-id2}). 
Thus, $\M^{\mu\nu\lambda}_{1(m)} $ is the matrix element of the pseudoscalar operator $\bar \psi \gam^5 \psi $ instead of the axial-vector operator $\bar \psi \gam^\mu \gam^5 \psi $. 
Explicitly, it is identified with the matrix element 
of the pseudoscalar condensate as 
\begin{eqnarray}
\label{eq:PS_condensate}
 i  q_\mu [\M^{\mu\nu\lambda}_{1(m)} +  \M^{\mu\nu\lambda}_{2(m)} ]
   \epsilon^\ast_\nu (p) \epsilon^\ast_\lambda (k)
=  2 im  \int d^4 x e^{-iq x} \langle p,k |  \bar \psi \gam^5 \psi | 0\rangle
\, ,
\end{eqnarray}
where the sum of the two triangle diagrams 
provides a factor of $ 2$. 
The presence of this term is consistent with the explicit symmetric breaking, 
$ \pd_\mu j_A^\mu = 2 i m\bar \psi \gam^5 \psi $, from the Dirac equation. 
Notice that the matrix element of the pseudoscalar condensate 
$ \langle \bar \psi \gam^5 \psi \rangle $ cannot be finite 
without insertions of the external photon legs. 
For example, the simplest one-loop integral 
without any insertion reads 
$ \langle \bar \psi \gam^5 \psi \rangle \propto \tr[ \gam^5 (\sla p + m)] = m \tr[\gam^5] =0 $; 
One has to take the mass term to mix 
the right and left chirality 
as they are mixed in the bilinear spinor 
$  \bar \psi \gam^5 \psi  $. 
It turns out that the lowest nonzero contribution 
needs two external photon legs. 
That is, the matrix element (\ref{eq:PS_condensate}) is 
the leading nonvanishing contribution 
and its diagrammatic representation still 
has the triangle form (cf. Fig.~\ref{fig:triangle}). 
This pseudoscalar triangle diagram (\ref{eq:PS-condensate-1}) 
does not have a divergence and 
can be computed quite straightforwardly. 

\begin{figure} 
	\begin{center}
\includegraphics[width=0.9\hsize]{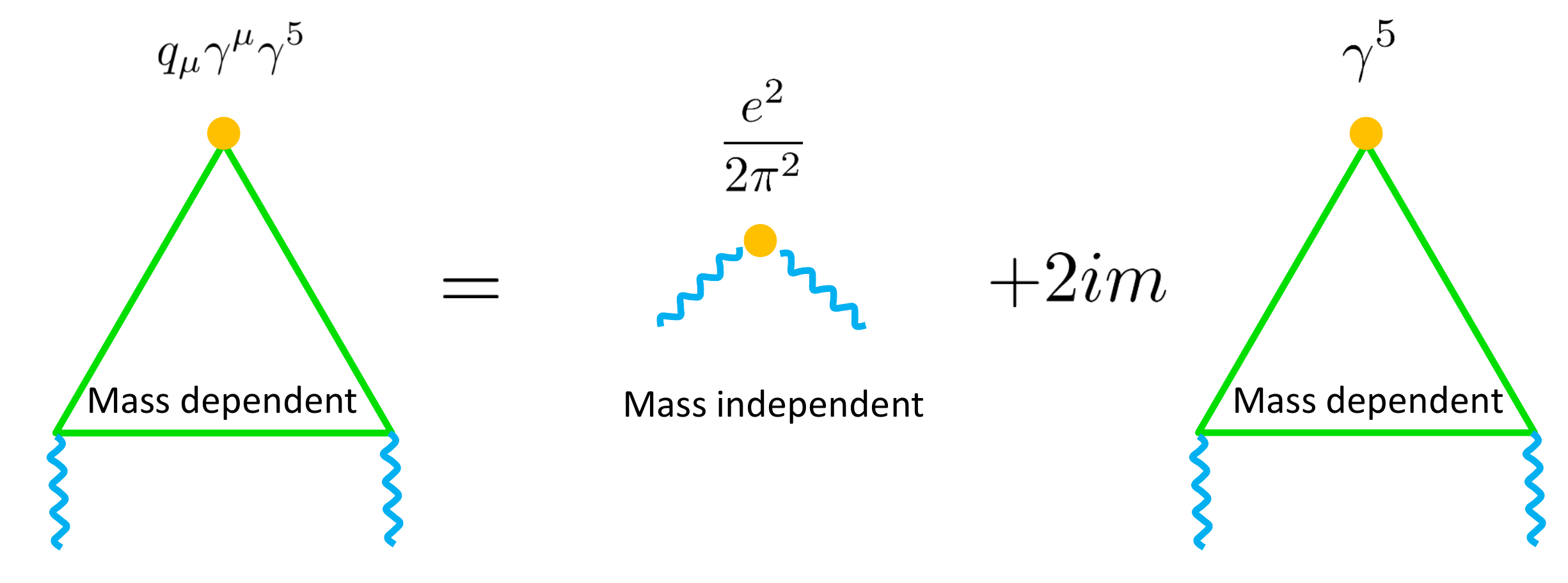}
	\end{center} 
\vspace{-0.7cm}
\caption{Splitting the massive triangle diagram into the anomalous and pseudoscalar-condensate terms.}
\label{fig:triangle}
\end{figure}

Summarizing the above, we obtain the AWI as the sum of 
the anomalous term (\ref{eq:anomalous}) 
and the pseudoscalar triangles (\ref{eq:PS_condensate}) as 
\cite{Adler:1969gk, Bell:1969ts} 
\begin{eqnarray}
\label{eq:anomaly-massive}
\langle p,k | \pd_\mu j^\mu_A | 0\rangle  
\=  \langle p,k | -   \frac{ e^2  }{ 8\pi^2 }   F^{\mu\nu} \tilde F_{\mu\nu} 
+ 2 im   \bar \psi \gam^5 \psi | 0\rangle
\\
\= - \frac{ e^2  }{ 8 \pi^2 }   \langle p,k |   F^{\mu\nu} \tilde F_{\mu\nu}  | 0\rangle  
\Big[ \, 1 +    
   \int_0^1dx \int_0^{1-x} dy   \frac{ 2m^2 }{  x (1-x) k^2  +  y (1-y)  p^2 + 2 x y  \, p\cdot k - m^2}
      \, \Big] 
      \nn
      \, .
\end{eqnarray}
We emphasize that both the mass-independent anomalous term 
and the mass-dependent pseudoscalar condensate stem 
from the same triangle diagrams with the massive fermion loops (cf. Fig.~\ref{fig:triangle}). 
The remaining integral is only 
a function of normalized photon momenta 
$ \bar p^\mu = p^\mu /m $  and $ \bar k^\mu = k^\mu/m $, 
so that the magnitude of the mass correction depends 
on the relative magnitudes 
between the photon momenta and the fermion mass. 

We shall examine the remaining integral. 
In the massless limit, the second term simply vanishes, 
reproducing the triangle anomaly in the strictly massless theory. 
On the other hand, in the large-mass limit such that $  \bar p^\mu , \bar k^\mu \ll 1  $, we have 
\begin{eqnarray}
\label{eq:offset}
\langle p,k | \pd_\mu j^\mu_A | 0\rangle  
 \to - \frac{ e^2  }{ 8 \pi^2 } \langle p,k |   F^{\mu\nu} \tilde F_{\mu\nu}  | 0\rangle  
\Big[ \, 1 +  \int_0^1dx \int_0^{1-x} dy \frac{ 2m^2    }{  - m^2}    \, \Big]
   = 0
   \, .
\end{eqnarray}
The axial-vector current is {\it conserved} in this limit. 
As shown just below, a vanishing momentum limit $ \bar p^\mu , \, \bar k^\mu\to0 $ 
corresponds to the constant electromagnetic fields. 
Therefore, one finds that the axial-vector current carried by massive fermions is conserved 
in the constant and perturbatively weak electromagnetic fields.\footnote{ 
In the above, we only discussed the perturbatively weak 
electromagnetic fields with the triangle diagrams. 
We could have a contribution of the Schwinger pair production 
when there is a strong electric field 
even if it is a constant field. 
}

When the two photons are on-shell ($ p^2 = k^2 = 0 $), 
we have $ 2 p \cdot k = q^2 $ according to the momentum conservation. 
Therefore, the integral is only a function of the normalized momentum $ q^2/m^2 $. 
Considering the neutral-pion decay $ \pi^0 \to 2\gamma $, 
which is one of the representative consequences of the chiral anomaly, 
we put $ q^2 = m_\pi^2 $ to find a ratio to the current quark mass 
$ m^2/q^2 \sim (5/140)^2 \sim 0.001$. 
Therefore, the quark-mass correction to the decay rate is expected to be quite small. 
The integral can be written with the polylogarithmic functions in this limit if necessary.

\subsubsection*{Limit of a constant magnetic field}

Finally, we examine one more limit in Eq.~(\ref{eq:anomaly-massive}) where one of the photon momentum is zero while the other is finite. 
We take $ p^\mu \to 0 $, and then $ q^\mu = k^\mu $ led by the momentum conservation. 
One can show that the vanishing momentum limit corresponds to 
a constant electromagnetic field as follows. 
For simplicity, we may consider a parallel electromagnetic field in the $  z$ direction. 
Other configurations can be generated by the Lorentz transformation. 
Such a parallel field can be described by a linear gauge configuration 
\begin{eqnarray}
\label{eq:constantf-field}
A^\mu = ( - c_0 z E_0 (\bx_\perp) , - d_1 y  B_0(t,z) ,  d_2 x  B_0(t,z)  , -c_3 t E_0 (\bx_\perp) )
\, ,
\end{eqnarray}
with constants such that $ c_0 + c_3 =1 $ and $ d_1 + d_2 =1 $. 
The gauge transformation does not change the following observations 
since we consider the strengths of the electric and magnetic fields 
\begin{subequations}
\begin{eqnarray}
\tilde E_z(q^\mu) \=  \tilde E_0(\bq_\perp) 
[c_3 \omega \delta'( \omega)  \delta(q_z) + c_0 q_z \delta'(q_z)  \delta(\omega)]
= (2\pi)^2 \delta(\omega) \delta(q_z) \tilde E_0 (\bq_\perp) 
\, ,
\\
\tilde B_z(q^\mu) \=  \tilde B_0(\omega,p_z) 
[d_2 q_x \delta'( q_x)  \delta(q_y) + d_1 q_y  \delta(q_x) \delta'(q_y)  ]
= (2\pi)^2 \delta(q_x) \delta(q_y)  \tilde B_0 (\omega,q_z)
\, .
\end{eqnarray}
\end{subequations}
where the tildes on the fields stand for the Fourier spectrum. 
When the electric field is independent of $ t $ and $  z$, 
its Fourier spectrum corresponds to vanishing $ \omega $ and $ q_z $ 
in spite of the facts that $ q^\mu $ is the momentum of the gauge field 
and that the gauge field itself depends on $ t $ and $  z$. 
In a similar manner, when the magnetic field is independent of $ x $ and $  y$, 
its Fourier spectrum corresponds to vanishing $ q_x $ and $ q_y $. 
When the electromagnetic field is independent of all the spacetime coordinate, 
one gets the four-dimensional delta functions  
\begin{subequations}
\begin{eqnarray}
\tilde E_z(q) \=  (2\pi)^4  \delta^{(4)}(q) E
\, ,
\\
\tilde B_z(q) \=  (2\pi)^4  \delta^{(4)}(q) B
\, ,
\end{eqnarray}
\end{subequations}
where $ E $ and $B  $ are constant strengths.

When $ p^\mu \to 0 $, the AWI (\ref{eq:anomaly-massive}) reads 
\begin{eqnarray}
\label{eq:anomaly-massive-constant-0}
\left. {\rm F.T.} \ \langle p, k | \pd_\mu j^\mu_A | 0\rangle \right|_{p^\mu \to 0}
\=   \frac{q_f^2}{2\pi^2} \frac12 [ \tilde \bE(q)  \cdot  \bB +  \tilde \bB(q)  \cdot  \bE ]
\Big[ \, 1 -  I \Big( \frac{q^2}{4m^2} \Big)   \, \Big] V_4 
\, ,
\end{eqnarray}
with the constant field strengths $  \bE  $ and $  \bB  $ 
and the system volume $ V_4 = (2\pi)^4 \delta^{(4)}(0)  $. 
Remarkably, the integral agrees with the $ I $ function (\ref{eq:I}).\footnote{
To see the agreement, first notice that 
\begin{eqnarray}
  \int_0^1dx \int_0^{1-x} dy   \frac{ 2m^2 }{  x (1-x) q^2  - m^2}
=  \int_0^1dx   \frac{ 2m^2 (1-x) }{  x (1-x) q^2  - m^2}
= \int_0^1dx  \frac{ 2m^2 x }{  (1-x) x q^2  - m^2}
\, ,
\end{eqnarray}
where the last expression is obtained via a change of the integral variable. 
Then, summing the latter two expressions, we get 
\begin{eqnarray}
\int_0^1dx   \frac{ 2m^2 (1-x) }{  x (1-x) q^2  - m^2} 
= \frac12 \int_0^1dx  \frac{ 2m^2 }{  x (1-x) q^2  - m^2}
= - I( q^2/4m^2)
\, .
\end{eqnarray}
When one sends the other momentum $ k^\mu $ to zero
the integral reduces to the $ I $ function in the same manner. 
}
Adding the two cases where either $ p^\mu $ 
or $  k^\mu$ is sent to zero, 
one finds that the AWI in a constant 
and perturbatively weak magnetic field reads 
\begin{eqnarray}
\label{eq:anomaly-massive-constant}
{\rm F.T.} \ \langle p, k | \pd_\mu j^\mu_A | 0\rangle 
\=  \frac{q_f^2}{2\pi^2}  \tilde \bE(q)  \cdot  \bB \Big[ \, 1 -  I \Big( \frac{q^2}{4m^2} \Big)   \, \Big]  V_4 
\, ,
\end{eqnarray}
where we assumed that there is not a constant electric field, i.e., $\bE = 0  $. 
This expression agrees with Eq.~(\ref{eq:AWI}) in the constant and strong magnetic field 
when the electric field is homogeneous in the transverse plane, i.e., $ |\bq_\perp| \to 0 $.

\subsubsection{The lowest Landau level in effective (1+1) dimensions}

\label{anomaly-massive-vac}

We next examine the AWI in the effective (1+1) dimensions 
realized in the LLL. 
One can find more computational details in the appendix of Ref.~\cite{Hattori:2022wao}. 
The anomaly diagram here is the two-point function 
composed of the LLL fermion loop:  
\begin{eqnarray}
\label{eq:axial-correlator}
\int d^4 x \, e^{-i q x} \langle k | \pd_\mu j_A^\mu(x) | 0 \rangle
= \rho_B e^{- \frac{ |\bq_\perp|^2}{2|q_f B|} } 
(2\pi)^2  \delta^{(2)}( k_\para - q_\para) i q_\mu \Pi_{1+1 (A)}^{\mu\nu} (q_\parallel) \epsilon^{\ast}_\nu (q) 
\, .
\end{eqnarray}
The Landau degeneracy factor and the Gaussian come from 
the transverse-momentum integral as in the computation of the vector-current correlator (\ref{eq:VV}). 
The residual (1+1)-dimensional longitudinal part is given as 
\begin{eqnarray}
iq_\mu \Pi_{1+1 (A)}^{\mu\nu} (q_\parallel) 
= (-1) ( - i q_f )\int \!\! \frac{d^2p_\parallel}{(2\pi)^2} 
\frac{\tr [\, \sla q_\para \gam^5 i (\sla p_\parallel + m) \prj_+
\gam_\parallel^\nu i ( \, (\sla p_\parallel + \sla q_\parallel) + m\, ) \prj_+ \, ]}
{ (p_\parallel^2 -m^2)( \,  (p_\parallel + q_\parallel)^2 -m^2 \,)}
\, .
\end{eqnarray} 
The two-dimensional integral is (superficially) logarithmically divergent, 
and we use the dimensional regularization 
with a displacement from two dimensions, i.e., $2 \to d $. 
Similar to the identity (\ref{eq:anom-id2}), we here use another identity 
\begin{eqnarray}
 \sla q_\para\gam^5 = \gam^5  ( \sla p_\para - m)  + ( \sla q_\para + \sla p_\para - m )  \gam^5 
+ 2 ( m - \tilde {\sla p}_\para )  \gam^5
\, .
\end{eqnarray}
Plugging the above identity to the (1+1)-dimensional part, we have 
\begin{eqnarray} 
i q_\mu \Pi_{1+1 (A)}^{\mu\nu} (q_\parallel) \epsilon^{\ast}_\nu (q) 
\= - 2 q_f  \epsilon^{\ast}_\nu (q)  \int \!\! \frac{d^d p_\parallel}{(2\pi)^d} 
\frac{\tr [\,  \tilde {\sla p}_\para \gam^5 (\sla p_\parallel + m) \prj_+
\gam_\parallel^\nu  ( \, (\sla p_\parallel + \sla q_\parallel) + m\, ) \prj_+ \, ]}
{ (p_\parallel^2 -m^2)( \,  (p_\parallel + q_\parallel)^2 -m^2 \,)}
\nnb
&& + 2im \langle \bar \psi \gam^5 \psi \rangle
\label{eq:Pi-A-2}
\, ,
\end{eqnarray} 
where we have 
\begin{eqnarray}
\label{eq:PS-2dim}
 \langle \bar \psi \gam^5 \psi \rangle
 =   ( -1)  [ - i q_f  \epsilon^{\ast}_\nu (q)  ]    \int \!\! \frac{d^2p_\parallel}{(2\pi)^2} 
\frac{\tr [\,  \gam^5 i (\sla p_\parallel + m) \prj_+
  \gam^\nu_\para i ( \, (\sla p_\parallel + \sla q_\parallel) + m\, ) \prj_+ \, ] }
{ (p_\parallel^2 -m^2)( \,  (p_\parallel + q_\parallel)^2 -m^2 \,)}
\, .
\end{eqnarray}
This is the leading nonvanishing matrix element of 
the pseudoscalar condensate, which is represented 
by the diagram with one-insertion of the external photon leg. 
The first term in Eq.~(\ref{eq:Pi-A-2}) 
will result in the anomalous term.

We shall first examine the anomalous term. 
Introducing the Feynman parameter and performing the spinor trace to the first term in Eq.~(\ref{eq:Pi-A-2}), 
one arrives at a momentum integral 
\begin{eqnarray}
  \int \!\! \frac{d^d p_\parallel}{(2\pi)^d}  \frac{  \tilde p_\para^{2}  } { ( p_\para^2 - \Delta_\para )^2 }
= \frac{d-2}{d} \times  \frac{(-1)i}{(4\pi)^{d/2}} \frac{d}{2}
 \frac{ \Gamma(- \frac{d-2}{2}) } {\Gamma(2)} \Delta_\para^{ \frac{d-2}{2}}
 =   \frac{ i}{ 4\pi } 
 \, ,
\end{eqnarray}
where we used $ (d-2) \Gamma(- \frac{d-2}{2}) = - 2 +  \order( d-2) $. 
As in the integral (\ref{eq:mom-integral-triangle}) 
for the triangle diagrams, 
the above integral is finite because of the overall dimensional factor and all the mass dependence goes away on the rightmost side after the limit ($d \to 2 $) is taken. 
the correlator (\ref{eq:Pi-A-2}) results in 
\begin{eqnarray}
\label{eq:axial-Ward-app}
i q_\mu \Pi_{1+1 (A)}^{\mu\nu} (q_\parallel)  \epsilon^{\ast}_\nu (k) 
= - \frac{ q_f}{\pi}   \langle k |  \tilde E_\para | 0 \rangle   +  2im \langle \bar \psi \gam^5 \psi \rangle
\, ,
\end{eqnarray}
where the momentum conservation in Eq.~(\ref{eq:axial-correlator}) is understood 
(with the delta function suppressed). 
The electric field is defined as 
$\tilde  E_\para = - i   \epsilon^{\mu\nu}_\para q_\mu  A_\nu 
$ with $ \epsilon^{\mu\nu}_\para = \epsilon^{\mu\nu12} $ as below Eq.~(\ref{eq:Ward-ids}).

One can straightforwardly compute the matrix element of 
the pseudoscalar condensate (\ref{eq:PS-2dim}) to find that 
\begin{eqnarray}
\label{eq:PS-condensate-LLL}
2im \langle \bar \psi \gam^5 \psi \rangle
=  \frac{s_f  q_f  }{\pi } \tilde E_\para  I \Big( \frac{q_\para^2}{4m^2} \Big)
\, .
\end{eqnarray}
Notice that we have the $ I $ function 
defined in Eq.~(\ref{eq:I}). 
As discussed below Eq.~(\ref{eq:PS_condensate}) 
for the triangle diagrams, 
one needs insertion of external photon legs 
to have a nonvanishing matrix element of 
$\langle \bar \psi \gam^5 \psi \rangle $. 
In the present case, one needs to include 
one external photon leg 
that results in $ \tilde E_\para$ in addition to 
the resummed magnetic-field legs.

Combining the anomalous term and pseudoscalar condensate 
in the above, we obtain  
\begin{eqnarray}
\label{eq:AWI-vacuum}
{\rm F.T.} \ \pd_\mu j_A^\mu  =    \frac{q_f^2}{2\pi^2} \tilde \bE(q) \cdot \bB
 \Big[ \, 1 -   I \Big( \frac{q_\para^2}{4m^2} \Big) \, \Big]    e^{ - \frac{|\bq_\perp|^2}{2|q_fB|}} 
   \, .
\end{eqnarray}
This result reproduces the AWI (\ref{eq:AWI}) obtained 
from $ \Pi_\para $ in Eq.~(\ref{eq:vac_massive}). 
As in the discussions about the triangle diagrams, 
we emphasize that both the mass-independent anomalous term 
and the mass-dependent pseudoscalar condensate stem 
from a single diagram with the massive fermion loop. 
The massive fermion loop vanishes as a whole in the large-mass limit, $\lim_{x\to0}  I(x) =1$. 
This clearly explains the conservation of 
the axial current observed below Eq.~(\ref{eq:AWI}). 
Notice again that this expression (\ref{eq:AWI-vacuum}) agrees with the AWI (\ref{eq:anomaly-massive-constant}) 
from the triangle diagrams in a weak and constant magnetic field 
when the electric field is homogeneous in the transverse plane ($ |\bq_\perp| =0 $).

%

\subsection{Thermal contributions at finite temperature and density}


\label{sec:VP_therm}

When there are thermally populated particles and antiparticles, 
their polarization can contribute to the polarization tensor $ \Pi^{\mu\nu} $. 
Below, we discuss such a medium contribution 
to $\Pi_\para $ in Eq.~(\ref{eq:Pi_LLL}) within the LLL 
in strong magnetic fields. 
We write it as $\Pi_\para = \Pi_\para^\vac + \Pi_\para^\temp $ 
where the vacuum contribution discussed 
in Appendix~\ref{sec:VP_vac} and the medium contribution discussed below are denoted as $\Pi_\para^\vac $ 
and $\Pi_\para^\temp$, respectively. 
To compute the medium contribution, 
one can use either the real-time formalism \cite{Dolan:1973qd, Smilga:1991xa, Baier:1991gg, Fukushima:2015wck, Hattori:2022uzp} 
or the imaginary-time formalism \cite{Kao:1998yt}). 
The equivalence between those results is commented below. 
Since all the computational details are summarized 
in the appendix of Ref.~\cite{Hattori:2022uzp}. 
We here only quote the essential results.

In the strictly massless case, 
one can show that the medium correction to 
the polarization tensor exactly vanishes, i.e., 
$ \Pi_\para^\temp =0 $. 
This result can be also confirmed by the use of 
the bosonization technique 
(see, e.g., Refs.~\cite{Fukushima:2011jc, Fukushima:2015wck}). 
This result implies the absence of medium corrections 
to the AWI (\ref{eq:anomaly-LLL-2}) in the massless case.

Even in the massive case, one finds a partial cancellation. 
As anticipated from the complete cancellation 
in the massless case, the residual terms are 
all proportional to the fermion mass squared $ m^2 $. 
After some computational efforts 
(see Ref.~\cite{Hattori:2022uzp}), 
the medium contribution is obtained as \cite{Fukushima:2015wck} 
\begin{eqnarray}
\Pi_\para^{\temp } (\omega, q_z)
=  \pi m_B^2 \, e^{- \frac{ \vert \bq_\perp \vert ^2}{2 \vert q_fB\vert} } 
\, m^2 \!\! \int_{-\infty}^\infty \frac{d p_z}{2\pi \epsilon_p} 
\frac{ ( q_\parallel^2 + 2 q_z p_z) \,   [n_+(\epsilon_p)  + n_-(\epsilon_p)] } 
{  q_\parallel^2  ( p_z - \frac{1}{2} q_z) ^2 - \frac{\omega^2}{4 }  (q_\parallel^2 -4m^2) }
\label{eq:Pi_medium}
\, ,
\end{eqnarray} 
where $\epsilon_p = \sqrt{p_z^2+m^2}$ 
and $  n_\pm (p^0) = [\, e^{( p^0 \mp \mu)/T} + 1 \, ]^{-1}$ 
with temperature $ T $ and chemical potential $ \mu $. 
The LLL contribution does not yield tensor structures 
other than $P_\para^{\mu\nu} $. 
In fact, there is no other transverse tensor structure that can be constructed in the (1+1) dimensions, 
whereas the projection operator is usually split 
into two components in a medium due to the absence of the Lorentz symmetry. 
Instead, the absence of the Lorentz symmetry manifests itself in 
the remaining $ p_z $ integral that separately depends on the external gluon/photon energy $\omega$ 
and momentum $q_z$ in separate forms 
in addition to the boost-invariant form $  q_\para^2 = \omega^2 - q_z^2$. 
After some algebra, the above expression reduces to 
the one shown in Eq.~(7) of Ref.~\cite{Kao:1998yt}.\footnote{ 
In Ref.~\cite{Kao:1998yt}, the imaginary-time formalism was applied to the computation of 
the polarization tensor in the massive Schwinger model. 
To confirm the agreement, notice that the numerator in the integrand can be rewritten as  
$\{ (\epsilon_\mp - \omega)^2 -\epsilon_\pm^2 \} + \{ (\epsilon_\mp + \omega)^2 -\epsilon_\pm^2 \} 
= 2 ( \omega^2  \mp  2 p_z^\prime q_z  ) = 2 (  q_\para^2  +   2 p_z  q_z   ) $
where $ \epsilon_\pm = \sqrt{ ( p_z^\prime \pm q_z/2)^2 + m^2} $ is defined in Ref.~\cite{Kao:1998yt}. 
We utilized shifts of the integral variable $ p_z^\prime = \mp p_z \pm \frac12 q_z $. 
The denominator can be arranged in the same way.  
}

Noticing that the integrand in Eq.~(\ref{eq:Pi_medium}) 
has poles, one can rearrange it in a more transparent form 
\begin{eqnarray}
\Pi_\para^{\temp }(\omega, q_z) 
=   \frac{ 2 \pi  m^2 m_B^2 }{ q_\para^2\sqrt{1- 4m^2/q_\para^2 }  } e^{ - \frac{|\bq|^2}{2|q_fB|}}
 \int_{-\infty}^\infty \frac{d p_z}{2\pi \epsilon_p} 
[ n_+(\epsilon_p)  + n_-(\epsilon_p)]
\Big( \frac{ \epsilon_p^+ s_+ }{ p_z - p_z^+ } - \frac{ \epsilon_p^- s_-}{ p_z - p_z^- } \Big)
\label{eq:Pi_medium-1}
\, .
\end{eqnarray} 
With the infinitesimal imaginary part $ \omega \to \omega + i \epsilon $ for the retarded correlator, 
the pole positions are found to be 
\begin{eqnarray}
\label{eq:poles}
p^{\pm}_z  = \frac{1}{2} \big( \, q_z \pm \omega \sqrt{1- 4m^2/q_\para^2 } \ \big) \pm i   \epsilon
\, .
\end{eqnarray}
Accordingly, we define corresponding energies  
\begin{eqnarray}
\label{eq:poles-energy}
\ep_p^\pm &\equiv& \sqrt{ (p^{\pm}_z)^2 + m^2 }
= \frac{1}{2} \Big \vert  \omega \pm  q_z   \sqrt{ 1 -  4m^2/q_\para^2 } \ \Big \vert
\, .
\end{eqnarray}
Note that those are complex-valued quantities 
when $ 1- 4m^2/q_\para^2 <0$, meaning that there is 
no pole contribution to the integral in this kinematical region. 
We also introduced sign functions 
\begin{eqnarray}
\label{eq:energy-signs}
s_\pm &:=& \sgn \Big(\omega \pm q_z    \sqrt{1- 4m^2/q_\para^2 } \Big)
\, ,
\end{eqnarray}
for $ 1- 4m^2/q_\para^2 > 0 $. 
When $  1 -  4m^2/q_\para^2 \leq 0 $ and thus $ \ep_p^\pm $ is complex-valued, 
we define that $ s_\pm =1 $ just to maintain the original form 
$ \epsilon_p^\pm s_\pm = \frac{1}{2}  \Big (  \omega \pm   q_z   \sqrt{ 1 -  4m^2/q_\para^2 } \Big)$.

\subsubsection{Thermally induced imaginary parts}
 
We closely look into the medium effects contained in 
the imaginary part of the integral expression (\ref{eq:Pi_medium-1}). 
Since the imaginary part is related to the squared amplitude of on-shell processes, 
it helps us to intuitively understand how the interplay between the effects of the strong magnetic field 
and medium work in the relevant processes.

The denominator of the integrand in Eq~(\ref{eq:Pi_medium-1}) 
has a definite sign as long as $ 1- 4m^2/q_\para^2  <0 $, that is, $0 < q_\parallel^2 < 4m^2$. 
In this kinematics, the regular $p_z$-integral does not generate any imaginary part. 
However, when $q_\parallel^2 \leq0$ or $q_\parallel^2 \geq 4m^2$, 
the integral picks up the pole contributions (\ref{eq:poles}), 
implying existence of an imaginary part in these regimes. 
By the use of the formula $ 1/(x \pm i \epsilon) = P(1/x) \mp i\pi \delta (x) $, 
one can extract the imaginary part of the polarization tensor:  
We obtain the total imaginary part 
\begin{eqnarray}
\label{eq:imag-sum}
\Im m \Pi_\para \= \Im m \Pi_\para^{\vac} + \Im m \Pi_\para^{\temp} 
\\
\=  -  \pi m_B^2 \, e^{- \frac{ \vert \bq_\perp \vert ^2}{2 \vert q_fB\vert} } 
\,  \frac{ 2  m^2   }{  q_\parallel^2 \sqrt{ 1 -4m^2 /q_\parallel^2 } }  
\Big[  (1- N_+) \theta( q_\para^2 - 4m^2)  \sgn(\omega) 
-  N_-   \theta( - q_\para^2 )  \Big]
\nn
\, ,
\end{eqnarray}
where we defined $ s_\pm = \theta(q_\para^2 -4m^2) \sgn(\omega) \pm \theta(-q_\para^2)$ and 
$ N_\pm =  \frac12 \{ n_+(\epsilon_p^+)  +  n_- (\epsilon_p^+) \}
\pm  \frac12\{ n_+(\epsilon_p^-)  +  n_- (\epsilon_p^-) \} $. 
The above result is symmetric under the sign flip of the chemical potential $ \mu \to - \mu $ as expected. 
One can combine the two kinematical regions 
$q_\para^2 <0 $ and $ q_\para^2 - 4m^2 > 0 $ to get 
another expression by hyperbolic functions: 
\begin{eqnarray}
\label{eq:imag-total-density}
\Im m \Pi_\para  
=  -  \pi m_B^2 \, e^{- \frac{ \vert \bq_\perp \vert ^2}{2 \vert q_fB\vert} } 
\frac{ 2  m^2   }{  q_\parallel^2 \sqrt{ 1 -4m^2 /q_\parallel^2 }  }
\cdot \frac12 \sum_{c=\pm}
\frac{   \sinh( \frac{ \omega }{2T})  \, \theta (1-  4m^2/q_\para^2 )  }
{ \cosh( \frac{\omega}{2T}  ) + \cosh \big( \, \frac{q_z }{2T}  \sqrt{1-  4m^2/q_\para^2}
+ c \frac{\mu}{T} \, \big)  }
\, .
\end{eqnarray} 
At a vanishing chemical potential $ \mu \to 0 $, 
this expression essentially agrees with those in Ref.~\cite{Kao:1998yt} (up to a typo there) 
and in Ref.~\cite{Baier:1991gg} where the authors compute the Feynman causal correlator 
instead of the retarded correlator. 
One can immediately find that 
the above result (\ref{eq:imag-sum}), 
and thus (\ref{eq:imag-total-density}) as well, 
is an odd function of $ \omega $, 
which originates from a general property of the spectral density (see, e.g., Ref.~\cite{Kapusta:2006pm}).

Now, we shall examine the physical processes that give rise to 
the imaginary part in the regions $ q_\para^2 > 4m^2 $ and $ q_\para^2 < 0 $ separately. 
Below, we refer those regions ``time-like'' and ``space-like'' as in the (1+1)-dimensional case. 
Notice that the ``time-like'' condition $ q_\para^2 > 4m^2 $ can be compatible with 
the light-like on-shell condition $ q^2 = q_\para^2 + q_\perp^2 = 0 $ in the four dimensions.

\cout{

One can confirm that the zero-temperature limit of Eq.~(\ref{eq:imag-total}) 
agrees with the vacuum expression (\ref{eq:vac_final}) as follows. 
In this limit, we have 
\begin{eqnarray}
\label{eq:zero-T}
\lim_{T \to 0} \Im m \Pi_\para  
= -  \pi m_B^2 \, e^{- \frac{ \vert \bq_\perp \vert ^2}{2 \vert q_fB\vert} } 
 \frac{ 2  m^2   }{  q_\parallel^2 \sqrt{ 1 -4m^2 /q_\parallel^2 } }  
\lim_{T \to 0}   \frac{   \{\theta(\omega) - \theta(-\omega) \} \theta (1-  4m^2/q_\para^2 )  }
{ 1  + \exp \big [ \,  \frac{ 1 }{2T}  \big ( |q_z| \sqrt{1-  4m^2/q_\para^2} - |\omega| \big ) \, \big ]  }
\, .
\end{eqnarray} 
The absolute values of $ \omega,q_z $ come from the hyperbolic cosine functions. 
The limit depends on the sign on the shoulder of the exponential. 
As expected, the limit is vanishing in the space-like region $ q_\para^2 < 0 $, 
while it reproduces the finite value of the imaginary part 
in vacuum in the time-like region $ q_\para^2 > 4m^2 $ [see Eq.~(\ref{eq:vac_massive})]. 

}

\subsubsection*{Finite temperature}

Here, we consider the zero-density case $ \mu = 0  $, 
and both the fermion and antifermion distribution functions reduce to the same form 
$ n_\pm (p^0) \to n(p^0) = [  e^{p^0/T} + 1 ]^{-1}  $. 
We first focus on the high-temperature limit where $ T \gg \omega, q_z $. 
Observe that the leading term in the high-temperature expansion of $ N_+ $ exactly cancels the vacuum contribution in Eq.~(\ref{eq:imag-sum}), i.e., 
\begin{eqnarray}
1- N_+  \sim   \frac{1}{4T}  ( \epsilon_p^+ + \epsilon_p^-)  =   \frac{|\omega|}{4T} 
\, .
\end{eqnarray}
Consequently, the total imaginary part is highly suppressed 
in the time-like region by a factor of  $ |\omega|/(4T) \ll 1$. 
One can interpret this result in the following way. 
In vacuum, the imaginary part indicates a pair creation from an incident photon/gluon. 
However, in the presence of a heat bath, its inverse process is also activated 
because thermal fermions and antifermions can annihilate into a photon/gluon.\footnote{
Note that kinematics of the 1-to-2 and 2-to-1 processes is allowed in a magnetic field 
even for an on-shell photon, though it is forbidden in the four dimensions without a magnetic field. 
} 
Due to the detail balance between these processes, 
the net imaginary part in the time-like regime is suppressed at high temperature. 
One can confirm this observation with a simple identity 
\begin{eqnarray}
\label{eq:N+---0}
1 - N_+  =    [ 1- n(\ep_p^+) ] [ 1-  n(\ep_p^-)]  - n(\ep_p^+)  n(\ep_p^-) 
\, .
\end{eqnarray}
The first term corresponds to the pair-creation channel $ \gam \to f \bar f $ 
that is subject to the Pauli-blocking effect in the final state, 
while the second term corresponds to the pair-annihilation channel $ f \bar f  \to  \gam $. 
Here, we also confirm the energy conservation 
$ |\omega| =  \ep_p^+ + \ep_p^-$ 
according to Eq.~(\ref{eq:poles-energy}). 
The net pair creation occurs with a significant rate only in 
a sufficiently high energy regime $\omega , q_z \gg T  $, 
where the thermal contribution is exponentially suppressed.
In such cases, the vacuum contribution stands as the dominant contribution to the imaginary part.

On the other hand, one finds a new medium-induced channel in the space-like regime, 
that is, a contribution from the Landau damping proportional to $N_-  $. 
The Landau damping is purely a medium effect where a medium fermion (antifermion) 
is scattering off a space-like photon. 
In case of vacuum, the imaginary parts in the space-like regime 
cancel out in the final expression of the polarization tensor 
in all order of the Landau levels \cite{Hattori:2012je}. 
One finds an identity 
\begin{eqnarray}
- N_- =    n(\ep_p^-)  [ 1- n(\ep_p^+) ] - n(\ep_p^+) [ 1-  n(\ep_p^-)] 
\, ,
\end{eqnarray} 
where the first (second) term corresponds to the gain (loss) term 
for the phase-space volume at $ p^0 = \ep_p^+ $. 
Namely, this identity shows the detail balance 
between the scatterings $ f + \gam^\ast \to f $ 
and $ f \to f  + \gam^\ast $. 
The Pauli blocking effect appears in the final states 
on the right-hand side of the above identity. 
We confirm the energy conservation 
$  \omega + \ep_p^- = \ep_p^+$ 
according to Eq.~(\ref{eq:poles-energy}).

\subsubsection*{Finite density}

At nonzero density $ (\mu \not = 0) $, 
the fermion and antifermion distribution functions should be distinguished. 
Nevertheless, we can interpret the pair creation/annihilation and the Landau damping 
in a similar way to the zero-density case discussed above.

Similar to the zero-density case (\ref{eq:N+---0}), 
one finds an identity 
\begin{eqnarray}
\label{eq:id-density-1}
1 - N_+ 
 &=& \sum_{\a=\pm}  \frac12 \left[ \ \{ 1 - n_+(\ep_p^\a) \}\{ 1-  n_-(\ep_p^{-\a}) \} 
 -  n_+(\ep_p^\a)  n_-(\ep_p^{-\a})  \ \right]
 \, .
\end{eqnarray}
This expression has the same form as in Eq.~(\ref{eq:N+---0}) 
up to the difference between the fermion and antifermion distribution functions. 
In the pair creation/annihilation channel, the fermion and antifermion can have different energies 
$ \ep_p^+ $ and $  \ep_p^- $ when the photon momentum is nonzero, i.e., $ q_z \not = 0 $. 
Thus, there are two kinematical windows depending on either a fermion or antifermion takes $ \ep_p^+ $. 
Clearly, interchanging $ \ep_p^+ $ and $  \ep_p^- $ brings one kinematics to the other. 
A finite chemical potential distinguishes those two cases, 
whereas they are degenerate at zero density (\ref{eq:N+---0}).

In the space-like region, the Landau damping of the fermions and antifermions occurs 
with different magnitudes. 
Separating the fermion and antifermion distribution functions, 
we find an identity 
\begin{eqnarray}
\label{eq:id-density-2}
- N_- 
&=&  \sum_{c=\pm} \frac12 \left [ \   n_{c}(\ep_p^-) \{ 1 - n_{c}(\ep_p^+) \}
 - n_{c}(\ep_p^+) \{1-  n_{c}(\ep_p^-) \}  \ \right]
\, .
\end{eqnarray}
The sum over $ c = + $ ($  - $) is for the Landau damping of fermions (antifermions), 
assuming that $ \mu > 0 $ without lose of generality.

Notice that, at zero temperature, the fermion distribution functions reduce to step functions, and the Pauli blocking factors vanish, i.e., $1-  n_{\pm} \to 0 $ 
below the Fermi surface. 
That is, the Pauli blocking at zero temperature  
requires photons/gluons to be energetic enough 
for the above on-shell processes to occur, 
leading to threshold shifts. 
This can be also confirmed on the basis of 
the analytic result of the integral (\ref{eq:Pi_medium-1}) 
that can be performed exactly at zero temperature \cite{Hattori:2022uzp}.

\cout{

At zero temperature, the distribution functions reduce 
to the step functions $ n_\pm(p^0) = \theta( \pm \mu - p^0 ) $. 
Thus, there is no annihilation channel, and only the pair-creation channel is left as 
\begin{eqnarray}
\label{eq:pair-zero-T}
\left. 1 - N_+ \right|_{T=0} 
 &=& \frac12 [\,  \theta (\ep_p^+ -  \mu  )  + \theta ( \ep_p^- -  \mu  )  \,]
 \, .
\end{eqnarray}
Either fermion or antifermion is free of the Pauli-blocking effect 
since its positive-energy state is completely vacant. 
However, the other one needs to find an unoccupied state 
above the Fermi surface to avoid the Pauli-blocking, resulting in the above step functions. 
On the other hand, the Landau damping occurs 
between an occupied initial state and an unoccupied final state. 
Thus, there are only fermion contributions (when we assume $ \mu > 0 $ as above). 
The gain and loss terms read  
\begin{eqnarray}
- \left.  N_- \right|_{T=0} 
&=&   \frac12 [ \,  \theta (  \mu - \ep_p^- ) \theta (\ep_p^+ -  \mu  ) 
-  \theta (  \mu - \ep_p^+ ) \theta (\ep_p^- -  \mu  )   \,]
\, .
\end{eqnarray} 
It is a simple task to confirm the zero-density limit $ (\mu \to 0) $, 
where we find that 
\begin{subequations}
\begin{eqnarray}
&&
\lim_{\mu \to 0} \left. 1 - N_+ \right|_{T=0} = 1
\, ,
\\
&&
\lim_{\mu \to 0} \left.  N_- \right|_{T=0} = 0
\, .
\end{eqnarray}
\end{subequations}
These limits reproduce the vacuum expression (\ref{eq:vac_final}). 

}

\cout{

\subsection{Thermal contributions at finite temperature and density}

When there are thermally populated particles and antiparticles, 
their polarization can contribute to the polarization tensor $ \Pi^{\mu\nu} $. 
We discuss such a medium contribution from the LLL fermions in the strong magnetic fields.

\subsubsection{Self-energy from the real-time formalism}

\label{sec:VP_therm}

We use the real time formalism to compute the medium contribution 
at finite temperature/density \cite{Dolan:1973qd, Smilga:1991xa, Baier:1991gg, Fukushima:2015wck}. 
The equivalent computation can be performed by the imaginary time formalism 
as well (see, e.g., Ref.~\cite{Kao:1998yt}). 
Below, we will explicitly confirm the agreement between the results 
from these formalisms in the literature.

The vacuum LLL propagator is given in Eq.~(\ref{eq:prop-LLL}). 
In the r-a basis of the real-time formalism, this propagator is extended to the following forms  
\begin{subequations}
\label{eq:S-all}
\begin{eqnarray}
\label{eq:Sra}
 S_{ra}(p) &=& 2 i\, e^{- \frac{ \vert \bm p_\perp \vert^2}{ |q_f  B|} } \prj_+ \,
  { \slashed p_\parallel + \mq  \over p_\parallel^2-\mq^2}
  \Big|_{p^0\to p^0+i\varepsilon}
  \,, 
\\
\label{eq:Sar}
 S_{ar}(p) &=& 2 i\, e^{- \frac{ \vert \bm p_\perp \vert^2}{ |q_f  B|} } \prj_+ \,
  {\slashed p_\parallel+\mq \over p_\parallel^2-\mq^2}
  \Big|_{p^0\to p^0-i\varepsilon} 
\,, 
\\
\label{eq:Srr}
 S_{rr}(p) &=& \big[{1\over 2} - n_+(p^0)\big]\bigl[S_{ra}(p)  -S_{ar}(p)\bigr]
\, ,
\\
 S_{aa} (p) &=& 0
 \, ,
\end{eqnarray}
\end{subequations}
where $  n_\pm (p^0) = [\, e^{( p^0 \mp \mu)/T} + 1 \, ]^{-1}$ 
with temperature $ T $ and chemical potential $ \mu $. 
By the use of these propagators, the medium contribution to 
the retarded correlator is written down as 
\begin{eqnarray}
\Pi_R^{ \mu\nu} &=& \M_1^{ \mu\nu} + \M_2^{ \mu\nu} 
\, ,
\end{eqnarray}
where 
\begin{subequations}
\begin{eqnarray}
\label{eq:med1}
i \M_1^{ \mu\nu}  &=& - (ig)^2 \int \frac{d^4p}{(2\pi)^4} \tr[ \gam^\mu S_\ar(p+q) \gam^\nu S_\rr(p) ]
\, ,
\\  
\label{eq:med2}
i \M_2^{ \mu\nu}  &=& - (ig)^2 \int \frac{d^4p}{(2\pi)^4} \tr[ \gam^\mu S_\rr(p+q) \gam^\nu S_\ra(p) ]
\, . 
\end{eqnarray}
\end{subequations} 
Note that one of the propagators $   S_{rr}(p) $ is proportional to the spectral density 
\begin{eqnarray}
\rho (p) = S_\ra(p) - S_\ar(p) 
= e^{-\frac{|\bp_\perp|^2}{ |q_f B| }} ( \slashed p_\parallel + m)   \prj_+
\cdot  \frac{2\pi}{\epsilon_p}  \left[ \delta( p^0  - \epsilon_p) - \delta( p^0  + \epsilon_p) \right]
\, ,
\end{eqnarray}
where $\epsilon_p = \sqrt{p_z^2+m^2}$. 
Having computed the vacuum part in the previous subsection, 
we drop the vacuum contributions that are independent of the temperature and chemical potential. 
Then, the medium-induced part in $   S_{rr}(p) $ reads 
\begin{eqnarray}
 S_{rr}(p) = - n_+(p^0)  e^{ - \frac{|\bp_\perp|^2}{ |q_f B| }} ( \slashed p_\parallel + m)   \prj_+
\cdot  \frac{2\pi}{\epsilon_p}  \left[ \delta( p^0  - \epsilon_p) - \delta( p^0  + \epsilon_p) \right]
\, .
\end{eqnarray}
The integral for the transverse momentum $  p_\perp$ is 
again just the Gaussian integral (\ref{eq:I_Gaussian}), which results in the density of states. 
As in the vacuum case, it is clear already at this stage that 
the polarization tensor, which is a dimension-two quantity 
in the four dimensions, has the overall factor of $ q_f B$, 
so that the remaining longitudinal part should be given by 
a function of dimensionless combinations of $ q_\para^2$, $ m $, $ q^0 $, $ T $, and $ \mu $. 
Especially, this implies that the thermal contribution 
cannot appear with the overall factor of $ (gT)^2$ due to the dimensional reason, 
showing a clear contrast to the four-dimensional thermal field theory. 



Inserting the propagators (\ref{eq:S-all}) into Eq.~(\ref{eq:med1}), we have 
\begin{eqnarray}
  \M_1^{\mu\nu}
&=& -   g^2 \frac{ |q_f B| }{8\pi} e^{- \frac{ \vert \bq_\perp \vert ^2}{2 |q_f B| } }
\int \frac{d p_z}{2\pi} \frac{ 1 }{ \epsilon_p}
\biggl[\, 
n_+(\epsilon_p) \frac{\tr[ \gam^\mu_\parallel  ( \slashed p_+ + \slashed q_\parallel )
\gam^\nu_\parallel   \slashed p _{+}   ] + 4m^2 g^{\mu\nu}_\parallel }
{ (p_++q_\parallel)^2 - m^2 -i \sgn(\epsilon_p+q^0) \varepsilon }
\nonumber
\\
&&\hspace{4.3cm} + n_-(\epsilon_p)
\frac{\tr[ \gam^\mu_\parallel   ( \slashed p_- + \slashed q_\parallel)
\gam^\nu_\parallel   \slashed p _{-}  ]
+ 4m^2 g^{\mu\nu}_\parallel  }
{ (p_-+q_\parallel)^2 - m^2 -i \sgn(-\epsilon_p+q^0) \varepsilon }
\, \biggl]
\, ,
\end{eqnarray}
where $p_{ \pm } = ( \pm \epsilon_p , 0, 0, p_z)$. 
We just consumed the delta functions with the $ p^0 $ integral 
and used a relation $ n_\pm (-x) = 1- n_\mp (x)   $ to drop the vacuum part. 
The same manipulation enables us to get the other contribution 
\begin{eqnarray}
  \M_2^{\mu\nu}
&=& -  g^2 \frac{ |q_f B| }{8\pi} e^{- \frac{ \vert \bq_\perp \vert ^2}{2  |q_f B|  } }
\int \frac{d p_z}{2\pi} \frac{ 1 }{ \epsilon_p}
\biggl[\, 
n_+(\epsilon_p) \frac{\tr[ \gam^\mu_\parallel \slashed p _{+} 
\gam^\nu _\parallel ( \slashed p_+ - \slashed q_\parallel) ] 
+ 4m^2 g^{\mu\nu}_\parallel }
{ (p_+-q_\parallel)^2 - m^2  +  i \sgn(\epsilon_p-q^0) \varepsilon }
\nonumber
\\
&&\hspace{4.3cm} 
+ n_-(\epsilon_p) \frac{\tr[ \gam^\mu_\parallel \slashed p _{-} 
\gam^\nu_\parallel ( \slashed p_- - \slashed q_\parallel ) ] 
+ 4m^2 g^{\mu\nu}_\parallel }
{ (p_- - q_\parallel)^2 - m^2  +  i \sgn(-\epsilon_p-q^0) \varepsilon }
\, \biggl]
\, ,
\end{eqnarray}
where we shifted the integral variables as 
$ p_\para^\mu \to p_\para^{\prime \mu} =  p_\para^\mu + q_\para^\mu $. 
Notice that changing the integral variable $p_z \to - p_z$ 
induces a change $ p_{ \pm} \to (\pm \epsilon_p, 0, 0, -p_z) =  - p_{ \mp} $ 
and also that the spinor trace is symmetric, i.e., $\tr[\gam^\mu \slashed p_1 \gam^\nu \slashed p_2]
= \tr[\gam^\mu \slashed p_2 \gam^\nu \slashed p_1]$ for arbitrary vectors $ p_{1,2}^\mu $. 
$ \tr[ \gam^\mu \gam^\a \gam^\nu \gam^\b ] = 4 ( g^{\mu\a} g^{\nu\b} - g^{\mu\nu} g^{\alpha\beta} 
+ g^{\mu\b} g^{\nu\b} ) $. 
It follows from those observations that $  \M_2^{\mu\nu}  $ has the same form as $   \M_1^{\mu\nu} $ 
up to simultaneous replacements $ n_\pm (\epsilon_p) \to n_\mp (\epsilon_p) $. 
Therefore, the sum of the two contributions reads 
\begin{eqnarray}
\label{eq:split-M}
\Pi_\temp^{\mu\nu} =    \M_1^{\mu\nu} +  \M_2^{\mu\nu} 
=  -  g^2 \frac{ |q_f B| }{2\pi} e^{- \frac{ \vert q_\perp \vert ^2}{2 |q_f B| } } \M_\para^{\mu\nu}
\, ,
\end{eqnarray}
where the fermion distribution functions are factorized in the additive form as 
\begin{eqnarray}
\M^{\mu\nu}_\para \= 
\frac14 \int \frac{d p_z}{2\pi} \frac{ n_+(\epsilon_p) +n_-(\epsilon_p)  }{ \epsilon_p}
\nnb 
&& \times
 \biggl[\, 
\frac{\tr[ \gam^\mu_\parallel ( \slashed p_+ - \slashed q_\parallel )
\gam^\nu_\parallel  \slashed p _{ + }  ] 
+ 4m^2 g^{\mu\nu}_\parallel }{ (p_+ - q_\parallel)^2 - m^2   }
+ 
\frac{\tr[ \gam^\mu_\parallel ( \slashed p_- - \slashed q_\parallel )
\gam^\nu _\parallel   \slashed p _{-}  ] 
+ 4m^2 g^{\mu\nu}_\parallel } { (p_- -q_\parallel)^2 - m^2    }
\, \biggl]
\label{eq:M_para}
\, .
\end{eqnarray}
The explicit forms of the denominators are given as 
$ ( p_\pm - q_\parallel)^2 -m ^2 =  2 ( \mp \omega \epsilon_p + q_z p_z + \frac{1}{2} q_\parallel^2 ) $ 
with $q_\parallel^\mu = (\omega, 0, 0, q_z)$. 
Putting the whole integrand over a common denominator, 
the denominator reads 
\begin{eqnarray}
\{( p_+ - q)_\parallel^2 -m ^2\} \{( p_- - q)_\parallel^2 -m ^2\} 
&=& - 4 \{ q_\parallel^2 ( p_z - \frac{1}{2} q_z) ^2
- \frac{1}{4} \omega^2 (q_\parallel^2 -4m^2) \}
\, .
\end{eqnarray}
As long as $0 < q_\parallel^2 < 4m^2$, this denominator has a definite sign, and the $p_z$ integral is regular. 
When $q_\parallel^2 \geq 4m^2$ or $q_\parallel^2 \leq0$, 
the denominator changes the sign, implying emergence of an imaginary part from the singularity.

Next, we perform the spinor trace in the numerators. 
We examine the diagonal and off-diagonal Lorentz components separately. 
In the diagonal components, the trace is carried out as 
\begin{eqnarray}
T_\pm ^\diag =
\frac14 \tr[ \gam^\mu_\parallel ( \slashed p_\pm - \slashed q_\parallel  )
\gam^\nu_\parallel  \slashed p _{ \pm}  ]
= ( \epsilon_p^2 + p_z^2  - q_z p_z )  \mp \omega \epsilon_p 
\, .
\end{eqnarray}
After the reduction to common denominator, the numerator is arranged as 
\begin{eqnarray}
&& \hspace{-1cm}
T_+^\diag  \{( p_- - q)_\parallel^2 -m ^2\} + T_-^\diag  \{( p_+ - q)_\parallel^2 -m ^2\} 
\nnb
\= 8 q_z p_z \{ (p_z- \frac{1}{2} q_z )^2 - \frac{1}{4} \omega^2 \}
+ 4m^2 \{ q_z p_z - \omega^2 + \frac{1}{2} q_\parallel^2 \} 
\, .
\end{eqnarray}
As we will see shortly, the mass-independent part vanishes identically 
when integrated with respect to $p_z$. 
Likewise, the trace for the off-diagonal (0,3) component is carried out as 
\begin{eqnarray}
T_\pm^{03} = \frac14 \tr[ \gam^0_\parallel ( \slashed p_\pm - \slashed q_\parallel  )
\gam^3_\parallel  \slashed p _{ \pm}  ] 
=  - \omega p_z  \pm ( 2 \epsilon_p p_z -  q_z \epsilon_p )
\, ,
\end{eqnarray}
leading to the corresponding numerator 
\begin{eqnarray}
&& \hspace{-1cm}
T_+^{03}  \{( p_- - q)_\parallel^2 -m ^2\}  + T_-^{03}  \{( p_+ - q)_\parallel^2 -m ^2\} 
\nnb
&= &
8 \omega p_z \{ (p_z- \frac{1}{2} q_z )^2 - \frac{1}{4} \omega^2 \}
+ 4 m^2 \{ 2 \omega p_z - \omega q_z \}
\, .
\end{eqnarray}
The trace should be symmetric, i.e., $ T_\pm^{03} = T_\pm^{30} $, 
because there is no room for the completely antisymmetric tensor to appear without $ \gam^5 $.

Now, one can show that the medium correction to 
the polarization tensor vanishes in the massless case. 
Both the numerator and denominator are 
proportional to $ (p_z- \frac{1}{2} q_z )^2 - \frac{1}{4} \omega^2  $ to cancel each other. 
What remains is a simple integral 
\begin{eqnarray}
\left.  \M_\para \right|_{m=0} 
&\propto& \int_{-\infty}^\infty \frac{d p_z}{2\pi} \frac{ n_+(\epsilon_p) + n_-(\epsilon_p) }{ \epsilon_p} p_z = 0
\, ,
\end{eqnarray}
where $\epsilon_p =  |p_z| $ at $ m=0 $. Clearly, this integral is vanishing.\footnote{
The correct dispersion relation of the massless LLL fermions is 
$\epsilon_p =  \pm p_z $ without the symbol of absolute value (see Sec.~\ref{sec:massless}). 
Nevertheless, the spectral function $ \rho(p) \propto \left[ \delta( p^0  - p_z) + \delta( p^0 + p_z) \right] /p^0  $ 
is still an even function of $ p_z $, allowing a replacement $  p_z \to |p_z| $ in $ \epsilon_p  $. 
Thus, the conclusion still holds with the correct dispersion relation. 
} 
This fact can be confirmed by the use of 
the bosonization technique as well (see, e.g., Refs.~\cite{Fukushima:2011jc, Fukushima:2015wck}). 

%

Even in the massive case, one finds a partial cancellation between the numerator and denominator. 
As anticipated from the complete cancellation in the massless case, 
the residual terms are all proportional to the fermion mass $ m^2 $. 
Including the terms proportional to $ 4m^2 g^{\mu\nu}_\parallel  $ in Eq.~(\ref{eq:M_para}) as well, 
we finally arrive at~\cite{Fukushima:2015wck} 
\begin{eqnarray}
\label{eq:722}
\M_\para^{\mu\nu} 
= -  m^2 P_\para^{\mu\nu} 
\!\! \int_{-\infty}^\infty \frac{d p_z}{2\pi \epsilon_p} 
\frac{ ( q_\parallel^2 + 2 q_z p_z) \,   [n_+(\epsilon_p)  + n_-(\epsilon_p)] } 
{  q_\parallel^2  ( p_z - \frac{1}{2} q_z) ^2 - \frac{\omega^2}{4 }  (q_\parallel^2 -4m^2) } 
\, .
\end{eqnarray}
Notice that the diagonal and off-diagonal components are combined 
in the form of $ P_\para^{\mu\nu}  $, indicating manifest transversality of the polarization tensor, 
serving as a consistency check of the calculation. 
While the transverse projection operator is usually split 
into two components in a medium due to the absence of the Lorentz symmetry, 
there is no other transverse tensor structure that can be constructed in the (1+1) dimensions. 
Instead, the absence of the Lorentz symmetry manifests itself in 
the remaining $ p_z $ integral that separately depends on the external gluon/photon energy $\omega$ 
and momentum $q_z$ in separate forms 
in addition to the boost-invariant form $  q_\para^2 = \omega^2 - q_z^2$. 
Matching the above result to the form in Eq.~(\ref{eq:selfenergy}) together with 
the transverse factors in Eq.~(\ref{eq:split-M}), the medium correction to $ \Pi_\para $ is obtained as 
\begin{eqnarray}
\Pi_\para^{\temp }
=  \pi m_B^2 \, e^{- \frac{ \vert \bq_\perp \vert ^2}{2 \vert q_fB\vert} } 
\, m^2 \!\! \int_{-\infty}^\infty \frac{d p_z}{2\pi \epsilon_p} 
\frac{ ( q_\parallel^2 + 2 q_z p_z) \,   [n_+(\epsilon_p)  + n_-(\epsilon_p)] } 
{  q_\parallel^2  ( p_z - \frac{1}{2} q_z) ^2 - \frac{\omega^2}{4 }  (q_\parallel^2 -4m^2) }
\label{eq:Pi_medium}
\, .
\end{eqnarray}
After some algebra, the above expression reduces the one shown in Eq.~(7) of Ref.~\cite{Kao:1998yt} 
up to an overall factor of $ - \tr[t^a t^b]  \rho_B \exp (- \frac{ \vert \bq_\perp \vert ^2}{2 |q_fB|} ) $. 
There, the imaginary-time formalism was applied to the computation of 
the polarization tensor in the massive Schwinger model, 
and the differences come from the color trace 
and the Landau degeneracy factor in the magnetic field.\footnote{ 
To confirm the agreement, notice that the numerator in the integrand can be rewritten as  
$\{ (\epsilon_\mp - \omega)^2 -\epsilon_\pm^2 \} + \{ (\epsilon_\mp + \omega)^2 -\epsilon_\pm^2 \} 
= 2 ( \omega^2  \mp  2 p_z^\prime q_z  ) = 2 (  q_\para^2  +   2 p_z  q_z   ) $
where $ \epsilon_\pm = \sqrt{ ( p_z^\prime \pm q_z/2)^2 + m^2} $ is defined in Ref.~\cite{Kao:1998yt}. 
We utilized shifts of the integral variable $ p_z^\prime = \mp p_z \pm \frac12 q_z $. 
The denominator can be arranged in the same way.  
}

Notice that the integrand in Eq.~(\ref{eq:Pi_medium}) has poles. 
With the infinitesimal imaginary part $ \omega \to \omega + i \epsilon $ for the retarded correlator, 
the pole positions are found to be 
\begin{eqnarray}
\label{eq:poles}
p^{\pm}_z  = \frac{1}{2} \big( \, q_z \pm \omega \sqrt{1- 4m^2/q_\para^2 } \ \big) \pm i   \epsilon
\, .
\end{eqnarray}
Accordingly, we define energies  
\begin{eqnarray}
\ep_p^\pm &\equiv& \sqrt{ (p^{\pm}_z)^2 + m^2 }
= \frac{1}{2} \Big \vert  \omega \pm  q_z   \sqrt{ 1 -  4m^2/q_\para^2 } \ \Big \vert
\, .
\end{eqnarray}
They are complex-valued quantities when $ 1- 4m^2/q_\para^2 <0$, 
meaning that there is no pole contribution to the integral in this kinematics. 
Then, the integrand in Eq.~(\ref{eq:Pi_medium}) can be rearranged 
in a simple and convenient form 
\begin{eqnarray}
\tilde \Pi_\para^{\temp }(\omega, q_z) 
=   \frac{ 2 \pi  m^2  }{ q_\para^2\sqrt{1- 4m^2/q_\para^2 }  }
 \int_{-\infty}^\infty \frac{d p_z}{2\pi \epsilon_p} 
[ n_+(\epsilon_p)  + n_-(\epsilon_p)]
\Big( \frac{ \epsilon_p^+ s_+ }{ p_z - p_z^+ } - \frac{ \epsilon_p^- s_-}{ p_z - p_z^- } \Big)
\label{eq:Pi_medium-1}
\, ,
\end{eqnarray}
where we used $ q_\parallel^2 + 2 q_z p_z^\pm =    2 \omega \epsilon_p^\pm s_\pm$ 
and $p_z^+ - p_z^- =  \omega  \sqrt{1- 4m^2/q_\para^2 }  $.
We also introduced sign functions 
\begin{eqnarray}
\label{eq:energy-signs}
s_\pm &:=& \sgn \Big(\omega \pm q_z    \sqrt{1- 4m^2/q_\para^2 } \Big)
\, ,
\end{eqnarray}
for $ 1- 4m^2/q_\para^2 > 0 $. 
When $  1 -  4m^2/q_\para^2 \leq 0 $ and thus $ \ep_p^\pm $ is complex-valued, 
we define that $ s_\pm =1 $ just to maintain the original form 
\begin{eqnarray}
\label{eq:original}
\epsilon_p^\pm s_\pm 
= \frac{1}{2}  \Big (  \omega \pm   q_z   \sqrt{ 1 -  4m^2/q_\para^2 } \Big)
\, .
\end{eqnarray}

\subsubsection{Thermally induced imaginary parts}
 
We closely look into the medium effects contained in 
the imaginary part of the integral expression (\ref{eq:Pi_medium-1}). 
Since the imaginary part is related to the squared amplitude of on-shell processes, 
it helps us to understand how the interplay between the effects of the strong magnetic field 
and medium work in the relevant processes.

The denominator of the integrand in Eq~(\ref{eq:Pi_medium-1}) 
has a definite sign as long as $ 1- 4m^2/q_\para^2  <0 $, that is, $0 < q_\parallel^2 < 4m^2$. 
In this kinematics, the regular $p_z$-integral does not generate any imaginary part. 
However, when $q_\parallel^2 \leq0$ or $q_\parallel^2 \geq 4m^2$, 
the integral picks up the pole contributions (\ref{eq:poles}), 
implying existence of an imaginary part in these regimes. 
By the use of the formula $ 1/(x \pm i \epsilon) = P(1/x) \mp i\pi \delta (x) $, 
one can extract the imaginary part of the polarization tensor: 
\begin{eqnarray}
\Im m \Pi_\para^{\temp }
\= \pi m_B^2 \, e^{- \frac{ \vert \bq_\perp \vert ^2}{2 \vert q_fB\vert} } \, \frac{ m^2 }{ q_\parallel^2  }
\Im m \int_{-\infty}^\infty \frac{d p_z}{2\pi}  
\frac{ n_+(\epsilon_p)  +  n_- (\epsilon_p)   }{ \epsilon_p} 
\left[ \, \frac{  1 } { p_z - p_z^+} - \frac{ 1} { p_z - p_z^- } \, \right]
\frac{  q_\parallel^2 + 2 q_z p_z } { p_z^+ - p_z^-}
\nnb
\=  \pi m_B^2 \, e^{- \frac{ \vert \bq_\perp \vert ^2}{2 \vert q_fB\vert} } 
\, \frac{ 2 m^2 } {   q_\parallel^2    \sqrt{ 1-4m^2/q_\para^2}  } 
\left[   \, s_+  \frac{n_+(\epsilon_p^+)  +  n_- (\epsilon_p^+)}{2}  
+  s_-   \frac{n_+(\epsilon_p^-)  +  n_- (\epsilon_p^-)}{2}   \, \right]
\label{eq:ImPi-med}
\, .
\end{eqnarray}

Next, we shall examine the sign function $s _\pm $ introduced just above. 
Since the polarization tensor is an even function of $ q_z  $, 
one can hereafter replace $ q_z $ by its absolute value. 
Note also that $ \omega^2 - (q_z  \sqrt{1 - 4m^2/q_\para^2})^2 = q_\para^2 [1 +  4m^2 (q_z/q_\para^2)^2] $, 
which shows a relative magnitude of the two terms in $ s_\pm $.  
The sign $  s_+$ is negative only when $  \omega <0$ and $ q_\para^2 > 4m^2 $, otherwise it is positive. 
Also, the sign $  s_-$ is positive only when $  \omega >0$ and $ q_\para^2 > 4m^2 $, otherwise it is negative. 
Summarizing the above cases, one finds 
\begin{eqnarray}
\label{eq:signs}
s_\pm \= \theta(q_\para^2 -4m^2) \sgn(\omega) \pm \theta(-q_\para^2)
\, .
\end{eqnarray}
Plugging this expression into Eq.~(\ref{eq:ImPi-med}), 
one can reorganize the imaginary part as 
\begin{eqnarray}
\label{eq:med_final}
\Im m  \Pi_\para^{\temp }
=  \pi m_B^2 \, e^{- \frac{ \vert \bq_\perp \vert ^2}{2 \vert q_fB\vert} } 
\,  \frac{ 2  m^2   }{  q_\parallel^2 \sqrt{ 1 -4m^2 /q_\parallel^2 } }  
\Big[   N_+ \theta( q_\para^2 - 4m^2)   \sgn(\omega) 
+  N_-   \theta( - q_\para^2 )  \Big]
\, ,
\end{eqnarray}
where  
\begin{eqnarray}
N_\pm : =   \frac{n_+(\epsilon_p^+)  +  n_- (\epsilon_p^+)}{2}  
\pm   \frac{n_+(\epsilon_p^-)  +  n_- (\epsilon_p^-)}{2} 
\, .
\end{eqnarray}
Since $ N_+ $ ($ N_- $) is an even (odd) function of $ \omega $, 
the imaginary part $ \Im m  \Pi_\para^{\temp } $ is an odd function of $ \omega $. 
$ N_- $ is a negative quantity when $ \epsilon_p^+ > \epsilon_p^- $. 
The imaginary part of the vacuum contribution (\ref{eq:vac_massive}) is given as 
\begin{eqnarray}
\Im m  \Pi_\para^{\vac} 
=  - \pi m_B^2 \, e^{- \frac{ \vert \bq_\perp \vert ^2}{2 \vert q_fB\vert} } 
 \frac{ 2  m^2   }{  q_\parallel^2 \sqrt{ 1 -4m^2 /q_\parallel^2 } }  
 \theta(q_\para^2-4m^2)  \sgn(\omega) 
\label{eq:vac_final}
\, .
\end{eqnarray}
Therefore, we obtain the total imaginary part 
\begin{eqnarray}
\label{eq:imag-sum}
\Im m \Pi_\para \= \Im m \Pi_\para^{\vac} + \Im m \Pi_\para^{\temp} 
\\
\=  -  \pi m_B^2 \, e^{- \frac{ \vert \bq_\perp \vert ^2}{2 \vert q_fB\vert} } 
\,  \frac{ 2  m^2   }{  q_\parallel^2 \sqrt{ 1 -4m^2 /q_\parallel^2 } }  
\Big[  (1- N_+) \theta( q_\para^2 - 4m^2)  \sgn(\omega) 
-  N_-   \theta( - q_\para^2 )  \Big]
\nn
\, .
\end{eqnarray}
We shall examine the physical processes that give rise to 
the imaginary part in the regions $ q_\para^2 > 4m^2 $ and $ q_\para^2 < 0 $ separately. 
Below, we refer those regions ``time-like'' and ``space-like'' as in the (1+1)-dimensional case. 
Notice that the ``time-like'' condition $ q_\para^2 > 4m^2 $ can be compatible with 
the light-like on-shell condition $ q^2 = q_\para^2 + q_\perp^2 = 0 $ in the four dimensions. 
This result is symmetric under the sign flip of the chemical potential $ \mu \to - \mu $ as expected.

\subsubsection*{Finite temperature}

Here, we consider the zero-density case $ \mu = 0  $, 
and both the fermion and antifermion distribution functions reduce to the same form 
$ n_\pm (p^0) \to n(p^0) = [  e^{p^0/T} + 1 ]^{-1}  $. 
We first focus on the high-temperature limit where $ T \gg \omega, q_z $. 
By the use of Eq.~(\ref{eq:signs}), we then obtain 
\begin{eqnarray}
N_+  \sim  1 - \frac{1}{4T}  ( \epsilon_p^+ + \epsilon_p^-)  =   1 - \frac{|\omega|}{4T} 
\, .
\end{eqnarray}
Observe that the first term in $ N_+ $ exactly cancels the vacuum contribution in Eq.~(\ref{eq:imag-sum}), i.e., 
\begin{eqnarray}
1- N_+  \sim   \frac{1}{4T}  ( \epsilon_p^+ + \epsilon_p^-)  =   \frac{|\omega|}{4T} 
\, .
\end{eqnarray}
Consequently, the total imaginary part is highly suppressed 
in the time-like region by a factor of  $ |\omega|/(4T) \ll 1$. 
One can interpret this result in the following way. 
In vacuum, the imaginary part indicates a pair creation from an incident photon/gluon. 
However, in the presence of a heat bath, its inverse process is also activated 
because thermal fermions and antifermions can annihilate into a photon/gluon.\footnote{
Note that kinematics of the 1-to-2 and 2-to-1 processes is allowed in a magnetic field 
even for an on-shell photon, though it is forbidden in the four dimensions without a magnetic field. 
} 
Due to the balance between these processes, 
the net imaginary part in the time-like regime is suppressed at high temperature. 
One can confirm this observation with a simple identity 
\begin{eqnarray}
\label{eq:N+---0}
1 - N_+  =    [ 1- n(\ep_p^+) ] [ 1-  n(\ep_p^-)]  - n(\ep_p^+)  n(\ep_p^-) 
\, .
\end{eqnarray}
The first term corresponds to the pair-creation channel $ \gam \to f \bar f $ 
that is subject to the Pauli-blocking effect in the final state, 
while the second term corresponds to the pair-annihilation channel $ f \bar f  \to  \gam $. 
According to the sign functions (\ref{eq:signs}), we confirm 
the energy conservation $ |\omega| =  \ep_p^+ + \ep_p^-$. 
The net pair creation occurs with a significant rate only in 
a sufficiently high energy regime $\omega , q_z \gg T  $, 
where the thermal contribution is exponentially suppressed.
In such cases, the vacuum contribution stands as the dominant contribution 
to the imaginary part [see Eq.~(\ref{eq:zero-T}) for the zero-temperature limit].

On the other hand, one finds a new medium-induced channel in the space-like regime, 
that is, a contribution from the Landau damping proportional to $N_-  $. 
The Landau damping is purely a medium effect where a medium fermion (antifermion) 
is scattering off a space-like photon. 
In case of vacuum, the imaginary parts in the space-like regime 
cancel out in the final expression of the polarization tensor 
in all order of the Landau levels \cite{Hattori:2012je}. 
Using again Eq.~(\ref{eq:signs}), we have 
\begin{eqnarray}
- N_- \sim  \frac{1}{4T}  ( \epsilon_p^+ - \epsilon_p^-)  =  \frac{\omega}{4T} 
\, .
\end{eqnarray}
Similar to the time-like region, 
the imaginary part is again suppressed in the space-like region in the high temperature limit. 
This is due to a balance between the gain and loss terms as seen in an identity 
\begin{eqnarray}
- N_- =    n(\ep_p^-)  [ 1- n(\ep_p^+) ] - n(\ep_p^+) [ 1-  n(\ep_p^-)] 
\, ,
\end{eqnarray} 
where the Pauli blocking effect appears in the final states. 
The first (second) term corresponds to the gain (loss) term 
for the phase-space volume at $ p^0 = \ep_p^+ $. 
Namely, this identity shows the balance between the scatterings
 $ f + \gam^\ast \to f $ and $ f \to f  + \gam^\ast $. 
According to the sign functions (\ref{eq:signs}), we confirm 
the energy conservation $  \omega + \ep_p^- = \ep_p^+$.

Combining the expressions in the time-like and space-like regions, 
we obtain the high-temperature expansion of the total imaginary part  
\begin{eqnarray}
\label{eq:imag-total-expansion}
\Im m \Pi_\para \sim - \pi m_B^2 \, e^{- \frac{ \vert \bq_\perp \vert ^2}{2 \vert q_fB\vert} } 
 \frac{ 2  m^2   }{  q_\parallel^2 \sqrt{ 1 -4m^2 /q_\parallel^2 } }  
\cdot
\frac{\omega}{4T} \theta (1-  4m^2/q_\para^2 )
\, .
\end{eqnarray}
Also, summing Eqs.~(\ref{eq:med_final}) and  (\ref{eq:vac_final}) without the high-temperature expansion, 
we can straightforwardly obtain the complete form 
\begin{eqnarray}
\label{eq:imag-total}
\Im m \Pi_\para  
=  -  \pi m_B^2 \, e^{- \frac{ \vert \bq_\perp \vert ^2}{2 \vert q_fB\vert} } 
\frac{ 2  m^2   }{  q_\parallel^2 \sqrt{ 1 -4m^2 /q_\parallel^2 } }  
\cdot
\frac{   \sinh( \frac{ \omega }{2T})  \, \theta (1-  4m^2/q_\para^2 )  }
{ \cosh( \frac{\omega}{2T}  ) + \cosh \big( \, \frac{q_z }{2T}  \sqrt{1-  4m^2/q_\para^2} \, \big)  }
\, .
\end{eqnarray}
Since all the terms in Eqs.~(\ref{eq:med_final}) and (\ref{eq:vac_final}) are odd functions of $ \omega $, 
so is the final result in Eq.~(\ref{eq:imag-total}), 
which originates from a general property of the spectral density (see, e.g., Ref.~\cite{Kapusta:2006pm}). 
This expression essentailly agrees with those in Ref.~\cite{Kao:1998yt} (up to a typo there) 
and in Ref.~\cite{Baier:1991gg} where the authors compute the Feynman causal correlator 
instead of the retarded correlator. 
One can confirm that the zero-temperature limit of Eq.~(\ref{eq:imag-total}) 
agrees with the vacuum expression (\ref{eq:vac_final}) as follows. 
In this limit, we have 
\begin{eqnarray}
\label{eq:zero-T}
\lim_{T \to 0} \Im m \Pi_\para  
= -  \pi m_B^2 \, e^{- \frac{ \vert \bq_\perp \vert ^2}{2 \vert q_fB\vert} } 
 \frac{ 2  m^2   }{  q_\parallel^2 \sqrt{ 1 -4m^2 /q_\parallel^2 } }  
\lim_{T \to 0}   \frac{   \{\theta(\omega) - \theta(-\omega) \} \theta (1-  4m^2/q_\para^2 )  }
{ 1  + \exp \big [ \,  \frac{ 1 }{2T}  \big ( |q_z| \sqrt{1-  4m^2/q_\para^2} - |\omega| \big ) \, \big ]  }
\, .
\end{eqnarray} 
The absolute values of $ \omega,q_z $ come from the hyperbolic cosine functions. 
The limit depends on the sign on the shoulder of the exponential. 
As expected, the limit is vanishing in the space-like region $ q_\para^2 < 0 $, 
while it reproduces the finite value of the imaginary part 
in vacuum in the time-like region $ q_\para^2 > 4m^2 $ [see Eq.~(\ref{eq:vac_massive})]. 

\subsubsection*{Finite density}

At nonzero density $ (\mu \not = 0) $, 
the fermion and antifermion distribution functions should be distinguished. 
Nevertheless, we can interpret the pair creation/annihilation and the Landau damping 
in a similar way to the zero-density case.

Similar to the finite-temperature case (\ref{eq:N+---0}), one finds an identity 
\begin{eqnarray}
1 - N_+ 
\= 1-  \frac12 \left[ \,  \{ n_+(\epsilon_p^+)  +  n_- (\epsilon_p^-)\}
+ \{  n_+(\epsilon_p^-)   +  n_- (\epsilon_p^+) \}   \, \right]
\nnb
 &=& \sum_{\a=\pm}  \frac12 [ \, \{ 1 - n_+(\ep_p^\a) \}\{ 1-  n_-(\ep_p^{-\a}) \} 
 -  n_+(\ep_p^\a)  n_-(\ep_p^{-\a})  \,]
 \, .
\end{eqnarray}
This expression has the same form as in Eq.~(\ref{eq:N+---0}) 
up to the difference between the fermion and antifermion distribution functions. 
In the pair creation/annihilation channel, the fermion and antifermion can have different energies 
$ \ep_p^+ $ and $  \ep_p^- $ when the photon momentum is nonzero, i.e., $ q_z \not = 0 $. 
Thus, there are two kinematical windows depending on either a fermion or antifermion takes $ \ep_p^+ $. 
Clearly, interchanging $ \ep_p^+ $ and $  \ep_p^- $ brings one kinematics to the other. 
A finite chemical potential distinguishes those two cases, 
whereas they are degenerate at zero density (\ref{eq:N+---0}).

In the space-like region, the Landau damping of the fermions and antifermions occurs 
with different magnitudes. 
Therefore, separating the fermion and antifermion distribution functions, 
we find an identity 
\begin{eqnarray}
- N_- 
\= - \frac12 \left[ \,  \{ n_+(\epsilon_p^+)  -  n_+ (\epsilon_p^-) \}
+ \{  n_- (\epsilon_p^+ )   -  n_- (\epsilon_p^-) \}   \, \right]
\nnb
&=&  \sum_{c=\pm} \frac12 [ \,   n_{c}(\ep_p^-) \{ 1 - n_{c}(\ep_p^+) \}
 - n_{c}(\ep_p^+) \{1-  n_{c}(\ep_p^-) \}  \,]
\, .
\end{eqnarray}
The sum over $ c = + $ ($  - $) is for the Landau damping of fermions (antifermions), 
assuming that $ \mu > 0 $ without loosing generality.

At zero temperature, the distribution functions reduce 
to the step functions $ n_\pm(p^0) = \theta( \pm \mu - p^0 ) $. 
Thus, there is no annihilation channel, and only the pair-creation channel is left as 
\begin{eqnarray}
\label{eq:pair-zero-T}
\left. 1 - N_+ \right|_{T=0} 
 &=& \frac12 [\,  \theta (\ep_p^+ -  \mu  )  + \theta ( \ep_p^- -  \mu  )  \,]
 \, .
\end{eqnarray}
Either fermion or antifermion is free of the Pauli-blocking effect 
since its positive-energy state is completely vacant. 
However, the other one needs to find an unoccupied state 
above the Fermi surface to avoid the Pauli-blocking, resulting in the above step functions. 
On the other hand, the Landau damping occurs 
between an occupied initial state and an unoccupied final state. 
Thus, there are only fermion contributions (when we assume $ \mu > 0 $ as above). 
The gain and loss terms read  
\begin{eqnarray}
- \left.  N_- \right|_{T=0} 
&=&   \frac12 [ \,  \theta (  \mu - \ep_p^- ) \theta (\ep_p^+ -  \mu  ) 
-  \theta (  \mu - \ep_p^+ ) \theta (\ep_p^- -  \mu  )   \,]
\, .
\end{eqnarray} 
It is a simple task to confirm the zero-density limit $ (\mu \to 0) $, 
where we find that 
\begin{subequations}
\begin{eqnarray}
&&
\lim_{\mu \to 0} \left. 1 - N_+ \right|_{T=0} = 1
\, ,
\\
&&
\lim_{\mu \to 0} \left.  N_- \right|_{T=0} = 0
\, .
\end{eqnarray}
\end{subequations}
These limits reproduce the vacuum expression (\ref{eq:vac_final}).

Similar to Eq.~(\ref{eq:imag-total}), one can express the full form of 
the imaginary part with the hyperbolic functions. 
The complete form at finite temperature and density is found to be 
\begin{eqnarray}
\label{eq:imag-total-density}
\Im m \Pi_\para  
=  -  \pi m_B^2 \, e^{- \frac{ \vert \bq_\perp \vert ^2}{2 \vert q_fB\vert} } 
\frac{ 2  m^2   }{  q_\parallel^2 \sqrt{ 1 -4m^2 /q_\parallel^2 }  }
\cdot \frac12 \sum_{c=\pm}
\frac{   \sinh( \frac{ \omega }{2T})  \, \theta (1-  4m^2/q_\para^2 )  }
{ \cosh( \frac{\omega}{2T}  ) + \cosh \big( \, \frac{q_z }{2T}  \sqrt{1-  4m^2/q_\para^2}
+ c \frac{\mu}{T} \, \big)  }
\, .
\end{eqnarray}
The chemical potential $ \mu $ appears only in the denominator. 
One can easily confirm that this expression reduces to Eq.~(\ref{eq:imag-total}) when $ \mu \to 0 $. 

}

\subsubsection{Photon masses from the polarization effects}

\label{sec:photon_mass}

We discuss physical applications of 
the in-medium polarization tensor given in Eq.~(\ref{eq:Pi_medium-1}). 
The first application is the in-medium gluon/photon masses \cite{Hattori:2022uzp} 
and another is the AWI at finite temperature 
and/or density \cite{Hattori:2022wao} in the next subsection. 
According to the resummed propagator (\ref{eq:Photon-prop-LLL}), 
the pole position of the parallel mode is given as 
\begin{eqnarray}
\label{eq:pole}
q^2 -  \Pi_\para (q_\para; q_\perp) = 0
\, .
\end{eqnarray}
The other mode is not modified by the polarization of the LLL fermions as already discussed there. 
In the massless limit, there is no medium modification on the polarization tensor (\ref{eq:Pi-massless}), 
so that the gluon/photon dispersion relation is immediately obtained as 
\begin{eqnarray}
\label{eq:massless-dispersion}
\omega^2 = |\bq|^2 + m_B^2 \, e^{-\frac{\vert \bq_\perp \vert^2}{2 \vert q_fB\vert}} 
\, .
\end{eqnarray} 
Computing the gluon/photon dispersion relation with massive fermions is more involved due to 
the momentum dependence of the polarization tensor; 
one needs to solve Eq.~(\ref{eq:pole}) with respect to the frequency. 
In general, the polarization tensor exhibits different limiting behaviors 
in the vanishing frequency and momentum limits; those limits do not commute with each other. 
Thus, there are two different gluon/photon masses (see, e.g., Ref.~\cite{Bellac:2011kqa}). 
They are called the Debye screening mass and the plasma frequency, 
which we discuss below in order.

It should be emphasized that the medium contribution~(\ref{eq:Pi_medium}) 
gives rise to finite gluon/photon masses even in the massive case, 
in contrast to the absence of a static mass 
in the vacuum contribution [see discussions below Eq.~(\ref{eq:I})]. 
This is simply because the thermal fermions populated 
above the mass gap respond to the gluon/photon fields. 
From this observation, the masses may be sizeable when $m/T, \, m/\mu \lesssim 1  $, 
and approach $ m_B $ in the high-temperature or -density limit, $ m/T \to 0 $ or $ m/\mu \to 0  $.

The Debye screening mass is the inverse of the screening length 
of a static (heavy) charge embedded in medium. 
In the static limit, the pole position is given as $   |\bq|^2 + \Pi_\para (\omega=0) = 0$, 
and the polarization tensor cuts off the long-range propagation of spatial gluons/photons. 
The Debye screening mass is defined as 
\begin{eqnarray}
m_D^2 
\=  \lim_{q_z/m \to 0}  \Pi_\para^{\temp }( \omega=0  , q_z; q_\perp=0) 
\end{eqnarray}
We have first taken the vanishing frequency limit 
and the vanishing transverse-momentum limit $(  q_\perp=0)$. 
The ordering of the limits does not matter for $ q_\perp $ in the present case. 
There is no vacuum contribution to $ m_D^2 $ 
as long as the fermion mass $m  $ is finite 
according to the property of the $  I$ function in Eq.~(\ref{eq:I-limits}); 
Massive fermions are not excited by photons with an infinitesimal energy. 
Inserting the medium contribution (\ref{eq:Pi_medium-1}), we have 
\begin{eqnarray}
m_D^2  
\=  -  m_B^2   \lim_{\bar q_z \to 0}   \frac{1}{ 2 \bar q_z} 
 \prj \int_{-\infty}^\infty \frac{d \bar p_z}{ \bar \epsilon_p} 
\frac{ n_+(m \bar \epsilon_p)  + n_-(m \bar \epsilon_p) } { \bar p_z - \bar q_z }
\label{eq:Debye-sc}
\, .
\end{eqnarray} 
We introduced a dimensionless variable $ \bar q_z = q_z/(2m) $, 
and scaled the integral variable as $ p_z \to p_z' = p_z/m $ accordingly. 
Other variables are also normalized as 
$ \bar \epsilon_p = \epsilon_p/m $, $ \bar T = T/m $, and $ \bar \mu = \mu/m $. 
The integral should be understood as the Cauchy principal value denoted with $ \prj $. 
The Debye mass is determined by the first derivative of 
the integral value with respect to $  \bar q_z$. 
The integral value is dominated by the infrared contribution 
such that $ \bar p_z \sim \bar q_z \to 0 $ due to 
the effective dimensional reduction of the integral measure.

We first examine the high-temperature or -density limit. 
The fermion distribution functions provide a cutoff scale at temperature $ T $ or chemical potential $ \mu $ as 
\begin{eqnarray}
m_D^2 
= - m_B^2   \lim_{\Lambda/m\to \infty }   \lim_{ \bar q_z \to 0}   \frac{1}{ 2 \bar q_z} 
\prj \int_{- \Lambda/m}^{\Lambda/m} \frac{d \bar p_z}{ \bar \epsilon_p} 
\frac{ 1 } { \bar p_z - \bar q_z }
= m_B^2
\label{eq:Debye-sc-infinite}
\, ,
\end{eqnarray}
where the cut-off $ \Lambda $ is given by 
a large value of either $ T $ or $ \mu $. 
In this limit, the Debye mass agrees with the Schwinger mass. 
Note that the hierarchy $ T/\mu \gg1 $ ($ T/\mu \ll 1 $) needs to be satisfied for the above replacement 
in addition to the high-temperature limit $\bar T \to \infty  $ (high-density limit $ \bar \mu \to \infty$).

\begin{figure} 
\begin{center}
\includegraphics[width=0.55\hsize]{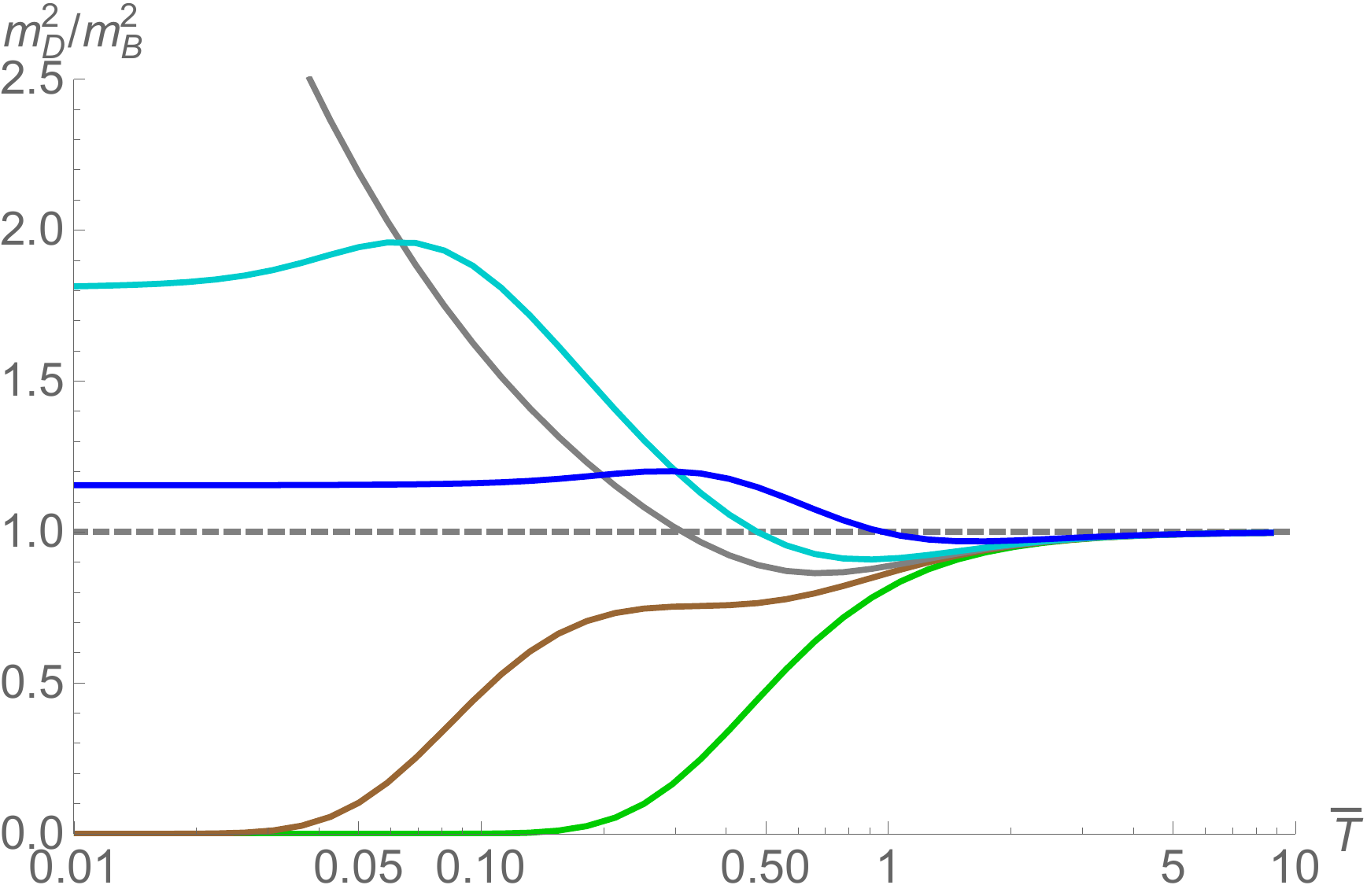} 
\end{center}
\caption{Temperature dependence of the Debye screening mass 
at $\bar  \mu = \{ 0, \, 0.8, \, 1,\, 1.2, \, 2 \} $ shown in \{green, brown, gray, blue, cyan\}. 
}
\label{fig:photon-mass}
\end{figure}

In Fig.~\ref{fig:photon-mass}, we show numerical results for the temperature dependence 
of the Debye screening mass at several values of density. 
We take $\bar  \mu = \{ 0, \, 0.8, \, 1,\, 1.2, \, 2 \} $ for the lines shown in 
\{green, brown, gray, blue, cyan\}. 
One can confirm that the Debye mass approaches the Schwinger mass when $ \bar T \gg1  $ 
for all the shown values of $  \bar \mu$. 
As we decrease temperature, the lines splits near $ \bar T =1 $ depending on the value of $ \bar \mu $. 
The Debye mass approaches zero when $\bar \mu <1  $ due to the absence of thermal excitations, 
while it increases when $\bar \mu \geq 1  $. 
This contrast behavior stems from a simple fact that the density of occupied low-energy states 
increases near the Fermi surface as we decrease temperature in the latter case. 
The reader is referred to Ref.~\cite{Hattori:2022uzp} for more results and discussions.

In short, the Debye screening mass is dominantly induced by 
the fermions in the low-energy regime, and is enhanced when those states are occupied. 
Notice also that the Debye mass is scaled by the Schwinger mass $ m_B^2 $. 
The Debye screening mass only depends on the magnetic-field strength 
through the Landau degeneracy factor in $ m_B^2 $. 
The density of degenerate states increases as we increase the magnetic-field strength, 
and thus naturally enhances the Debye screening mass.

Next, we discuss the plasma frequency that appears in 
the dispersion relation of an on-shell gluon/photon in medium, 
which can be regarded as a collective excitation composed of a gluon/photon and oscillating plasma. 
The dispersion relations may be no longer gapless in the absence of the Lorentz symmetry. 
To find the magnitude of the energy gap at $ |\bq| =0 $, one should take the vanishing momentum limit first. 
Therefore, the plasma frequency $  \omega_p$ is determined by the following equation 
\begin{eqnarray}
\label{eq:plasma-frequency}
\omega_p^2 =  \Pi_\para(\omega_p, 0) 
=  m_B^2   \bigg[ \, 1- I \big( \frac{\omega_p^2}{4m^2} \big)
+ 2  m^2 \prj \int_{-\infty}^\infty \frac{d p_z}{ \epsilon_p} 
\frac{  n_+(\epsilon_p)  + n_-(\epsilon_p) } {   (2 \epsilon_p)^2 -  \omega_p^2  }
 \,  \bigg] 
 \, .
\end{eqnarray}
The right-hand side is still a function of the frequency $ \omega_p $, 
so that one needs to solve the above equation explicitly. 
Note that the first two terms in the brackets come from the vacuum contribution. 
One needs to maintain those terms in addition to the medium contribution 
because the medium contributions does not necessarily dominate over the vacuum contribution 
due to the effective dimensional reduction. 
This contrasts to the four dimensional case where  the medium contribution, which is 
proportional to $ (q_f T)^2 $ or $ (q_f \mu)^2 $, governs the polarization effects 
at the high temperature/density regime.

One can show that Eq.~(\ref{eq:plasma-frequency}) always has a solution as follows \cite{Hattori:2022uzp}. 
When $ \omega_p \to 0 $, the vacuum contribution exactly vanishes 
and the integral for the medium contribution gives $   \Pi_\para $ a positive value. 
On the other hand, the behavior of the right-hand side is dominated by 
the $ I $ function that gives $   \Pi_\para $ a divergence to negative infinity as $  \omega_p^2 \to 4m^2$. 
This divergence is associated with the pair creation threshold as we discussed repeatedly. 
Therefore, one can find a solution in the range $ 0 < \omega_p^2 <  4m^2 $ 
as long as temperature is finite. 
At zero temperature case, 
the upper bound is shifted with a chemical potential 
since the Pauli-blocking effect induces the threshold shift 
as we discussed below Eq.~(\ref{eq:id-density-2}).

In the left panel of Fig.~\ref{fig:PF}, we show the normalized plasma frequency 
$ \bar \omega_p^2 =\omega_p^2/(4m^2) $ 
as a function of the normalized magnetic-field strength $  \bar m_B^2$. 
We take a set of temperature $ \bar T = \{ 0.5, 1, 3, 5, 10\} $ 
at vanishing chemical potential $ \bar \mu = 0 $. 
All the curves increase as we increase the magnetic-field strength $  \bar m_B^2$. 
The plasma frequency approaches the Schwinger mass, shown with the light green line, 
and saturates at $ \bar \omega_p^2=1 $ as we increase temperature. 
The former behavior is naturally expected from that in the massless limit $ \bar T \to \infty$. 
The latter behavior may be more subtle, implying that an infinitesimal mass makes 
the curve approach $ \bar \omega_p^2=1 $ rather than the Schwinger mass. 
The deviation from the Schwinger mass gets larger as we increase $  \bar m_B^2$, 
and the plasma frequency is strongly suppressed in such a strong-field limit, 
meaning that the fermion-mass effect becomes sizable in the photon dispersion relation.

In the right panel of Fig.~\ref{fig:PF}, 
we show the plasma frequency as a function of $ \bar m_B^2 $ at $ \bar \mu = \{ 1.1 , \ 2 , \ 3 \} $. 
The plasma frequency approaches the Schwinger mass as we increase $  \bar \mu $. 
As we increase $ \bar m_B^2 $, the plasma frequency saturates 
at $\bar \omega_p^2 =  \bar \mu^2  $ [$ \omega_p^2 =  (2 \mu)^2  $] at zero temperature, 
whereas we have seen that the saturation value is always given by 
$\bar \omega_p^2 =  1 $ [$ \omega_p^2 =  (2 m)^2  $] at finite temperature. 
This difference originates from the shift of the pair-creation threshold induced by the Pauli blocking effect.

Here is one caveat; For the LLL approximation to work, 
all the physical scales should be smaller than the magnetic-field strength, 
i.e., $ m^2, \ \omega_p^2, \ T^2 , \ \mu^2 ,  \ll q_fB$. 
Otherwise, the higher Landau levels come into play a role. 
Noticing that $ \bar m_B^2 = \a_\EM \rho_B/m^2 \sim  |q_fB|/m^2 \times 10^{-3} $, 
the LLL approximation should work for the lines shown in Fig.~\ref{fig:PF}, 
except for the regime where $  \bar m_B^2 $ is too small.

\begin{figure}
\begin{minipage}{0.48\hsize} 
	\begin{center} 
\includegraphics[width=\hsize]{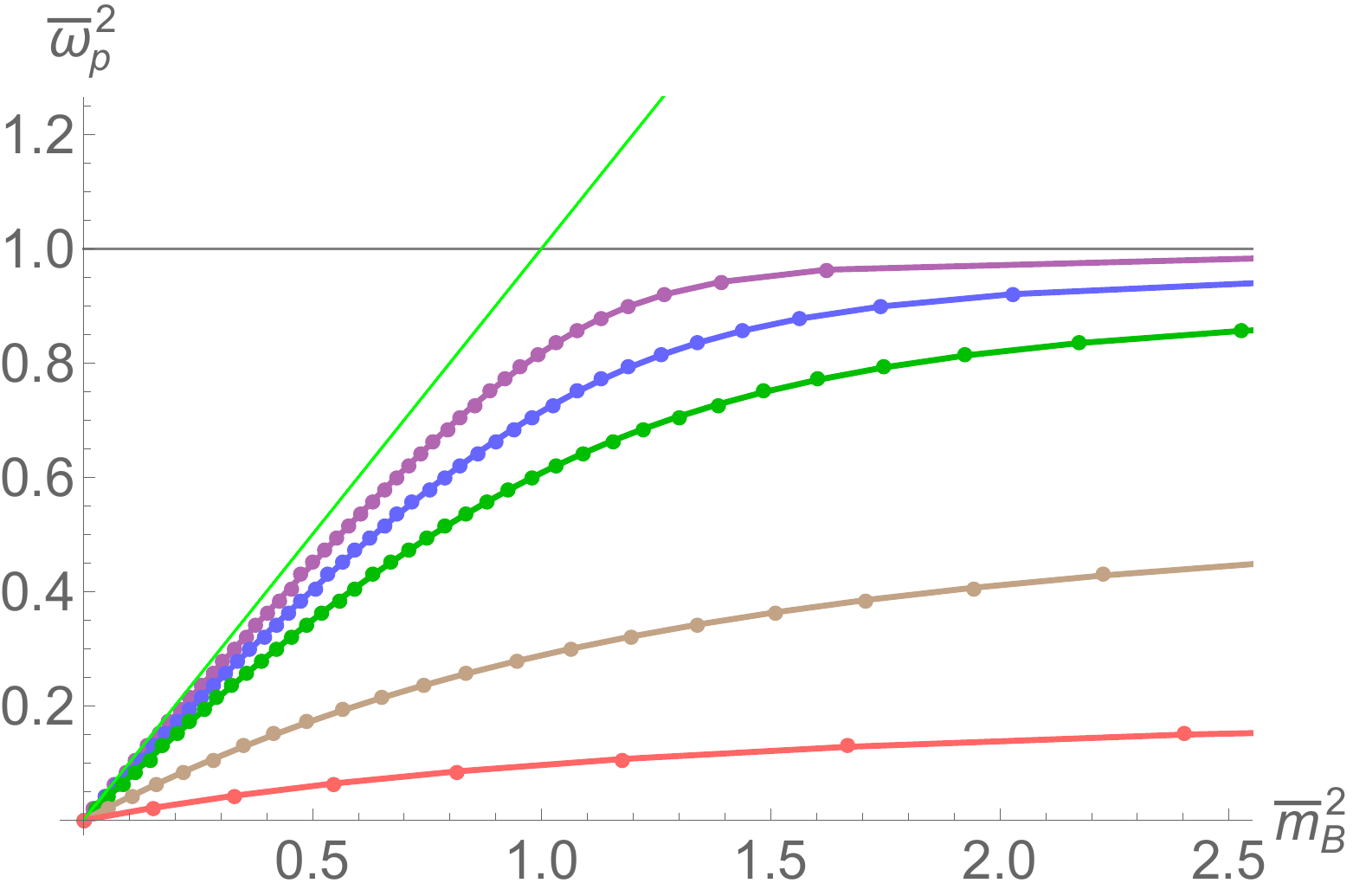}
	\end{center}
\end{minipage}
\begin{minipage}{0.48\hsize}
	\begin{center}
\includegraphics[width=\hsize]{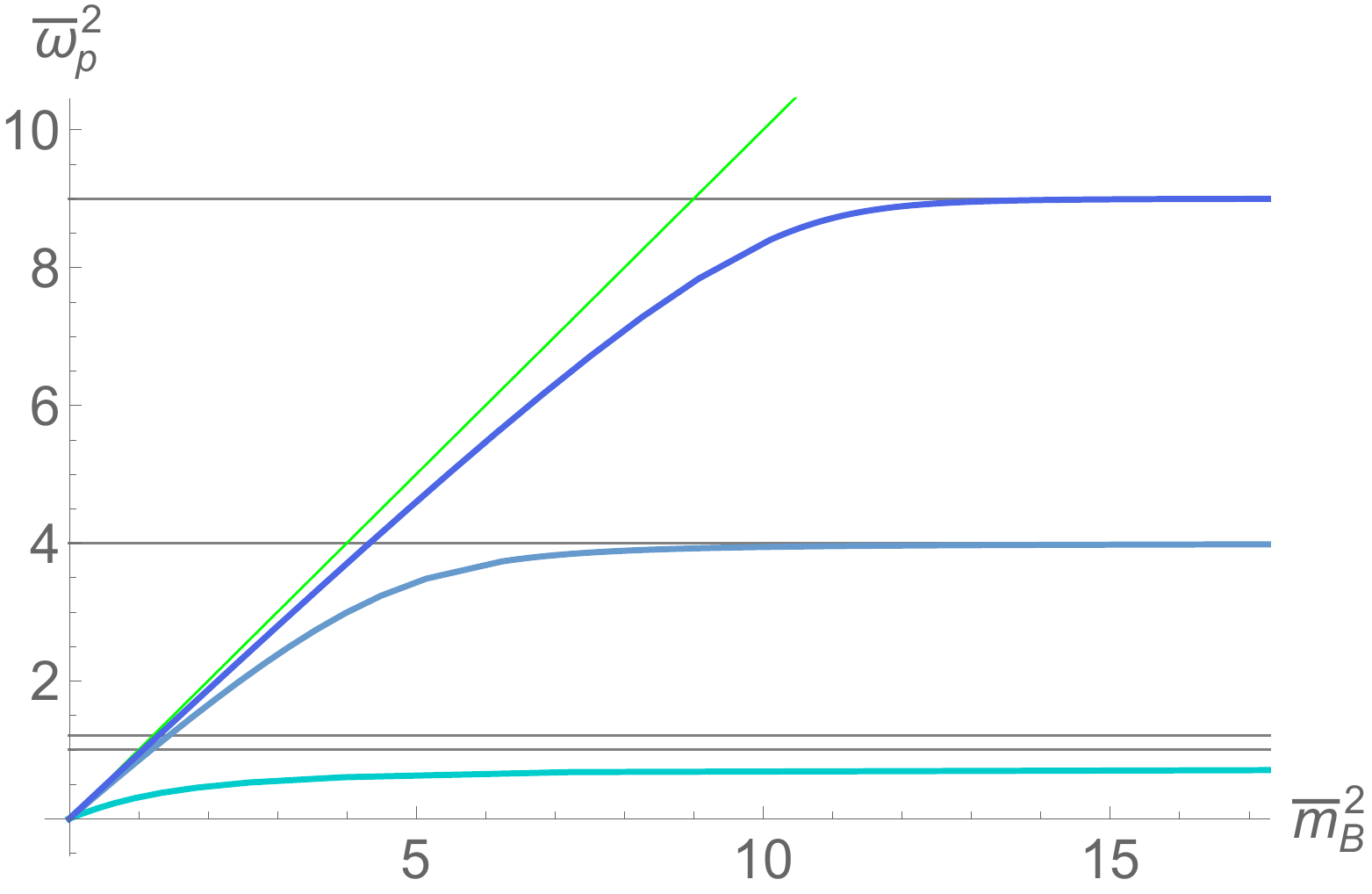}
	\end{center}
\end{minipage}
\caption{
The plasma frequency as a function of the normalized magnetic-field strength $ \bar m_B^2 $ (left). 
We take $\bar \mu=0  $ and $ \bar T = \{ 0.5, \ 1, \ 3, \ 5, \ 10\} $ from bottom to top. 
The plasma frequency as a function of the normalized magnetic-field strength $ \bar m_B^2 $ (right). 
We take $ \bar \mu = \{ 1.1, \ 2 , \ 3 \} $ from bottom to top.  
}
\label{fig:PF}
\end{figure}

\subsubsection{Axial Ward identity at finite temperature and density}

\label{sec:AWI-medium}

We now come back to the AWI (\ref{eq:anomaly-LLL-1}). 
In case of zero temperature, the axial current is conserved 
in the massive case $ m^2 \gg q_\para^2 $ where $  \Pi_\para \to 0$. 
This is because there are no fermions that respond to the soft photon. 
In contrast, even a soft photon can perturb the system 
when there are thermal fermions at finite temperature 
as already discussed below Eq.~(\ref{eq:Pi_medium}). 
In this case, the spectral flow of thermal fermions,
which is induced by a constant electric field, 
can create the axial charge due to the helicity flip 
during the acceleration \cite{Hattori:2022wao}. 
Also, a gauge field $ A^\mu $ at finite frequency can induce 
helicity flop of massive fermions by scattering 
and can decay into a pair of fermion and antifermion 
with the same helicity, both of which contribute to 
creation of the axial charge.

Including both the vacuum contribution (\ref{eq:vac_massive}) 
and the thermal contribution (\ref{eq:Pi_medium-1}), 
we obtain the in-medium AWI 
\begin{subequations}
\label{eq:AWI-thermal}
\begin{eqnarray} 
&& {\rm F.T.} \ \pd_\mu j_A^\mu = 
 \frac{ q_f^2}{2\pi^2} \tilde \bE(q) \cdot \bB  \, e^{-\frac{|\bq_\perp|^2}{2|q_fB|}}  
 +2im \langle \bar \psi \gam^5 \psi \rangle
\, ,
\end{eqnarray}
\end{subequations}
where 
\begin{eqnarray}
\label{eq:PS-condensate-LLL-thermal}
2im \langle \bar \psi \gam^5 \psi \rangle
\=  \frac{ q_f^2}{2\pi^2} \tilde \bE(q) \cdot \bB  \, e^{-\frac{|\bq_\perp|^2}{2|q_fB|}}  
\nnb
&& \times  \bigg[ \, - I \Big( \frac{q_\para^2}{4m^2} \Big)
 +
 m^2  \int_{-\infty}^\infty \frac{d p_z}{ \epsilon_p}  
 \frac{    n_+(\epsilon_p)  + n_-(\epsilon_p)   }{ q_\para^2\sqrt{1- 4m^2/q_\para^2 }  } 
\Big( \frac{ \epsilon_p^+ s_+ }{ p_z - p_z^+ } - \frac{ \epsilon_p^- s_-}{ p_z - p_z^- } \Big) \, \bigg]
\, .
\end{eqnarray}
This is an extension of the vacuum expression (\ref{eq:AWI-vacuum}). 
The first term in Eq.~(\ref{eq:AWI-thermal}) 
is the anomalous term 
as in the vacuum massless case (\ref{eq:anomaly-LLL-2}). 
The remaining terms are proportional to the fermion mass 
and are identified with matrix elements of the pseudoscalar condensate (\ref{eq:PS-condensate-LLL-thermal}) 
as confirmed with explicit computation of 
the anomaly diagrams in Eq.~(\ref{eq:PS-condensate-LLL}). 
The first and second terms in Eq.~(\ref{eq:PS-condensate-LLL-thermal}) are the vacuum and thermal contributions, respectively. 
Note that the expression between the square brackets is dimensionless and should be a function of dimensionless combinations of the parameters.

In the ``massless'' limit $q_\para^2 / m^2 \to \infty  $, 
there is no thermal correction, and the AWI takes the same form 
as in vacuum with massless fermions (\ref{eq:AWI-vacuum}). 
While we here does not discuss radiative corrections, 
the absence of temperature correction to the anomalous term was 
shown to all orders in the coupling constant in the strictly massless case \cite{Itoyama:1982up}. 
Note that the above limit, $q_\para^2 / m^2 \to \infty  $, 
needs to be defined with a nonzero energy $ \omega $ and/or momentum $ q_z $ of the electric field $ \tilde E_\para $ along the magnetic-field direction. 
If $q_\para^2 $ goes to zero faster than $m^2 $, 
it is rather the adiabatic limit discussed below. 
This is the case where the electric field is 
homogeneous over the Compton length $1/m $ 
and includes the case of a constant electric field (see Sec.~\ref{sec:triangle-constant}).


 \begin{figure}[t]
\begin{minipage}{0.5\hsize} 
	\begin{center} 
\includegraphics[width=\hsize]{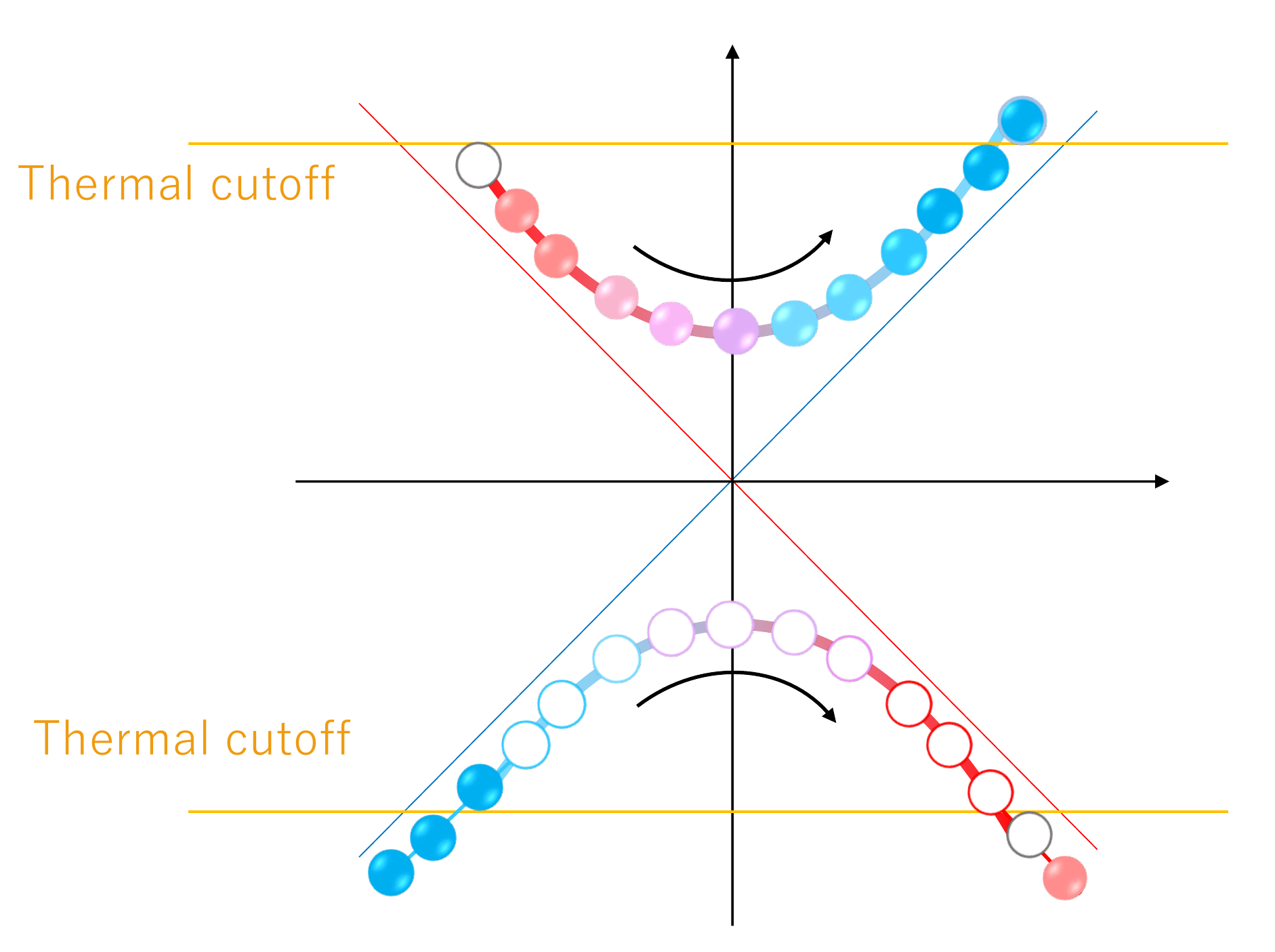}
	\end{center}
\end{minipage}
\begin{minipage}{0.45\hsize}
	\begin{center}
\includegraphics[width=\hsize]{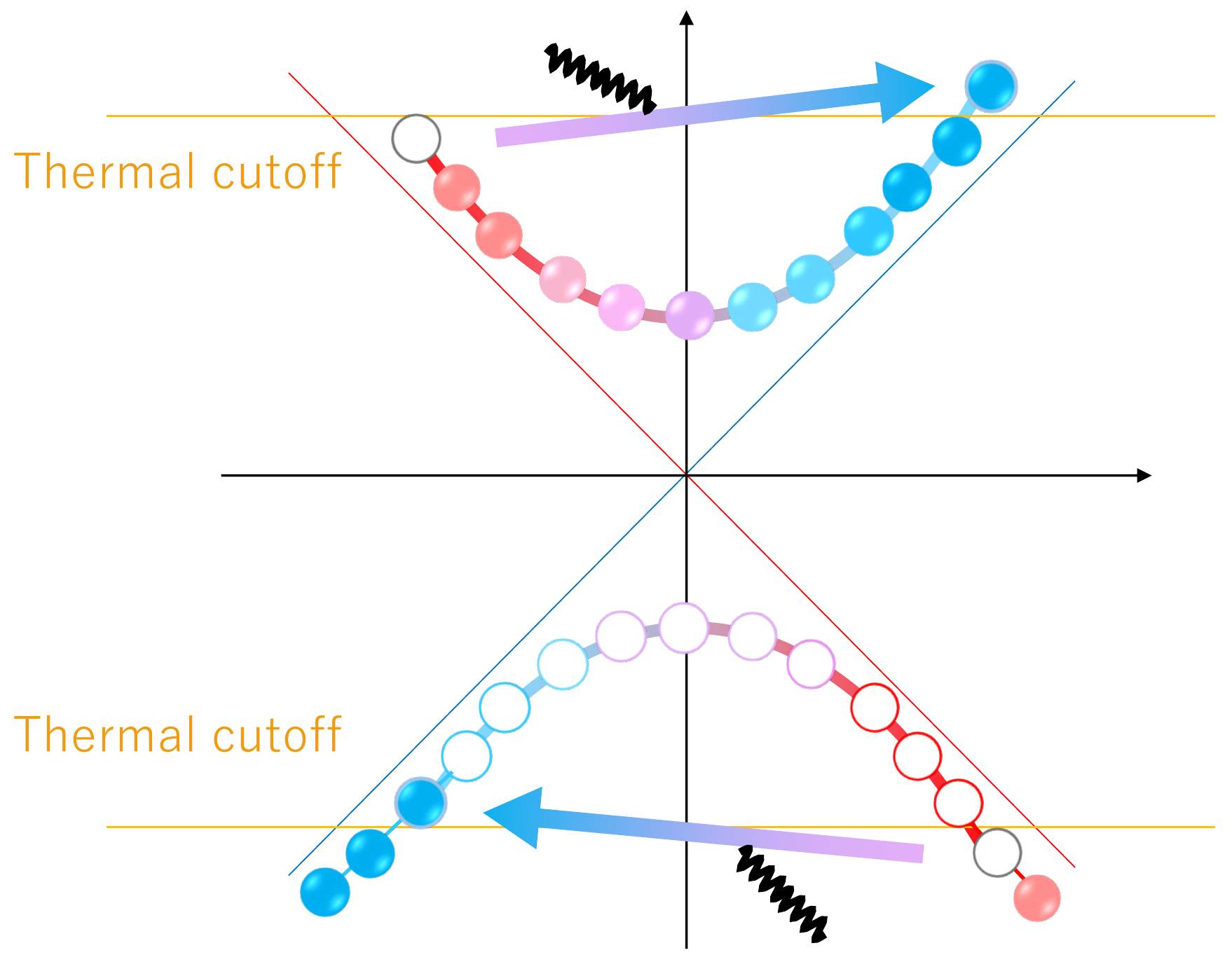}
	\end{center}
\end{minipage}
\caption{Axial-charge creation by the thermal spectral flow  (left) and by the Landau damping (right) in the plane of 
fermion dispersion relation. 
Those processes are regarded as transitions from 
the left to right half planes (and in the inverse direction) 
that are accompanied by helicity flip.  
 }
\label{fig:AWI-thermal}
\end{figure}

One can recognize a qualitative difference between the vacuum and in-medium cases 
in the adiabatic limit $q_\para^2 / m^2 \to 0  $. 
As explicitly shown around Eq.~(\ref{eq:constantf-field}), this limit corresponds to constant electric fields.  
The axial current is conserved in vacuum since a massive fermion and antifermion pair 
cannot be created in an adiabatic process. 
The spectral flow of the negative-energy states is always there, 
but does not give rise to axial-charge creation in the infrared regime 
because the spectral flow along the parabolic dispersion curve turns back 
to the bottom of the Dirac sea \cite{Ambjorn:1983hp} 
(see the left panel in Fig.~\ref{fig:AWI-thermal}). 
In contrast, the spectral flow can occur to the thermal particles in medium 
and give rise to a nonzero divergence of the axial current. 
This is because acceleration of thermal (on-shell) particles by electric fields 
induces a momentum flip and thus a helicity flip when the spectral flow goes through 
the bottom of the parabolic dispersion curve for the positive-energy states. 
What remains in the adiabatic limit is only the thermal contribution to the pseudoscalar condensate, 
which, therefore, is responsible for the thermal spectral flow. 
Interestingly, the AWI nevertheless takes the same form 
as the anomalous term in the high-temperature or density limit ($ m/T , \ m/ \mu \to0 $). 
Namely, one finds that 
\begin{eqnarray}
\label{eq:AWI-thermal-adiabatic-2}
\lim_{\omega/m\to 0 } \lim_{q_z /m\to 0 } {\rm F.T.} \ \pd_\mu j_A^\mu 
= \lim_{q_z/m\to 0 } \lim_{\omega /m\to 0 } {\rm F.T.} \ \pd_\mu j_A^\mu \=
 \frac{ q_f^2}{2\pi^2} \tilde \bE(q_\perp) \cdot \bB  \, e^{-\frac{|\bq_\perp|^2}{2|q_fB|}}  
\, ,
\end{eqnarray}
where the integrals can be performed as in the previous subsection for the photon masses. 
One gets the same results irrespective of the order of limits. 
In the thermal spectral flow, the axial-charge imbalance appears near the thermal cutoff 
of the order of temperature or density instead of the infrared regime. 
Therefore, effects of the finite curvature along the dispersion curve 
is negligible in the high temperature or density limit defined as $ m/T , \ m/\mu \to 0 $.

Diabatic processes can also give rise to a nonzero divergence of the axial-vector current. 
In the present case, those processes have been identified with 
the pair creation and the Landau damping. 
We already discussed the contribution from the pair creation in vacuum. 
A net chirality carried by a pair of massive fermion 
and antifermion belonging to different chirality sectors. 
The pair creation rate is captured by the imaginary part of the polarization tensor 
inspected in the preceding subsection. 
The Landau damping can also contribute because a fermion or antifermion 
can be back-scattered during this process, implying occurrence of the helicity flip. 
Those contributions were numerically studied in Ref.~\cite{Hattori:2022wao}.

Overall, one can only get simple results in the adiabatic limit 
that can be understood with the spectral-flow picture. 
However, the flow pattern crucially depends on the fermion mass, temperature, and density. 
Once diabatic processes are activated, 
fermions do not follow such a collective flow. 
Then, one needs to investigate 
the frequency and wavelength dependences of the AWI 
with the explicit form of the response function, 
which do not exhibit universal behaviors. 
The AWI takes the simple universal form in the massless case
only because those diabatic processes are kinematically prohibited. 
In a sense, this situation is similar to that in quantum Hall systems and topological insulators 
where clean edge currents can be measured 
thanks to the absence of 
non-universal metallic currents in the bulk; 
Noises are muted by inherent mechanisms. 
The massless nature strongly restricts possible processes to those that 
occur in the right- and left-handed sectors independently.

\cout{
It is also worth mentioning that mass effects on the axial-charge generation 
in a parallel electric and magnetic field was discussed in Ref.~\cite{Ambjorn:1983hp} 
and more recently in Ref.~\cite{Copinger:2018ftr}, suggesting the absence of 
the axial-charge generation over a finite mass gap 
in an equilibrium state. 
The spectral flow on a parabolic dispersion relation turns back toward 
the bottom of the Dirac sea in an {\it adiabatic} process 
in contrast to the spectral flow on a gapless dispersion relation for the massless fermion \cite{Nielsen:1983rb}. 
Note that the Schwinger mechanism and the Laudau-Zener transition 
mentioned in Sec.~\ref{sec:tunneling} are {\it diabatic} processes. 
}

\subsection{Gluon/photon self-energy in a weak magnetic field}

\label{sec:selfenergy-weak_B}

In this subsection, we discuss the vacuum polarization, 
or the photon/gluon self-energy, in the weak field limit 
that can be addressed with perturbative insertion of 
the external-field lines to the self-energy diagram. 
Such computation was well-investigated in 
a somewhat different context in the computation of 
the Wilson coefficients for the operator product expansion 
(see Ref.~\cite{Reinders:1984sr} for a review). 
There is an established algorithm useful 
for analytic computation of 
the self-energy diagrams \cite{Nikolaev:1982rq}. 
We will summarize the flow chart of the computation and 
the result of the leading-order correction below.

The zeroth-order diagram is a familiar vacuum polarization tensor without any insertion of the external field. 
The computation and properties of this diagram is 
given in standard textbooks, e.g., in Ref.~\cite{Peskin:1995ev}, 
so that we do not discuss this diagram. 
The diagrams with one insertion of the external-field line vanish according to the Furry's theorem. 
Therefore, the perturbative corrections by the external field start at the second-order diagrams 
shown in Fig.~\ref{fig:selfenergy2}. 
Corresponding to those diagrams, one may write the leading-order correction as 
\begin{eqnarray}
\Pi^{\mu\nu} (q) = \Pi_{(1)}^{\mu\nu} (q)  + 2  \Pi_{(2)}^{\mu\nu} (q) 
\, .
\end{eqnarray}
The latter two diagrams provide the same contributions when the quark pair 
carry an identical mass and charge, resulting in a factor of two. 
These terms can be written down by the use of the fermion propagators (\ref{eq:fermion1})-(\ref{eq:fermion2}) 
with the insertions of one and two external fields \cite{Reinders:1984sr, Nikolaev:1982rq} as 
\begin{subequations}
\begin{eqnarray}
i \Pi_{(1)}^{\mu\nu} (q)  &=& g^2 \tr[t^a t^b]
\int \frac{d^4p}{(2\pi)^4} \tr[ \, \gam^\mu S_1 (p) \gam^\nu S_1 (p-q) \, ]
\, ,
\\
 i \Pi_{(2)}^{\mu\nu} (q) &=&  g^2 \tr[t^a t^b]
\int \frac{d^4p}{(2\pi)^4} \tr[ \, \gam^\mu S_2 (p) \gam^\nu S_0 (p-q) \, ]
\, .
\end{eqnarray}
\end{subequations}
Note that the product of the coupling constant and the color trace, $  g^2 \tr[t^a t^b]$, 
for the gluon self-energy should be replaced by $ q_f^2 \tr[ \id]_{\rm color} $ 
for the photon self-energy with a quark loop. 
Inclusion of leptons needs a modification of the electric charge and 
removal of the color matrix as obvious.

\begin{figure}[t]
     \begin{center}
              \includegraphics[width=0.8\hsize]{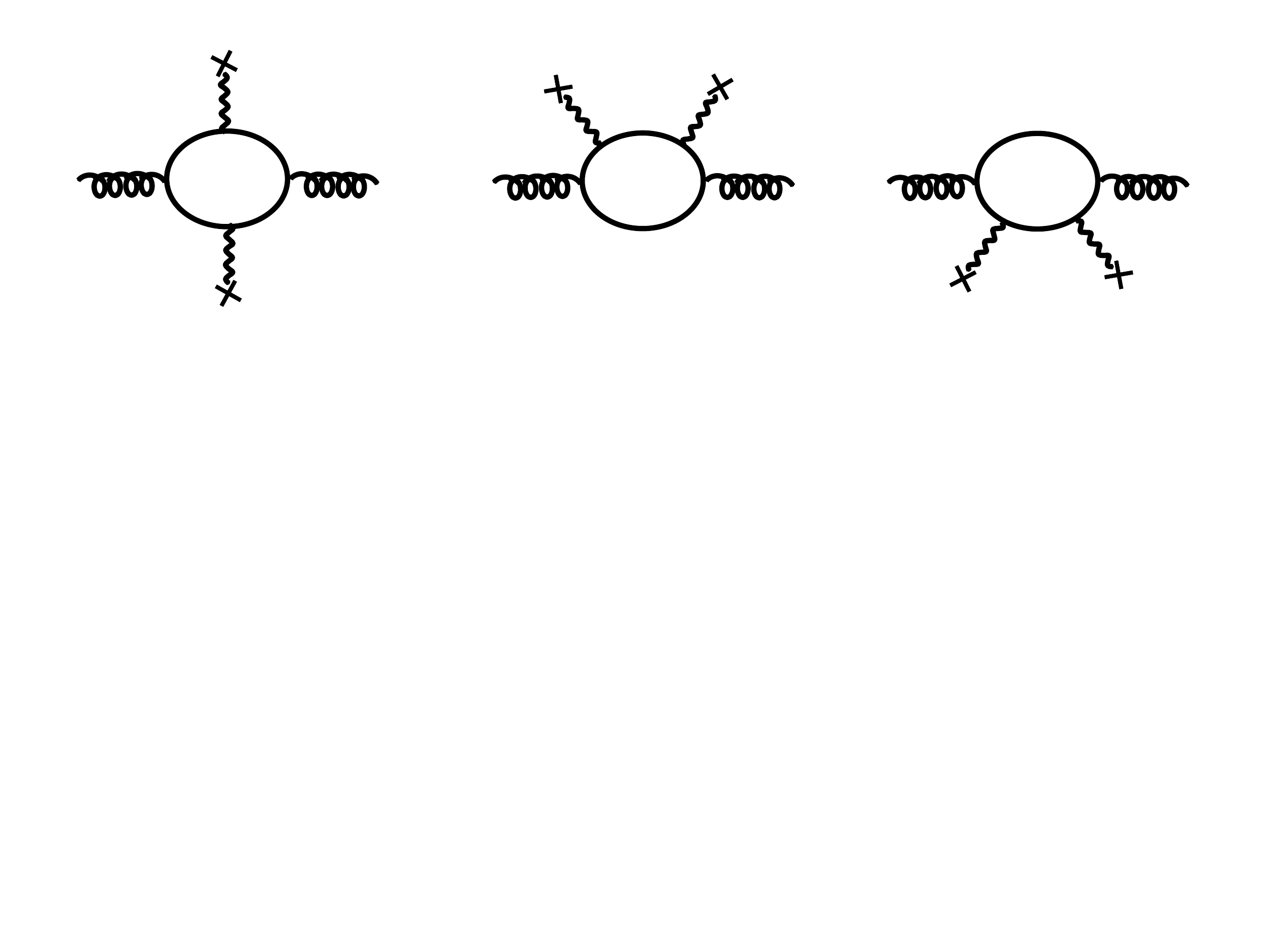}
     \end{center}
\vspace{-0.5cm}
\caption{Leading-order perturbative corrections to the gluon/photon self-energy 
with two insertions of an external field (wavy lines with crosses). }
\label{fig:selfenergy2}
\end{figure}

At the second order, one can decompose the field strength tensor as 
\begin{eqnarray}
q_f^2 F_{\mu\nu} F_{\alpha\beta} = 
\frac{1}{12} ( g_{\mu\alpha} g_{\nu\beta} - g_{\mu\beta}g_{\nu\alpha}) H
+ \frac{1}{2} ( g_{\mu\alpha} H_{\nu\beta} - g_{\mu\beta} H_{\nu\alpha} 
+ g_{\nu\beta} H_{\mu\alpha} - g_{\nu\alpha} H_{\mu\beta}   ) 
- \frac{2}{4!} \epsilon_{\mu\nu\alpha\beta} \tilde H
\, ,
\end{eqnarray}
where the scalar, the symmetric and traceless, and the antisymmetric parts 
are, respectively, given by 
\begin{subequations}
\begin{eqnarray}
&&
H = q_f^2 F_{\mu\nu} F^{ \mu\nu} 
\, ,
\\
&&
H_{\mu\nu} =  q_f^2  \left( F_{\mu\sigma} F_\nu^{ \ \sigma}  
- \frac{1}{4} g_{\mu\nu}  F_{\lambda \sigma} F^{ \lambda \sigma}  \right)
\, ,
\\
&&
\tilde H = q_f^2  F_{\mu\nu}  \tilde F^{ \, \mu\nu} 
\,  .
\end{eqnarray}
\end{subequations}
Accordingly, the self-energy can be also decomposed into the scalar and symmetric-tensor parts as 
\begin{eqnarray}
 \Pi^{\mu\nu} = \frac{g^2}{\pi^2} \tr[t^a t^b]  \left( \Pi^{\mu\nu} _s + \Pi^{\mu\nu} _t \right)
\label{eq:2insertions}
\, ,
\end{eqnarray}
with 
\begin{subequations}
\label{eq:PI-perturbative}
\begin{eqnarray}
 \Pi^{\mu\nu} _s&=& H \left[ \ 
 - \frac{1}{12} \phi^{(1) {\mu\nu} } -   m^2 \phi^{(2){\mu\nu}}
\ \right]
\label{eq:PI_scalar}
\, ,
\\
 \Pi^{\mu\nu}_t &=&   H^{\alpha\beta}
\left[ \ 
\frac{1}{2} \psi^{(1){\mu\nu}}_{\alpha\beta} + 
4 \left( \psi^{(21){\mu\nu}}_{\alpha\beta} - \psi^{(22){\mu\nu}}_{\alpha\beta} \right)
\ \right]
\label{eq:twist2}
\, .
\end{eqnarray}
\end{subequations}
The antisymmetric part does not contribute in usual environments 
where the correlator is symmetric in the Lorentz indices. 
More specifically, one needs an odd number of $ \gamma^5 $ so that 
the spinor trace yields an antisymmetric tensor.\footnote{
Possible examples are the presence of an axial chemical potential, 
or a mixing between the currents which have different parities. 
}
The coefficients in Eqs.~(\ref{eq:PI_scalar}) and (\ref{eq:twist2}) are obtained by performing the loop integrals: 
\begin{subequations}
\begin{eqnarray}
\phi^{(1){\mu\nu}} &=& 
-i\frac{ (4\pi)^2 }{32}  \int \frac{d^4p}{(2\pi)^4} 
\frac{1}{A^2 B^2 } 
\tr[ \ 
\gam^\mu  \{ \sigma^{\delta\gamma} , \slashed p \} 
\gam^\nu \{ \sigma_{\delta\gamma}, (\slashed p - \slashed q) \} 
+ (2m)^2 \gam^\mu  \sigma^{\delta\gamma} \gam^\nu \sigma_{\delta\gamma}  
\ ]
\, ,
\\
\phi^{(2){\mu\nu}} &=& 
-i \frac{ (4\pi)^2 }{4} \int \frac{d^4p}{(2\pi)^4} 
\frac{1}{A^4 B^1 }
\tr[ \ 
\gam^\mu \gam^\nu p^2 
+ \gam^\mu  \slashed p \gam^\nu (\slashed p - \slashed q) 
\ ]
\, ,
\end{eqnarray}
\end{subequations}
and 
\begin{subequations}
\begin{eqnarray}
\psi^{(1){\mu\nu}}_{\alpha\beta} &=& 
i \frac{ (4\pi)^2 }{16}  \int \frac{d^4p}{(2\pi)^4} 
\frac{1}{A^2 B^2 } 
\tr[ \ 
\gam^\mu  \{ \sigma_{\alpha\lambda} , \slashed p \} 
\gam^\nu  \{ \sigma_{\beta}^{\ \; \lambda}, (\slashed p - \slashed q) \} 
+ (2m)^2 \gam^\mu  \sigma_{\alpha\lambda} \gam^\nu \sigma_{\beta}^{\ \; \lambda}  
\ ]
\, ,
\nonumber
\\
\\
\psi^{(21){\mu\nu}}_{\alpha\beta} &=& 
-i \frac{ (4\pi)^2 }{8} \int \frac{d^4p}{(2\pi)^4} 
\frac{1}{A^3 B^1 } 
\tr[ \ 
\gam^\mu ( p_\alpha \gamma_\beta + p_\beta \gamma_\alpha ) 
\gam^\nu   (\slashed p - \slashed q) \} 
\ ]
\, ,
\\
\psi^{(22){\mu\nu}}_{\alpha\beta} &=& 
-i \frac{ (4\pi)^2 }{4}  \int \frac{d^4p}{(2\pi)^4} 
\frac{ p_\alpha p_\beta }{A^4 B^1 } 
\tr[ \ 
m^2 \gam^\mu \gam^\nu 
+\gam^\mu  p  \gam^\nu (\slashed p - \slashed q)  
\ ]
\, ,
\end{eqnarray}
\end{subequations}
where $A= p^2 -m^2  $ and $  B = (p-q)^2 -m^2  $. 
One can arrange the above expressions into compact ones 
by the use of a nice algorithm \cite{Nikolaev:1982rq}. 
We briefly summarize the procedure in the following.

As in the standard procedure \cite{Peskin:1995ev}, 
the propagators are combined with the help of the Feynman parameter: 
\begin{eqnarray}
\frac{ 1 }{A^m B^n } 
&=& \frac{ \Gamma(n+m) }{ \Gamma(m) \Gamma(n) }
\int \frac{ x^{m-1} (1-x)^{n-1} \ dx }{ [ x( p^2 -m ^2 ) + (1-x) \{ (p-q)^2 - m^2 \} ]^{m+n} }
\, .
\end{eqnarray}
Then, replacing an integral variable as $\ell = p - (1-x)q $, 
the loop integral can be performed as 
\begin{eqnarray}
\int \frac{d^4\ell}{(2\pi)^4}  \frac{ (\ell^2)^r }{ (\ell^2 - \Delta )^s } 
&=& i \frac{(-1)^{r-s}}{(4\pi)^2} 
\frac{ \Gamma(r+2) \Gamma(s-r-2) }{ \Gamma(s) } \; \Delta^{r-s+2} 
\, ,
\end{eqnarray}
where we introduced 
$\Delta = m^2 - x(1-x) q^2$. 
After the integration, one will obtain an expression with single integrals 
with respect to the Feynman parameter, of which the general form is given by 
\begin{eqnarray}
I^m_n (N) 
= \int_0^1 \frac{x^m (1-x)^n}{ [m^2+x(1-x)Q^2]^N } \, dx 
\label{eq:eqeq}
\, ,
\end{eqnarray}
with integers $ m$, $n $, $N $. 
By using recursive relations\footnote{
From a simple identify $  x^n (1-x)^{n-1} = x^{n-1} (1-x)^{n-1} - x^{n-1} (1-x)^{n}$, 
we have $  I^n_{n-1} (N) = I_{n-1} (N) - I^{n-1}_{n} (N) $. 
Thus, noting that $I^n_m(N)  = I^m_n(N) $, the first recursive equation is readily obtained. 
The other relations are also obtained in a similar way. 
} 
\begin{subequations}
\begin{eqnarray}
I^n_{n-1} (N) &=& \frac{1}{2} I_{n-1}(N) 
\, ,
\\
I^n_{n-2}(N)  &=& - I_{n-1} (N) + \frac{1}{2} I_{n-2}(N) 
\, ,
\\
I^n_{n-3} (N) &=& - \frac{3}{2} I_{n-2} (N) + \frac{1}{2} I_{n-3} (N) 
\,  ,
\end{eqnarray}
\end{subequations}
one can reduce the difference between the upper and lower indices, 
and arrive at the expression given by 
$ I_n (N) \equiv I^n_n (N)  $, that is, 
\begin{eqnarray}
I_n (N)  
= \int_0^1 \frac{x^n(1-x)^n}{ [m^2-x(1-x)q^2]^N } \, dx 
=: \frac{1}{(m^2)^N} J_N^n
\, ,
\end{eqnarray}
where we introduced a dimensionless integral $ J_N^n $. 
Furthermore, 
one can immediately get another recursive relation 
\begin{eqnarray}
J_N^n &=& y^{-1} ( \, J_{N-1}^{n-1} - J_{N}^{n-1} \, )
\, .
\end{eqnarray}
By the use of this relation, one can reduce the order $ n $ in the numerator, 
and finally reach the integrals that do not have the Feynman parameters in the numerators, 
i.e., 
\begin{eqnarray}
J_N := J^0_N = \int_0^1 \, [\, 1 + x(1-x) y \,]^{-N} \, dx
\, ,
\end{eqnarray}
with $y=-q^2/m^2$. 
Then, the integral with an arbitrary index $  N$ can be obtained recursively as 
\begin{eqnarray}
J_N  
&=& \frac{(2N-3)!!}{(N-1)!} \left( \frac{a-1}{2a} \right)^{N-1}  \left[ \, 
J_1 
+ \sum_{k=0}^{N-2} \frac{ k!}{(2k+1)!!} \left( \frac{a-1}{2a} \right)^{-k}
\, \right]
\, ,
\end{eqnarray}
where $ a = 1 + 4/y $, and 
\begin{eqnarray}
J_1 = \left( \frac{a-1}{2a} \right)  \sqrt{a} \ln \frac{\sqrt{a}+1}{\sqrt{a}-1}
\, .
\end{eqnarray}


Now, performing a lengthy but 
straightforward calculation outlined above, 
we obtain the leading-order correction (\ref{eq:2insertions}) 
to the vacuum polarization as 
\begin{subequations}
\begin{eqnarray}
\Pi^{\mu\nu}_{s}
&=& 
\frac12 H w P_0^{\mu\nu}
\, ,
\\
\Pi^{\mu\nu}_{t} &=& \frac12 H_{\alpha\beta} 
[ \   v  P_0^{\mu\nu} q^\alpha q^\beta
+ 
u \, \{ - g^{\mu\nu} q^\alpha q^\beta + 
 g^{\mu\alpha} q^\nu q^\beta + g^{\nu\alpha} q^\mu q^\beta 
- g^{\mu\alpha} g^{\nu\beta} q^2 \} 
\ ]  
\, ,
\end{eqnarray}
\end{subequations}
\cout{
NOTE: 
\begin{subequations}
\label{eq:}
\begin{eqnarray} 
\frac{\alpha_s}{\pi} \cdot M
\= \frac{g^2}{4 \pi^2} \cdot 
\langle G^{a\mu\nu} G^{a\mu\nu} \rangle
\nnb
\= \frac{2}{4} \cdot \frac{g^2}{\pi^2} 
\cdot \tr[t^a t^b] \langle G^{a\mu\nu} G^{a\mu\nu} \rangle
\\
\frac{\alpha_s}{\pi} \cdot (2 M^{\a\b})
\= \frac{g^2}{4 \pi^2} 
\cdot \langle G^{a\mu\a} G^{a\nu}_{\ \, \a}\rangle_{\rm TS}
\nnb
\= \frac{2}{4} \cdot \frac{g^2}{\pi^2} 
\cdot \tr[t^a t^b] \langle G^{a\mu\a} G^{b\nu}_{\ \, \a}\rangle_{\rm TS}
\end{eqnarray}
\end{subequations}

\begin{eqnarray}
[- E^2 g_\para^{\mu\nu} + B^2 g_\perp^{\mu\nu} 
- \frac12 (B^2-E^2) g^{\mu\nu}] q_\mu q_\nu
\= - E^2 q_\para^2 + B^2 q_\perp^2 
- \frac12 (B^2-E^2) ( q_\para^2 + q_\perp^2 )
\nnb
\= - \frac12 (E^2 + B^2) ( q_\para^2 -  q_\perp^2 )
\end{eqnarray}

}
where $P_0^{\mu\nu} =  (g^{\mu\nu} -q^\mu q^\nu/q^2) $ and 
\begin{subequations}
\begin{eqnarray}
w &=&  \frac{1}{12 q^2} ( 1 - 3J_2 + 2 J_3 ) 
\, ,
\\
v &=& 
\frac{1}{q^4} \left[ \;  - \frac{2}{3}  +2  J_1 - 2 J_2 + \frac{2}{3} J_3 \; \right]
\, ,
\\
u &=& 
 \frac{1}{q^4} \left[ \; \frac{1}{2} + \left( 1-\frac{1}{3} y \right) J_1 - \frac{3}{2} J_2 \; \right]
 \, .
\end{eqnarray}
\end{subequations}
The tensor part $ \Pi_t^{\mu\nu} $ has been obtained 
as the Wilson coefficient for the twist-two gluon condensate in medium \cite{Klingl:1998sr} 
(up to the differences in color factors). 
Choosing the Lorentz frame where 
the electric and magnetic fields are parallel/antiparallel 
to each other, we find the leading correction to 
the gluon/photon self-energy (\ref{eq:2insertions}) as 
\begin{subequations}
\begin{eqnarray}
\Pi^{\mu\nu}_{s} &=& q_f^2  (B^2-E^2) w P_0^{\mu\nu}
\, ,
\\
\Pi^{\mu\nu}_{t} &=& 
- \frac{1}{4} q_f^2  (B^2+E^2) [ \   v  (q_\para^2-q_\perp^2) P_0^{\mu\nu} 
-
2u (q_\para^2 P_\para^{\mu\nu} - q_\perp^2 P_\perp^{\mu\nu} )
\ ]  
\, .
\end{eqnarray}
\end{subequations}
Here, without the medium effect, the self-energy contains the transversal tensor structures discussed 
around Eq.~(\ref{eq:selfenergy}) 
in the beginning of this section. 
The results at $B \not = 0 , \, E=0 $ were used to investigate charmonium spectroscopy 
in an external magnetic field by means of the QCD sum rule~\cite{Cho:2014exa, Cho:2014loa}. 


\section{Chiral magnetic/separation effect with massive fermions}

\label{sec:massive}


We discuss mass effects on the anomalous transport phenomena in a magnetic field.  
Here, we focus on a straightforward extension of the computation in Sec.~\ref{sec:Ritus-CME} by the use of the mode expansion of fermion fields in the Ritus basis. 
This extension assumes existence of constant axial and vector chemical potentials in the thermal equilibrium. 
Then, we will reach a somewhat misleading conclusion that there is no mass effect on the CME. 
Does this really mean that the same amount of the CME current is transported no matter how heavy the fermion is? 
In reality, this should not be taken literary since the axial charge density is not a conserved quantity due to 
the explicit chiral symmetry breaking by a fermion mass 
and the CME current is implicitly, but possibly significantly, 
dependent on a fermion mass. 
The axial charge will be damped out  
in an equilibrium due to the helicity flipping processes 
(unless there is an external pumping effect, 
which may realize a steady state). 
One should keep these caveats in mind when computing the CME current with an axial chemical potential $ \mu_A$ 
as a static parameter.  
On the other hand, the vector charge and 
associated vector chemical potential $\mu_V $ 
are well-defined equilibrium quantity 
protected by a $ U(1)_V $ symmetry, 
so that we will find a simpler and more robust conclusion on the mass correction to the CSE. 


Computation of the currents goes almost in parallel to 
that in the massless case performed in Sec.~\ref{sec:Ritus-CME}. 
Here, we should use the V/A basis to compute the currents 
since the chirality is no longer a good quantum number of massive fermions. 
We focus on the spatial component along the magnetic field 
\begin{eqnarray}
\label{eq:current-definition-VA}
 j^\mu_{V/A} (x) = 
  \langle \bar \psi(x) \gam_{V/A}^\mu \psi(x) \rangle
 \, ,
 \end{eqnarray} 
where $\gam_{V}^3 = \gam^3 $ and $\gam_{A}^3 = \gam^3 \gam^5 $.
The convention of $ \gam^5 $ is given below Eq.~(\ref{eq:metrics}).

To organize the mode expansion similar to 
Eq.~(\ref{eq:Ritus-mode}), we first find 
solutions for the Dirac equation with the chemical potentials. 
In addition to the vector chemical potential, 
we introduce an axial chemical potential as a temporal component of the axial gauge field 
that is coupled to the right- and left-handed helicity modes with opposite signs 
(Note that this procedure itself does not justify the existence of the static axial chemical potential). 
The Dirac operator with nonzero vector and axial chemical potentials read [cf. Eq.~(\ref{Dirac_Ritus})]
\begin{eqnarray}
(i \sla D + \mu_V \gam^0 + \mu_A  \gam^0 \gam^5 -m)  
= ( i \sla \partial _\para  + \mu_V \gam^0  + \mu_A \gam^0 \gam^5  -m) 
- \sqrt{2n|q_fB|} \gam^1 (\hat a \prj_+ + \hat a^\dagger \prj_-)
\, .
\end{eqnarray}
As usual, the origin of energy is shifted by the vector chemical potential $ \mu_V $. 
On the other hand, the longitudinal momentum is shifted by the axial chemical potential $ \mu_A$ 
with spin-dependent signs as implied by an identity $\gam^0 \gam^5 \prj_\pm  = \pm s_f   \gam^3 \prj_\pm$. 
Since the LLL has the unique spin state, 
we simply have a shift of the momentum when the Dirac operator acts on the Ritus basis in the LLL: 
\begin{eqnarray}
(i \sla D + \mu_V \gam^0  + \mu_A \gam^0 \gam^5 -m)  \psi_{\LLL} (x) 
= e^{- i p_\para x} \Ritus_{n=0, \qn}   \{ (p^0+\mu_V) \gam^0 - (p^3 - s_f \mu_A ) \gam^3 - m  \}  u
\, ,
\end{eqnarray}
where we assumed the ansatz (\ref{eq:Dirac-Ritus}) for 
positive-energy solutions. 
On the other hand, the spin degeneracy in the hLLs 
is resolved by the inclusion of $ \mu_A $ as 
\begin{eqnarray}
(i \sla D + \mu_V \gam^0 + \mu_A \gam^0 \gam^5 -m)  \psi_\hLL (x) 
&=&
e^{- i p_\para x} \Ritus_{n,\qn}   (\sla p_n - m) \, u 
+  (\mu_V \gam^0 + \mu_A \gam^0 \gam^5)   \psi_\hLL (x) 
\nn
\\
&=& e^{-i p_\para \cdot x} \Ritus_{n,\qn}  \left[ \prj_+ ( \sla p_{n}^+  - m  )  +  \prj_-  ( \sla p_{n}^-  - m  )  \right] u
\, ,
\end{eqnarray}
where we used $ \prj_\pm \prj_\mp =0 $ and 
defined the shifted momenta $ p_{n}^{\pm} = (p^0+\mu_V, \sqrt{2n|q_fB|}, 0, p^3 \mp s_f \mu_A ) $. 
Correspondingly, one can decompose the spinor basis as 
$u = u^+ \prj_+ + u^- \prj_- $ where 
the spinors $ u^\pm $ satisfy 
distinct ``free'' Dirac equations 
$ ( \sla p_{n}^{\pm}  - m  ) u^\pm =0 $. 
Similarly, the negative-energy solutions satisfy 
the above equations with the sign of momentum flipped; 
The spin basis $ v^\pm$ is provided by the solutions for 
$ ( \bar{\sla p}_{n}^{\pm} + m  ) v^\pm =0 $ with 
$ \bar p_{n}^{\pm} = (p^0 - \mu_V, -\sqrt{2n|q_fB|}, 0, p^3 \pm s_f \mu_A ) $. 
One can then perform the the mode expansion 
in the same manner as in Eq.~(\ref{eq:Ritus-mode}) 
but with the distinct dispersion relations for 
the two spin states.

Now, we are ready to compute the current.  
Inserting the mode expansion as in Eq.~(\ref{eq:current-RL-0}), 
we get the currents expressed by the Ritus basis. 
This amounts to replacement of the gamma matrices 
and the dispersion relations as 
\begin{eqnarray}
j^3_{V/A} (q=0) \= \frac{1}{L_x L_y} \int d^3 x \, e^{ - i 0 \cdot \bx}  j^3_{V/A} (x) 
 \nnb
 \=   \frac{1}{L_x L_y}   \int d^3 x  \sum_{\kappa=\pm}  \sum_{\kappa'=\pm}
  \sum_{n=0}^{\infty} \sum_{n'=0}^{\infty}
  \int \frac{d p_z}{2\pi}   \int  \frac{dp_y dp'_y}{(2\pi)^2} 
\label{eq:current-VA-0}
\\
&&
\times \Big[ \  
\frac{1}{\sqrt{4 \epsilon_n^\kappa \epsilon_{n'}^\kappa}}
\langle a_{p_{n'}, p'_y}^{\kappa' \dagger}  a_{p_n, p_y}^{\kappa} \rangle  
  \bar u ^{\kappa'} (p_{n'}^{\kappa'}) 
  \Ritus_{n',p_y'}^\dagger(x_\perp)  
\gam_{V/A}^3 \Ritus_{n,p_y} (x_\perp)   u ^\kappa (p_n^\kappa)
   e^{  i (\epsilon_{n'}^\kappa - \epsilon_{n}^\kappa ) t} 
  \nnb
&& 
  +  
  \frac{1}{\sqrt{4 \bar \epsilon_n^\kappa \bar \epsilon_{n'}^\kappa}}
  \langle b_{\bar{p}_{n'}, p'_y}^{\kappa'}  b_{\bar{p}_{n}, p_y}^{\kappa \dagger}  \rangle 
 \bar v ^{\kappa'} (\bar p_{n'}^{\kappa'})    \Ritus_{n',p_y'}^\dagger(x_\perp)
\gam_{V/A}^3 \Ritus_{n,p_y} (x_\perp)  v ^\kappa (\bar p_n^\kappa)   e^{ - i (\epsilon_{n'}^\kappa - \epsilon_{n}^\kappa ) t} 
\ \Big]
\nn
 \, ,
 \end{eqnarray}  
with $L_{x,y}  $ being the system lengths in the transverse directions. 
The temporal component of the momentum is given by 
the shifted Landau levels as 
$(p_n^\kappa)^0 = \sqrt{ (p_z \mp s_f \mu_A)^{ 2} + m^ 2+ 2 n |q_f B| } = :\epsilon_n^\kappa $ 
and $(\bar p_n^\kappa)^0= \sqrt{ (p_z \pm s_f \mu_A)^{ 2} + m^2 + 2 n |q_f B| } =: \bar \epsilon_n^\kappa $. 
We use the fact that  $  \prj_\pm$ commutes 
with $ \gam^3 $ and $ \gam^5 $ 
and then the orthogonal relation (\ref{eq:orthogonal-p}) to find that 
\begin{eqnarray}
\label{eq:current-VA}
j^3_{V/A} (q=0) \=  \frac{|q_fB|}{2\pi}
 \sum_{\kappa=\pm}  \sum_{\kappa'=\pm} \sum_{n=0}^{\infty}  
  \int \frac{d p_z}{2\pi} 
\\
&&
\times \Big[  \frac{1}{2 \epsilon_n } \langle a_{p_{n}, p_y}^{\kappa' \dagger}  a_{p_n, p_y}^{\kappa} \rangle  
  \bar u ^{\kappa'} (p_{n}^{\kappa'}) \gam_{V/A}^3 I_n  u ^\kappa (p_n^\kappa) 
 +  \frac{1}{2 \bar \epsilon_n }  \langle  b_{\bar{p}_n, p_y}^{\kappa'} b_{\bar{p}_{n}, p_y}^{\kappa \dagger}  \rangle 
 \bar v ^{\kappa'} (\bar p_{n}^{\kappa'}) \gam_{V/A}^3  I_n  v ^\kappa (\bar p_n^\kappa)  
 \Big]
 \nn
 \, .
 \end{eqnarray} 
The Landau degeneracy factor is reproduced from the $ y $ integral in the Landau gauge (see the last part of Sec.~\ref{sec:Landau-g}). 
The thermal expectation value is assumed to be diagonal 
in the spin basis, i.e., $ \langle a_{p_{n}, p_y}^{\kappa' \dagger}  a_{p_n, p_y}^{\kappa} \rangle 
= \delta_{\kappa\kappa'} f(\epsilon_n^\kappa - \mu_V) $, 
where $ f(\epsilon) = 1/[\exp(\epsilon /T) +1] $ is the Fermi-Dirac distribution function. 
Likewise, we have $ \langle  b_{\bar{p}_n, p_y}^{\kappa'} b_{\bar{p}_{n}, p_y}^{\kappa \dagger}  \rangle 
= \delta_{\kappa\kappa'} \{1-f(\bar \epsilon_n^\kappa + \mu_V )\}$. 
Then, we arrive at  
\begin{eqnarray}
\label{eq:current-VA-2}
j^3_{V/A} (q=0) 
 \=  \frac{|q_fB|}{2\pi} \sum_{\kappa=\pm} \sum_{n=0}^{\infty}    \int \frac{d p_z}{2\pi} 
\Big[ 
\frac{ \tr[ \gam^3_{V/A}  I_n  ( \sla p_{n}^\kappa + m) \prj_\kappa ]}
{2 \epsilon_n^\kappa }  f(\epsilon_n^\kappa - \mu_V) 
\nnb
&&
+ \frac{ \tr[  \gam^3_{V/A}  I_n (\bar{\sla p}_{n}^\kappa - m )\prj_\kappa  ]}  
 {2 \bar \epsilon_n^\kappa }
 \{1-f(\bar \epsilon_n^\kappa + \mu_V )\} \Big] 
 \, .
 \end{eqnarray}  
The mass term in the numerator vanishes due to the spinor trace,  
meaning that the chirality cannot be flipped by 
the mass term in those tadpole diagrams. 
Therefore, mass dependences could appear only through 
the energy levels $ \epsilon^\pm, \, \bar \epsilon^\pm $ if any.  
Below, we examine the vector and axial-vector currents separately.

\cout{

For the LLL, we have 
\begin{eqnarray}
\label{eq:current-VA-LLL}
j^3_{V/A} (q=0) 
 \=  \frac{|q_fB|}{2\pi}  \int \frac{d p_z}{2\pi} 
\Big[ 
\frac{ \tr[ \gam^3_{V/A}  \sla p_{n}^+  \prj_+ ]}
{2 \epsilon_n^+ }  f(\epsilon_n^+ - \mu_V) 
+ \frac{ \tr[  \gam^3_{V/A} \bar{\sla p}_{n}^+ \prj_+  ]}  
 {2 \bar \epsilon_n^+ } 
 \{1-f(\bar \epsilon_n^+ + \mu_V )\} \Big] 
 \, .
 \end{eqnarray} 
As for the hLLL, we have 
\begin{eqnarray}
\label{eq:current-VA-hLL}
j^3_{V/A} (q=0) 
 \=  \frac{|q_fB|}{2\pi} \sum_{\kappa=\pm} \sum_{n=0}^{\infty}    \int \frac{d p_z}{2\pi} 
\Big[ 
\frac{ \tr[ \gam^3_{V/A} \sla p_{n}^\kappa  \prj_\kappa ]}
{2 \epsilon_n^\kappa }  f(\epsilon_n^\kappa - \mu_V) 
 +  \frac{ \tr[  \gam^3_{V/A} \bar{\sla p}_{n}^\kappa \prj_\kappa  ]}  
 {2 \bar \epsilon_n^\kappa }
 \{1-f(\bar \epsilon_n^\kappa + \mu_V )\} \Big] 
 \, .
 \end{eqnarray} 
}

Carrying out the spinor trace, we find the vector current  
\begin{eqnarray}
\label{eq:current-V-0}
j^3_{V} (q=0) 
 \=  \frac{|q_fB|}{2\pi} \sum_{\kappa=\pm} \sum_{n=0}^{\infty}    \int \frac{d p_z}{2\pi} \Big[ 
\frac{ p_z -\kappa s_f \mu_A }{\epsilon_n^\kappa } 
f(\epsilon_n^\kappa - \mu_V) 
 + \frac{ p_z +\kappa s_f \mu_A }{\bar \epsilon_n^\kappa } 
 \{1-f(\bar \epsilon_n^\kappa + \mu_V )\}  \Big] 
 \nnb
 \=  \frac{|q_fB|}{2\pi} \sum_{\kappa=\pm} \sum_{n=0}^{\infty}    \int_{-\Lambda}^\Lambda \frac{d p_z}{2\pi} 
\frac{ p_z +\kappa s_f \mu_A }{\bar \epsilon_n^\kappa } 
 \, .
\end{eqnarray} 
After shifting the integral variable in the terms 
proportional to the distribution functions, one finds that 
those contributions vanish in contrast 
to the massless case (\ref{eq:current-final}). 
The vacuum contribution has a subtle structure 
since the integrand approaches a constant in the UV region. 
Introducing the cut-off $\Lambda $, we find that 
\begin{eqnarray}
\label{eq:current-V}
j^3_{V} (q=0) 
&=& \frac{|q_fB|}{2\pi} \sum_{\kappa=\pm} \sum_{n=0}^{\infty}   
\frac{1}{2\pi} \lim_{\Lambda\to\infty}  
\left[ \sqrt{ (\Lambda+\kappa s_f \mu_A)^2 + m^2 }
-\sqrt{ (\Lambda- \kappa s_f \mu_A)^2 + m^2 } \right] 
 \nn
\\
&=& \frac{q_fB}{2\pi^2} \mu_A
\label{eq:JV_LLL}
 \, .
\end{eqnarray} 
The final result stems from the LLL contribution: 
There is no net contribution from the hLL 
due to cancellation between the spin states $\kappa = \pm $. 
This result is independent of the fermion mass 
and have the same form as in the massless case (\ref{eq:current-final})~\cite{Fukushima:2009ft, Fukushima:2012vr, Zyuzin:2012tv}. 
However, let us reiterate that this conclusion is 
obtained under a strong assumption for 
the existence of a {\it constant} axial chemical potential, 
which is not always a good approximation in real systems. 
Also, the origins of the same results are different 
as mentioned above. 
The terms proportional to the distribution functions 
do not provide any contribution to the current 
in the massive case. 
In contrast, the current in the massless case 
are transported by thermally populated fermions 
as seen in Sec.~\ref{sec:Ritus-CME}, 
while the explicit temperature dependence finally 
goes away because of the cancellation 
between the particle and antiparticle contributions 
in the integral (\ref{eq:current-final}).

Next, we examine the axial-vector current. 
Carrying out the spinor trace in Eq.~(\ref{eq:current-VA-2}), 
we find that 
\begin{eqnarray}
\label{eq:current-A-0}
j^3_{A} (q=0) 
 \= s_f \frac{|q_fB|}{2\pi} \sum_{\kappa=\pm} \sum_{n=0}^{\infty}    \int  \frac{d p_z}{2\pi} 
\kappa \Big[ f(\epsilon_n^\kappa - \mu_V) 
+  \{1-f(\bar \epsilon_n^\kappa + \mu_V )\}  \Big] 
\nnb
 \= \frac{q_fB}{2\pi}  
 \int  \frac{d p_z}{2\pi} \left[ 
 f(\epsilon_0 - \mu_V) - f(\epsilon_0 + \mu_V ) \right] 
 \, ,
\end{eqnarray} 
where $\epsilon_0 = \sqrt{ p_z^2 + m^2} $. 
We dropped the divergent vacuum term which depends 
on neither $ T $, $ \mu_V$, nor $\mu_A  $ and 
shifted the integral variable $ p_z $ to 
absorb the dependence on $  \mu_A$, 
which is allowed in the convergent integrals. 
The particle and antiparticle contributions have opposite signs 
because their spins are aligned in the opposite directions 
in the LLL. 
The hLL contributions are canceled out 
between the two spin states in each hLL. 

Wrapping up the computation in the massive case, 
the net vector and axial-vector currents are 
transported by fermions in the LLL; 
there are no contributions from the hLLs 
to either vector or axial-vector current 
due to the cancellations. 
This fact is common to the massless case. 
We have obtained 
\begin{subequations}
\begin{eqnarray}
\bj_{V} &=&  q_f \frac{\mu_A } {2\pi^2} \bB
\, ,
\\
\bj_{A} &=&  q_f \frac{ \lambda } {2\pi^2} \bB
\label{eq:JA-final}
\, .
\end{eqnarray}
\end{subequations}
While the vector current has the same form 
as in the massless case (\ref{eq:CME-CSE}), 
the axial current is subject to the mass correction 
encoded in an integral 
\begin{eqnarray}
 \lambda = \frac{1}{2}   \int_{-\infty}^\infty \!\! dp_z
\left[ \, f(  \ep_0 - \mu_V  ) -  f ( \ep_0 + \mu_V ) \, \right]
\, ,
\end{eqnarray}
which agrees with the results from the Kubo formula calculation 
without the explicit Landau quantization~\cite{Lin:2018aon}. 
Note again that the origins of the vector current are different 
in the massless and massive cases~\cite{Fukushima:2009ft, Fukushima:2012vr}. 
The vector current in the massive case
originates from the vacuum contribution in Eq.~(\ref{eq:JV_LLL}) 
with vanishing finite-temperature contributions, 
while the finite-temperature contributions in the massless case result in the temperature-independent result (\ref{eq:current-final}).

\begin{figure}[t]
\begin{minipage}{0.5\hsize} 
	\begin{center} 
             \includegraphics[width=0.9\hsize]{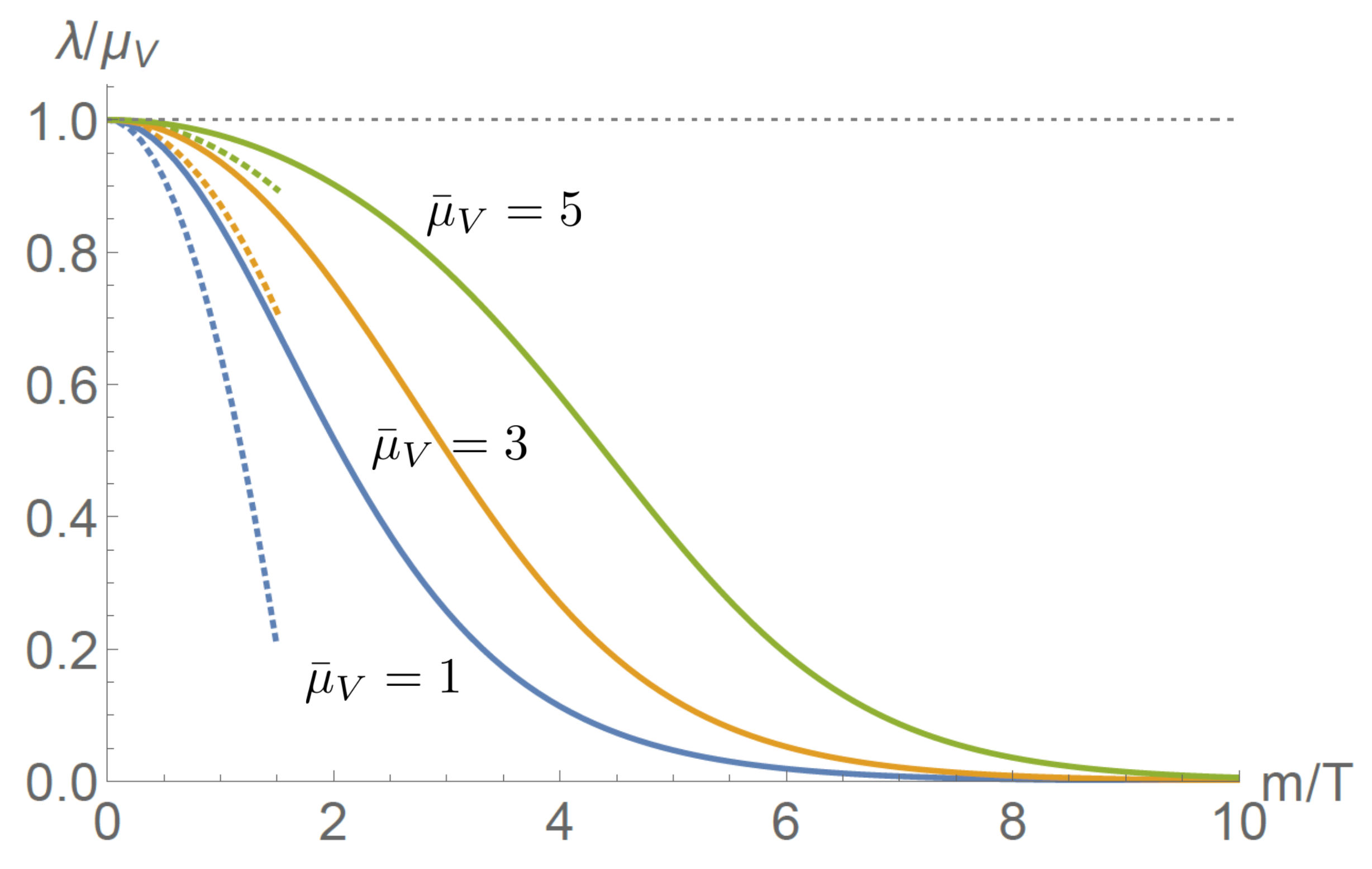}
	\end{center}
\end{minipage}
\begin{minipage}{0.5\hsize}
	\begin{center}
             \includegraphics[width=0.9\hsize]{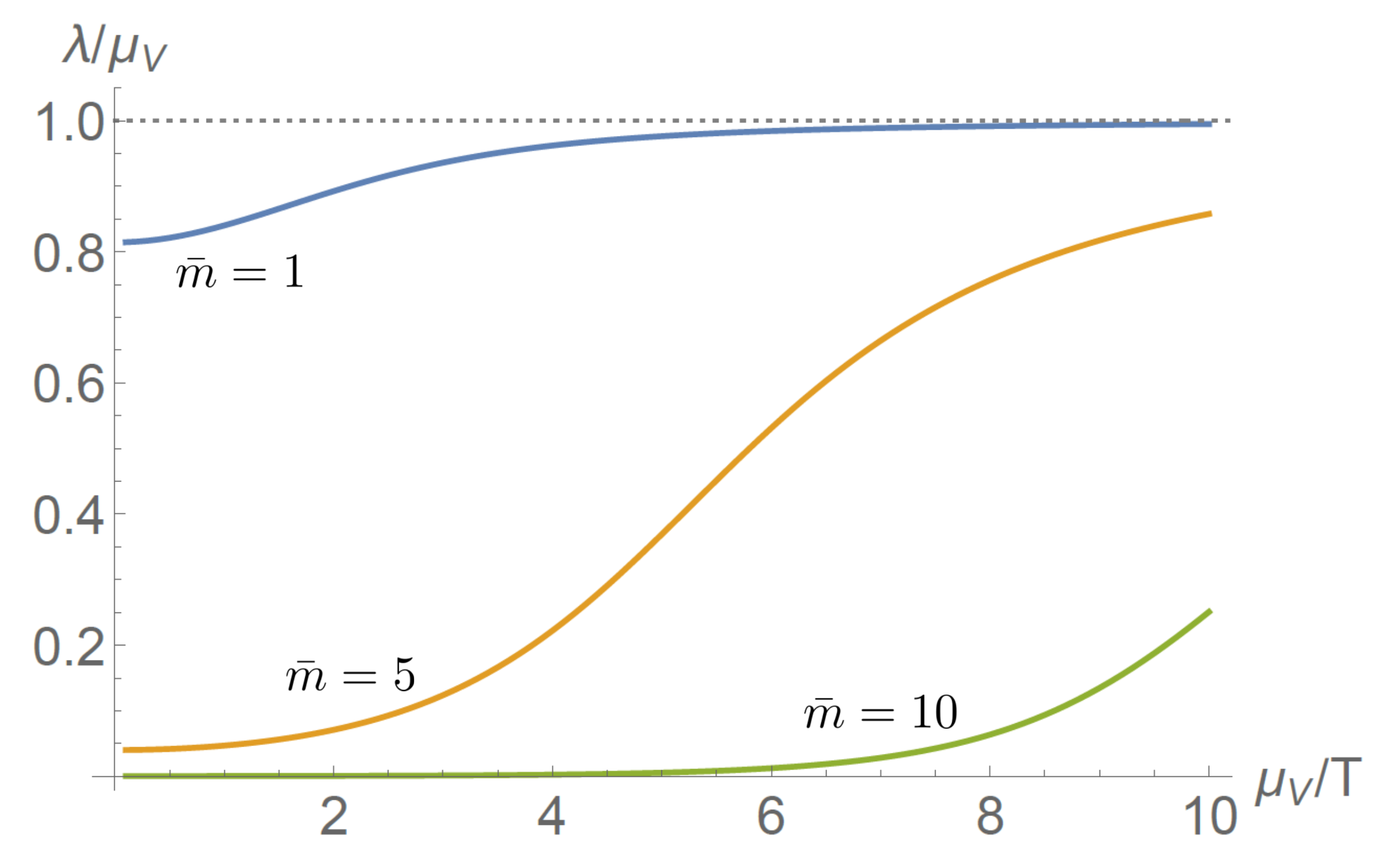}
	\end{center}
\end{minipage}
\caption{Mass corrections to the integral $  \lambda$ for the axial current 
with $\bar \mu_V =  \mu_V/T  $ and $ \bar m = m/T  $. 
Solid lines show the results of numerical integration, 
while the dotted lines show those of the small-mass expansion in $ \order(\bar m^2) $.}
\label{fig:lambda}
\end{figure}

In the rest of this appendix, we more closely look into the mass corrections to the axial-vector current (\ref{eq:JA-final}). 
In the massless limit, one can get an analytic result of the integral
\begin{eqnarray}
  \lambda = \mu_V
  \, ,
 \end{eqnarray} 
which reproduces the previous result in the massless case (\ref{eq:CME-CSE}). 
One should expect sizeable mass effects 
when the mass is larger than both temperature and chemical potential. 
Otherwise, the mass scale may be negligible as compared to either of them. 
Numerical results on mass corrections to the integral $ \lambda $ are shown in Fig.~\ref{fig:lambda}, 
where we introduce normalized variables, $ \bar \mu_V = \mu_V/T $ and $ \bar m = m/T $. 
In the left figure, we confirm that the curves start at $  \lambda = \mu_V$ where $ \bar m =0 $ 
and approach zero as we increase $ \bar m $, 
indicating a suppression of the axial current by mass effects 
(at a given temperature and chemical potential). 
Dotted curves show a small-mass expansion (\ref{eq:small-mass}) shown just below. 
In the right figure, when the normalized mass is small $ \bar m \lesssim 1 $, 
the integral $ \lambda $ exhibits an almost linear dependence on $ \bar \mu_V $ like the massless case. 
As we increase the mass, the integral $ \lambda $ is significantly suppressed when 
the chemical potential is smaller than the mass $ \bar \mu_V  \lesssim \bar m$. 
However, the integral $ \lambda $ again approaches a linear dependence on $ \bar \mu_V $ 
as the chemical potential exceeds the mass scale, $ \bar m \ll\bar \mu_V $.

One can find some analytic forms of the integral $  \lambda$. 
At zero temperature, one can easily perform the integral. 
There is no contribution from antiparticles (particles) when $ \mu_V >0 $ ($ \mu_V < 0 $), 
so that the direction of the current depends on the sign of $ \mu_V $. 
The result is immediately obtained as~\cite{Metlitski:2005pr, Gorbar:2013upa, Lin:2018aon}
\begin{eqnarray}
\bj_A = q_f \frac{\sqrt{ \mu_V^2 - m^2 }  }{2\pi^2} \bB  \ \theta(\mu_V^2 - m^2) \, \sgn(\mu_V) 
\, ,
\end{eqnarray}
where $ \theta(x) $ is the step function. 
At finite temperature, a small-mass expansion was shown in Ref.~\cite{Lin:2018aon}, which is reproduced as 
\begin{eqnarray}
\bj_A 
&=&  q_f \frac{ \mu_V}{2\pi^2}  \bB \left[ \, 1 
- \frac{\bar m^2}{2 \bar \mu_V} \int_{0}^\infty \! \frac{d \bar p_z}{\bar p_z} 
\frac{ \partial \ }{ \partial \bar \mu_V} \left( \, 
 \frac{1}{  \e^{ \bar p_z   - \bar \mu_V } +1 } +  \frac{1}{  \e^{ \bar p_z   + \bar \mu_V } +1 } 
\, \right)
+ {\mathcal O}(\bar m^4) 
\, \right]
\nn
\\
&=&   q_f \frac{ \mu_V}{2\pi^2}  \bB \left[ \, 1  + 
\frac{\bar m^2}{2\bar \mu_V} \lim_{s\to0} \frac{ \partial \ }{ \partial \bar \mu_V} \Gamma(s) \left\{ \, 
{\rm Li}_{s} ( - \e^{-\bar \mu_V} ) + {\rm Li}_{s} ( - \e^{\bar \mu_V} )
\, \right\}
+ {\mathcal O}(\bar m^4)
\, \right]
\nn
\\
&=&q_f \frac{ \mu_V}{2\pi^2}  \bB \left[ \, 1 - 
\frac{\bar m^2}{2\bar \mu_V} \,  \frac{ \partial \ }{ \partial s} \left.  \left\{ \, 
{\rm Li}_{s} ( - \e^{- \bar \mu_V} ) - {\rm Li}_{s} ( - \e^{\bar \mu_V} )
\, \right\} \right|_{s=-1}
+ {\mathcal O}(\bar m^4)
\, \right]
\label{eq:small-mass}
\, ,
\end{eqnarray}
where $ \Gamma(x) $ and $  {\rm Li}_{s} (x) $ are gamma and polylogarithmic functions, respectively.\footnote{
To get the above result, use the following formulas~\cite{Polylogarithm}: 
$ \int_0^\infty dp \, p^s /( \e^{p-\mu} + 1) = - \Gamma(s+1) {\rm Li}_{s+1} ( - \e^\mu) $, 
$\frac{ \partial \ }{\partial x} {\rm Li}_{s} (x) = x^{-1} {\rm Li}_{s-1} (x) $, 
and $ {\rm Li}_{m} (x) + (-1)^{m}  {\rm Li}_{m} (1/x)=0 $ for a negative integer $ m$. 
}
Note that each integral in the leading mass correction has a singularity at $ p_z=0 $ 
which manifests itself as a divergence of the gamma function in the intermediate line. 
Nevertheless, the expansion coefficient in the last line is finite 
thanks to a cancellation of the divergences originating form the particle and antiparticle contributions. 
The leading mass correction (\ref{eq:small-mass}) is shown 
by the dotted curves in Fig.~\ref{fig:lambda}.

\cout{

\section{Notes on relativistic magnetohydrodynamics}

\label{sec:suppMHD}

We provide supplemental notes on the relativistic magnetohydrodynamics discussed in Sec.~\ref{sec:MHD}. 
Here, we discuss choices of hydrodynamic variables and frames, decomposition of the viscous tensor, 
inequalities from the second law of thermodynamics.

\subsection{Conserved charges in magnetohydrodynamics}

\label{sec:time}

First, we shall see that an electric field $ \bE $ and a charge-density fluctuation $ \delta n $ 
in a neutral plasma are damped out in a finite time. 
The charge neutrality implies that the total charge-density fluctuation in a volume vanishes, $ \int d^3x \, \delta n(t,\bx) = 0 $. 
An electric field $ \bE$, induced by the inhomogeneous charge density, 
obeys the Gauss's law $ \nabla \cdot \bE  = \delta n/\epsilon$ with $ \epsilon $ being a dielectric constant. 
Charges will be redistributed via an Ohmic current $ {\bm j}  = \sigma \bE $. 
Plugging the Gauss's law into the current conservation $  \partial_\mu j^\mu =0 $, 
we find 
\begin{eqnarray}
\frac{\partial \delta n(t,\bx)}{\partial t} 
= - \nabla \cdot {\bm j} (t,\bx)
=  - \frac{ \sigma } { \epsilon }\delta n(t, \bx)
\, .
\end{eqnarray}
Thus, the charge-density fluctuation is damped out in time as 
\begin{eqnarray}
\delta n(t,\bx) = \delta n(0,\bx) \exp( - \frac{ \sigma } { \epsilon } t )
\, .
\end{eqnarray}
Plugging this solution back to the Gauss's law, we naturally find that 
the electric field is also damped out in the same time scale 
as the inhomogeneity of the charge density is damped out in a finite time scale $ t \sim \epsilon/ \sigma $. 
Therefore, neither of them are conserved quantities 
and are included in a set of slow variables to construct hydrodynamic equations for a neutral plasma. 
Nevertheless, they can be induced in the dynamics, 
and, if any, should be expressed as functions of  hydrodynamic variables, i.e., constitutive equations. 
On the other hand, the magnetic field persists in a plasma. 
This difference arises form the presence of the current term in the Maxwell equation, 
which breaks the duality between the electric and magnetic fields.

\subsection{Frame choices}

\label{sec:T1}

Next, we discuss ambiguities in the definitions of the flow velocity and thermodynamic quantities beyond the ideal order. 
Settling those ambiguities, we assume a specific form of the energy-momentum tensor in Eq.~(\ref{eq:T1_mn}). 

The most general tensor structures of the first-order dissipative corrections $ T_{(1) {\rm dis}}^{\mu\nu}  $ 
can be written as\footnote{
Since the energy-momentum tensor is even under the parity and charge-conjugation transformations, 
a term proportional to $u^{(\mu} b^{\nu)}  $ requires its coefficient be odd under those transformations. 
Therefore, this term is not allowed in those systems where the parity and charge-conjugation symmetries are preserved. 
}  
\begin{eqnarray}
T_{(1) {\rm dis}}^{\mu\nu} = \delta \epsilon u^\mu u^\nu 
+ \delta p_\para (-b^\mu b^\nu) + \delta p_\perp \varXi^{\mu\nu} 
+  h^{(\mu} u^{\nu)} + f^{(\mu} b^{\nu)} + \tau^{\mu\nu} 
\label{eq:T1_mn0}
\, ,
\end{eqnarray}
where $ \{ \delta \epsilon , \delta P_\para , \delta P_\perp,  h^\mu, f^{\mu} , \tau^{\mu\nu} \} \sim {\mathcal O} (\partial^1) $. 
The off-equilibrium pressure correction, which is written as $ \delta P \Delta^{\mu\nu} $ 
in the absence of a magnetic field, is split into two terms due to the preferred orientation in a magnetic field. 
The fourth term is proportional to the familiar heat current $ h^\mu $. 
Those tensors are transverse to both $  u^\mu$ and $b^\mu  $, i.e., $ u_\mu h^\mu =   b_\mu h^\mu =0$, 
$ u_\mu f^\mu =   b_\mu f^\mu =0$, 
and $   u_\mu \tau^{\mu\nu} = b_\mu \tau^{\mu\nu} =  0$ 
with the symmetric property $  \tau^{\mu\nu}  =  \tau^{\nu\mu}  $. 
They are all transverse to $  u^\mu$ and $b^\mu  $. 
If they have longitudinal components, the heat current, for example, 
can be decomposed into the longitudinal and transverse parts 
as $ h^\mu = (\varXi^{\mu\alpha} + u^\mu u^\alpha - b^\mu b^\alpha) h_\alpha 
=\varXi ^{\mu\alpha} h_\alpha  + \{ \, (u^\alpha h_\alpha) u^\mu  +  (b^\alpha h_\alpha) b^\mu\, \} $. 
Only the first term provides an independent structure to the energy-momentum tensor, 
and the other terms between the braces can be absorbed by the terms existing in Eq.~(\ref{eq:T1_mn0}). 


In an off-equilibrium state, the local thermodynamic quantities as well as the flow velocity are not uniquely defined, 
but are subject to ambiguities of the order of gradients~\cite{landau1959course:fluid, 
Hiscock:1985zz, Bhattacharya:2011tra, Kovtun:2012rj, Minami:2012hs, Jensen:2012jh, Hernandez:2017mch}. 
The gradient corrections vanish as the system approaches an equilibrium state, 
and those quantities reduce to the same equilibrium values 
which may be defined unambiguously by the expectation values of the microscopic current operators 
in equilibrium statistical mechanics. 
We shall, therefore, express the coefficients by the energy-momentum tensor as 
\begin{eqnarray}
&&
\epsilon + \delta \epsilon = u_\alpha u_\beta  T^{\alpha\beta}
, \quad
p_\parallel + \delta  p_\parallel = -b_\alpha b_\beta  T^{\alpha\beta} 
, \quad
p_\perp  + \delta  p_\perp = \frac{1}{2} \varXi_{\alpha\beta}   T^{\alpha\beta} 
,
\nn
\\
&&
h^\mu =  \frac{1}{2} u_{(\alpha} \varXi^{\mu}_{ \beta)}    T^{\alpha\beta} 
, \quad  
f^\mu =  - \frac{1}{2}  b_{(\alpha} \varXi^{\mu}_{ \beta)}  T^{\alpha\beta} 
, \quad
 \tau^{\mu\nu} =( \varXi^{  (\mu} _\alpha \varXi^{\nu) }_\beta - \varXi^{\mu\nu}\varXi_{\alpha\beta} )   T^{\alpha\beta} 
.
\end{eqnarray}
The energy-momentum tensor may be equally parametrized by either set of variables 
before and after the redefinition $ T \to T + \delta T$ and $u^\mu \to u^\mu + \delta u^\mu$ 
with $ \delta T, \delta u^\mu \sim \order(\partial^1) $. 
Under these shifts, the zeroth-order thermodynamic quantities are transformed as $ \epsilon(T) \to \epsilon(T+\delta T) 
\sim \epsilon(T) + (\partial \epsilon/\partial T) \delta T $ and so on, 
while their derivative corrections as $ \delta \epsilon(T) \to \delta \epsilon(T+\delta T) 
\sim \delta \epsilon(T) + \order(\partial^2) $. 
Therefore, up to the first order in gradient, we find 
\begin{subequations}
\label{eq:redef-scalar}
\begin{eqnarray}
&&
\epsilon + \delta \epsilon \to  
\epsilon  + \frac{\partial \epsilon}{\partial T} \delta T+ \delta \epsilon + {\mathcal O} (\partial^2)
,
\\
&&
p_\parallel + \delta  p_\parallel \to
p_\para + \frac{\partial p_\para}{\partial T} \delta T + \delta p_\para+ {\mathcal O} (\partial^2)
,
\\
&&
p_\perp  + \delta  p_\perp \to 
p_\perp + \frac{\partial p_\perp}{\partial T} \delta T + \delta p_\perp+ {\mathcal O} (\partial^2)
,
\end{eqnarray}
\end{subequations}
and 
\bseq
\label{eq:redef-tensor}
\begin{eqnarray}
&&
h^\mu 
\to  h^\mu + (\epsilon+p)\delta u^\mu + {\mathcal O} (\partial^2)
,
\\
&&
f^\mu \to 
f^\mu +  (\epsilon+ p) \delta b^\mu  + {\mathcal O} (\partial^2)
,
\\
&&
\tau^{\mu\nu} = 0 + {\mathcal O} (\partial^2)
.
\end{eqnarray}
\eseq
The redefinition of the flow vector do not affect on the scalar thermodynamic quantities at the first order in gradient. 
We will neglect the second-order terms which are beyond the current working accuracy.

By using the transforms of the scalar quantities (\ref{eq:redef-scalar}), 
one can eliminate the off-equilibrium quantities $ \delta \epsilon $,  $ \delta p_\para $, or $\delta p_\perp $. 
It is customary to eliminate the off-equilibrium energy density by setting $ \delta T 
= - (\partial \epsilon/\partial T)^{-1}\delta\epsilon $ 
so that the equilibrium energy density $ \epsilon $ is maintained. 
Then, the pressure remains to have the off-equilibrium contributions, corresponding to the bulk viscosities. 
This is a general structure: If there are $ n $ conserved quantities, such as energy density and an electric charge ($ n=2 $), 
one can maintain their equilibrium quantities by the redefinitions of the $ n $ thermodynamic conjugate parameters, 
temperature and a chemical potential for the electric charge, leaving the off-equilibrium components of the pressure.

Similarly, one can eliminate a dissipative current with a suitable choice of the first-order term 
arising from the redefinition of the flow velocity in Eq.~(\ref{eq:redef-tensor}). 
One traditional choice is the Landau-Lifshitz frame \cite{landau1959course:fluid, Minami:2012hs} 
with $  \delta u^\mu = -h^\mu /(\epsilon+p) $. 
After setting the ambiguities as explained above, 
we reach the first-order corrections to the energy-momentum tensor~(\ref{eq:T1_mn}) in the Landau-Lifshitz frame. 


\subsection{Viscous coefficients and inequalities} 

\label{sec:MHD-viscosities}

The viscous tensor in a magnetic field has been investigated in classic works~\cite{hooyman1954coefficients, Landau10:kinetics}, 
and recently in Refs.~\cite{Huang:2011dc, Critelli:2014kra, Hernandez:2017mch, Grozdanov:2016tdf} for relativistic systems. 
Those authors obtained the same number of the independent viscous coefficients, 
so that their viscous tensors are expected to be equivalent. 
This point was elaborated especially in Ref.~\cite{Hernandez:2017mch}. 
Here, we shall provide complementary notes on comparison among the viscous tensors in the literature, 
and clarify discrepancies in the inequalities which should be satisfied by the viscous coefficients.

The construction of the independent tensor structures are elaborated 
by Huang, Sedrakian, and Rischke (HSR) in Sec.~2.2 of Ref.~\cite{Huang:2011dc}, 
providing useful building blocks and a relativistic extension 
of the nonrelativistic theory in a classic textbook~\cite{Landau10:kinetics}. 
Therefore, the result by HSR serves as a good starting point for comparisons to other works. 
Since the technical machinery has been already detailed there, 
we shall start with the result given in Eq.~(39) of Ref.~\cite{Huang:2011dc}: 
\begin{eqnarray}
T_\one^{\mu\nu}&=& 
-3\zeta_\parallel^{\rm HSR} b^\mu b^\nu \theta_\para 
+ \frac{3}{2}\zeta_\perp^{\rm HSR} \varXi^{\mu\nu} \theta_\perp
+2\eta_0 ^{\rm HSR}  \left(w^{\mu\nu}-\frac{1}{3}\Delta^{\mu\nu}\theta\right) 
- \eta_1^{\rm HSR}  \left( b^\mu b^\nu + \frac{1}{2}\varXi^{\mu\nu} \right)
\left( \theta_\para - \frac{1}{2}\theta_\perp\right) 
\nn\\
&&
+2\Bigl\{\, 
-\eta_2^{\rm HSR}   b^\alpha b^{(\mu} \varXi^{\nu) \beta}
-\eta_3^{\rm HSR}  \varXi^{\alpha (\mu} \bar b^{\nu)\beta} 
+\eta_4^{\rm HSR}  b^{\alpha}b^{(\mu} \bar b^{\nu)\beta} 
\, \Bigr\} w_{\alpha\beta}
\label{eq:T1-HSR}
\, .
\end{eqnarray}
Here, we used our notations introduced in Sec.~\ref{sec:MHD}. 
Especially, note that $ \theta = \theta_\perp + \theta_\para $. 
Nevertheless, the above expression and definitions of the viscous coefficients 
are an exact duplicate 
of the result by HSR up to simple arrangements. 

The tensors in the first two terms in Eq.~(\ref{eq:T1-HSR}) have nonvanishing traces, 
while the others are traceless. 
Therefore, the two $\zeta^{\rm HSR} $'s and the five $\eta^{\rm HSR} $'s 
are called the bulk and shear viscosities in HSR~\cite{Huang:2011dc}, respectively (see also Ref.~\cite{Landau10:kinetics}). 
The fourth term with $ \eta_1^{\rm HSR}  $ is proportional to the expansion/compression rates $ \theta_{\para, \perp} $. 
However, this term is a traceless part of the viscous tensor, 
so that the fluid volume does not change as the expansion in the parallel/perpendicular direction 
to the magnetic field is compensated by the compression in the perpendicular/parallel direction. 
In this sense, this term was not regarded as a bulk viscosity in HSR~\cite{Huang:2011dc}.\footnote{
The authors thank X.-G. Huang for discussions about these points.} 
We will see that this term generates the cross viscosity $ \zeta_\times $ discussed in Sec.~\ref{sec:MHD}. 
In terms of the symmetries, one may add a term proportional to 
$\bar b^{\mu\alpha} \bar b^{\nu\beta} w_{\alpha\beta}$, 
which is, however, not independent of the other tensors involved in Eq.~(\ref{eq:T1-HSR}). 
The simple reason is that this tensor contains two antisymmetric tensors 
which can be identically decomposed into products of the metric tensors. 
Therefore, one should not include this tensor as an independent term. 

Now, by using the identity (\ref{eq:w_mn}), one can immediately arrange Eq.~(\ref{eq:T1-HSR}) in the following form 
\begin{eqnarray}
T_\one^{\mu\nu}&=& 
(  \zeta_\para \theta_\para - \zeta_\times \theta_\perp ) (- b^\mu b^\nu)
+ (  \zeta_\times \theta_\para  - \zeta_\perp \theta_\perp ) \varXi^{\mu \nu}
\label{eq:Tmunu-dis2}
\\
&&
+  \Bigl\{\, 
- \eta_\para b^\alpha b^{(\mu} \varXi^{\nu) \beta}
+ \frac{\eta_\perp}{2}  ( \varXi^{ \alpha (\mu} \varXi^{\nu) \beta} - \varXi^{\mu\nu}\varXi^{\alpha\beta} )
+ \xi_\Hall^\para b^{\alpha}b^{(\mu} \bar b^{\nu)\beta} 
+ \xi_\Hall^\perp \varXi^{\alpha (\mu} \bar b^{\nu)\beta} 
\, \Bigr\} w_{\alpha\beta}
\nn
\, .
\end{eqnarray}
Referring to our conventions in Eqs.~(\ref{eq:T1_mn}) and (\ref{eq:T1-viscosities}), 
we have identified the correspondences between the viscous coefficients introduced 
in Sec.~\ref{sec:MHD} and in HSR~\cite{Huang:2011dc} as 
\begin{eqnarray}
&&
\zeta_\para = 3\zeta_\para^{\rm HSR}  + \left( \frac{4}{3} \eta_0^{\rm HSR}  + \eta_1^{\rm HSR}   \right)
, \quad
\zeta_\perp = \frac{3}{2} \zeta_\perp ^{\rm HSR} + \frac{1}{4} \left( \frac{4}{3} \eta_0 ^{\rm HSR} + \eta_1 ^{\rm HSR}  \right)
, \quad 
\zeta_\times = - \frac{1}{2} \left( \frac{4}{3} \eta_0 ^{\rm HSR} + \eta_1 ^{\rm HSR}  \right)
, 
\nn\\
&&
\eta_\perp = 2 \eta_0 ^{\rm HSR} 
, \quad
\eta_\para = 2(\eta_0^{\rm HSR}  + \eta_2^{\rm HSR} )
, \quad
\xi_\Hall^\para= 2\eta_4^{\rm HSR} 
, \quad
\xi_\Hall^\perp = -2\eta_3^{\rm HSR} 
\label{eq:correspondences}
.
\end{eqnarray}
It is now clear that the cross viscosity $ \zeta_\times  $ was generated 
from the traceless terms proportional to $ \eta_0 $ and $ \eta_1  $ in Eq.~(\ref{eq:T1-HSR}). 
%
%
%
%
The correspondences between the viscous coefficients in HSR~\cite{Huang:2011dc} and 
Hernandez and Kovtun (HK)~\cite{Hernandez:2017mch} are available in Eq.~(B.1) of HK~\cite{Hernandez:2017mch}. 
Now, putting Eq.~(\ref{eq:correspondences}) together, the list of correspondences is expanded as 
\begin{eqnarray}
&&
\eta_\perp^{\rm HK} = \eta_0^{\rm HSR} = \frac{1}{2} \eta_\perp
 \, , 
\nn \\
&&
\tilde \eta_\perp^{\rm HK} = - 2 \eta_3^{\rm HSR} = \xi_\Hall^\perp
 \, , 
 \nn \\
 &&
 \eta_\para^{\rm HK} =  \eta_0^{\rm HSR} +  \eta_2^{\rm HSR} = \frac{1}{2} \eta_\para
 \, , 
\nn \\
&&
\tilde \eta_\para^{\rm HK} = - \eta_4^{\rm HSR} = - \frac{1}{2}\xi_\Hall^\para
 \, , 
\nn \\
 &&
 \eta_1^{\rm HK} = - \frac{1}{2} \eta_0 ^{\rm HSR} - \frac{3}{8} \eta_1^{\rm HSR}  - \frac{3}{4} \zeta_\perp ^{\rm HSR} 
 = - \frac{1}{2} ( \zeta_\perp -  \zeta_\times)
\, , 
\\
 &&
\eta_2^{\rm HK} = \frac{3}{2} \eta_0 ^{\rm HSR} + \frac{9}{8} \eta_1^{\rm HSR}  
+ \frac{3}{4} \zeta_\perp ^{\rm HSR} + \frac{3}{2} \zeta_\para ^{\rm HSR} 
= \frac{1}{2} ( \zeta_\para - 2  \zeta_\times +\zeta_\perp )
 \, , 
 \nn \\
 &&
\zeta_1^{\rm HK} = \zeta_\perp ^{\rm HSR} = \frac{1}{3}(2 \zeta_\perp +  \zeta_\times)
 \, , 
\nn \\
&&
\zeta_2^{\rm HK} = \zeta_\para ^{\rm HSR} -\zeta_\perp ^{\rm HSR} 
= \frac{1}{3} ( \zeta_\para  +  \ \zeta_\times - 2  \zeta_\perp)
\nn
\, .
\end{eqnarray}
The superscripts ``HK'' and ``HSR'' place remarks on the viscous coefficients 
introduced in HK~\cite{Hernandez:2017mch} and HSR~\cite{Huang:2011dc}, respectively, 
while those without superscripts are introduced in this paper. 
We note that the viscous coefficients in the current paper agree with those 
in Eqs.~(3.9), (3.10), (3.13), and (3.14) of Grozdanov, Hofman, and Iqbal (GHI)~\cite{Grozdanov:2016tdf} 
up to conventions in the signs and factors. 
Further correspondences to the conventions in Ref~\cite{Critelli:2014kra} are discussed in 
the appendix of HK~\cite{Hernandez:2017mch}. 
Compared with Eq.~(A5) in Ref.~\cite{Li:2017tgi}, one also finds correspondences to 
the viscosities, which are relevant for the hydrodynamic modes in an external nondynamical magnetic field, 
\begin{eqnarray}
\eta^{\rm LY} = \eta_0 ^{\rm HSR}  + \eta_2^{\rm HSR} = \eta_\para
, \quad
\zeta^{\rm LY} = 3 \zeta_\para ^{\rm HSR} + \frac{4}{3} \eta_0 ^{\rm HSR} + \eta_1 ^{\rm HSR} = \zeta_\para
, \quad
\zeta^{\prime \, {\rm LY} }= - \frac{2}{3} \eta_0 ^{\rm HSR} - \frac{1}{2} \eta_1 ^{\rm HSR} = \zeta_\times
,
\end{eqnarray}
where the superscripts LY denote the viscosities introduced in Ref.~\cite{Li:2017tgi}. 
It is now clear that all of them quantify the responses to the flow perturbations along the external magnetic field.


With the above identifications,
the involved inequalities in Eq.~(B.2) of HK~\cite{Hernandez:2017mch} can be rewritten in simple forms 
\begin{eqnarray}
\label{eq:inequalities-HK}
\eta_\perp \geq 0
\, , \ \ \ 
\eta_\para \geq 0
\, , \ \ \ 
\zeta_\perp \geq 0
\, , \ \ \ 
\zeta_\para + \zeta_\perp + 2 \zeta_\times \geq 0
\, , \ \ \ 
\zeta_\para \zeta_\perp - \zeta_\times^2 \geq 0
\, .
\end{eqnarray}
As mentioned below Eq.~(\ref{eq:ineqalities}), 
one may remove one inequality $  \zeta_\para \geq0$ from the list in Eq.~(\ref{eq:ineqalities}), 
or putting it differently, one can prove that $  \zeta_\para \geq0$ 
when the two inequalities $ \zeta_\perp \geq 0$ and $\zeta_\para \zeta_\perp - \zeta_\times^2 \geq 0 $ are satisfied. 
Therefore, an essential difference from Eq.~(\ref{eq:ineqalities}) 
is only the existence of the second-last inequality, $ \zeta_\para + \zeta_\perp + 2 \zeta_\times \geq 0 $. 
This inequality can be, however, deduced from the others in Eq.~(\ref{eq:inequalities-HK}), as follows. 
We can immediately show that 
$ ( \zeta_\para +  \zeta_\perp)^2 - (2  \zeta_\times)^2 
=  (  \zeta_\para - \zeta_\perp)^2 
+ 4( \zeta_\para   \zeta_\perp -  \zeta_\times^2) \geq 0 $. 
Since we have $ \zeta_{\para,\perp} \geq0 $, 
we find that $   \zeta_\para  + \zeta_\perp + 2  \zeta_\times \geq 0$, regardless of the sign of $   \zeta_\times$. 
This proof suggests that the fourth inequality in Eq.~(\ref{eq:inequalities-HK}), 
and thus in Eq.~(B.2) of HK~\cite{Hernandez:2017mch}, is redundant. 
The minimal list of the inequalities is given by 
\begin{eqnarray}
\label{eq:inequalities-HK-final}
\eta_\perp \geq 0
\, , \ \ \ 
\eta_\para \geq 0
\, , \ \ \ 
\zeta_\perp \geq 0
\, , \ \ \ 
\zeta_\para \zeta_\perp - \zeta_\times^2 \geq 0
\, .
\end{eqnarray}
The inequality $ \zeta_\perp \geq0 $ can be alternatively replaced by the other one, $ \zeta_\para \geq0  $.

}

\bibliographystyle{apsrev4-1_r}
\bibliography{bib_review}

\end{document}